\documentclass[11pt,a4paper,twoside,openright]{scrreprt}
\pdfoutput=1

\usepackage{CLICdp}

\usepackage{CLICdp_definitions}
\newcommand{\tabt}[1]{\multicolumn{1}{c}{#1}}

\usepackage{csquotes}

\usepackage{tikz}
\tikzset{every picture/.style={/utils/exec={\sffamily}}}
\usepackage{pgfplots}
\pgfplotsset{compat=1.12}
\usepackage[colorinlistoftodos,prependcaption,textsize=tiny]{todonotes}
\usepackage{tikz}
\usetikzlibrary{shapes,arrows,positioning}
\usepackage[european]{circuitikz}
\usepackage{isotope}
\usepackage[eulergreek]{sansmath}

\usepackage{threeparttable}
\usepackage{standalone}
\usepackage{gincltex}

\usepackage{amssymb}

\title{Detector technologies for CLIC}

\clicdpdraft{2019}{002}  

\date{\today}

\collName{Contributors}
\addauthor{
A.C.~Abusleme~Hoffman\\
\emph{Pontificia Universidad Cat\'olica de Chile, Santiago, Chile}
}

\addauthor{
G. Par\`es\\
\emph{CEA-Leti, Grenoble, France}
}

\addauthor{
T.~Fritzsch,
M.~Rothermund\\
\emph{Fraunhofer-Institut f\"ur Zuverl\"assigkeit und Mikrointegration IZM, Berlin, Germany}
}

\addauthor{
H.~Jansen,
K.~Kr\"uger,
F.~Sefkow,
A.~Velyka\\
\emph{DESY, Hamburg, Germany}
}

\addauthor{
J.~Schwandt\\
\emph{University of Hamburg, Hamburg, Germany}
}

\addauthor{
I.~Peri\'c\\
\emph{Karlsruhe Institute of Technology (KIT), Karlsruhe, Germany}
}

\addauthor{
L.~Emberger,
C.~Graf,
A.~Macchiolo$^{1}$,
F.~Simon,
M.~Szalay,
N.~van~der~Kolk$^{2}$\\
\emph{Max-Planck-Institut f\"ur Physik, Munich, Germany}
% ...
}

\addauthor{
G.~Casse,
J.~Vossebeld\\
\emph{University of Liverpool, Liverpool, United Kingdom}
}

\addauthor{
T.~Coates,
F.~Salvatore\\
\emph{University of Sussex, Brighton, United Kingdom}
}

\addauthor{
J.~Repond,
L.~Xia\\
\emph{Argonne National Laboratory, Argonne, USA}
}

\addauthor{
C.~Kenney,
A.~Tomada\\
\emph{SLAC National Accelerator Laboratory, Menlo Park, USA}
}

\addauthor{
\begin{flushleft}This work was carried out in the framework of the CLICdp Collaboration.
\end{flushleft}
}

\addauthor{
\begin{flushleft}
{$^{1}$}Now at University of Zurich, Zurich, Switzerland\\
{$^{2}$}Now at Nikhef/Utrecht University, Amsterdam/Utrecht, The Netherlands\\
% ...
\end{flushleft}
}

\abstract{The Compact Linear Collider (CLIC) is a high-energy high-luminosity linear electron--positron collider under development.
It is foreseen to be built and operated in three stages, at centre-of-mass energies of \SI{380}{\GeV}, \SI{1.5}{\TeV} and \SI{3}{\TeV}, respectively.
It offers a rich physics program including direct searches as well as the probing of new physics through a broad set of precision measurements of Standard Model processes, particularly in the Higgs-boson and top-quark sectors.
The precision required for such measurements and the specific conditions imposed by the beam dimensions and time structure put strict requirements on the detector design and technology. This includes low-mass vertexing and tracking systems with small cells, highly granular imaging calorimeters, as well as a precise hit-time resolution and power-pulsed operation for all subsystems. A conceptual design for the CLIC detector system was published in 2012. Since then, ambitious R\&D programmes for silicon vertex and tracking detectors, as well as for calorimeters have been pursued within the CLICdp, CALICE and FCAL collaborations, addressing the challenging detector requirements with innovative technologies. 
 This report introduces the experimental environment and detector requirements at CLIC and reviews the current status and future plans for detector technology R\&D.}

\graphicspath{
  {./logos/}
  {./sections/IntroductionConclusion/}
  {./sections/VertexTracking/}
  {./sections/Calorimetry/}
  {./sections/FCAL/}
  {./sections/Electronics/}
  {./sections/Appendix/}
}

\DeclareSIUnit\year{yr}
\DeclareSIUnit\neq{\text{\ensuremath{n_{\textup{eq}}}}}

\newcommand{\apsq}{Allpix\textsuperscript{2}\xspace}

\begin{document}

\pagenumbering{roman}
\begin{titlepage}
\noindent
\setlength{\unitlength}{1mm}
\begin{center}
\begin{tikzpicture}[overlay,font=\rmfamily]
\node[anchor=west] at (5,0) {CERN-2019-001};
\node at (0,-4) {\includegraphics[width=15cm]{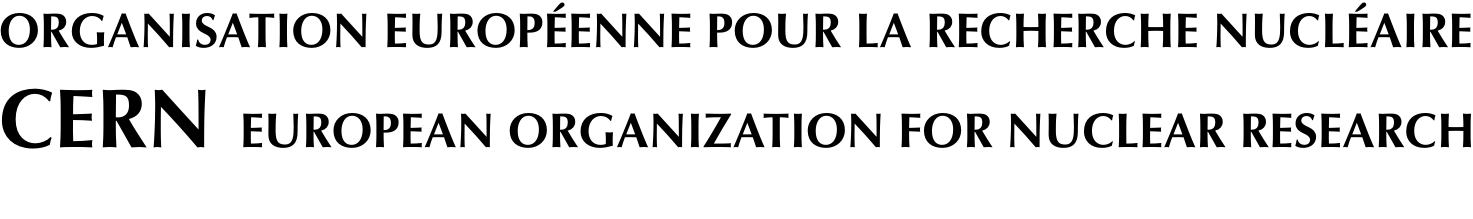}};
\node at (0,-11) {\includegraphics[width=11.25cm]{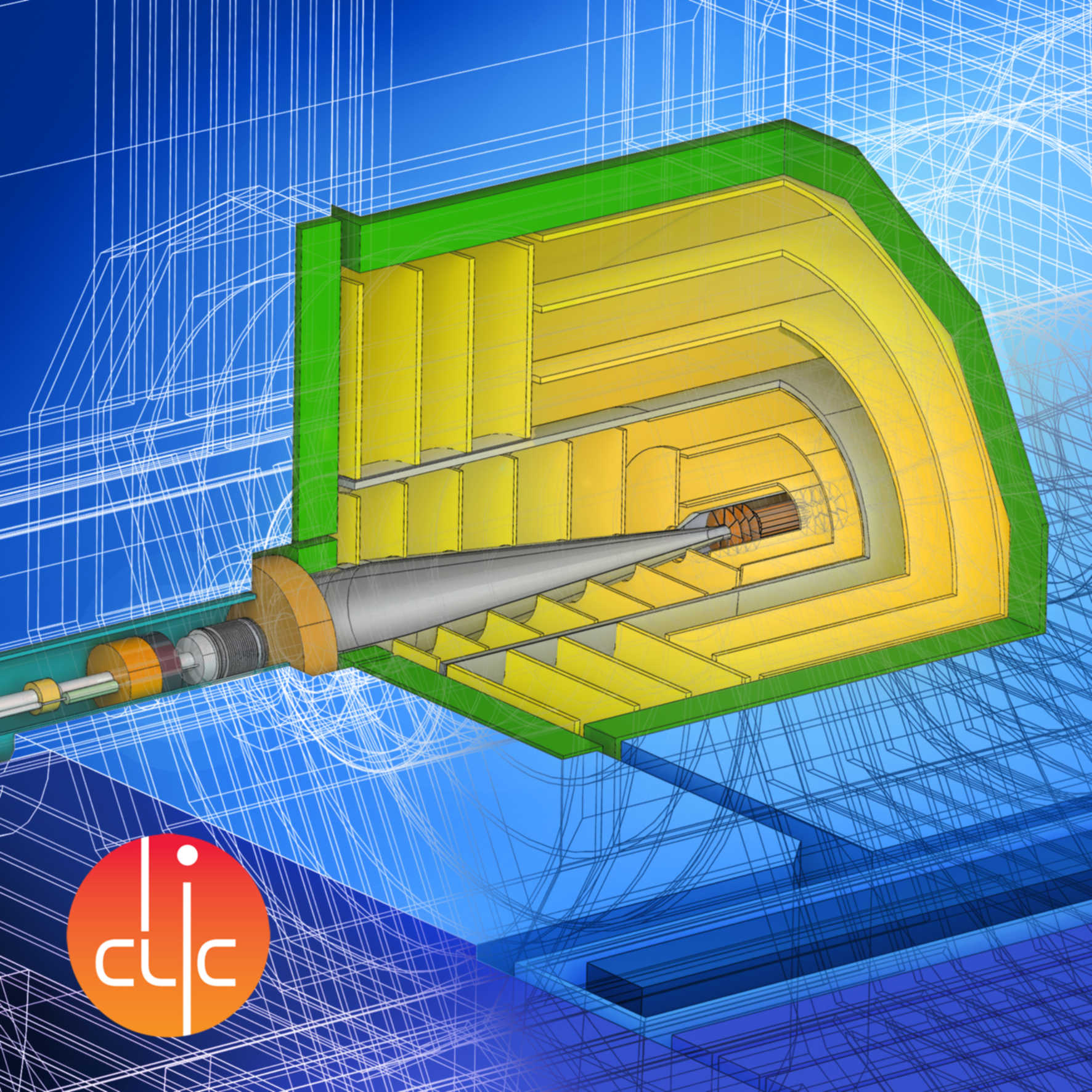}};
\node at (0,-19) {\huge\bfseries {\scshape Detector Technologies for CLIC}};
\node at (0,-23) {GENEVA};
\node at (0,-23.5) {2019};
\end{tikzpicture}
\end{center}
\newpage

\thispagestyle{empty}
\mbox{}
\vfill

\begin{flushleft}
CERN Yellow Reports: Monographs\\
Published by CERN, CH-1211 Geneva 23, Switzerland\\[3mm]

\begin{tabular}{@{}l@{~}l}
ISBN & 978--92--9083--535--6 (Print)\\
ISBN & 978--92--9083--536--3 (Online)\\
ISSN & 2519--8068 (Print)\\
ISSN & 2519--8076 (Online)\\
DOI  & \texttt{\href{http://dx.doi.org/10.23731/CYRM-2019-001}{http://dx.doi.org/10.23731/CYRM-2019-001}}
\end{tabular}\\[3mm]
Accepted for publication by the CERN Report Editorial Board (CREB) on 2 May 2019\\
Available online at \url{http://publishing.cern.ch/} and \url{http://cds.cern.ch/}\\[3mm]

Copyright \copyright{} CERN, 2019\\[1mm]
\raisebox{-1mm}{\includegraphics[height=12pt]{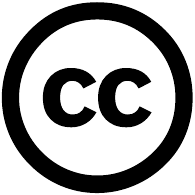}}
 Creative Commons Attribution 4.0\\[1mm]
Knowledge transfer is an integral part of CERN's mission.\\[1mm]
CERN publishes this volume Open Access under the Creative Commons Attribution 4.0 license\\
(\url{http://creativecommons.org/licenses/by/4.0/}) in order to permit its wide dissemination and use.\\
The submission of a contribution to a CERN Yellow Report series shall be deemed to constitute the contributor's agreement to this copyright and license statement. Contributors are requested to obtain any clearances that may be necessary for this purpose.\\[5mm]

Image credit for all images in this volume, unless otherwise noted: CLICdp.\\[5mm]

This volume is indexed in: CERN Document Server (CDS), INSPIRE.\\[5mm]


This volume should be cited as:\\[1mm]
Detector Technologies for CLIC, edited by D.\ Dannheim, K.\ Kr\"{u}ger, A.\ Levy, A.\ N\"{u}rnberg, E.\ Sicking, CERN--2019--001 (CERN, Geneva, 2019),  \texttt{\href{http://dx.doi.org/10.23731/CYRM-2019-001}{http://dx.doi.org/10.23731/CYRM-2019-001}} \\[3mm]

\end{flushleft}

\cleardoublepage

\thispagestyle{empty}
\vspace*{3cm}
\begin{center}
  \large{\bfseries\sffamily Abstract}
\end{center}
\begin{quotation}
\noindent\MyAbstract
\end{quotation}
\vfill

\begin{center}
{\large {\bfseries\sffamily Corresponding editors}} 
\vspace*{0.25cm}

{
Dominik Dannheim (CERN), 
Katja Kr\"{u}ger (DESY Hamburg),
Aharon Levy (Tel Aviv University), 
Andreas N\"{u}rnberg (KIT), 
Eva Sicking (CERN)
}
\end{center}
\end{titlepage}

\clearpage
\textcolor{white}{ }
\thispagestyle{empty}
\newpage

 \begin{center}
   \Large\bfseries\sffamily\MyCollName
 \end{center}

{
\setcounter{footnote}{0}\def\@currentlabel{}
\begingroup\def\thefootnote{\arabic{footnote}}
\def\@makefnmark{\hbox{$^{\@thefnmark)}$}}
\large
{
\MyAuthors
\par}
\endgroup

\vspace{-5mm}
\normalsize

\tableofcontents

\cleardoublepage

\pagenumbering{arabic}


\chapter{Introduction}\label{chap:intro}
The Compact Linear Collider (CLIC) is a high-luminosity high-energy electron--positron collider proposed for the post High-Luminosity Large Hadron Collider (HL-LHC) phase. CLIC is foreseen to be built and operated in three centre-of-mass energy stages from 380~GeV up to 3~TeV~\cite{CLIC-summary-report-2018}. The physics programme of CLIC includes precision measurements of the properties of the Higgs boson and the top quark, as well as searches for physics beyond the Standard Model (BSM).

Initial studies, described in the CLIC Detector and Physics Conceptual Design Report (CDR) in 2012, demonstrated the feasibility of performing precision measurements at CLIC~\cite{cdrvol2}. They were based on the two detector concepts \clicild and \clicsid, derived from previous studies for the International Linear Collider~\cite{ildloi:2009,sidloi:2009}.
This report on Detector Technologies for CLIC summarises the studies since the publication of the CDR. A single detector concept, CLICdet, has been developed~\cite{AlipourTehrani:2254048,CLICdet-performance} and a broad technology R\&D programme is being pursued.
In view of the time scales involved and the limited resources, the developments focus on areas where the CLIC requirements are the most challenging: the silicon vertex and tracking detectors and the high-granularity calorimeter systems.

The detector design and technology choices are driven by the physics programme and match the expected experimental conditions at CLIC, as briefly described in \cref{chap:overview-backgrounds}. The detector concepts and hardware R\&D focus on the most challenging 3~TeV case. Only small modifications for the inner detector region are anticipated for the initial stage at 380 GeV. 

The studies for the vertex and tracking detectors, presented in \cref{chap:vtxtrk}, are carried out by the CLIC detector and physics collaboration (CLICdp)~\cite{CLICdp-collaboration} and are closely linked to other detector development projects. Both hybrid and monolithic pixel detector concepts are under consideration and various technology demonstrators have been built and characterised in view of the CLIC requirements. Detailed simulations of the detector performance are validated with existing prototypes and used to optimise the design of future sensors and readout chips. Detector integration aspects are addressed through the conceptual design and prototyping of low-mass support structures, air-flow cooling systems, power-delivery and power-pulsing concepts, as well as detector assembly scenarios.

The R\&D for the main electromagnetic and hadronic calorimeters, presented in \cref{chap:calorimeters}, is performed within the CALICE collaboration, developing calorimeters for high-energy \epem experiments~\cite{Adloff:2012dla}. Highly granular sampling calorimeter prototypes with silicon- and scintillator-based active layers have been built and their performance assessed in beam tests. Recent studies aim at demonstrating the scalability of the proposed technological solutions in view of mass production, detector assembly and integration.
CLIC-specific aspects related to the performance at high centre-of-mass energies and to the required time resolution for the suppression of beam-induced background particles have been addressed in dedicated studies.

  Radiation-hard and compact sandwich calorimeters for the luminosity measurement and the tagging of forward-going electrons and photons are developed within the FCAL collaboration~\cite{FCAL_website}, as described in \cref{chap:fcal}. Detectors and readout systems of increasing size and sophistication have been developed and tested in recent years, demonstrating the feasibility of forward calorimetry at CLIC.

 The detector occupancies and data volumes resulting from the desired high granularities and expected experimental conditions are presented in \cref{chap:electronics-daq}. These estimates form crucial input for future data-acquisition and data-storage concepts.
\cref{chap:conclusions} concludes this report, giving a summary of the results obtained and an outlook on future developments.


\chapter{CLIC accelerator and detector overview}\label{chap:overview-backgrounds}
This chapter gives an overview of the CLIC accelerator and detector and of the expected experimental conditions at CLIC.
The most important parameters of the CLIC accelerator, which affect the detector design and technology choices, are presented in \cref{sec:clic-machine-parameters}. The physics requirements on the detector performance are listed in \cref{sec:det-requirements}.
\cref{sec:clic-detector-concept} briefly introduces the overall concept of the CLIC detector. The vertex- and tracking detectors and calorimeters will then be described in more detail in the subsequent chapters.
The expected rates and energy depositions from beam-induced background particles in the different detector regions are summarised in \cref{sec:beam_induced_backgrounds}, followed by a discussion of the resulting radiation exposure in~\cref{sec:radiation}.

\section{CLIC accelerator}\label{sec:clic-machine-parameters}
The CLIC accelerator is foreseen to be built and operated in three centre-of-mass energy stages from \SI{380}{\giga\electronvolt} up to \SI{3}{\tera\electronvolt}, targeting different aspects of the CLIC physics programme~\cite{CLIC-summary-report-2018}. Following a preparation and construction phase for the \SI{380}{\giga\electronvolt} collider of approximately 15~years and upgrade periods of 2~years between subsequent stages, each stage would operate for 7--8~years, including a luminosity ramp-up phase of 2--3~years.

\cref{scd:clic_layout} shows a schematic layout of the CLIC accelerator complex for the 380~GeV stage. CLIC uses a novel two-beam acceleration scheme, in which a drive beam of rather low energy but high current is decelerated, and its energy is transferred to the low-current main beam, which is accelerated with gradients of up to \SI{100}{\mega\volt\per\meter}. The two-beam acceleration technique thus removes the need for RF power sources along the main linac and allows for reaching centre-of-mass energies of up to \SI{3}{\tera\electronvolt} with an overall length of the accelerator complex of approximately \SI{50}{\km}.

\begin{figure}[ht!]
\includegraphics[width=\textwidth]{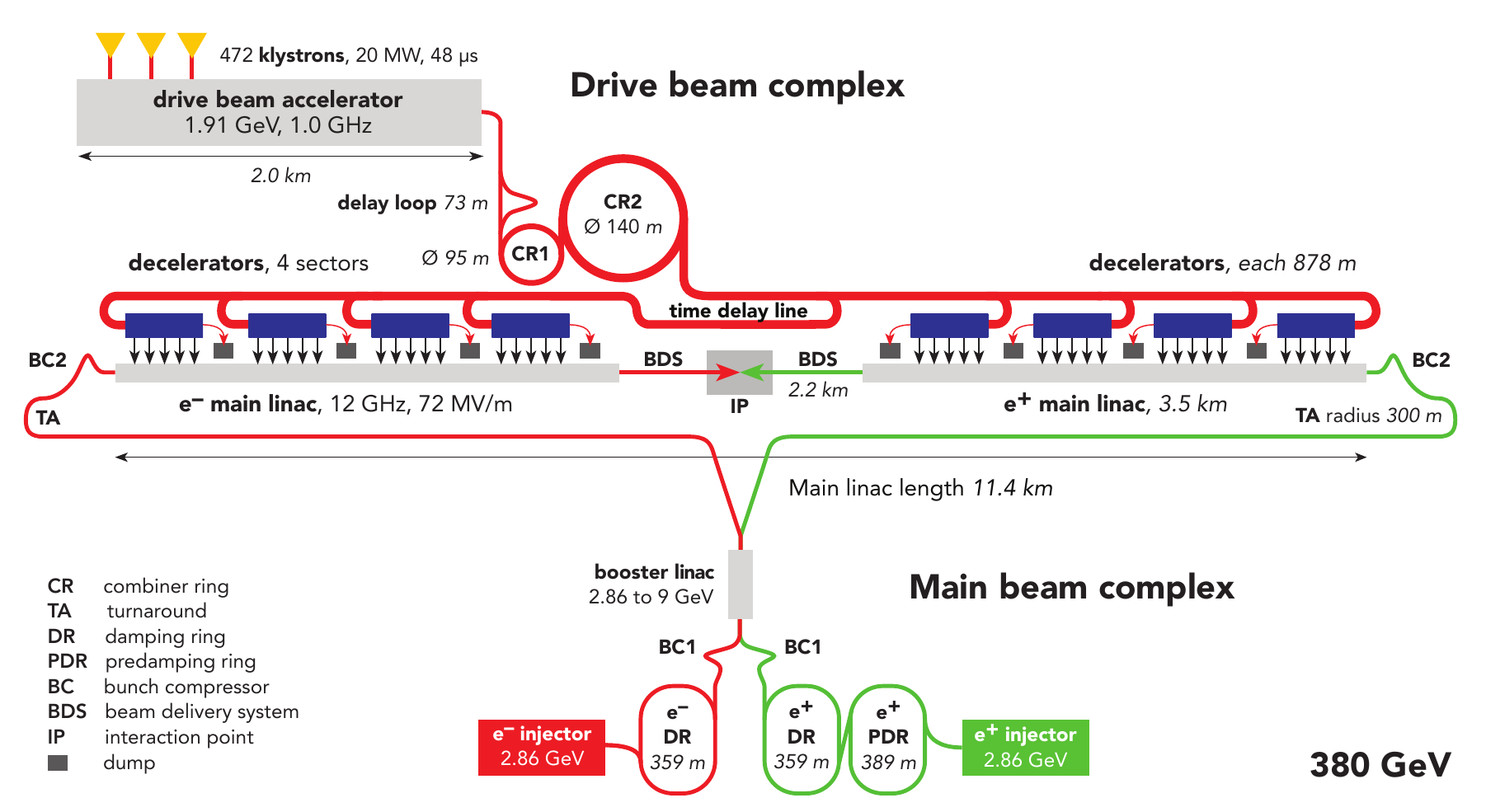}
\caption{Schematic layout of the CLIC accelerator complex for the 380~GeV stage~\cite{CLIC-summary-report-2018}.}
\label{scd:clic_layout}
\end{figure}

The main accelerator parameters constraining the design of the detector system are summarised in \cref{tab:CLIC-parameters} for the three centre-of-mass energy stages.
In order to reach its design luminosity of $1.5-6\times 10^{34}$cm$^{-2}$s$^{-1}$, CLIC will operate with very small bunch sizes (less than  \SI{150}{\nano\meter} in $x$ and \SI{3}{\nano\meter} in $y$ and less than \SI{100}{\micron} along the beam). Collisions occur at a beam crossing angle of $16.5-20~\si{\mrad}$ in bunch crossings (BX) every 0.5 ns for a train duration of 156--176~ns. The train repetition rate is 50 Hz, resulting in a very low duty cycle of less than \SI{0.001}{\percent}.

\begin{table}[ht!]
\caption{Selected parameters of the CLIC accelerator for the different centre-of-mass energy stages~\cite{CLIC-summary-report-2018}.}
\label{tab:CLIC-parameters}
\centering
\begin{tabular}{l l l l l}
\toprule
Parameter                  &   Unit &    Stage 1 &   Stage 2 &   Stage 3 \\
\midrule
Centre-of-mass energy               &\si{\GeV}                                     & 380     & 1500          & 3000\\
Bunch repetition rate                &\si{\Hz}                                     & 50      & 50            & 50\\
Number of bunches per train         &                                              & 352     & 312           & 312\\
Bunch separation                    &\si{\ns}                                      & 0.5     & 0.5           & 0.5\\
\midrule
Accelerating gradient               &\si{\mega\volt/\meter}                        & 72      & 72/100        & 72/100\\
\midrule
Total luminosity                    &\SI{e34}{\per\centi\meter\squared\per\second} & 1.5     & 3.7           & 5.9 \\
Luminosity above \SI{99}{\percent} of $\sqrt{s}$ & \SI{e34}{\per\centi\meter\squared\per\second} & 0.9     & 1.4           & 2\\
Equivalent run time at full luminosity per year  & \SI{e7}{\second} & 1.2 &  1.2 &  1.2  \\
Total integrated design luminosity per year& \si{\per\fb}                                  & 180     & 444           & 708 \\
Years of operation (ramp-up + nominal conditions)                 &                                               & $3+5$   & $2+5$         & $2+6$ \\
\midrule
Main linac tunnel length            &\si{\km}                                      & 11.4    & 29.0          & 50.1\\
Crossing angle at interaction point (IP) &\si{\mrad}                                    & 16.5    & 20            & 20 \\
Number of particles per bunch       &\num{e9}                                      & 5.2     & 3.7           & 3.7\\
Bunch length  $\sigma_z$            &\si{\um}                                      & 70      & 44            & 44\\
IP beam size $\sigma_x/\sigma_y$    &\si{\nm}                                      & 149/2.9 & $\sim$ 60/1.5 & $\sim$ 40/1\\
\bottomrule
\end{tabular}
\end{table}

The short train duration and the fact that on average less than one interesting physics collision event is expected per train implies that trigger-less readout of all subdetectors, once per train, can be implemented. The power consumption of the detectors, and therefore the material required for cooling infrastructure, is reduced by switching off parts of the frontend electronics during the 20 ms gaps between trains ({\em power pulsing}). Hence lower material budgets can be reached for the vertex and tracking systems as well as a lower abundances of inactive light material in the calorimeter systems. Both are of importance for the physics performance of the CLIC detector.

\section{Detector requirements}\label{sec:det-requirements}
The demands for precision physics~\cite{CLIC-summary-report-2018} lead to challenging performance targets for the CLIC detector system:
\begin{itemize}
\item track-momentum resolution for high-momentum tracks of $\sigma_{\pT}/\pT^2 \leq \SI{2e-5}{\per\GeV}$ in the central detector;
\item impact-parameter resolution of $\sigma_{d_0}^2 = (\SI{5}{\um})^2 + (\SI{15}{\um\GeV})^2/(p^2\sin^{3}\theta)$, to allow accurate reconstruction and enable flavour tagging with clean $\PQb$-, $\PQc$-, and light-quark jet separation;
\item jet-energy resolution for light-quark jets of $\sigma_E/E \leq \SI{3.5}{\percent}$ for jet energies in the range \SI{100}{\GeV} to \SI{1}{\TeV} ($\leq \SI{5}{\percent}$ at \SI{50}{\GeV});
\item detector coverage for electrons and photons to very low polar angles (\SI{\sim10}{\mrad}) to assist with background rejection.
\end{itemize}
These physics performance goals have to be met in the challenging experimental conditions provided by the CLIC accelerator. The resulting subdetector requirements are discussed in the following chapters.

\section{CLIC detector concept}\label{sec:clic-detector-concept}
The CLIC detector concept \mbox{CLICdet}~\cite{AlipourTehrani:2254048} is optimised for precision measurements based on Particle Flow Analysis (PFA) reconstruction.
The PFA approach allows for the distinguishing of individual particles within jets and combines the information from the high-precision tracking system with the energy measurement in the calorimeter, resulting in an optimal jet-energy measurement~\cite{thomson:pandora,MARSHALL2013153}.
The inner region of \mbox{CLICdet}, surrounding the conical beam pipe, comprises an all-silicon vertex and tracking detector with a central barrel part and two endcap sections. The tracker is surrounded by highly-granular electromagnetic (silicon-tungsten ECAL) and hadronic (scintillator-steel HCAL) sampling calorimeters.
A superconducting solenoid surrounding the calorimeters provides a magnetic field of \SI{4}{\tesla}.
Beyond the solenoid, \mbox{CLICdet} contains an iron yoke interleaved with detectors for muon identification.
Forward calorimeters located close to the beam pipe, called LumiCal and BeamCal, provide luminosity measurements and forward electron-tagging.
 CLICdet was optimised for operation at $\roots=\SI{3}{\TeV}$. As background rates at $\roots=\SI{380}{\GeV}$ are lower, some modifications to the inner detector layers are anticipated for the first energy stage~\cite{cdrvol2}.
A quarter-view of the longitudinal cross section and a transverse cross section of \mbox{CLICdet} are shown in \cref{fig:clicdet}. 
More details on the layout of the tracking detectors and calorimeters are given in the following chapters.

\begin{figure}
\begin{subfigure}[T]{.49\linewidth}
    \includegraphics[width=\linewidth]{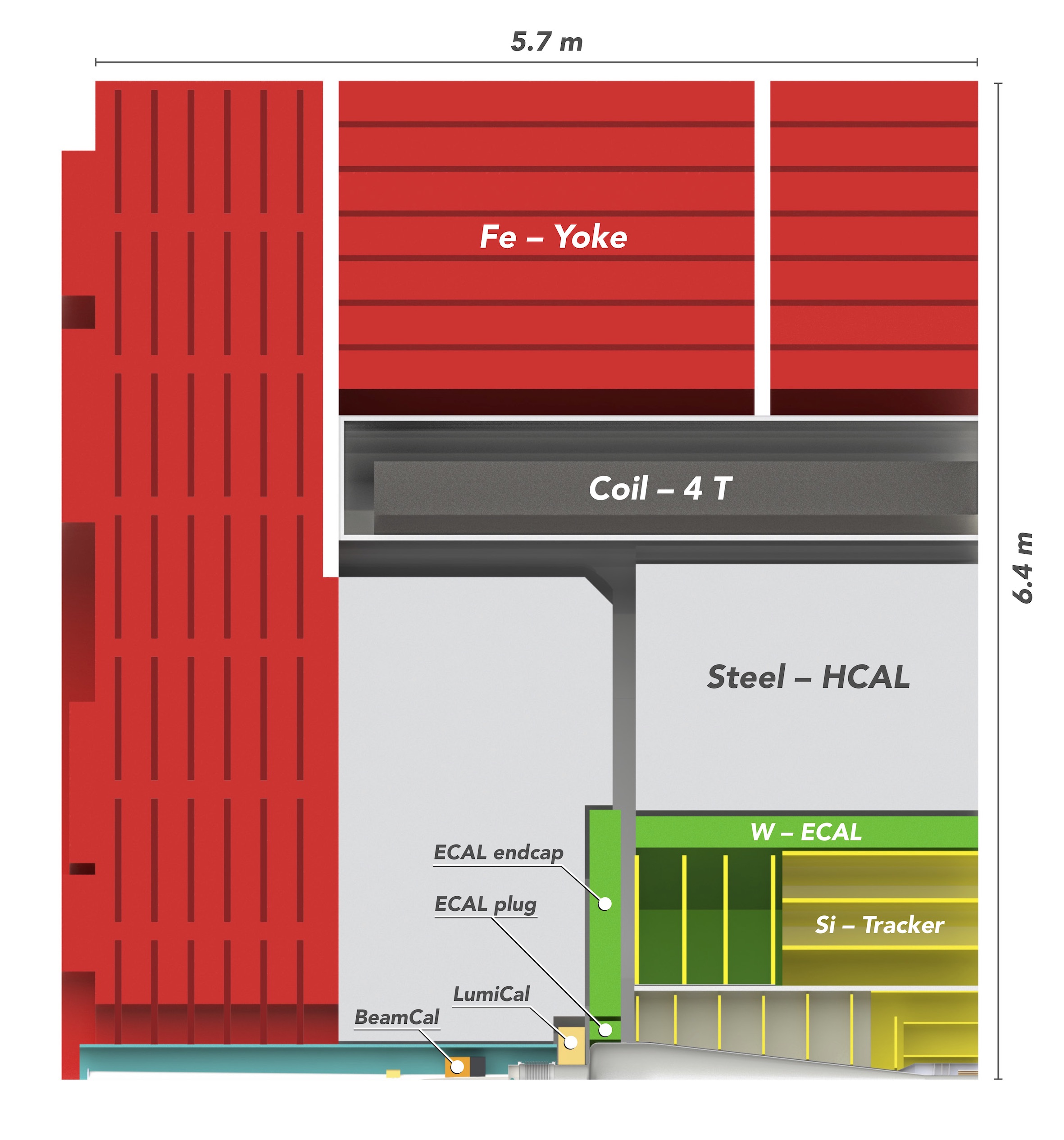}
    \caption{}\label{fig:clicdet_quarterview}
   \end{subfigure}
   \hfill
  \begin{subfigure}[T]{.49\linewidth}
  \includegraphics[width=\linewidth]{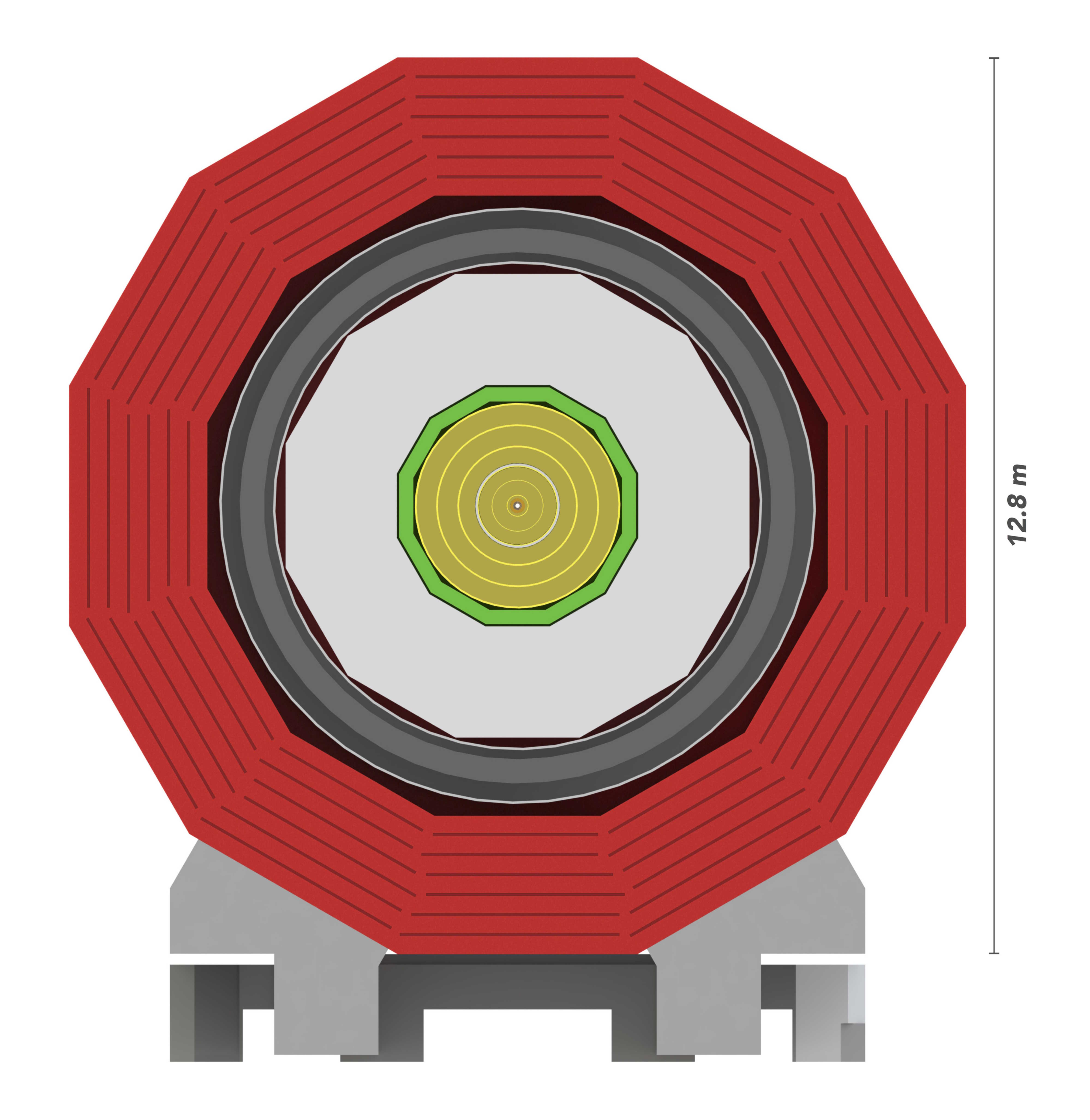}
  \caption{}\label{fig:clicdet_3dcorner}
\end{subfigure}
  \caption{\subref{fig:clicdet_quarterview} Longitudinal cross section of the top left quadrant and \subref{fig:clicdet_3dcorner} transverse (XY) cross section of CLICdet.}\label{fig:clicdet}
\end{figure}

\section{Beam-induced backgrounds}\label{sec:beam_induced_backgrounds}
The very small bunch sizes lead to strong electromagnetic radiation (Beamstrahlung) from the electron and positron bunches in the field of the opposite beam. The creation of the Beamstrahlung photons reduces the available centre-of-mass energy of the e$^+$e$^-$ collisions and interactions involving Beamstrahlung photons result in high rates of lepton pairs and hadrons being produced.
Pile-up rejection algorithms based on hit time stamping at the 1--10~ns level are therefore needed to separate physics from background events.

Most of the particles originating from Beamstrahlung events are produced at very low polar angles and are thus confined within the conical beam pipe by the axial magnetic field. The dominant backgrounds in the detectors are incoherently produced electron--positron pairs and \gghad events, shown in \cref{fig:bg-theta} for the 380~GeV and the 3~TeV stages. The rate of incoherent pairs reaching the central detector is approximately 5 times higher at 3~TeV, compared to the 380~GeV case. For particles from \gghad, the increase corresponds to a factor of almost 20~\cite{CLICdet-performance}. 

\begin{figure}[h]
\begin{center}
\begin{subfigure}{.49\textwidth}
  \centering
  \includegraphics[width=\textwidth]{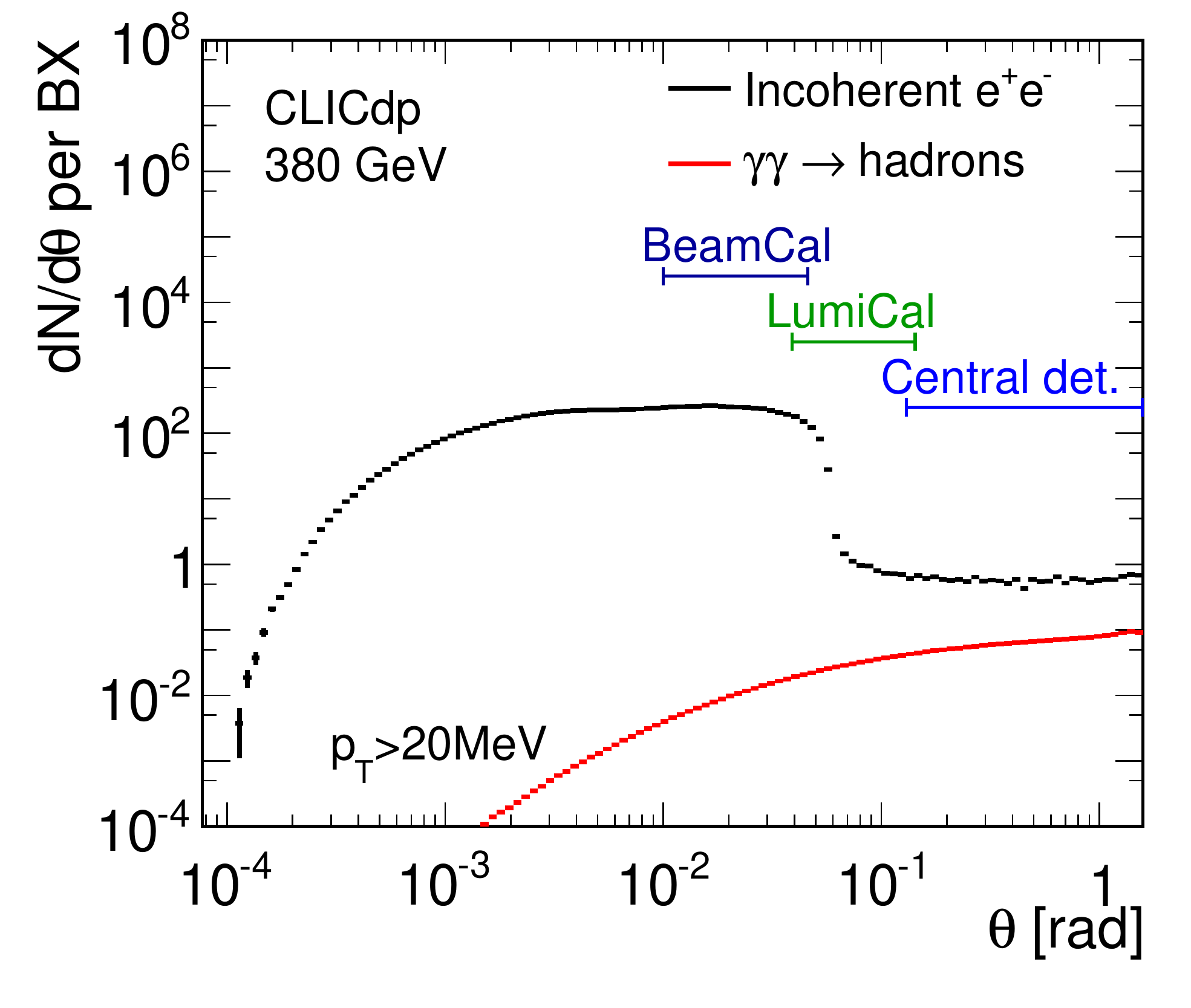}
  \caption{}
  \label{fig:bg-theta-380gev}
\end{subfigure}
\begin{subfigure}{.49\textwidth}
  \centering
  \includegraphics[width=\textwidth]{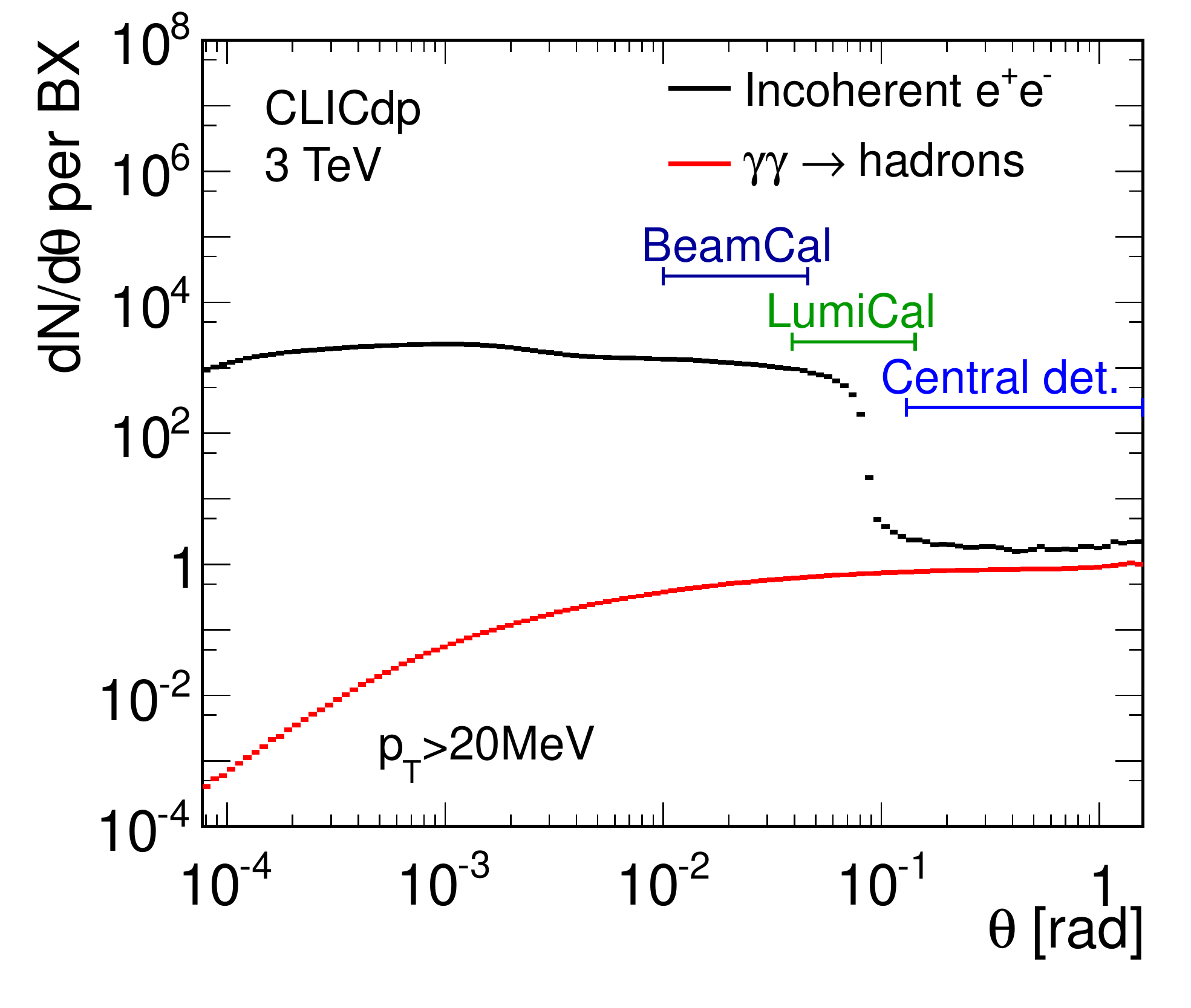}
  \caption{}
  \label{fig:bg-theta-3tev}
\end{subfigure}
 \caption{Polar angle distribution of beam-induced backgrounds for \subref{fig:bg-theta-380gev} 380~GeV and \subref{fig:bg-theta-3tev} 3~TeV centre-of-mass energy.
  Generated particle distributions for $p_{\text{T}}>20\,$MeV are shown, including a \SI{2}{\GeV} c.m.\ threshold for \gghad and excluding safety factors for the simulation uncertainties. The polar-angle acceptance range of the forward calorimeters and the central detector region are indicated with horizontal lines.}
\label{fig:bg-theta}
\end{center}
\end{figure}

The electron--positron pairs are predominantly produced at very small transverse momenta and low polar angles. The detector occupancies in the innermost layers can therefore be reduced to an acceptable level by a careful design optimisation of the inner- and forward-detector regions. The central beam-pipe walls are placed outside the high-rate region, in order to reduce the production of secondary particles in the material of the central beam pipe (0.6~mm beryllium). The inner detectors are shielded from back-scattered particles originating in the forward region by the 4.8~mm thick steel walls of the conical beam pipe sections. The beam-pipe walls point to the interaction point (IP) at an angle of approximately 7\si{\degree}, which defines the forward acceptance of the tracking system.

The particles produced in \gghad interactions, on the other hand, show a harder transverse momentum spectrum and a more central polar-angle distribution, resulting in sizeable  rates and energy depositions from background particles reaching the outer detector layers~\cite{Barklow:1443518}.

\geant-based~\cite{geant4,geant4-2,geant4-3} full detector simulations of beam-induced background events have been performed for the 380~GeV and 3~TeV stages, in order to estimate hit rates in the inner detectors and energy depositions in the calorimeters, as well as the expected radiation exposure of these detectors. The simulations include showering and back-scattering effects~\cite{CLICdet-performance,ThesisAndre,CLICdet-background}.

\cref{tab:tracker-backgrounds} summarises the expected hit rates from incoherent-pair and \gghad background particles reaching the CLICdet tracking region. The highest rates are observed in the innermost layers of the vertex detector, reaching up to 8 hits per mm$^{2}$ per bunch train. A strong radial dependence of the hit rates is observed, leading to a spread of approximately two orders of magnitude in each detector region. As expected from the higher production rates at 3~TeV, the hit rates are increased by a factor of 5 (incoherent pairs) to 20 (\gghad) with respect to the 380~GeV stage. Even in the outer tracking layers the hit rates are dominated by incoherent pair events, for which significant contributions from secondary particles are observed, despite the implemented shielding and layout-optimisation measures.

The high background hit rates constrain the technology choices for the vertex and tracking detectors in terms of pixel size and hit-time resolution, as discussed in more detail in \cref{chap:vtxtrk}.

\begin{table}[ht!]
\centering
  \caption{
    Expected background hit rates in the CLICdet tracking detectors at 380~GeV and 3~TeV. Safety factors for the simulation uncertainties of five for the incoherent pairs, and two for the \gghad events are included and the corresponding contributions are added to obtain the total hit rates. The minimum and maximum rate per detector region is given, averaged over the azimuthal angle and using \SI{10}{\mm} radial bins for the endcap regions.}\label{tab:tracker-backgrounds}
  \begin{tabular}{l S S S S }
    \toprule
    Energy stage    & \multicolumn{2}{c}{\SI{380}{GeV}} & \multicolumn{2}{c}{\SI{3}{TeV}}                                                               \\\cmidrule(r){2-3}\cmidrule(l){4-5}
    Subdetector     & \tabt{Minimum}                & \tabt{Maximum}                    & \tabt{Minimum}                & \tabt{Maximum}                \\
                    & \tabt{Hits[1/mm$^{2}$/train]} & \tabt{Hits[1/mm$^{2}$/train]}      & \tabt{Hits[1/mm$^{2}$/train]} & \tabt{Hits[1/mm$^{2}$/train]} \\ \midrule
    Vertex barrel   & 0.2                           & 3.2                               & 0.6                          & 8.8                           \\
    Vertex endcaps  & 0.1                           & 2.7                               & 0.2                          & 8.8                           \\
    Tracker barrel  & 0.0003                        & 0.03                              & 0.002                        & 0.1                           \\
    Tracker endcaps & 0.0004                        & 0.1                               & 0.002                        & 0.6                           \\
    \bottomrule
  \end{tabular}
\end{table}

\cref{tab:calo-backgrounds} summarises the energy depositions from backgrounds in the sensitive volumes of the calorimeters at 380~GeV and at 3~TeV, integrated over a full bunch train. Both the direct hits and the hits from back-scattered particles originating in the forward calorimeters occur predominantly in the forward regions, leading to very large energy depositions of up to 12~TeV per bunch train at 3~TeV in the endcap calorimeters, largely dominated by incoherent pairs in the HCAL endcaps. The total energies released in ECAL and HCAL at 380~GeV are a factor of four lower than at 3~TeV.
The energy deposits in BeamCal, which is located closest to the beam line, reach up to 270 TeV at 3~TeV, three orders of magnitude more than in LumiCal and largely dominated by incoherent pairs. The total energy deposits at 380~GeV are \mbox{\numrange{5}{7}} times lower than at 3~TeV for the forward calorimeters.
Highly granular readout cells and nanosecond timing of hits are required in all calorimeters to separate background hits from physics hits, as discussed in more detail in \cref{chap:calorimeters} and \cref{chap:fcal}.

\begin{table}[ht!]
\centering
  \caption{
    Expected energy deposits from background particles in the sensitive volumes of the CLICdet calorimeters at 380~GeV and 3~TeV.
    The numbers correspond to the summed energy deposits for an entire CLIC bunch
    train, excluding safety factors for the simulation uncertainties.
    \label{tab:calo-backgrounds}}
    \begin{tabular}{l S S S S }
    \toprule
    Energy stage      & \multicolumn{2}{c}{\SI{380}{GeV}}                   & \multicolumn{2}{c}{\SI{3}{TeV}}                             \\\cmidrule(r){2-3}\cmidrule(l){4-5}
    Subdetector       & \tabt{Incoherent pairs}           & \tabt{\gghad{}} & \tabt{Incoherent pairs} & \tabt{\gghad{}} \\
                      & \tabt{[\si{GeV\per{}train}]}      &  \tabt{[\si{GeV\per{}train}]} &  \tabt{[\si{GeV\per{}train}]}            &  \tabt{[\si{GeV\per{}train}]}   \\ \midrule
    ECAL barrel       & 3.6                               & 2.1             & 14                    & 52             \\
    ECAL endcaps + plugs & 11.1                              & 9.4             & 39                    & 252            \\ \midrule
    HCAL barrel       & 0.05                              & 0.18            & 0.22                    & 5.0             \\ 
    HCAL endcaps      & 2874                              & 7.0             & 11790                   & 312             \\ \midrule 
    Total ECAL+HCAL   & 2889                              & 19            & 11840                   & 621            \\ \midrule
    LumiCal           & 68.5                              & 4.5             & 283                     & 193             \\
    BeamCal           & 54730                             & 5.6             & 270600                  & 540             \\ \bottomrule
  \end{tabular}
\end{table}

\section{Radiation exposure}\label{sec:radiation}
Due to the absence of large QCD backgrounds in lepton collisions and due to the small interaction rates in linear colliders, the radiation exposure of the main detectors at CLIC is expected to be much smaller than the one of the current LHC detectors. Sizeable levels of total ionizing radiation dose (TID) and non-ionizing energy loss (NIEL) will only be present in the forward calorimeters LumiCal and BeamCal, due to their exposure to the strongly forward-peaked beam-induced backgrounds discussed in the previous section.

The radiation levels from incoherent pairs and \gghad events in the silicon tracking layers and in the BeamCal at 3~TeV were estimated using a full \geant-based simulation of the energy loss inside the sensor volumes and of the hit densities scaled with displacement damage factors~\cite{Dannheim:1443516,ThesisAndre}. This study used the CLIC\_ILD detector model~\cite{lcd:muennichsailer2011}, with the same 4~T magnetic field and a similar vertex-detector geometry as CLICdet. A similar study has been performed to estimate the radiation exposure of the CLIC\_ILD ECAL at \SI{3}{\TeV}~\cite{RadiationNoteSailer}. The current running scenario assumes a $40\%$ increased integrated luminosity performance of the accelerator (\cref{tab:CLIC-parameters}), compared to the assumptions for the earlier radiation level estimates. The numbers given in the following take this increased luminosity into account. Moreover, safety factors for the simulation uncertainties of five for the incoherent pairs, and two for the \gghad events are included~\cite{cdrvol2}.

Most of the NIEL damage in the vertex and tracking layers originates from \gghad events. The expected 1-MeV neutron-equivalent fluence for the inner vertex-detector layers is approximately
\SI{6e10}{\neq\per\centi\meter\squared\per\year}. 
The TID is dominated by incoherent pairs, resulting in a maximum ionising dose of
\SI{300}{\gray\per\year} 
in the inner vertex-detector layers.

 A total fluence of approximately
\SI{2e11}{\neq\per\centi\meter\squared\per\year},   
is expected for the ECAL endcaps, with similar contributions from pairs and \gghad events. The ionising dose in the ECAL endcaps reaches up to
\SI{6}{\gray\per\year} 
for the \gghad events and  
\SI{3}{\gray\per\year} 
for the incoherent pairs. The rates fall steeply with increasing radius.

The radiation levels in the tracking detectors and main calorimeters are several orders of magnitude below the levels expected in the corresponding regions of the ATLAS and CMS detectors during LHC run~1--3~\cite{ATLAS-ITKpixel-TDR,ATLAS-LAr-TDR,CMS-tracker-TDR,HGCALTDR}. No dedicated R\&D is therefore pursued for the radiation tolerance of the sensor and readout technologies for the CLIC tracking detectors and main calorimeters.

The radiation exposure of the BeamCal is dominated by incoherent pair background. The TID reaches up to
\SI{7}{\mega\gray\per\year} at \SI{3}{\tera\electronvolt}  
and the NIEL up to
\SI{1.4e14}{\neq\per\centi\meter\squared\per\year}  
in the innermost cells. Radiation-hard sensor and readout technology is therefore required in this detector region, as discussed in \cref{chap:fcal}.

No dedicated studies were performed for the HCAL and LumiCal, nor for the \SI{380}{\giga\electronvolt} stage. From the hit rates and energy deposits discussed above, it is expected that the HCAL endcaps will have a much larger radiation exposure than the ECAL. An optimised detector layout with additional shielding will help to mitigate the effect of the beam-induced backgrounds in this region, as discussed in \cref{sec:detector-optimisation-high-occupancy}. The LumiCal layers, on the other hand, are expected to be exposed to several orders of magnitude lower radiation levels than BeamCal, due to the larger radius. The radiation levels at 380~GeV are expected to be much lower than at \SI{3}{\tera\electronvolt}, in view of the lower rates of beam-induced backgrounds (\cref{fig:bg-theta}) and corresponding energy depositions (\cref{tab:calo-backgrounds}).


\chapter{Vertex and tracking detector}\label{chap:vtxtrk}
This chapter gives an overview of the developments for the vertex and tracking detector at CLIC. The requirements are introduced in \cref{sec:vtx-trk-requirements}, followed in \cref{sec:vtx-trk-concept} by a description of the detector concept matching these requirements. An overview of the sensor and readout R\&D is given in \cref{sec:sensor-readout-overview}. \cref{sec:vtx-trk-sim-characterisation} presents the simulation and characterisation infrastructure aiding the detector R\&D. For the vertex detector region, fine-pitch hybrid readout ASICs, described in \cref{sec:vtx-trk-hybrid-asics}, are studied in bump-bonded assemblies with thin passive sensors (\cref{sec:vtx-trk-hybrid-assemblies}) and in capacitively coupled assemblies with active CMOS sensors (\cref{sec:capacitively_coupled_hvcmos}). A variety of integrated CMOS sensor technologies, described in \cref{sec:vtx-trk-monolithic-cmos,sec:hrcmos,sec:soi}, are under development for the large-area tracker region. Detector integration aspects are discussed in \cref{sec:vtx-trk-integration}. The chapter concludes in \cref{sec:vtx-trk-summary} with a summary of the presented results and an outlook on plans for future R\&D.

\section{Requirements}\label{sec:vtx-trk-requirements}
A precise measurement of the displaced decay vertices of heavy-quark flavour hadrons and tau-leptons is needed to meet the precision physics requirements at CLIC. For the CLIC tracking system, a transverse impact parameter resolution of $\sigma(d_0)=5\oplus 15/(p[\si{\giga\electronvolt}]\sin^{\frac{3}{2}}\theta)\si{\micron}$ is aimed for~\cite{cdrvol2}. Taking into account that the inner detector radius is constrained to $\geq$31~mm due to background hits from incoherent pairs (see \cref{sec:beam_induced_backgrounds}), this requirement translates into a single point resolution of \SI{3}{\micron} for the vertex detector~\cite{lcd-note-2011-031}. Small, squared pixels with \SI[product-units = brackets-power]{25x25}{\micron\squared} size and the readout of charge information per pixel are needed to reach this resolution target. The pixel area in the innermost layers is also constrained by the high background rates, which lead to occupancies of a few percent in the innermost layers for \SI[product-units = brackets-power]{25x25}{\micron\squared} pixels (see \cref{sec:datarate}). 

The main requirement for the tracker is a transverse momentum resolution for high-$p_\text{T}$ tracks ($\geq$\SI{100}{\giga\electronvolt}) of $\sigma_{p_\text{T}} /p_T^2\leq\SI{2e-5}{\per\giga\electronvolt}$ \cite{cdrvol2}. The total tracker radius is \SI{1.5}{\m}. To achieve the required momentum resolution in a \SI{4}{\tesla} solenoid field, the single point resolution of the tracking detector layers has to be smaller than \SI{7}{\micron}~\cite{Nurnberg:2261066}. This requires the cell length in the R$\varphi$ direction to be in the range of \SIrange{30}{50}{\micron}. The cell length in the beam direction (for the barrel layers) and in the radial direction (for the endcap layers) is limited by the detector occupancy to \SIrange{1}{10}{\mm} at \SI{50}{\micron}, depending on the detector layer.

To achieve the stringent requirements on impact parameter and momentum resolution, the material budget in the detector is limited to \SI{0.2}{\percent{}X_0} per detection layer in the vertex and to \SIrange{1}{2}{\percent{}X_0} per layer in the main tracking detector. Low-power detector elements deploying power-pulsing features and low-mass services and supports are therefore necessary. In particular in the vertex detector, the strict material budget does not allow for liquid cooling of the detector. In order to enable cooling through forced air flow, the average power dissipation in the vertex detector should therefore not exceed \SI{50}{\milli\watt\per\cm\squared}. For the tracker a leak-less water-cooling system is foreseen, allowing for an average power dissipation of approximately \SI{150}{\milli\watt\per\cm\squared}. Dedicated studies on low-mass support structures for the vertex and tracking detectors (\cref{sec:support_structures}) and on power pulsing (\cref{sec:power}) and air-flow cooling (\cref{sec:cooling}) of the vertex detector have been performed, to ensure that these requirements can be met.

The distinct beam structure of the CLIC accelerator with its low duty cycle allows for the full readout of the \SI{156}{\ns} bunch train in the \SI{20}{\ms} gap between bunch trains. Therefore, no trigger system is foreseen. To suppress hits from out-of-time background as outlined in \cref{sec:beam_induced_backgrounds}, a precise hit timing of about \SI{5}{\ns} is required. It is assumed that only the first hit per readout cell and bunch train will be time tagged.

The expected hit occupancies from background particles of 1-3\% per bunch train in the inner layers have a similar effect as random noise hits from the detector. Moreover, as only one time stamp per detector cell is read out per bunch train, such background and noise hits will also deteriorate the detection efficiency for physics hits reconstructed within the correct time window~\cite{Nurnberg:2261066}. As a goal, the intrinsic detector inefficiency and noise occupancies in the silicon detectors should both be an order of magnitude below the corresponding inefficiency and hit occupancy caused by beam-induced background particles. This goal translates into a required hit efficiency of 99.7-99.9\% and a maximum tolerable noise rate below \SI{6.5}{\kilo\hertz} per pixel for operation at CLIC with \SI{156}{\ns} bunch duration and \SI{50}{\hertz} train repetition rate. A much smaller noise rate is typically required for efficient detector operation with particle sources and in test beams with high duty cycles.

Only moderate radiation tolerance is required for the vertex and tracking-detector region.
The expected radiation exposure from non-ionising energy loss is about \SI[per-mode=reciprocal]{6e10}{\neq\per\cm\squared\per\year} for the inner vertex layers.
The maximum total ionising dose in this region is about \SI{300}{\gray\per\year}.
More details on the expected background hit rates and radiation exposure can be found in \cref{sec:radiation} and the references therein.

\section{Detector concept}\label{sec:vtx-trk-concept}

The vertex detector consists of three double layers in the barrel region, ranging in radius from \SIrange{31}{60}{\mm} and discs on each side of the detector, as illustrated in \cref{fig:vertex_layout}. The limited material budget does not allow for liquid cooling within the vertex detector volume, hence the detector is cooled by forced air flow. To allow for better air-flow through the detector, the discs are arranged in a spiral geometry. It has been demonstrated in \geant-based full simulation studies on dijet-events, that the spiral endcap geometry has only a minor impact on the flavour-tagging performance of the detector~\cite{Tehrani:FlavorTaggingSpirals, AlipourTehrani:1742993}.

\begin{figure}[ht]
  \centering
                 \includegraphics[width=.8\linewidth]{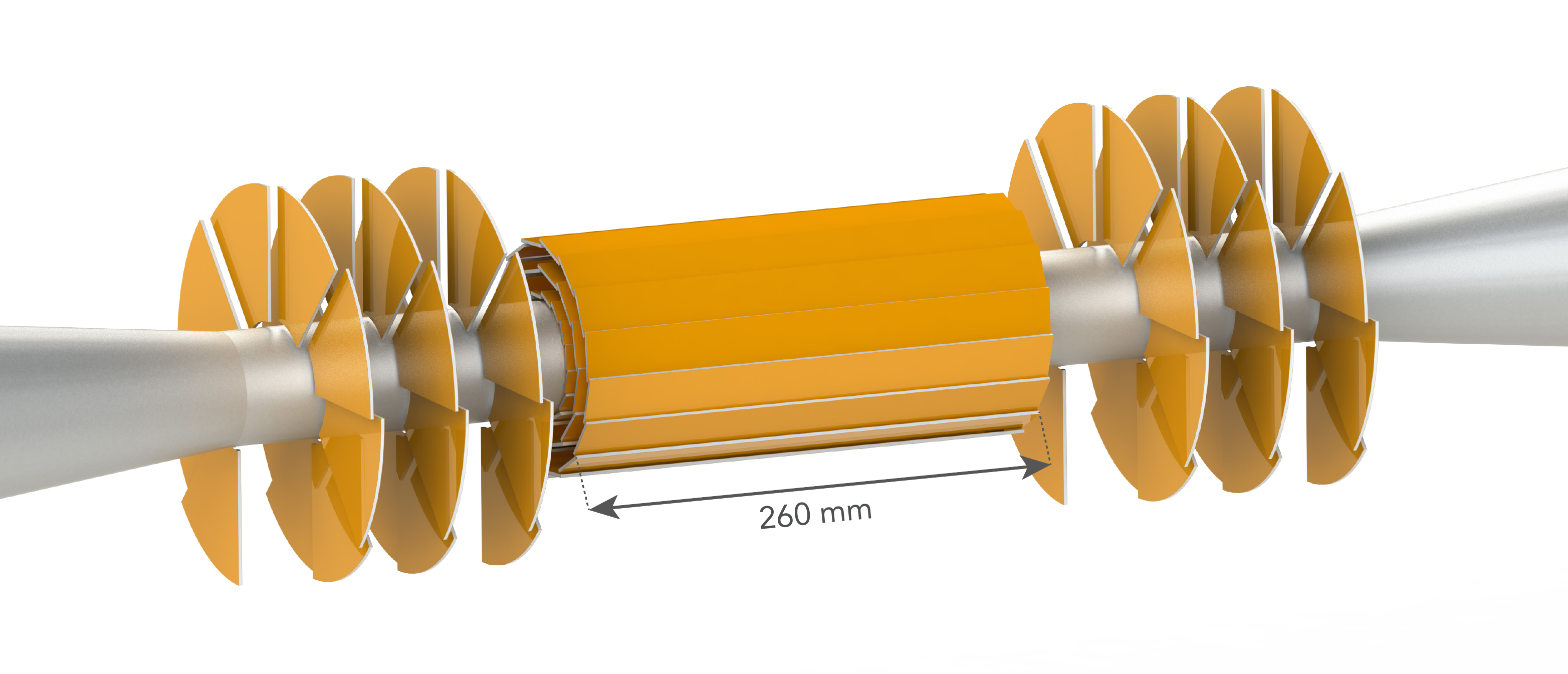}
                 \caption{Schematic view of the CLIC vertex detector consisting of three double layers in the barrel and three double-layer forward petals in a spiralling arrangement to facilitate air cooling~\cite{CLIC-summary-report-2018}.}\label{fig:vertex_layout}
  \end{figure}

The main tracker is divided into inner and outer parts by a cylindrical support tube. The radius of this tube, which also supports the third inner barrel layer, has been chosen to be sufficiently large to allow for an extended coverage of the forward discs. It also serves as a support for the vacuum beam pipe (see \cref{sec:beam-pipe-design}). The tracker consists of 6 layers in the barrel (with radii from 127 to \SI{1486}{\mm}) and 7 inner discs and 4 outer discs on each detector side (extending up to $z=\SI{2190}{\mm}$), as illustrated in~\cref{fig:tracker}.

\begin{figure}[t]
         \centering
         \includegraphics[width=\linewidth]{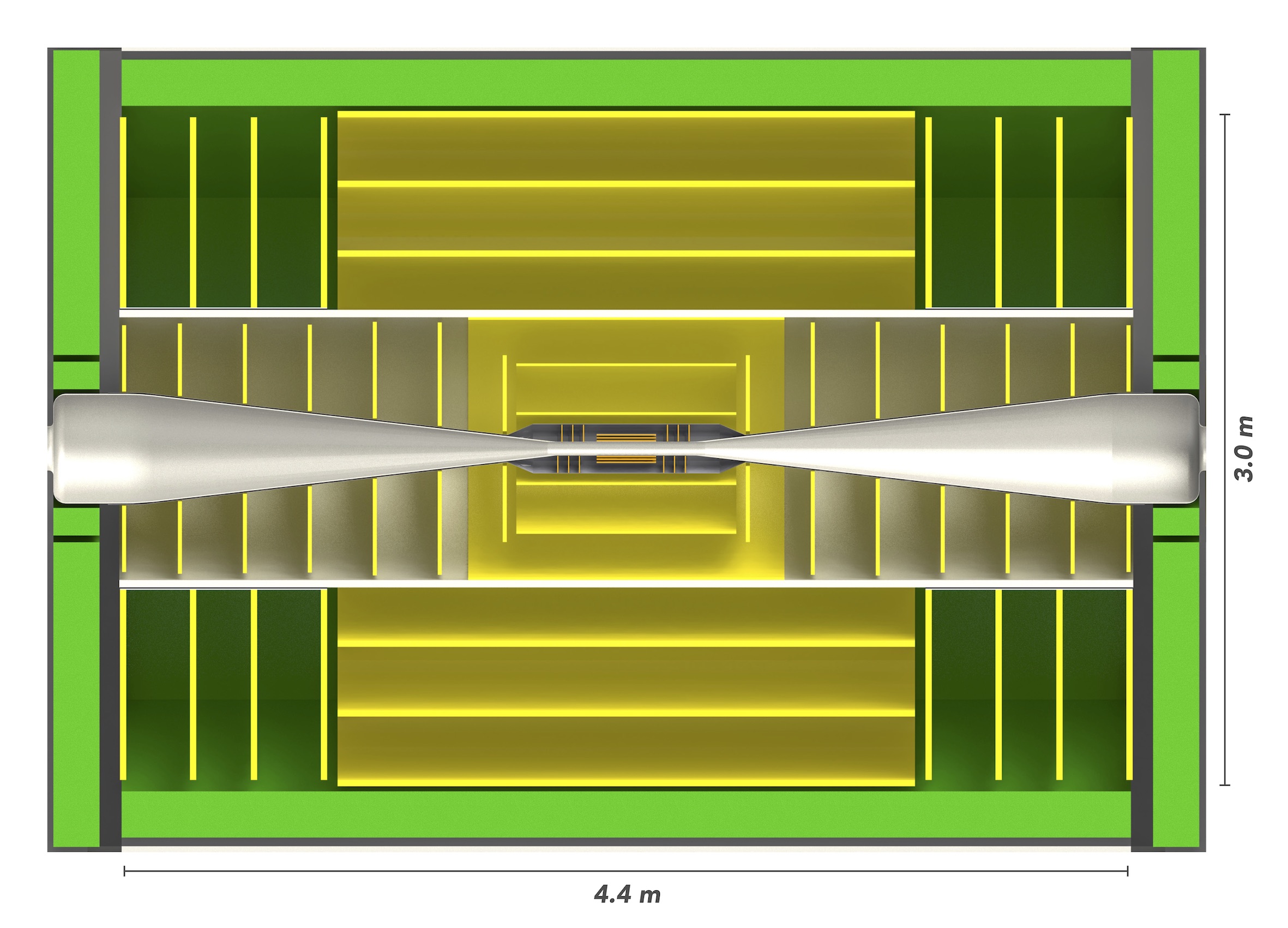}
         \caption{Top view of the tracking system. Starting from the centre: the beam pipe, shown in grey, consists of a cylindrical central part and two conical sections. The vertex-detector layers, shown in orange, are surrounded by an envelope for air cooling, shown in dark grey. The main tracker layers are depicted in yellow and are subdivided into outer and inner parts by a cylindrical support tube shown in light grey. The tracker is surrounded by the electromagnetic calorimeter, depicted in green.} \label{fig:tracker}
\end{figure}

\cref{fig:tracker1} presents the detector model as implemented in simulations, showing the detailed layout of sensitive material, supports as well as the routing of services. The main parameters of the vertex and tracking detectors and the beam pipe are summarised in \cref{tab:vertex-tracker-parameters}.
\cref{fig:material_budget} shows the material budget of the different regions inside the tracking system, including the beam pipe, supports and cables, as a function of the polar angle.

\begin{figure}[ht]
  \centering
  \includegraphics[scale=0.4]{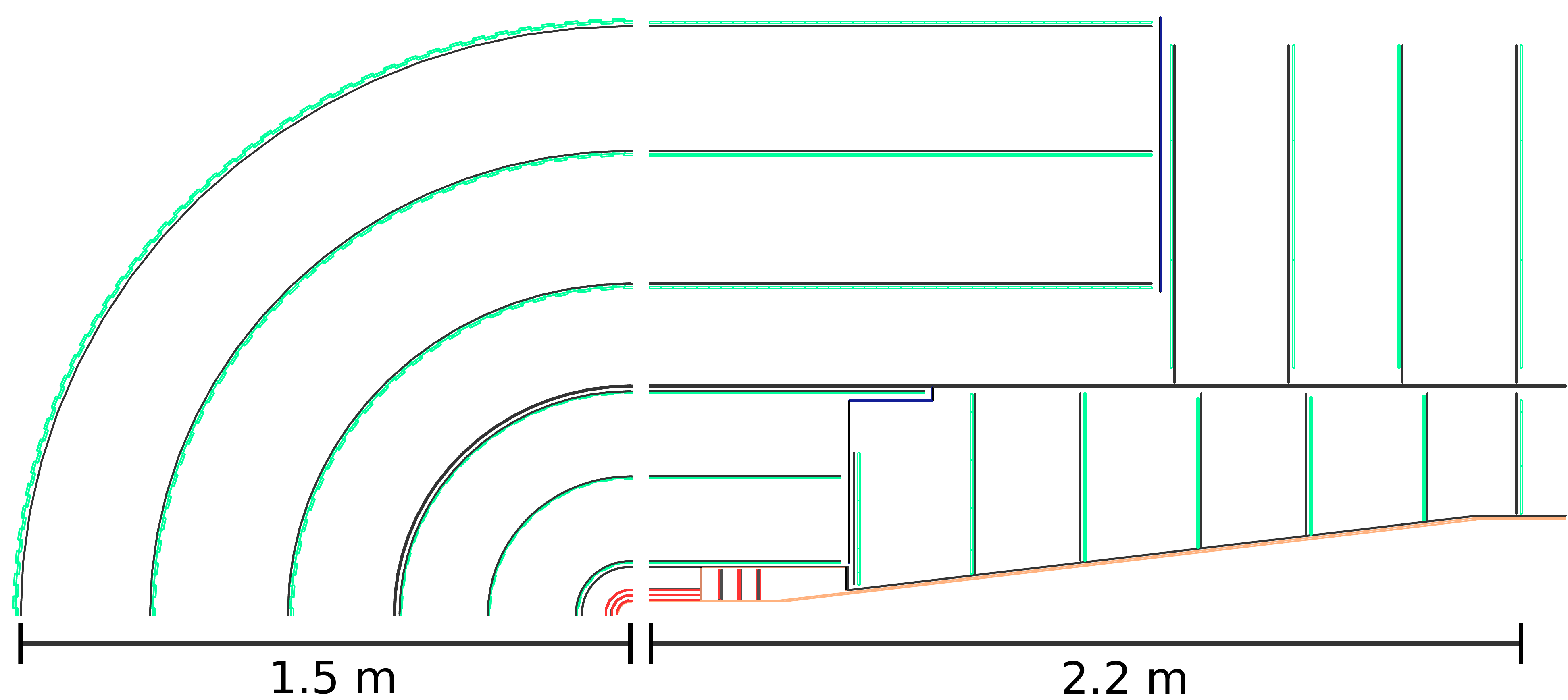}
  \caption{Layout of the tracking system in the $xy$  plane (left) and the $xz$  plane (right), as implemented in the simulation model~\cite{CLICdet-performance}. The vertex detector is shown in the centre (in red). Tracker sensors are shown in green, support material in black. The orange line indicates the vacuum tube. The blue lines represent additional material (e.g.\ cables) in the critical transition regions between barrel and endcap layers.}
  \label{fig:tracker1}
\end{figure}

\begin{table}
\centering
\caption{Main parameters of the CLIC beam pipe and vertex and tracking detectors.}
\label{tab:vertex-tracker-parameters}
\begin{tabular}[htpb]{l l l}
\toprule
Parameter& \multicolumn{2}{c}{Value} \\
\midrule
Central beam pipe inner radius & \multicolumn{2}{c}{29.4~mm} \\
Central beam pipe thickness and material  & \multicolumn{2}{c}{0.6~mm beryllium (0.17\%$X_0$)}\\
Conical beam pipe thickness and material  &  \multicolumn{2}{c}{4.8~mm steel}\\
\midrule
 & Vertex region & Tracker region \\
\midrule
Material budget per layer & 0.2\%$X_{0}$ &  $\approx$1\%$X_{0}$ \\
Max. cell size & \SI[product-units = brackets-power]{25x25}{\micron\squared} & $(0.050\times$1--10~mm)$^{2}$ \\
\midrule
Number of barrel layers & \SI{3x2}{} & 6 \\
Full length of barrel layers & 260~mm & 964--2528~mm\\
Radius barrel layers & 31--60~mm & 127-1486~mm \\
\midrule
Number of endcap layers & 6 & 7 \\
Position of endcap layers in $z$ & 160--299~mm &  524--2190~mm \\
\midrule
Total sensor area & 0.84~m$^{2}$ & 137~m$^{2}$ \\
\bottomrule
\end{tabular}
\end{table}

\begin{figure}[ht]
   \centering
   \includegraphics[width=.5\linewidth]{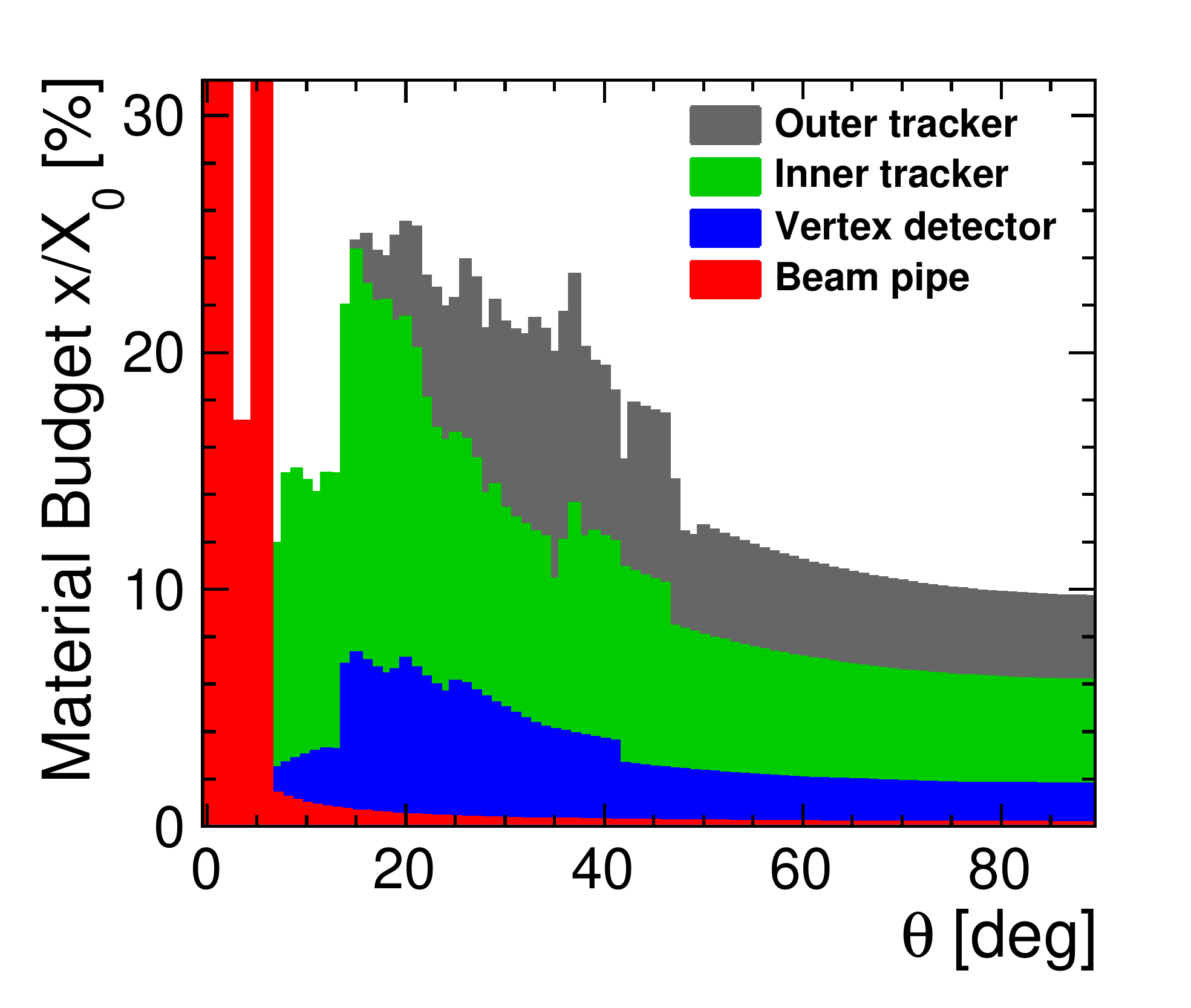}
   \caption{Stacked material budget of the different regions inside the tracking system, as a function of the polar angle~\cite{CLICdet-performance}. Contributions from sensitive layers, cables, supports and cooling are included in the respective regions.
}
   \label{fig:material_budget}
\end{figure}

\section{Overview of sensor and readout ASIC developments}\label{sec:sensor-readout-overview}
The challenging requirements for the sensors and readout ASICs of the vertex and tracking detector in terms of high spatial and temporal measurement precision and minimal material content have inspired a broad technology R\&D programme in this domain.
Starting from tests with existing pixel-detector readout ASICs and sensors, a progressively more specialised development programme is pursued, aiming at simultaneously fulfilling all of the requirements at CLIC. Synergies are exploited with other detector R\&D projects, such as the HL-LHC upgrades for the LHC experiments~\cite{Garcia_Sciveres_2018} or the pixel-detector developments for the Mu3e experiment~\cite{Vilella_2018}. In several areas they have led to common developments and testing efforts.

\cref{tab:sensor-asic-overview} gives an overview of the various test assemblies produced as technology demonstrators for the vertex and tracking detectors, which are described in detail in the following sections.
\begin{table}[h]
\footnotesize
  \centering
\begin{threeparttable}
  \caption{\label{tab:sensor-asic-overview} Summary of investigated pixel-detector assemblies and technologies.}
  \begin{tabular}{c  *6{c}}
    \toprule
Test       &y Type & Coupling & CMOS     & Pixel                & Target      &  Presented  \\
assembly   &      &          & feature  & dimensions           & application &  in        \\
           &      &          & size     &                      &             &  section   \\
           &      &          & [nm]     & $[\upmu\text{m}^2]$  &             &    \\
    \midrule
Timepix/Timepix3 + Si sensor & hybrid planar & bump-bonded & 250/130 & $55\,\times\,55$ & sensor R\&D & \ref{sec:vtx-trk-hybrid-asics}, \ref{sec:vtx-trk-hybrid-assemblies}   \\
 CLICpix(2) + Si sensor & hybrid planar & bump-bonded & 65 & $25\,\times\,25$ & vertex &  \ref{sec:vtx-trk-hybrid-asics}, \ref{sec:vtx-trk-hybrid-assemblies}  \\
 CLICpix(2) + CCPDv3/C3PD & hybrid HV-CMOS & capacitive & 65, 180 & $25\,\times\,25$ & vertex & \ref{sec:capacitively_coupled_hvcmos}  \\
 ATLASpix simple & HV-CMOS & monolithic & 180 & $40\,\times\,130$ & tracker & \ref{sec:vtx-trk-monolithic-cmos} \\
 ALICE Investigator & HR-CMOS & monolithic & 180 & $28\,\times\,28$ & vertex/tracker & \ref{sec:hrcmos}\\
 CLICTD (in production) & HR-CMOS & monolithic & 180 & $30\,\times\,300$ & tracker & \ref{sec:clictd} \\
 Cracow SOI & SOI & monolithic  & 200 & $30\,\times\,30$ & vertex/tracker &  \ref{sec:soi}\\
 CLIPS (produced) & SOI & monolithic  & 200 & $20\,\times\,20$ & vertex/tracker & \ref{sec:clips}\\
     \bottomrule
 \end{tabular}
\end{threeparttable}
\end{table}

Initial studies for the vertex detector focussed on assessing the feasibility of small-pitch hybrid detectors with very thin (down to \SI{50}{\micron}) sensors.
The hybrid approach allows for a separate optimisation of readout ASICs and sensors.
The general-purpose Timepix (\SI{250}{\nm} feature size) and Timepix3 (\SI{130}{\nm} feature size) hybrid readout ASICs with \SI{55x55}{\micron} pitch were used as test vehicles, bump bonded to sensors with various thicknesses and advanced design features (slim edge and active edge).
The results of these performance assessments were used to improve the simulation tools and optimise future dedicated ASIC and sensor designs for CLIC.
While the Timepix and Timepix3 assemblies with very thin sensors (\SIrange{50}{100}{\micron}) show an excellent detection efficiency and time resolution, the spatial resolution for perpendicular tracks is severely impacted by the lack of charge sharing between neighbouring pixels between neighbouring pixels separated by a \SI{55}{\micron} pitch. This is due to the reduced number of charge carrier that are created in the thin sensor which at the same time have undergo less diffusion before being collected.

A smaller pixel pitch is therefore necessary, in order to reach a better spatial resolution.
The availability of a more advanced CMOS process with \SI{65}{\nm} feature size enabled the development of the CLICpix and CLICpix2 readout ASICs with \SI{25x25}{\micron} pitch, targeting specifically the requirements of the CLIC vertex detector.
Bump bonding at this small pitch remains a challenge, and the spatial resolution target of \SI{3}{\micron} for the vertex detector has not yet been reached with \SI{50}{\micron} thin planar sensors.
New sensor designs with enhanced lateral drift are therefore under study, with the aim of increasing the charge sharing and thereby improving the position resolution for a given readout pitch and sensor thickness.

An alternative hybrid detector concept is under study, which is based on capacitive coupling through a thin layer of glue between CLICpix/CLICpix2 readout ASICs and active CCPDv3/C3PD sensors implemented in a \SI{180}{\nm} High-Voltage CMOS process.
Tightly controlled glue-assembly procedures, as well as dedicated simulation and calibration efforts are required for this technology, in order to cope with the complex signal-transfer chain.
Similar position resolution values are reached as for the thin planar sensor assemblies, meaning that they currently do not meet the CLIC requirements.

The High-Voltage CMOS process is also suited for building fully monolithic depleted sensors with larger (elongated) pixels.
The ATLASpix HV-CMOS sensor with \SI{30x300}{\micron} pitch was designed as a technology demonstrator for the ATLAS HL-LHC upgrade, but also targets the CLIC tracker requirements.
Studies performed with the CLICdp test-beam setup show sufficient timing precision, while the spatial resolution is limited by the pixel pitch to values significantly above the required \SI{7}{\micron}.
A new design with an adapted pixel geometry is planned in order to reach this resolution value.

An alternative depleted CMOS sensor technology with very small collection electrodes implemented on a High-Resistivity (HR) substrate has been chosen for the HL-LHC upgrade of the ALICE Inner Tracking System (ITS).
Promising CLICdp test-beam results with the Investigator analogue test chip have led to the design of the fully monolithic CLICTD demonstrator chip with sub-segmented macro pixels of \SI{30x300}{\micron} pitch, targeting the requirements of the CLIC tracker.

The Silicon-On-Insulator (SOI) technology allows for producing thin monolithic sensors consisting of a CMOS readout layer separated from a fully depleted high-resistivity sensor layer through an insulator oxide layer.
The Cracow SOI developments are based on a \SI{200}{\nm} SOI process.
Generic technology demonstrator sensors with a pixel pitch of \SI{30x30}{\micron} and rather thick sensor layers (300 and \SI{500}{\micron}) have been tested successfully.
The CLIPS demonstrator sensor with \SI{20x20}{\micron} pitch and a snap-shot time and energy measurement concept has recently been produced and specifically targets the CLIC vertex-detector requirements.

\section{Simulation and characterisation infrastructure}\label{sec:vtx-trk-sim-characterisation}

Development of new detectors, especially in not yet established technologies, requires dedicated infrastructure. This ranges from detailed device simulations for gauging new detector designs, through data acquisition and readout systems for each new detector, to reference beam telescopes and corresponding reconstruction and analysis frameworks, which facilitate the measurement of basic device characteristics and figures of merit such as tracking efficiency or spatial resolution. This section gives a brief overview of simulation and characterisation tools developed in the context of the CLIC vertex and tracker R\&D and in close collaboration with other experiments.

\subsection{Monte Carlo detector simulations with \apsq}
\label{sec:apx_2}

Detailed simulations of silicon detectors are a crucial tool for understanding their performance.
The complex characteristics of novel devices, such as highly non-uniform electric fields in the sensor, remain a challenge to detector simulations.
Nevertheless, a better understanding of the device performance from simulations can significantly improve new detector designs as well as drastically reduce the cost and time required for the development of a novel device.
Advanced tools for simulation such as finite-element Technology Computer Aided Design (TCAD)~\cite{synopsys-tcad} exist, but are very demanding on computing time and do not easily allow integration with other tools in order to facilitate a Monte Carlo approach, an essential method in high-energy physics given the stochastic nature of particle interactions.

In order to support the CLIC vertex and tracking detector R\&D activities, \apsq~\cite{allpix-squared, allpix-squared-webpage} has been developed. It is a comprehensive and modular open-source framework for Monte Carlo detector simulations combining detailed device descriptions with simplified models of charge transport. 

The main novelty of the \apsq framework is the possibility of easily combining TCAD-simulated electric fields with a \geant{}~\cite{geant4,geant4-2,geant4-3} simulation of particle interactions with matter, including stochastic effects such as Landau fluctuations and the production of secondary particles.
This allows detector performance parameters, such as resolution and efficiency, to be directly assessed with high precision.

The electrostatic finite element TCAD simulation of the complex field configuration in the sensor is converted from the adaptive mesh used in TCAD to a regular mesh for fast interpolation and lookup of field values during charge transport.
The charge carriers deposited by the initial \geant{} simulation are then transported through the sensor along these field lines using a Runge-Kutta-Fehlberg integration method~\cite{fehlberg}.
This approach has shown to be especially useful for novel technologies under consideration for the vertex and tracking detector, such as CMOS pixel sensors with complex implantation profiles (see \cref{sec:hrcmos-simulation-results}).
With event simulation rates of several tens of \SI{}{\hertz}, this allows high-statistics samples to be gathered, which are necessary for detailed studies of the detector behaviour.

The simulation chain is arranged with the help of configuration files containing key-value pairs with physical units, and an extensible system of modules which implement separate simulation steps such as the initial energy deposition and charge carrier creation, the transport of the charge carriers through the silicon, or the shaping and amplification of the signal collected at the electrodes by the front-end electronics of the detector.
The flexibility enables simulations of different detectors in the same setup, such as a beam telescope for reference tracks together with the actual device under test.
An example for such a simulation chain involving different modules employed for different detectors is shown in \cref{fig:simulation:complex}.

\begin{figure}[ht]
  \centering
  \includegraphics[width=\textwidth]{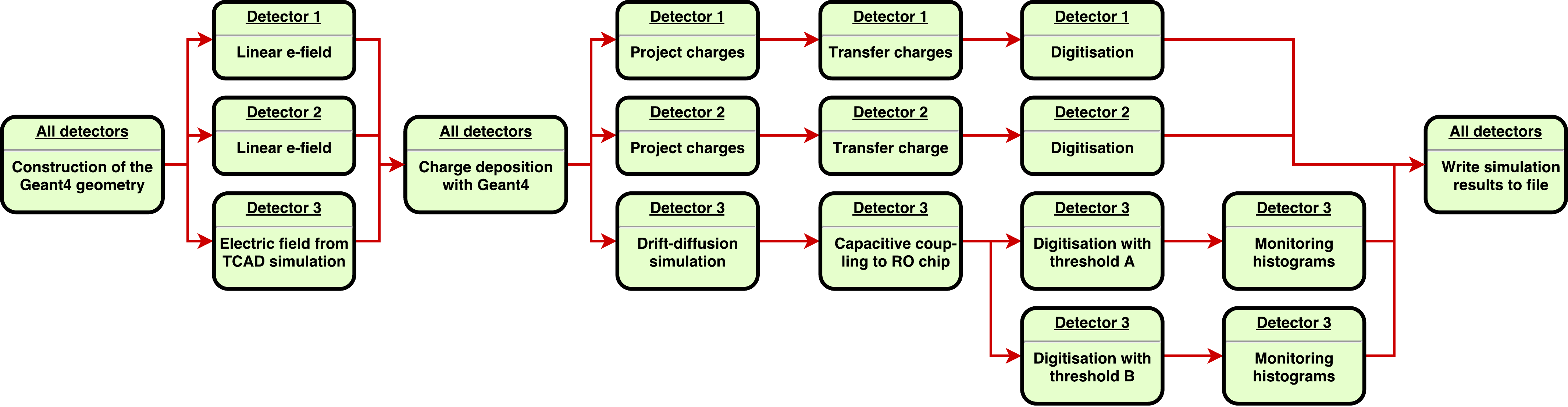}
  \caption{Setup of an \apsq simulation chain with three detectors, running different simulation modules~\cite{allpix-squared}.}
  \label{fig:simulation:complex}
\end{figure}


Different physics models for charge transport with varying accuracy and computation speed have been implemented to cover various use cases, such as the detailed simulation of the detector under investigation, combined with a fast simulation of reference detectors in the same setup.
This approach allows the reproduction of full detector setups as e.g.\ found in test-beam measurement campaigns, including the beam telescope used for reference tracks.
For the detailed drift-diffusion model implemented in \apsq, where charge carriers are propagated through the silicon sensor step by step, while updating mobility and velocity depending on the electric field at the given position, the framework can produce line graphs depicting the drift path of individual charge carriers.
An example for such a graph is shown on \cref{fig:simulation:linegraph}, where electrons and holes drift to different electrodes under the influence of the applied electric field.
This representation can help in understanding the behaviour of charge carriers in the sensor, especially with more complex electric fields simulated with TCAD.
\begin{figure}[ht]
  \begin{subfigure}[T]{0.7\textwidth}
    \includegraphics[width=0.75\linewidth]{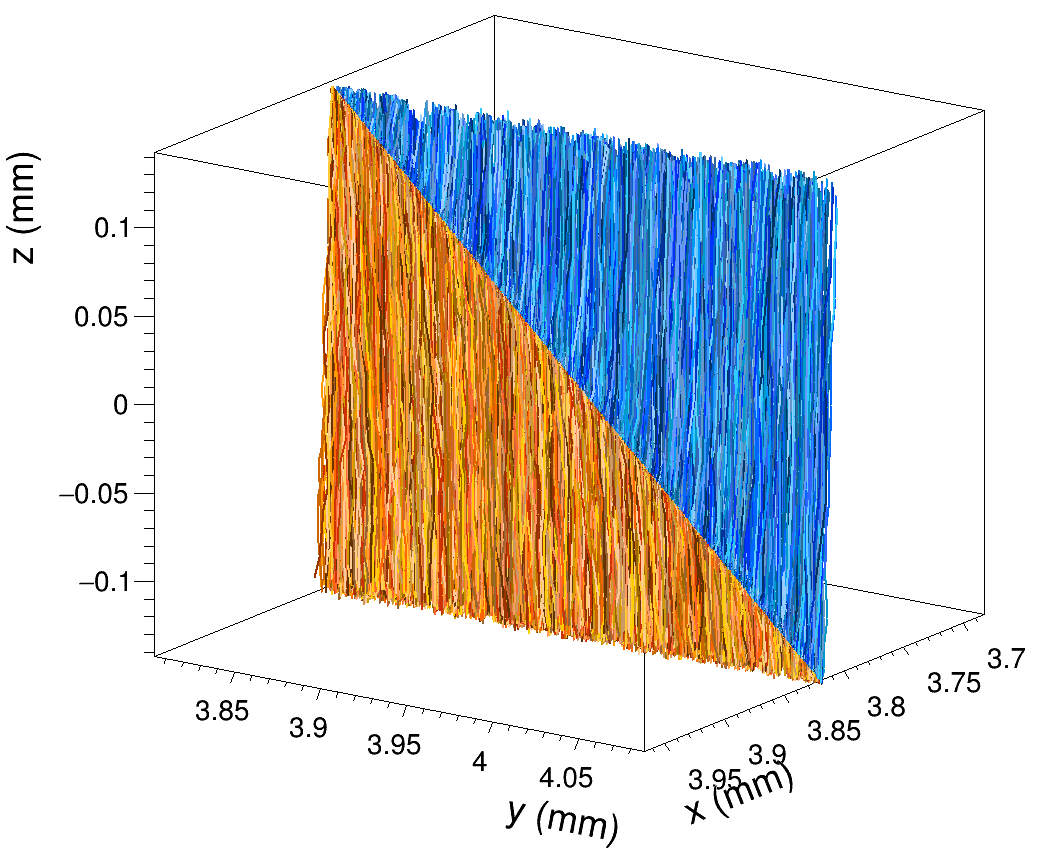}
    \caption{}\label{fig:sim-lg-3d}
  \end{subfigure}
  \begin{subfigure}[T]{0.2\textwidth}
    \includegraphics[width=0.5\linewidth]{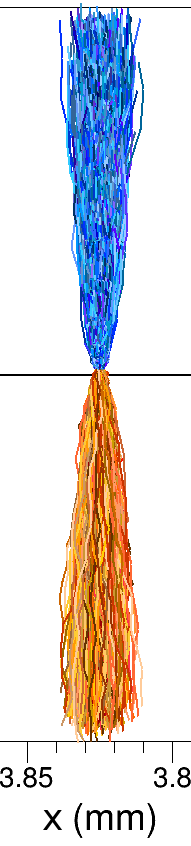}
    \caption{}\label{fig:sim-lg-2d}
  \end{subfigure}
 \caption{Line graph generated by \apsq illustrating the charge transport through the silicon sensor. \subref{fig:sim-lg-3d} The ionizing radiation crosses the detector at an angle of \SI{45}{\degree} and electrons (visualised in azure) and holes (visualised in orange) drift towards their respective collection electrodes. \subref{fig:sim-lg-2d} depicts a projection along the particle trajectory, showing the additional component of diffusion.}\label{fig:simulation:linegraph}
\end{figure}

The \apsq framework has seen continuous development and extension of its functionality since its first release. It is disseminated together with a comprehensive and continuously updated user manual~\cite{clicdp-apsq-manual}.

\subsection{CaRIBOu, a flexible pixel-detector readout system}
\label{sec:caribou}
Developing new detectors requires the design of an adequate readout system, a task usually requiring significant personnel and financial resources.
In order to ease this task, and to facilitate the testing of many different detectors within the CLIC vertex and tracking detector R\&D, a flexible open-source readout system, CaRIBOu~\cite{caribou, caribou_fiergolski, caribou-webpage}, has been designed, which only requires minimal adaptation to new detectors.

\begin{figure}[ht]
  \centering
  \includegraphics[width=.95\linewidth]{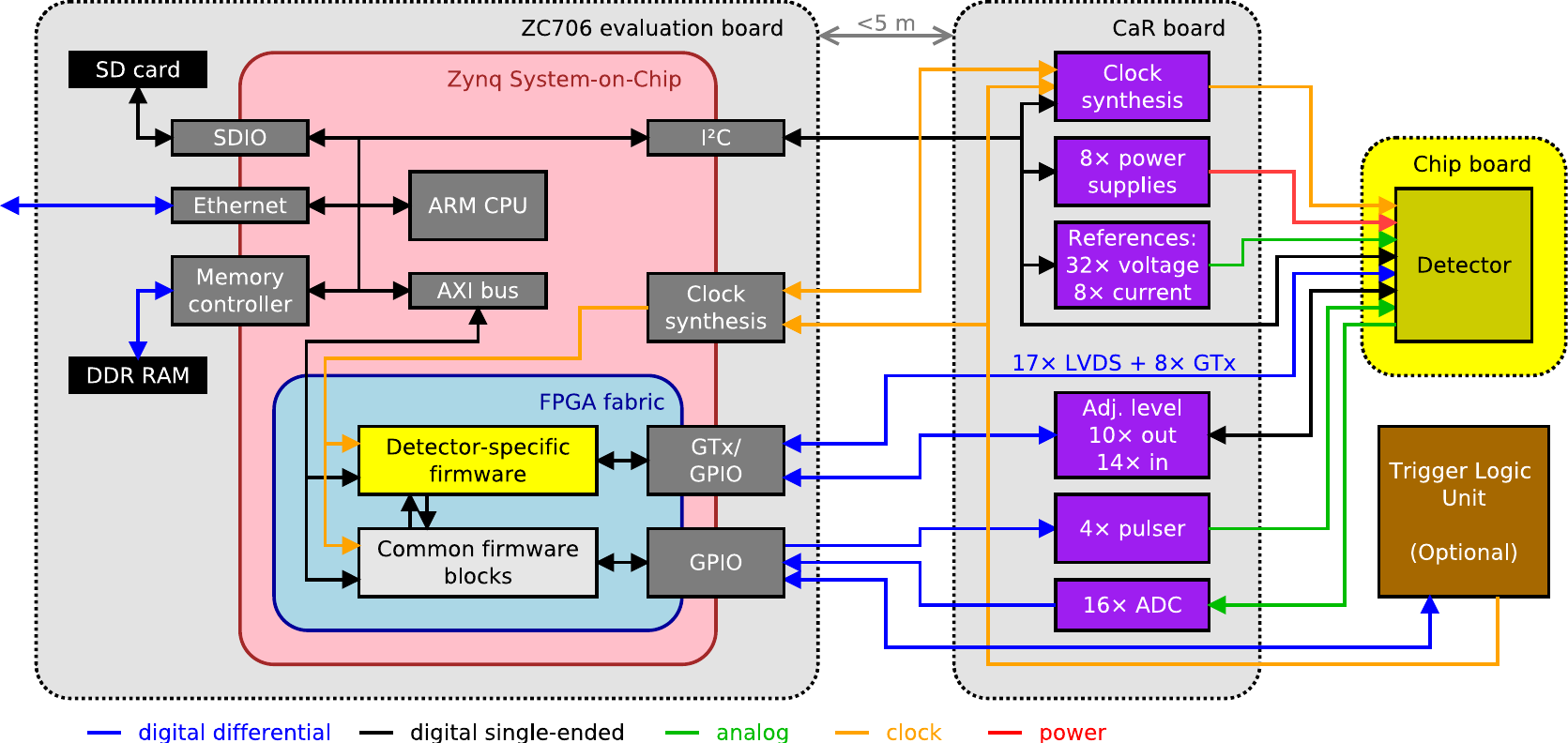}
    \caption{Schematic representation of the CaRIBOu hardware architecture, consisting of a ZC706 evaluation board, a universal CaR board and an application-specific chip board.}
  \label{fig:caribou-hardware-schematic}
\end{figure}

\cref{fig:caribou-hardware-schematic} shows a schematic representation of the hardware components included in the CaRIBOu system.
At the detector-side of the readout chain, a custom designed \emph{chip board} implements the routing from a specific detector to the interface connector of the CaRIBOu system.
The chip board is connected to the \emph{CaR board} (Control and Readout board) which provides the hardware environment for various target ASICs, including programmable power supplies with monitoring, programmable voltage and current references, analogue-to-digital converters (ADCs), an I$^2$C interface bus, as well as a number of general-purpose communication links including high-speed full-duplex serial links up to 12.5~Gbps (GTx), differential links for up to 1.2~Gbps (LVDS) and single-ended general-purpose inputs and outputs (GPIO) with adjustable voltage level.
Furthermore, it is equipped with a clock generator which can be used to generate stable clock signals for use in both the detector and the firmware blocks.
Furthermore, it can receive an external clock and triggers from a trigger logic unit (TLU) to synchronise with external devices and other readout systems.
The board is connected to the chip board via a 320 pin SEARAY connector and is therefore re-usable for different devices and setups.
The core of the CaRIBOu system is a Xilinx Zynq System-on-Chip (SoC) device hosted on the ZC706 evaluation board.
It combines a dual-core ARM Cortex-A9 CPU and a Kintex-7 Field Programmable Gate Array (FPGA) fabric connected through a silicon interposer.

For operation in radiation environments, the connection between the Xilinx system and the CaR board can be established through an optional FMC cable (up to approximately 5~m) which allows for a remote placement of the SoC system.
The general-purpose links are routed to the FPGA fabric inside the Zynq SoC. The user can write a detector-specific firmware block (IP core) handling these links or use some of the available IP cores that implement most of the standard communication protocols. 
Several CaRIBOu-specific IP cores are provided with the system. 
They provide an interface to components on the CaR board as well as some generic data handling usable for any detector, such as ring buffers for data reception from the detector interface to the ARM processing system via an Advanced eXtensible Interface (AXI) bus.
New firmware modules can be written and added to the existing design in order to customise the functionality to match the requirements of a specific detector.

The system is directly connected to the ethernet network and does not require an additional desktop PC for operation.
The ARM CPU of the system runs a full Linux-based operating system, implemented using the Yocto project, which is widely used in industry for embedded-system solutions~\cite{yocto}.
The user logs into the system using secure shell (SSH).

\begin{figure}[ht]
  \centering
  \includegraphics[width=.8\linewidth]{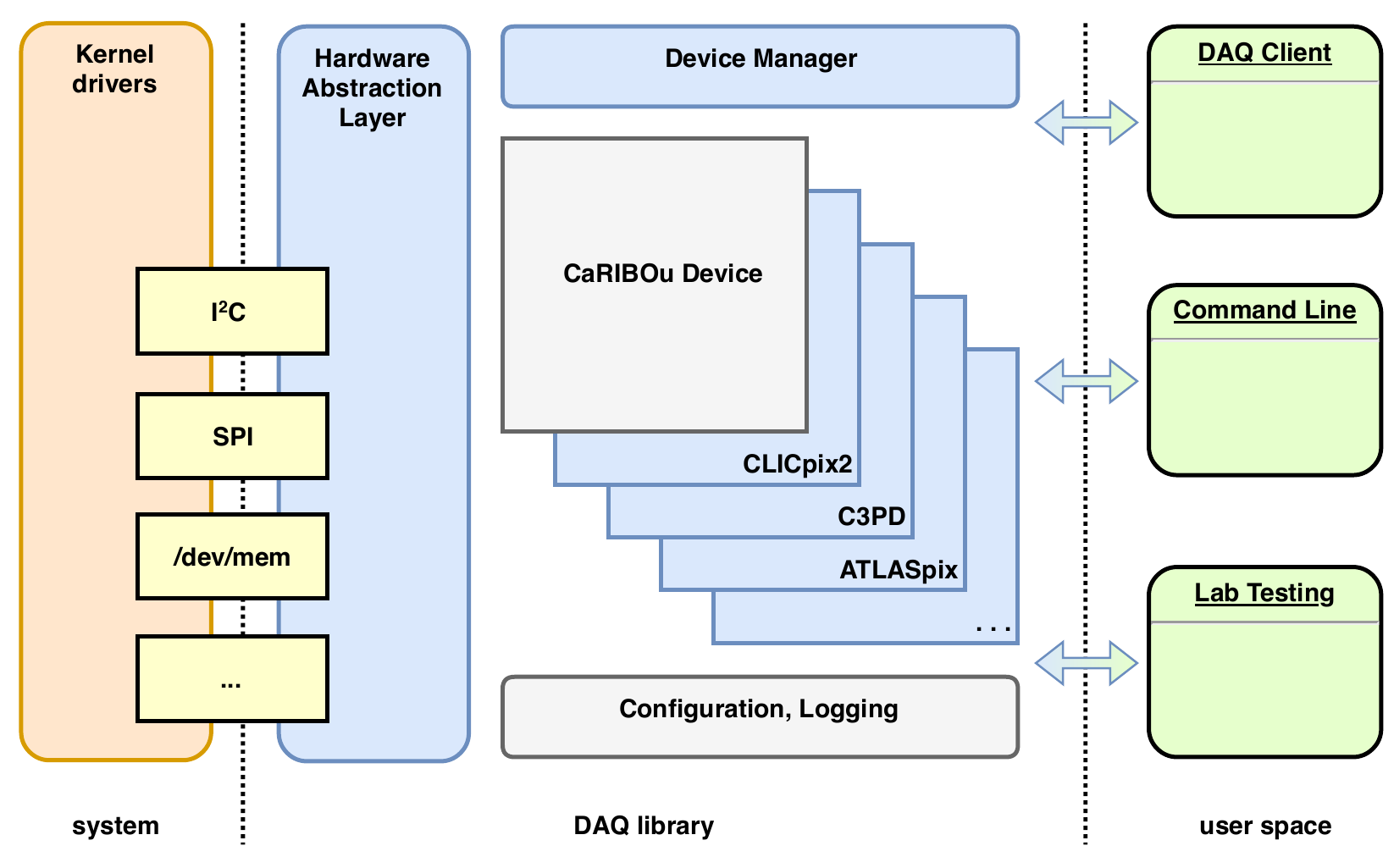}
    \caption{Schematic representation of the Peary software architecture, comprising kernel modules, a hardware abstraction layer and individual detector modules which are represented through a generic API.}
  \label{fig:caribou-software-schematic}
\end{figure}

A flexible data acquisition software called \emph{Peary} allows the user to access the attached detectors and to control the different periphery components of the systems, such as voltage regulators or ADCs.
The different components of the software are shown in \cref{fig:caribou-software-schematic}.
Kernel modules are provided to access different hardware blocks as well as registers in the FPGA via a memory-mapped device.
The hardware abstraction layer provides convenient functions to control different parts of the readout system, such as voltage regulators, while shielding much of the complexity of enabling and configuring the respective I$^2$C devices from the user.
The individual detectors are supported via \emph{CaRIBOu devices}, which register required periphery components, implement a common set of functions and can export additional individual functions for control and readout to the environment.
The interaction with the user is then established through a common application programming interface, which is independent of the detector controlled.
This unified communication between end-user interfaces such as a command-line tool, scripts, or a central data acquisition control software, and the actual hardware attached to the system significantly reduces the effort required to implement support for new devices.

A photograph of the individual hardware components from a CLICpix2 laboratory measurement setup is shown in \cref{fig:components}.
Currently, CaRIBOu supports the devices CLICpix2, C3PD, FEI4, H35Demo and MuPix/ATLASpix, several of which will be discussed in more detail in the following sections. Various new devices, such as CLICTD or CLIPS, are currently being integrated into the system.

\begin{figure}[ht]
  \centering

      \begin{tikzpicture}
        \node[anchor=south west,inner sep=0] at (0,0)(image){  \includegraphics[width=.8\linewidth]{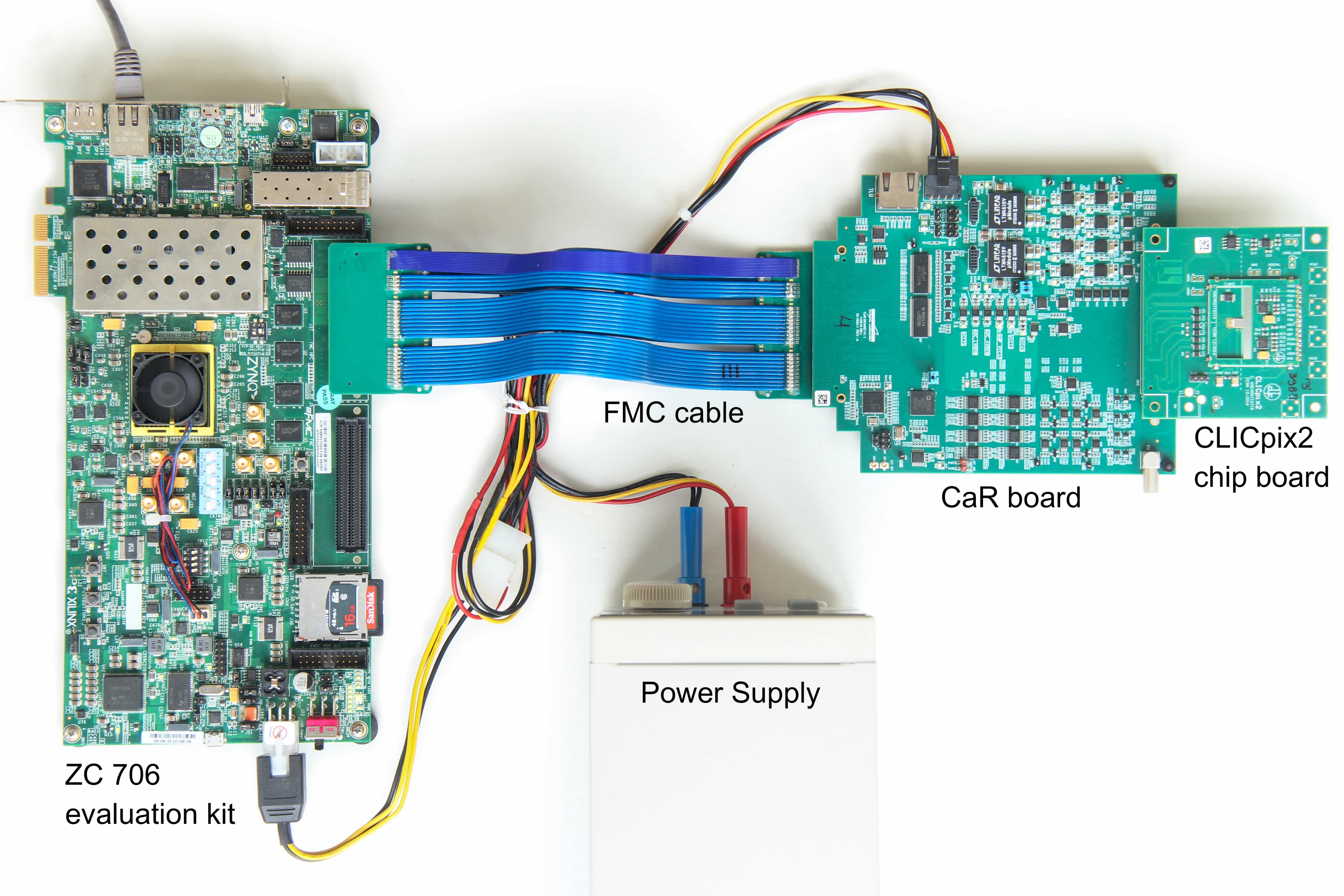}};
    \begin{scope}[x={(image.south east)},y={(image.north west)}]
      \draw[<->,black,thick](0.64,0.4)--(0.875,0.4) node[below,pos=0.5]{\footnotesize \SI{13}{\cm}};
    \end{scope}
    \end{tikzpicture}
    \caption{CaRIBOu setup for laboratory measurements.}
  \label{fig:components}
\end{figure}

\subsection{Beam telescope for test-beam measurements}
\label{sec:timepix3_telescope}
Efficient detector R\&D with particle beams requires the use of a high-performance reference tracking system. To this end, a dedicated beam telescope for studies within the CLICdp collaboration has been set up in the H6 beam line at the CERN SPS North Area test-beam facility. The architecture of the system is based on the LHCb Timepix3 telescope~\cite{lhcb-timepix3-telescope}.

The telescope consists of seven planes of Timepix3 detector assemblies for reference tracking mounted inside a light-tight enclosure, as shown in \cref{fig:timepix3-telescope}. \SI{300}{\micron} thick p-on-n planar sensors are bump-bonded to the Timepix3 readout ASICs, which are described in more detail in \cref{sec:timepix3}. The planes are slightly rotated with respect to the beam axis in order to increase the average cluster size and to optimise the spatial resolution. The setup is usually operated in a \SI{120}{\GeV} \PGpp{} beam with parallel tracks and a narrow beam profile (few mm RMS in both transverse directions), resulting in a track extrapolation accuracy of approximately \SI{2}{\micron} on the device under test (DUT) in the centre of the telescope~\cite{ThesisNilou}. Exploiting the precise hit-time measurement of Timepix3, a track-impact time resolution of about \SI{1}{\ns} is achieved~\cite{calib_timing_tpx3}. In typical SPS data-taking conditions, the telescope system is capable of recording approximately \SI{2e5} particle tracks per second without efficiency loss. This rate is limited by the number of links used for the off-chip data transfer and by high local occupancies due to the narrow beams and the bunched SPS extraction scheme. As an external timing reference and for DUTs requiring an external trigger, a scintillator-based trigger system, consisting of three scintillators read out by photo-multiplier tubes (PMTs) and connected in coincidence, is available.

The device under test is mounted on an x/y linear movement and rotation stage, allowing for automatic position and angular scans. To facilitate parasitic operation in parallel to other users in the same beam line, the telescope box can be moved horizontally and vertically by several \si{cm} with a remote-controlled motion stage.

The data acquisition from the telescope planes is based on the SPIDR (Speedy PIxel Detector Readout) readout system~\cite{SPIDR}. Two detector planes are read out by one commercial FPGA development board (Xilinx VC707) hosting a Virtex 7 FPGA. Data frames received from the Timepix3 detectors are encapsulated in User Datagram Protocol (UDP) frames and sent via an optical 10~Gigabit link to the readout PC for storage.

The data streams from the telescope planes and DUTs can be synchronised during the reconstruction by matching telescope time stamps with DUT hit time stamps obtained from a common reference clock. In this mode, the Timepix3 ASICs and DUTs receive a common \SI{40}{\mega\hertz} clock and a synchronous reset signal at the beginning of each run.

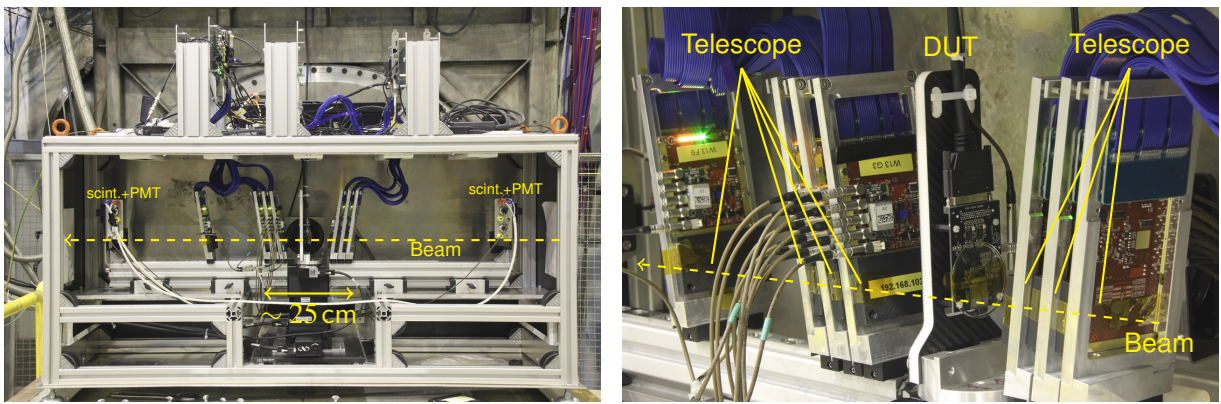
\begin{figure}[ht]
  \begin{subfigure}[T]{0.49\textwidth}
    \begin{tikzpicture}
    \node[anchor=south west,inner sep=0] at (0,0)(image){	\includegraphics[width=\linewidth]{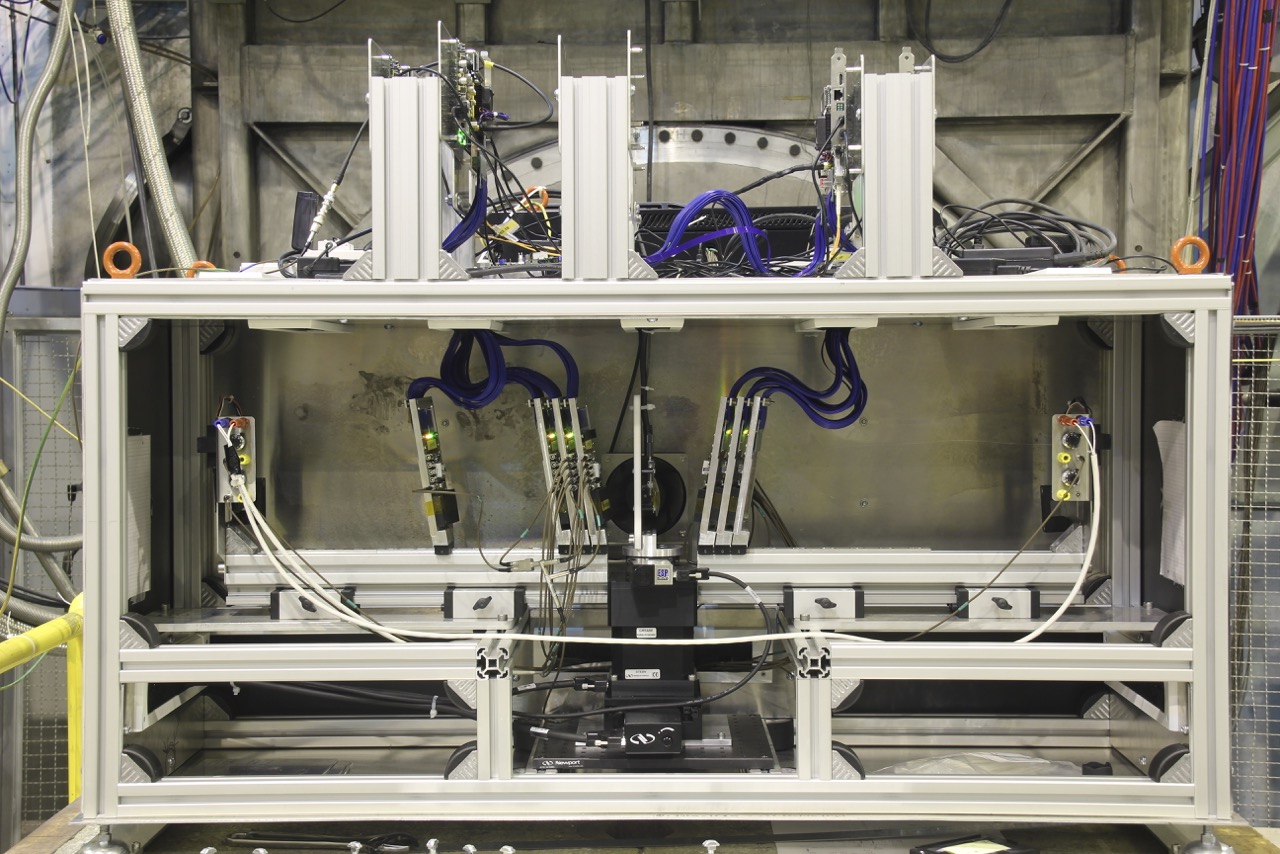}};
    \begin{scope}[x={(image.south east)},y={(image.north west)}]
      \draw[<->,yellow,thick](0.435,0.275)--+(0.15,0) node[below,pos=0.5]{$\sim\SI{25}{\cm}$};
      \draw[->,yellow,thick, dashed](0.93, 0.41)--(0.1, 0.41) node [pos=0.25,below,yshift=1mm]{\scriptsize Beam};
      \node[yellow] at (0.2,0.53)(s1){\tiny scint.+PMT};
      \node[yellow] at (0.84,0.54)(s2){\tiny scint.+PMT};
    \end{scope}
    \end{tikzpicture}
  \end{subfigure}
  ~
  \begin{subfigure}[T]{0.49\textwidth}
    \begin{tikzpicture}
    \node[anchor=south west,inner sep=0] at (0,0)(image){	\includegraphics[width=\linewidth]{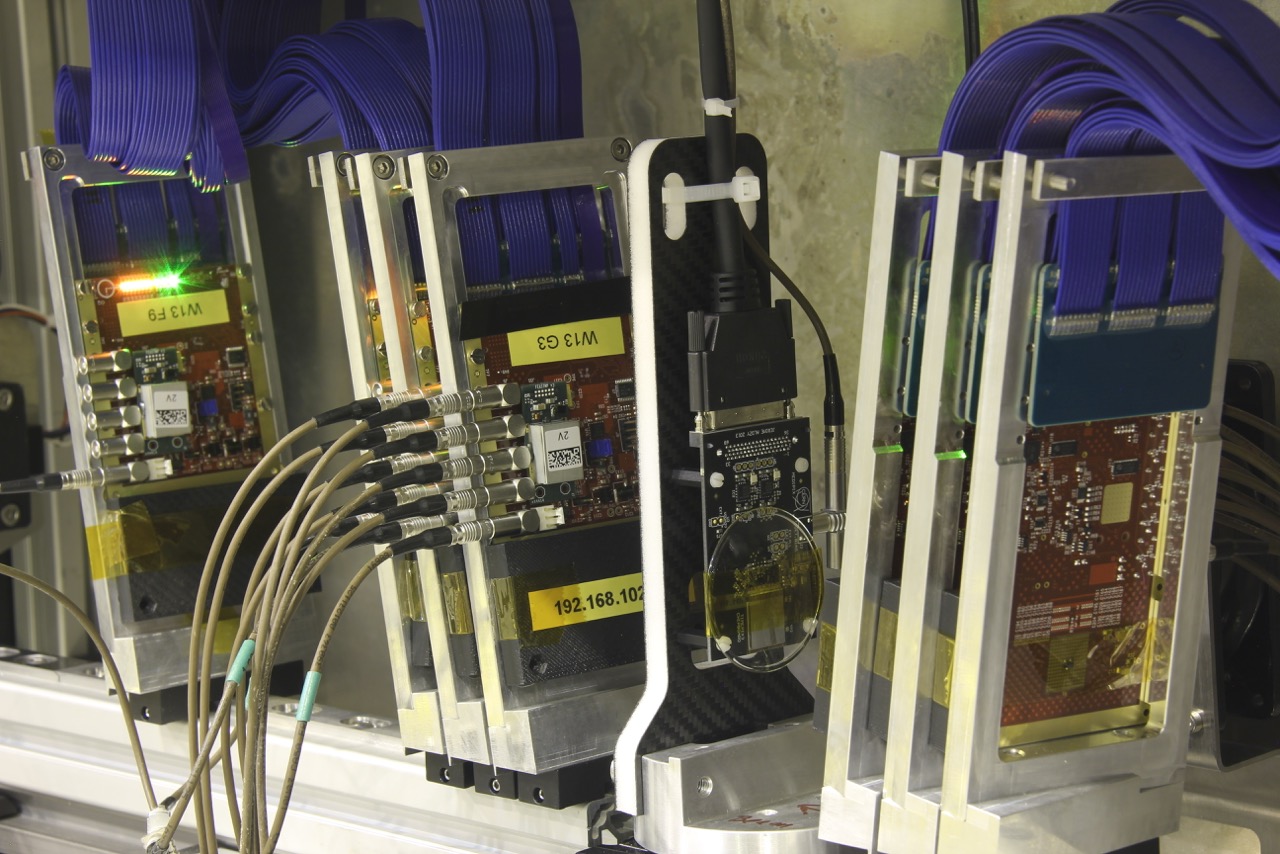}};
    \begin{scope}[x={(image.south east)},y={(image.north west)}]

    \draw[->, thick, dashed, yellow](0.9,0.2)--(0.025,0.35) node [pos=0,below]{Beam};
    \node[yellow] at (0.2,0.9)(t1){Telescope};
    \node[yellow] at (0.55,0.9)(dut){DUT};
    \node[yellow] at (0.85,0.9)(t2){Telescope};

      \draw[yellow,thick] (0.15,0.35)--(t1.south);
    \draw[yellow,thick] (0.3,0.35)--(t1.south);
      \draw[yellow,thick] (0.35,0.325)--(t1.south);
    \draw[yellow,thick] (0.4,0.3)--(t1.south);

    \draw[yellow,thick] (0.675,0.3)--(t2.south);
      \draw[yellow,thick] (0.725,0.275)--(t2.south);
    \draw[yellow,thick] (0.8,0.25)--(t2.south);
    \end{scope}
    \end{tikzpicture}
  \end{subfigure}
  \caption{The CLICdp Timepix3 telescope setup operated in the SPS H6 beam line at CERN. Seven planes of Timepix3 detectors, a device under test and a scintillator+PMT trigger setup are mounted inside a light-tight enclosure. Four SPIDR FPGA readout boards are located on top of the enclosure.}\label{fig:timepix3-telescope}
\end{figure}

\subsection{The Corryvreckan reconstruction and analysis framework}
\label{sec:corryvreckan}

The reconstruction and analysis of test-beam data from novel detector prototypes and reference telescopes requires a lightweight and flexible software framework which provides a balance between adaptability and ease of use. Especially, the tracking and alignment algorithms have to be able to adapt to different experimental conditions.
The \emph{EUTelescope} software framework~\cite{eutelescope,eutelescope-conf} has been used for many of the earlier test-beam data analyses described in the following sections. EUTelescope is based on a trigger-based event model, which relies on receiving one event per trigger per detector. This event model has been adapted for the CLIC vertex and tracker test-beam analyses, in order to combine devices with different frame-based and data-driven readout schemes.
Most of the recent studies use the new \emph{Corryvreckan}\footnote{Corryvreckan was named after a maelstrom in the Inner Hebrides off the west coast of mainland Scotland.} software framework~\cite{corryvreckan-gitlab}, which was conceived as a flexible, easy-to-use and to maintain reconstruction and analysis toolkit. Built on a time-based event model, it supports a straightforward combination of devices with different frame-based and data-driven readout schemes.

Corryvreckan is designed in a modular way, comprising a central component which deals with parsing of configuration files, the central event loop and coordinate transformations, and individual modules which implement the individual steps of the reconstruction process.
Several so-called \emph{event loader} modules allow the reading and decoding of data from a variety of devices such as ATLASpix, CLICpix, CLICpix2, Timepix and Timepix3, which will be described in detail in the following sections.
In addition, data files recorded with the EUDAQ framework~\cite{eudaq} can be processed directly.
The reconstruction and analysis workflow is configured via two configuration files; one describing the detector setup including the position and orientation of all detectors as well as their role (i.e.\ reference detector or device under test); the other describing the modules to be executed together with their parameters.

One important feature of the framework is its capability of using timing information from individual detectors for reconstruction.
Among the available modules are clustering and tracking algorithms which act on four coordinates, taking into account timing information available e.g.\ for the detectors of the reference beam telescope described in \cref{sec:timepix3_telescope}.
The untriggered data stream is read in and split into chunks in time for processing.
The clustering algorithm then searches for pixel hits which are close to each other both in space and time.
This allows a drastic reduction in the combinatorics and allows the recovery of individual clusters even in high-rate environments.
Tracks are then built from individual clusters using the available time information in a similar fashion.

The modular structure and the flexibility of the event definition also allows for the combination of data from detectors with different readout architectures.
One example is the correlation of a frame-based device, where pixel hits are only recorded during a certain time frame and read out afterwards, and a data-driven device where hits are read out and stored as they are detected.
In this case, Corryvreckan uses the start and stop time signals of the frame-based device as delimiters, and only pixel hits from the data-driven detectors which fall into this frame are accepted and used for the reconstruction.
This allows Corryvreckan to correctly compute the efficiency of the frame-based device despite its deadtime outside the acquisition frame.

By reducing the dependency on external data handling frameworks, the framework is very lightweight and fast.
For a typical dataset recorded by the CLICdp Timepix3 telescope and a device under test in the CERN SPS beam, the framework is capable of reconstructing several thousand tracks per second on a standard desktop PC.
During test-beam campaigns, this speed can be exploited by using the framework for online monitoring purposes.
The \emph{OnlineMonitor} module provides an interactive graphical user interface which presents a configurable set of plots taken from any module in the reconstruction chain.
These plots are updated continuously during the run and allow the user to directly gauge the performance of all involved detectors not only based on hit maps or correlation plots, but also directly from tracking results.
It is therefore capable of providing the figures of merit for the detector performance, such as efficiency or resolution, already during data taking and thus allows for the detection of possible problems, such as a misconfiguration of the device under test, immediately.

The Corryvreckan framework features a comprehensive and continuously updated user manual~\cite{corryvreckan-manual}.

\section{Hybrid readout ASICs}\label{sec:vtx-trk-hybrid-asics}
This section introduces the hybrid pixel readout ASICs used in the context of the CLIC vertex-detector R\&D. Initial studies have focussed on assemblies of existing ASICs (Timepix and Timepix3) bump bonded to thin planar sensors. The CLIC vertex-detector requirements are specifically targeted by the CLICpix and CLICpix2 ASICs.

\subsection{Timepix}
Timepix~\cite{Timepix_paper} is a pixelated readout ASIC designed by the Medipix2 collaboration~\cite{medipix2}. It is implemented in a 250~nm CMOS process with a matrix size of \SI{256x256}{pixels} at a pixel pitch of \SI{55x55}{\micron}. The readout is based on a global shutter signal. The pixel matrix is sensitive as long as the shutter signal is active, and the full matrix is read out after closure of the shutter.

The analogue pixel front-end is based on the Krummenacher architecture~\cite{KRUMMENACHER1991527} for signal amplification and shaping, followed by a discriminating stage, using a threshold voltage with \SI{4}{bit} local adjustment for hit detection. Each pixel incorporates a \SI{14}{bit} counter that can be operated in one of three modes.  The Time-over-Threshold (ToT) mode is used for hit energy measurement. The counter is incremented as long as the discriminator output surpasses the threshold. The Time-of-Arrival (ToA) mode is used for hit time determination. The counter is incremented starting from the time when a particle hit is detected until the shutter is closed. In the hit counting mode the counter is incremented each time the discriminator output surpasses the threshold.

For the Timepix planar-sensor studies presented in \cref{sec:vtx-trk-hybrid-assemblies}, the FITPix USB readout system~\cite{fitpix} was used to provide low voltage to the Timepix ASIC and for control and data readout.

\subsection{Timepix3}\label{sec:timepix3}
The Timepix3 hybrid readout ASIC is implemented in a \SI{130}{\nm} CMOS process. It builds on the Timepix experience and includes advanced features, such as simultaneous time-of-arrival and time-over-threshold measurements with high precision~\cite{timepix3_paper}. The matrix size of \SI{256x256}{pixels} and the pixel area of \SI{55x55}{\micron} are identical to Timepix, such that the same sensor types can be used for both ASICs. Timepix3 is used in a wide range of particle tracking, imaging and dosimetry applications.
The analogue front-end contains a preamplifier with Krummenacher leakage current compensation feedback circuitry, a 4-bit digital-to-analogue converter (DAC) for local threshold tuning and a discriminator.
In addition to the simultaneous ToT/ToA readout, the timing precision has been improved with respect to Timepix by a factor of 6, and a zero-suppressed data-driven readout scheme is implemented to reduce dead times at low occupancy. Time-over-threshold is measured with \SI{10}{bit} precision. The hit arrival time is obtained with a step size of \SI{1.5625}{\ns} and a dynamic range of \SI{18}{bits}, using a combination of a global \SI{40}{\mega\hertz} clock and local \SI{640}{\mega\hertz} oscillators.
Power-pulsing features are included in the ASIC, allowing for switching dynamically between nominal power and shutdown modes in the analogue domain, and for gating the system clock and the clock of the pixel matrix. Dedicated power-pulsing tests with Timepix3 assemblies are described in \cref{sec:timepix3-pp}.

This feature set, especially the accurate timing and energy resolution, make Timepix3 a suitable test vehicle for investigating pixelated silicon sensors for the CLIC vertex and tracking detectors. Planar silicon pixel sensors of various thickness and with active-edge processing have been studied using Timepix3, as described in more detail in \cref{sec:thin_planar_sensors,sec:active_edge_sensors}. A dedicated beam telescope based on Timepix3 has been built and is operated at the CERN SPS, as described in \cref{sec:timepix3_telescope}.

For data acquisition and readout, the SPIDR readout system~\cite{SPIDR} is used. The SPIDR system was designed as a general readout platform for pixelated readout ASICs, such as Timepix3. Using a \SI{10}{gigabit} ethernet connection, Timepix3 ASICs can be read out at their maximal data rate of \SI{80}{\mega hits\per\second} per ASIC. The system is based on VC707 development boards from Xilinx. The Timepix3 ASIC is mounted on a separate chip carrier board, which is connected to the system on the FMC port. One FPGA development board is capable of reading two Timepix3 ASICs simultaneously. Fast inputs for clock signals and synchronisation signals for running multiple systems via a common timing control unit are available and have been used in the telescope setup described in \cref{sec:timepix3_telescope}. A separate TDC channel can be used to provide reference time stamps for applications that require a trigger.

\subsection{CLICpix}\label{sec:CLICpix}
The CLICpix hybrid readout ASIC is a technology demonstrator and targets the CLIC vertex-detector requirements. It has been designed and fabricated in a \SI{65}{\nm} CMOS process~\cite{CLICpixValerio,Valerio:1610583}. \cref{tab:clicpix_specifications} summarises the most important ASIC specifications. The pixel matrix consists of \num{64x64} pixels at \SI{25x25}{\micron} pixel size. The main features include simultaneous 4-bit measurements of Time-over-Threshold and Time-of-Arrival with \SI{10}{\ns} accuracy, on-chip data compression and power pulsing capability.

\begin{table}
  \centering
  \caption{CLICpix and CLICpix2 design features}\label{tab:clicpix_specifications}
  \begin{tabular}{l c c}
    \toprule
    & CLICpix & CLICpix2\\ \midrule
ASIC size & \SI{1.85x3}{\mm} & \SI{3.35x4.06}{\mm} \\
Active area & \SI{1.6x1.6}{\mm}  & \SI{3.2x3.2}{\mm} \\
Matrix size & \num{64x64} pixels & \num{128x128} pixels \\
CMOS technology & \SI{65}{\nm} & \SI{65}{\nm} \\
Pixel pitch & \SI{25x25}{\micron} & \SI{25x25}{\micron} \\
ToT counter depth & 4 bit & 5 bit \\
ToA counter depth & 4 bit & 8 bit \\
ToA bin size & \SI{10}{\ns} & \SI{10}{\ns}\\
ENC (w/o) sensor & $\sim\SI{55}{\Pem{}}$ & $\sim\SI{67}{\Pem{}}$ \\
Minimum threshold (6$\sigma$ margin) & 550 \Pem{} & 440 \Pem{} \\
Acquisition mode & frame-based & frame-based\\
Readout mode & single column readout & 1/2/4/8 parallel column readout\\
Data encoding & - & 8\,bit\,/\,10\,bit\\
Readout system & $\upmu$ASIC & CaRiBOu\\
Voltage reference & external & bandgap\\
Testpulse & external & internal\\Analogue power diss./pixel & \SI{7}{\micro\watt} & \SI{6.6}{\micro\watt}\\
Amplifier gain & \SI{44}{\milli\volt\per\kilo\Pem{}} & \SI{33}{\milli\volt\per\kilo\Pem{}}\\
Slow control & custom & SPI\\
Clock speed & \multicolumn{2}{c}{\SI{100}{\mega\hertz} (acquisition) and \SI{320}{\mega\hertz} (readout)}\\
Data type & \multicolumn{2}{c}{Zero compression (pixel, super-pixel and column skipping)} \\
Power saving & \multicolumn{2}{c}{Clock gating (digital part), power gating (analogue part)} \\   
\bottomrule
  \end{tabular}
\end{table}

The ASIC front-end is sketched in \cref{fig:clicpix_frontend_sketch}. Current pulses coming from the sensor or from a test-pulse capacitor are amplified and shaped by the preamplifier and Krummenacher feedback network and compared to a global threshold. This threshold is locally adjusted with a 4-bit DAC to compensate for pixel-to-pixel threshold mismatch. The result of the comparison is used in the pixel logic as an enable signal for the counting clocks of both the ToT and ToA counters. Local state machines are implemented in order to decide when to stop counting: for the ToA counter, it is tied to a global shutter signal, which is synchronously distributed to the whole pixel matrix and used as a timing reference. For the ToT measurement, the counting will stop as soon as the discriminator signal goes below the threshold.

\begin{figure}[t]
  \centering
  \includegraphics[width=.8\linewidth]{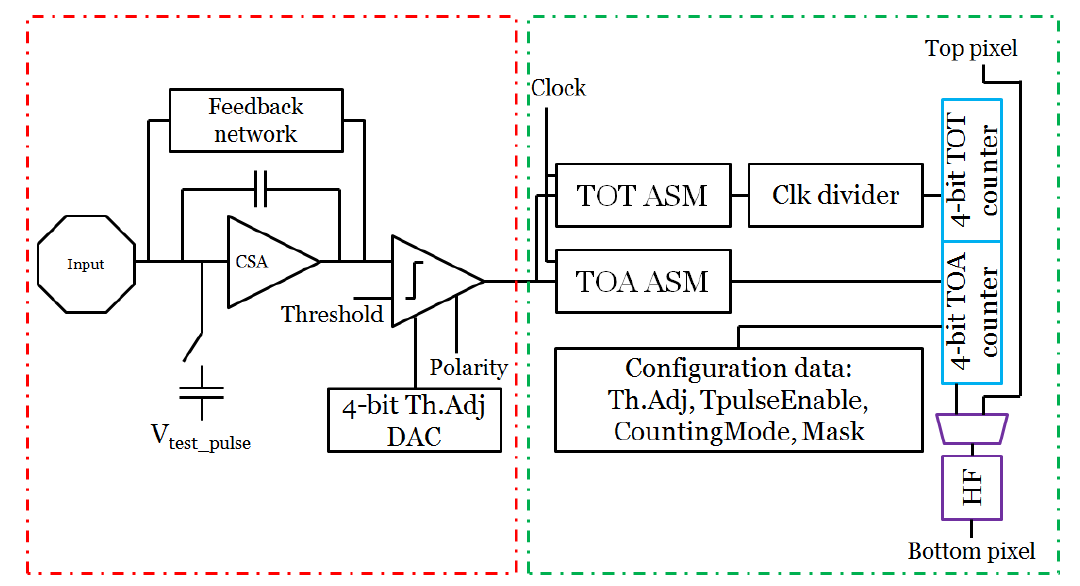}
  \caption{Block diagram of the pixel circuits in CLICpix~\cite{CLICpixValerio}. The left part is the analogue front-end, with the preamplifier, the discriminator and the threshold equalisation DAC. The right part shows the digital front-end, with Asynchronous State Machines (ASM) generating enabling signals for the ToT and ToA counters and configuration latches. }\label{fig:clicpix_frontend_sketch}
\end{figure}

The CLICpix ASIC is read out by the $\upmu$ASIC custom readout system, based on a SPARTAN6 FPGA board and a modular interface card~\cite{uasic}. Bump-bonded and capacitively coupled detector assemblies have been produced and successfully tested, as described in \cref{sec:finepitch_bump_bonding,sec:thin_planar_sensors,sec:capacitively_coupled_hvcmos}. A bare ASIC mounted on a readout board is depicted in  \cref{fig:clicpix_photo}.

\subsection{CLICpix2}\label{sec:CLICpix2}
To overcome several limitations observed with the first generation CLICpix ASIC, a second iteration (CLICpix2~\cite{clicpix2-manual}) was designed in the same \SI{65}{\nm} CMOS process as CLICpix. \cref{fig:clicpix2} shows pictures of CLICpix and CLICpix2 at a similar scale, mounted and wire-bonded on the respective readout boards. For CLICpix2 the pixel matrix area has been increased by a factor of four to \num{128x128} pixels, and the counter depths have been increased to 8 bit for the ToA and 5 bit for the ToT counters. Both are beneficial for better testability in test-beam environments. Further improvements include a better noise isolation and the removal of a cross-talk issue observed in CLICpix, a more sophisticated input-output protocol with parallel column readout and 8\,bit\,/\,10\,bit encoding, a band gap voltage reference and test pulse DACs inside the ASIC. The readout of CLICpix2 has been implemented in the CaRIBOu readout system~\cite{caribou_fiergolski}, which is described in \cref{sec:caribou}.
A second iteration (CLICpix2~\cite{clicpix2-manual}) was designed in the same \SI{65}{\nm} CMOS process as CLICpix. \cref{fig:clicpix2} shows pictures of CLICpix and CLICpix2 at a similar scale, mounted and wire-bonded on the respective readout boards. For CLICpix2 the pixel matrix area has been increased by a factor of four to \num{128x128} pixels, and the counter depths have been increased to 8 bit for the ToA and 5 bit for the ToT counters. Both are beneficial for better testability in test-beam environments. Further improvements include a better noise isolation and the removal of a cross-talk issue observed in CLICpix, a more sophisticated input-output protocol with parallel column readout and 8\,bit\,/\,10\,bit encoding, a band gap voltage reference and test pulse DACs inside the ASIC. The readout of CLICpix2 has been implemented in the CaRIBOu readout system~\cite{caribou_fiergolski}, which is described in \cref{sec:caribou}.

\begin{figure}[ht]
  \begin{subfigure}[T]{.49\linewidth}
    	\includegraphics[width=\linewidth]{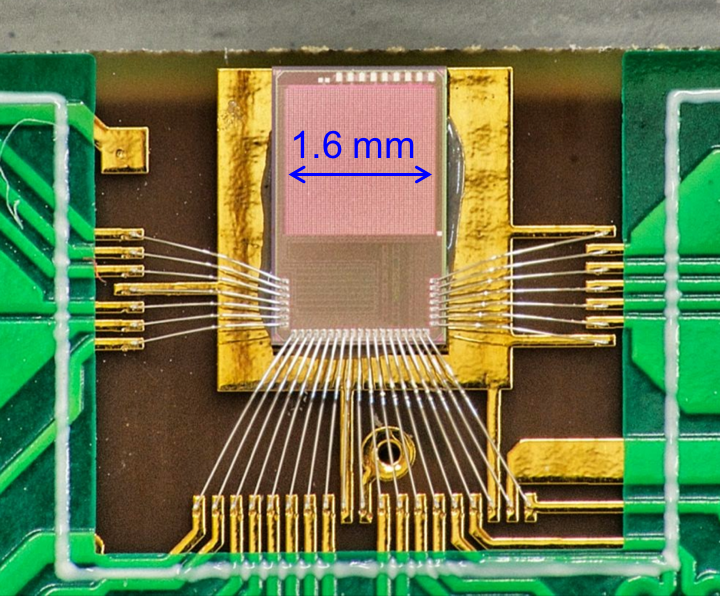}
      \caption{}\label{fig:clicpix_photo}
  \end{subfigure}
  \hfill
  \begin{subfigure}[T]{.49\linewidth}
  	\begin{tikzpicture}
	 			\node[anchor=south west,inner sep=0] (image) at (0,0) { 	\includegraphics[width=\linewidth,clip,trim=11cm 9.35cm 10.5cm 4.1cm]{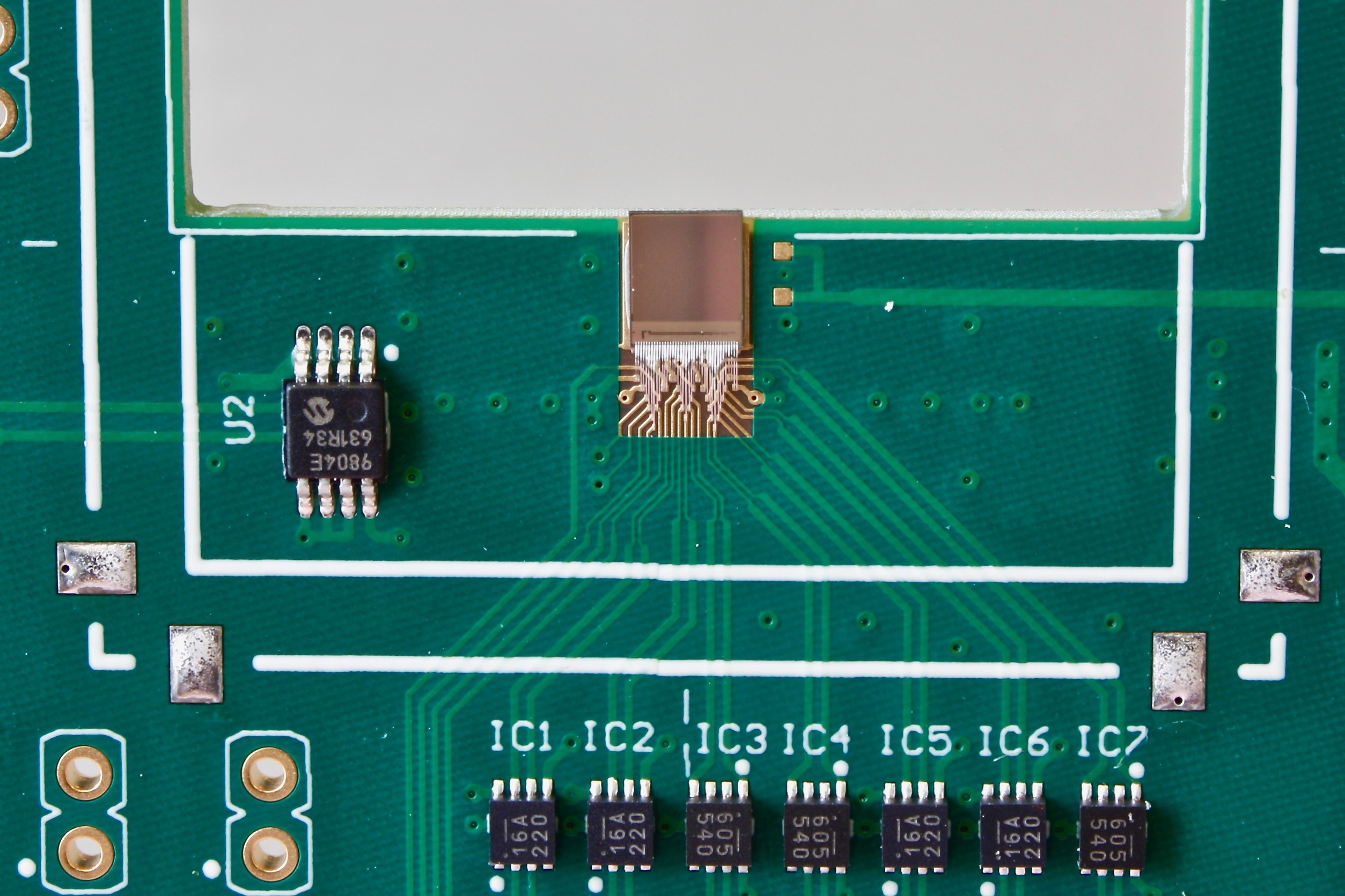}};
	 		\begin{scope}[x={(image.south east)},y={(image.north west)}]

          \draw[<->, ultra thick, blue](0.3,0.8) -- (0.71,0.8) node[pos=0.5, above]{\large\textsf{\SI{3.2}{\mm}}};
          \draw[<->, ultra thick, blue](0.225,0.98) -- (0.225,0.35) node[pos=0.5, rotate=90, above]{\large\textsf{\SI{4}{\mm}}};
	 			\end{scope}
	 	\end{tikzpicture}
    \caption{}\label{fig:clicpix2_photo}
  \end{subfigure}
  \caption{\subref{fig:clicpix_photo} CLICpix and \subref{fig:clicpix2_photo} CLICpix2 ASICs glued and wire-bonded to readout PCBs. Both pictures have a similar scale of approximately 10:1.}\label{fig:clicpix2}
\end{figure}

\section{Hybrid passive sensor assemblies}\label{sec:vtx-trk-hybrid-assemblies}
Hybrid passive pixel detectors are widely used for vertexing in modern high-energy collider experiments. A planar pixelated p-n junction on a high-resistivity sensor wafer acts as the sensitive material, while highly customised front-end ASICs perform the charge amplification, digitisation and signal transmission to the back-end electronics. The sensor electrodes are connected to the amplification stages in the ASIC through a grid of solder bump bonds. The separation of sensor and ASIC in two silicon dies allows the best suited technology for both components to be chosen. However, the need for very small pixels makes the bump-bonding interconnect process difficult and costly. It is therefore one of the limiting factors for the pixel pitch. Novel sensor-fabrication techniques like active-edge (\cref{sec:active_edge_sensors}) and Enhanced Lateral Drift sensors (\cref{sec:ELAD}) aim at further optimising charge collection, resolution and material budget.

\subsection{Fine-pitch bump bonding}\label{sec:finepitch_bump_bonding}
A variety of fine-pitch interconnect processes have been explored to produce hybrid assemblies for the thin-sensor studies presented in \cref{sec:thin_planar_sensors} and \cref{sec:active_edge_sensors}.
Initial assemblies with \SI{55}{\micron}-pitch Timepix and Timepix3 ASICs and thin slim-edge and active-edge sensors were made using established full-wafer lithographic under-bump metalisation (UBM) and bump deposition processes, followed by die-to-die flip-chip and reflow.
Assemblies with \SI{25}{\micron}-pitch CLICpix and CLICpix2 dies from Multi-Project-Wafer (MPW) productions required the development of dedicated carrier-wafer single-chip bump-bonding processes.

\subsubsection{Bump bonding of Timepix assemblies} \label{sec:tpx-bump-bonding}
Slim-edge p-in-n and n-in-p sensors with a thickness of \SIrange{100}{300}{\micron} produced at Micron Semiconductor~\cite{Micron} were hybridised at the Fraunhofer Institute for Reliability and Microintegration IZM~\cite{IZM} to Timepix ASICs of \SI{700}{\micron} native thickness and to Timepix ASICs thinned down to \SI{100}{\micron}~\cite{Fritzsch:2014zya}.
The processing steps for the sensors are:
\begin{enumerate}
\item Sputter deposition of the plating base TiW/Cu;
\item Preparation of carrier wafers and bonding on carrier wafers (for $\leq$\SI{200}{\micron} sensors);
\item Deposition and patterning of the resist layer by mask lithography;
\item Electroplating of UBM (Cu), stripping of resist and plating base;
\item Removal from carrier wafer (for $\leq$\SI{200}{\micron} sensors);
\item Cleaning;
\item Dicing (performed at Micron Semiconductor);
\end{enumerate}
The Timepix wafer processing consists of the following steps:
\begin{enumerate}
\item Mechanical grinding of the Timepix wafer to \SI{100}{\micron} (for Timepix wafer thinned to \SI{100}{\micron});
\item Preparation of glass carrier wafers and temporary bonding on glass carrier wafers (for Timepix wafer thinned to \SI{100}{\micron});
\item Sputter deposition of the plating base TiW/Cu, deposition and patterning of the resist layer by mask lithography;
\item Electroplating of UBM and bump metallisation (Cu-SnAg), stripping of resist and plating base;
\item Automated optical bump inspection;
\item Dicing;
\end{enumerate}
The final bonding of sensor and ASIC dies, after cleaning, inspection and sorting, is performed in a pick-and-place process using a flip-chip bonder tool, followed by re-flow soldering.
For the \SI{100}{\micron} Timepix ASICs, a laser debonding of the carrier chips is performed after reflow.
\cref{fig:100-on-100-TPX} shows a photograph of an assembly consisting of a \SI{100}{\micron} Timepix and a \SI{100}{\micron} p-in-n sensor.

Timepix assemblies with \SI{50}{\micron} p-in-n active-edge sensors were produced at Advacam~\cite{Advacam}, using SnPb solder bumps and a similar process flow as for the hybridisation of Timepix3 (see \cref{sec:tpx3-bump-bonding}). A photograph of one of the assemblies is shown in \cref{fig:50umTPX}.

Measurements with radioactive sources, x-ray photons and in test beams (see \cref{sec:thin_planar_sensors}) showed an excellent bond yield for all Timepix assemblies. Typically fewer than 50 isolated unresponsive channels out of 65536 channels were observed, except for a few assemblies with masked columns on the readout ASICs~\cite{timepix_calibration_sophie}.

\begin{figure}[ht]
  \begin{subfigure}[T]{0.51\textwidth}
    \begin{tikzpicture}
      \node[anchor=south west,inner sep=0] (image) at (0,0) {\includegraphics[width=\linewidth]{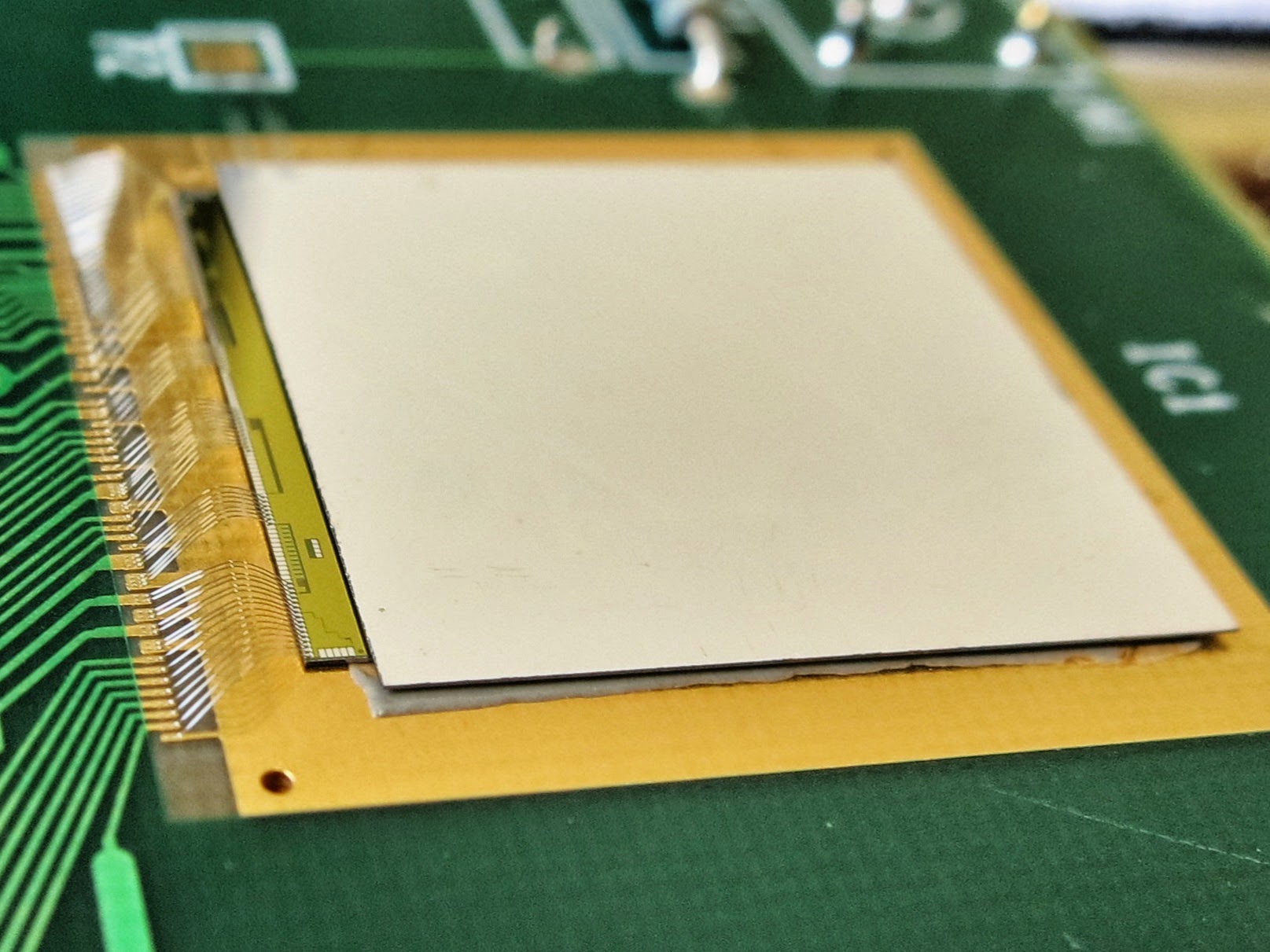}};
      \begin{scope}[x={(image.south east)},y={(image.north west)}]
         \draw[<->, ultra thick, blue](0.32,0.32) -- (0.94,0.375) node[pos=0.5, rotate=5, above]{\large\textsf{\SI{14}{\mm}}};
      \end{scope}
    \end{tikzpicture}
    \caption{}\label{fig:100-on-100-TPX}
  \end{subfigure}
  \hfill
  \begin{subfigure}[T]{0.46\textwidth}
    \begin{tikzpicture}
	 			\node[anchor=south west,inner sep=0] (image) at (0,0) { 	\includegraphics[width=\linewidth]{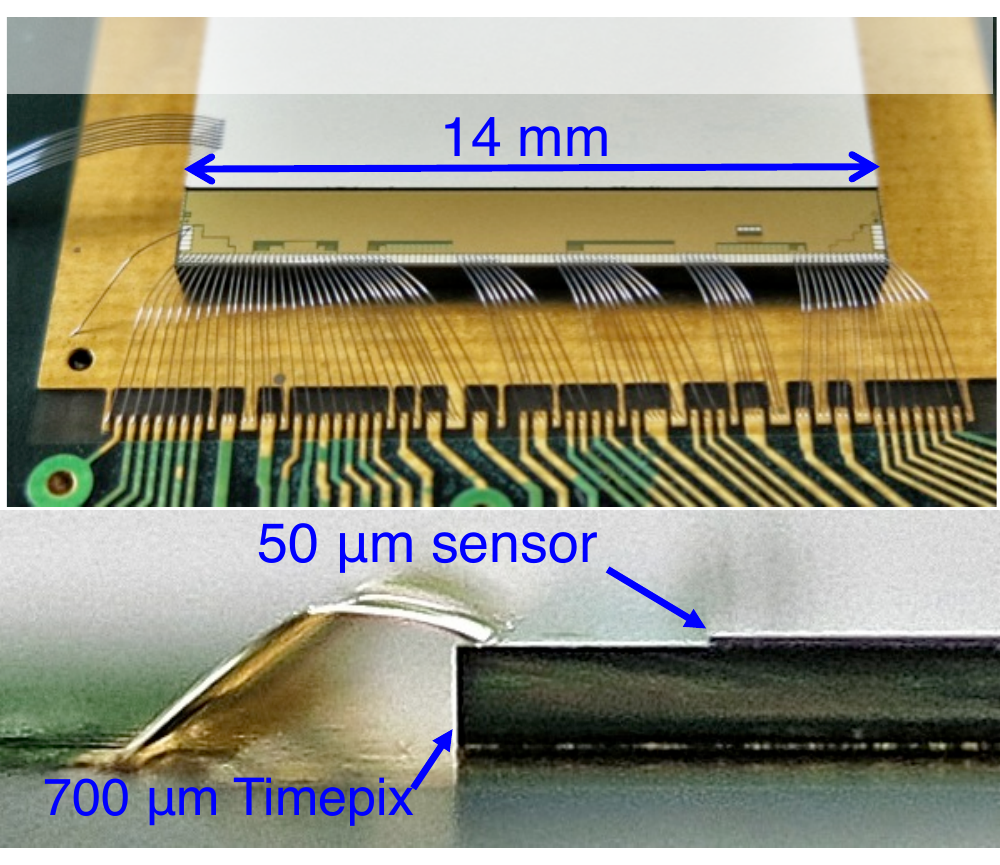}};
	 		\begin{scope}[x={(image.south east)},y={(image.north west)}]
          \fill[white] (0,0.41) rectangle(1,0.45);
	 			\end{scope}
	 	\end{tikzpicture}
    \caption{}\label{fig:50umTPX}
  \end{subfigure}
  \caption{\SI{100}{\micron} thick Micron slim-edge sensor on a \SI{100}{\micron} thick Timepix ASIC, hybridised at \subref{fig:100-on-100-TPX} Fraunhofer IZM, and \subref{fig:50umTPX} \SI{50}{\micron} thick Advacam active-edge sensor on a \SI{700}{\micron} thick Timepix ASIC hybridised at Advacam.}\label{fig:tpx-assemblies}
\end{figure}

\subsubsection{Bump bonding of Timepix3 active-edge sensor assemblies} \label{sec:tpx3-bump-bonding}
Sensors of n-in-p type with activated cut edge and thicknesses from \SIrange{50}{150}{\micron} were produced in a carrier-wafer process and hybridised to Timepix3 ASICs at Advacam~\cite{Advacam}.
The active-edge technology used is based on Deep Reactive Ion Etching (DRIE) of a trench around the active sensor matrix and subsequent implantation of the exposed sensor edges~\cite{Wu_2012}, resulting in singularised sensors without a dedicated dicing step.
The sensor front-side patterning and UBM layer deposition is therefore performed as part of the sensor production, prior to the DRIE processing.
Different UBM deposition processes were explored, including thin-film deposition of Ni/Au or Pt, as well as electroplating of Cu/Au or SnPb.
The singularised sensors are removed from the carrier wafer and the sensor back side is metalised.
The Timepix3 wafer processing consists of UBM and SnPb solder deposition, followed by dicing.
Sensors and ASICs are bonded in a pick-and-place process using a flip-chip bonder tool and subsequent re-flow soldering.
A photograph of one of the resulting assemblies is shown in \cref{fig:timepix3-assembly}.

An excellent pixel response yield was achieved for all tested Timepix3 assemblies, with less than 50 isolated unresponsive or noisy readout-ASIC channels and less than 20 unresponsive or unconnected sensor channels out of 65536 channels per assembly~\cite{timepix3_calibration_flo}.

\begin{figure}[ht]
  \centering
  \begin{tikzpicture}
    \node[anchor=south west,inner sep=0] (image) at (0,0) { \includegraphics[width=.75\linewidth]{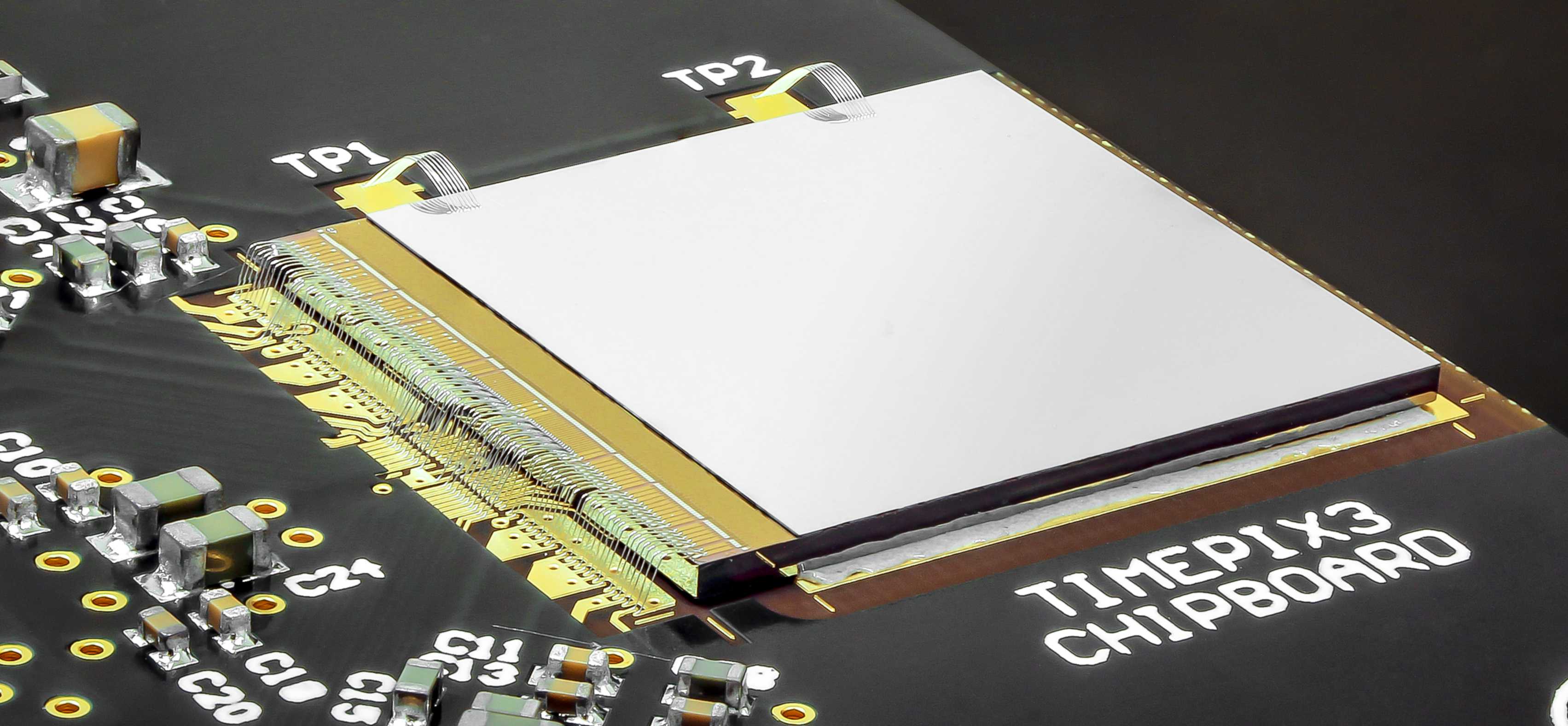}};
      \begin{scope}[x={(image.south east)},y={(image.north west)}]
        \draw[<->, thick, blue](0.49,0.3) -- (0.89,0.53) node[pos=0.5, rotate=17, above]{\large\textsf{\SI{14}{\mm}}};
      \end{scope}
  \end{tikzpicture}
  \caption{Assembly of a \SI{50}{\micron} thin active-edge sensor bump-bonded to a Timepix3 ASIC, wire-bonded to a readout PCB.}\label{fig:timepix3-assembly}
\end{figure}

\subsubsection{Bump bonding of CLICpix assemblies} \label{sec:clicpix-bump-bonding}
Commercially available bump-bonding processes usually require the processing of full wafers for the UBM and solder deposition.
The \SI{25}{\micron}-pitch CLICpix dies were however received from a Multi-Project-Wafer (MPW) production, without access to the ASIC wafers.
A dedicated single-chip Indium bump-bonding process was therefore developed at SLAC, based on carrier-wafers for patterning and metal deposition steps on the sensors and ASICs~\cite{SLAC_bumpbonding}.
Slim-edge n-in-p sensors from Micron with \SI{200}{\micron} thickness and active edge n-in-p sensors from Advacam with \SI{50}{\micron} and \SI{150}{\micron} thickness were used for the test assemblies. All sensors were provided without wafer-level UBM.
The carrier-wafer processing for both the sensors and the ASICs consists of the following steps:
\begin{enumerate}
\item Align sensors / ASICs on carrier wafer;
\item Spin photoresist;
\item Expose with contact aligner;
\item Deposit indium layer in evaporator;
\item Lift-off indium, resulting in \SI{4}{\micron} high bumps.
\end{enumerate}
Example pictures of the resulting UBM and indium bumps on a CLICpix ASIC and a Micron sensor are shown in \cref{fig:ubm}, demonstrating a good uniformity of the bumps.

\begin{figure}[ht]
		\begin{subfigure}[T]{0.49\textwidth}
			 \includegraphics[width=\linewidth]{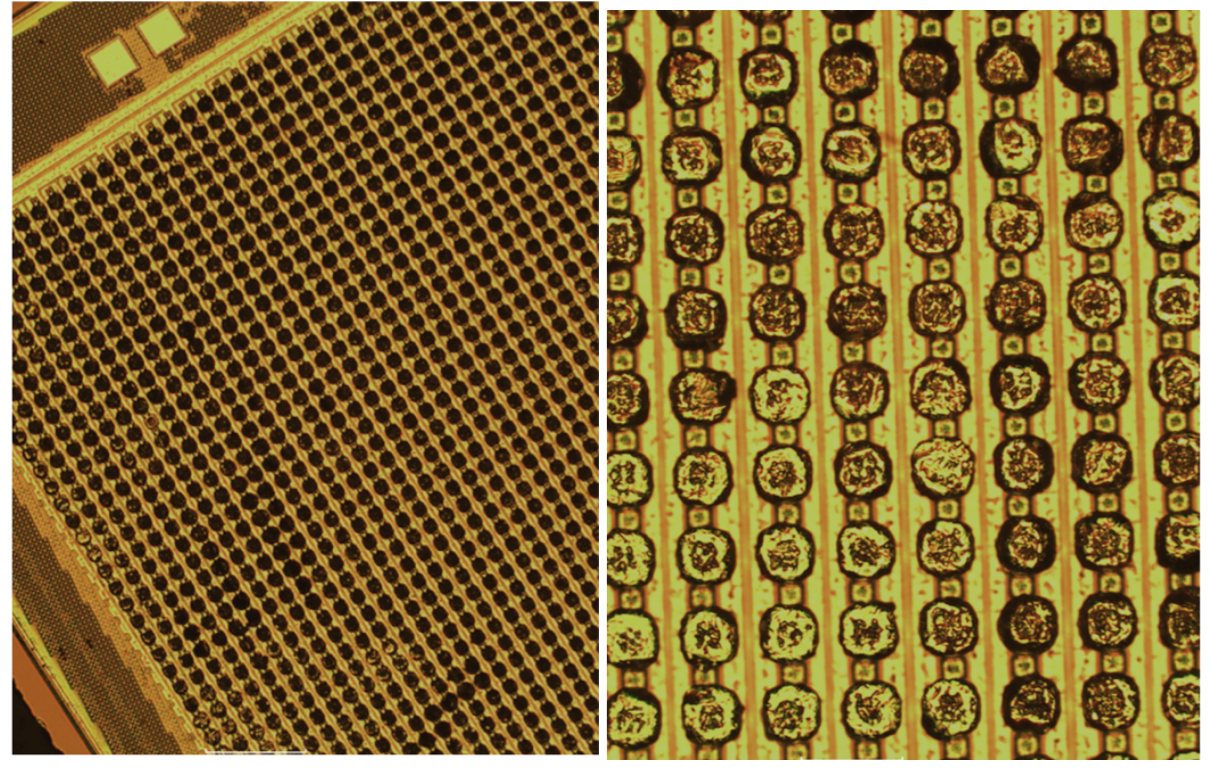}
			 \caption{}\label{fig:ubm_clicpix}
	\end{subfigure}%
		\begin{subfigure}[T]{0.49\textwidth}
			 \includegraphics[width=\linewidth]{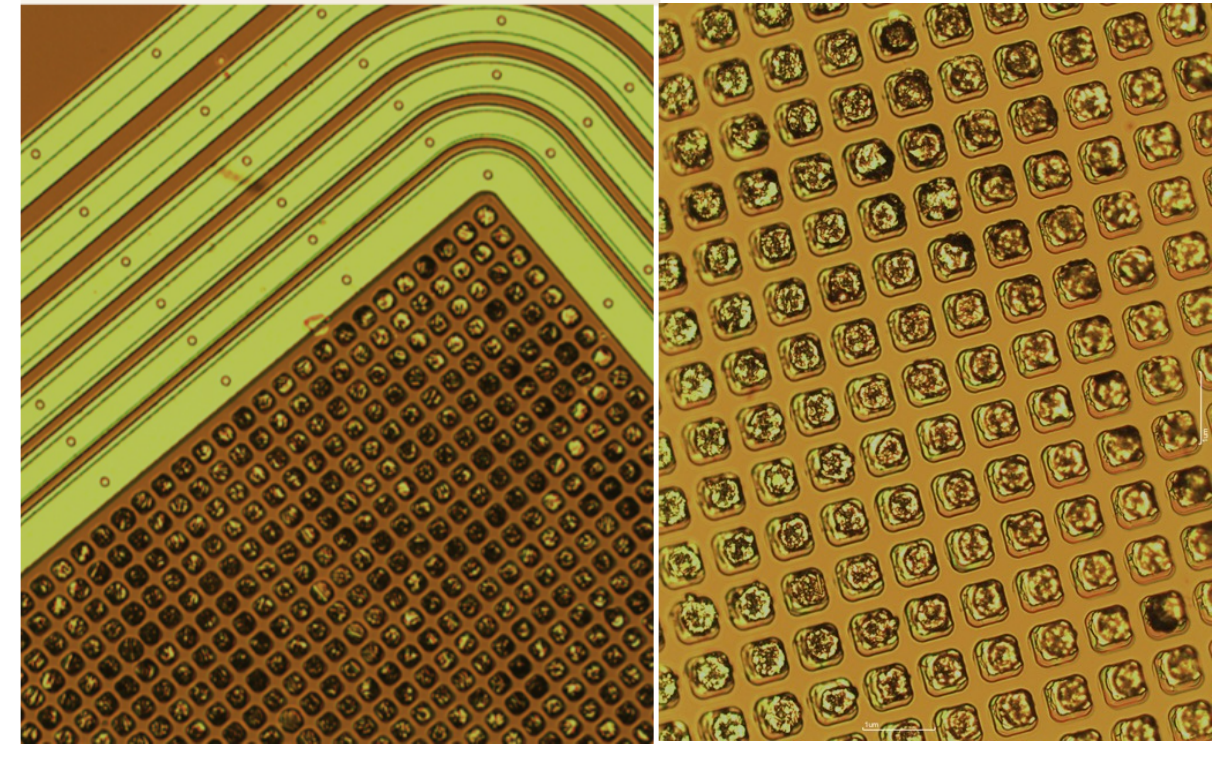}
			 \caption{}\label{fig:ubm_sensor}
	\end{subfigure}%
	\caption{UBM and indium bumps deposited on the \subref{fig:ubm_clicpix} CLICpix readout ASIC and \subref{fig:ubm_sensor} the sensor prior to the flip-chip process. Image credit: SLAC.}\label{fig:ubm}
\end{figure}

The final bump bonding is performed in a flip-chip component placer at relatively low temperature (\SI{170}{\celsius}) and force (\SI{2}{\newton}).
An example of a \SI{50}{\micron} sensor assembly wire-bonded to a readout PCB is shown in \cref{fig:50um_assembly_photo}.
\begin{figure}[ht]
	\centering
  \begin{tikzpicture}
      \node[anchor=south west,inner sep=0] (image) at (0,0) { 		\includegraphics[width=.75\linewidth]{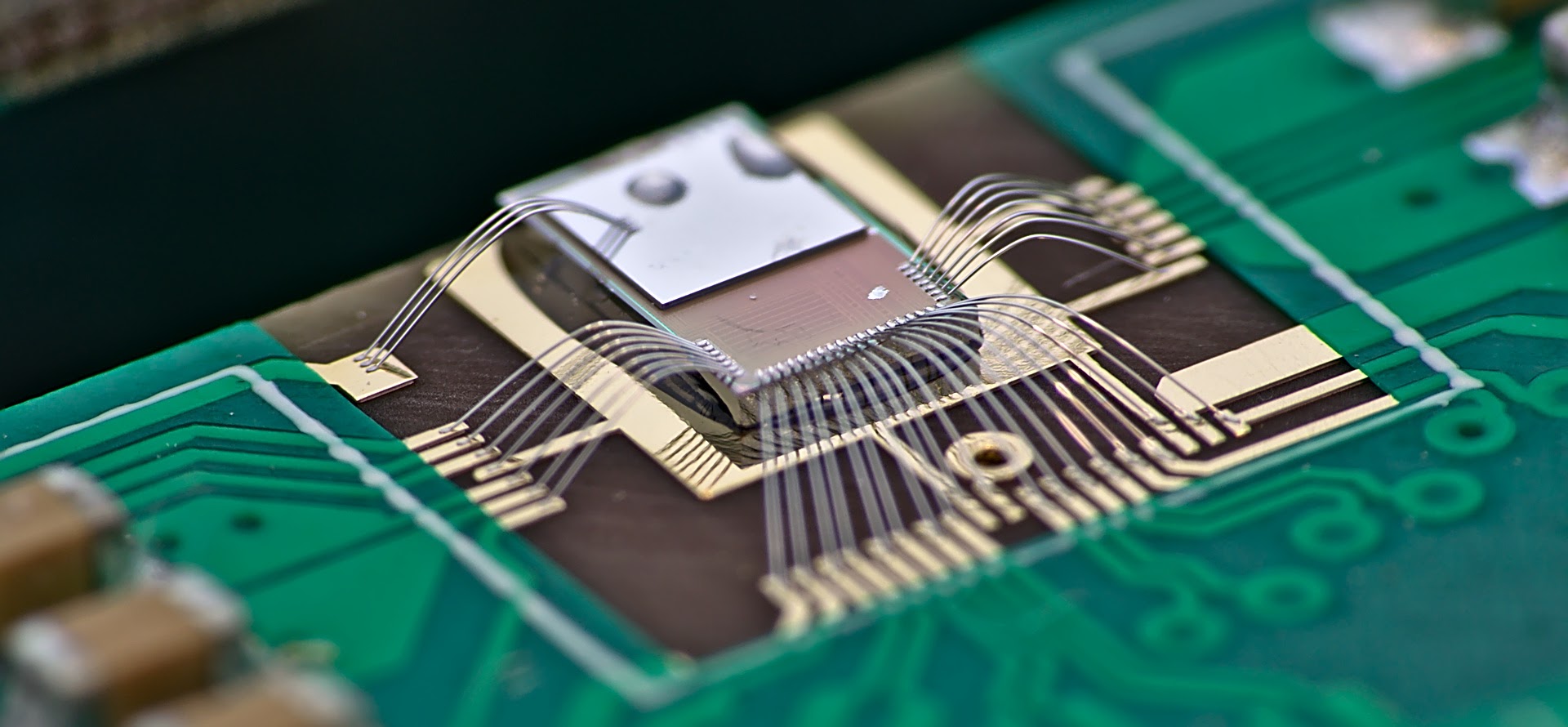}};
    \begin{scope}[x={(image.south east)},y={(image.north west)}]
        \draw[white, ultra thick, <->](0.315,0.75)--(0.47,0.89)node[pos=0.5,above,rotate=24]{\SI{1.6}{\mm}};
      \end{scope}
  \end{tikzpicture}
	\caption{Assembly of a \SI{50}{\micron} thin planar sensor (top) bump-bonded to a CLICpix ASIC (bottom), wire-bonded to a readout PCB.}\label{fig:50um_assembly_photo}
\end{figure}

The quality of the resulting sensor-to-ASIC interconnection was investigated using source and test-beam measurements~\cite{Buckland_thesis}.
Missing bump connections, as well as shorts between bumps, were identified based on the response to ionisation signals in the sensors and a comparison with electrical test-pulse measurements of the ASIC response.
\cref{fig:seed_subpixel} shows the signal response of a fully depleted \SI{200}{\micron} CLICpix sensor assembly to minimum ionising particles, operated in a test-beam setup at a bias voltage of \SI{50}{\volt}.
Regions of weak signal response, mostly visible on the top side, correspond to missing bond connections.
Regions with shorts between neighbouring bonds can be identified by a pair-wise reduction of the signal response to approximately half of the typical values.
For all tested assemblies, the number of shorted pixels was below 2\%.
Pixels with weak or no sensor signal were found in 2-40\% of the cases.
Larger regions of unconnected pixels were found to be correlated with missing bumps visible in the pictures taken prior to the flip-chip step.
The number of pixels with good sensor signals ranged from 40\% up to 97\%.

\begin{figure}[ht]
	\centering
	\begin{subfigure}[T]{.45\textwidth}
		\begin{tikzpicture}
	 			\node[anchor=south west,inner sep=0] (image) at (0,0) { 	\includegraphics[width=\linewidth]{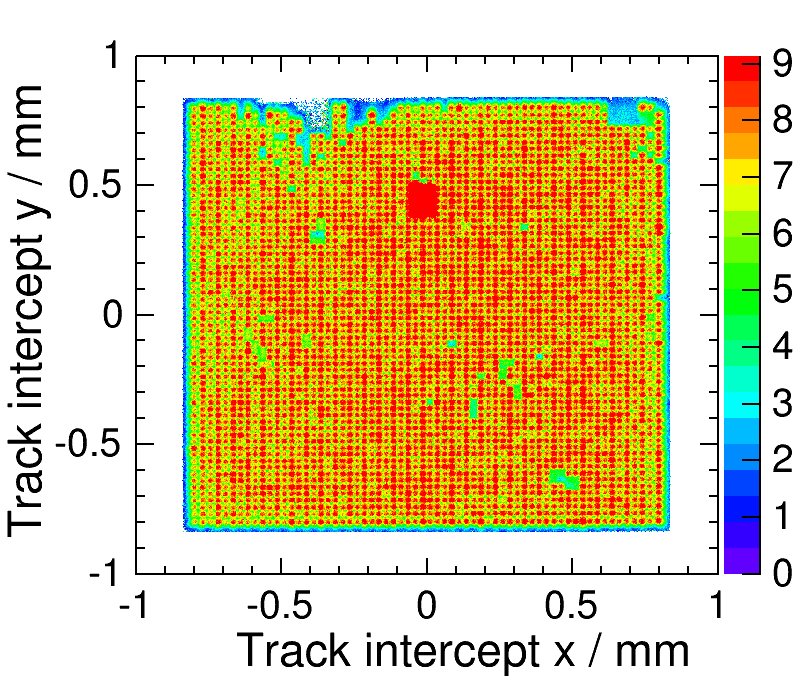}};
	 		\begin{scope}[x={(image.south east)},y={(image.north west)}]
					\draw[thick] (0.29,0.46)rectangle(0.41,0.57);
          \node[anchor=south west, rotate=-90] at (1,0.95){\sffamily Avg. seed signal};
	 			\end{scope}
	 	\end{tikzpicture}
		\caption{}\label{fig:clicpix_seed}
	\end{subfigure}
	\qquad
	\begin{subfigure}[T]{.45\textwidth}
    \begin{tikzpicture}
        \node[anchor=south west,inner sep=0] (image) at (0,0) { 		\includegraphics[width=\linewidth]{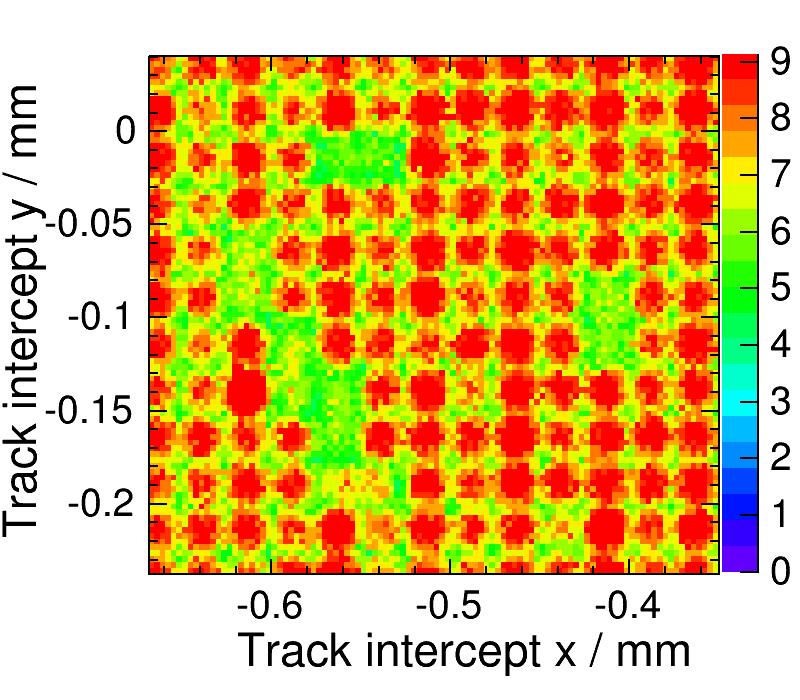}};
      \begin{scope}[x={(image.south east)},y={(image.north west)}]
          \node[anchor=south west, rotate=-90] at (1,0.95){\sffamily Avg. seed signal};
        \end{scope}
    \end{tikzpicture}
		\caption{}\label{fig:clicpix_seed_zoom}
	\end{subfigure}
	\caption{\subref{fig:clicpix_seed}: Average signal of the leading pixel in a cluster as a function of the track intercept position for a CLICpix \SI{200}{\micron} sensor assembly operated in a test-beam telescope setup, showing regions with defective interconnections (missing bumps and shorts) as explained in the text. \subref{fig:clicpix_seed_zoom}: zoom into the highlighted region with several shorted bumps.}\label{fig:seed_subpixel}
\end{figure}
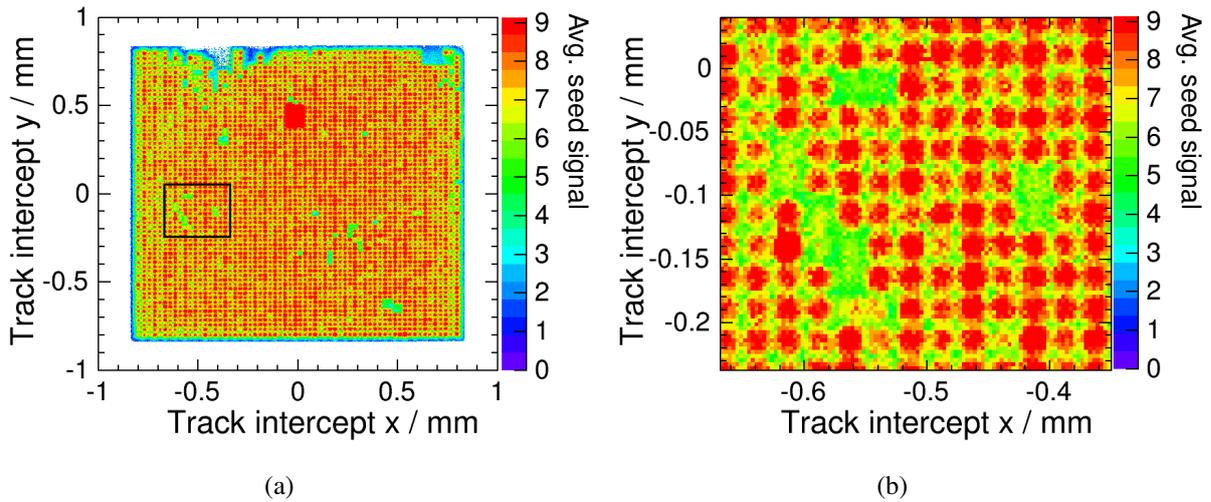

\subsubsection{Bump bonding of CLICpix2 assemblies}
A single-chip bump-bonding process based on carrier-wafers for CLICpix2 ASICs and active-edge sensors from Advacam~\cite{Advacam} and FBK-CMM~\cite{FBK} is currently under development at IZM~\cite{IZM}.
The \SI{100}{\micron} and \SI{150}{\micron} Advacam sensors received thin-film Ni/Au UBM during the sensor production, as described in \cref{sec:tpx3-bump-bonding} for the Timepix3 active-edge sensors.
The \SI{130}{\micron} thick FBK-CMM sensors received Ti/W/Cu UBM at IZM, as described in \cref{sec:tpx-bump-bonding} for the Micron sensors.
The carrier-wafer processing steps for the ASIC UBM and SnAg bump deposition include:
\begin{enumerate}
\item Preparation of carrier wafers with mask alignment marks and bond layer;
\item Bond single-chip dies on carrier wafer;
\item Bump deposition: sputtering of plating base, resist lithography, Cu+SnAg-galvanic, resist removal and etching of plating base outside bumps, reflow of bumps;
\item Removal of ASICs from carrier wafer.
\end{enumerate}
\cref{fig:clicpix2_bumps} shows the resulting uniform distribution of SnAg bumps on one of the CLICpix2 ASICs.
The flip-chip assembly of the ASICs and the sensors is performed in a pick-and-place process using a flip-chip bonder tool, followed by a further reflow step.
\cref{fig:clicpix2-assembly} shows a photograph of a CLICpix2 assembly with a \SI{130}{\micron} active-edge sensor, wire-bonded to a readout PCB.
\begin{figure}[ht]
	\centering
        \begin{tikzpicture}
	  \node[anchor=south west,inner sep=0] (image) at (0,0) { 	\includegraphics[width=.7\linewidth]{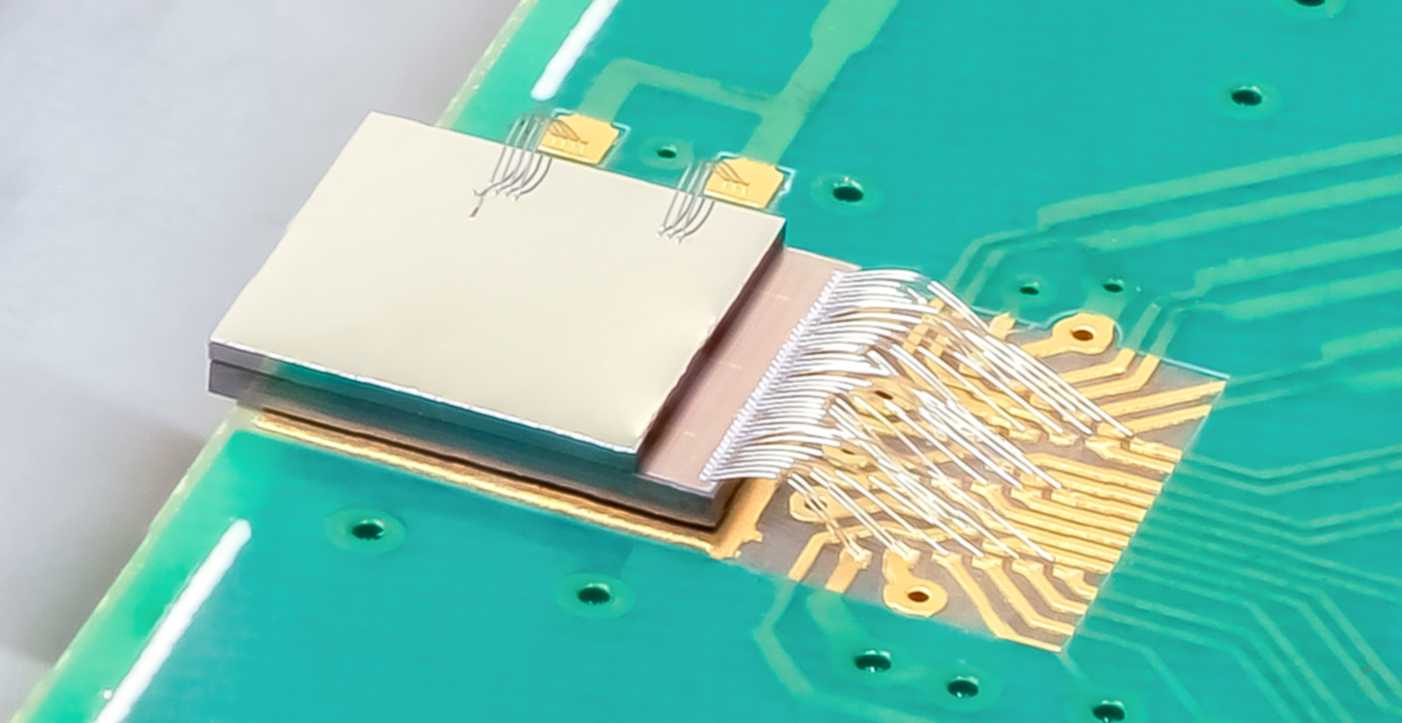}};
	  \begin{scope}[x={(image.south east)},y={(image.north west)}]
             \draw[<->, thick, blue](0.14,0.55) -- (0.25,0.86) node[pos=0.5, rotate=55, above]{\large\textsf{\SI{3.4}{\mm}}};
	  \end{scope}
	\end{tikzpicture}
	\caption{Assembly of a \SI{130}{\micron} thick active-edge sensor bump-bonded to a CLICpix2 ASIC, wire-bonded to a readout PCB.}\label{fig:clicpix2-assembly}
\end{figure}

X-ray images of the resulting assemblies are taken for quality control purposes.
The bump connectivity was investigated in more detail for one of the assemblies by performing a destructive cross-section and taking light microscope and Scanning-Electron-Microscope (SEM) pictures at various positions inside the assembly.
An example microscope picture is shown in \cref{fig:clicpix2_cross-section}, demonstrating a good solder connection between the two dies in the corresponding region.
\begin{figure}[ht]
  \centering
  \begin{subfigure}[T]{.41\textwidth}
    \includegraphics[width=\linewidth]{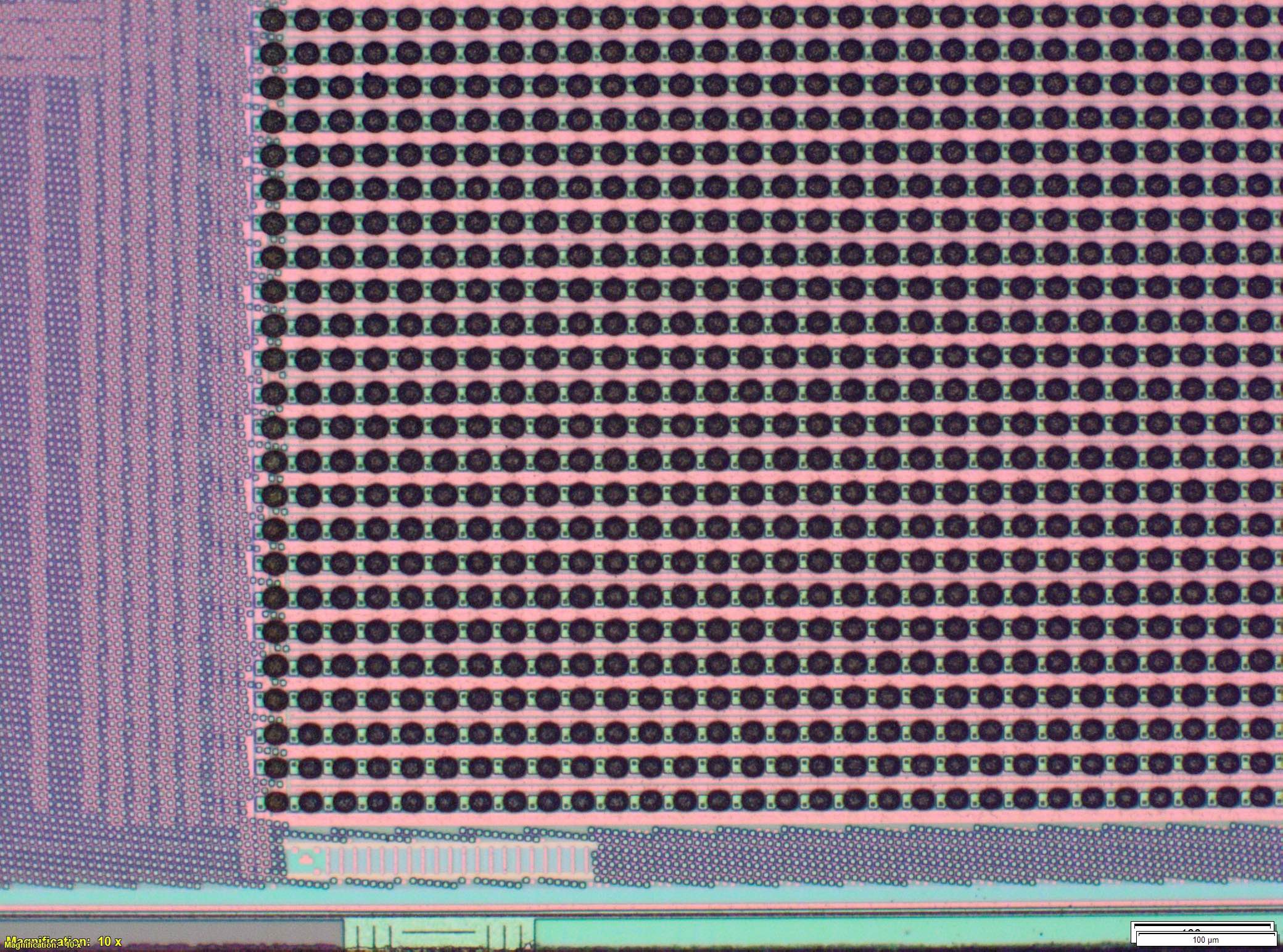}
    \caption{}\label{fig:clicpix2_bumps}
  \end{subfigure} \hfill
  \begin{subfigure}[T]{.56\textwidth}
    \includegraphics[width=\linewidth, clip, trim=0 0 0 7cm]{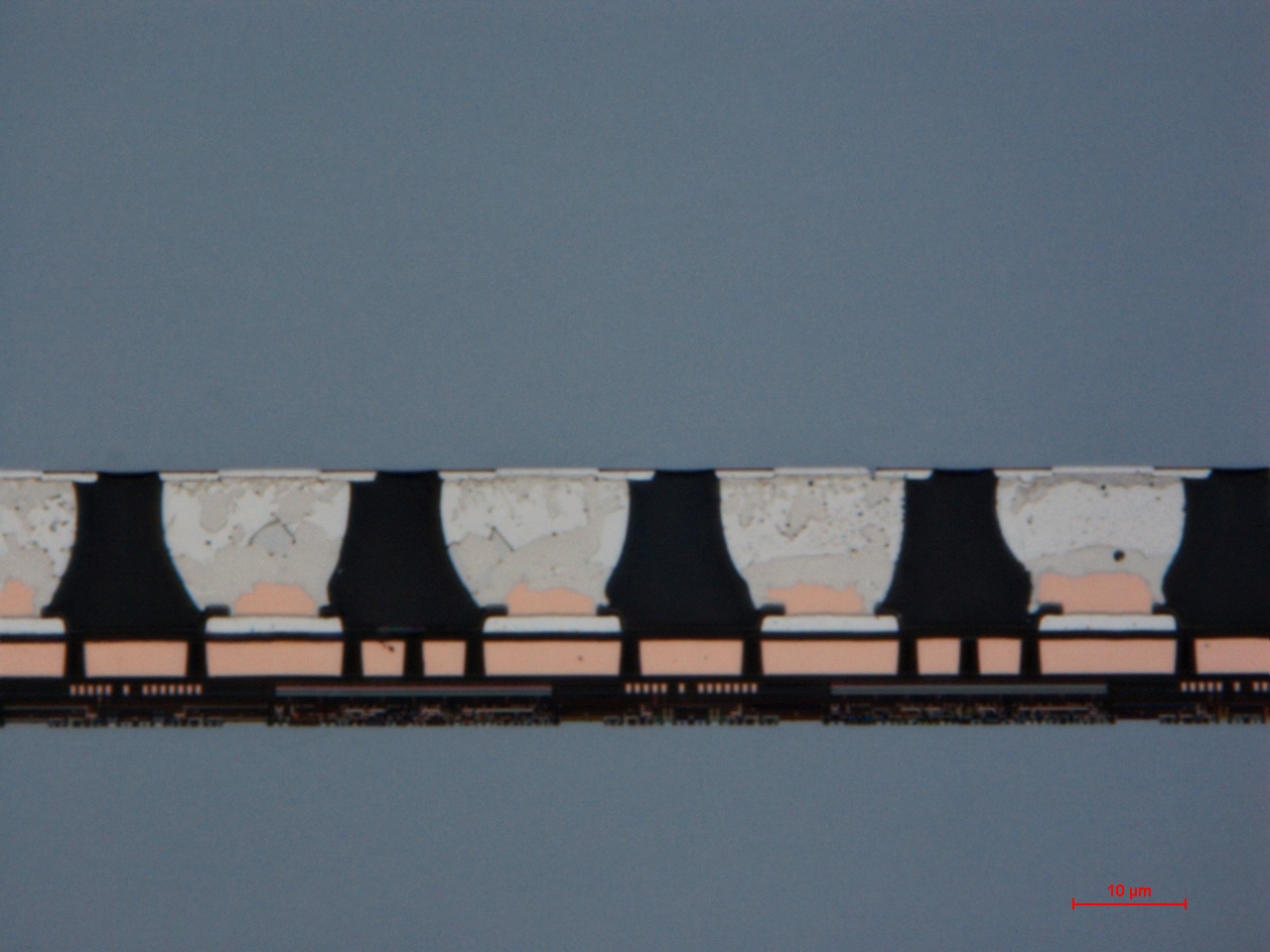}
    \caption{}\label{fig:clicpix2_cross-section}
  \end{subfigure}
  \caption{SnAg solder bumps on CLICpix2 ASIC \subref{fig:clicpix2_bumps} and cross-section through CLICpix2 and Advacam active-edge sensor after bump-bonding \subref{fig:clicpix2_cross-section}. Image credit Fraunhofer IZM.}
 \end{figure}

Several assemblies were operated successfully and exposed to radioactive sources and test beams.
Preliminary results indicate a good interconnect quality for the sensors with Ti/W/Cu UBM.
\cref{fig:clicpix2-Sr90} shows the pixel hit rate in a \isotope[90]{Sr} source exposition of a \SI{130}{\micron} FBK sensor assembly.
An interconnect yield of 99.7\% is observed.
The 55 missing bumps (0.3\% of all pixels), visible also on the x-ray picture of the assembly (\cref{fig:clicpix2-xray}), are located in a continuous region at the ASIC corner and were lost when lifting the ASIC from the tape.
Further studies are ongoing to investigate the significantly lower interconnect yield observed for the assemblies made with thin-film UBM sensors.
\begin{figure}[ht]
  \centering
  \begin{subfigure}[T]{.50\textwidth}
    \includegraphics[width=\linewidth]{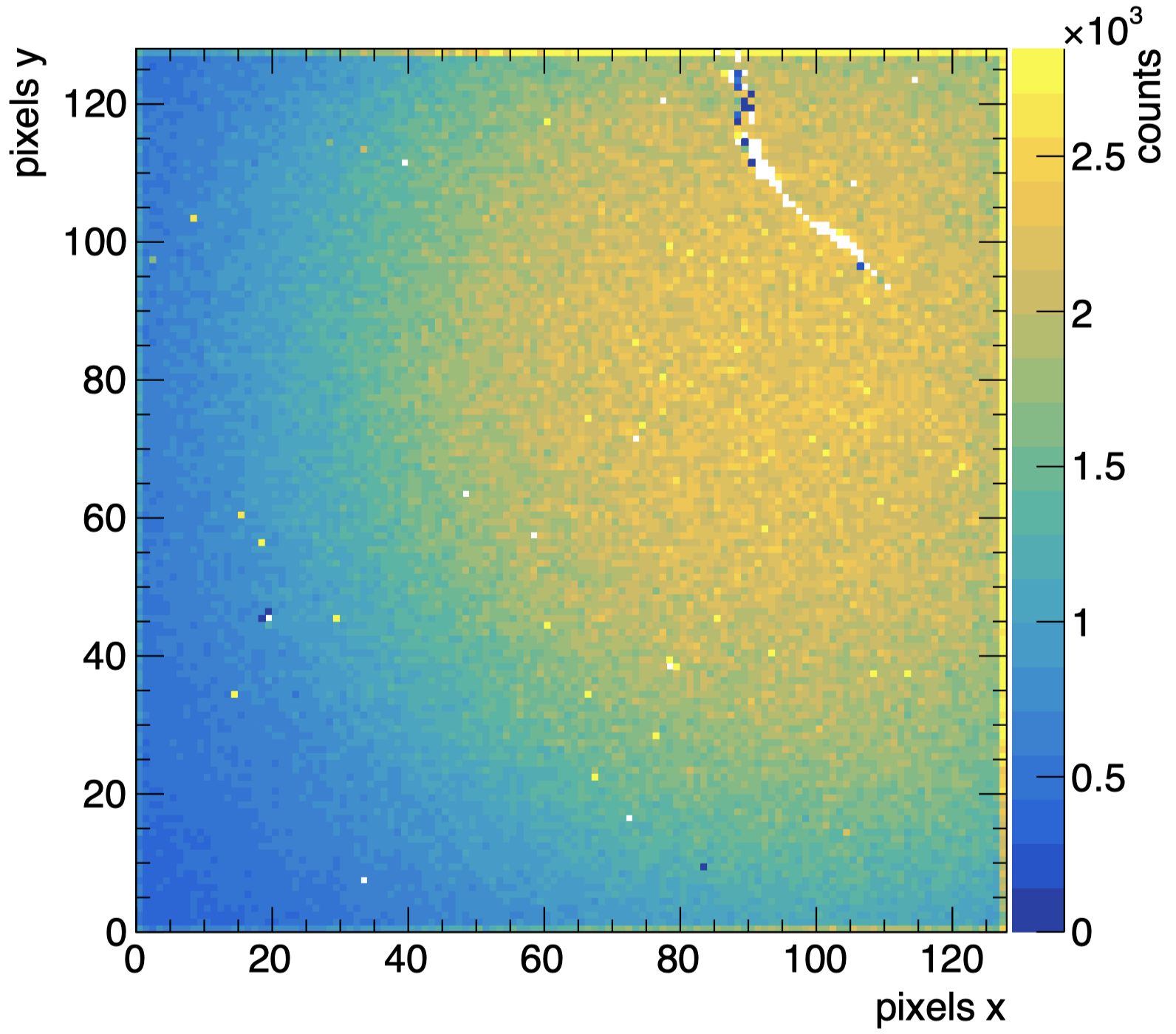}
    \caption{}\label{fig:clicpix2-Sr90}
  \end{subfigure} \hfill
  \begin{subfigure}[T]{.42\textwidth}
    \includegraphics[width=\linewidth]{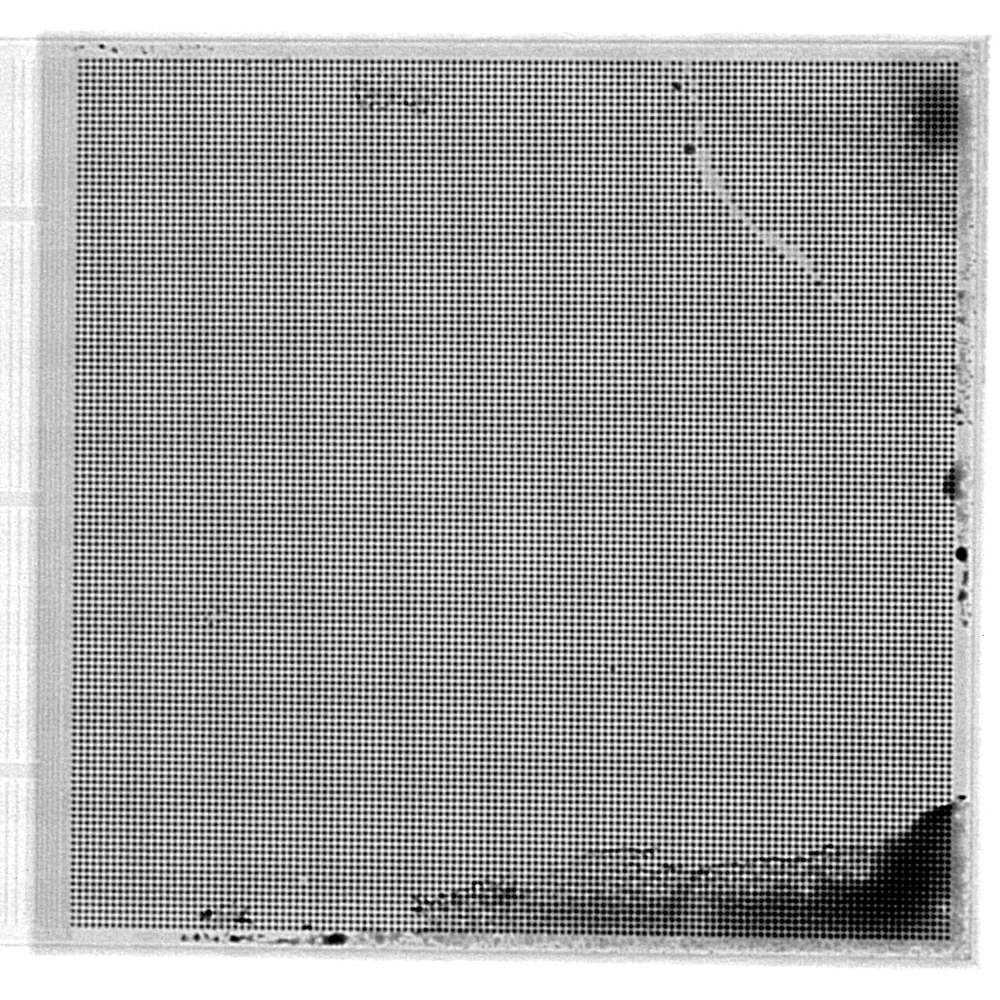}
    \caption{}\label{fig:clicpix2-xray}
  \end{subfigure}
  \caption{\subref{fig:clicpix2-Sr90} \isotope[90]{Sr} source measurement of a CLICpix2 assembly bump bonded to a \SI{130}{\micron} thick active-edge sensor. Most of the missing 55 solder bumps are located in the notch in the top right corner and were lost when lifting the ASIC from the tape. \subref{fig:clicpix2_cross-section} The notch is also visible in the x-ray picture taken after the flip-chip process (image credit IZM). The dark areas in the x-ray picture are caused by residues in the packaging material and are not related to assembly defects.}
 \end{figure}

\subsection{Measurements with thin planar sensor assemblies}\label{sec:thin_planar_sensors}

Timepix and Timepix3 readout ASICs have been used to study planar passive pixel sensors in laboratory and test-beam campaigns. Despite not matching all requirements for application in the CLIC vertex detector, the study gave valuable insight into charge collection, charge sharing, single-hit point and time resolution and detection efficiency of the technology. It thus helped in setting the specifications for the dedicated CLICpix ASICs and in validating the simulation tools used for the optimisation of future sensors and readout ASICs.

\subsubsection{Energy and time calibration of Timepix3 assemblies}
The energy and time measurement response of the Timepix and Timepix3 detector assemblies has been calibrated individually for each pixel in order to achieve a uniform response of the detector over the whole pixel matrix~\cite{timepix_calibration_sophie,timepix3_calibration_flo,calib_timing_tpx3}. This is necessary, in order to correct for systematic variations across the pixel matrix and for process-related variations in the ASIC. First, a relative calibration per pixel is performed using the test pulse circuitry of the readout ASIC. Owing to the use of a fixed threshold value in combination with a constant rise time, non-linear functions are required to parameterise the front-end response in terms of energy and time measurement as a function of the test-pulse amplitude (time walk). Following the test-pulse calibration, signals of known energy or known time-of-arrival are used to perform an absolute calibration of the device.

\paragraph*{Energy calibration}
For the energy measurement, the test pulse calibration results in a conversion of the time-over-threshold counts to the injected pulse amplitude in \si{\milli\volt}. Signals of known energy from x-ray fluorescence targets allow for an absolute energy calibration of the detector to be performed. The results are shown in \cref{fig:timepix3_energy_calibration} for three different Timepix3 detector assemblies with \SI{50}{\micron}, \SI{100}{\micron} and \SI{150}{\micron} thick planar sensors. All fluorescence peaks originating from the different target materials are reconstructed at the expected position, validating the calibration procedure. It should be pointed out that only the Iron and Indium peaks are used for the calibration, while the Copper, Lead and Zirconium measurements are only used for verification purposes.

\begin{figure}[ht]
  \centering
  \begin{tikzpicture}
\node[anchor=south west,inner sep=0] at (0,0)(image){  \includegraphics[width=.7\linewidth]{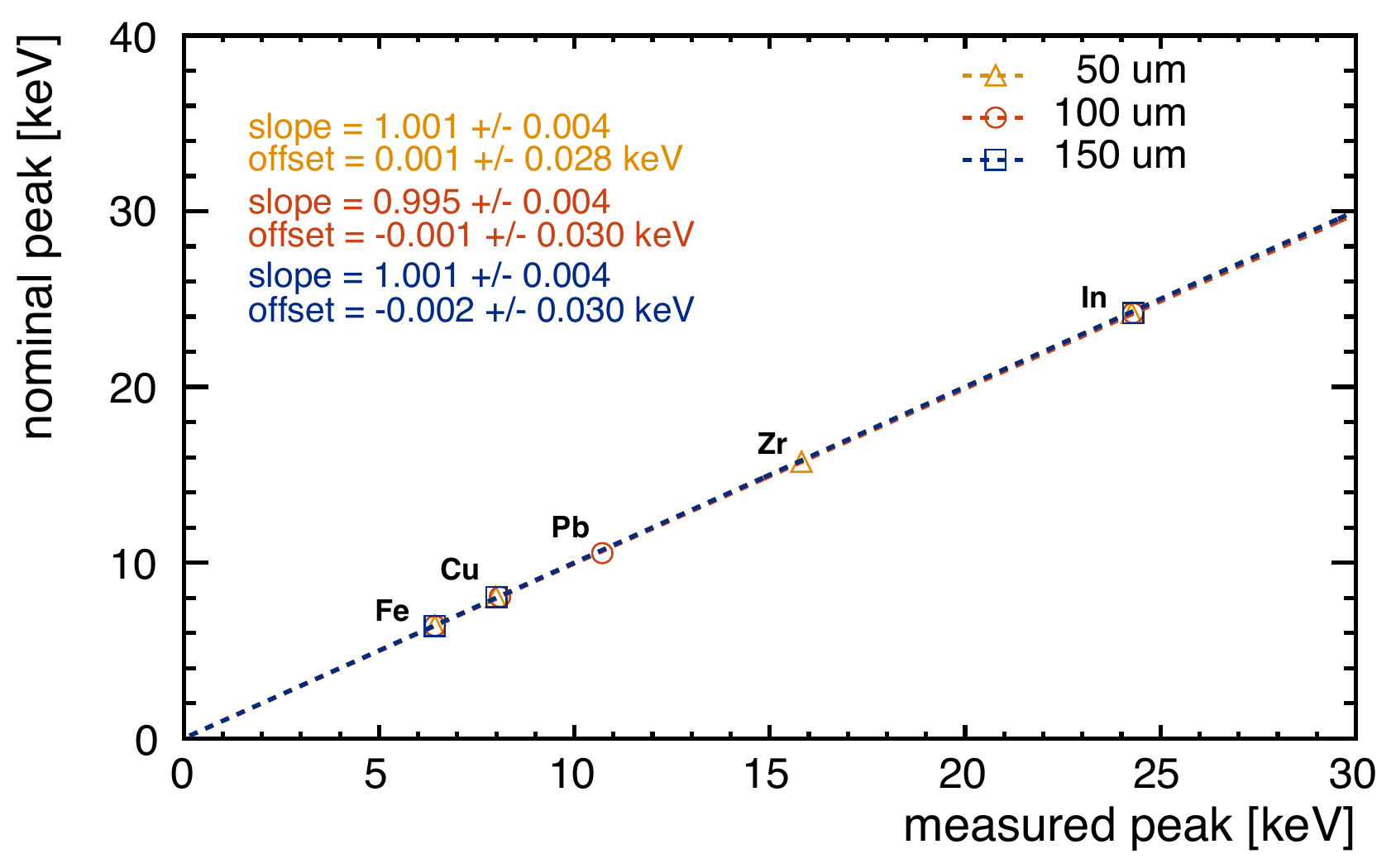}};
  \begin{scope}[x={(image.south east)},y={(image.north west)}]
   \node[anchor=north east] at (0.85,0.35){CLICdp};
\end{scope}
\end{tikzpicture}
  \caption{Measured peak energy vs. nominal peak energy obtained from an x-ray fluorescence measurement for three different Timepix3 detector assemblies with \SI{50}{\micron}, \SI{100}{\micron} and \SI{150}{\micron} thick planar sensors~\cite{timepix3_calibration_flo}. Fe and In points are used for calibration, while the Cu, Pb and Zr points are added for verification purposes. Only a sub-set of all available target materials was used in each case. The results of linear fits to the data points are also given.}\label{fig:timepix3_energy_calibration}
\end{figure}

\paragraph*{Time calibration}
Similar to the energy calibration, the time calibration is done in two steps by a combination of test pulses for the non-linear time walk and test-beam data to calibrate the delay~\cite{calib_timing_tpx3}. \cref{fig:timepix3_timing_rms_corrections} shows an example of the timing residuals at various stages of the calibration.

\begin{figure}[ht]
\begin{subfigure}[T]{.49\linewidth}  
  \begin{tikzpicture}
\node[anchor=south west,inner sep=0] at (0,0)(image){  \includegraphics[width=\linewidth]{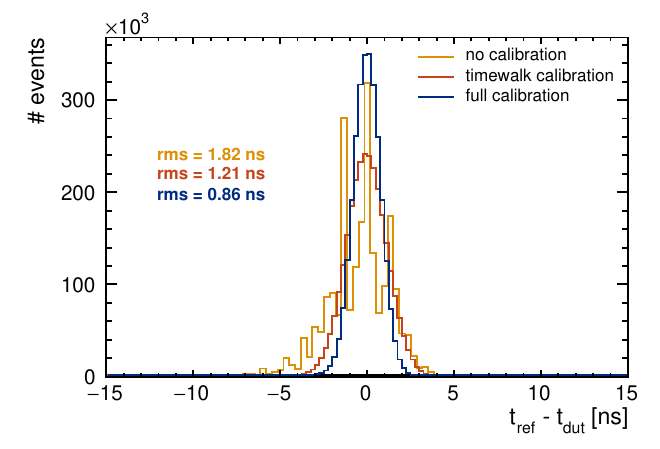}};
  \begin{scope}[x={(image.south east)},y={(image.north west)}]
   \node[anchor=north east] at (0.85,0.35){CLICdp};
\end{scope}
\end{tikzpicture}
  \caption{}\label{fig:timepix3_timing_rms_corrections_50um}
\end{subfigure}
\hfill
\begin{subfigure}[T]{.49\linewidth}  
  \begin{tikzpicture}
\node[anchor=south west,inner sep=0] at (0,0)(image){  \includegraphics[width=\linewidth]{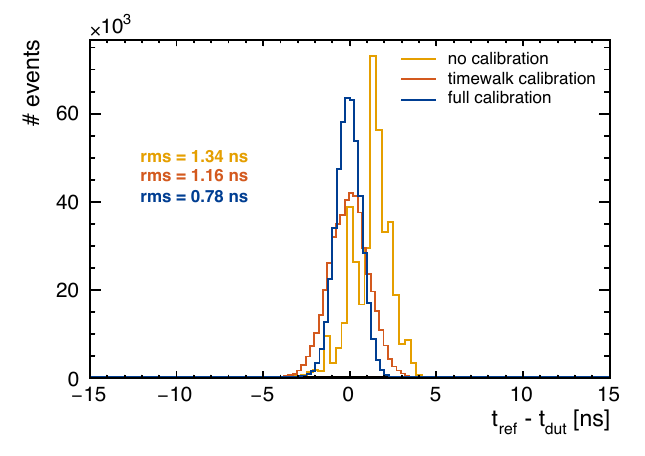}};
  \begin{scope}[x={(image.south east)},y={(image.north west)}]
   \node[anchor=north east] at (0.85,0.35){CLICdp};
\end{scope}
\end{tikzpicture}
  \caption{}\label{fig:timepix3_timing_rms_corrections_150um}
\end{subfigure}
  \caption{Timing residual between the reconstructed hit on a Timepix3 assembly with a \subref{fig:timepix3_timing_rms_corrections_50um} \SI{50}{\micron} and \subref{fig:timepix3_timing_rms_corrections_150um} \SI{150}{\micron} thin sensor and the scintillator reference. The raw distribution and two calibration stages are shown~\cite{calib_timing_tpx3}.}\label{fig:timepix3_timing_rms_corrections}
\end{figure}

To parameterise the time-walk effect of the front-end, the time-of-arrival of analogue test pulses with varying amplitude is compared to the arrival time of a digital reference pulse, which is not affected by time walk. The test pulse injection is taking place relative to the same clock that is measuring the arrival time, and thus any systematic variation in the clock phase between pixels remains undetected. To compensate for that effect, external signals with known arrival time have to be injected into the detector. This has been performed in beam tests using the Timepix3 based reference telescope described in~\cref{sec:timepix3_telescope}. The telescope is capable of reconstructing a particle hit on the device under test with about \SI{2}{\micron} spatial and \SI{1}{\ns} temporal precision. The coincidence between three additional scintillators is used to achieve a more precise timestamp of the order of several hundred picoseconds of the particle impact.

\paragraph*{Time resolution}
\cref{fig:timepix3_timeres_thickness} summarises the time resolution obtained in beam tests with \SI{120}{\giga\electronvolt} pions for 3 different sensor thicknesses operated \SI{5}{\volt} above the full depletion voltage. A significant improvement in the time resolution can be achieved when using a time-walk correction for the \SI{50}{\micron} sensor. The relative improvement is reduced for the thicker sensors due to the larger energy deposits. After time-walk correction, all sensors yield approximately the same resolution at their nominal operating~voltage.

The electric field in the sensor has a strong influence on the time resolution. The investigated \SI{50}{\micron}, \SI{100}{\micron} and \SI{150}{\micron} thick sensors deplete around \SI{10}{\volt}, \SI{15}{\volt} and \SI{25}{\volt} respectively. \cref{fig:timepix3_timeres_bias} shows the obtained time resolution for different bias voltages for the \SI{50}{\micron} and \SI{150}{\micron} sensor. The resolution however is improving beyond that voltage due to stronger electric fields at higher bias voltages. This is due to an increase in the drift velocity that in turn results in a faster signal creation on the input of the charge sensitive amplifier and in a bias towards single-pixel clusters with higher pixel signals. The best time resolution achieved in this study at significant over-depletion is \SI{720+-40}{\ps}.

\begin{figure}[ht]
  \begin{subfigure}[T]{.49\linewidth}
    \begin{tikzpicture}
  \node[anchor=south west,inner sep=0] at (0,0)(image){  \includegraphics[width=\linewidth]{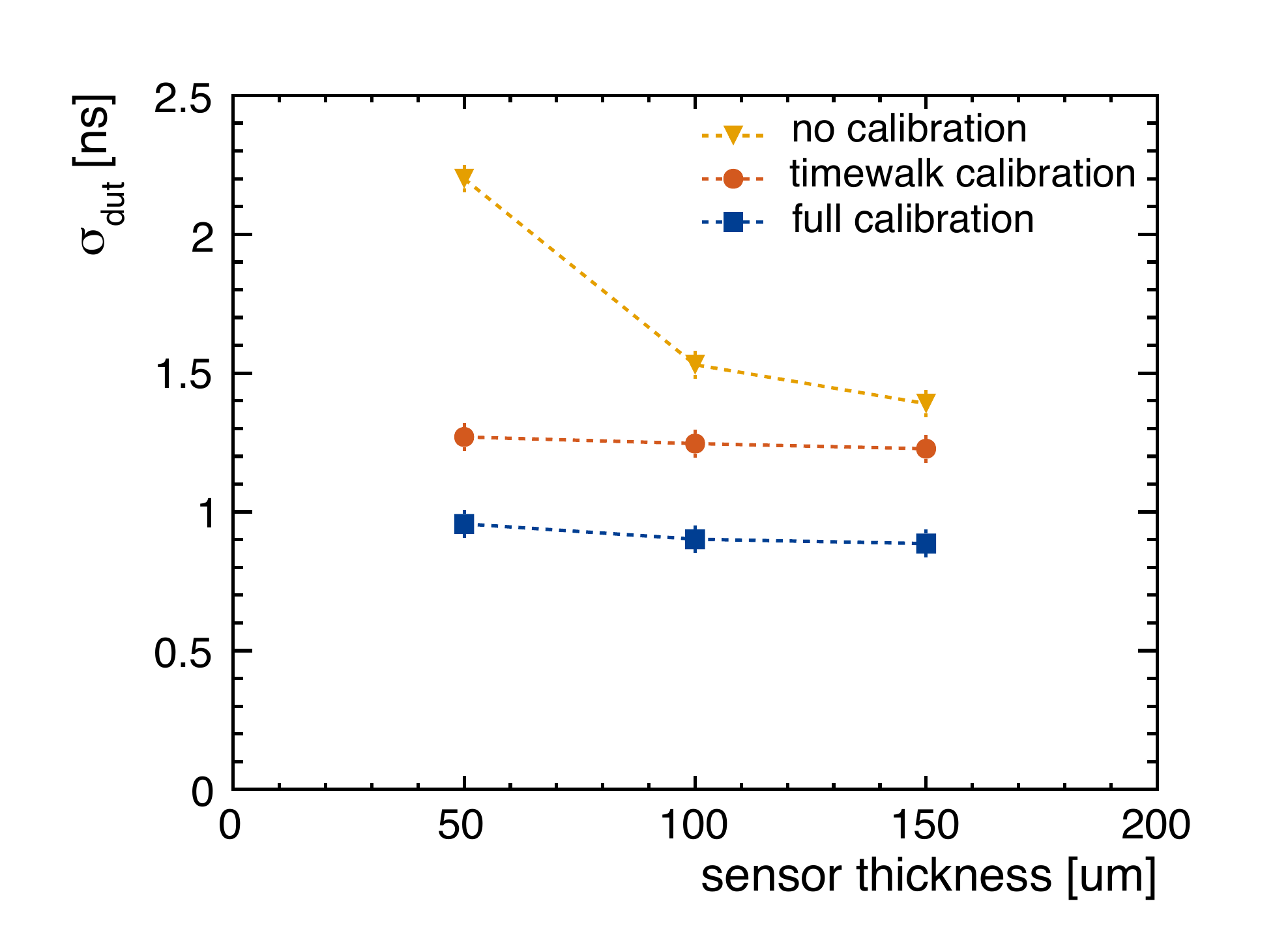}};
    \begin{scope}[x={(image.south east)},y={(image.north west)}]
     \node[anchor=north east] at (0.85,0.3){CLICdp};
  \end{scope}
  \end{tikzpicture}
    \caption{}\label{fig:timepix3_timeres_thickness}
  \end{subfigure}
  \hfill
  \begin{subfigure}[T]{.49\linewidth}
    \begin{tikzpicture}
  \node[anchor=south west,inner sep=0] at (0,0)(image){  \includegraphics[width=\linewidth]{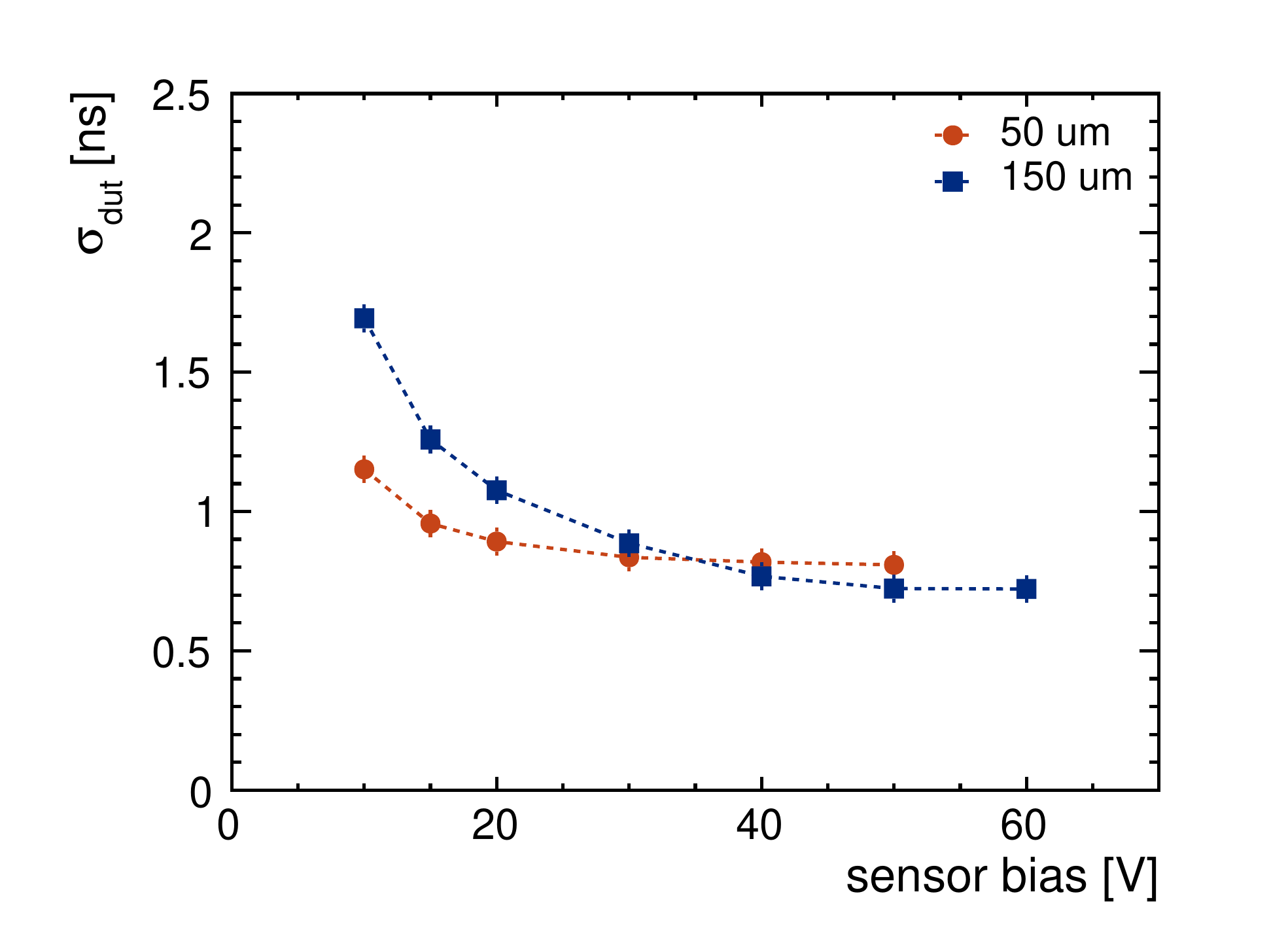}};
    \begin{scope}[x={(image.south east)},y={(image.north west)}]
     \node[anchor=north east] at (0.85,0.3){CLICdp};
  \end{scope}
  \end{tikzpicture}
    \caption{}\label{fig:timepix3_timeres_bias}
  \end{subfigure}
  \caption{ \subref{fig:timepix3_timeres_thickness} Time resolution of Timepix3 detector assemblies with various sensor thicknesses with and without time-walk and delay corrections, biased at 5~V over-depletion. \subref{fig:timepix3_timeres_bias} Improvement of time resolution with increasing bias voltage for two different sensor thicknesses~\cite{calib_timing_tpx3}. The resolution of the reference time stamp from an organic scintillator readout by a PMT of \SI{0.3}{\ns} has been unfolded.}\label{fig:timepix3_timeres}
\end{figure}

\subsubsection{Impact of sensor thickness on tracking performance}\label{sec:timepix3_planar_study}
The very low material budget for the vertex detector constrains the sensor thickness in hybrid assemblies not to exceed \SI{50}{\micron} per single detection layer.
A systematic study of the influence of the sensor thickness on the tracking performance was therefore performed, to assess the performance of such thin active layers.

\paragraph*{Spatial resolution}
The cluster size is defined as the number of adjacent pixels exceeding the charge threshold after a particle hit. \cref{fig:timepix_cs_vs_thickness} illustrates the relative occurrence of different cluster size categories for sensors of various thicknesses from \SIrange[range-phrase={~to~},range-units=repeat]{50}{500}{\micron} bump-bonded to Timepix ASICs and operated in a \SI{5.6}{\giga\electronvolt} \Pem{} beam at perpendicular incident~\cite{TimepixTestbeamNote}. In thicker sensors, the drifting charge diffuses further and gives rise to a larger fraction of multi-pixel clusters. As expected for squared pixels, the cluster size along both pixel axes is similar.

\begin{figure}[ht]
  \begin{subfigure}[T]{.49\linewidth}
    \begin{tikzpicture}
  \node[anchor=south west,inner sep=0] at (0,0)(image){  \includegraphics[width=\linewidth]{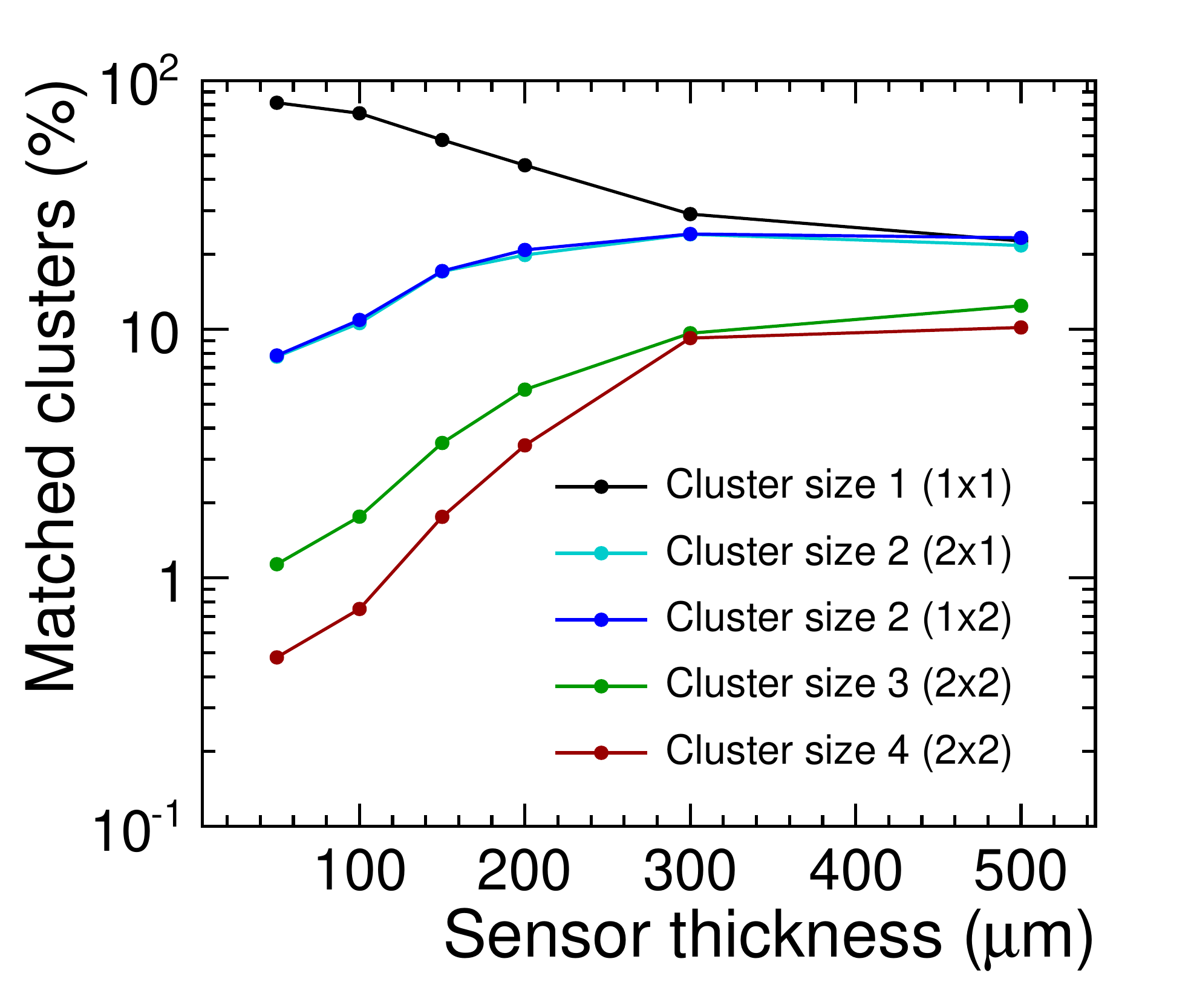}};
    \begin{scope}[x={(image.south east)},y={(image.north west)}]
     \node[anchor=north east] at (0.85,0.875){CLICdp};
  \end{scope}
  \end{tikzpicture}
    \caption{}\label{fig:timepix_cs_vs_thickness}
   \end{subfigure}
   \hfill
  \begin{subfigure}[T]{.49\linewidth}
  \includegraphics[width=\linewidth]{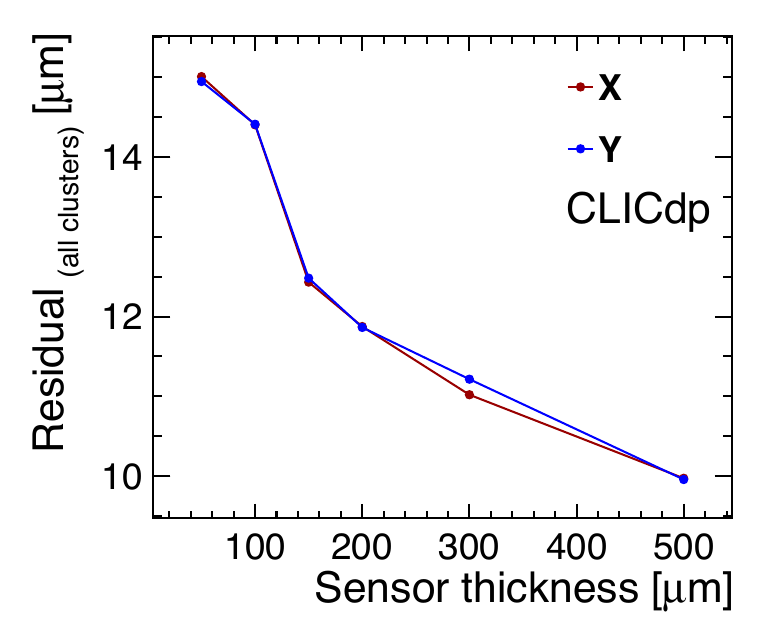}
  \caption{}\label{fig:timepix_resolution_vs_thickness}
\end{subfigure}
\caption{Test-beam results of Timepix assemblies with sensors of different thicknesses: \subref{fig:timepix_cs_vs_thickness} Fraction of different cluster sizes and \subref{fig:timepix_resolution_vs_thickness} single-point x and y residuals as a function of the sensor thickness~\cite{TimepixTestbeamNote}. The average of the RMS calculated per run in a range of $\pm\SI{40}{\micron}$ around 0 is used. The track resolution is not unfolded.}
\end{figure}

The spatial resolution of the sensor is closely linked to the cluster size.
For multi-pixel clusters, the hit position can be determined more precisely by means of charge weighting, using e.g.\ the $\eta$-correction algorithm~\cite{Belau}, and thus the resolution improves with the increasing fraction of multi-pixel clusters.
For that reason, the trend of rising cluster size with sensor thickness is directly reflected in the distributions of the residuals between the reconstructed sensor-hit position and the track impact point obtained from the telescope.
\cref{fig:timepix_resolution_vs_thickness} summarises the obtained resolution as a function of the sensor thickness, demonstrating the superior resolution of the thicker sensors.
For \SI{50}{\micron} sensor thickness, where more than 90\% of all clusters contain only one pixel, the resolution is only slightly better than the binary resolution of pitch$/\sqrt{12}$, which amounts to $\sim\SI{16}{\micron}$.
Examples of the residual distributions are shown in \cref{fig:timepix_residual_thicknesses_50um,fig:timepix_residual_thicknesses_100um,fig:timepix_residual_thicknesses_200um,fig:timepix_residual_thicknesses_500um}.
For the thinner sensors, a narrow peak from more precisely reconstructed multi-pixel clusters can clearly be distinguished from the broader distribution of single-pixel clusters. 
The RMS calculated in a range of $\pm\SI{40}{\micron}$ is used as a measure of the width of the distributions.

\begin{figure}[ht]
  \begin{subfigure}[T]{0.24\linewidth}
    \includegraphics[width=\linewidth, clip, trim=0cm 0 12cm 0]{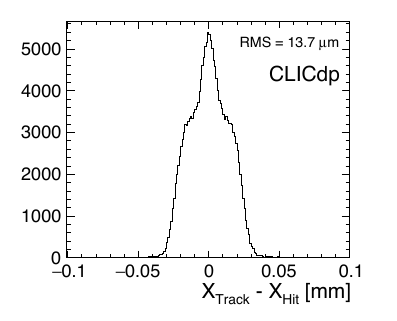}
    \caption{}\label{fig:timepix_residual_thicknesses_50um}
  \end{subfigure}
  \hfill
  \begin{subfigure}[T]{0.24\linewidth}
    \includegraphics[width=\linewidth, clip, trim=4cm 0 8cm 0]{figures/timepix_residual_thicknesses.pdf}
    \caption{}\label{fig:timepix_residual_thicknesses_100um}
  \end{subfigure}
  \hfill
  \begin{subfigure}[T]{0.24\linewidth}
    \includegraphics[width=\linewidth, clip, trim=8cm 0 4cm 0]{figures/timepix_residual_thicknesses.pdf}
    \caption{}\label{fig:timepix_residual_thicknesses_200um}
  \end{subfigure}
  \hfill
  \begin{subfigure}[T]{0.24\linewidth}
    \includegraphics[width=\linewidth, clip, trim=12cm 0 0cm 0]{figures/timepix_residual_thicknesses.pdf}
    \caption{}\label{fig:timepix_residual_thicknesses_500um}
  \end{subfigure}
  \caption{Example single-point x residuals for Timepix assemblies with fully depleted silicon sensors of different thicknesses~\cite{TimepixTestbeamNote}: \subref{fig:timepix_residual_thicknesses_50um} \SI{50}{\micron} p-in-n, \subref{fig:timepix_residual_thicknesses_100um} \SI{100}{\micron} p-in-n, \subref{fig:timepix_residual_thicknesses_200um} \SI{200}{\micron} n-in-p and \subref{fig:timepix_residual_thicknesses_500um} \SI{500}{\micron} n-in-p. The RMS is calculated within a range of $\pm\SI{40}{\micron}$ around 0.}
\end{figure}

The limited resolution of the track impact point onto a sensor under study contributes in quadrature to the width of the residual distribution. The resolution of the reference telescope used here is $\sim\SI{2}{\micron}$. This is significantly smaller than the resolution of the sensor under study, and its contribution is thus neglected.

\paragraph*{Detection efficiency}
The most probable charge deposit of a minimal ionising particle in a \SI{50}{\micron} thin sensor is around \SI{3300}{\Pem{}}.
To efficiently detect a particle hit, the threshold of the front-end has to be significantly lower than that, in order to account for statistical fluctuations in the energy deposit and for charge sharing between adjacent pixels.
The detection efficiency of the sensor is defined as the number of detected hits that can be associated to a reconstructed particle track divided by the total number of reconstructed particle tracks through the sensitive detector area.
This has been studied on planar-sensor assemblies using Timepix3 ASICs~\cite{ThesisNilou}.
For the track to hit matching, a radius of \SI{100}{\micron} around the predicted track impact point has been chosen.
\cref{fig:eff_timepix3} shows the detection efficiency as a function of the threshold for Timepix3 assemblies with sensor thicknesses of \SI{50}{\micron}, \SI{100}{\micron} and \SI{150}{\micron}, obtained in a \SI{120}{\giga\electronvolt} pion beam.
For all three sensors, the resulting detection efficiency is above \SI{99}{\percent} for thresholds around \SI{1000}{\Pem{}} and below.
Several pixels (less than 50, corresponding to less than \SI{0.1}{\percent{} of the matrix}) have been masked manually due to their abnormally high dark count rate.
These known inefficient pixels have not been taken into account when estimating the efficiency result.
A similar analysis of test-beam data taken with thin-sensor Timepix assemblies resulted in efficiencies of more than 99.7\% at nominal operation thresholds for sensor thicknesses down to \SI{50}{\micron}~\cite{TimepixTestbeamNote}.

\begin{figure}[ht]
  \begin{subfigure}[T]{.45\linewidth}
    \begin{tikzpicture}
  \node[anchor=south west,inner sep=0] at (0,0)(image){  \includegraphics[width=\linewidth]{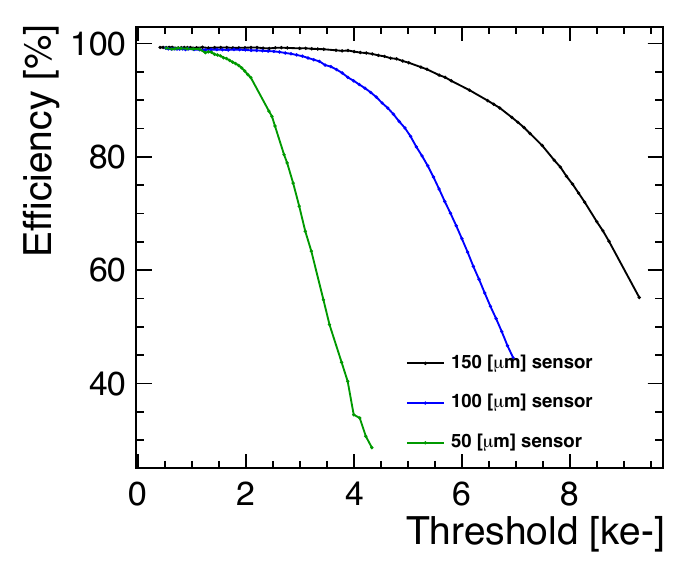}};
    \begin{scope}[x={(image.south east)},y={(image.north west)}]
     \node[anchor=north east] at (0.9,0.9){CLICdp};
  \end{scope}
  \end{tikzpicture}
    \caption{}\label{fig:eff_timepix3_full}
  \end{subfigure}
  ~
  \begin{subfigure}[T]{.45\linewidth}
    \begin{tikzpicture}
  \node[anchor=south west,inner sep=0] at (0,0)(image){  \includegraphics[width=\linewidth]{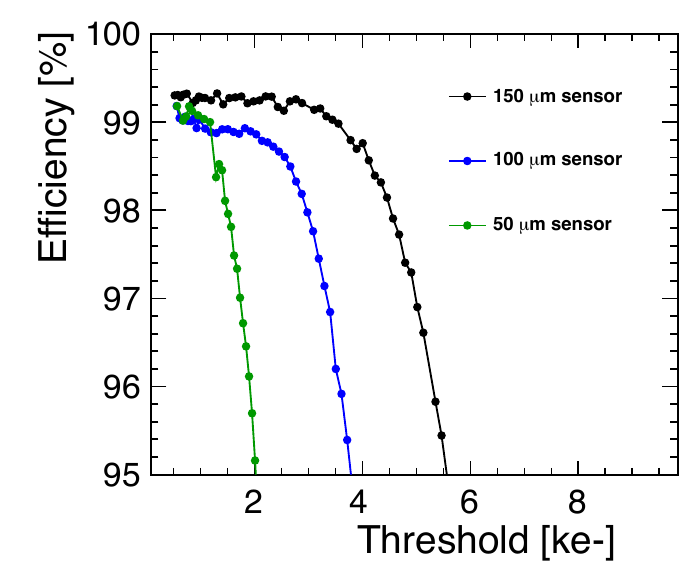}};
    \begin{scope}[x={(image.south east)},y={(image.north west)}]
     \node[anchor=north east] at (0.9,0.35){CLICdp};
  \end{scope}
  \end{tikzpicture}
    \caption{}\label{fig:eff_timepix3_zoom}
  \end{subfigure}
  \caption{\subref{fig:eff_timepix3_full} Global detection efficiency as a function of the threshold for Timepix3 assemblies with sensor thicknesses of \SI{50}{\micron}, \SI{100}{\micron} and \SI{150}{\micron}. \subref{fig:eff_timepix3_zoom} enlarged scale showing the efficiencies between \SI{95}{\percent} and \SI{100}{\percent}~\cite{ThesisNilou}.}\label{fig:eff_timepix3}
\end{figure}

The window of threshold values required to maintain high efficiency decreases as the sensor thickness is reduced. This implicates the necessity of low-noise readout electronics, allowing for a reduction of the detection threshold while simultaneously maintaining a low noise occupancy in the detector.

\subsubsection{Comparison of planar-sensor performance with simulations}\label{sec:allpix_validation}
The Timepix3 planar sensor performance measured in test-beam experiments has been compared to simulations using the \apsq simulation framework (\cref{sec:apx_2})~\cite{allpix-squared}.
The Geant4-based simulation includes the CLICdp Timepix3 telescope and the tested assemblies with thicknesses of \SI{50}{\micro\meter} and \SI{100}{\micro\meter}.
The reconstruction and analysis of both measured and simulated data is performed with the Corryvreckan framework (\cref{sec:corryvreckan})~\cite{corryvreckan-gitlab}.
It reproduces the results from an earlier study of the same data set~\cite{ThesisNilou}, which used the EUTelescope reconstruction framework~\cite{eutelescope-conf} and compared the data to Allpix simulations~\cite{benoit-thesis,allpix-twiki,allpix-github}. 

A comparison between the simulated and measured cluster size is depicted in \cref{fig:ap2data_cs}.
The cluster size and hence the charge sharing depend on the pixel geometry as well as the drift and diffusion in the sensor and the detection threshold.
The excellent agreement observed between data and simulation for both sensor thicknesses thus indicates a correct modelling of these dependencies in the simulation.

\begin{figure}[ht]
  \begin{subfigure}[T]{.45\linewidth}
    \begin{tikzpicture}
  \node[anchor=south west,inner sep=0] at (0,0)(image){  \includegraphics[width=\linewidth]{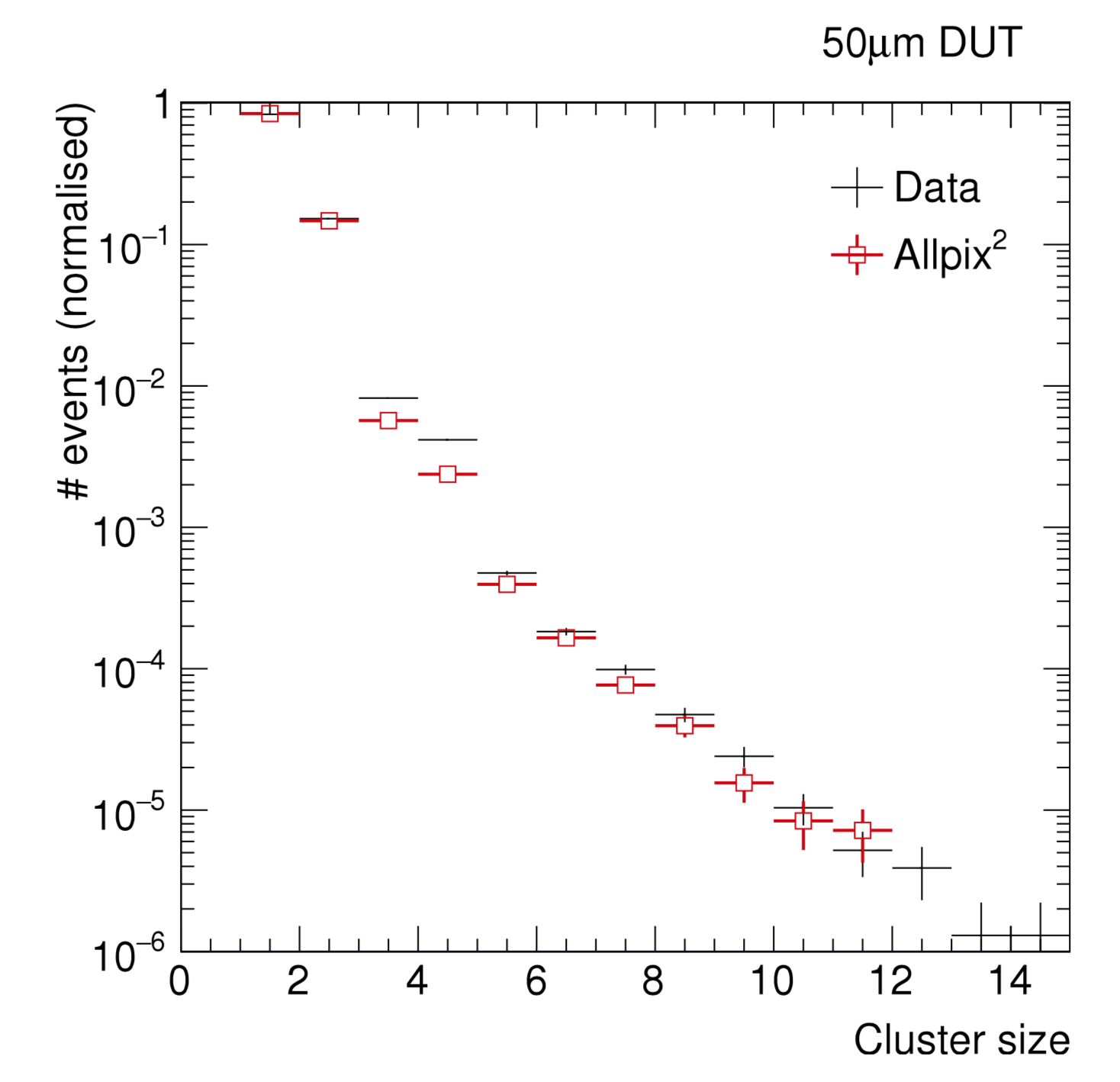}};
    \begin{scope}[x={(image.south east)},y={(image.north west)}]
     \node[anchor=north east] at (0.9,0.65){CLICdp};
  \end{scope}
  \end{tikzpicture}
    \caption{}\label{fig:ap2data_cs_50}
  \end{subfigure}
  \hfill
  \begin{subfigure}[T]{.45\linewidth}
    \begin{tikzpicture}
  \node[anchor=south west,inner sep=0] at (0,0)(image){  \includegraphics[width=\linewidth]{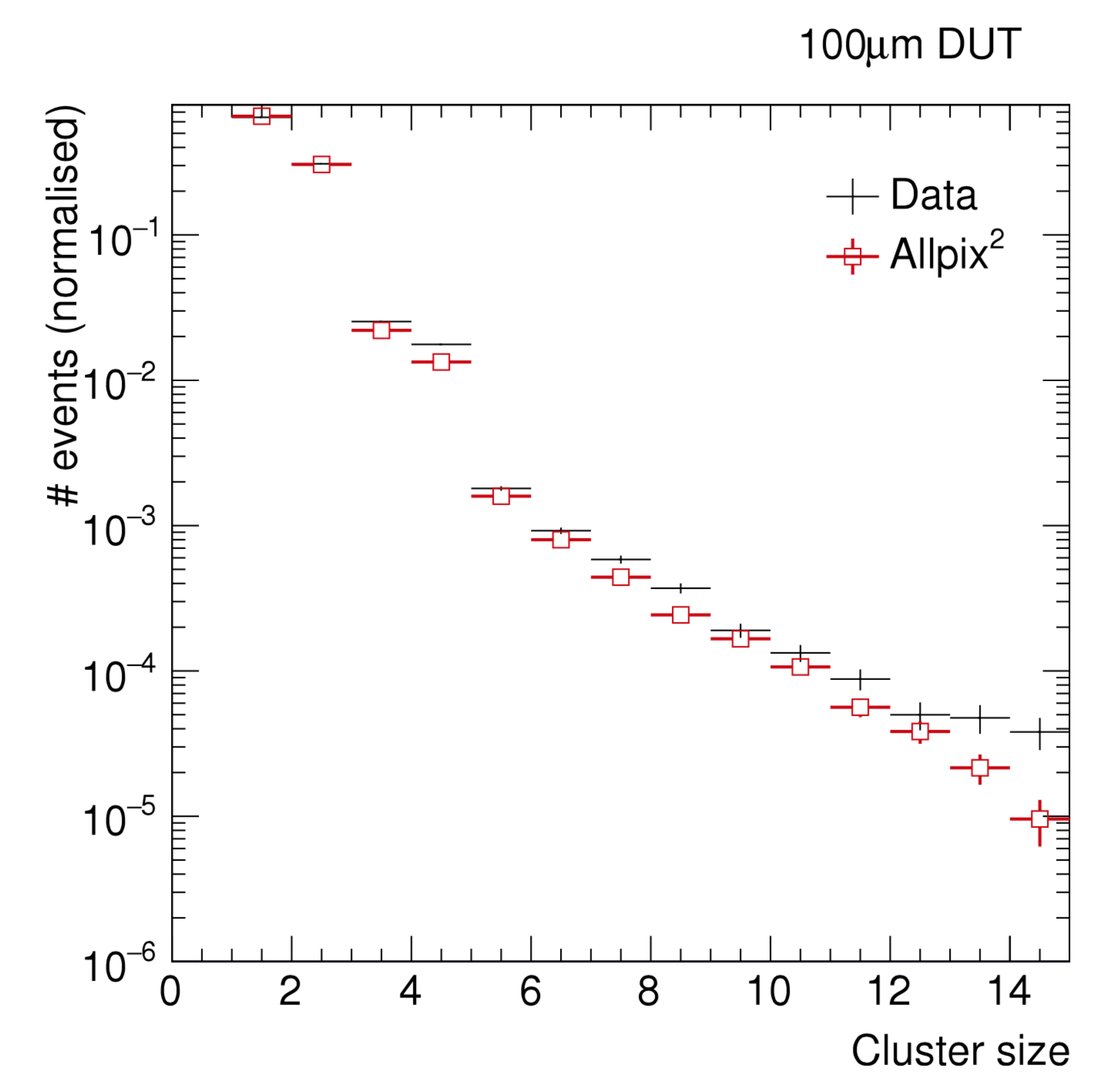}};
    \begin{scope}[x={(image.south east)},y={(image.north west)}]
     \node[anchor=north east] at (0.9,0.65){CLICdp};
  \end{scope}
  \end{tikzpicture}
    \caption{}\label{fig:ap2data_cs_100}
  \end{subfigure}
  \caption{Cluster size distributions for Timepix3 sensor data and \apsq simulations for a sensor thickness of \subref{fig:ap2data_cs_50} \SI{50}{\micro \meter} and  \subref{fig:ap2data_cs_100} \SI{100}{\micro \meter}~\cite{allpix-squared}.}
  \label{fig:ap2data_cs}
\end{figure}

\cref{fig:ap2data_res} shows the comparison between data and simulation for the spatial residual distributions.

\begin{figure}[ht]
  \begin{subfigure}[T]{.45\linewidth}
    \begin{tikzpicture}
  \node[anchor=south west,inner sep=0] at (0,0)(image){  \includegraphics[width=\linewidth]{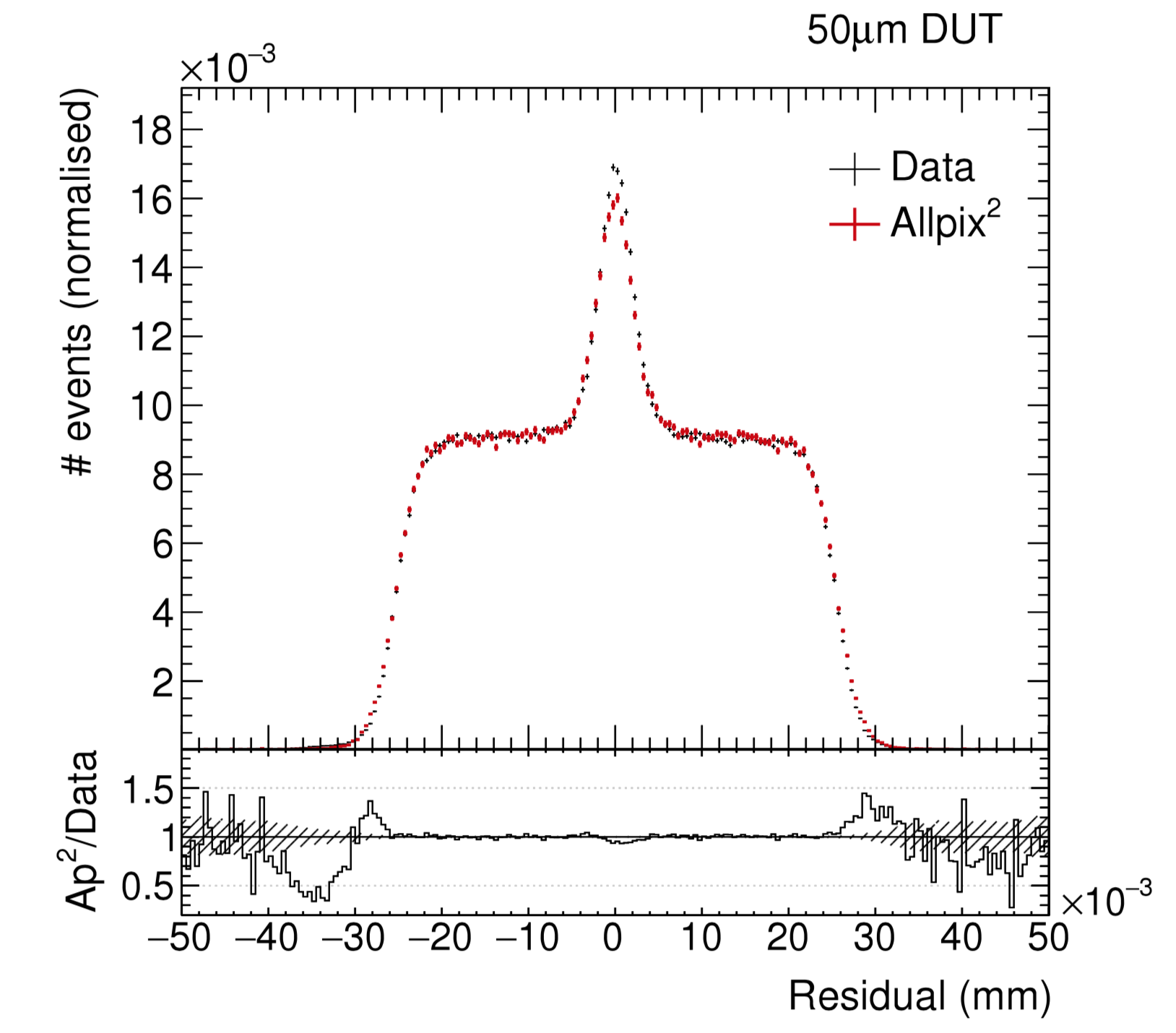}};
    \begin{scope}[x={(image.south east)},y={(image.north west)}]
     \node[anchor=north east] at (0.875,0.7){CLICdp};
  \end{scope}
  \end{tikzpicture}
    \caption{}\label{fig:ap2data_res_50}
  \end{subfigure}
  \hfill
  \begin{subfigure}[T]{.45\linewidth}
    \begin{tikzpicture}
  \node[anchor=south west,inner sep=0] at (0,0)(image){  \includegraphics[width=\linewidth]{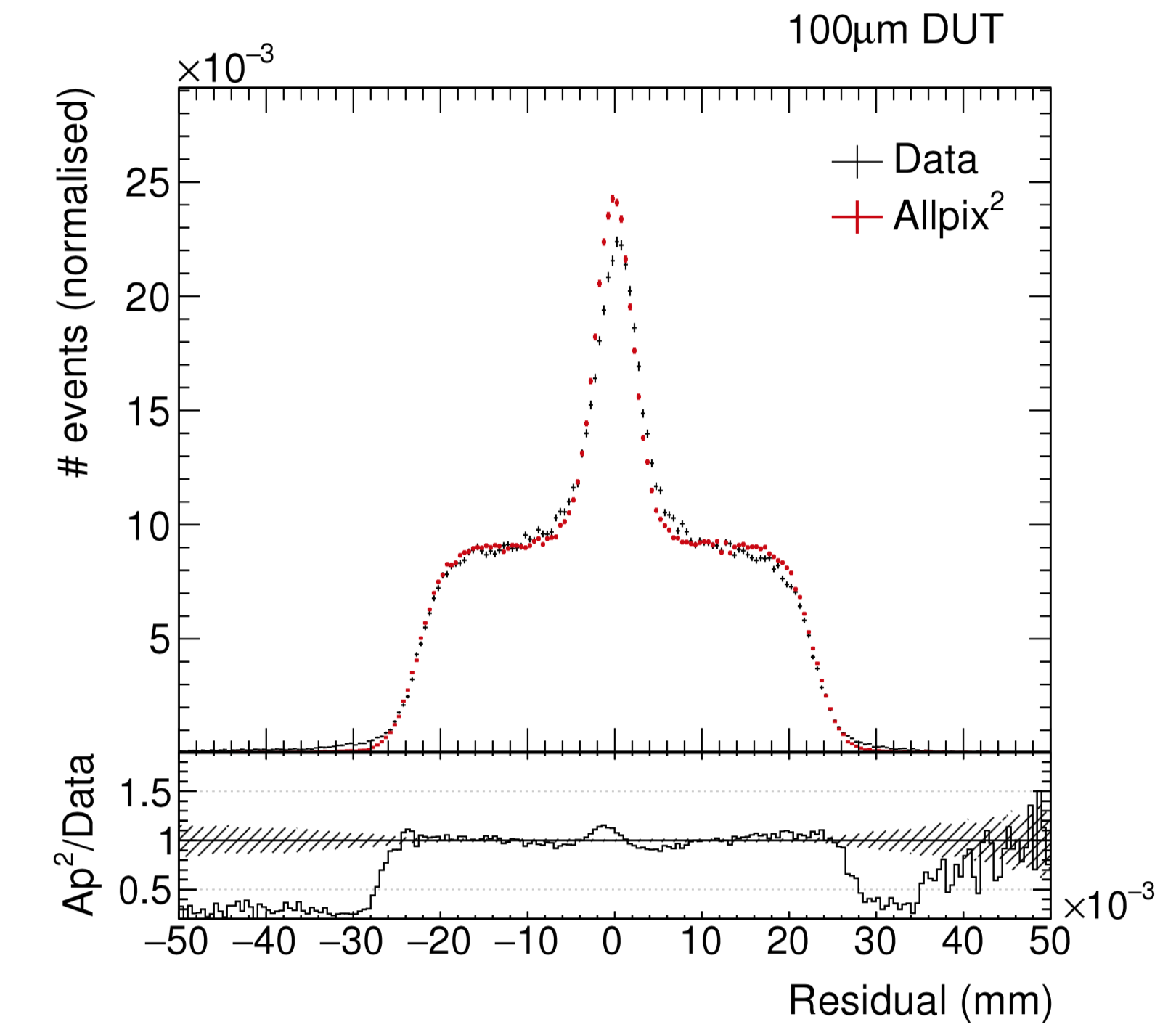}};
    \begin{scope}[x={(image.south east)},y={(image.north west)}]
     \node[anchor=north east] at (0.875,0.7){CLICdp};
  \end{scope}
  \end{tikzpicture}
    \caption{}\label{fig:ap2data_res_100}
  \end{subfigure}
  \caption{Residual distributions for Timepix3 sensor data and \apsq simulations for a sensor thickness of \subref{fig:ap2data_res_50} \SI{50}{\micro \meter} and \subref{fig:ap2data_res_100} \SI{100}{\micro \meter}~\cite{allpix-squared}.}
  \label{fig:ap2data_res}
\end{figure}

For both sensor thicknesses a good agreement between data and simulation has been found, which further shows that the reference is being modelled well in the simulation.

\subsubsection{Power-pulsing tests with Timepix3 assemblies}\label{sec:timepix3-pp}
Power-pulsing capabilities of the analogue and digital pixel front-end have been implemented in the Timepix3 ASIC, before CLIC specific front-end ASICs became available. 
Laboratory and test-beam measurements have been performed with a Timepix3 ASIC bump bonded to a \SI{50}{\micro \meter} thin n-on-p planar sensor.

For the analogue power pulsing, the three most power consuming nodes (the preamplifier and two of the discriminators) are multiplexed between the ON and OFF states, configurable through 3 independent DACon and DACoff values.
In the digital domain, the system and pixel matrix clocks are gated, such that they do not get distributed to the pixel matrix during the power-off state. This reduces the power consumption in the digital state-machines in the pixel and further reduces the power consumption of the chip, when combined with power pulsing in the analogue chip domain. 

The noise rate and hit efficiency of the ASIC in a power pulsing cycle has been investigated in laboratory measurements. The delay between the time when the power is enabled and the time when the Timepix3 shutter opens has been varied between \SIrange[range-phrase={~and~},range-units=repeat]{5}{10}{\micro\second}, as illustrated in \cref{fig:power_pulsing}.

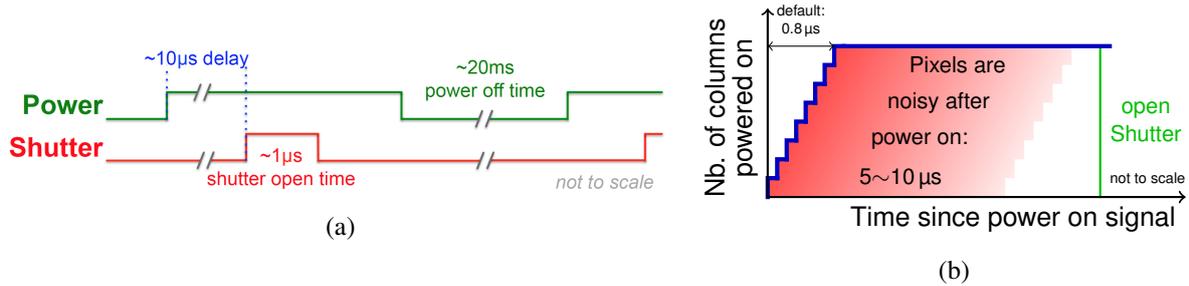
\begin{figure}[ht]
  \begin{subfigure}[T]{.575\linewidth}
  \includegraphics[width=\linewidth]{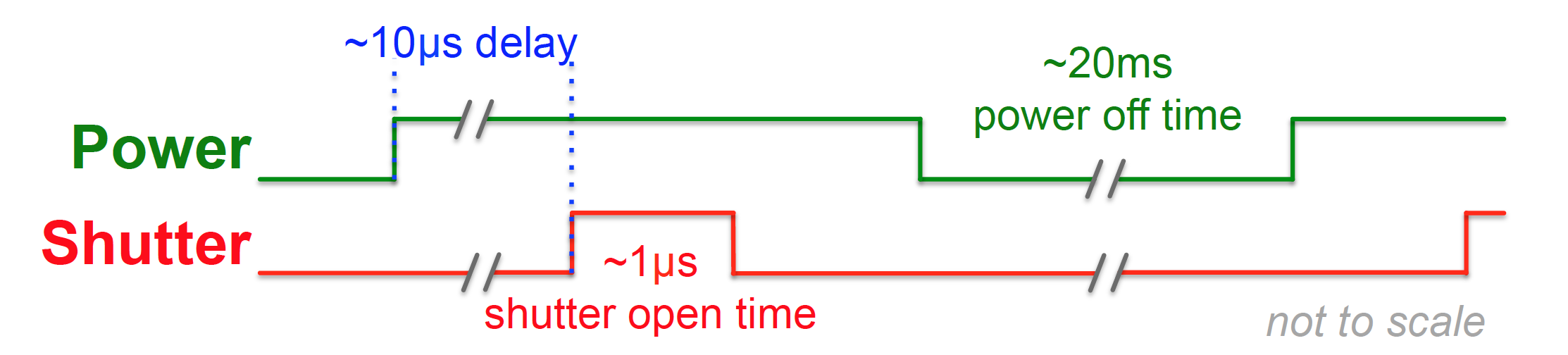}
  \caption{}\label{fig:power_pulsing}
\end{subfigure}
\hfill
\begin{subfigure}[T]{.425\linewidth}
  \begin{tikzpicture}[font=\sffamily,line cap=rect]

  \shade[left color=red!80!white, right color=white, shading=axis,shading angle=75] (0,0)|-++(0.125,0.25)
  |-++(0.125,0.25)
  |-++(0.125,0.25)
  |-++(0.125,0.25)
  |-++(0.125,0.25)
  |-++(0.125,0.25)
  |-++(0.125,0.25)
  |-++(0.125,0.25)
  --++(3,0)
  |-++(-0.125,-0.25)
  |-++(-0.125,-0.25)
  |-++(-0.125,-0.25)
  |-++(-0.125,-0.25)
  |-++(-0.125,-0.25)
  |-++(-0.125,-0.25)
  |-++(-0.125,-0.25)
  |-++(-0.125,-0.25)
    ;
  \foreach \i in {0,.25,0.5,...,1.75}
  {
        \draw[blue!75!black,ultra thick](\i/2,\i)--++(0,0.25)--+(0.125,0);
  }

  \draw[green!75!black,thick](4.375,0)--++(0,2)node[right,pos=0.5]{\footnotesize\shortstack{open\\Shutter}};

  \draw[<->](0,2)--+(.875,0) node[above,pos=0.5]{\tiny \shortstack{default:\\\SI{0.8}{\micro\second}}};

  \draw[blue!75!black,ultra thick](1,2)--++(3.5,0);

  \node[] at (2.5,1.75){\footnotesize Pixels are};
  \node[] at (2.25,1.25){\footnotesize noisy after};
  \node[] at (2,.75){\footnotesize power on:};
  \node[] at (1.75,0.25){\footnotesize \SIrange[range-phrase=$\sim$,range-units=single]{5}{10}{\micro\second}};

\node at (5,.25){\tiny not to scale};

\draw[->,thick](0,0)--(0,2.5)node[above,rotate=90,anchor=south east]{\shortstack{Nb. of columns\\powered on}};
\draw[->,thick](0,0)--(5.5,0) node[below,anchor=north east]{Time since power on signal};
\end{tikzpicture}
  \caption{}\label{fig:timepix3_pp_switch_on_sketch}
\end{subfigure}
  \caption{\subref{fig:power_pulsing} Illustration of the power pulsing duty cycle for CLIC. \subref{fig:timepix3_pp_switch_on_sketch} Detailed illustration of the power-on sequence to activate different column groups in the pixel matrix (blue line). The red shaded area corresponds to the approximate period in which the respective group of columns exhibits an increased noise level. After that period, the frontend is stable and the shutter is opened.}\label{fig:pp_scheme}
\end{figure}

The results of the laboratory measurements are summarised in \cref{fig:tpx_pp_lab_results}. To determine the minimal delay time between the power-on transition and the shutter-open signal, noise and signal measurements have been performed while the ASIC has been operated in a power-pulsed mode with a duty cycle of \SI{50}{\percent}.
\cref{fig:pp_hitcount} shows the number of detected hits within \SI{100}{\second} as a function of the shutter-open delay, measured with and without a \isotope[90]{Sr} source placed above the sensor. The measurement has been performed in two different operation modes: using analogue power pulsing only (APP) and using analogue and digital power pulsing simultaneously (ADPP).
Similar results have been measured for both operation modes.
Without using the \isotope[90]{Sr} source, the number of noise hits is monotonically decreasing
and is close to zero for a shutter open delay of \SI{8}{\micro\second} and longer. At shorter delay times, the  outputs of the front-end amplifiers are not yet settled into a steady signal, and thus may cross the threshold discriminator level. Noise hits are injected into the digital part of the detector.
The measurement with the \isotope[90]{Sr} source shows a stabilisation of the hit count at around \SI{6}{\micro\second} after the power has been switched on. At shorter delay times, the detector is dominated by noise effects.

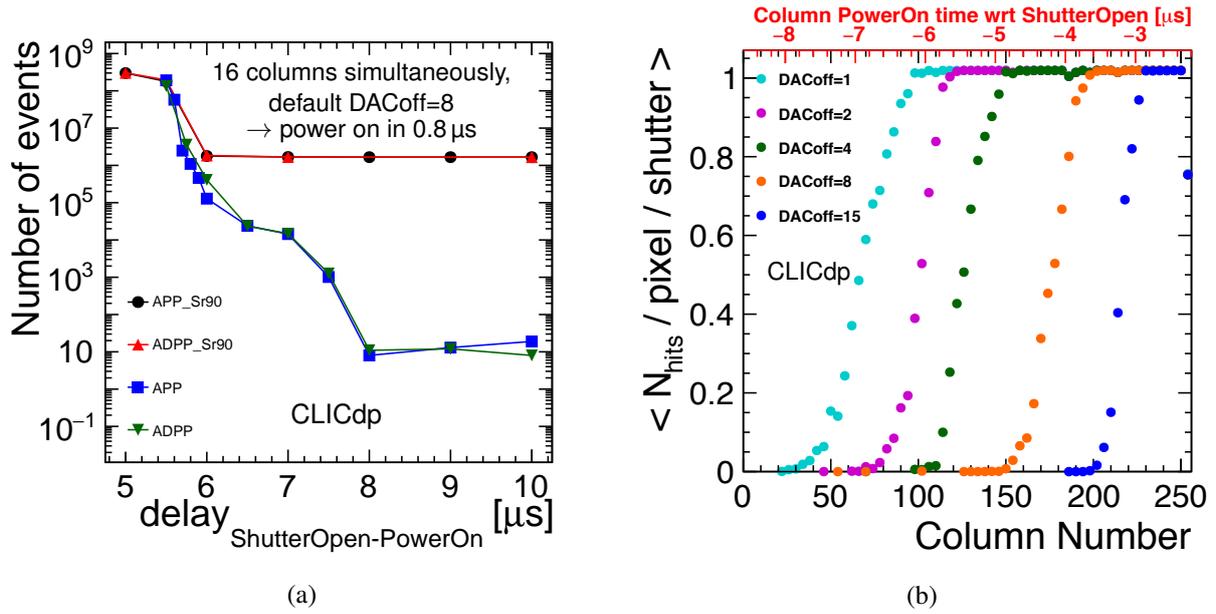
\begin{figure}[ht]
\begin{subfigure}[T]{.49\linewidth}
  \begin{tikzpicture}
\node[anchor=south west,inner sep=0] at (0,0)(image){  \includegraphics[width=\linewidth]{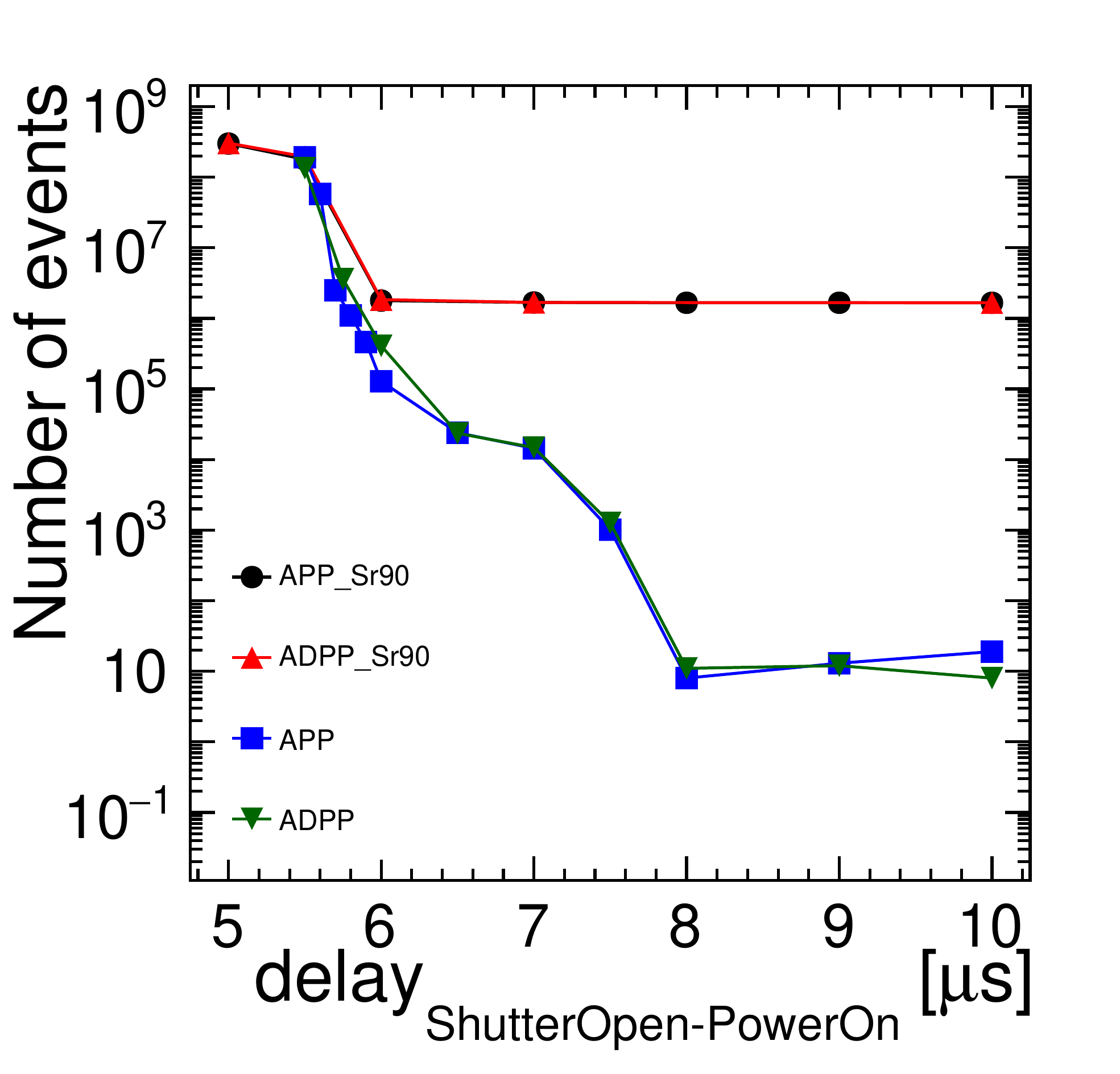}};
  \begin{scope}[x={(image.south east)},y={(image.north west)}]
   \node[anchor=north east] at (0.65,0.3){CLICdp};
   \node[anchor=north] at (0.6,0.9){\small\shortstack{16 columns simultaneously,\\default DACoff=8\\$\rightarrow$ power on in \SI{0.8}{\micro\second}}};
\end{scope}
\end{tikzpicture}
  \caption{}\label{fig:pp_hitcount}
\end{subfigure}
  \hfill
  \begin{subfigure}[T]{.49\linewidth}
    \begin{tikzpicture}
  \node[anchor=south west,inner sep=0] at (0,0)(image){  \includegraphics[width=\linewidth]{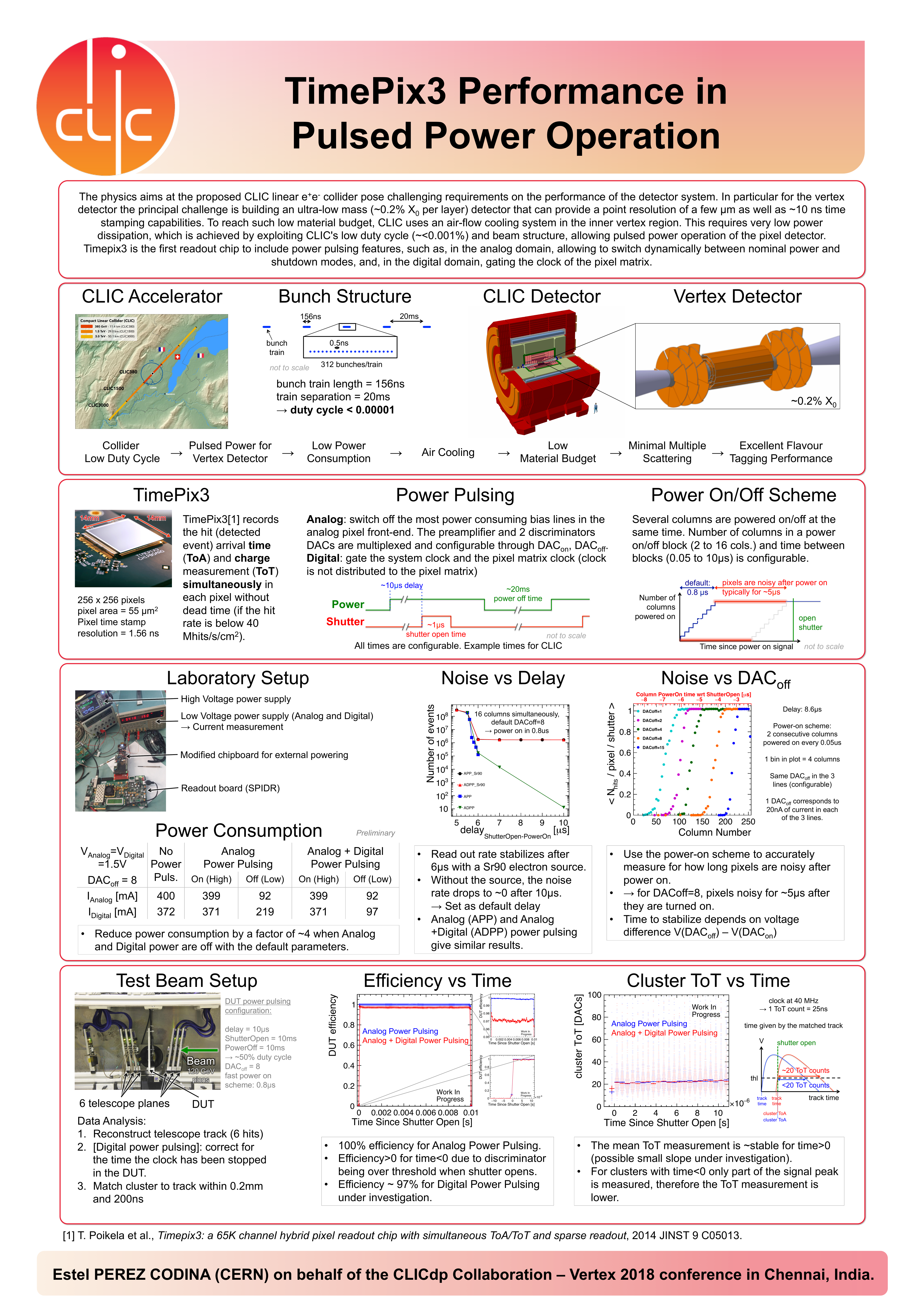}};
    \begin{scope}[x={(image.south east)},y={(image.north west)}]
     \node[anchor=north west] at (0.225,0.55){\small CLICdp};
  \end{scope}
  \end{tikzpicture}
    \caption{}\label{fig:pp_column_turnon}
  \end{subfigure}
  \caption{\subref{fig:pp_hitcount} Detected number of hits within \SI{100}{\second} as a function of the time delay between switching to the power-on state and the shutter open of the Timepix3 ASIC using analogue power-pulsing only (APP) and analogue and digital power-pulsing simultaneously (ADPP). The noise rate has initially been determined without particle illumination (blue and green curves). To estimate the hit efficiency, the detector has been illuminated by a \isotope[90]{Sr} source (red and black curves). The duty cycle of the pulsed power operation is \SI{50}{\percent{}}, with a shutter-open time of \SI{1}{\ms}. \subref{fig:pp_column_turnon} Noise occupancy as a function of the pixel column number~\cite{pp_estel}. The power is switched on individually for each double-column with a delay of \SI{50}{\ns}. For high column numbers, the delay between power-on and shutter-open is reduced, and thus the noise occupancy is increased. The measurement is shown for different settings of \mbox{DACoff}, i.e.\ for different current levels in the power-off state.}\label{fig:tpx_pp_lab_results}
\end{figure}

The transition from the low-power to the operational state happens consecutively in a fraction of columns at a time (configurable from 2 to 16 columns) and with configurable delay from one group to the next, as sketched in \cref{fig:timepix3_pp_switch_on_sketch}. This prevents spikes in power consumption during the transition to the powered state. \cref{fig:pp_column_turnon} illustrates the noise occupancy per pixel and frame as a function of the pixel column number. The delay between the start of the power-on sequence and the shutter open signal is \SI{8.6}{\micro\second}. The various curves correspond to various settings of the DAC that is defining the power consumption in the off-state. To best demonstrate the effect, only two consecutive columns are powered on every \SI{0.05}{\micro\second}, resulting in a sizeably different time delay to the shutter opening point for each double-column in the ASIC. As demonstrated before, the delay has to be of the order of \SI{10}{\micro\second} to effectively suppress noise injection into the pixel front-end. This is the reason why columns with high number show high noise occupancy while columns with low number are already in a stable state and show no noise hits. In addition, it can be seen that the settling time to recover from the power-off state is increasing, as the biasing of the front-end (DACoff) is decreased.

The power consumption of the ASIC has been measured separately for the power-on and power-off states. When fully powered, a consumption of about \SI{1.1}{\watt} has been measured, and a reduction of the power consumption by a factor of approximately 4 has been achieved when the analogue and digital power of the ASIC are switched off. Due to the low duty cycle operation in the CLIC detector, the power consumption in the power-off state would be the dominant contribution to the total power budget.

Results from test-beam measurements obtained using the Timepix3 telescope as reference system for particle tracking show a stable hit efficiency and ToT distribution under power-pulsed operation~\cite{pp_estel}. For these measurements the delay between the transition to power-on and the shutter opening has been fixed to \SI{10}{\micro\second}. Both analogue and analogue+digital power pulsing have been studied. 

\subsubsection{Test-beam performance of CLICpix planar-sensor assemblies}\label{sec:clicpix_planar_resolution}
The performance of \SI{25}{\micron}-pitch hybrid detectors with thin planar sensors is assessed using CLICpix. Assemblies with \SI{50}{\micron} and \SI{200}{\micron} thick sensors bump-bonded to CLICpix ASICs (see \cref{sec:clicpix-bump-bonding}) were exposed to a \SI{120}{\giga\electronvolt} \PGpp{} beam in the H6 beamline at the CERN SPS.
The \SI{200}{\micron} assemblies have been tested in the AIDA telescope~\cite{Rubinski2014,Jansen2016}, the \SI{50}{\micron} assembly (shown in \cref{fig:50um_assembly_photo}) in the Timepix3-based telescope (see \cref{sec:timepix3_telescope}).

 \cref{fig:clicpix_200um_30V_signal} shows the MPV of the cluster signal of the \SI{200}{\micron} CLICpix assembly, obtained as a function of the applied bias voltage. The deposited energy reaches a plateau value at full depletion around \SI{30}{\volt}, with a most probable energy deposition of approximately \SI{16000}{\Pem{}}.

The large average cluster signal in this assembly leads to a significant fraction of multi-pixel clusters, shown in \cref{fig:biasscan_fraction} as a function of the applied bias voltage.
At low bias voltages, the active volume is increasing. This leads to a longer drift time of the charge carriers and a larger lateral diffusion before they are collected at the readout electrodes.
Diffusion increases the probability to lift the signal in neighbouring pixels above the detection threshold of the readout ASIC and thus the probability to obtain a multi-pixel cluster increases.
As soon as the full bulk volume is depleted, a further increase of the voltage increases the electric field in the sensor, reducing the drift time and thereby the lateral diffusion, and the cluster size decreases again.
The maximal fraction of multi-pixel clusters is therefore reached around the full depletion voltage of \SI{30}{\volt}.

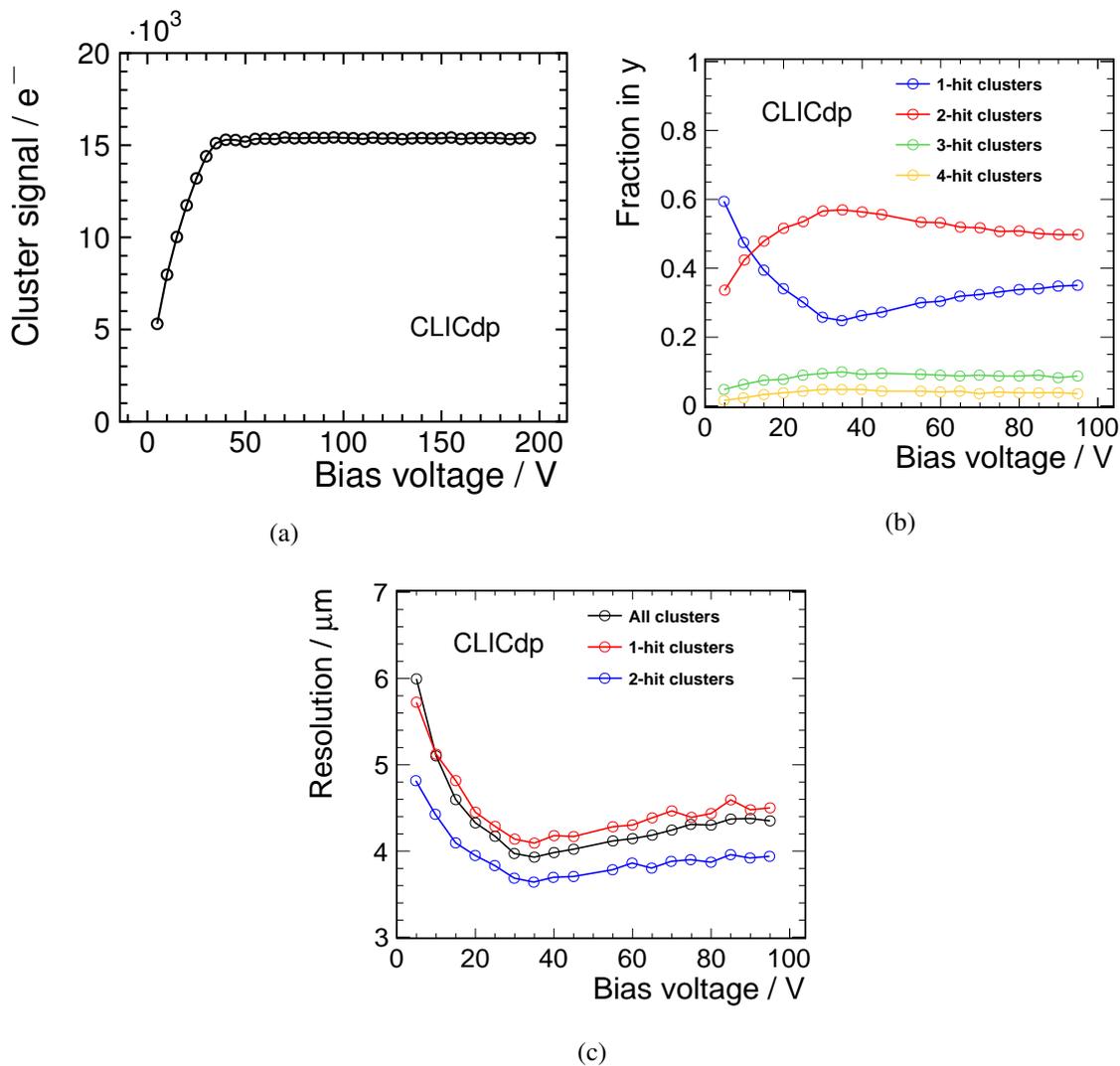
\begin{figure}[!htb]
  \centering
  \begin{subfigure}[T]{0.49\linewidth}
    \centering
    \begin{tikzpicture}
  \begin{axis}[
    /pgf/number format/1000 sep={},
    scaled y ticks=base 10:-3,
    axis line style={black,thick},
    ticklabel style={black,font={\sansmath\sffamily\large}},
		xlabel= Bias voltage / \si{\volt},
		ylabel = Cluster signal / \Pem{},
    ylabel near ticks,
    xlabel near ticks,
		label style={black,font={\sansmath\sffamily\Large}},
    xlabel style={at={(rel axis cs:1,0)}, anchor=north east,yshift=-.4cm},
    ylabel style={at={(rel axis cs:0,1)}, anchor=south east,yshift=.85cm},
    width=7.5cm,
    height=6.5cm,
    ymin=0,
    ymax=20000,
    minor y tick num={4},
    minor x tick num={4},
    every tick/.style={black,thick},
    every axis/.append style={font=\footnotesize},
    legend style={draw=none,text=black, fill=none},
    legend style={text=black},
    legend style={at={(0.95,0.05)},anchor=south east,nodes=right,font={\sansmath\sffamily\footnotesize}},
    ]

		\addplot[mark=o,thick,color=black] plot[] table[
		x=V,
		y expr={\thisrow{Clustersignal}*1.33}
		]{sections/VertexTracking/figures/clicpix/clustersignal_200um.dat}; 

   \node at (rel axis cs:0.75,0.25) {\normalsize CLICdp};
  \end{axis}
\end{tikzpicture}
    \caption{}\label{fig:clicpix_200um_30V_signal}
  \end{subfigure}
  \hfill
  \begin{subfigure}[T]{0.49\textwidth}
    \centering
  
    \begin{tikzpicture}
  \node[anchor=south west,inner sep=0] at (0,0)(image){  \includegraphics[width=\linewidth]{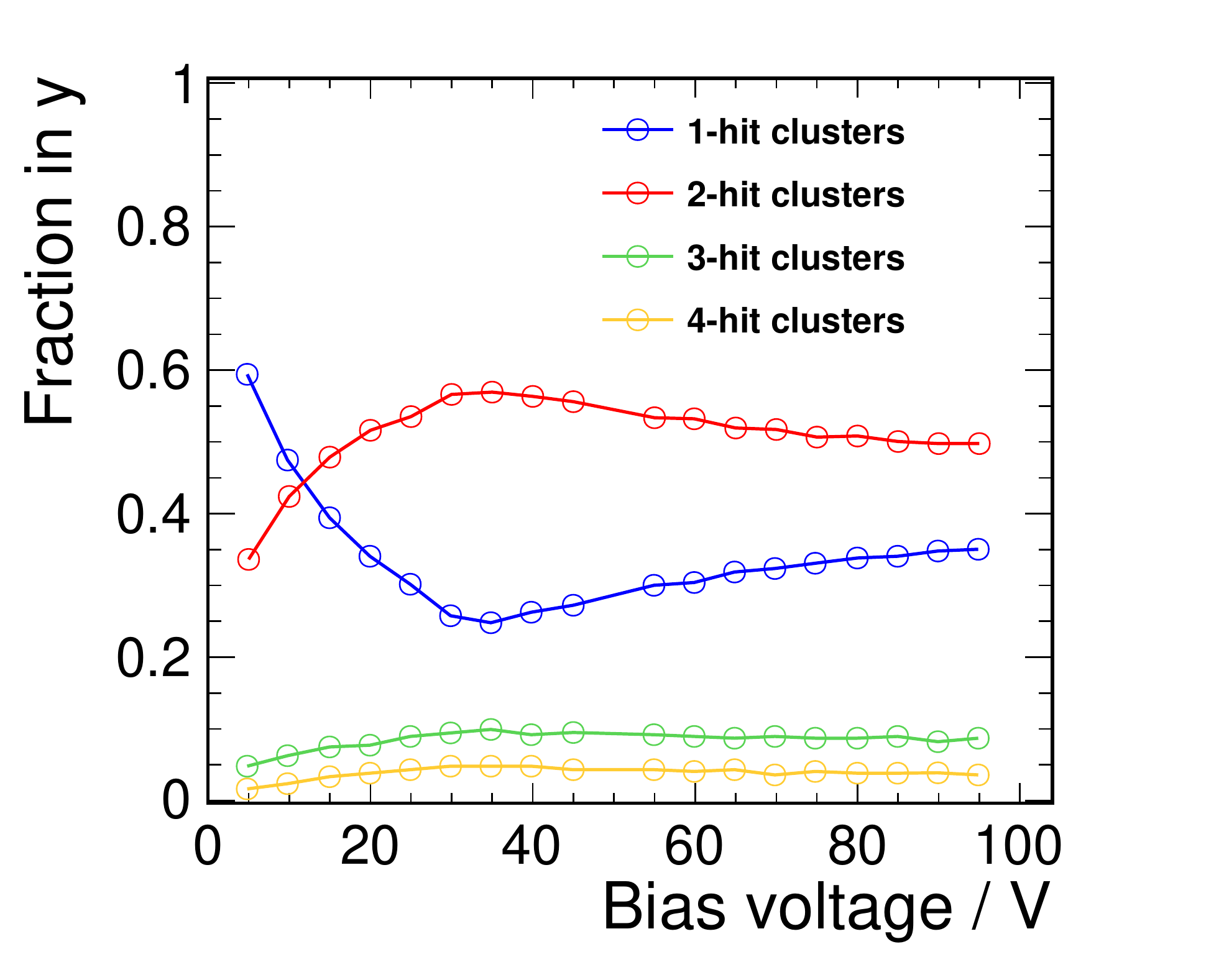}};
    \begin{scope}[x={(image.south east)},y={(image.north west)}]
     \node[anchor=north west] at (0.25,0.85){CLICdp};
  \end{scope}
  \end{tikzpicture}
    \caption{}\label{fig:biasscan_fraction}
  \end{subfigure}
  
  \begin{subfigure}[T]{.49\textwidth}
    \begin{tikzpicture}
  \node[anchor=south west,inner sep=0] at (0,0)(image){	\includegraphics[width=\linewidth]{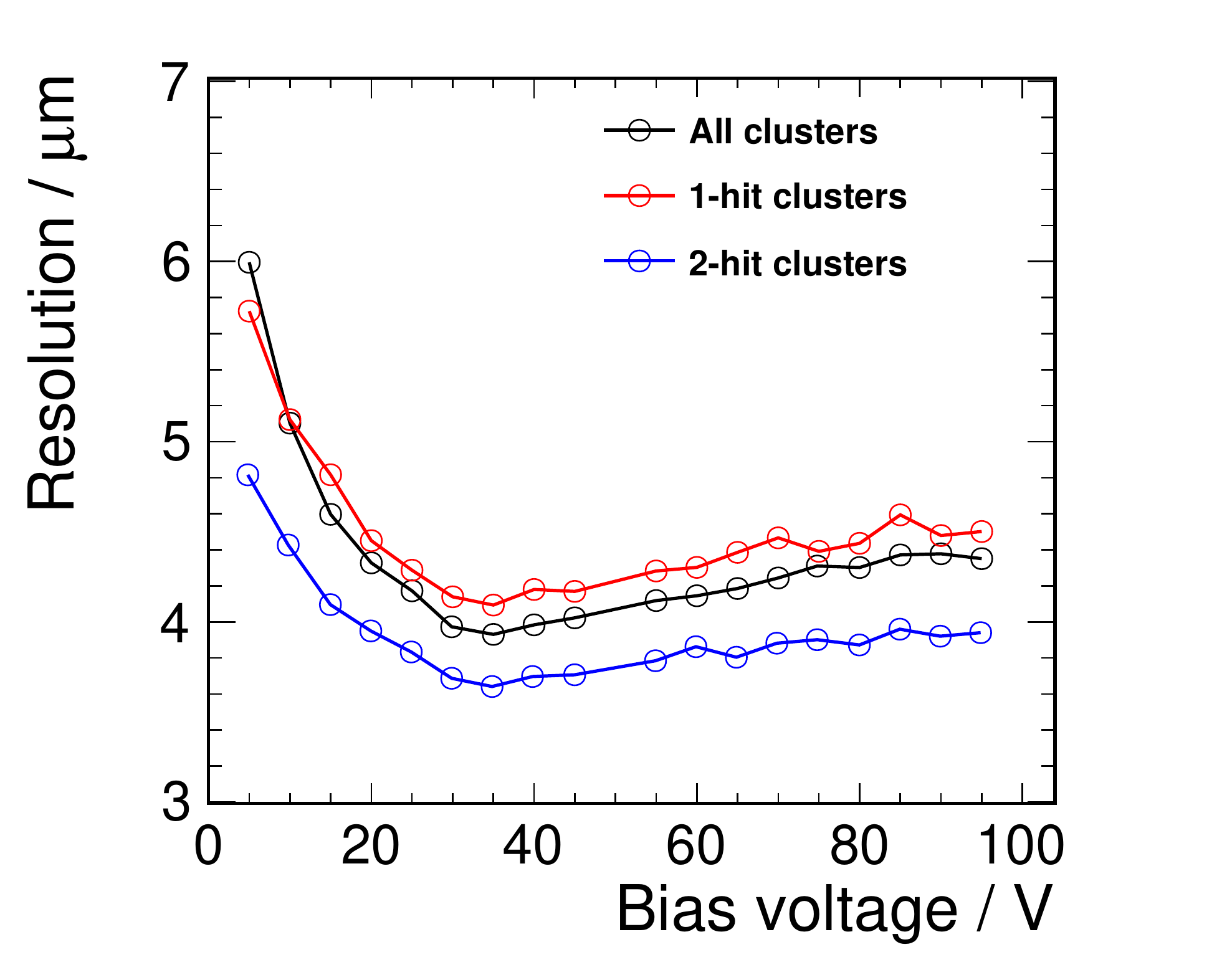}};
    \begin{scope}[x={(image.south east)},y={(image.north west)}]
     \node[anchor=north west] at (0.25,0.85){CLICdp};
  \end{scope}
  \end{tikzpicture}
  	\caption{}\label{fig:biasscan_resolution}
  	\end{subfigure}
  \caption{\subref{fig:clicpix_200um_30V_signal} Most-probable-value of the cluster signal in the \SI{200}{\micron} planar CLICpix assembly, \subref{fig:biasscan_fraction} fractional cluster multiplicity and \subref{fig:biasscan_resolution} spatial resolution as a function of the applied sensor bias voltage.}\label{fig:clicpix_200um_cluster-signal-fraction}
\end{figure}
\cref{fig:biasscan_resolution} shows the position resolution as a function of the applied bias voltage, obtained as the width of a Gaussian fit to the respective residual distributions. The most accurate position reconstruction is obtained for the voltage corresponding to the maximum in the two-pixel cluster distribution (\cref{fig:biasscan_fraction}). Therefore the spatial resolution of the device reaches a minimum of about \SI{4}{\micron} at this voltage. This number includes the uncertainty of the track-position prediction from the telescope of about \SI{2}{\micron}~\cite{Jansen2016}. When including the non-Gaussian tails, the RMS within a range covering 99.7\% (mean $\pm3\sigma$) of the distribution (RMS99.7) increases to \SI{6.9}{\micron}.
When unfolding the track uncertainty from the measurement result, the obtained spatial resolution of the \SI{200}{\micron} assembly with \SI{25}{\micron} pixel pitch is around \SI{3.5}{\micron}, close to fulfilling the requirement on the spatial resolution for the vertex detector at CLIC. The material content of about \SI{0.2}{\percent X_0} of the sensor alone is taking the full available budget for the detection layer, leaving no margin for the ASIC, support and services. Thus even thinner sensors down to \SI{50}{\micron} need to be considered.

The position reconstruction for multi-pixel clusters profits from charge weighting of the individual pixel contributions.
In \cref{fig:clustersize_res_50um_200um}, the distributions of the spatial residuals and the cluster size for the \SI{50}{\micron} assembly are compared to the corresponding distributions obtained with the \SI{200}{\micron} thick sensor. A Gaussian fit to the residuals obtained with the \SI{50}{\micron} thin sensor results in a resolution of \SI{7.9}{\micron}, the RMS99.7 is \SI{8.6}{\micron}. This is close to the binary expectation for single-pixel hits at \SI{25}{\micron} pixel pitch. As already seen during the investigation of the Timepix3 detector assemblies, the cluster multiplicity is reduced in thin sensors, due to the shorter drift path and reduced diffusion of the drifting charge cloud. Even at the small pixel size of \SI{25}{\micron}, the available signal in the \SI{50}{\micron} thin sensor and the amount of diffusion is not large enough to result in a sizeable fraction of multi-pixel clusters. The mean cluster size is \num{1.1}, and thus, the spatial resolution is still dominated by the pixel geometry.

\begin{figure}[ht]
  \centering
\begin{subfigure}[b]{0.49\linewidth}
  \centering
  \begin{tikzpicture}
\node[anchor=south west,inner sep=0] at (0,0)(image){\includegraphics[width=\linewidth]{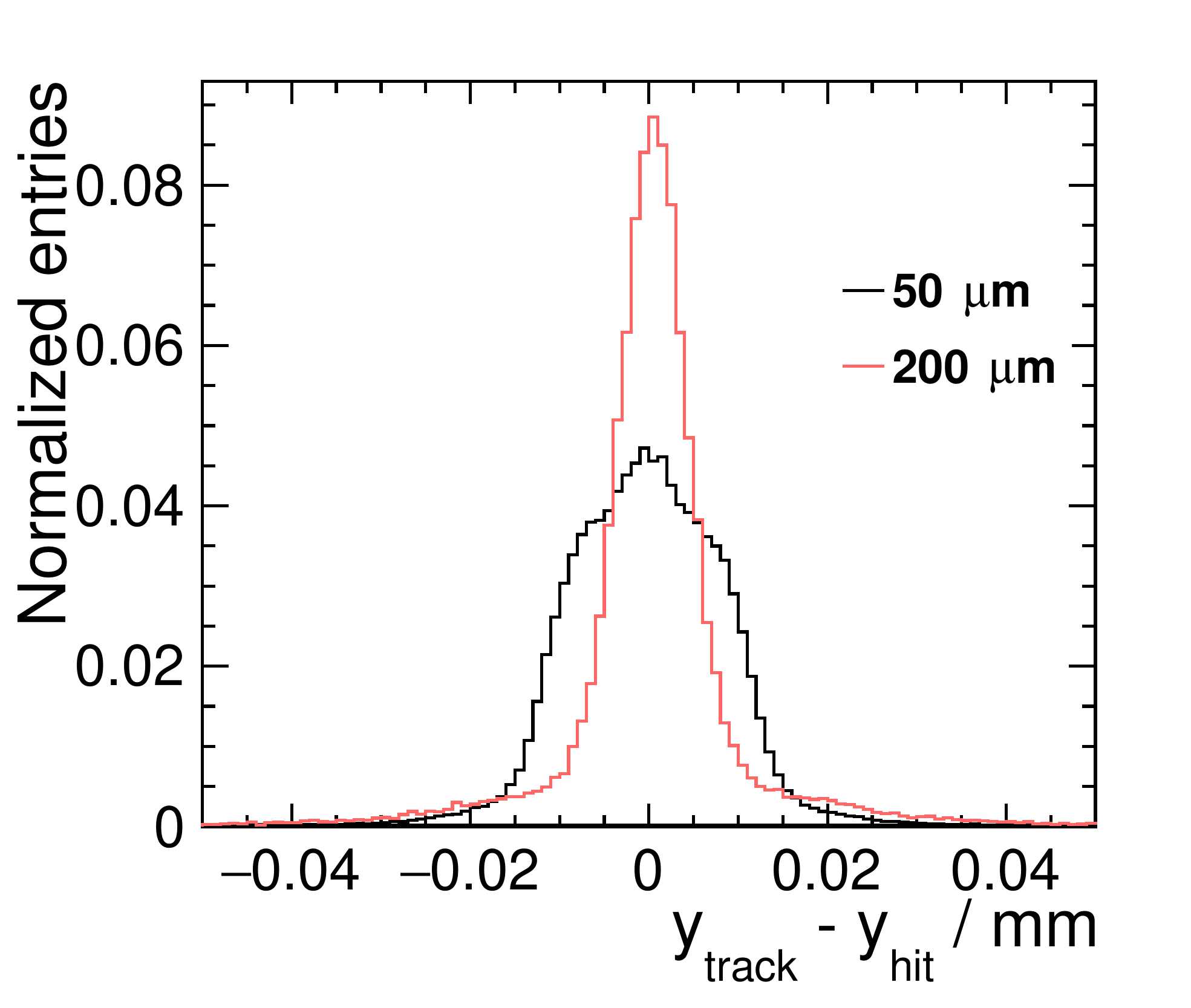}};
  \begin{scope}[x={(image.south east)},y={(image.north west)}]
   \node[anchor=north east] at (0.45,0.7){\textcolor{black}{\shortstack[l]{RMS99.7=\\ \SI{8.6}{\micron}}}};
   \node[anchor=north east] at (0.45,0.55){\textcolor{red}{\shortstack[l]{RMS99.7=\\\SI{6.9}{\micron}}}};

 \node[anchor=north east] at (0.875,0.85){CLICdp};

\end{scope}
\end{tikzpicture}
  \caption{}\label{fig:res_200_50}
\end{subfigure}
\hfill
\begin{subfigure}[b]{0.49\linewidth}
  \centering
  \begin{tikzpicture}
\node[anchor=south west,inner sep=0] at (0,0)(image){\includegraphics[width=\linewidth]{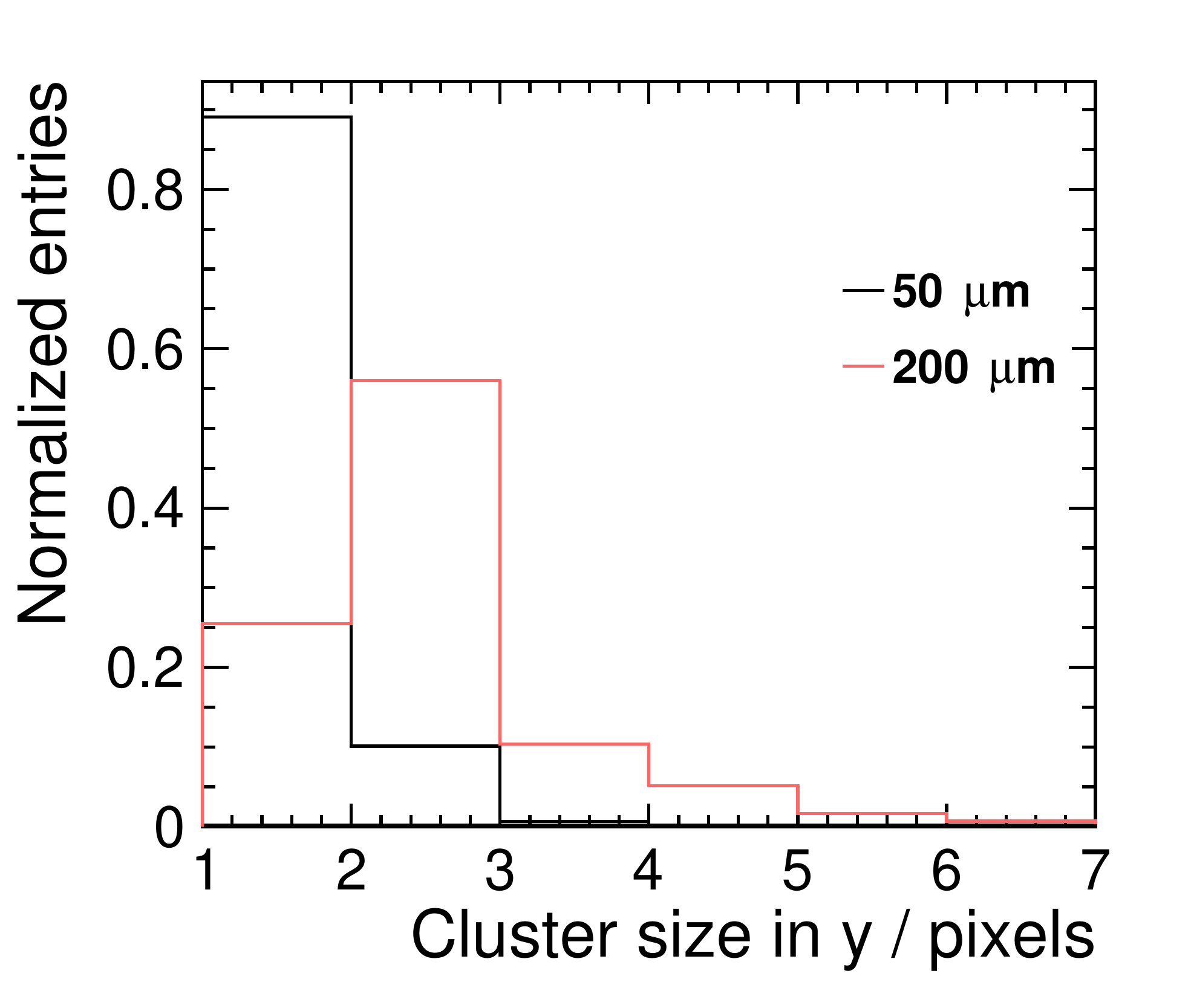}};
  \begin{scope}[x={(image.south east)},y={(image.north west)}]
   \node[anchor=north east] at (0.85,0.55){\textcolor{black}{mean=1.1}};
   \node[anchor=north east] at (0.85,0.45){\textcolor{red}{mean=2.0}};

 \node[anchor=north east] at (0.875,0.875){CLICdp};

\end{scope}
\end{tikzpicture}
\caption{}\label{fig:cs_200_50}
\end{subfigure}
	\caption{Comparison of the distributions of \subref{fig:res_200_50} spatial residual and  \subref{fig:cs_200_50} cluster size of the \SI{50}{\micron} and \SI{200}{\micron} thin sensor assembly.}\label{fig:clustersize_res_50um_200um}
\end{figure}

The use of a beam telescope with good tracking resolution allows the study of the sensor response as a function of the track intercept on the sensor surface with sub-pixel resolution. A geometrical representation of the sharing of charge is shown in \cref{fig:clicpix_inpixel_cs_200um,fig:clicpix_inpixel_cs_50um}, where the in-pixel hitmap for different cluster sizes is shown for the \SI{200}{\micron} thick assembly and the \SI{50}{\micron} thin assembly. In the \SI{200}{\micron} thick sensor, only tracks passing close to the pixel centre produce single pixel clusters, while tracks passing further towards the pixel edges and corners result in higher cluster multiplicities. Clusters formed of two or more pixels allow for an interpolation of the track impact position, and thus lead to the good observed spatial resolution. For the thin assembly, single pixel hits occur almost across the full pixel cell, restricting the spatial resolution close to the binary limit.

\begin{figure}[!htb]
	\centering
	\foreach{\i} in {1,2,3,4}
	{
	\begin{subfigure}[b]{.23\linewidth}
    		\begin{tikzpicture}
 	  	\node[anchor=south west,inner sep=0] at (0,0)(image){	\includegraphics[width=\linewidth]{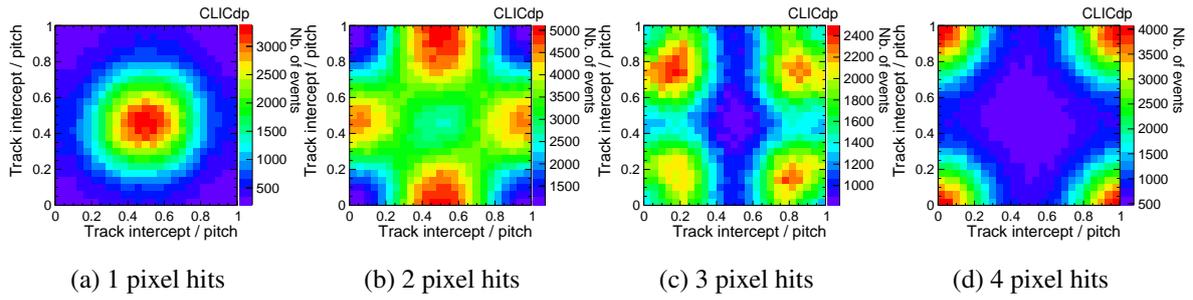}};
  		  \begin{scope}[x={(image.south east)},y={(image.north west)}]
        \node[anchor=south west,rotate=-90]at(0.975,0.95){\tiny\sffamily Nb. of events};
 	   	   \node[anchor=south west] at (0.65,0.9){\tiny CLICdp};
 		\end{scope}
 		\end{tikzpicture}
		\caption{\i~pixel hits}
		\end{subfigure}
}
		\caption{Track intercept within the pixel cell of the \SI{200}{\micron} planar CLICpix assembly for different cluster sizes. 
    }\label{fig:clicpix_inpixel_cs_200um}
	\end{figure}
  
  \begin{figure}[!htb]
    \centering
    \foreach{\i} in {1,2,3,4}
    {
    \begin{subfigure}[b]{.23\linewidth}
          \begin{tikzpicture}
        \node[anchor=south west,inner sep=0] at (0,0)(image){	\includegraphics[width=\linewidth]{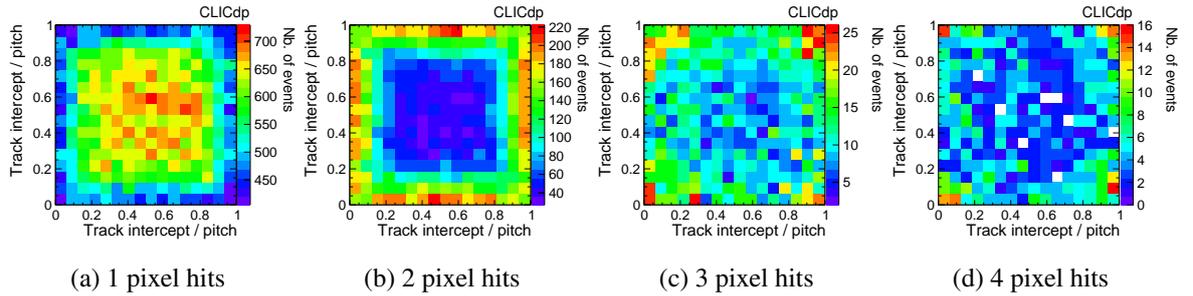}};
          \begin{scope}[x={(image.south east)},y={(image.north west)}]
            \node[anchor=south west,rotate=-90]at(0.95,0.95){\tiny\sffamily Nb. of events};
           \node[anchor=south west] at (0.65,0.9){\tiny CLICdp};
      \end{scope}
      \end{tikzpicture}
      \caption{\i~pixel hits}
      \end{subfigure}
  }
      \caption{Track intercept within the pixel cell of the \SI{50}{\micron} planar CLICpix assembly for different cluster sizes. 
      }\label{fig:clicpix_inpixel_cs_50um}
    \end{figure}
    
The detection efficiency as a function of the threshold is summarised in \cref{fig:clicpix_thresholdscan} for both sensor thicknesses. The efficiency in the \SI{50}{\micron} thin sensor is around \SI{95}{\percent}, at the lowest achievable threshold for that particular detector assembly of around \SI{1300}{\Pem{}}. Increasing the threshold from the nominal value, the detection efficiency of the assembly is degrading almost immediately, due to the low signal level. The most probable ionisation signal in a \SI{50}{\micron} silicon sensor is approximately \SI{3300}{\Pem{}}. It is thus desirable to reach threshold values below \SI{1000}{\Pem{}} for better efficiency. This emphasises the need for a low noise in the readout ASIC, in order to be able to set the detection threshold as low as possible and thus to allow for fully efficient operation of the thin sensor. The studies on Timepix3 detector assemblies summarised in \cref{sec:timepix3_planar_study} have demonstrated a fully efficient operation of \SI{50}{\micron} thin sensors at a threshold of \SI{600}{\Pem{}}.
The four times larger signal in the \SI{200}{\micron} sensor leads to a larger threshold range in which the assembly is operated at full efficiency. Up to a threshold of \SI{3000}{\Pem{}}, the efficiency is constant at \SI{99.7}{\percent{}}. 
    
The efficiency of the \SI{50}{\micron} thin sensor assembly as a function of the track intercept within the pixel cell is illustrated in \cref{fig:clicpix_inpixel_efficiency_50um}. It shows a drop in efficiency along the pixel edges, and most prominently at the pixel corners. This can be explained by the comparatively high operation threshold of \SI{1300}{\Pem{}}, which, in the presence of charge sharing, leads to a fraction of events where the charge for all of the pixels falls below the threshold.

\begin{figure}[!htb]
  \begin{subfigure}[b]{.49\linewidth}
    \begin{tikzpicture}[scale=0.94]
        \begin{axis}[
          /pgf/number format/1000 sep={},
          axis line style={black,thick},
          ticklabel style={black,font={\sansmath\sffamily\Large}},
          xlabel= Threshold / \si{\Pem{}},
          ylabel = Efficiency,
          ylabel near ticks,
          xlabel near ticks,
          label style={black,font={\sansmath\sffamily\Large}},
          xlabel style={at={(rel axis cs:1,0)}, anchor=north east,yshift=-.4cm},
          ylabel style={at={(rel axis cs:0,1)}, anchor=south east,yshift=.85cm},
          width=8cm,
          height=7cm,
          xmin=0,
          xmax=4500,
          ymin=0,
          ymax=1.025,
          minor y tick num={4},
          minor x tick num={4},
          every tick/.style={black,thick},
          every axis/.append style={font=\footnotesize},
          legend style={draw=none,text=black},
          legend style={text=black},
          legend style={at={(0.05,0.05)},anchor=south west,nodes=right,font={\sansmath\sffamily\footnotesize}},
          ]
  
          \addplot[mark=o,thick,color=black, only marks] plot[] table[]{sections/VertexTracking/figures/clicpix/efficiency_50um_planar.csv}; 
          \addlegendentry{\SI{50}{\micron} sensor};
  
          \addplot[mark=square,thick,color=red!80!white, only marks] plot[] table[x=thr, y=eff]{sections/VertexTracking/figures/clicpix/eff_vs_thr_clicpix_200umPlanar_best.txt}; 
          \addlegendentry{\SI{200}{\micron} sensor};
      %
         \node[anchor=south,text=black,font=\small\sffamily,align=center] at (rel axis cs:0.8,0.65){CLICdp};
  
        \end{axis}
      \end{tikzpicture}
    \caption{}\label{fig:clicpix_thresholdscan}
		\end{subfigure}
    \hfill
    \begin{subfigure}[b]{.49\linewidth}
      \begin{tikzpicture}
     \node[anchor=south west,inner sep=0] at (0,0)(image){
       \includegraphics[width=\linewidth]{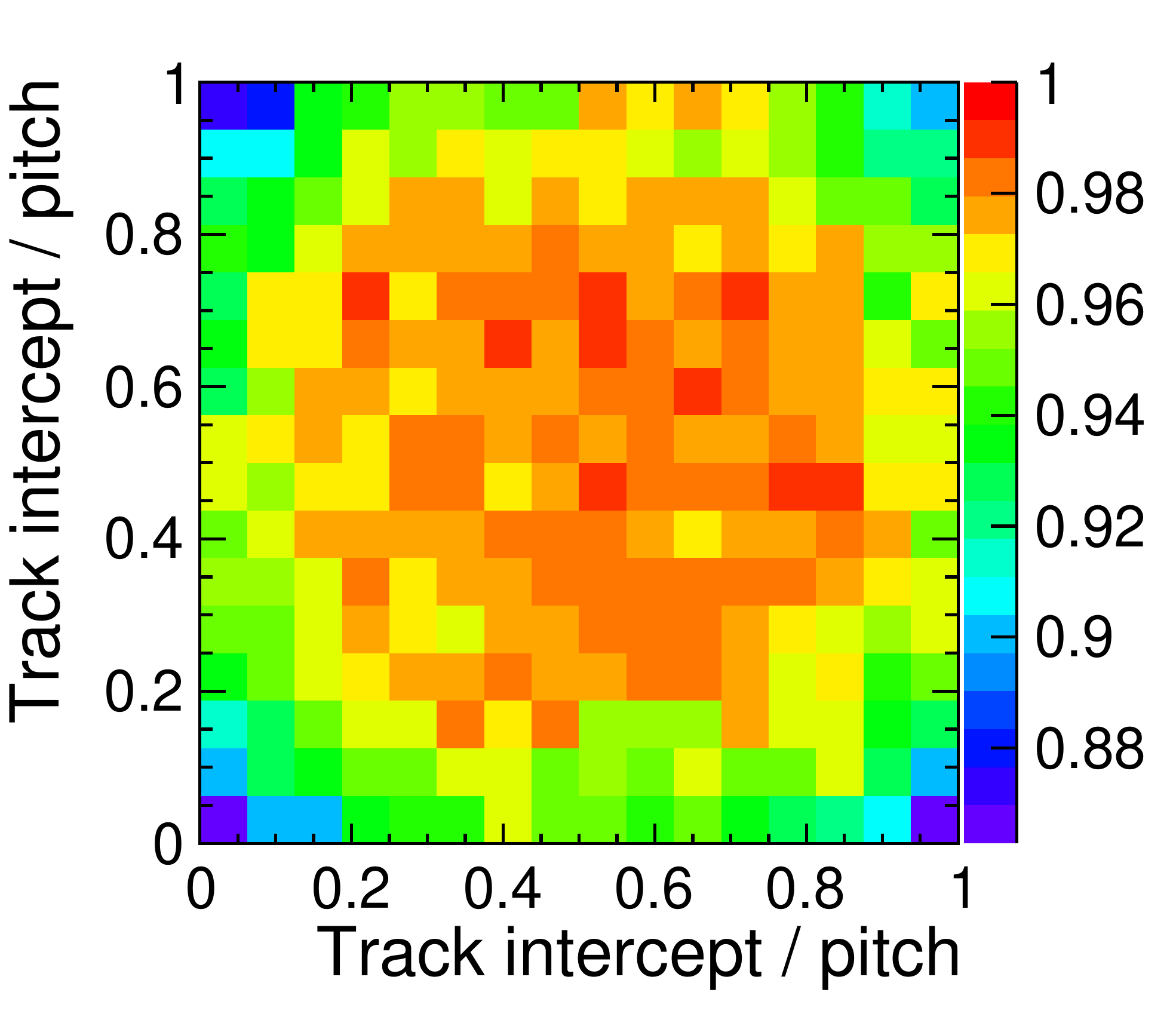}};
         \begin{scope}[x={(image.south east)},y={(image.north west)}]
          \node[anchor=south west,rotate=-90]at(0.975,0.95){\sffamily Efficiency};
          \node[anchor=south west] at (0.65,0.9){\scriptsize CLICdp};
     \end{scope}
     \end{tikzpicture}
     \caption{}\label{fig:clicpix_inpixel_efficiency_50um}
\end{subfigure}
  \caption{\subref{fig:clicpix_thresholdscan} Efficiency of the \SI{50}{\micron} and the \SI{200}{\micron} thick planar CLICpix assembly as a function of the threshold, and \subref{fig:clicpix_inpixel_efficiency_50um} efficiency within the pixel cell for the \SI{50}{\micron} thick sensor assembly operated at \SI{5}{\volt} bias and \SI{1300}{\Pem{}} detection threshold.}
\end{figure}

\subsubsection{Conclusions and outlook for thin planar-sensor studies}
The results obtained with thin planar passive pixel sensors bump-bonded to  Timepix, Timepix3, CLICpix and CLICpix2 readout ASICs demonstrate that individual requirements for application in the vertex-detector region, like spatial resolution, efficiency, timing performance and material budget, can be achieved by planar passive pixel detectors. It is however challenging to meet all requirements at the same time. A further reduction of the pixel pitch below \SI{25}{\micron} could result in better spatial resolution also at \SI{50}{\micron} thickness. However, the available CMOS feature sizes and logic densities, as well as the interconnect process between sensor and readout ASIC set limits on the achievable pixel pitch. The chosen \SI{65}{\nm} CMOS process technology and pixel architecture require a minimum pitch of approximately \SI{25}{\micron}.

\subsection{Active-edge sensors}\label{sec:active_edge_sensors}
To achieve a full coverage of the detector, inactive areas have to be minimised.
However, the low available material budget in the CLIC vertex detector disfavours overlaps between sensor tiles. One possible alternative is the use of sensors with active-edge technology. Being sensitive up to the physical edge of the silicon die, such sensors allow for seamless tiling of individual sensors.

The fabrication is based on a deep reactive ion etching (DRIE) process of the sensor edge~\cite{Wu_2012}.
A trench is formed and then implanted, thereby extending the back-side implantation to the sensor edge.
This allows the depleted area to be extended up to the physical edge of the sensor.
Using this technique, the physical edge can be moved very close to the pixel matrix, since no non-depleted material is needed to separate the depletion zone from the dicing area.

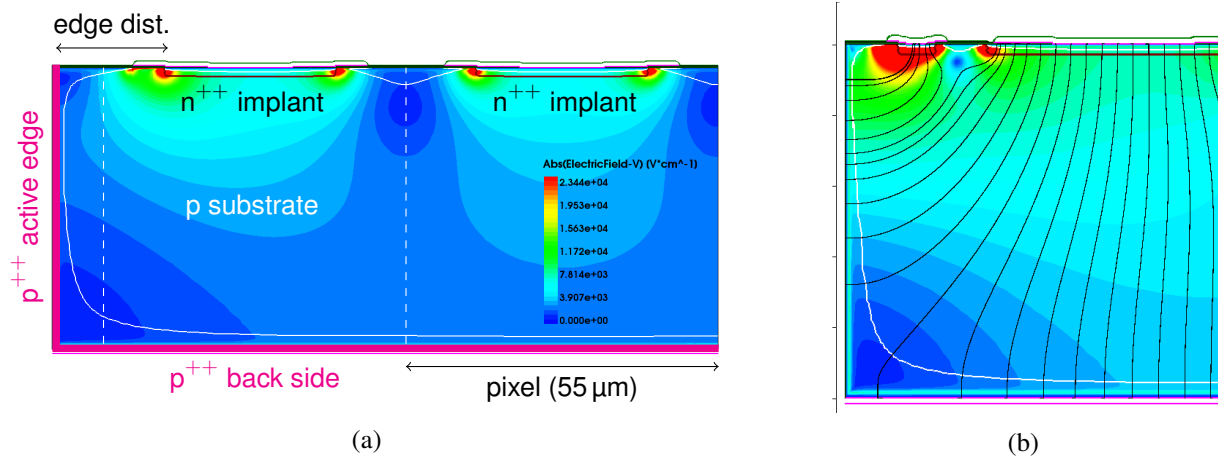
\begin{figure}[ht]
  \centering
    \begin{subfigure}[T]{.6\linewidth}
    \begin{tikzpicture}
    \node[anchor=south west,inner sep=0] (image) at (0,0)
    {\includegraphics[width=.925\linewidth]{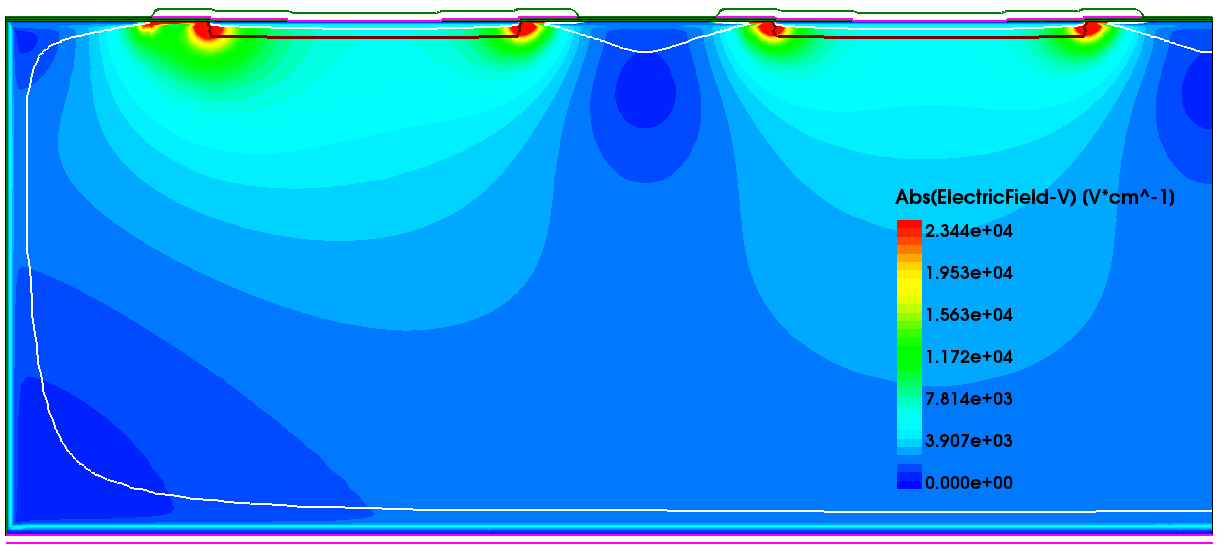}};
    \begin{scope}[x={(image.south east)},y={(image.north west)}]
       \draw[white,dashed](0.5275,0.05) -- ++(0,0.91);
       \draw[white,dashed](0.08,0.05) -- ++(0,0.91);
       \draw[<->](0.5275,-0.025)--(0.99,-0.025) node[below, pos=0.5]{pixel (\SI{55}{\micron})};
       \node[white] at (0.3,0.5){p substrate};
       \node at (0.3,0.85){$\text{n}^{++}$ implant};
       \node at (0.7625,0.85){$\text{n}^{++}$ implant};
       \draw[<->](0.01,1.025) -- (0.175,1.025) node[above, pos=0.5]{edge dist.};
       \draw[line width=1mm, magenta](0.99,0.035)--++(-0.98,0) node[below, pos=.7]{$\text{p}^{++}$ back side} -- ++(0,0.935) node[rotate=90, above, pos=0.5] {$\text{p}^{++}$ active edge};
    \end{scope}
  \end{tikzpicture}
  \caption{}\label{fig:active_edge_sim}
\end{subfigure}
\hfill
  \begin{subfigure}[T]{.32\linewidth}
    \begin{tikzpicture}
      \node[anchor=south west,inner sep=0] (image) at (0,0)
      {\includegraphics[width=\textwidth,clip=true,trim=3.5cm 4cm 20cm 1cm]{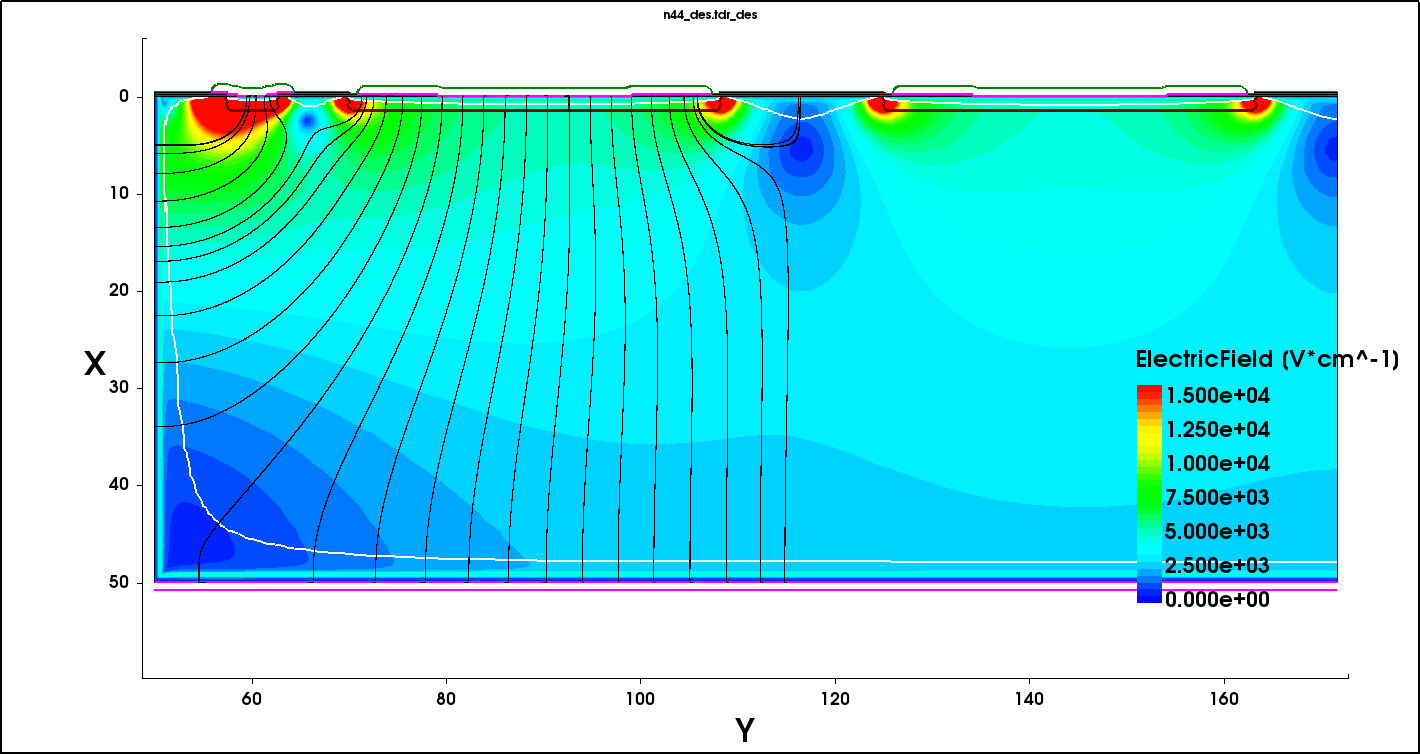}};
      \begin{scope}[x={(image.south east)},y={(image.north west)}]
      \end{scope}
    \end{tikzpicture}
    \caption{}
    \label{fig:field_gndgr}
  \end{subfigure}

  \caption{\subref{fig:active_edge_sim}: Electric field distribution in a  \SI{50}{\micron} thin planar sensor at \SI{-15}{\volt} bias in the 2D cross section. The dashed white lines indicate the regular \SI{55}{\micron} pixel grid. \subref{fig:field_gndgr} Electric field lines close to the sensor edge. In both figures, the solid white line indicates the border of the depletion volume.}\label{fig:active_edge}
\end{figure}

\cref{fig:active_edge} illustrates a finite-element simulation of the edge region of a planar sensor fabricated with active edge technology. The depleted area extends up to a few micron from the physical edge of the sensor. In this particular example, the edge region has been chosen to be significantly smaller than the pixel pitch.

Due to the extension of the back-side contact to the edge, the potential gradient between the sensor edge (bias voltage) and the outermost pixels (ground potential) makes this type of sensor potentially prone to early breakdown. To smoothen the potential gradient, additional guard rings can be placed between the outermost pixel cell and the sensor edge, surrounding the full pixel matrix. These guard rings can either be connected to ground potential or be kept floating.

Experimental studies on thin active-edge sensors with various guard-ring and edge layouts have been performed using Timepix3 readout ASICs~\cite{active_edge_paper}. Besides a study of the high-voltage stability and breakdown behaviour, the sensor assemblies have been operated as DUT in the CLICdp Timepix3 reference telescope. The main focus of the study is on the signal collection and detection efficiency of the sensor in the edge region.

The experimentally measured charge collection in \SI{50}{\micron} thin sensors close to the edge is summarised in \cref{fig:signal_edge}. For better comparability to 2D device simulations, only particle tracks which pass the sensor within the central \SI{40}{\percent{}} of the pixel cell area are considered.
In devices without guard ring, the recorded signal is constant up to the physical edge of the sensor, as the pixel implants are the only available contacts towards which the charges can drift. This is illustrated by the signal distribution as a function of the track incident position close to the sensor edge, shown in \cref{fig:signal_nogr}. In devices with floating guard ring, a slight drop in signal is observed, as shown in \cref{fig:signal_floatgr}. This can be explained by the capacitive coupling of the guard ring to the surrounding implants. A grounded guard ring is capable of collecting signal, and thus is competing with the pixel implant. As a result, a significant drop in signal is observed for tracks passing the sensor after the last pixel, as visible in \cref{fig:signal_gndgr28,fig:signal_gndgr55}.

Exploiting the good pointing capabilities of the reference telescope used, the efficiency of the active edge region has been mapped in two dimensions (see \cref{fig:eff_edge}). All 256 pixel rows are mapped to a \num{2x2} pixel grid, showing the edge region of the sensor. As already deduced from the signal distributions in \cref{fig:signal_edge}, the device without guard ring and the device with floating guard ring stay fully efficient up to the physical edge of the silicon, whereas for both sensors with grounded guard ring a significant loss of efficiency can be noted. This is attributed to the fact that the grounded guard ring and the last pixel implant compete in collecting the ionisation charge. Most of the signal close to the sensor edge is collected by the guard ring and therefore lost for detection. This effect is even more pronounced in the area close to the centre between two pixel implants.

\begin{figure*}[ht]
  \begin{minipage}{\linewidth}
    \begin{subfigure}[T]{.24\linewidth}
      \begin{tikzpicture}
    \node[anchor=south west,inner sep=0] at (0,0)(image){	\includegraphics[page=5,width=\linewidth]{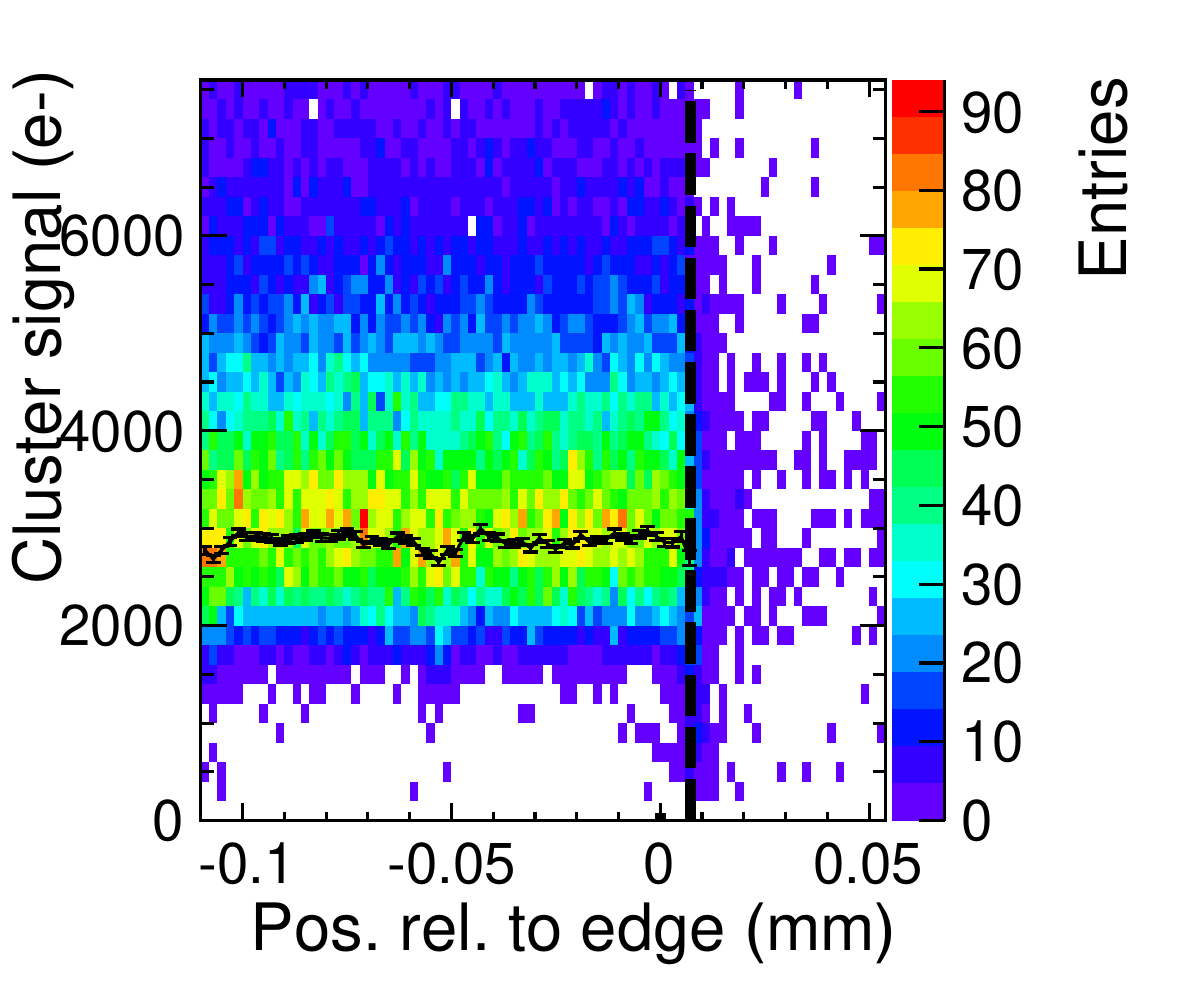}};
      \begin{scope}[x={(image.south east)},y={(image.north west)}]
       \node[anchor=south west] at (0.6,0.9){\tiny CLICdp};
  \end{scope}
  \end{tikzpicture}
      \caption{}
      \label{fig:signal_nogr}
    \end{subfigure}
    \begin{subfigure}[T]{.24\linewidth}
      \begin{tikzpicture}
    \node[anchor=south west,inner sep=0] at (0,0)(image){	\includegraphics[page=8,width=\linewidth]{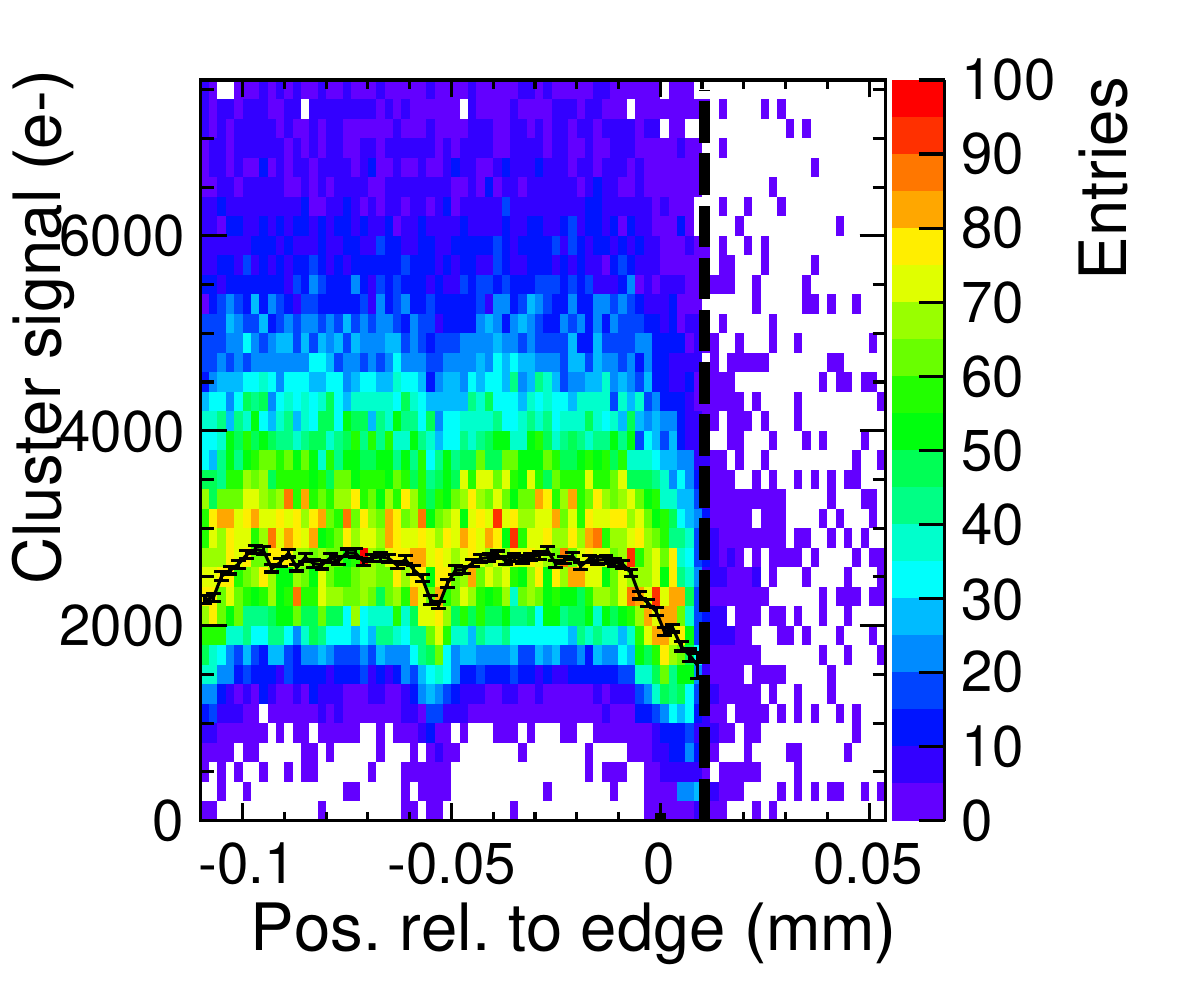}};
      \begin{scope}[x={(image.south east)},y={(image.north west)}]
       \node[anchor=south west] at (0.6,0.9){\tiny CLICdp};
  \end{scope}
  \end{tikzpicture}
      \caption{}
      \label{fig:signal_floatgr}
    \end{subfigure}
    \begin{subfigure}[T]{.24\linewidth}
      \begin{tikzpicture}
    \node[anchor=south west,inner sep=0] at (0,0)(image){	\includegraphics[page=11,width=\linewidth]{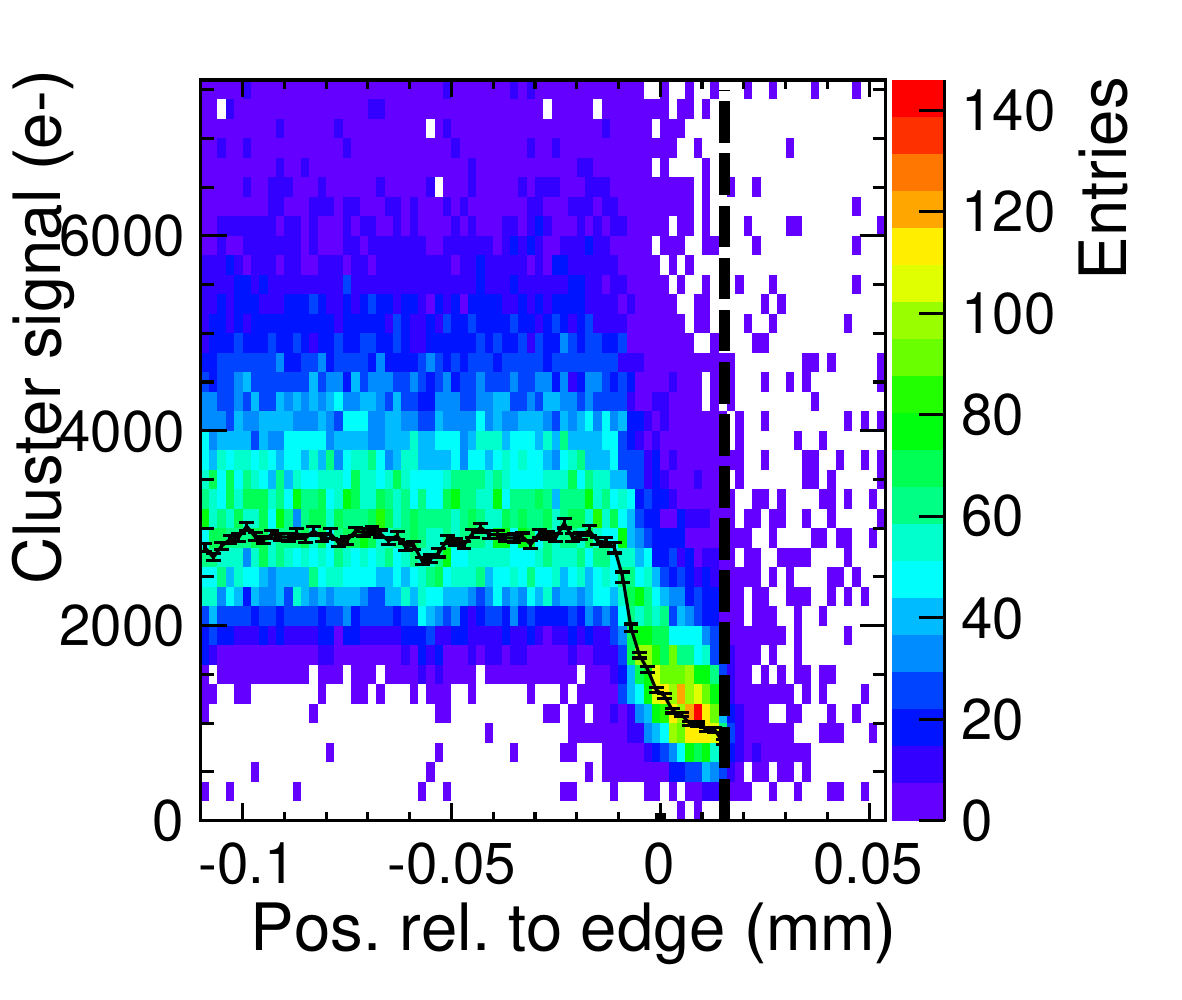}};
      \begin{scope}[x={(image.south east)},y={(image.north west)}]
       \node[anchor=south west] at (0.6,0.9){\tiny CLICdp};
  \end{scope}
  \end{tikzpicture}
      \caption{}
      \label{fig:signal_gndgr28}
    \end{subfigure}
    \begin{subfigure}[T]{.24\linewidth}
      \begin{tikzpicture}
    \node[anchor=south west,inner sep=0] at (0,0)(image){	\includegraphics[page=14,width=\linewidth]{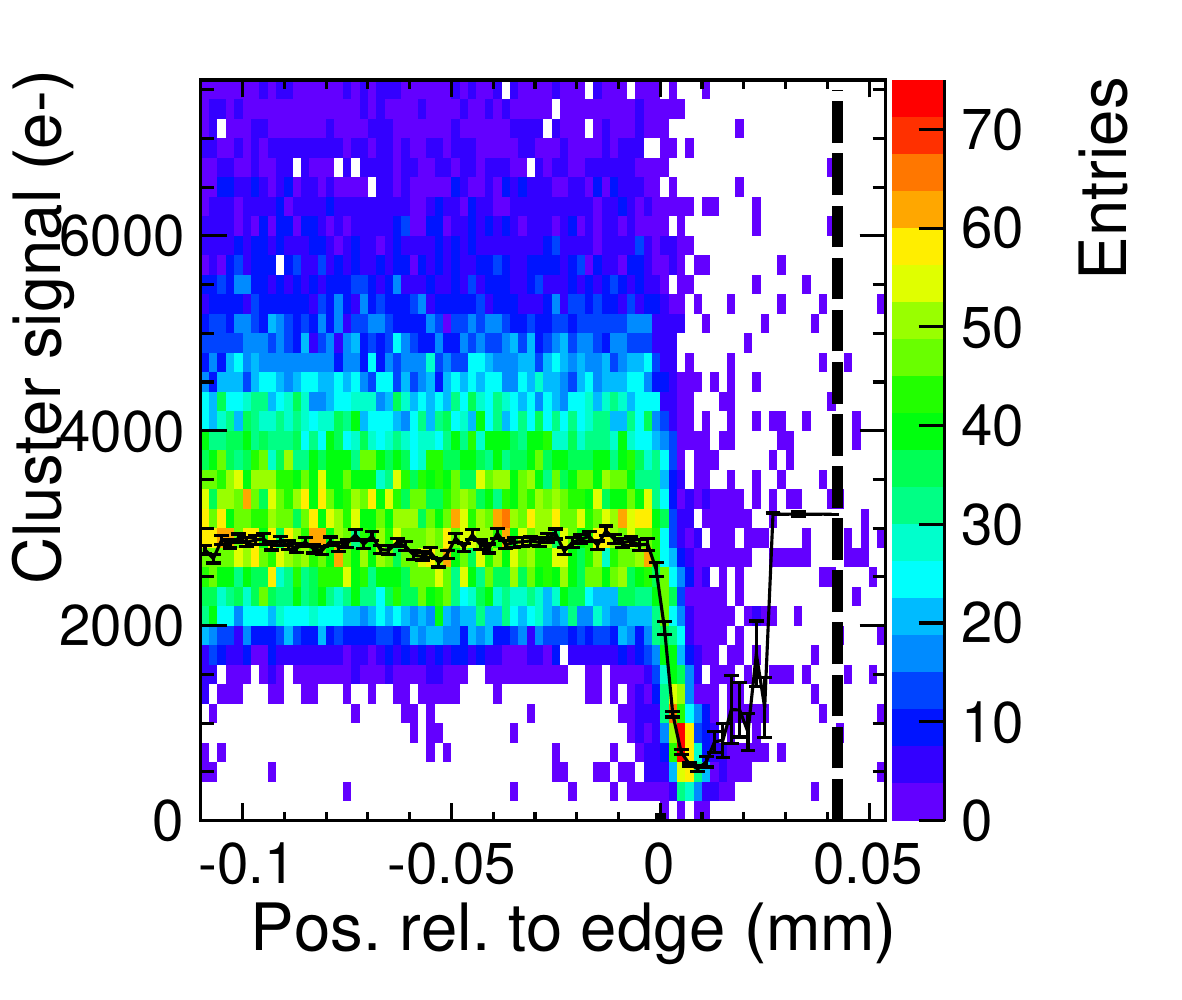}};
      \begin{scope}[x={(image.south east)},y={(image.north west)}]
       \node[anchor=south west] at (0.6,0.9){\tiny CLICdp};
  \end{scope}
  \end{tikzpicture}
      \caption{}
      \label{fig:signal_gndgr55}
      \end{subfigure}
    \caption{Experimentally obtained signal distribution close to the sensor edge for \SI{50}{\micron} thin sensors with different edge layout in planar Timepix3 assemblies~\cite{AlipourTehrani:2143800}: \subref{fig:signal_nogr} no guard ring, \subref{fig:signal_floatgr} floating guard ring, \subref{fig:signal_gndgr28} grounded guard ring with \SI{28}{\micron} edge and \subref{fig:signal_gndgr55} grounded guard ring with \SI{55}{\micron} edge. The vertical dashed line indicates the physical edge of the sensor. The solid black line indicates the most probable value of the cluster signal per x-bin.}\label{fig:signal_edge}
  \end{minipage}

  \begin{minipage}{\linewidth}
    \begin{subfigure}[T]{.24\linewidth}
      \begin{tikzpicture}
    \node[anchor=south west,inner sep=0] at (0,0)(image){	\includegraphics[page=3,width=\linewidth]{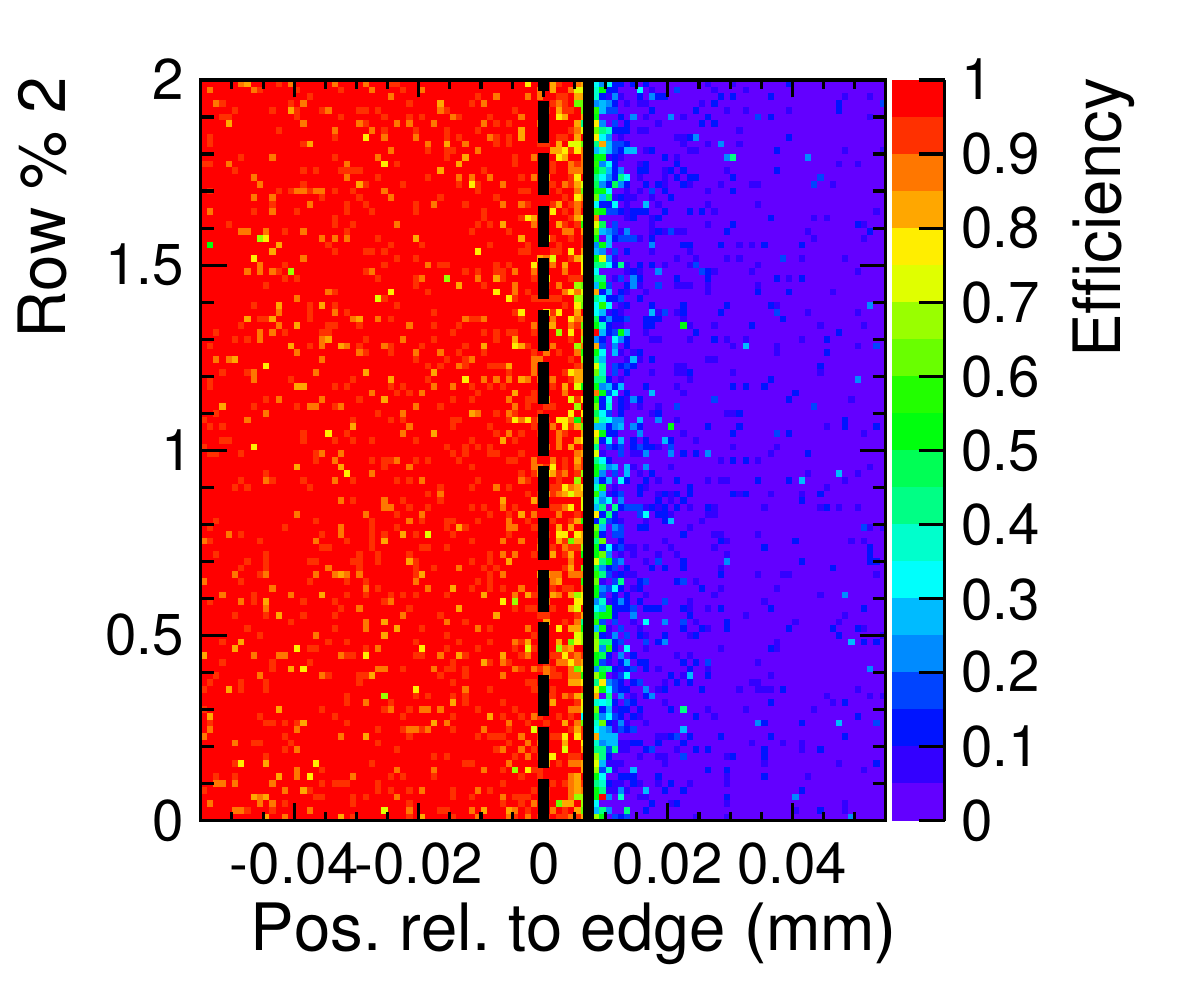}};
      \begin{scope}[x={(image.south east)},y={(image.north west)}]
       \node[anchor=south west] at (0.6,0.9){\tiny CLICdp};
  \end{scope}
  \end{tikzpicture}
      \caption{}
      \label{fig:eff_nogr}
    \end{subfigure}
      \begin{subfigure}[T]{.24\linewidth}
      \begin{tikzpicture}
    \node[anchor=south west,inner sep=0] at (0,0)(image){	\includegraphics[page=6,width=\linewidth]{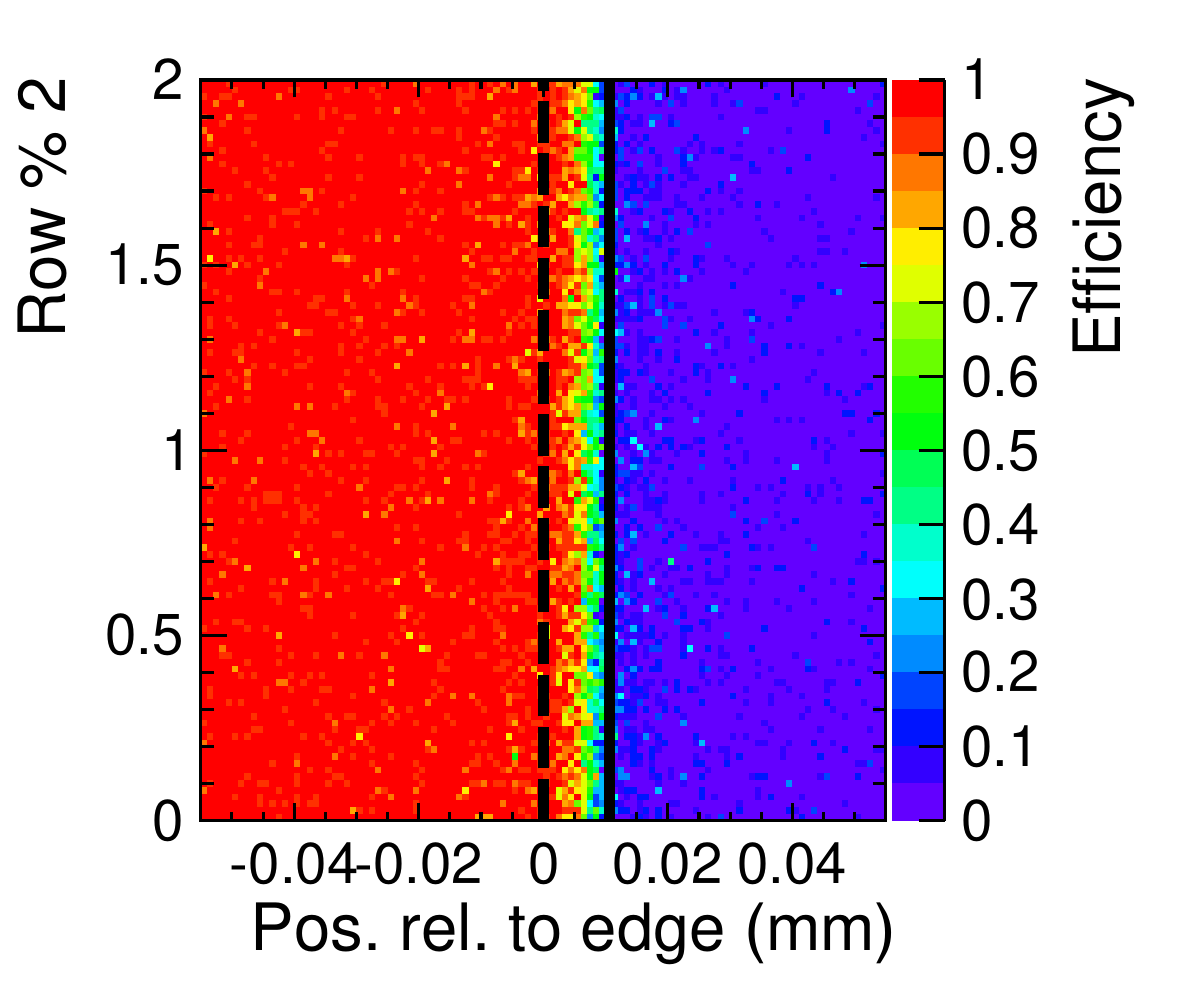}};
      \begin{scope}[x={(image.south east)},y={(image.north west)}]
       \node[anchor=south west] at (0.6,0.9){\tiny CLICdp};
  \end{scope}
  \end{tikzpicture}
      \caption{}
      \label{fig:eff_floatgr}
    \end{subfigure}
      \begin{subfigure}[T]{.24\linewidth}
      \begin{tikzpicture}
    \node[anchor=south west,inner sep=0] at (0,0)(image){	\includegraphics[page=9,width=\linewidth]{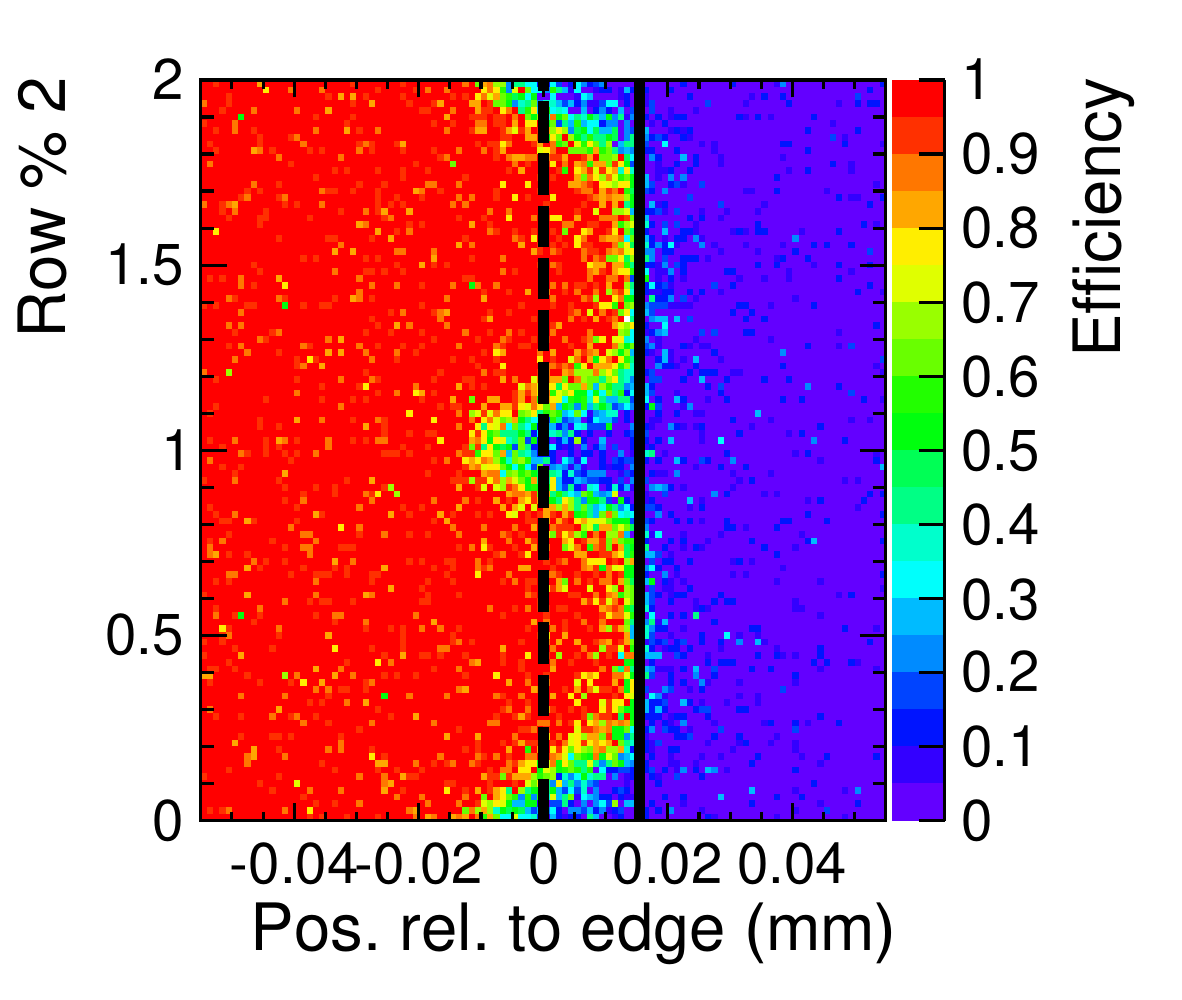}};
      \begin{scope}[x={(image.south east)},y={(image.north west)}]
       \node[anchor=south west] at (0.6,0.9){\tiny CLICdp};
  \end{scope}
  \end{tikzpicture}
      \caption{}
      \label{fig:eff_gndgr28}
    \end{subfigure}
      \begin{subfigure}[T]{.24\linewidth}
      \begin{tikzpicture}
    \node[anchor=south west,inner sep=0] at (0,0)(image){	\includegraphics[page=12,width=\linewidth]{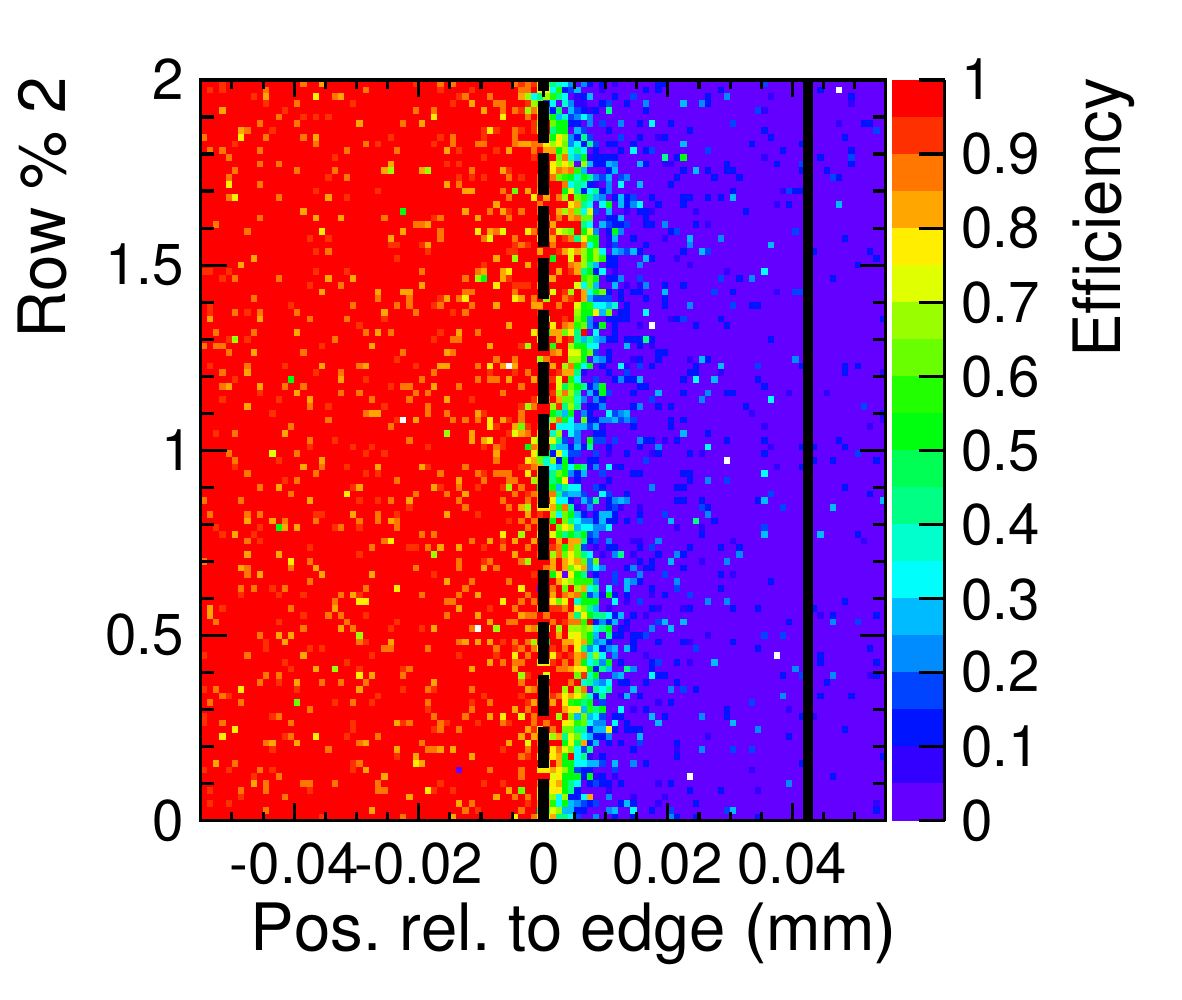}};
      \begin{scope}[x={(image.south east)},y={(image.north west)}]
       \node[anchor=south west] at (0.6,0.9){\tiny CLICdp};
  \end{scope}
  \end{tikzpicture}
      \caption{}
      \label{fig:eff_gndgr55}
      \end{subfigure}
    \caption{Experimentally obtained detection efficiency close to the sensor edge for \SI{50}{\micron} thin sensors with different edge layout in planar Timepix3 assemblies~\cite{AlipourTehrani:2143800}: \subref{fig:eff_nogr} No guard ring, \subref{fig:eff_floatgr} floating guard ring, \subref{fig:eff_gndgr28} grounded guard ring with \SI{28}{\micron} edge distance and \subref{fig:eff_gndgr55} grounded guard ring with \SI{55}{\micron} edge distance. The vertical dashed line indicates the end of the regular pixel matrix, the vertical solid line indicates the physical edge of the sensor.}
    \label{fig:eff_edge}
  \end{minipage}

  \end{figure*}

The results of the active-edge sensor study show that \SI{50}{\micron} thin planar pixel sensors can be operated fully efficiently up to the physical sensor edge.

\subsection{Sensors with enhanced lateral drift (ELAD)} \label{sec:ELAD}
The small diffusion length of the drifting charge carriers in very thin planar pixel sensors results in a low pixel multiplicity per particle hit, as shown in \cref{sec:timepix3_planar_study,sec:clicpix_planar_resolution}. The position resolution is therefore limited to pitch/$\sqrt{12}$. The pixel pitch for hybrid pixel detectors is however limited by the feature size of the readout electronics node and the available area for the readout logic in the ASIC, and furthermore the availability of appropriate interconnect techniques. For the CLIC studies on hybrid pixel detectors, pixel pitches smaller than \SI{25}{\micron} have so far been out of reach.

The goal of the development of enhanced lateral drift detectors (ELAD)~\cite{JANSEN2016242,elad_patent,elad_tipp2017} is to increase charge sharing in pixel sensors without decreasing the pixel pitch or increasing the sensor thickness. By locally engineering the electric field in the sensor bulk, a lateral drift component can be induced, and the ionisation charge can eventually be shared between neighbouring readout implants with a ratio of sharing that depends on the impact position. A lateral field component in the sensor is needed, realised by ion implants situated several tens of microns below the readout implants and acting as repulsive/attractive areas in the silicon bulk.

To investigate the feasibility of the concept, 2D TCAD finite element simulations of the charge collection in ELAD sensors with strip implants have been performed. A readout pitch of \SI{55}{\micron} has been chosen in the simulation, to match the Timepix3 footprint, and a total sensor thickness of \SI{150}{\micron} was selected for initial design studies. The aim is to achieve a cluster size of two pixels. To achieve an electric field profile with optimal charge sharing properties, extensive optimisations of the width and depth of the deep implants, the distance between implants within one layer and between layers, the position and shift to neighbouring layers, the number of layers, and most importantly the doping concentrations of the deep implants were performed.

Static simulations were carried out to optimise the electric field and depletion behaviour of the sensor at realistic production conditions, as well as to study possible early breakdown of the device. Transient simulations of the drifting charges after a particle incident show the increased lateral extent of the charge cloud during the drift, and thereby demonstrate the beneficial effect of the deep implants on the charge sharing. An example is illustrated in \cref{fig:elad_mip_transient_velyka}, where the electron current density in the ELAD sensor bulk is depicted after a particle hit for several snapshots in time. Electrons (in a p-type sensor) drifting through the implant region are guided towards the middle between two readout electrodes and cross the cell border, effectively sharing charges between two readout cells.

\begin{figure}[ht]
  \centering

  \begin{tikzpicture}
  \node[anchor=south west,inner sep=0] (image) at (0,0) {
    \includegraphics[width=.9\linewidth]{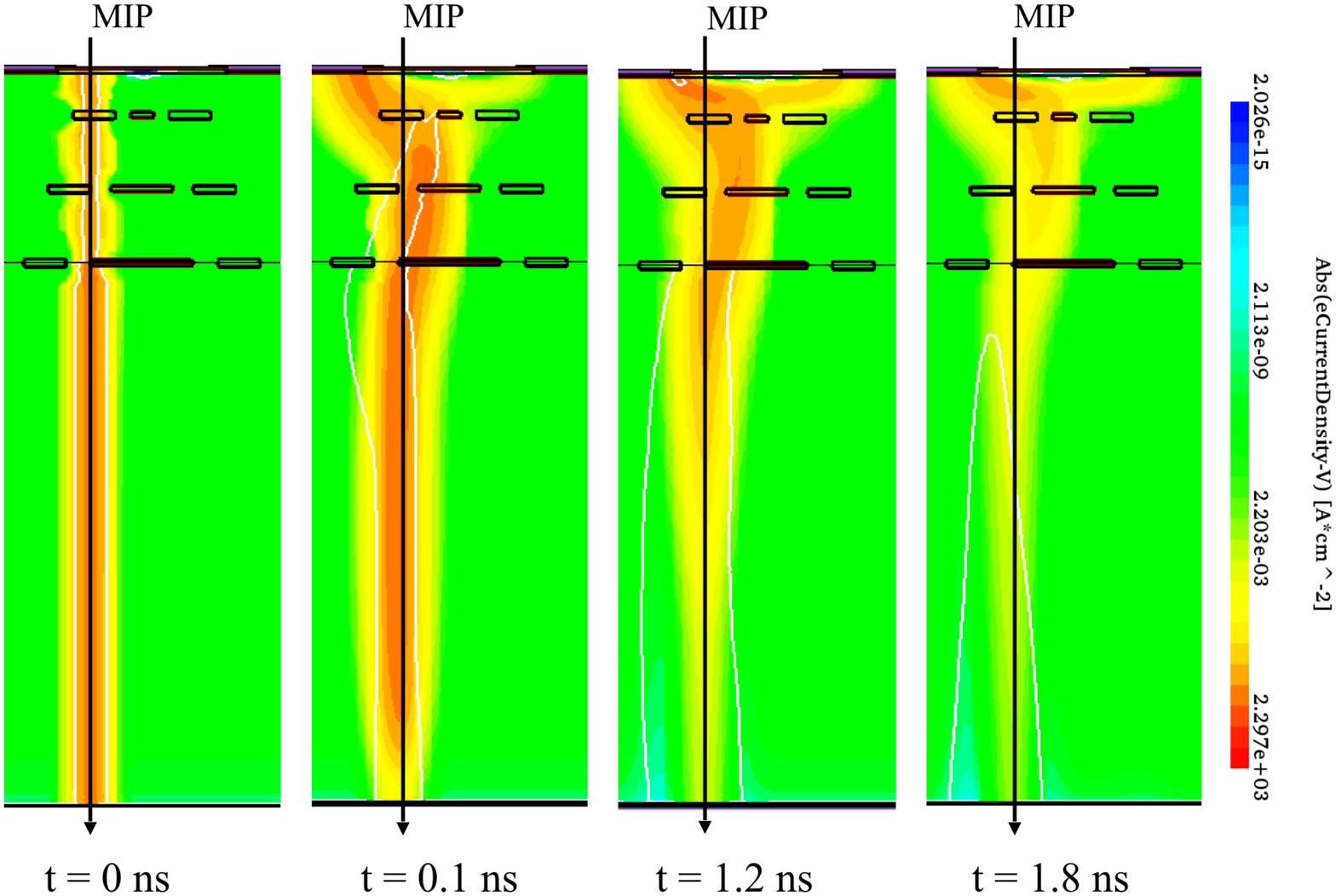}};
  \begin{scope}[x={(image.south east)},y={(image.north west)}]

    \draw[thick, <->](-0.02,0.125)--(-0.02,0.9) node[pos=0.5,rotate=90,above]{\SI{150}{\micron}};
  \end{scope}
\end{tikzpicture}
  \caption{TCAD transient simulation of the electron current density in an ELAD sensor after a minimum ionising particle hit at a distance of \SI{16.9}{\micron} from the left pixel implant. Charge drifting throughout the implant region is deflected and effectively shared between both readout implants.}\label{fig:elad_mip_transient_velyka}
\end{figure}

The collected charge on the two simulated strips as a function of the particle incident point is illustrated in \cref{fig:elad_eta_simulation_velyka}. The theoretical optimum for reconstructing the particle hit position with the highest accuracy corresponds to a linear splitting of the charge on both neighbouring strips, as indicated by the black dashed line. The simulation in a standard sensor without deep implants (black solid line) is far from that optimum. For most incident positions the majority of the charge is collected by an individual strip, and significant charge sharing between both electrodes occurs only close to the mid-gap position at x=\SI{27.5}{\micron}. With this distribution, the accuracy for reconstructing the exact particle hit point is mostly limited by the strip distance. The benefit of the ELAD design manifests itself in a linear charge sharing curve much closer to the theoretical optimum, as indicated by the blue solid curves. This offers the possibility to reconstruct the impact point by weighting the two signals and thus improving over the binary limit of strip distance divided by $\sqrt{12}$. This technique can also be applied in thinner sensors.

\begin{figure}[t]
  \centering
  \includegraphics[width=.9\linewidth]{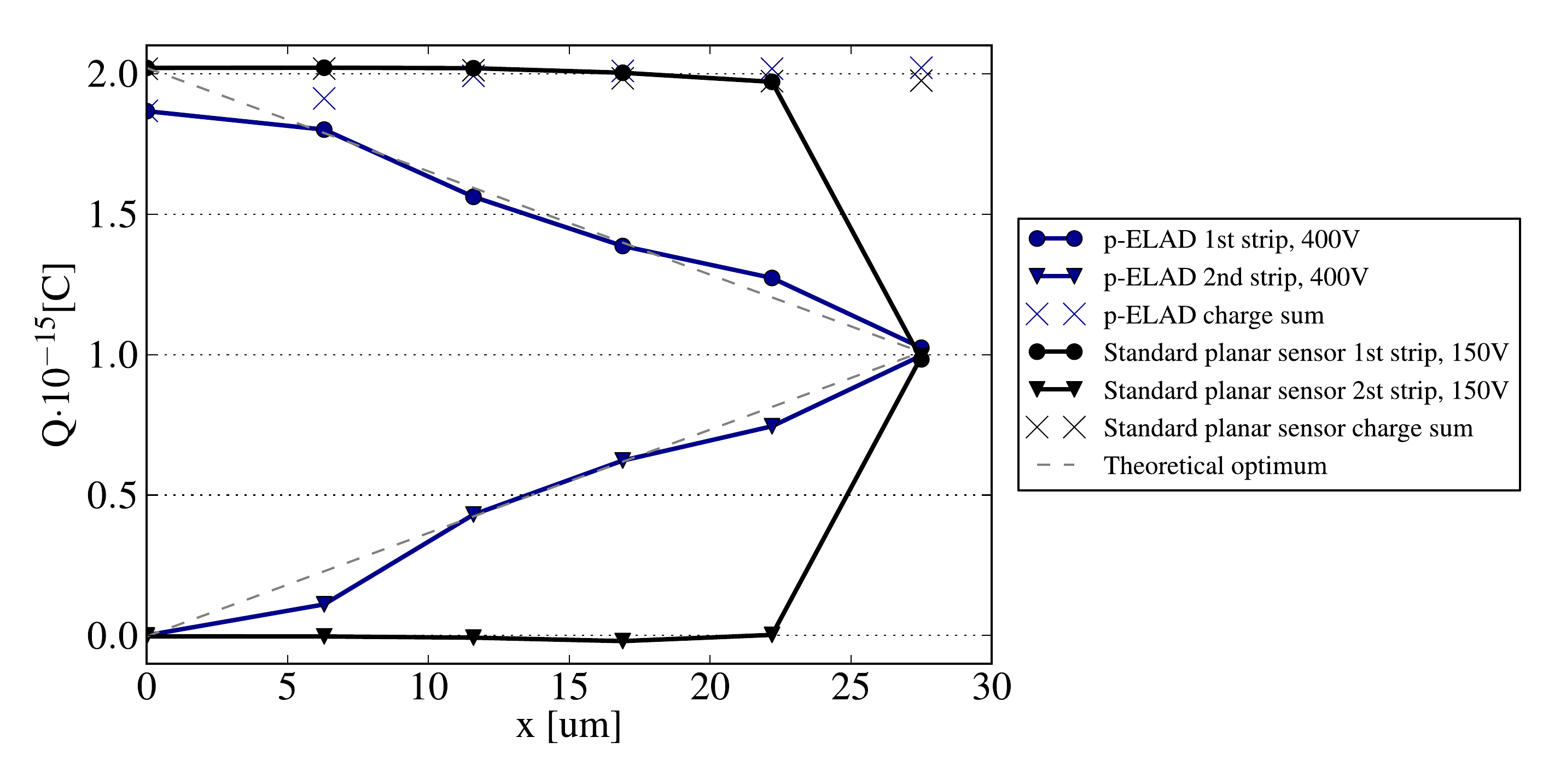}
    \caption{TCAD simulation of the collected charge on two neighbouring strips in an \SI{150}{\micron} thick ELAD detector as a function of the MIP incident position between the two strips. The centre of the left strip is at x=\SI{0}{\micron}, the simulation reaches up to the centre between both strips at x=\SI{27.5}{\micron}. The optimal case, where the charge is split linearly between both strips, is indicated by the dashed grey line. The simulation results for ELAD (dark blue solid lines) and standard strip sensors (black solid lines) are shown.}\label{fig:elad_eta_simulation_velyka}
\end{figure}

The potential detector-performance benefits of ELAD sensors, optimised using TCAD finite element simulations, have been evaluated making use of the \apsq detector simulation framework (cf. \cref{sec:apx_2}). The electric field in a standard pixel sensors and in a pixel sensor with ELAD strips were loaded in \apsq, and the charge drift and charge collection induced by traversing MIPs were simulated at perpendicular incidence. Residuals reflecting the difference between the Monte Carlo position of the MIP and the reconstructed position using a simple centre-of-gravity interpolation have been obtained.
An example of such distributions is shown in \cref{fig:elad_allpix} for the \SI{150}{\micron} thick sensor.
At a pitch of \SI{55}{\micron}, an optimal resolution of approximately \SI{7.6}{\micron} is achieved for a deep implant concentration of \SI{3e15}{\per\cubic\cm} and an operation voltage of \SI{300}{\volt}. This represents an improvement of a factor of two in comparison with the binary resolution. Similar performance gains are expected for sensors thinned to \SI{50}{\micron}.

\begin{figure}[ht]
  \centering
  \begin{tikzpicture}
  \node[anchor=south west,inner sep=0] (image) at (0,0) {
  \includegraphics{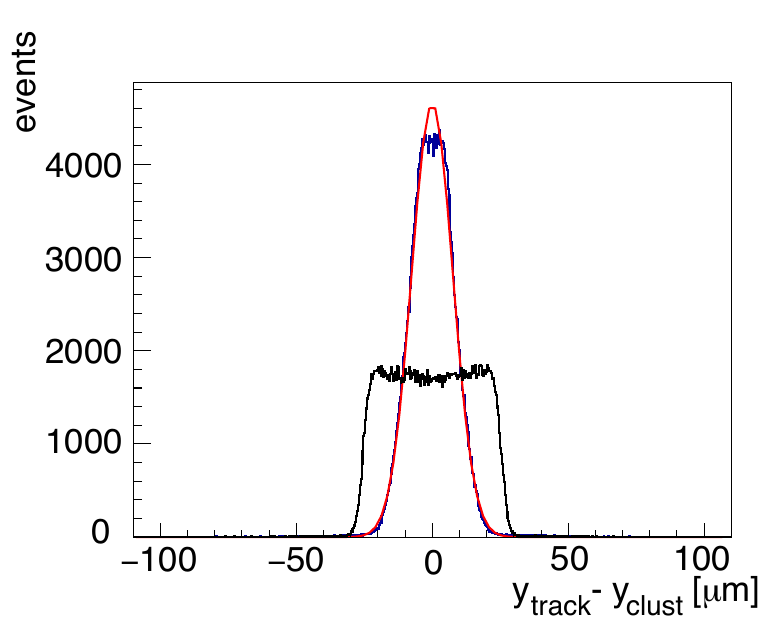}};
  \begin{scope}[x={(image.south east)},y={(image.north west)}]
    \draw[ultra thick, ->](0.75, 0.3)--(0.65,0.35) node[below, pos=0]{\shortstack{standard\\sensor}};
    \draw[ultra thick, red!80!black, ->](0.7, 0.65)--(0.6,0.6) node[above, pos=0]{ELAD};
  \end{scope}
\end{tikzpicture}
  
\caption{\apsq simulation results of the ELAD design: Comparison of the residual distribution in a \SI{150}{\micron} thick standard sensor and a \SI{150}{\micron} thick sensor with deep implants to enhance the lateral drift (ELAD). The deep implant concentration of the ELAD is \SI{3e15}{\per\cubic\cm} and both sensors are operated at \SI{300}{\volt}.}\label{fig:elad_allpix}
\end{figure}

The deep implants in the sensor bulk material makes the fabrication of ELAD devices more challenging than the fabrication of standard planar pixel sensors. The first technology demonstrator production foresees an alternating ion implantation and silicon epitaxial growth by chemical vapour deposition to create the deep implant structure. Process simulations show the feasibility of the procedure, without deteriorating the already implanted structures during the subsequent high temperature steps for epitaxial growth of the following layers and annealing for implant activation. Several layout variations of diodes, strip and pixel sensors have been implemented in the first production design of this novel detector type, in order to demonstrate the concept and to validate the layout and process optimisation. However, the producibility of the concept is not guaranteed.

\section{Capacitively coupled active High-Voltage CMOS sensors}\label{sec:capacitively_coupled_hvcmos}
The need for fine-pitch bump-bonding to interconnect the sensor to the readout ASIC in passive hybrid pixel detectors is a major cost driver and also one of the limiting factors for the pixel pitch. Instead of a direct coupling of the sensor output to the amplifier input via a small solder ball, the signal can also be transferred capacitively between the two dies. A thin layer of glue between the sensor and readout ASIC forms a capacitor between each pair of opposing pixels, resulting in a capacitively coupled pixel detector (CCPD)~\cite{PERIC2010576}. In order to maintain a good detection efficiency, the signal arriving at the readout front-end should not be smaller than the signal of a bump-bonded passive planar sensor. For that reason, a first stage of amplification is needed inside the sensor, as well as a sufficiently large coupling capacitance between the output pads of the active sensor and the input pads of the readout front-end.

\subsection{Fabrication process and demonstrator chips}
To study the prospects of CCPDs for the vertex detector at CLIC, a dedicated active sensor has been developed (CCPDv3)~\cite{CCPDv3_hynds}. The sensor offers a pixel matrix with \num{64x64} square pixels with \SI{25}{\micron} pixel pitch, and thus matches the footprint of the CLICpix ASIC. The sensor is fabricated in a commercial \SI{180}{\nm} High-Voltage CMOS (HV-CMOS) process. The basic pixel design is outlined in \cref{fig:capacitive_coupling_sketch}. A large deep n-well is placed in the p-substrate and covers most of the pixel area. A moderate reverse bias voltage ($\lesssim$\SI{90}{\volt}) can be applied to the substrate and thus a few microns around the n-well are depleted. Depending on the substrate resistivity, the depletion depth can vary between a few microns and several hundred microns. The n-well serves as collection diode for the signal and all active components are placed inside the deep n-well. They are thus being shielded from the high substrate voltage. Since the P-MOS transistors are placed directly in the collecting n-well, they are prone to injecting noise into the amplifier front-end. The front-end is therefore designed using as few P-MOS transistors as feasible. The fast drift signal collected in the depleted layer is transformed to a voltage signal by a transimpedance amplifier and sent to a metal readout pad, from where it couples capacitively to the corresponding input pad of the readout ASIC. The chip contains only the analogue front-end in each pixel, and offers no standalone readout circuitry, except for an analogue monitoring output of the amplifier response for a single pixel.

\begin{figure}[ht]
  \begin{subfigure}[T]{.49\linewidth}
    \includegraphics[width=\linewidth,clip,trim=0 0 0 2.2cm]{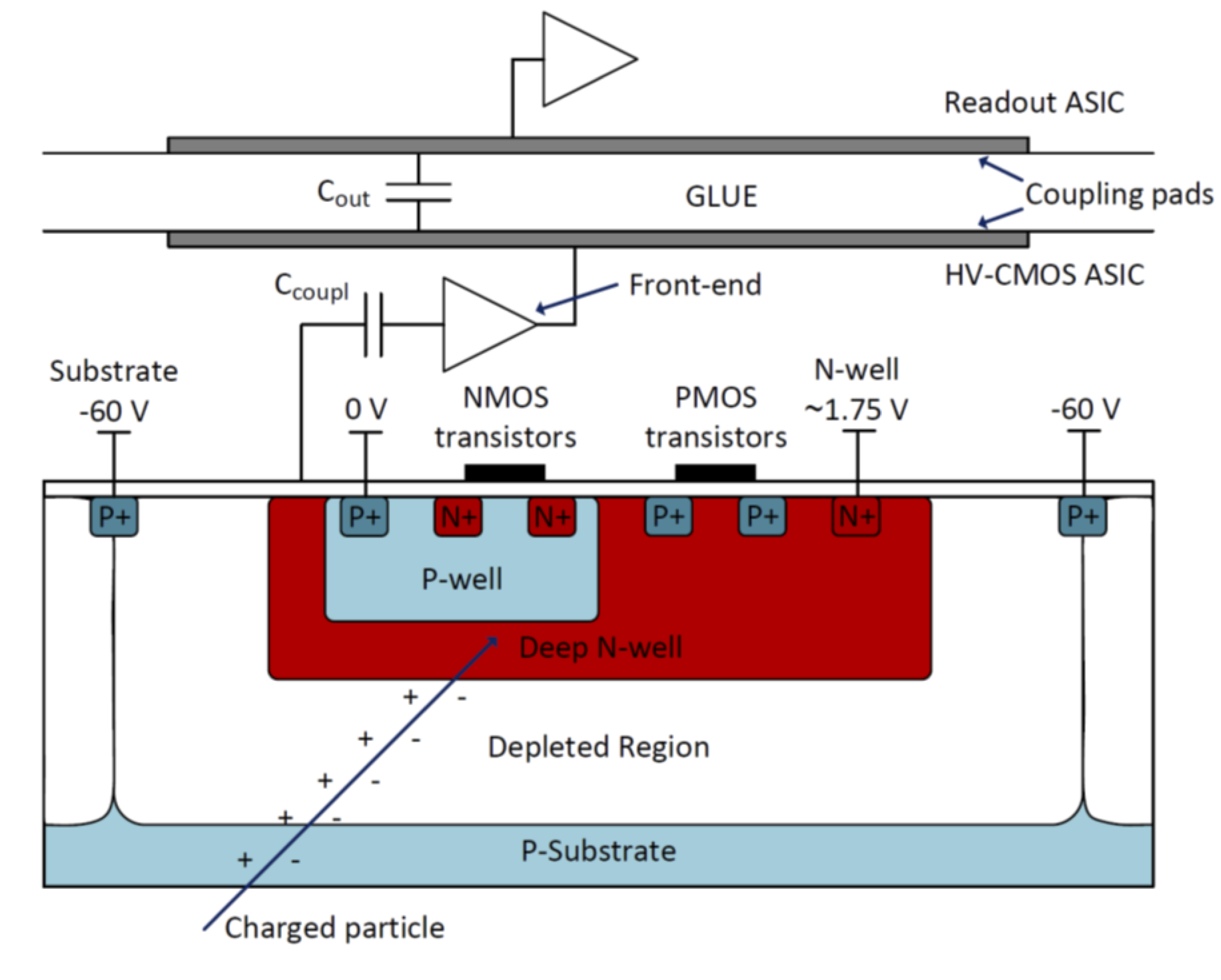}
    \caption{}\label{fig:capacitive_coupling_sketch}
  \end{subfigure}
  \hfill
  \begin{subfigure}[T]{.49\linewidth}
    \includegraphics[width=\linewidth]{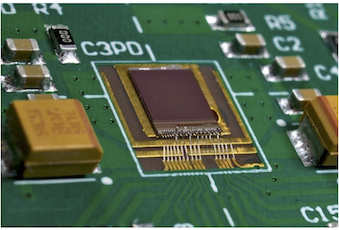}
    \caption{}\label{fig:c3pd_thinned}
  \end{subfigure}
  \caption{\subref{fig:capacitive_coupling_sketch} Schematic cross-section of a capacitively coupled active sensor and \subref{fig:c3pd_thinned} photograph of a \SI{50}{\micron} thin C3PD active sensor chip.}
\end{figure}

A second generation of the CCPDv3 produced in the same \SI{180}{\nm} HV-CMOS process, the CLIC capacitively coupled pixel detector (C3PD), matches the larger \num{128x128} pixel matrix footprint of the CLICpix2 ASIC and offers several advancements in the analogue front-end and digital periphery design for better testability~\cite{c3pd_standalone}. This includes a faster signal rise-time at the amplifier output, an improved test-pulse circuitry, an I$^2$C slow-control interface and power-pulsing functionality. \cref{fig:c3pd_thinned} shows a photograph of a bare C3PD sensor thinned to \SI{50}{\micron} and wire-bonded to a readout PCB. \cref{tab:ccpdv3-vs-c3pd} compares the main parameters of the CCPDv3 and C3PD sensors. 

\begin{table}
  \centering
  \caption{Comparison between CCPDv3 and C3PD active HV-CMOS sensor parameters.}\label{tab:ccpdv3-vs-c3pd}
  \begin{tabular}{l c c}
    \toprule
    & CCPDv3 & C3PD \\    
    \midrule
    Pixel pitch & \SI{25x25}{\micron} & \SI{25x25}{\micron} \\
    Matrix size & \num{64x64} & \num{128x128} \\
    RMS Noise & \SI{35}{\Pem{}} & \SI{40}{\Pem{}}\\
    Amplifier stages & 2 & 1\\
    Gain & \SI{250}{\milli\volt\per\kilo\Pem{}}& \SI{190}{\milli\volt\per\kilo\Pem{}}\\
    Rise-time & >\SI{100}{\ns}& \SI{20}{\ns}\\
    Power diss. per pixel & \SI{26}{\micro\watt} & $\SI{4.8}{\micro\watt}$\\
    Testpulse & full matrix & indiv. row/column\\
    Power pulsing & -- & analogue front-end\\
    \bottomrule
  \end{tabular}
\end{table}

\subsection{TCAD charge-collection simulation}\label{sec:hv-cmos-tcad}
Finite-element TCAD simulations have been performed to gain a deeper insight into the depletion and charge collection properties of the investigated HV-CMOS devices. A two-dimensional structure containing three pixels of the active HV-CMOS sensor has been built. The model was derived from the Graphic Database System (GDS) layout file of the CCPDv3 sensor. The entire structure is \SI{75}{\micron} wide and \SI{100}{\micron} thick, and the pixels have a pitch of \SI{25}{\micron}. In order to replicate a larger multi-pixel structure and not to be affected by edge effects, periodic boundary conditions were added to the physical edges of the model. Detailed comparisons between the two-dimensional model and a more realistic three-dimensional simulation have been made to confirm the validity of the applied simplifications~\cite{Buckland_thesis}.

Due to the sensors not operating at full depletion the exact depletion depth is an important quantity to simulate and measure, as it is related to the charge collection and timing performance. The depletion width of a p-n junction is proportional to the resistivity of the substrate and the square root of the applied bias voltage. The electric field of a sensor is another important quantity as it can give a better understanding of where a breakdown can occur and of the transport direction and magnitude of the charge carriers. From this, regions within the device that produce fast or slow charge collections can be identified. The value of the electric field for the three-pixel structure was simulated at a bias voltage of \SI{-60}{\volt}. A comparison of the electric field for different substrate resistivities between \SIrange[range-phrase={~and~},range-units=repeat]{10}{1000}{\ohm\cm} is shown in \cref{fig:hv_cmos_tcad_field_resistivity}. In all cases, the electric field penetrates the deepest under the deep n-well and has areas of low field under the p+ implants that make up the bias ring. This means that between pixels the charge collection will be the slowest. This is especially pertinent for the \SI{10}{\ohm\cm} case due to the very low field values. The high field value displayed in yellow is deepest for the \SI{10}{\ohm\cm} substrate compared to the higher resistivity options. As the resistivity increases, this area extends laterally, indicating a faster charge collection when the MIP passes in between two pixels, compared to lower resistivity models.

\begin{figure}[ht]
  \begin{subfigure}[T]{.49\linewidth}
    \includegraphics[width=\linewidth]{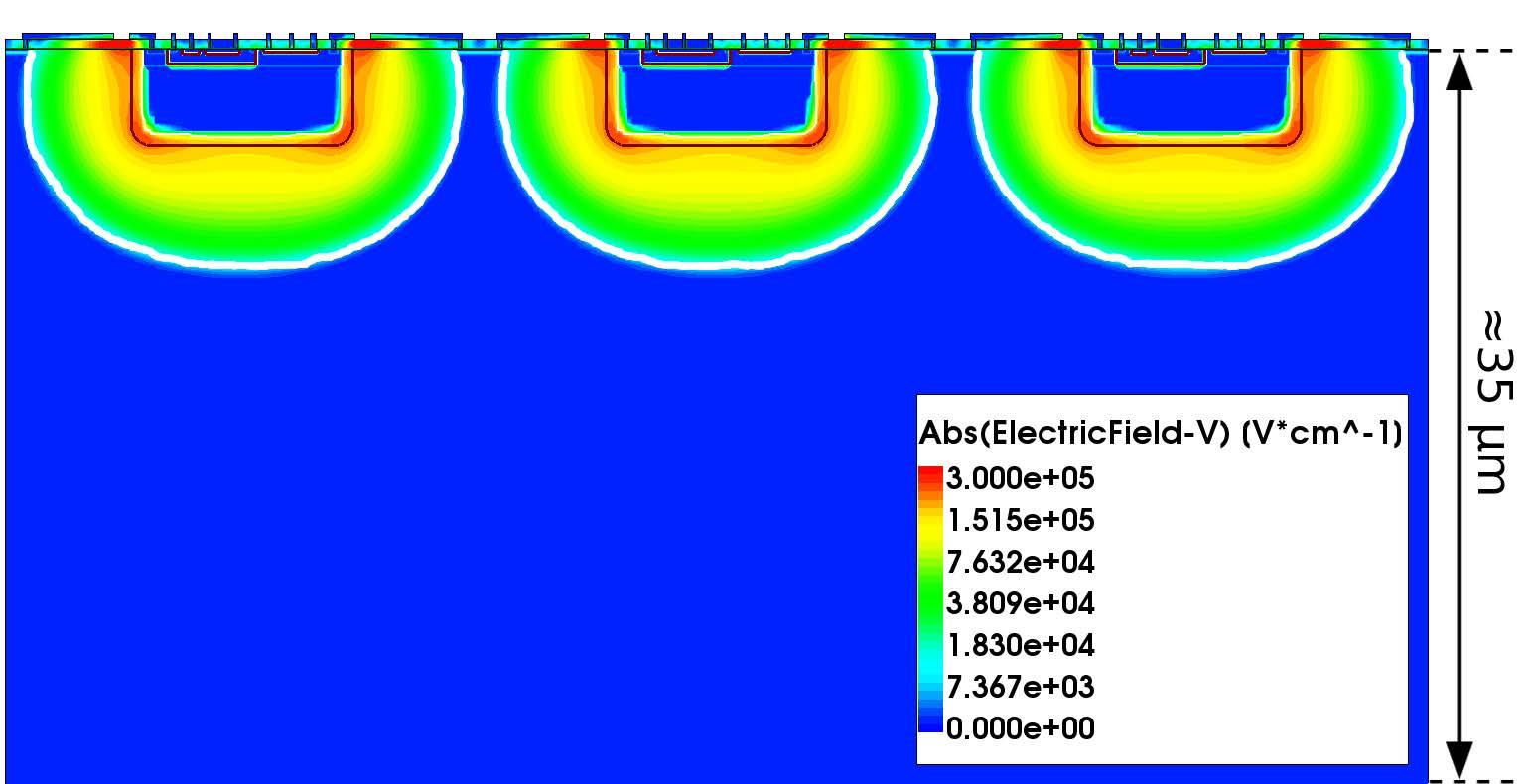}
    \caption{}\label{fig:hv_cmos_tcad_10ohmcm}
  \end{subfigure}
  \hfill
  \begin{subfigure}[T]{.49\linewidth}
    \includegraphics[width=\linewidth]{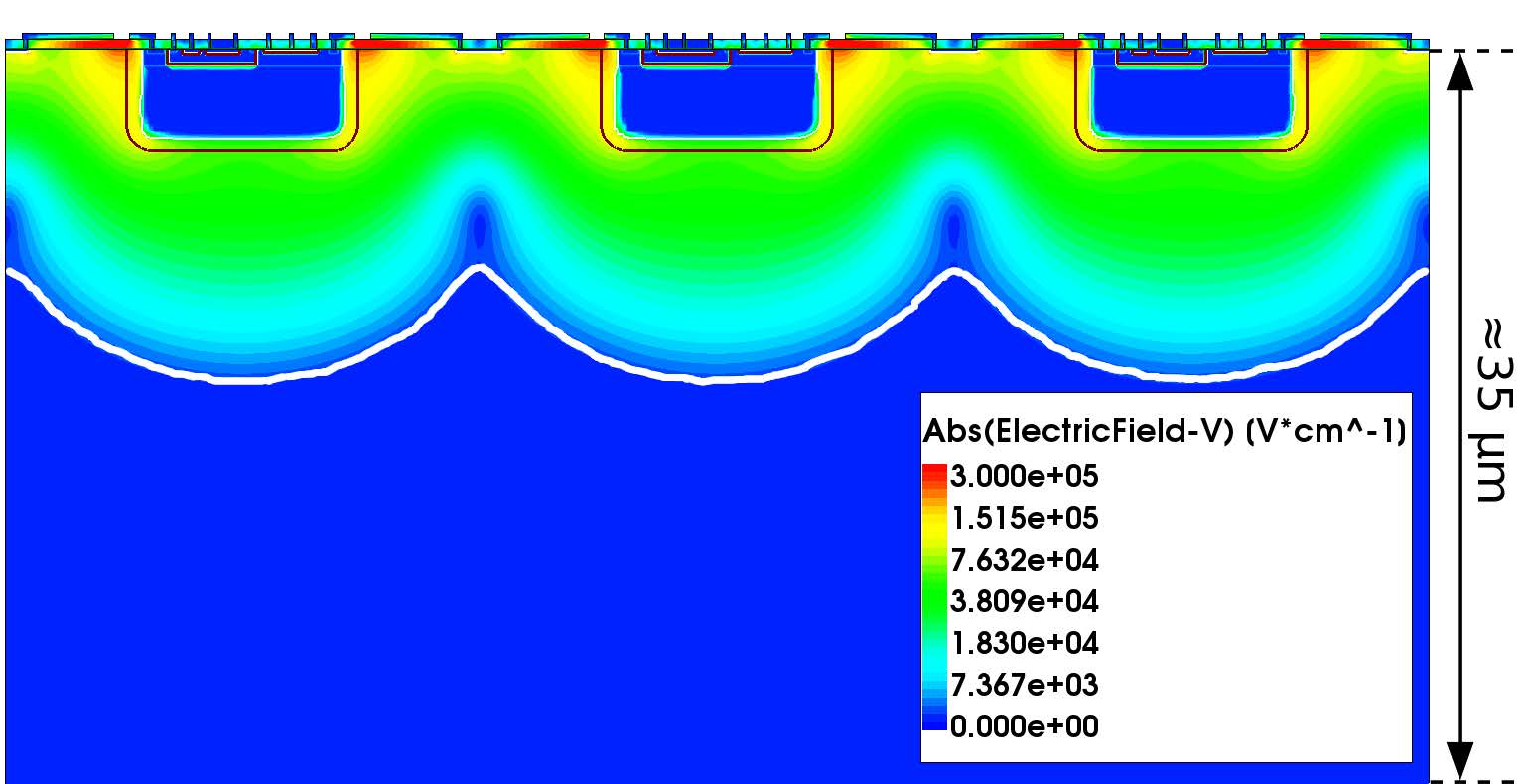}
    \caption{}\label{fig:hv_cmos_tcad_80ohmcm}
  \end{subfigure}

  \begin{subfigure}[T]{.49\linewidth}
    \includegraphics[width=\linewidth]{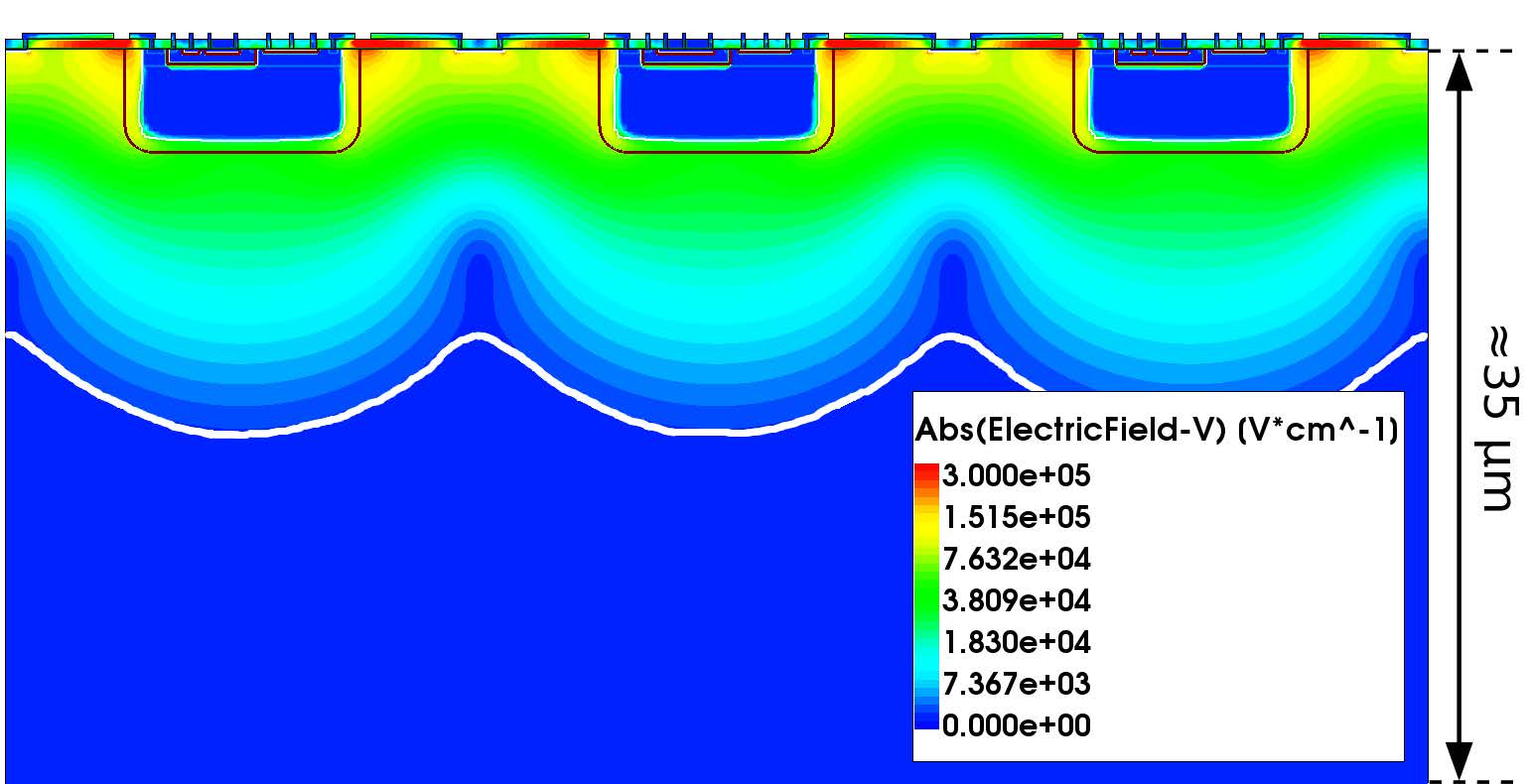}
    \caption{}\label{fig:hv_cmos_tcad_200ohmcm}
  \end{subfigure}
  \hfill
  \begin{subfigure}[T]{.49\linewidth}
    \includegraphics[width=\linewidth]{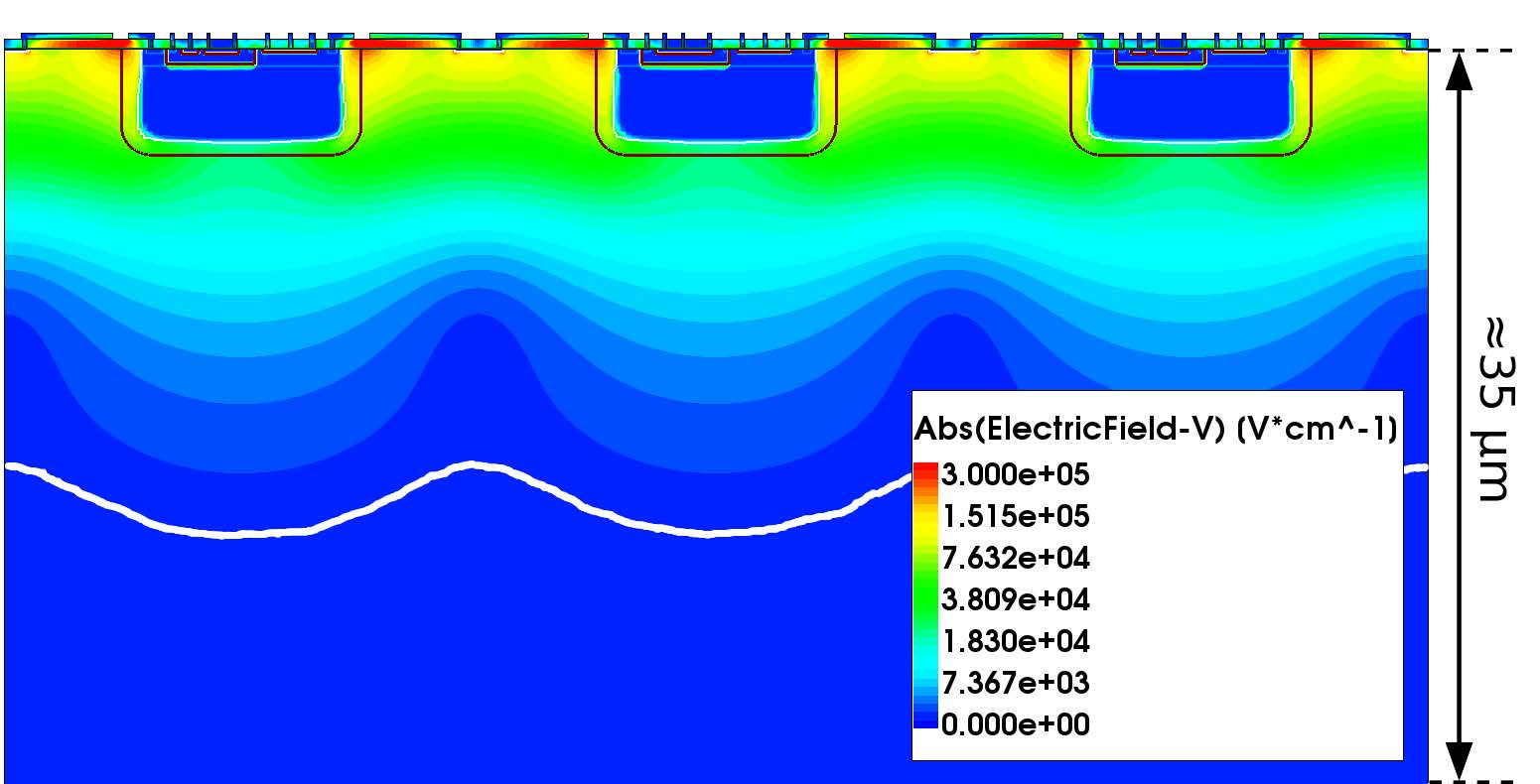}
    \caption{}\label{fig:hv_cmos_tcad_1000ohmcm}
  \end{subfigure}
  \caption{Simulated absolute value of the electric field at \SI{-60}{\volt} for four different values of substrate resistivity~\cite{Buckland_thesis}: \subref{fig:hv_cmos_tcad_10ohmcm} \SI{10}{\ohm\cm}, \subref{fig:hv_cmos_tcad_80ohmcm} \SI{80}{\ohm\cm}, \subref{fig:hv_cmos_tcad_200ohmcm} \SI{200}{\ohm\cm} and \subref{fig:hv_cmos_tcad_1000ohmcm} \SI{1}{\kilo\ohm\cm}. The white line represents the edge of the depletion region.}\label{fig:hv_cmos_tcad_field_resistivity}
\end{figure}

The collected charge of the simulations is found by integrating the current following a particle hit. The resulting charge transients are shown in \cref{fig:hv_cmos_tcad_charge} for various resistivities and on three different time scales. The lower resistivities initially have a slightly quicker charge collection in the sub-nano second range (\cref{fig:hv_cmos_tcad_500ps}). This is a consequence of the larger electric field value in the depletion region for the lower resistivity models. After \SI{0.5}{\ns}, the larger depletion region of the higher resistivities leads to larger charges collected. The \SI{1}{\kilo\ohm\cm} model has the highest collected charge, with around \SI{700}{\Pem{}} more charge collected than the \SI{200}{\ohm\cm} model and about \SI{1000}{\Pem{}} more than the \SI{10}{\ohm\cm} standard substrate after \SI{100}{\ns} (\cref{fig:hv_cmos_tcad_100ns}). Diffusion of charges from the undepleted bulk into the drift region leads to a significant increase of the collected charge up to several microseconds after the particle incidence (\cref{fig:hv_cmos_tcad_10us}). 

Besides higher resistivities, a further way to improve the performance is to apply the bias voltage from the back side instead of the top side. This makes the high values of the electric field extend more towards the bulk which will increase the speed and amount of charge collected. Especially for the higher bulk resistivity, the back-side bias contact is beneficial for the signal collection, as illustrated in \cref{fig:hv_cmos_tcad_charge}. However, back-side biasing requires an additional fabrication step to ion-implant and to apply the metallisation to the sensor back side for the contact. For most CMOS processes, this is a non-standard step and has to be applied in a separate post-processing after sensor fabrication is completed.

\begin{figure}[ht]
  \centering
  \begin{subfigure}[T]{.49\linewidth}
    \begin{tikzpicture}
  \node[anchor=south west,inner sep=0] at (0,0)(image){  \includegraphics[width=\linewidth]{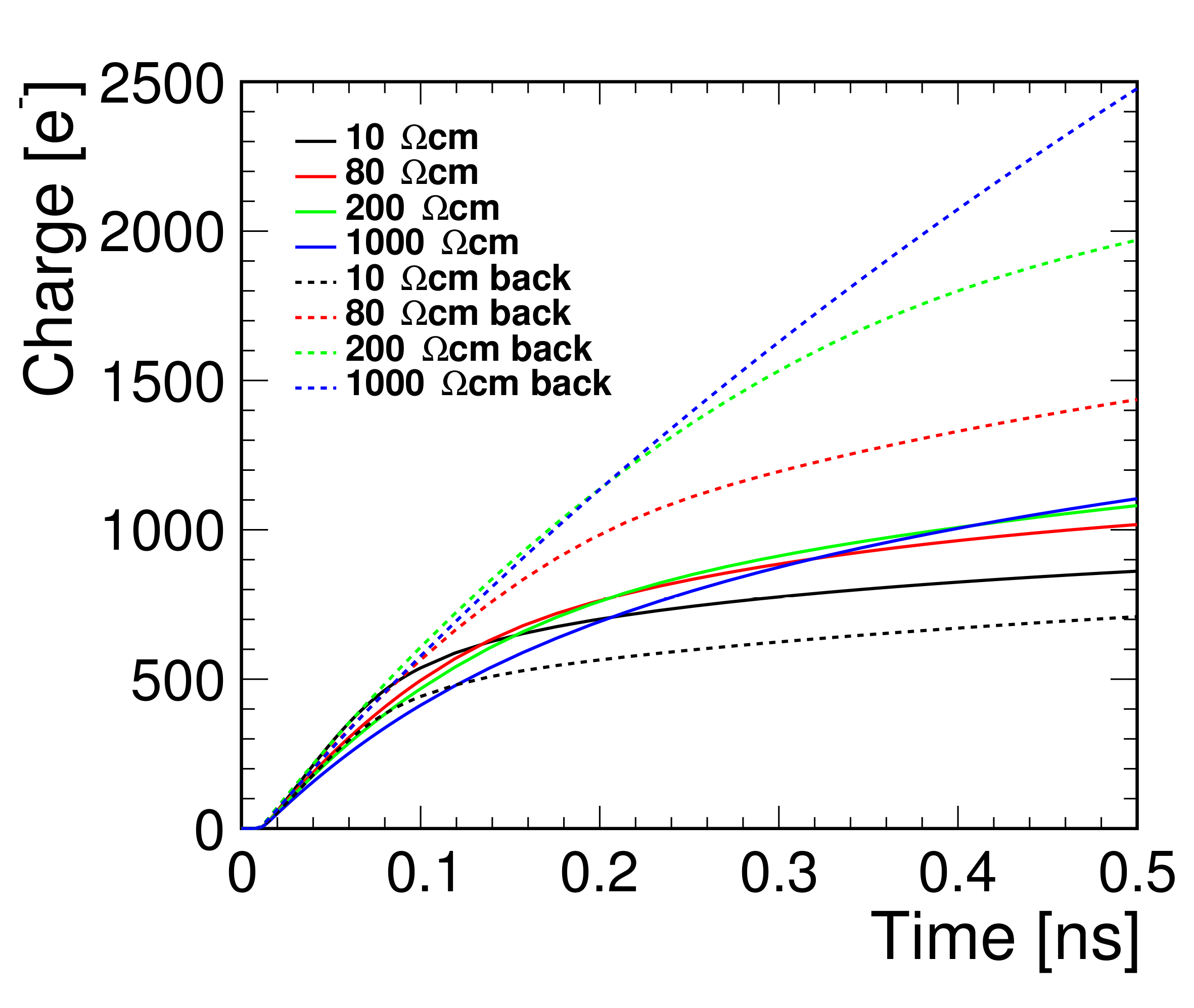}};
    \begin{scope}[x={(image.south east)},y={(image.north west)}]
     \node[anchor=north west] at (0.25,0.55){CLICdp};
  \end{scope}
  \end{tikzpicture}
    \caption{}\label{fig:hv_cmos_tcad_500ps}
  \end{subfigure}
  \hfill
  \begin{subfigure}[T]{.49\linewidth}
    \begin{tikzpicture}
  \node[anchor=south west,inner sep=0] at (0,0)(image){  \includegraphics[width=\linewidth]{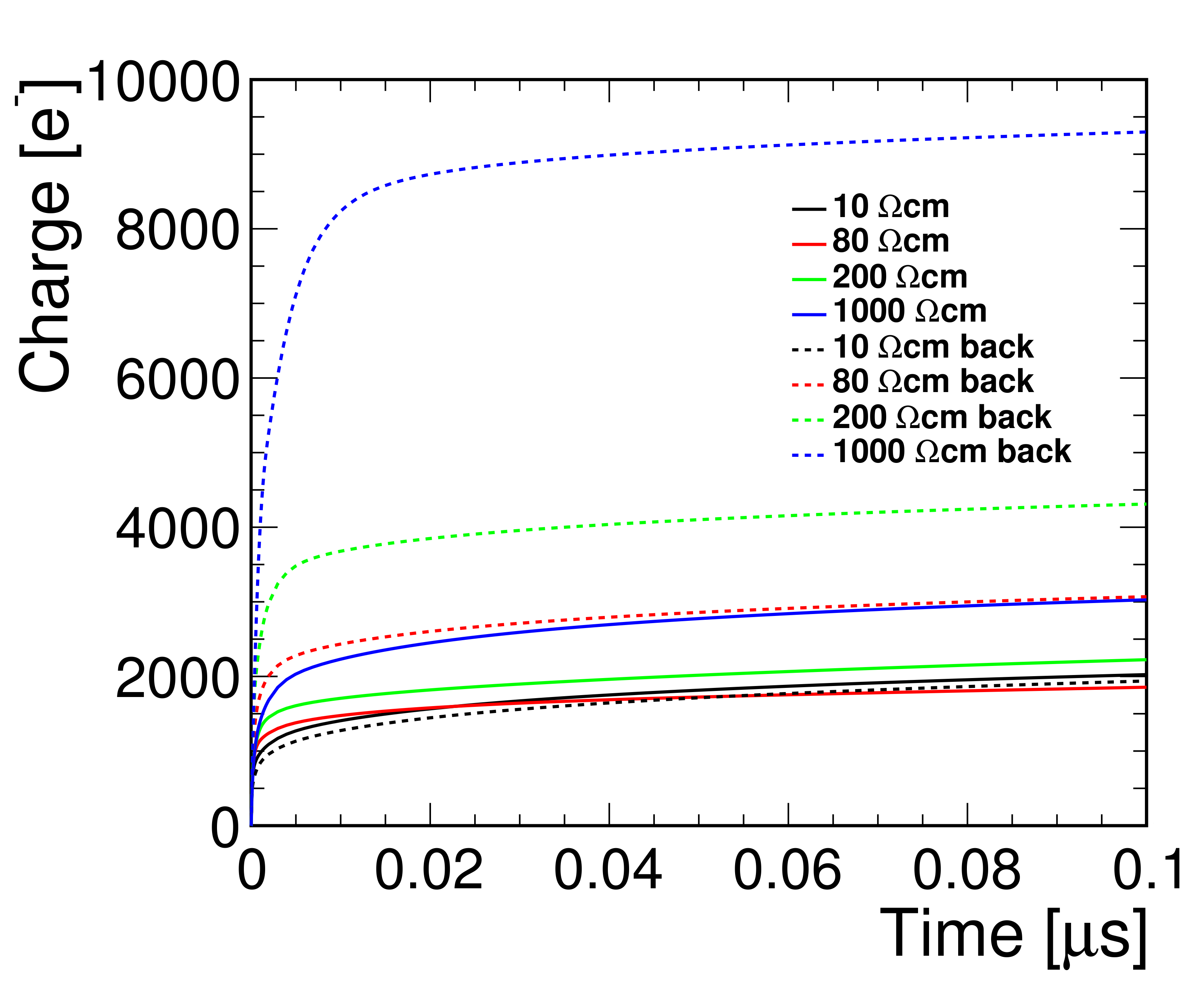}};
    \begin{scope}[x={(image.south east)},y={(image.north west)}]
     \node[anchor=north west] at (0.3,0.7){CLICdp};
  \end{scope}
  \end{tikzpicture}
    \caption{}\label{fig:hv_cmos_tcad_100ns}
  \end{subfigure}

  \begin{subfigure}[T]{.49\linewidth}
    \begin{tikzpicture}
  \node[anchor=south west,inner sep=0] at (0,0)(image){  \includegraphics[width=\linewidth]{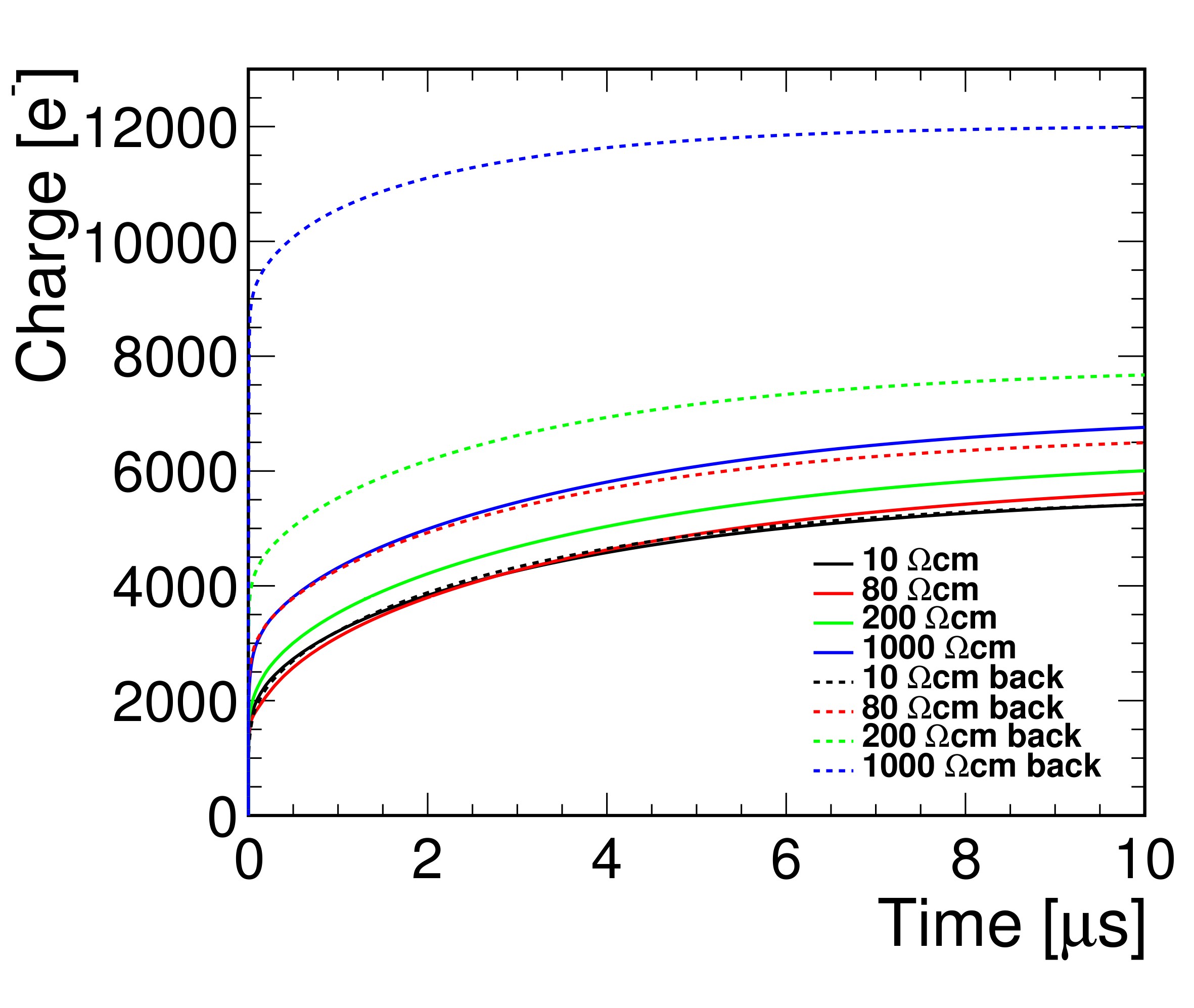}};
    \begin{scope}[x={(image.south east)},y={(image.north west)}]
     \node[anchor=north east] at (0.65,0.35){CLICdp};
  \end{scope}
  \end{tikzpicture}
    \caption{}\label{fig:hv_cmos_tcad_10us}
  \end{subfigure}
    \caption{The total charge collected for the three-pixel simulation model at different bulk resistivity values at \SI{-60}{\volt} bias over different time scales~\cite{Buckland_thesis}: \subref{fig:hv_cmos_tcad_500ps} \SI{0.5}{\ns}, \subref{fig:hv_cmos_tcad_100ns} \SI{100}{\ns} and \subref{fig:hv_cmos_tcad_10us} \SI{10}{\micro\second}. The high-resistivity material and the back-side biasing scheme both result in significantly higher collected charge compared to the standard resistivity of \SI{10}{\ohm\cm} and top-side biasing.}\label{fig:hv_cmos_tcad_charge}
\end{figure}

C3PD active high-voltage CMOS sensors have been fabricated on substrates with higher resistivity of \SI{80}{\ohm\cm}, \SI{200}{\ohm\cm} and \SI{1000}{\ohm\cm}. Compared to the standard substrate resistivity of \SIrange{10}{20}{\ohm\cm}, a larger depletion zone around the deep n-wells is formed, and a larger fraction of the deposited charge is collected by drift.
In addition, chips with post-processed back side will become available, allowing to apply the bias voltage to the back side of the substrate, and to deplete a larger fraction of the substrate. These chips, however, have not been tested, yet.

\subsection{C3PD standalone characterisation}
For a group of 9 pixels on C3PD the analogue amplifier output is routed to the periphery and can be monitored using an oscilloscope, four pixels at a time. Several C3PD chips have been tested in standalone mode, without being glued to a readout ASIC~\cite{c3pd_standalone}. Extensive optimisation on the settings of the on-chip DACs steering the front-end has been performed, to achieve a fast signal rise time, and high signal-to-noise ratio. \cref{fig:fe55_c3pd} shows an example set of pulses recorded from a detector being illuminated by an \isotope[55]{Fe}-source. Results show a rise time of the pixel output pulse of the order of \SI{20}{\ns}. The average charge gain obtained from the integral spectrum presented in \cref{fig:fe55_spectrum_c3pd} was measured to be \SI{190}{\milli\volt\per\kilo\Pem{}} with an RMS noise of \SI{40}{\Pem{}}. Samples of the C3PD chip have also been thinned to \SI{50}{\micron}, and have been successfully tested along with the standard \SI{250}{\micron} thick chips with no observed changes in their performance. Only small variations between devices in terms of rise time, power consumption, signal-to-noise ratio and leakage current have been observed.

\begin{figure}[ht]
  \begin{subfigure}[T]{0.48\linewidth}
    \begin{tikzpicture}
  \node[anchor=south west,inner sep=0] at (0,0)(image){  \includegraphics[width=\linewidth]{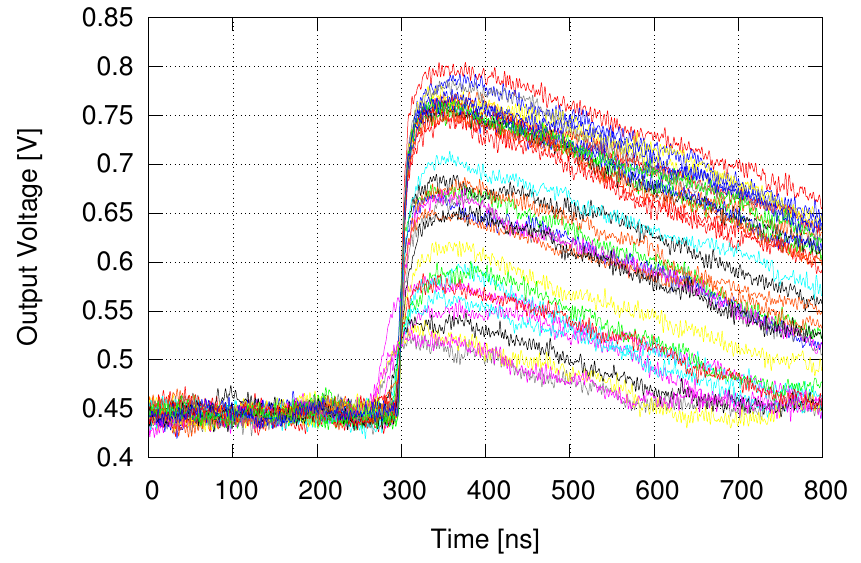}};
    \begin{scope}[x={(image.south east)},y={(image.north west)}]
     \node[anchor=north west] at (0.225,0.55){CLICdp};
  \end{scope}
  \end{tikzpicture}
    \caption{}\label{fig:fe55_c3pd}
  \end{subfigure}
  \hfill
  \begin{subfigure}[T]{0.48\linewidth}
    \begin{tikzpicture}
  \node[anchor=south west,inner sep=0] at (0,0)(image){  \includegraphics[width=\linewidth]{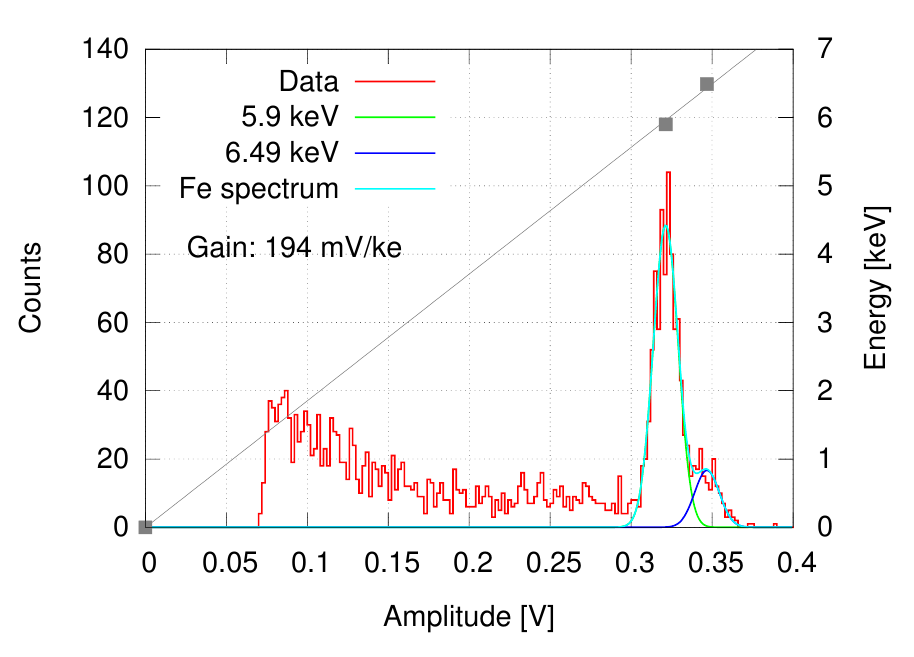}};
    \begin{scope}[x={(image.south east)},y={(image.north west)}]
     \node[anchor=north west] at (0.225,0.55){CLICdp};
  \end{scope}
  \end{tikzpicture}
    \caption{}\label{fig:fe55_spectrum_c3pd}
  \end{subfigure}
  \caption{\subref{fig:fe55_c3pd} Measured sample pulses from a \isotope[55]{Fe} source and \subref{fig:fe55_spectrum_c3pd} resulting amplitude spectrum and double Gaussian fit for one of the monitored C3PD pixels~\cite{c3pd_standalone}.}
\end{figure}

\subsection{Mechanical properties of capacitive coupled assemblies, chip-to-sensor alignment}\label{sec:ccpd_misalignment}
Capacitive coupling between the two silicon dies is achieved by bonding the two chips face-to-face using an epoxy adhesive. To avoid cross-coupling between neighbouring channels, a precise positioning of both dies with respect to each other is needed. For that reason, a flip-chip bonder tool is used for the gluing process. This ensures planarity, precise alignment and reproducibility of the final assembly.

To demonstrate the feasibility of the capacitive coupling, extensive studies on the assembly procedure and detector performance have been pursued. Mechanical assembly properties like sensor to ASIC alignment, glue layer thickness etc. have been extracted by means of destructive cross-sections performed on dummy material and active sensors. The flip-chipped assembly is embedded in an epoxy resin, and then ground to expose a cross-section perpendicular to the matrix surface at the desired depth. The cross-section is then inspected with an optical or electron microscope. An example of a scanning electron microscopy image showing a CCPDv3 and CLICpix assembly is shown in \cref{fig:ccpd_clicpix_crosssection}. The glue layer thickness is around \SI{0.2}{\micron}. The total distance between the CCPDv3 output pad and the CLICpix input pad including the passivation layers has been measured to be around \SI{3}{\micron}. The measured glue thickness was found to be insensitive to the applied bond force, which was varied between \SI{3}{\newton} and \SI{20}{\newton}. For the highest bond force of \SI{20}{\newton}, a deformation of the CLICpix aluminium pads was observed.

\begin{figure}[ht]
  \centering
\includegraphics[width=\linewidth]{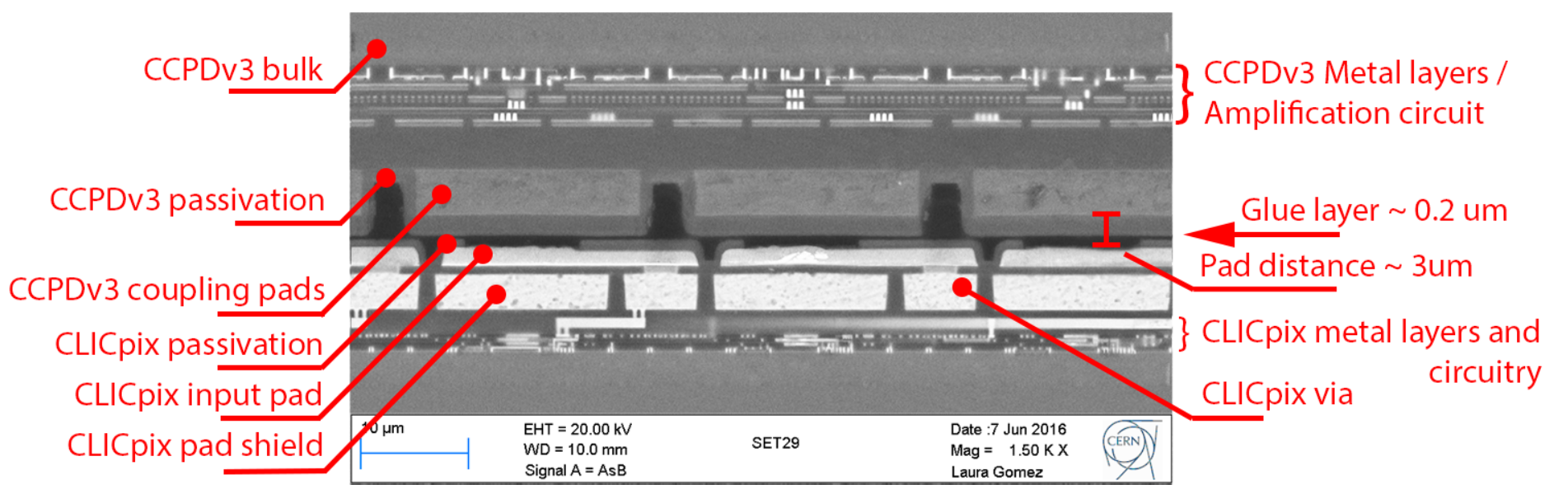}
\caption{Scanning electron microscopy of cross-section through CCPDv3 and CLICpix glue assembly fabricated with ideal alignment between the CCPD coupling pad and the CLICpix input pad.}\label{fig:ccpd_clicpix_crosssection}
\end{figure}

To investigate the influence of chip misalignment, several assemblies of CCPDv3 and CLICpix have been produced with an intentional offset between the two pixel pads. \cref{fig:ccpd_misalignment} illustrates the considered cases. In \cref{fig:ccpd_misalignment_perfect}, the ideal alignment is shown, where the centre-of-gravity between the sensor output pad and the input pad of the readout ASIC are exactly aligned to each other. This is expected to result in the highest coupling capacitance for matching channels and low cross-capacitance to neighbouring channels. In \cref{fig:ccpd_misalignment_quarter,fig:ccpd_misalignment_half} an intentional offset by \SI{6.25}{\micron} and \SI{12.5}{\micron}, corresponding to a quarter and the half of the pixel pitch, is introduced. The performance of the different assemblies has been evaluated in test beams. As an example, the detection efficiency as a function of the threshold is illustrated in \cref{fig:efficiency_ccpd_misalignment}. Several assemblies with ideal alignment are compared to a quarter pixel and a half pixel misaligned detector. Due to the misalignment, signal charge is coupled to several pixels, and the charge on an individual pixel is reduced. This results in a reduced detection efficiency and a faster drop with increasing threshold. While for the quarter pixel misaligned assembly the degradation is only moderate, for the half pixel misaligned detector the effect is strongly visible in the steep drop of the efficiency.

\begin{figure}[ht]
  \begin{subfigure}[T]{.32\linewidth}
  \includegraphics[width=\linewidth,clip,trim=0 0 23cm 0]{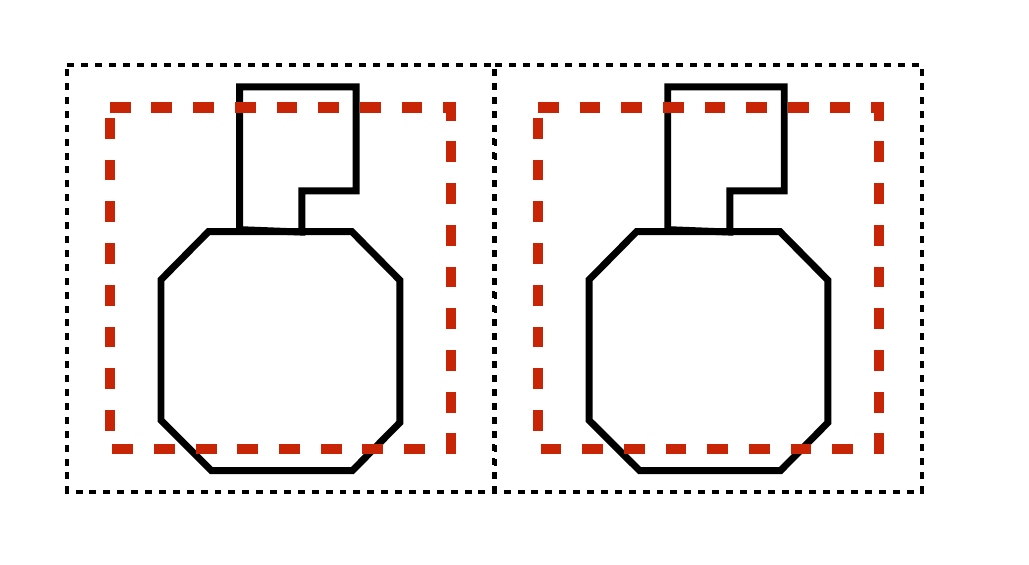}
  \caption{}\label{fig:ccpd_misalignment_perfect}
  \end{subfigure}
   \hfill
  \begin{subfigure}[T]{.32\linewidth}
  \includegraphics[width=\linewidth,clip,trim=11cm 0 12cm 0]{figures/misalignedPads.pdf}
  \caption{}\label{fig:ccpd_misalignment_quarter}
  \end{subfigure}
  \hfill
  \begin{subfigure}[T]{.32\linewidth}
  \includegraphics[width=\linewidth,clip,trim=22cm 0 0 0]{figures/misalignedPads.pdf}
  \caption{}\label{fig:ccpd_misalignment_half}
  \end{subfigure}
  \caption{Alignment of the CCPDv3 output pad (in red) and the CLICpix input pad and the neighbouring via (in black) for three cases: \subref{fig:ccpd_misalignment_perfect} perfect alignment of the gravity centres of the two pads, \subref{fig:ccpd_misalignment_quarter} quarter-pixel offset and \subref{fig:ccpd_misalignment_half} offset by half a pixel.}\label{fig:ccpd_misalignment}
\end{figure}

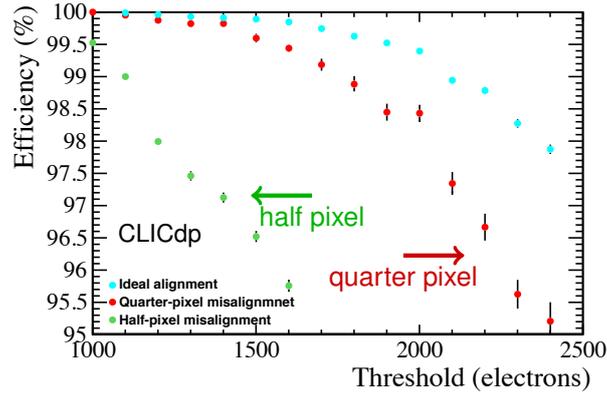
\begin{figure}[ht]
  \centering
  \begin{tikzpicture}
  \node[anchor=south west,inner sep=0] (image) at (0,0) {
  \includegraphics[width=.5\linewidth]{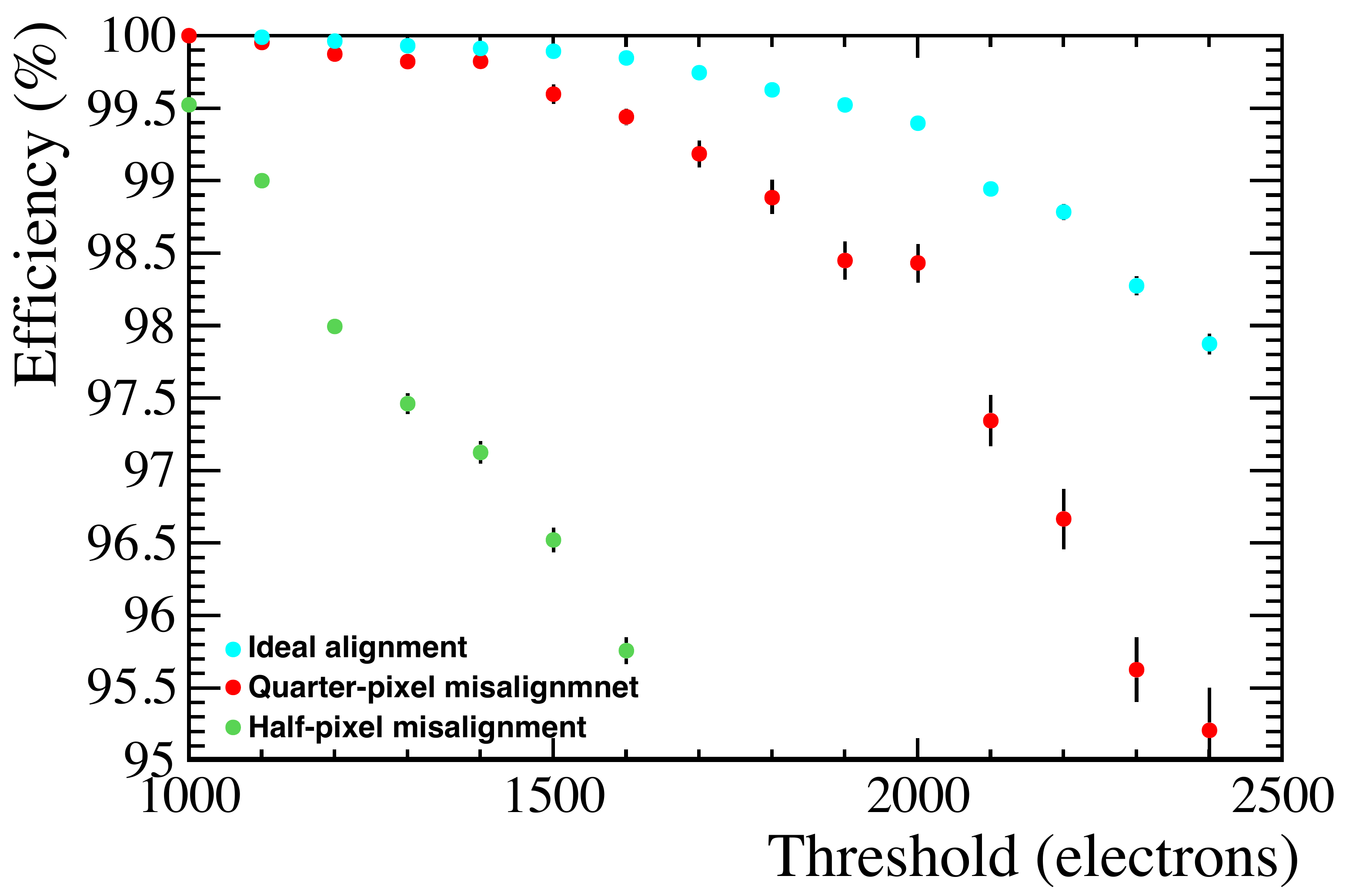}};
  \begin{scope}[x={(image.south east)},y={(image.north west)}]
    \draw[ultra thick, green!70!black, ->](0.5, 0.5)--(0.4,0.5) node[below, pos=0]{half pixel};
    \draw[ultra thick, red!80!black, ->](0.65, 0.35)--(0.75,0.35) node[below, pos=0]{quarter pixel};
    \node at (0.25,0.4){\small CLICdp};
  \end{scope}
\end{tikzpicture}
  \caption{Measured detection efficiency for minimum ionising particles as a function of the applied detection threshold for an ideally aligned CCPDv3+CLICpix sample (turquoise), a quarter-pixel misaligned sample (red) and a half-pixel misaligned sample (green) of \SI{10}{\ohm\cm} resistivity~\cite{StevenGreenThesis}.}\label{fig:efficiency_ccpd_misalignment}
\end{figure}

\subsection{Coupling simulation}\label{sec:coupling_simulation}
To better understand the coupling and cross-talk between neighbouring pixels, a three-dimensional finite element simulation of the capacitive coupling between the metal pads of the CCPDv3 active sensor and the CLICpix readout ASIC has been performed with the COMSOL Multiphysics software package~\cite{MateusComsol}. To simplify the calculations, a sub-matrix of \num{3x3} pixels of the sensor and readout ASIC have been modelled in three dimensions, as shown in \cref{fig:comsol_pixel_coordinates}. The thickness of the different chip layers as well as their material and dielectric properties are taken into account.

The electric field distribution in the system is calculated, and from this the capacitance between the various nodes is extracted. \cref{fig:ccpd_comsol_capacitance} illustrates the extracted capacitance between the central sensor pixel to all nine pixels on the readout ASIC for nominal alignment parameters. The highest capacitance of \SI{3.8}{\femto\farad} is found between the directly facing pixels.  A cross-coupling capacitance to neighbouring pixels of the order of \SI{0.5}{\percent{}} in row direction and up to \SI{3.7}{\percent{}} in column direction. The capacitance to corner pixels is negligible. The different cross-coupling to the various neighbouring pixels is attributed to the geometry of the CLICpix pixel pad and the connection via, as illustrated in \cref{fig:ccpd_misalignment}. This asymmetry in the cross-coupling has also been observed in test-beam data~\cite{CCPDv3_hynds}.

\begin{figure}[ht]
  \centering
  \begin{subfigure}[T]{.59\linewidth}
    \includegraphics[width=\linewidth]{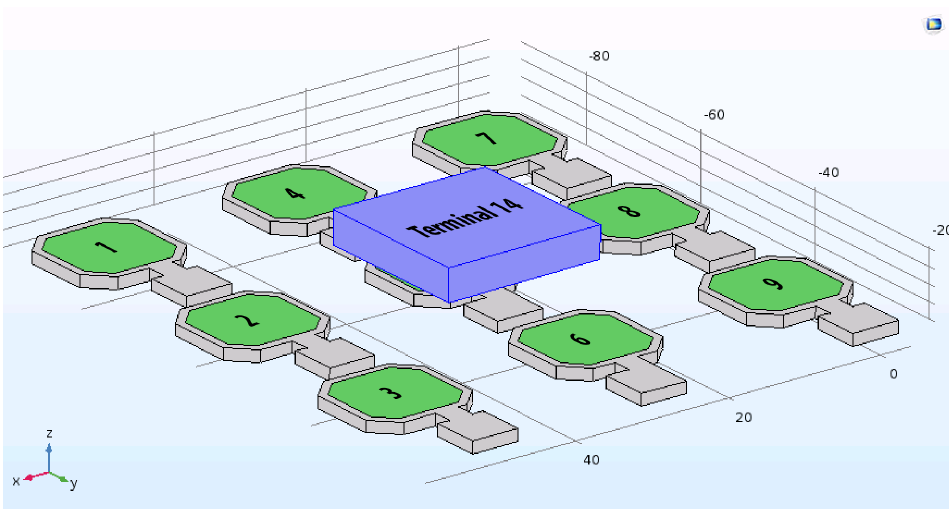}
    \caption{}\label{fig:comsol_pixel_coordinates}
  \end{subfigure}
  \hfill
  \begin{subfigure}[T]{.39\linewidth}
  \begin{tikzpicture}
\node[anchor=south west,inner sep=0] at (0,0)(image){	\includegraphics[width=\linewidth]{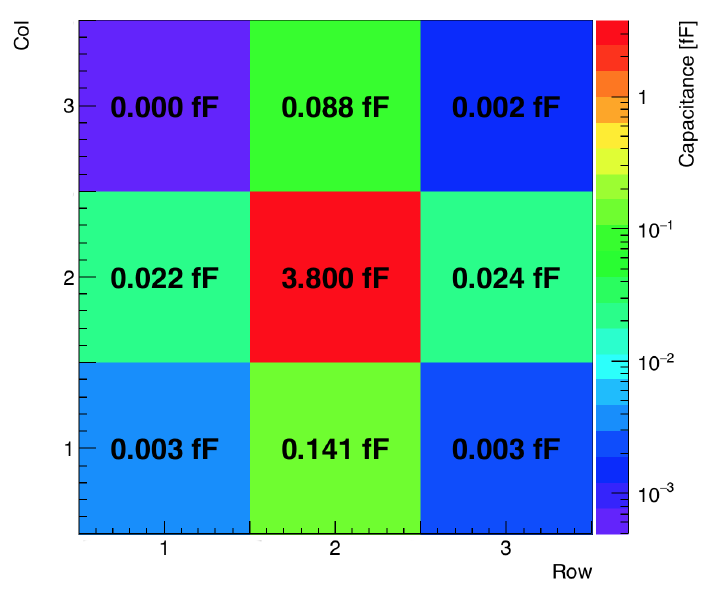}};
  \begin{scope}[x={(image.south east)},y={(image.north west)}]
   \node[anchor=south west] at (0.65,0.95){\scriptsize CLICdp};
\end{scope}
\end{tikzpicture}
  \caption{}\label{fig:ccpd_comsol_capacitance}
\end{subfigure}
\caption{Finite element simulation of the coupling capacitance between CCPDv3 and CLICpix pads~\cite{MateusComsol}: \subref{fig:comsol_pixel_coordinates} simulated geometry, with only a single sensor output pad shown for simplicity (blue box) and nine readout ASIC pixel pads (green octagons). Pad 5 is the directly matching pad on the readout ASIC, the remaining pads are the eight surrounding neighbouring channels. \subref{fig:ccpd_comsol_capacitance} extracted coupling capacitance of the central sensor output pad to the nine pads in the readout ASIC. The orientation corresponds to the view shown in \subref{fig:comsol_pixel_coordinates}.}\label{fig:ccpd_coupling_simulation_mateus}
\end{figure}

The coupling capacitance depends on the thickness of the glue layer and hence the distance between the metal pads. The simulation has been repeated for various distances between the metal pads from \SIrange{3}{100}{\micron}. The resulting capacitance values are shown in \cref{fig:ccpd_coupling_capacitance_comsol}. As the distance increases, the capacitance between the sensor and the readout ASIC pads decreases. The relative difference between the matching pixel and its neighbours decreases, as well. For a gap of \SI{20}{\micron}, the main capacitance drops by an order of magnitude with respect to the \SI{3}{\micron} gap, while the capacitance to the neighbour pixels is almost constant or initially even increases. This initial increase is attributed to the complex pad geometry leading to distance-dependent shielding effects. The steep decrease of the main coupling capacitance underlines the necessity of keeping the glue layer as thin and as homogeneous as possible, in order to achieve a sufficiently large signal and a homogeneous detector response over the pixel matrix.

To investigate the effects of misalignment between the two chips, the CLICpix ASIC was fixed in space and the CCPDv3 chip has been moved by $\pm\SI{13}{\micron}$  along the row and column direction. This covers the full range of possible mismatch between both chips, similar to the experimental study outlined in \cref{sec:ccpd_misalignment}. To maintain a full coverage of the two chips, the matrix has been increased to \num{5x5} pixels, where only the inner \num{3x3} pixels are analysed. Results of the scan in row direction are summarised in \cref{fig:ccpd_comsol_misalignment}. The colour bands indicate the uncertainty of the capacitance due to the $\sim\SI{2}{\micron}$ precision of the flip-chip alignment. A small misalignment does not affect the coupling of the main pixel, and even for the maximal misalignment, the capacitance drops only by a factor of two. However, as expected, a strong increase in the cross-coupling capacitance to the neighbouring pixels by several orders of magnitude is observed. Even for a slight misalignment of \SI{2}{\micron}, the cross coupling capacitance is increased by a factor of 3. This confirms the need for a precise alignment between the two pixel matrices to the micron-level, and confirms qualitatively the observed test-beam results in \cref{sec:ccpd_misalignment}.

\begin{figure}[ht]
  \begin{subfigure}[T]{.49\linewidth}
    \begin{tikzpicture}
  \node[anchor=south west,inner sep=0] at (0,0)(image){	 \includegraphics[width=\linewidth]{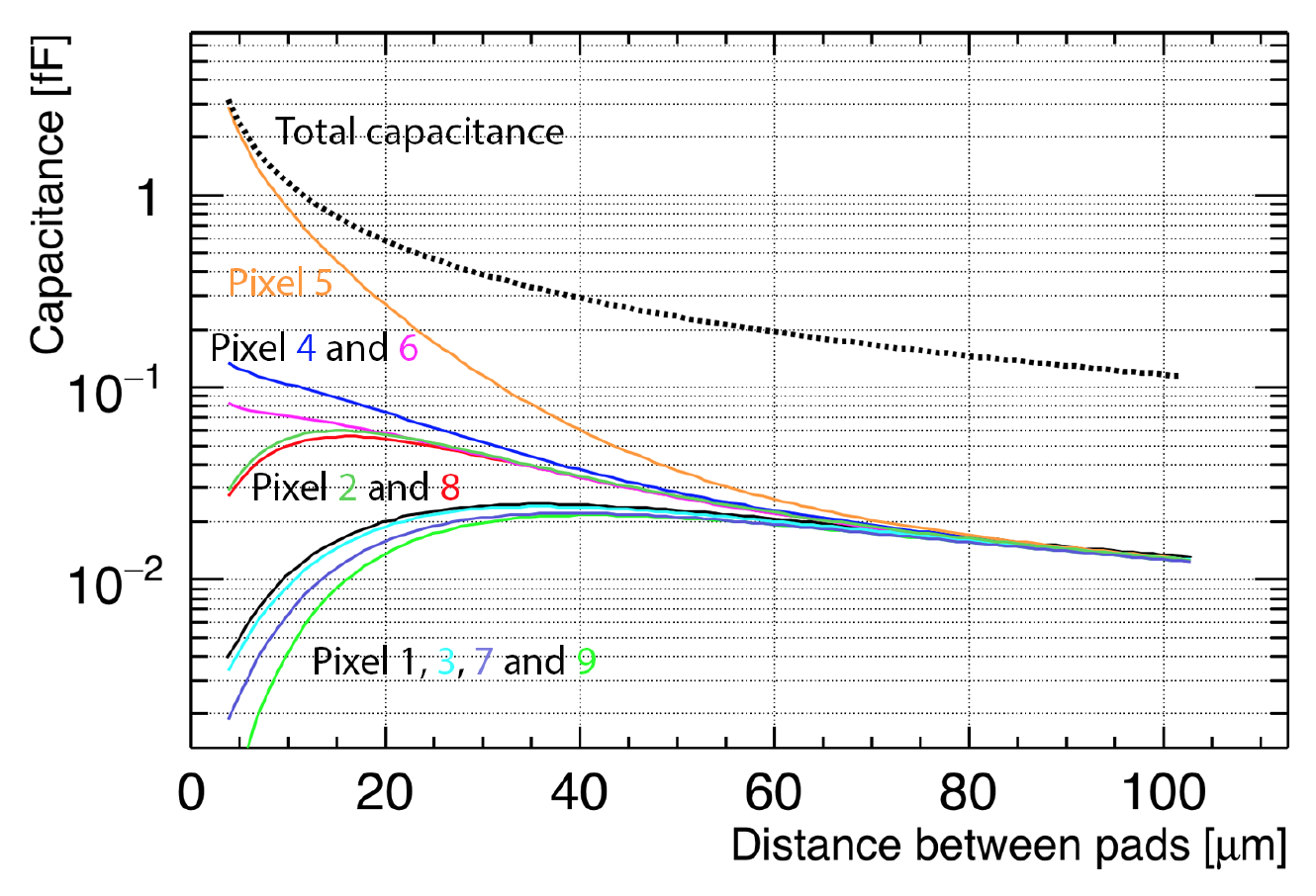}};
    \begin{scope}[x={(image.south east)},y={(image.north west)}]
     \node[anchor=north east] at (0.85,0.85){\small CLICdp};
  \end{scope}
  \end{tikzpicture}
    \caption{}\label{fig:ccpd_coupling_capacitance_comsol}
  \end{subfigure}
  \hfill
  \begin{subfigure}[T]{.49\linewidth}
    \begin{tikzpicture}
  \node[anchor=south west,inner sep=0] at (0,0)(image){	 \includegraphics[width=\linewidth]{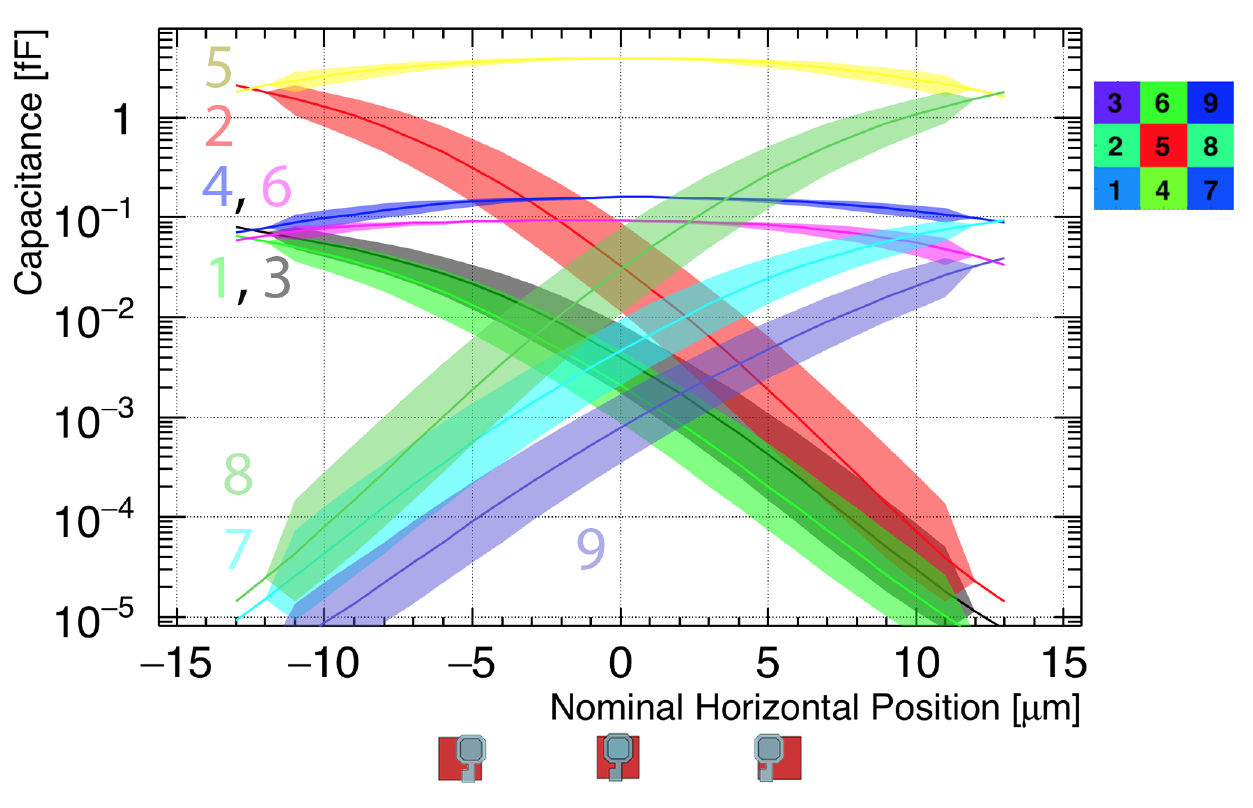}};
    \begin{scope}[x={(image.south east)},y={(image.north west)}]
     \node[anchor=south west] at (0.65,0.95){\scriptsize CLICdp};
  \end{scope}
  \end{tikzpicture}
    \caption{}\label{fig:ccpd_comsol_misalignment}
  \end{subfigure}
  \caption{Simulated coupling capacitance between the output pad on the CCPDv3 sensor and the input pad of the CLICpix readout ASIC~\cite{MateusComsol}. The coupling capacitance for the main pixel (number 5) and the cross-coupling capacitance to the eight surrounding neighbouring pixels is shown as a function of \subref{fig:ccpd_coupling_capacitance_comsol} the glue layer thickness and of \subref{fig:ccpd_comsol_misalignment} the lateral misalignment between sensor and readout ASIC.}\label{fig:ccpd_comsol_results}
\end{figure}

The capacitance simulation setup is also used to obtain the nominal coupling capacitance between the pads of the next generation sensor and readout ASIC (C3PD and CLICpix2) and to confirm the effect of a guard-ring mesh surrounding the C3PD pixel pads. For nominal alignment parameters, the coupling capacitance to the main CLICpix2 input pad is \SI{3.5}{\femto\farad}. The guard ring~\cite{MateusComsol} shields the C3PD pads from the neighbouring pixels in the CLICpix2 readout ASIC and thereby reduces the cross-coupling by almost an order of magnitude, while slightly increasing the main coupling capacitance. It also adds load to the C3PD amplifier, as the coupling capacitance to the guard ring itself is similar to the one to the main pixel of the readout ASIC.

\subsection{Calibration, glue uniformity, coupling capacitance}\label{sec:glue-quality}
Special care has been taken during the design phase of the second generation of the active sensor and the readout ASIC (C3PD and CLICpix2) to allow for a characterisation of the capacitive coupling between both chips in the assembly. To this end, both chips offer the possibility to inject test pulses into the respective amplifier input. Per pixel, a small capacitance can be switched between two voltage levels, and thereby a fixed amount of charge is injected into the front-end.

The ToT response to a test pulse injection into the CLICpix2 front-end for various test pulse amplitudes is used to calibrate the CLICpix2 amplifier stage. A nominal value of the test pulse capacitance of \SI{10}{\femto\farad} is assumed. Based on the assumption that the ToT response does not depend on the origin of the charge signal, whether test pulse or injected into the input pad, the amount of charge coupled into the CLICpix2 front-end due to a signal generated by C3PD can be determined.

To perform the characterisation of the coupling capacitance, the analogue output of several pixels on C3PD are made routed on pads in the chip periphery. The test pulse circuit of one pixel has been altered in the design to allow for a direct injection of a voltage signal into the C3PD output pad. Using this pixel, a voltage step with known amplitude can be injected into the glue layer. This results in a charge signal on the CLICpix2 input pad. Using the known relation between input charge and ToT response of CLICpix2, the amount of charge is measured. From this, and the monitored C3PD output amplitude, the unknown coupling capacitance can be determined.

This deduction is valid for the one pixel in C3PD for which the test pulse amplitude is routed directly to the output pad. For the other pixels with monitoring output, an additional calibration step between test pulse voltage and amplitude response is necessary. With this, the capacitance can be calculated for those pixels, as well. Under the assumption that this relation is common to all pixels in the C3PD matrix, the coupling capacitance through the glue layer can be calculated for all \num{128x128} pixels of the assembly.

\cref{fig:c3pd_glue_testpulse} illustrates the ToT response of the CLICpix2 front-end to a fixed test pulse charge injected into the sensor input for two capacitively coupled detector assemblies. The monitored pixels are located close to the upper right corner of the plots. For the assembly shown in \cref{fig:c3pd_glue_testpulse_ass7}, a coupling capacitance of \SI{2.95}{\femto\farad} in the upper right corner has been extracted. Given the slight gradient to higher ToT towards the lower end of the matrix, this is in good agreement to the simulated value of \SI{3.5}{\femto\farad} as summarised in \cref{sec:coupling_simulation}. The assembly shown in \cref{fig:c3pd_glue_testpulse_ass6} shows a strong non-uniformity of the test pulse response over the pixel matrix. The four dots of epoxy glue applied to the silicon die prior to the flip-chip assembly process are clearly visible as regions of stronger signals, surrounded by regions in which the lower dielectric constant of air, as opposed to the glue ,leads to a lower coupling capacitance. The extracted capacitance in the upper right corner is \SI{1.46}{\femto\farad}, attributed to the lack of glue in this region.
These results indicate that the gluing process has to be well controlled in order to achieve a homogeneous and high coupling capacitance throughout the detector assembly.
For full-size chips in the order of several \si{\cm} size, a good understanding and control of the gluing process will be essential.

\begin{figure}[ht]
  \begin{subfigure}[T]{0.45\linewidth}
\begin{tikzpicture}
\node[anchor=south west,inner sep=0] (image) at (0,0) {
\includegraphics[width=\linewidth]{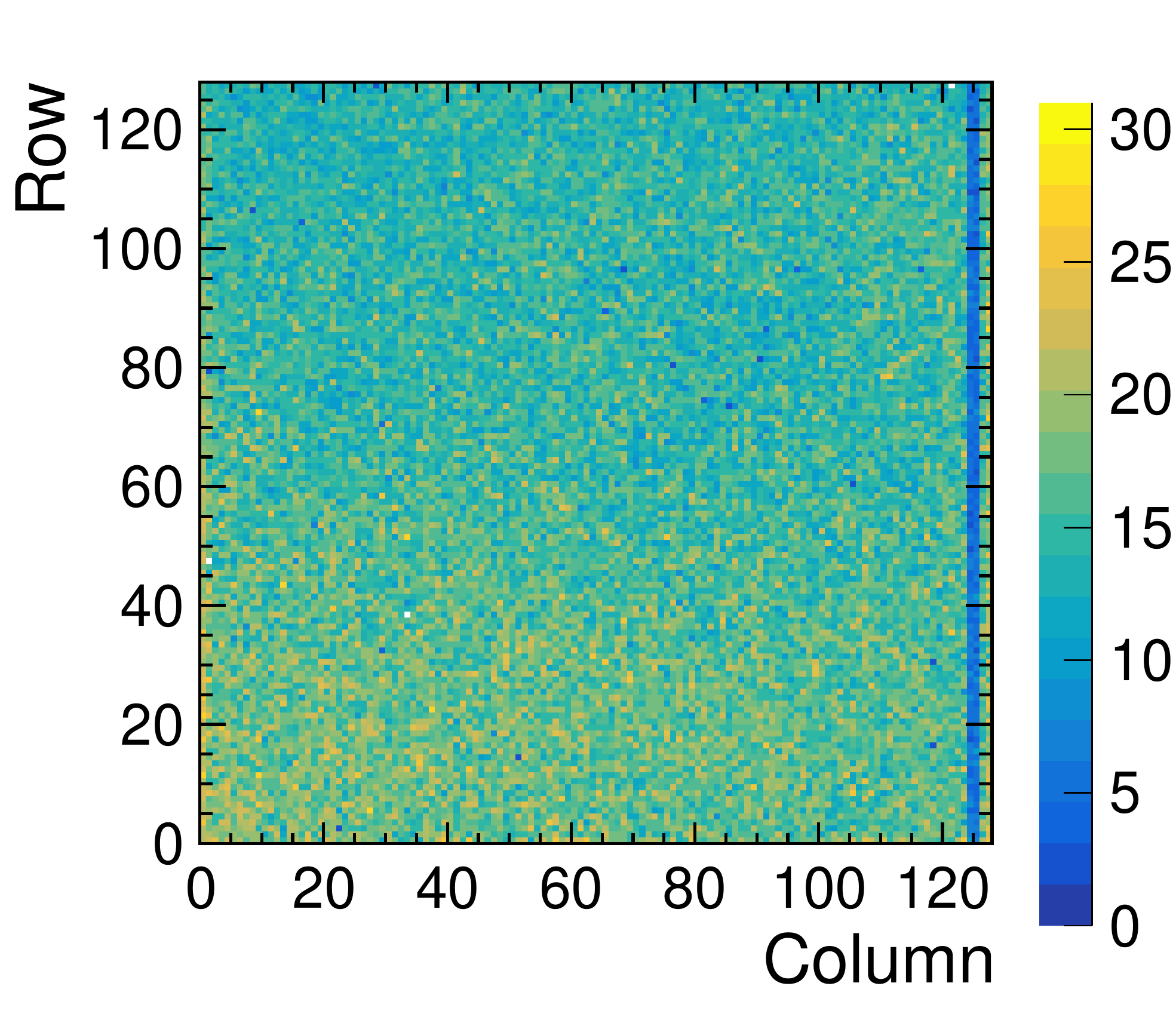}};
\begin{scope}[x={(image.south east)},y={(image.north west)}]
  \node[black,rotate=90,anchor=east] at (1.05,0.9){Mean ToT};
  \node[anchor=south east] at (0.85,0.925){\scriptsize CLICdp};
\end{scope}
\end{tikzpicture}
\caption{}\label{fig:c3pd_glue_testpulse_ass7}
\end{subfigure}
\hspace{.5cm}
\begin{subfigure}[T]{0.45\linewidth}
\begin{tikzpicture}
\node[anchor=south west,inner sep=0] (image) at (0,0) {
\includegraphics[width=\linewidth]{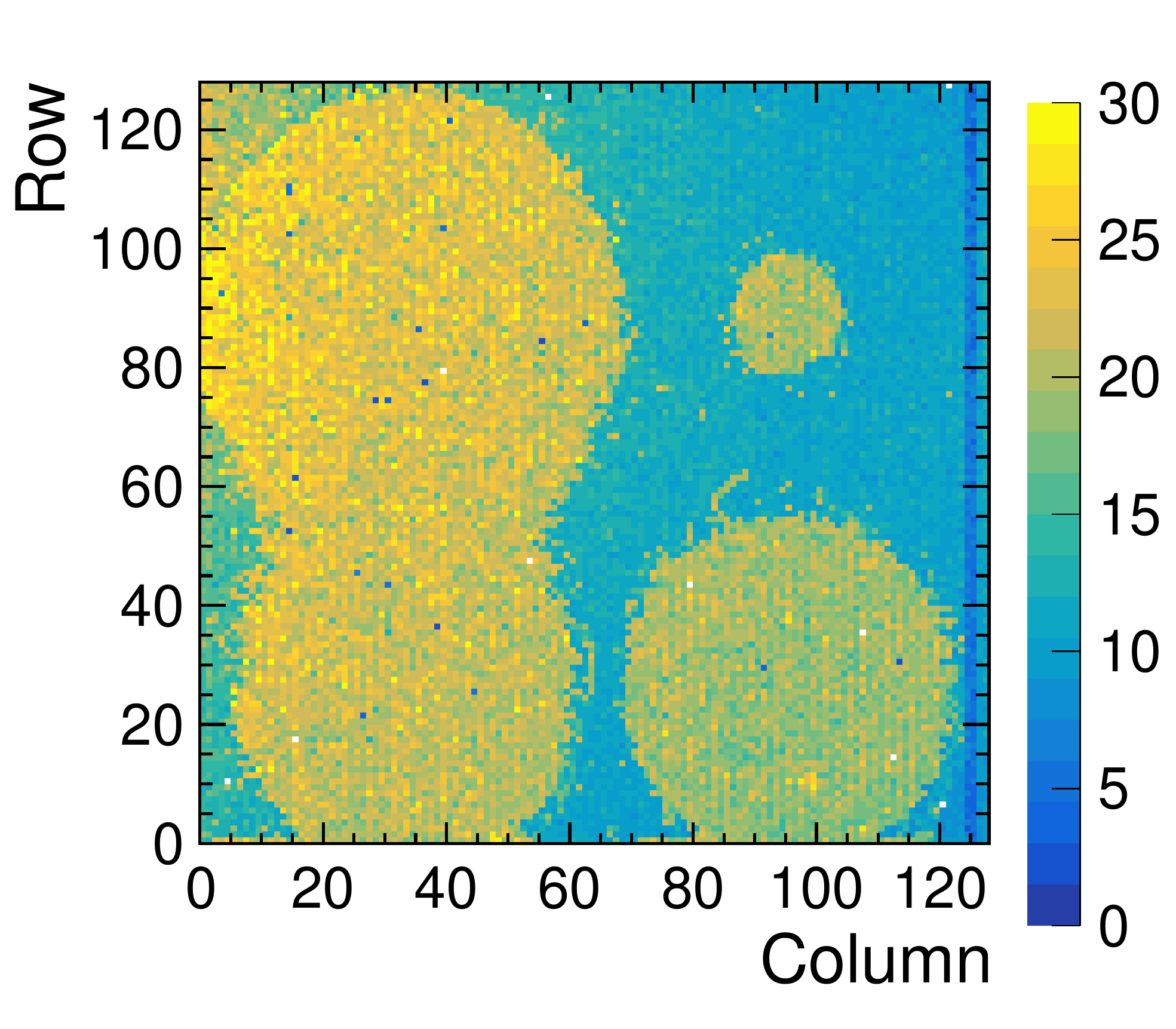}};
\begin{scope}[x={(image.south east)},y={(image.north west)}]
\node[black,rotate=90,anchor=east] at (1.05,0.9){Mean ToT};
\node[anchor=south east] at (0.85,0.925){\scriptsize CLICdp};

\end{scope}
\end{tikzpicture}
\caption{}\label{fig:c3pd_glue_testpulse_ass6}
\end{subfigure}
\caption{Testpulse response through the glue layer for two different CLICpix2 + C3PD capacitively coupled pixel-detector assemblies. The detector shown in \subref{fig:c3pd_glue_testpulse_ass7} has a rather uniform response over the pixel matrix, whereas the detector shown in \subref{fig:c3pd_glue_testpulse_ass6} clearly has a non-uniformity due to the glue application process. The reduced response in columns 124 and 125 originates from a different amplifier design in the C3PD front-end and is not related to the capacitive coupling.}\label{fig:c3pd_glue_testpulse}
\end{figure}
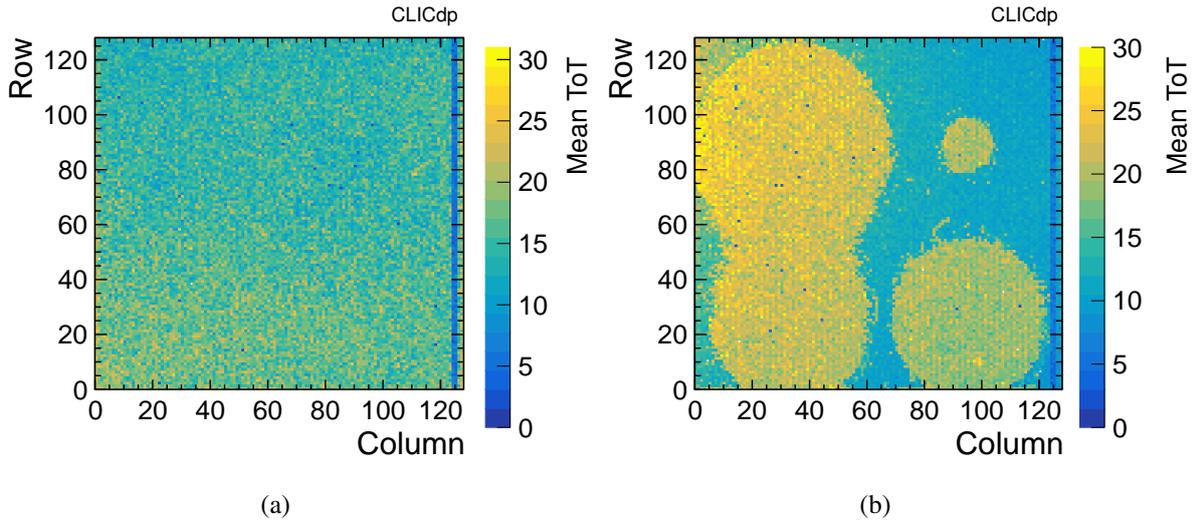

\subsection{Test-beam results}
Several capacitively coupled detector assemblies comprising CCPDv3 and CLICpix as well as C3PD and CLICpix2 have been tested in a particle beam at CERN. Detailed analyses of the tracking resolution, detection efficiency, cross-coupling, charge collection and timing have been performed~\cite{CCPDv3_hynds,buckland_ccpd,Nurnberg_Vertex2017}. The CLICpix and CCPDv3 hybrid detector assembly is more than \SI{99}{\percent{}} efficient at the nominal detection threshold of approximately \SI{1200}{\Pem} for a wide range of bias voltages and reaches close to \SI{100}{\percent{}} efficiency at the maximal applicable voltage of \SI{80}{\volt}, as illustrated in \cref{fig:ccpd_v3_efficiency_bias}. The single-point hit resolution at perpendicular particle incident is in the order of \SIrange{6}{8}{\micron} for C3PD (\cref{fig:ResDut_c3pd}) and CCPDv3, as expected from the fact that most hits lead to single-pixel clusters in the very thin depleted layers of the active sensors. For larger incidence angles the higher fraction of two-pixel clusters leads to a slightly improved resolution, as depicted in \cref{fig:ccpd_resolution_angle}. 
\begin{figure}[ht]
  \begin{subfigure}[T]{.49\linewidth}
    \includegraphics[width=\linewidth]{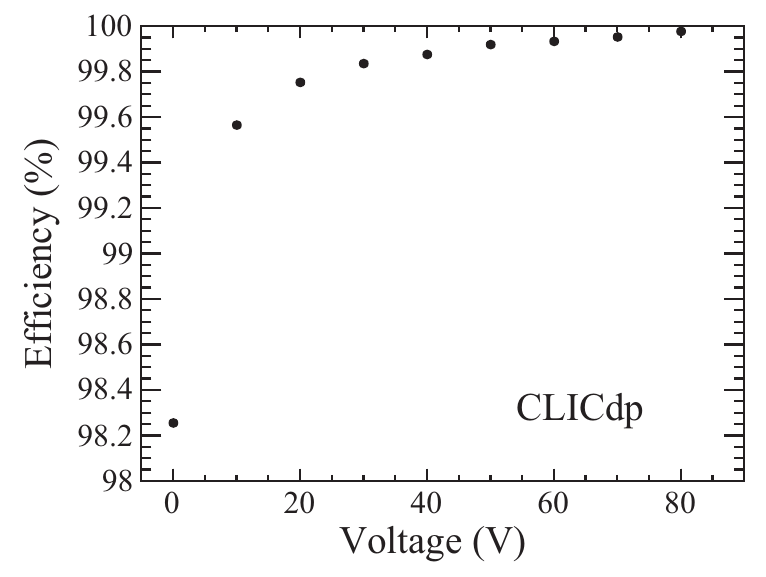}
    \caption{}\label{fig:ccpd_v3_efficiency_bias}
  \end{subfigure}
  \hfill
  \begin{subfigure}[T]{.49\linewidth}

    \begin{tikzpicture}
      \node[anchor=south west,inner sep=0] (image) at
      (0,0){  \includegraphics[width=\linewidth]{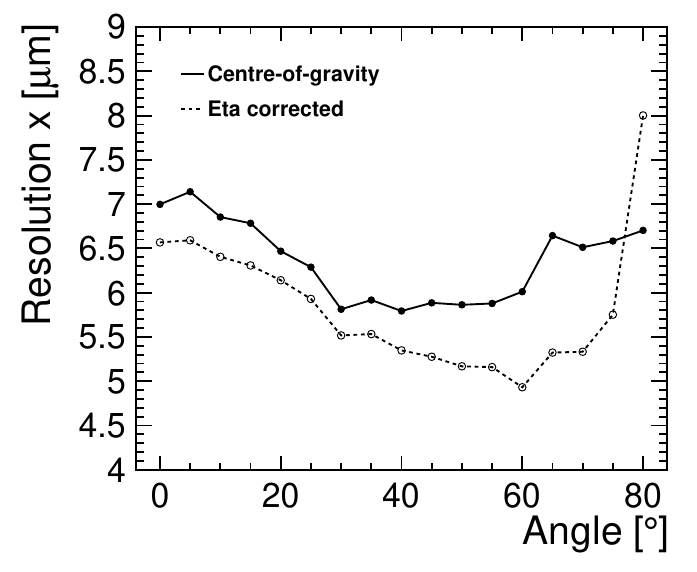}};
      \begin{scope}[x={(image.south east)},y={(image.north west)}]
        \node[above, color=black] at (0.75, 0.75) {\small{CLICdp}};
      \end{scope}
    \end{tikzpicture}
    \caption{}\label{fig:ccpd_resolution_angle}
  \end{subfigure}
  \caption{\subref{fig:ccpd_v3_efficiency_bias} Single hit efficiency versus bias voltage~\cite{CCPDv3_hynds} at the nominal detection threshold of approximately \SI{1200}{\Pem} and \subref{fig:ccpd_resolution_angle} single hit resolution in the x-direction as a function of the incidence angle and for two different cluster-position reconstruction algorithms~\cite{buckland_ccpd}, obtained in test-beam measurements with capacitively coupled hybrid assemblies of CLICpix and CCPDv3.}\label{fig:ccpd_testbeam_result}
\end{figure}

\cref{fig:c3pd_testbeam_summary} summarises results obtained with C3PD and CLICpix2 assemblies. The differences in cluster-signal and cluster-size distributions among individual detector assemblies are attributed to the varying glue-assembly quality discussed in \cref{sec:glue-quality}. As expected from the overall low cluster multiplicities, the width of the residual distributions of $\sim\SI{7}{\micron}$ is mostly determined by the square pixel pitch of \SI{25}{\micron} and not significantly affected by the difference in cluster signals.

\begin{figure}[ht]
  \begin{subfigure}[T]{.31\linewidth}
    \begin{tikzpicture}
  	\node[anchor=south west,inner sep=0] (image) at (0,0) {
  	\includegraphics[width=\linewidth]{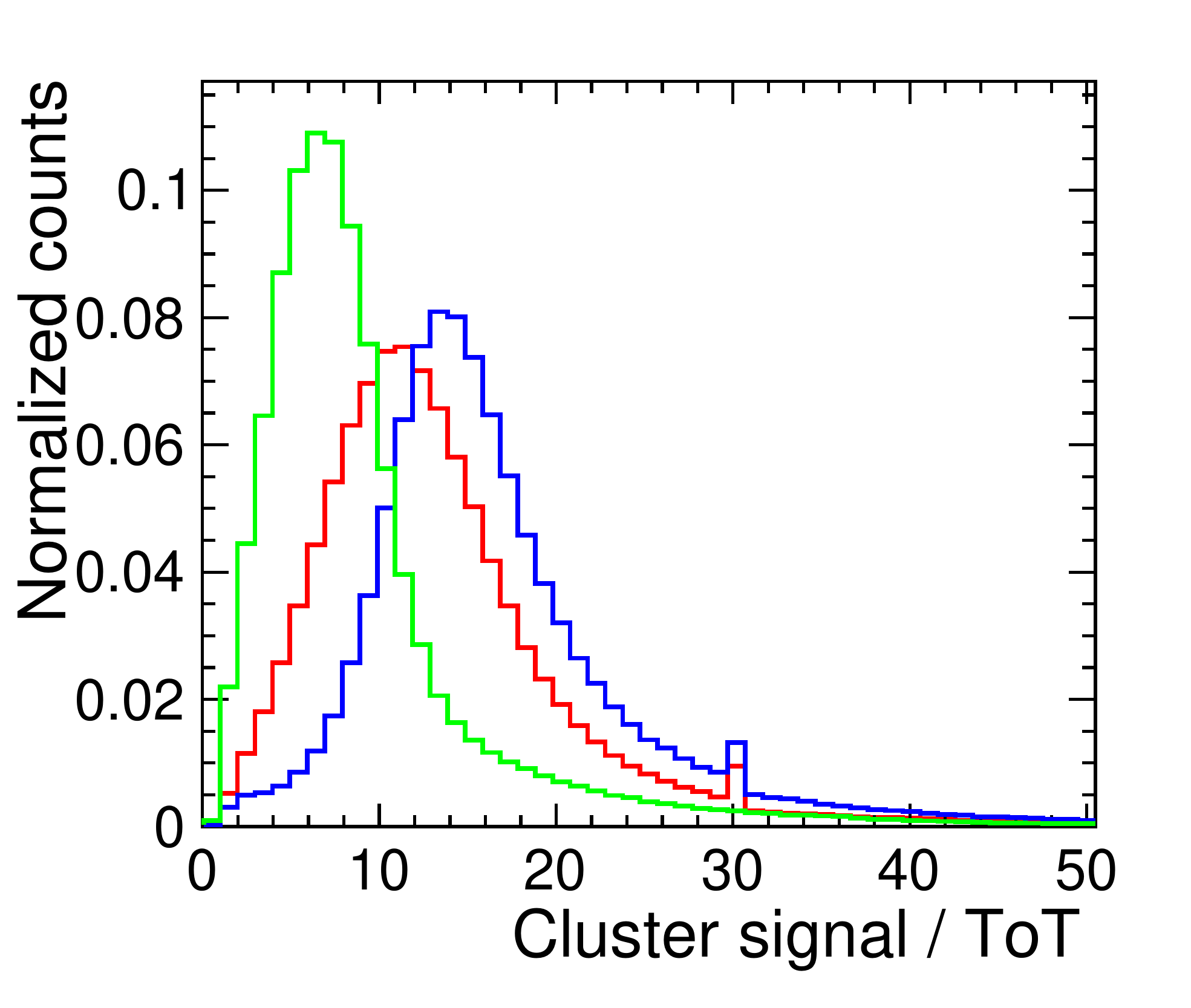}};
  	\begin{scope}[x={(image.south east)},y={(image.north west)}]
  		\node[red] at (0.75,0.8){Ass. 1};
  		\node[blue] at (0.75,0.7){Ass. 3};
  		\node[green] at (0.75,0.6){Ass. 5};
  		\node[anchor=south east, color=black] at (0.9, 0.25) {\scriptsize{CLICdp}};
  	\end{scope}
  \end{tikzpicture}
  \caption{}\label{fig:Clustersignal_c3pd}
  \end{subfigure}
  \hfill
  \begin{subfigure}[T]{.31\linewidth}
  \begin{tikzpicture}
	  \node[anchor=south west,inner sep=0] (image) at (0,0) {
	  \includegraphics[width=\linewidth]{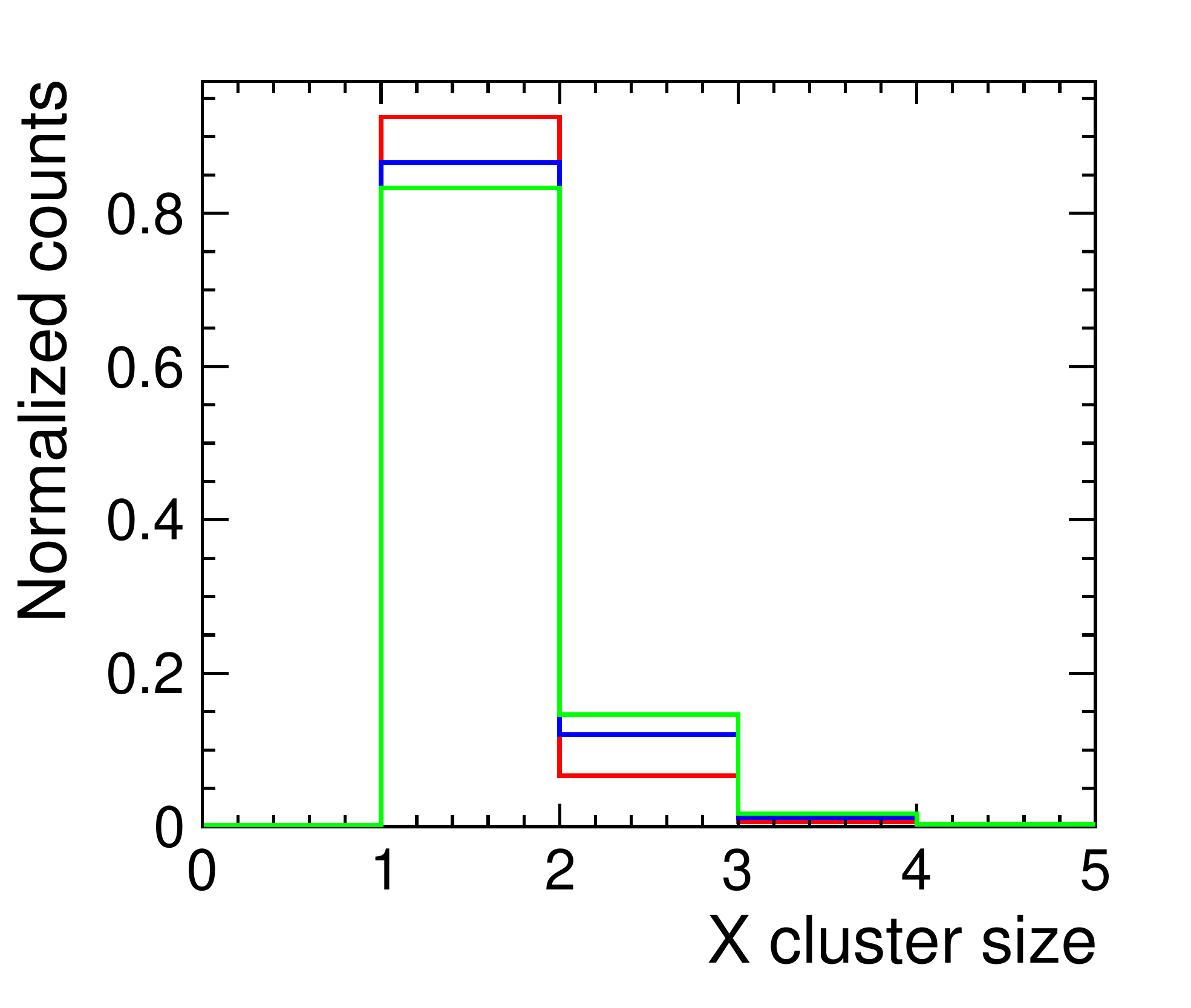}};
	  \begin{scope}[x={(image.south east)},y={(image.north west)}]
	    \node[red] at (0.75,0.8){Ass. 1};
	    \node[blue] at (0.75,0.7){Ass. 3};
	    \node[green] at (0.75,0.6){Ass. 5};
	    \node[anchor=south east, color=black] at (0.9, 0.25) {\scriptsize{CLICdp}};
	  \end{scope}
	\end{tikzpicture}
  \caption{}\label{fig:clustersize_x_c3pd}
  \end{subfigure}
  \hfill
    \begin{subfigure}[T]{.31\linewidth}
    \begin{tikzpicture}
    \node[anchor=south west,inner sep=0] (image) at (0,0) {
    \includegraphics[width=\linewidth]{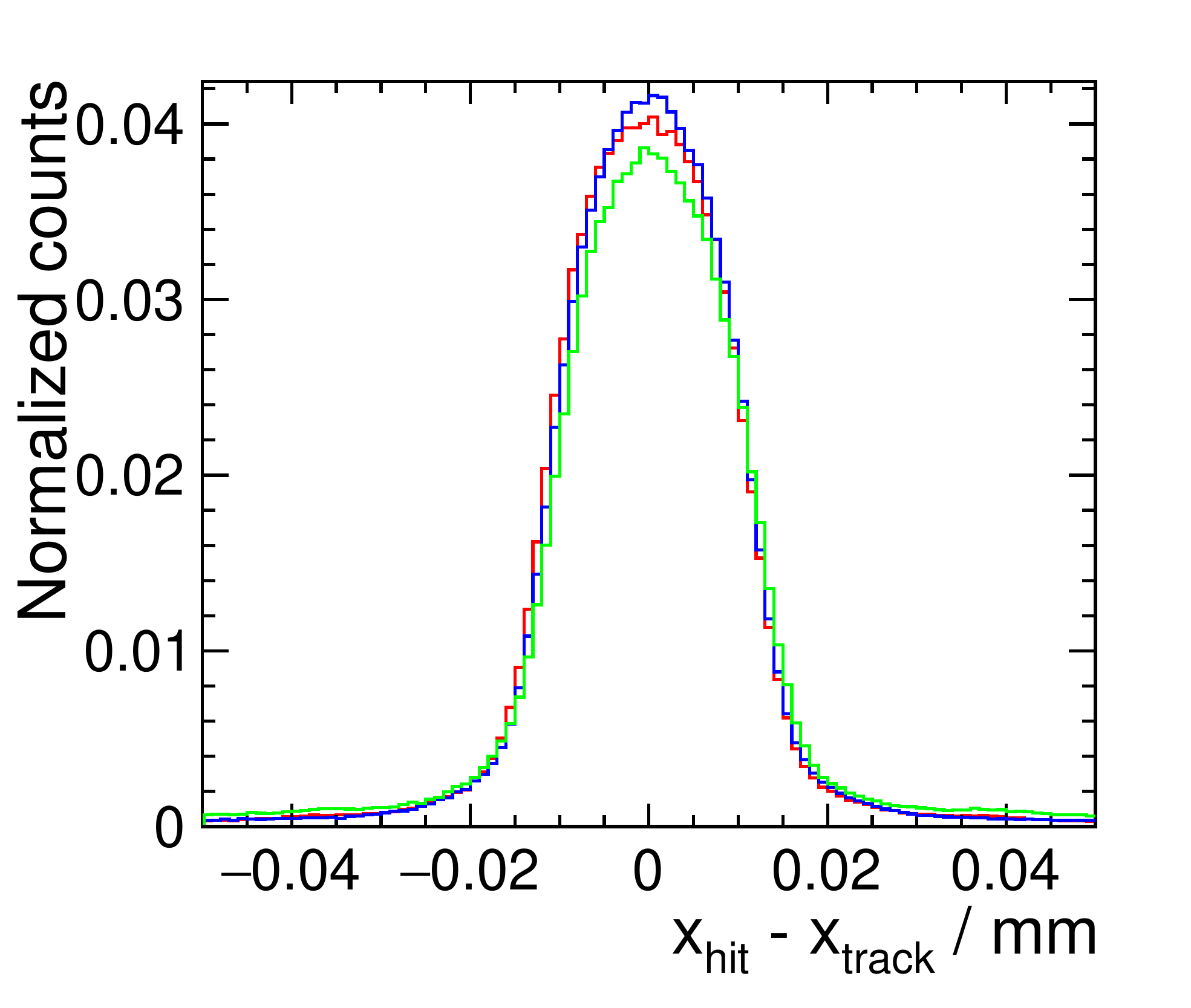}};
    \begin{scope}[x={(image.south east)},y={(image.north west)}]
      \node[red] at (0.75,0.8){Ass. 1};
      \node[blue] at (0.75,0.7){Ass. 3};
      \node[green] at (0.75,0.6){Ass. 5};
      \node[anchor=south east, color=black] at (0.9, 0.25) {\scriptsize{CLICdp}};
    \end{scope}
    \end{tikzpicture}
  \caption{}\label{fig:ResDut_c3pd}
  \end{subfigure}
  \caption{\subref{fig:Clustersignal_c3pd} cluster-signal, \subref{fig:clustersize_x_c3pd} cluster-size and \subref{fig:ResDut_c3pd} residual distribution, as measured in test beam for three capacitively coupled detector assemblies consisting of C3PD and CLICpix2~\cite{Nurnberg_Vertex2017}.}\label{fig:c3pd_testbeam_summary}
\end{figure}
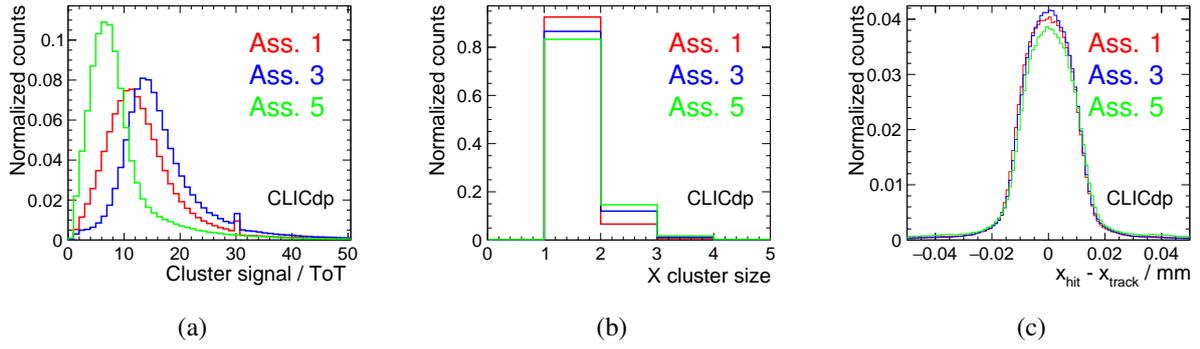

To determine the timing performance of the capacitively coupled assemblies in test-beam measurements, the time-of-arrival measurement by CLICpix2 is compared to the precise reference time of the particle track. CLICpix2 is operated with a \SI{100}{\mega\hertz} clock, and the time-stamp binning is thus \SI{10}{\ns}. \cref{fig:c3pd_timing_residual} depicts the time residual distribution, before and after applying a time-walk correction. A Gaussian fit to the non-corrected distribution yields a resolution of \SI{9}{\ns}.
The non-corrected distribution shows non-Gaussian tails towards later hit arrival times, as expected from time walk in the discriminator in the pixel front-end. The time-walk effect introduces a correlation between the hit arrival time and the hit energy measured in time-over-threshold counts (ToT). This dependence can be exploited to compensate for the effect. In \cref{fig:c3pd_timewalk} the extracted correction function for each ToT bin is shown in red. After correction, the time residual distribution has a Gaussian shape, and a fit results in a time resolution of \SI{7}{\ns}. In this result, the reference resolution has not been deconvolved.

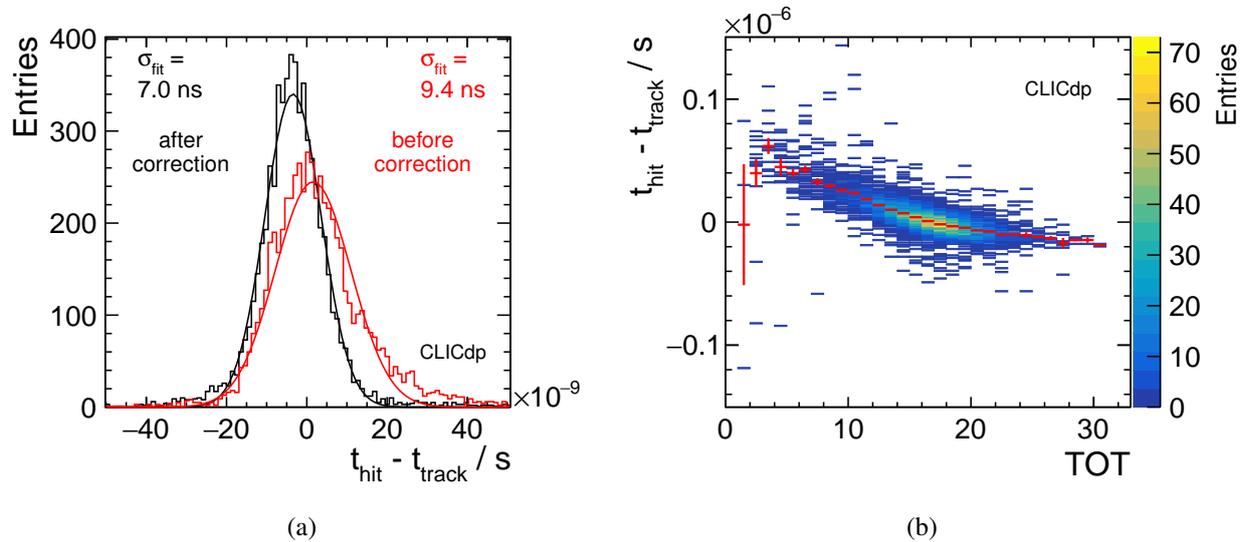
\begin{figure}[ht]
  \begin{subfigure}[T]{.49\linewidth}
\begin{tikzpicture}
\node[anchor=south west,inner sep=0] (image) at (0,0) {
\includegraphics[width=\linewidth]{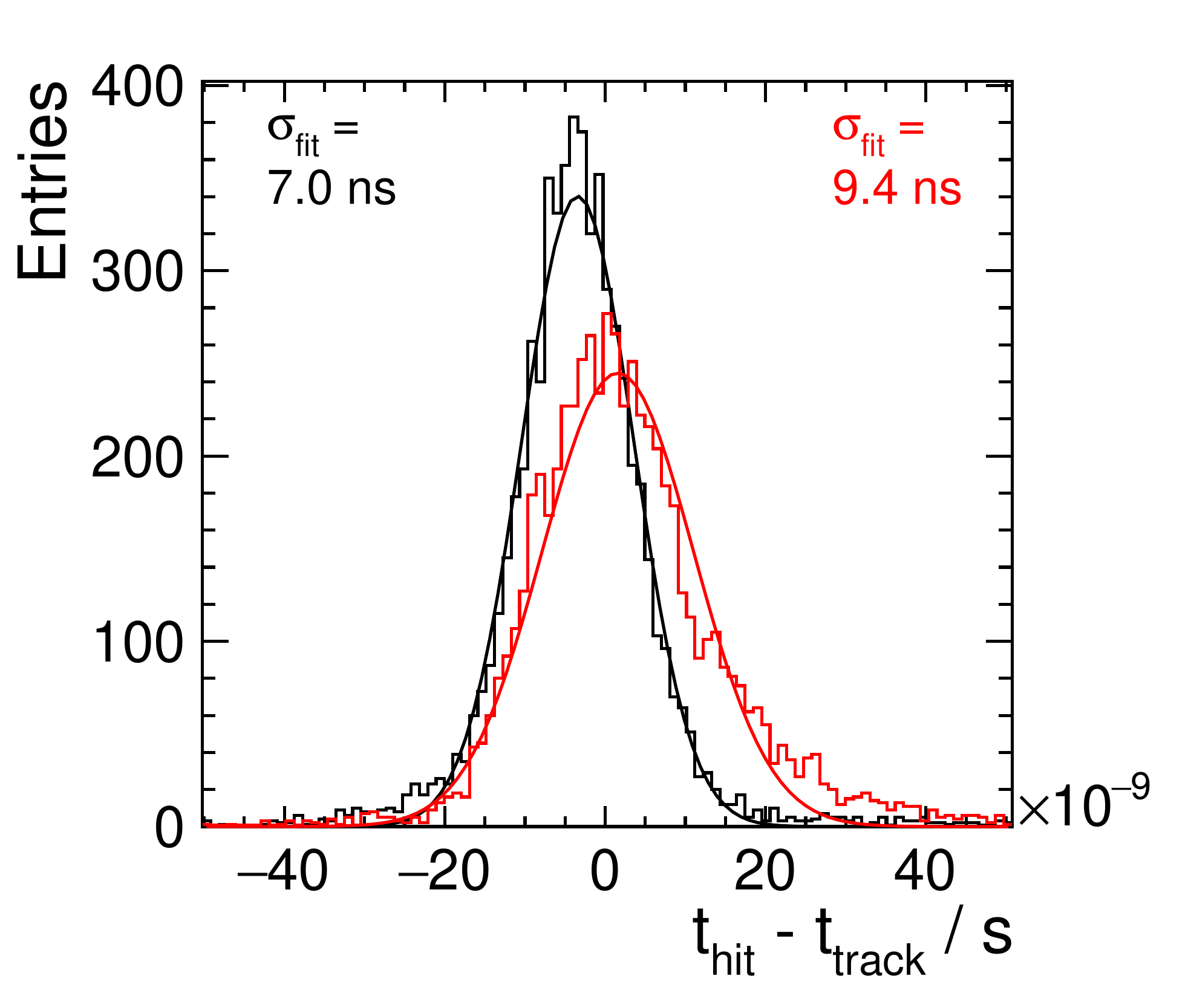}
};
\begin{scope}[x={(image.south east)},y={(image.north west)}]
  \node[anchor=south east, color=black] at (0.825, 0.25) {\scriptsize{CLICdp}};
  \node[anchor=north west,black] at (0.2,0.75){\footnotesize\shortstack{after\\correction}};
  \node[anchor=north east,red] at (0.8,0.75){\footnotesize\shortstack{before\\correction}};
\end{scope}
\end{tikzpicture}
\caption{}\label{fig:c3pd_timing_residual}
\end{subfigure}
\hfill
\begin{subfigure}[T]{.49\linewidth}
  \begin{tikzpicture}
  \node[anchor=south west,inner sep=0] (image) at (0,0) {
  \includegraphics[width=\linewidth]{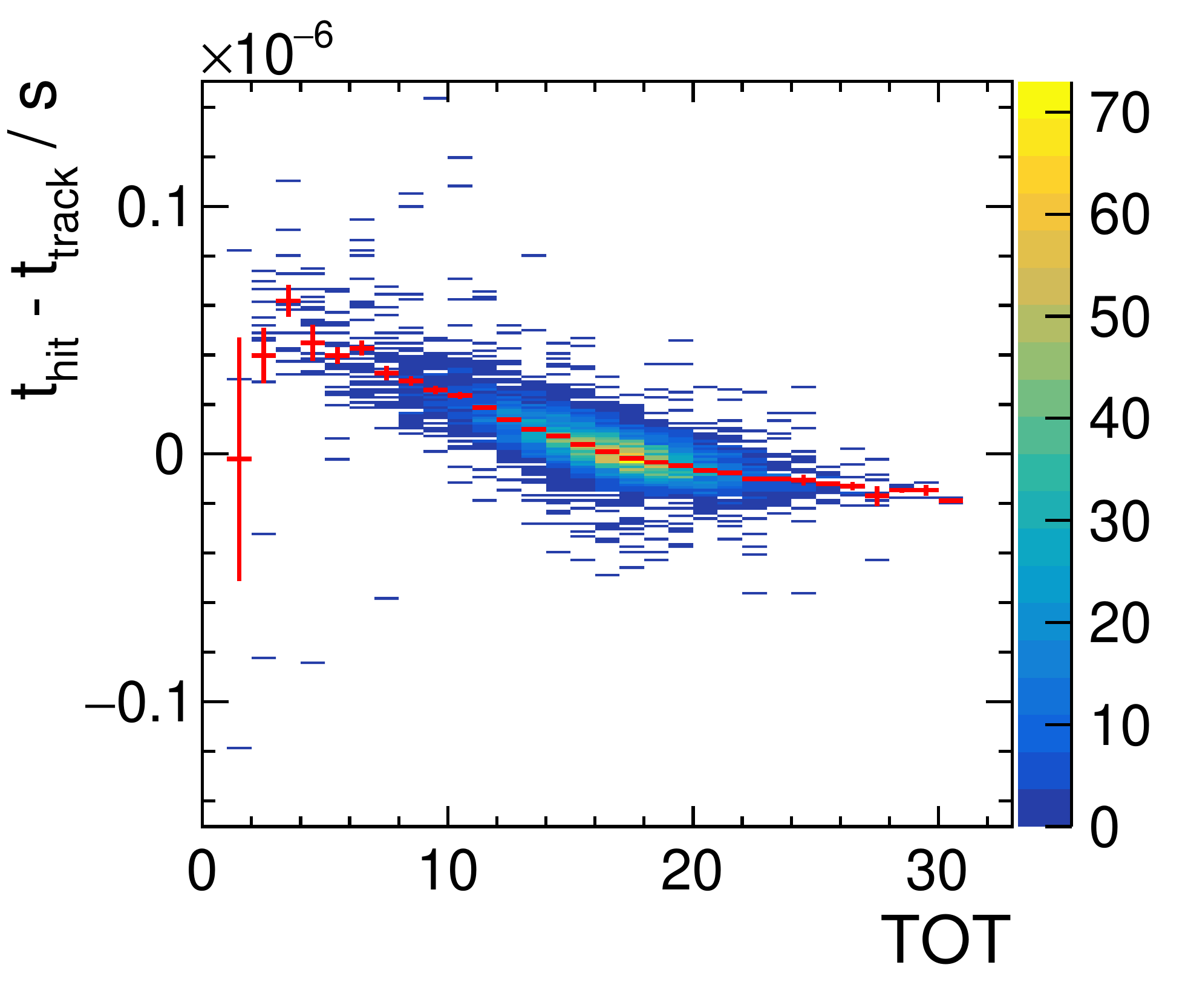}
  };
  \begin{scope}[x={(image.south east)},y={(image.north west)}]
    \node[black,rotate=90,anchor=east] at (1.01,0.925){\textsf{Entries}};
    \node[anchor=north east, color=black] at (0.8, 0.85) {\scriptsize{CLICdp}};
  \end{scope}
\end{tikzpicture}
\caption{}\label{fig:c3pd_timewalk}
\end{subfigure}
\caption{\subref{fig:c3pd_timing_residual} Timing residual distribution before and after time-walk correction and \subref{fig:c3pd_timewalk} illustration of the pulse-height dependence of the time measurement (time-walk) in a C3PD and CLICpix2 assembly~\cite{Nurnberg_Vertex2017}.}
\end{figure}

\section{Monolithic High-Voltage CMOS sensors}\label{sec:vtx-trk-monolithic-cmos}
The ATLAS CMOS collaboration and the Mu3e collaboration are developing several monolithic HV-CMOS sensors for the upgrade of the outer layer of the ATLAS pixel detector (ATLASpix) and the vertex detector of the Mu3e experiment (MuPix)~\cite{Vilella_2018}. One of these demonstrators is the self-triggered ATLASpix\_Simple sensor with an elongated pixel geometry~\cite{PERIC2018}, which has target specifications similar to the CLIC tracker requirements.
The most prominent differences are on the one hand the better spatial resolution and more stringent material and power-budget limits for CLIC, and on the other hand the relaxed radiation tolerance and readout rate, compared to ATLAS.
Several ATLASpix\_Simple sensors have been investigated in laboratory and test-beam measurements in view of the CLIC tracker requirements.

The ATLASpix\_Simple matrix was produced in a common reticle with other test chips, as shown in \cref{fig:reticle}. The main design parameters are presented in \cref{tab:atlaspix-overview-table}. The matrix contains \num{25x400} pixels of \SI{130x40}{\micron} size, giving an active matrix size of \SI{3.25x16}{\mm}. The chip is fabricated in the same commercial \SI{180}{\nm} HV-CMOS process as the active HV-CMOS sensors matching the CLICpix and CLICpix2 readout ASICs, as presented in \cref{sec:capacitively_coupled_hvcmos}. Substrates with different resistivity values have been used. The chip for which results are presented here was produced with a substrate resistivity of \SI{200}{\ohm\cm}. The collecting deep n-well covers most of the pixel area, as illustrated in \cref{fig:process}. The full pixel circuitry is embedded in this deep n-well, which also serves as collection diode for the signal. The pixel circuitry contains a charge sensitive amplifier followed by a discriminator stage. The discriminated front-end response for each pixel is routed to a dedicated readout cell in the chip periphery by a point-to-point metal connection. There, Time-of-Arrival and Time-over-Threshold is determined and passed on to the data serialiser circuitry. The process has been split into two subsets, one featuring an isolation of the P-MOS transistors by placing them in a separate, isolated n-well, rather than directly in the collecting deep n-well. The chip studied here does not have this additional shielding. To minimise noise injection into the collection well, the comparator is thus based on N-MOS transistors only. The reverse bias voltage to the substrate is applied on the front-side. The nominal operation voltage is \SI{-50}{\volt} to \SI{-60}{\volt}, resulting in an estimated depletion depth of approximately \SI{20}{\micron}, as simulated for the active sensors fabricated in the same technology (cf.~\cref{fig:hv_cmos_tcad_field_resistivity}).

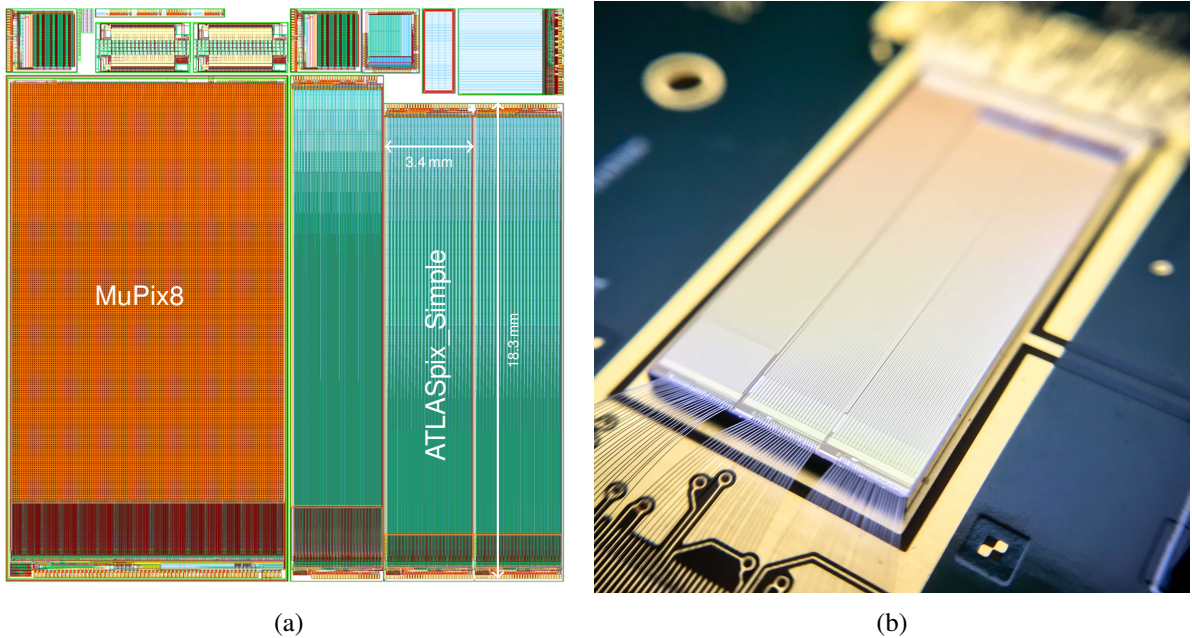
\begin{figure}[ht]
	\begin{subfigure}[b]{0.49\linewidth}\centering
	\begin{tikzpicture}
		 \node[anchor=south west,inner sep=0] (image) at (0,0) {
		 \includegraphics[width=\linewidth]{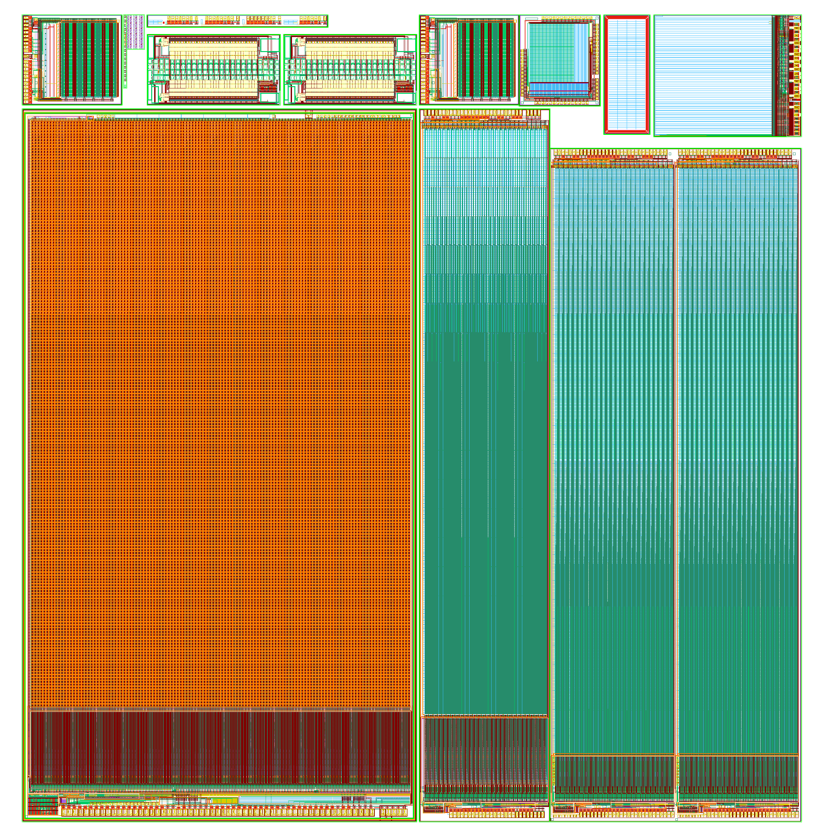}};
	\begin{scope}[x={(image.south east)},y={(image.north west)}]
				 \node[white] at (0.25,0.5){MuPix8};
				 \node[white,rotate=90] at (0.75,0.4){ATLASpix\_Simple};
				 \draw[<->,thick,white](0.85,0.024)--(0.85,0.824) node[below,pos=0.5,anchor=north,rotate=90]{\tiny\SI{18.3}{\mm}};
				 \draw[<->,thick,white](0.66,0.75)--(0.81,0.75) node[below,pos=0.5,anchor=north]{\tiny\SI{3.4}{\mm}};
				 \end{scope}
		 \end{tikzpicture}
		 \caption{}\label{fig:reticle}
	 \end{subfigure}
	 \begin{subfigure}[b]{.49\linewidth}
		\includegraphics[width=\linewidth]{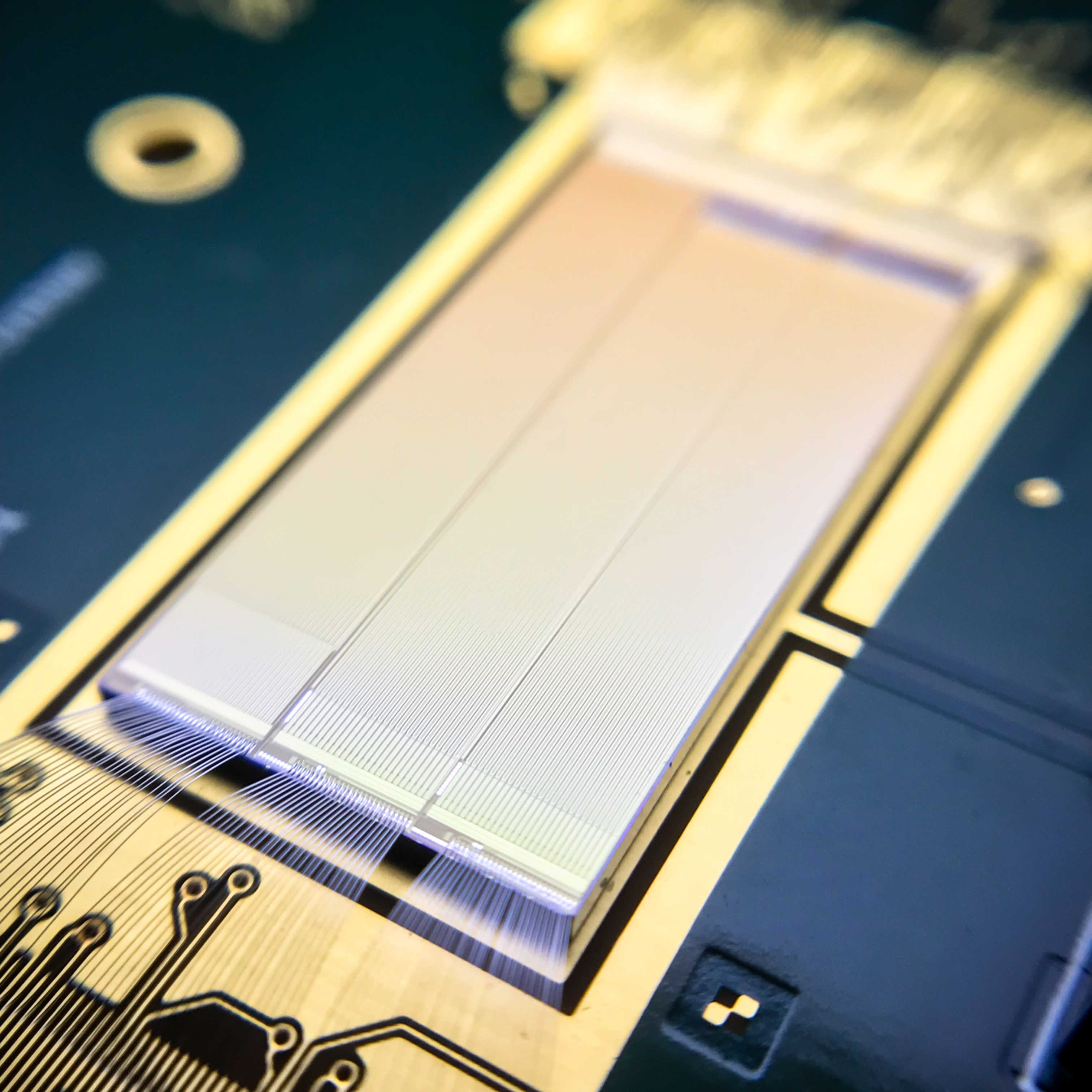}
		\caption{}\label{fig:photo}
	 \end{subfigure}
	\caption{\subref{fig:reticle} Reticle of the common MuPix/ATLASpix submission and \subref{fig:photo} photograph of the \mbox{ATLASpix} chip wire-bonded to a readout PCB.}\label{fig:atlaspix}
\end{figure}

\begin{figure}[ht]
	\includegraphics[width=\linewidth]{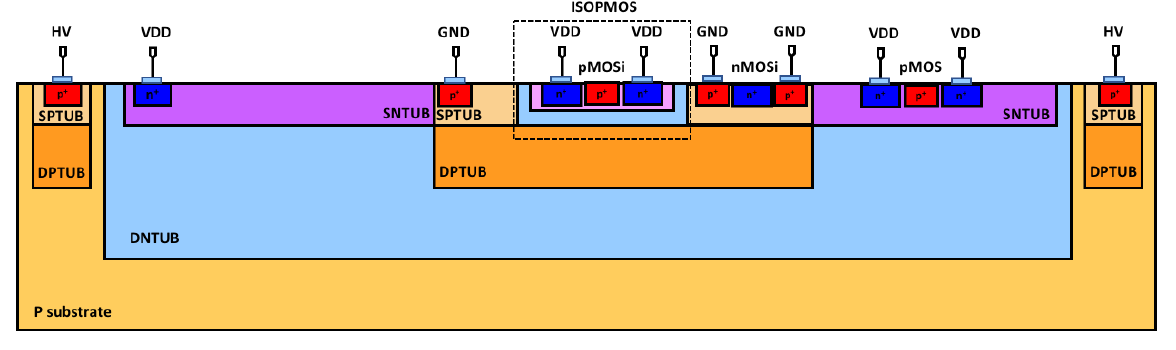}
	\caption{Schematic cross section of the \SI{180}{\nm} high voltage CMOS process used for the \mbox{ATLASpix\_Simple} monolithic sensor.}\label{fig:process}
\end{figure}

\begin{table}
  \centering
  \caption{Design and production parameters of the ATLASpix\_Simple demonstrator chip.}\label{tab:atlaspix-overview-table}
  \begin{tabular}{l c}
    \toprule
Parameter  & Value \\ \midrule
Process technology & \SI{180}{\nm} HV-CMOS \\
Substrate resistivities & 10, 50--100, 200-400, 700-2000~\SI{}{\ohm\cm}\\  
ASIC size & \SI{3.4x18.3}{\mm} \\ 
Sensitive area & \SI{3.25x16}{\mm}  \\
Matrix size & \num{25x400} pixels \\
Pixel pitch & \SI{130x40}{\micron} \\
ToT counter depth & 6.5~bit \\
ToA counter depth & 10 bit \\
ToA bin size & $\geq$\SI{12.5}{\ns} \\
Readout scheme & self triggered \\
Data output rate & 1.6~Gbps\\
Data encoding &  8\,bit\,/\,10\,bit\\
Power consumption & $<200$~mW/cm$^2$ \\ 
\bottomrule
  \end{tabular}
\end{table}

Several ATLASpix\_Simple sensors have been wire-bonded to readout PCBs (\cref{fig:photo}) and tested in the Timepix3 based telescope (\cref{sec:timepix3_telescope}) using \SI{120}{\giga\electronvolt} pion and muon beams. The CaRIBou readout system (\cref{sec:caribou}) was used to operate the sensor. The chip has been operated with a \SI{125}{\MHz} clock, resulting in a Time-of-Arrival bin size of \SI{16}{\ns}.

For most of the particle incidents on the sensor, only the directly hit pixel collects enough charge to surpass the threshold value (estimated to be around \SIrange{800}{1000}{\Pem{}}). \cref{fig:ATLASpix_mean_cluster_size_inpixel_CLICdpStyle} illustrates the distribution of pixel multiplicities mapped into a single pixel. Only along the pixel borders, charge is detected by two adjacent pixels simultaneously. For tracks hitting the majority of the pixel area, only the directly hit pixel detects the signal.

\begin{figure}[ht]
  \centering
\begin{tikzpicture}
\node[anchor=south west,inner sep=0] (image) at (0,0) {
\includegraphics[width=\linewidth]{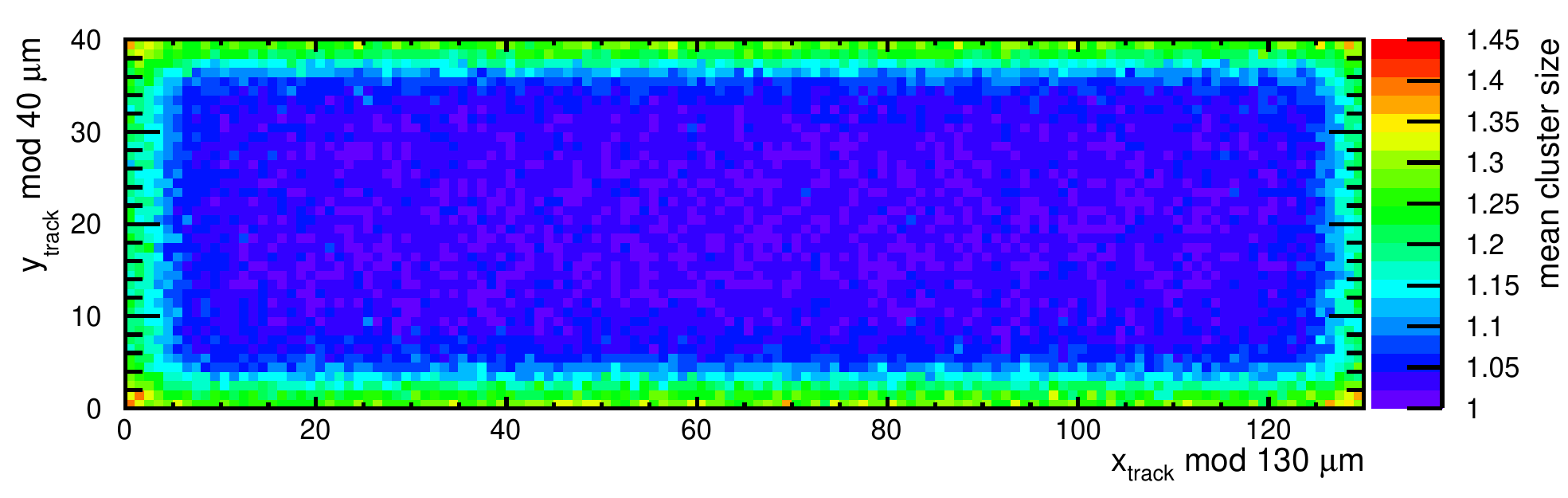}};
\begin{scope}[x={(image.south east)},y={(image.north west)}]
\node[anchor=south east] at (0.85,0.925){\scriptsize CLICdp};
\end{scope}
\end{tikzpicture}
\caption{Pixel multiplicities mapped into a single pixel, as measured in beam tests for an \mbox{ATLASpix\_Simple} sensor.}\label{fig:ATLASpix_mean_cluster_size_inpixel_CLICdpStyle}
\end{figure}

\cref{fig:atlaspix_res} shows the spatial residual distributions. Along both pixel axes, the distribution resembles a box distribution, reflecting the pixel pitch, with slight distortions coming from the track prediction resolution of the reference telescope. The box shape is closely linked to the low cluster multiplicity. The obtained RMS of the residual distribution is \SI{37.1}{\micron} along the long pixel dimension, and \SI{11.8}{\micron} along the short pixel dimension.

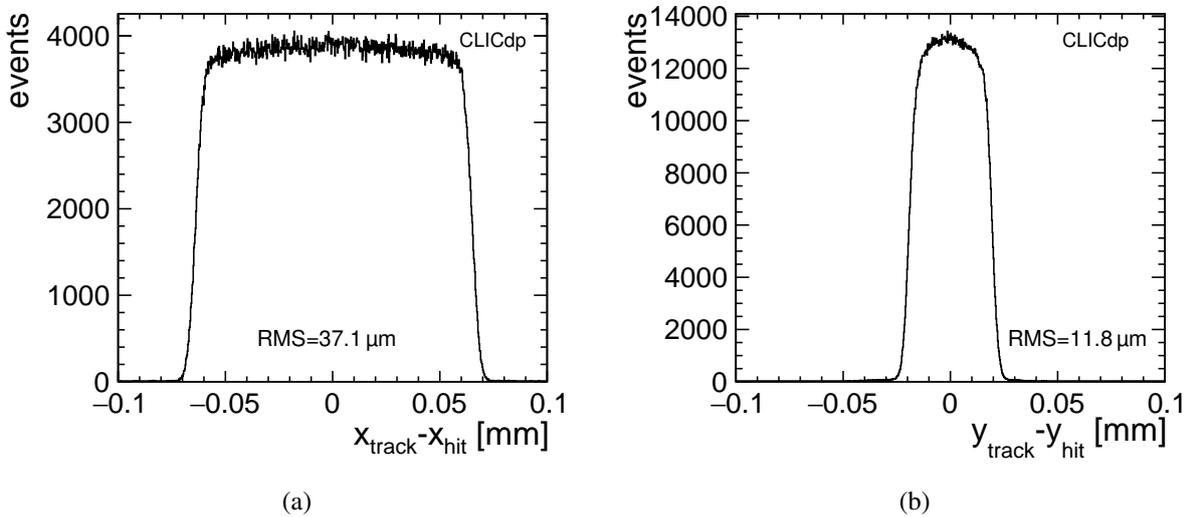
\begin{figure}[ht]
  \begin{subfigure}[b]{0.49\linewidth}
    \begin{tikzpicture}[font=\sffamily]
      \node[anchor=south west,inner sep=0] (image) at
      (0,0){	\includegraphics[width=\linewidth]{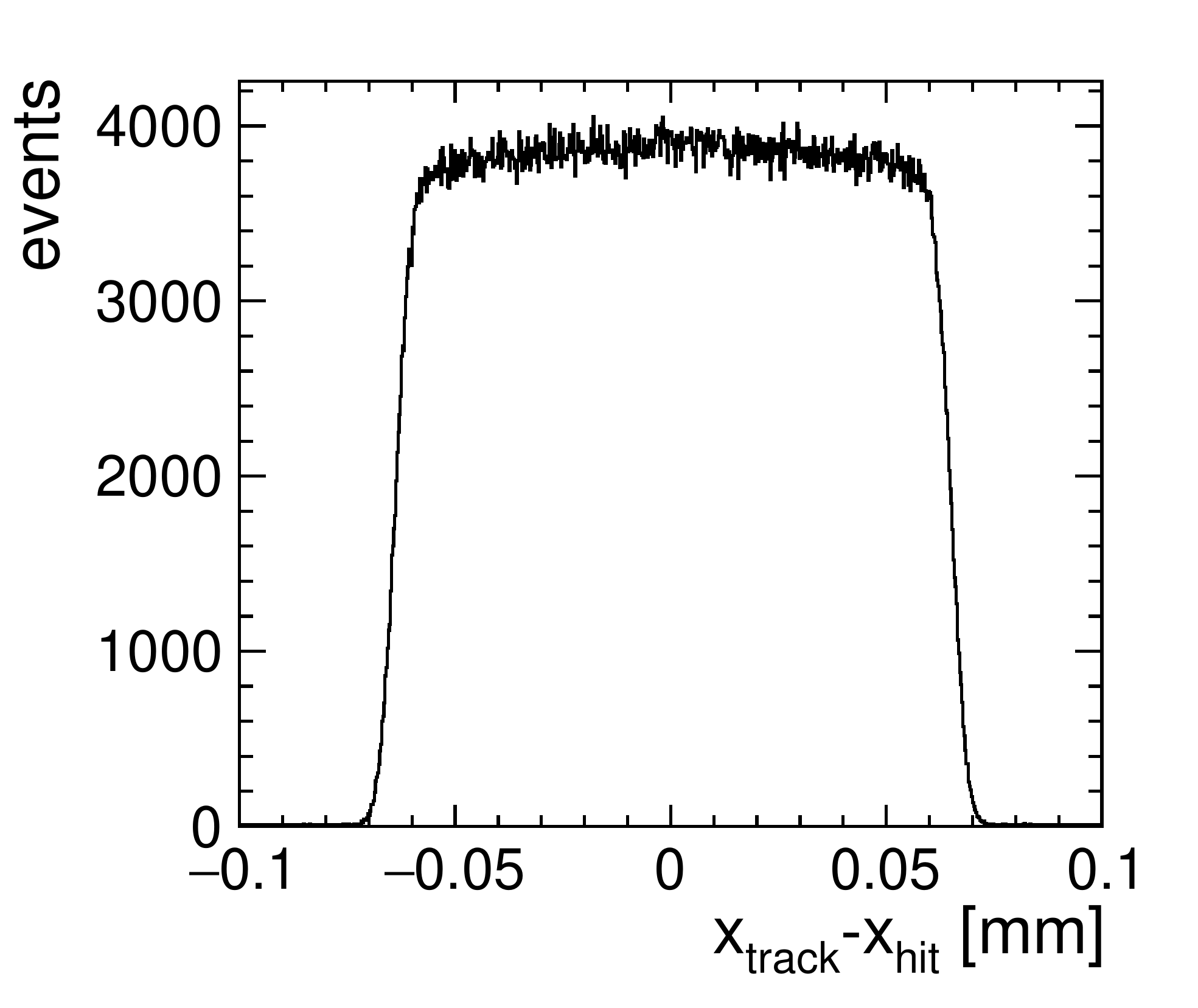}};
      \begin{scope}[x={(image.south east)},y={(image.north west)}]
        \node[anchor=north east, color=black] at (0.9, 0.9) {\scriptsize{CLICdp}};
        \node[anchor=south] at (0.55,0.225){\footnotesize RMS=\SI{37.1}{\micron}};
      \end{scope}
    \end{tikzpicture}
    \caption{}\label{fig:atlaspix_resx}
  \end{subfigure}
  ~
  \begin{subfigure}[b]{0.49\linewidth}
  \begin{tikzpicture}[font=\sffamily]
  	\node[anchor=south west,inner sep=0] (image) at
  	(0,0){	\includegraphics[width=\linewidth]{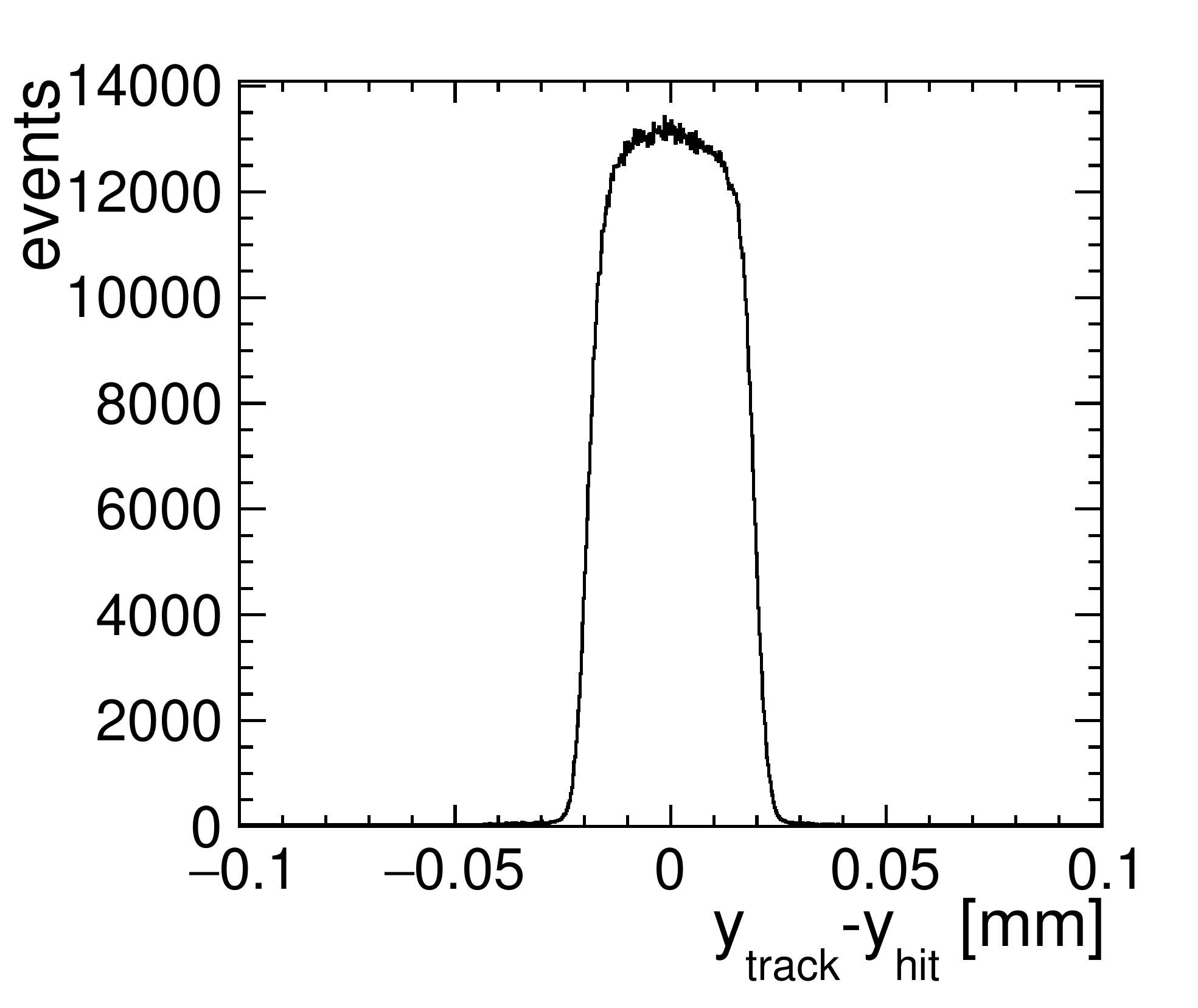}};
  	\begin{scope}[x={(image.south east)},y={(image.north west)}]
  		\node[anchor=north east, color=black] at (0.875, 0.9) {\scriptsize{CLICdp}};
      \node[anchor=south west] at (0.64,0.225){\footnotesize RMS=\SI{11.8}{\micron}};
  	\end{scope}
  \end{tikzpicture}
  \caption{}\label{fig:atlaspix_resy}
\end{subfigure}
  \caption{ATLASpix\_Simple residual distribution between the track impact point and the reconstructed hit on the detector along the \subref{fig:atlaspix_resx} long and \subref{fig:atlaspix_resy} short pixel dimension.}\label{fig:atlaspix_res}
\end{figure}

The overall detection efficiency of the studied sensor has been measured to be \SI{99.7}{\percent{}}. No obvious substructure within the pixel cell is observed, indicating no systematic efficiency degradation due to e.g.\ charge sharing or the electrode geometry. \cref{fig:atlaspix_thresholdscan} summarises the detection efficiency as a function of the applied threshold (\cref{fig:atlaspix_eff_threshold}) and bias voltage (\cref{fig:atlaspix_eff_biasvoltage}). The large operation window for efficient functioning of the detector below \SI{1500}{\Pem{}} threshold and above \SI{40}{\volt} bias is well separated from the noise region. Only below approximately \SI{600}{\Pem{}}, a total noise rate on the full chip of more than \SI{1}{\hertz} is measured. The total noise rate of about \SI{100}{\hertz} at \SI{500}{\Pem{}} threshold is still tolerable for operation in the CLIC detector. Above a threshold of \SI{1500}{\Pem{}} or a bias below \SI{40}{\volt}, the efficiency degrades due to small or shared signals not being detectable any more. The spatial resolution is not affected by the detection threshold. This is due to the fact that almost no charge is shared among neighbouring pixels, and thus the resolution is determined by the pixel geometry, approximately to $\text{pitch}/\sqrt{12}$.

\begin{figure}[ht]
    \begin{subfigure}[b]{0.49\linewidth}
  \begin{tikzpicture}[font=\sffamily]
    \node[anchor=south west,inner sep=0] (image) at
    (0,0){\includegraphics[width=\linewidth]{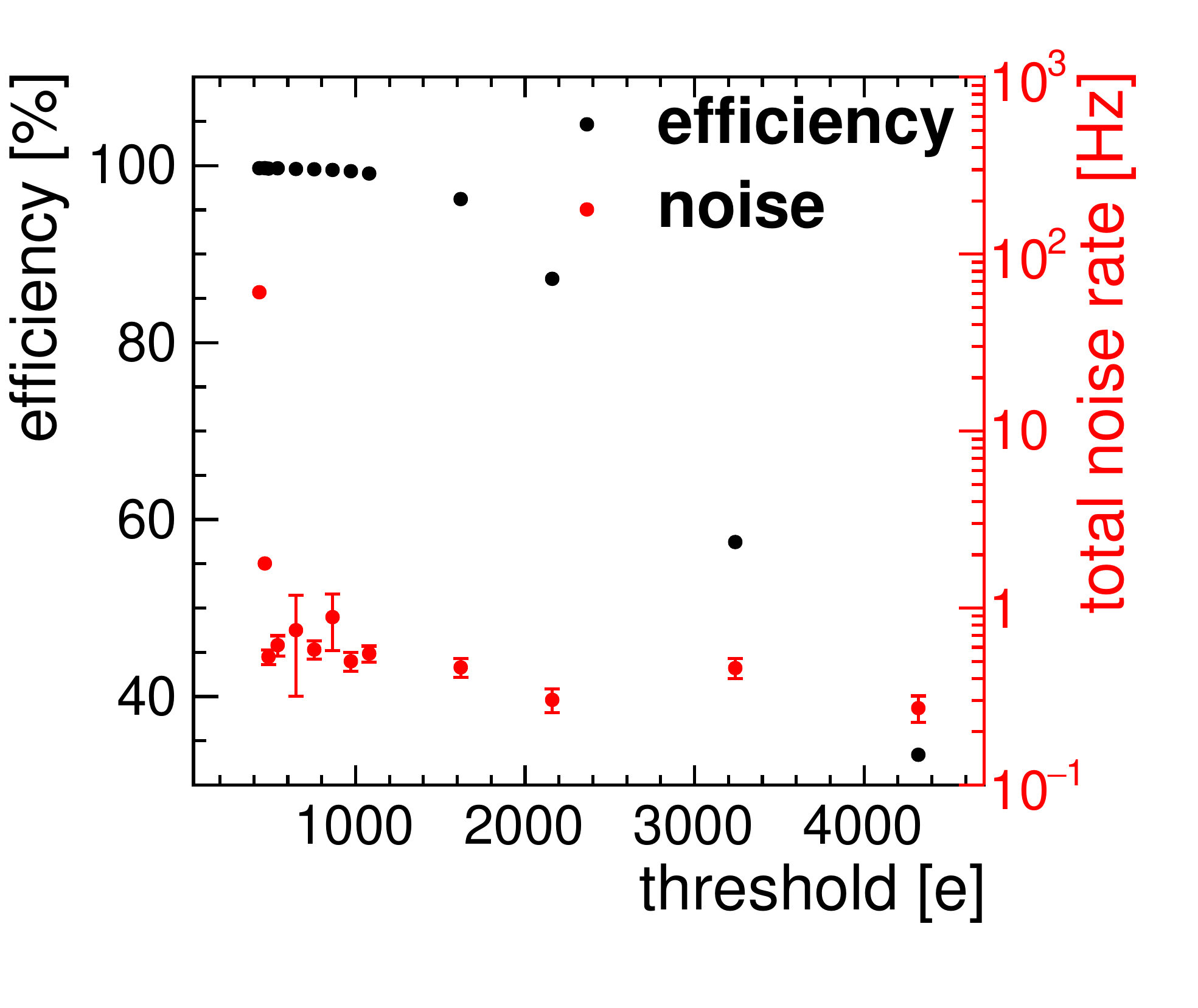}};
    \begin{scope}[x={(image.south east)},y={(image.north west)}]
      \node[anchor=east] at (0.8,0.65){\footnotesize CLICdp};
    \end{scope}
  \end{tikzpicture}
  \caption{}\label{fig:atlaspix_eff_threshold}
\end{subfigure}
~
\begin{subfigure}[b]{0.49\linewidth}
\begin{tikzpicture}[font=\sffamily]
  \node[anchor=south west,inner sep=0] (image) at
  (0,0){\includegraphics[width=\linewidth]{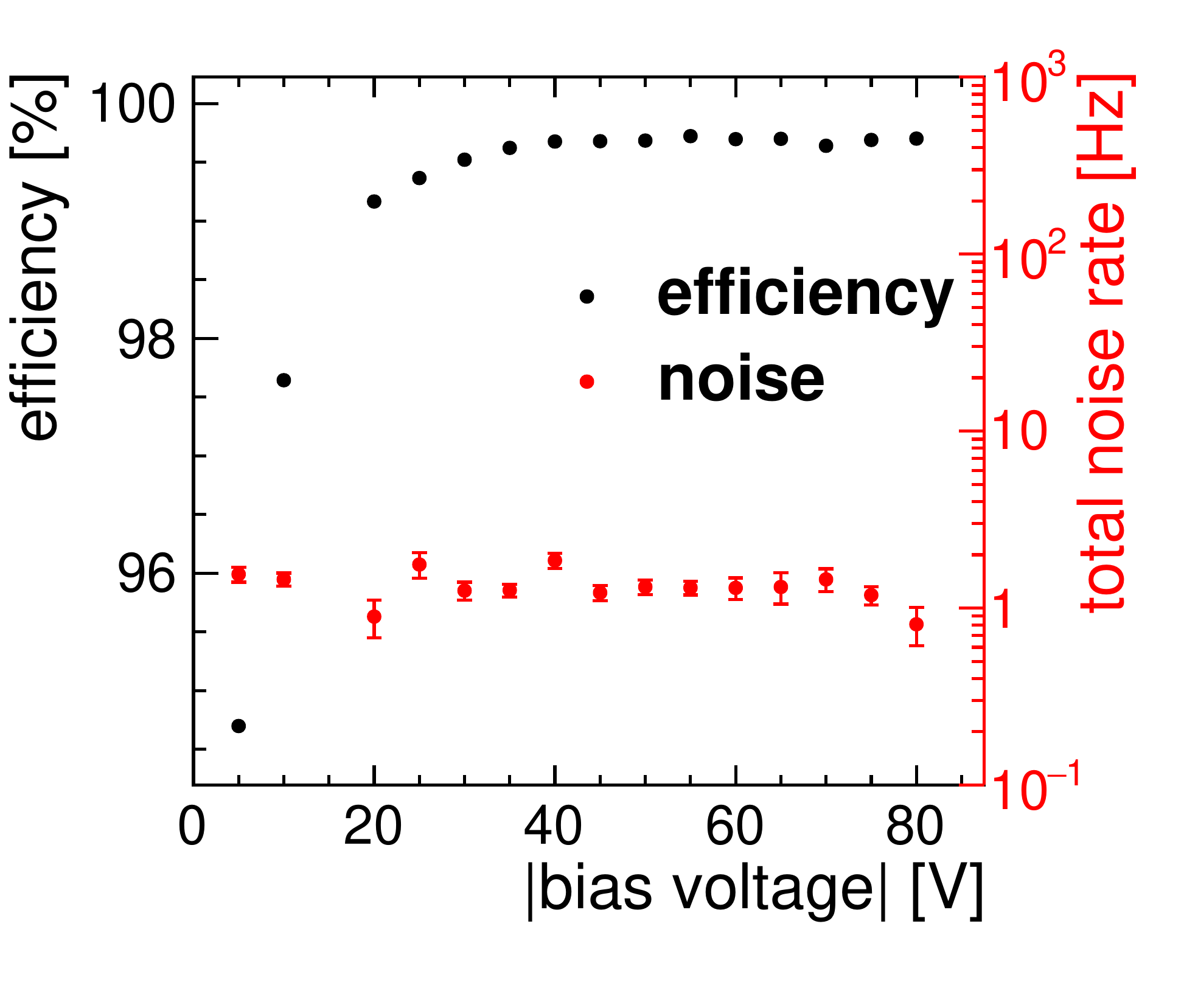}};
  \begin{scope}[x={(image.south east)},y={(image.north west)}]
      \node[anchor=east] at (0.8,0.525){\footnotesize CLICdp};
  \end{scope}
\end{tikzpicture}
\caption{}\label{fig:atlaspix_eff_biasvoltage}
\end{subfigure}
\caption{ATLASpix\_Simple efficiency and total noise rate as a function of \subref{fig:atlaspix_eff_threshold} the detection threshold and \subref{fig:atlaspix_eff_biasvoltage} the bias voltage.}\label{fig:atlaspix_thresholdscan}
\end{figure}

\cref{fig:atlaspix_timing_gaussbox} illustrates the measured timing residual between the track time stamp and the hit time reconstructed by the ATLASpix\_Simple detector. The sensor has been operated at \SI{-75}{\volt} reverse bias and the threshold was set to approximately \SI{500}{\Pem{}}.
 A row-by-row correction of the propagation delay along the metal lines to the periphery has been extracted and applied to the data. Due to the low operating threshold, time walk in the front-end does not contribute significantly to the resolution, and thus has not been corrected for here. The chip has been operated with a clock of \SI{125}{\mega\hertz}, which results in a time-of-arrival binning of \SI{16}{\ns}, due to a chip-internal down scaling of the clock frequency by a factor of two for the ToA measurement. The standard deviation of the residual distribution is \SI{6.8}{\ns}. Additionally, the histogram was fitted with a box function convolved with a Gaussian, to take the clock binning uncertainty into account.

With increasing threshold, time-walk of the detector front-end significantly degrades the timing resolution of the detector, as illustrated in \cref{fig:atlaspix_timingResolution_vs_thres}. The RMS of the distribution, as well as the Gaussian component of a convolution of a Gaussian and a \SI{16}{\ns} wide box-shaped distribution is shown. The RMS is close to \SI{7}{\ns} at \SI{480}{\Pem{}} threshold and increases to around \SI{16}{\ns} at \SI{2200}{\Pem{}} threshold. The RMS describes the resolution with which a single particle hit can be timed (including the clock binning), whereas the Gaussian component of the convolution describes an upper limit on the
intrinsic timing resolution of the detector from the charge collection process and the and analogue front-end. This intrinsic limit is close to \SI{5}{\ns} at \SI{480}{\Pem{}}.

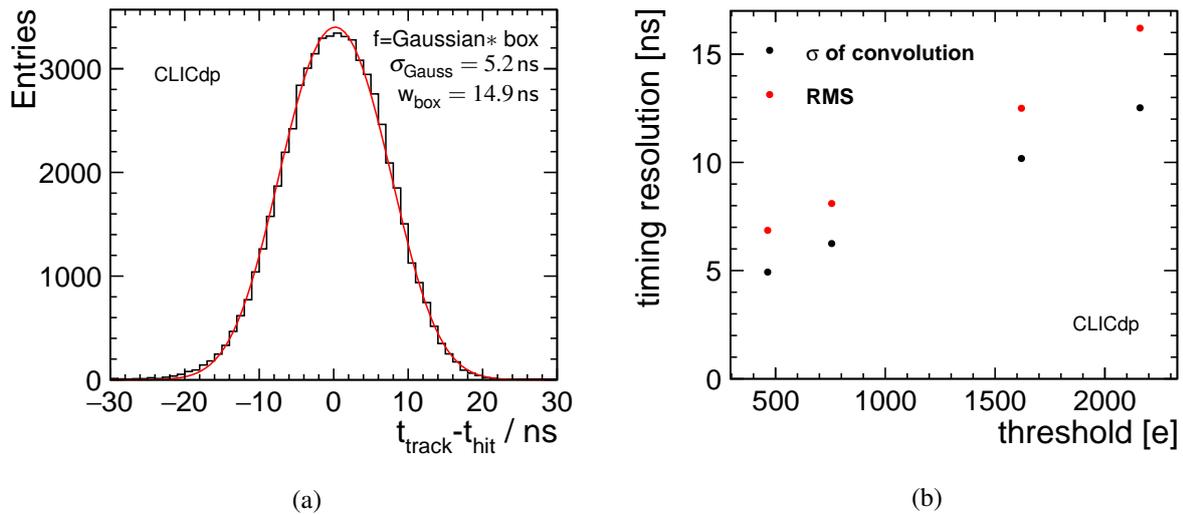
\begin{figure}[ht]
  \begin{subfigure}[T]{.49\linewidth}
  \begin{tikzpicture}
  \node[anchor=south west,inner sep=0] at (0,0)(image){\includegraphics[width=\linewidth]{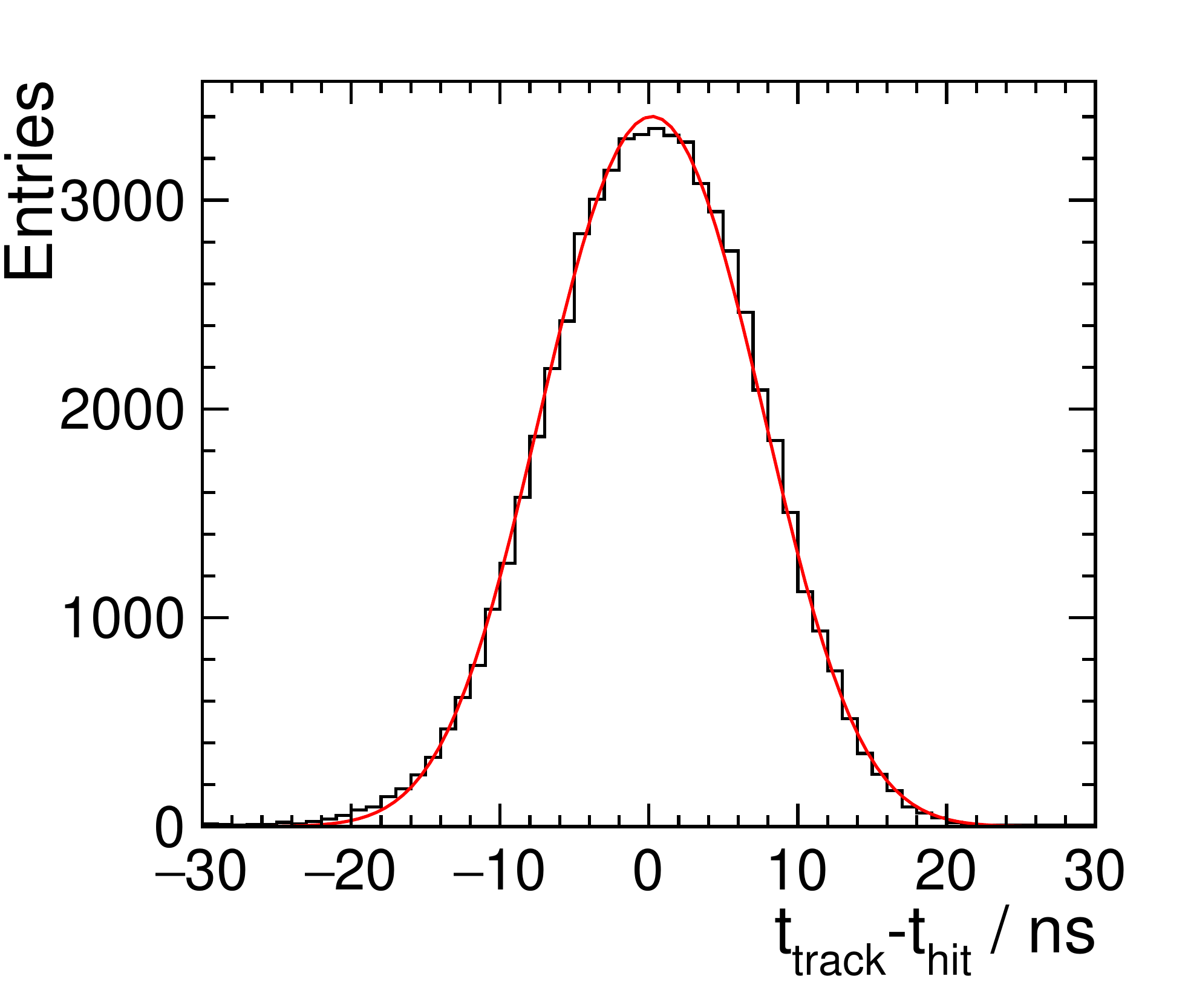}};
  \begin{scope}[x={(image.south east)},y={(image.north west)}]
    \node at (0.75,0.8){\footnotesize\shortstack[r]{f=Gaussian$\ast$ box\\$\sigma_\text{Gauss}=\SI{5.2}{\ns}$\\$\text{w}_\text{box}=\SI{14.9}{\ns}$}};
    \node[above, color=black] at (0.3, 0.75) {\sffamily\scriptsize{CLICdp}};
  \end{scope}
  \end{tikzpicture}
  \caption{}\label{fig:atlaspix_timing_gaussbox}
\end{subfigure}
\hfill
\begin{subfigure}[T]{.49\linewidth}
  \begin{tikzpicture}[font=\sffamily]
    \node[anchor=south west,inner sep=0] (image) at
    (0,0){\includegraphics[width=\linewidth]{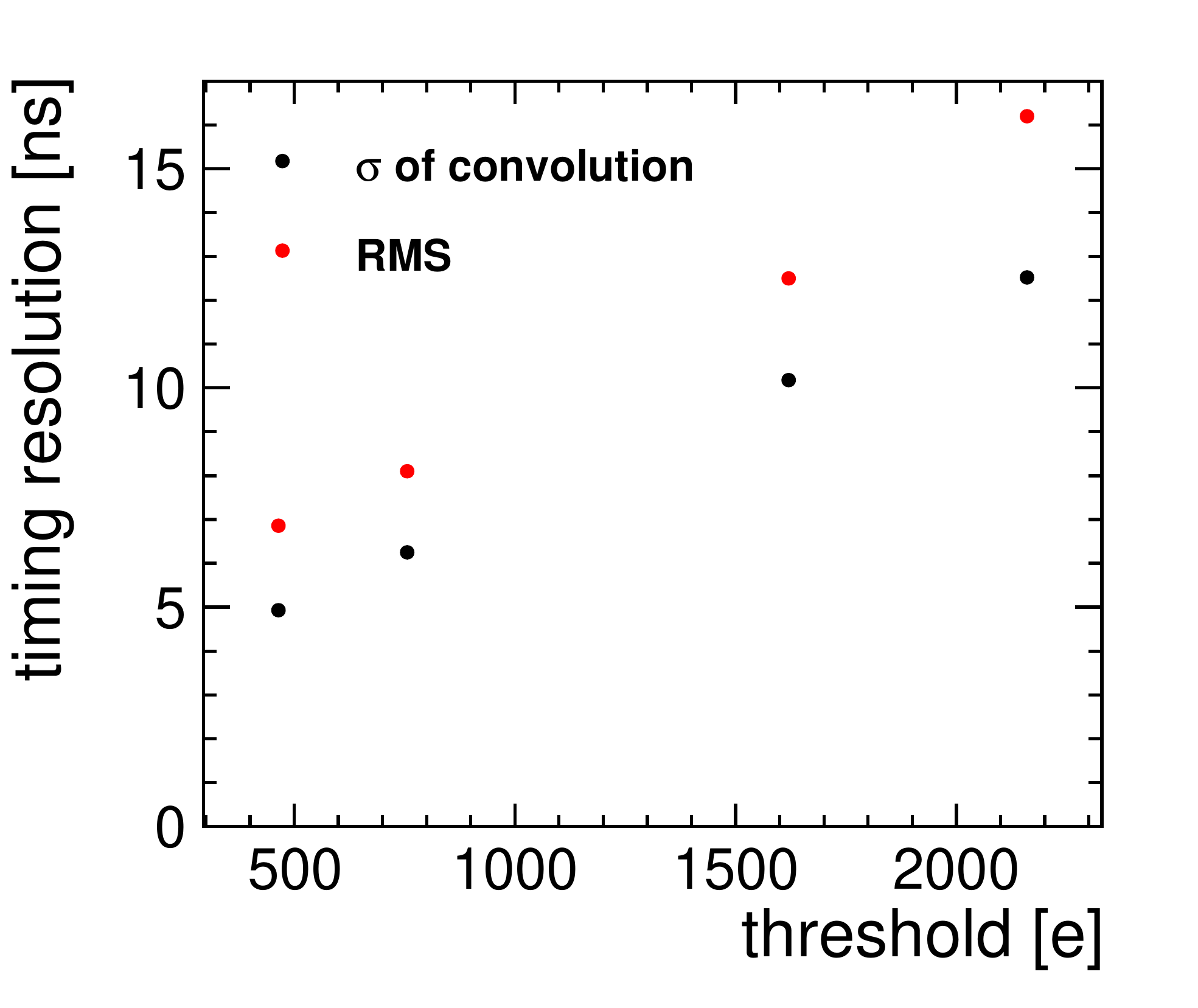}};
    \begin{scope}[x={(image.south east)},y={(image.north west)}]
      \node[above, color=black] at (0.8, 0.25) {\scriptsize{CLICdp}};
    \end{scope}
  \end{tikzpicture}
  \caption{}\label{fig:atlaspix_timingResolution_vs_thres}
\end{subfigure}
  \caption{\subref{fig:atlaspix_timing_gaussbox} Timing residual distribution between the telescope track and the hit on the ATLASpix\_Simple sensor, and \subref{fig:atlaspix_timingResolution_vs_thres} variation of the width of the timing residual distribution with the detection threshold .}\label{fig:atlaspix_timing_residual}
\end{figure}

For an application in the CLIC tracking detector, \SI{7}{\micron} resolution along the r$\varphi$ direction are aimed for (see \cref{sec:vtx-trk-requirements}). To achieve this, a modification of ATLASpix\_Simple with \SI{25}{\micron} wide pixels is under development. For this version the long pixel dimension has to be elongated to approximately \SI{200}{\micron}, to maintain the pixel area needed to fit the necessary transistors into the pixel n-well. This elongation is well within the pixel-size requirements for the CLIC tracker. All other requirements, such as the material budget, the detection efficiency, the timing resolution and the power consumption are already met by the current demonstrator chip.

\section{Monolithic CMOS sensors with a small collection electrode} \label{sec:hrcmos}
CMOS sensors with a small collection electrode and a high-resistivity substrate have been developed for particle-detector applications~\cite{alice_tdr} and recently tested in view of the CLIC vertex and tracking detector requirements~\cite{investigator_pub}. In these designs, the CMOS circuitry is placed in deep p-well rings, surrounding the pixel implants.
The P-MOS transistors are thereby physically separated from the collection electrode, resulting in a good noise isolation and allowing for full CMOS circuitry in the pixel.
Moreover, the size of the collection electrode can be minimised, leading to a very small sensor capacitance in the order of a few \si{\femto\farad} which is beneficial for achieving a very low noise and analogue power consumption~\cite{walter_power}.

\subsection{Fabrication process}\label{sec:towerjazz-process}
\cref{fig:towerjazz} outlines the studied \SI{180}{\nm} imaging process.
To create a depleted area around the collection electrode, a bias voltage can be applied to the p-wells and the substrate.
For the studied version of the Investigator this bias voltage is limited to maximally \SI{6}{\volt}, however further prototypes produced in this technology allow for a higher bias voltage.
Due to this bias voltage limit and the small junction area around the collection electrode, achieving a full lateral depletion of the pixel cell is not guaranteed, as illustrated in \cref{fig:towerjazz_standard}, where the junction is marked by the yellow line and the edge of the depleted area by the white line.
Thus, to achieve a sizeable depletion in these sensors, a high resistivity (\SIrange{1}{8}{\kilo\ohm\cm}) p-type epitaxial layer is used as sensing material.
The epitaxial layer is grown on a low resistivity p-type substrate material.
To improve the depletion behaviour, a recent process modification was introduced, based on a deep planar junction that is caused by an additional deep n-doped implantation, as depicted in \cref{fig:towerjazz_modified}~\cite{Snoeys_modified_process}.

\begin{figure}[ht]
	\begin{subfigure}[]{.49\linewidth}
		\includegraphics[width=\linewidth]{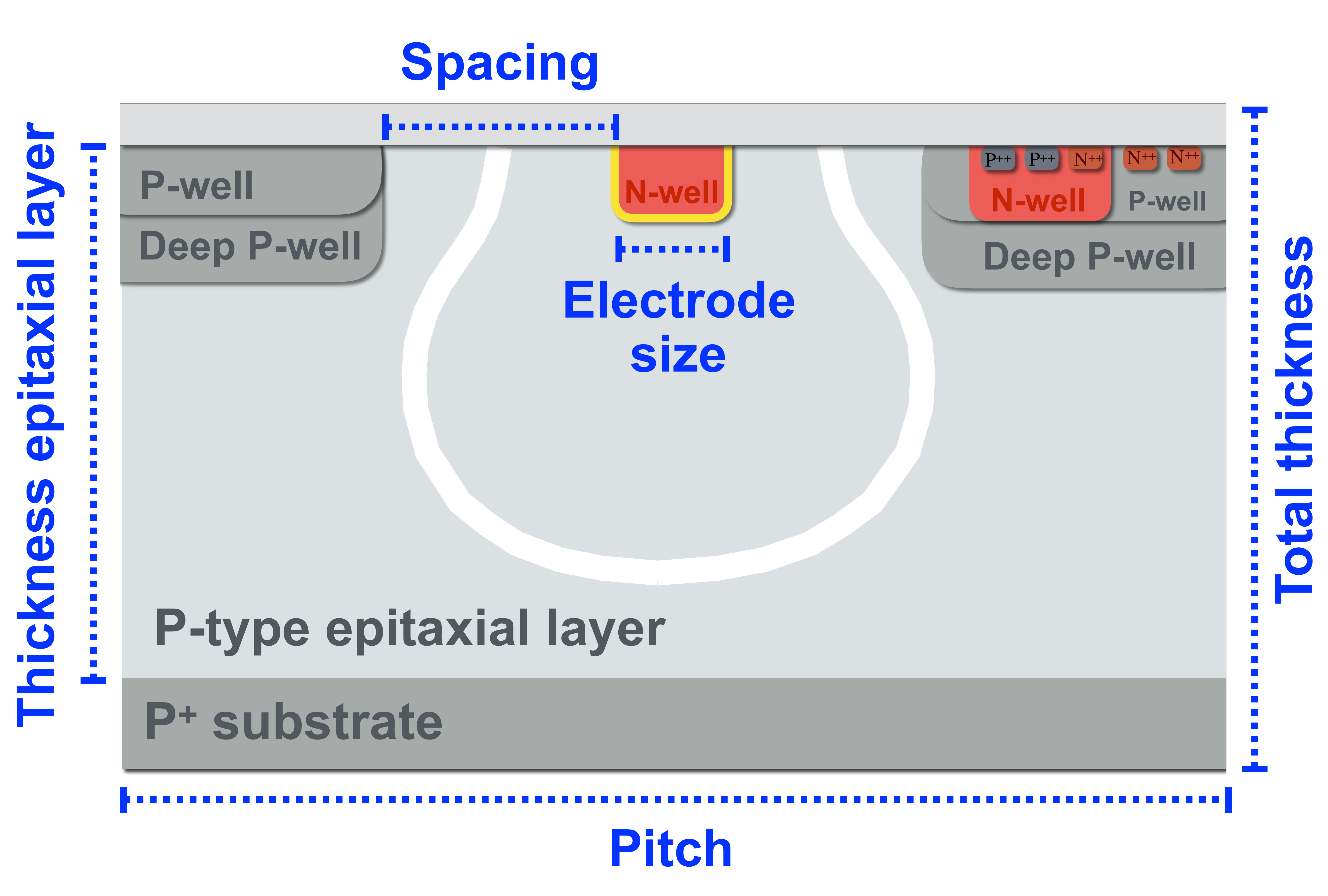}
		\caption{}\label{fig:towerjazz_standard}
	\end{subfigure}
	\hfill
	\begin{subfigure}[]{.49\linewidth}
		\includegraphics[width=\linewidth]{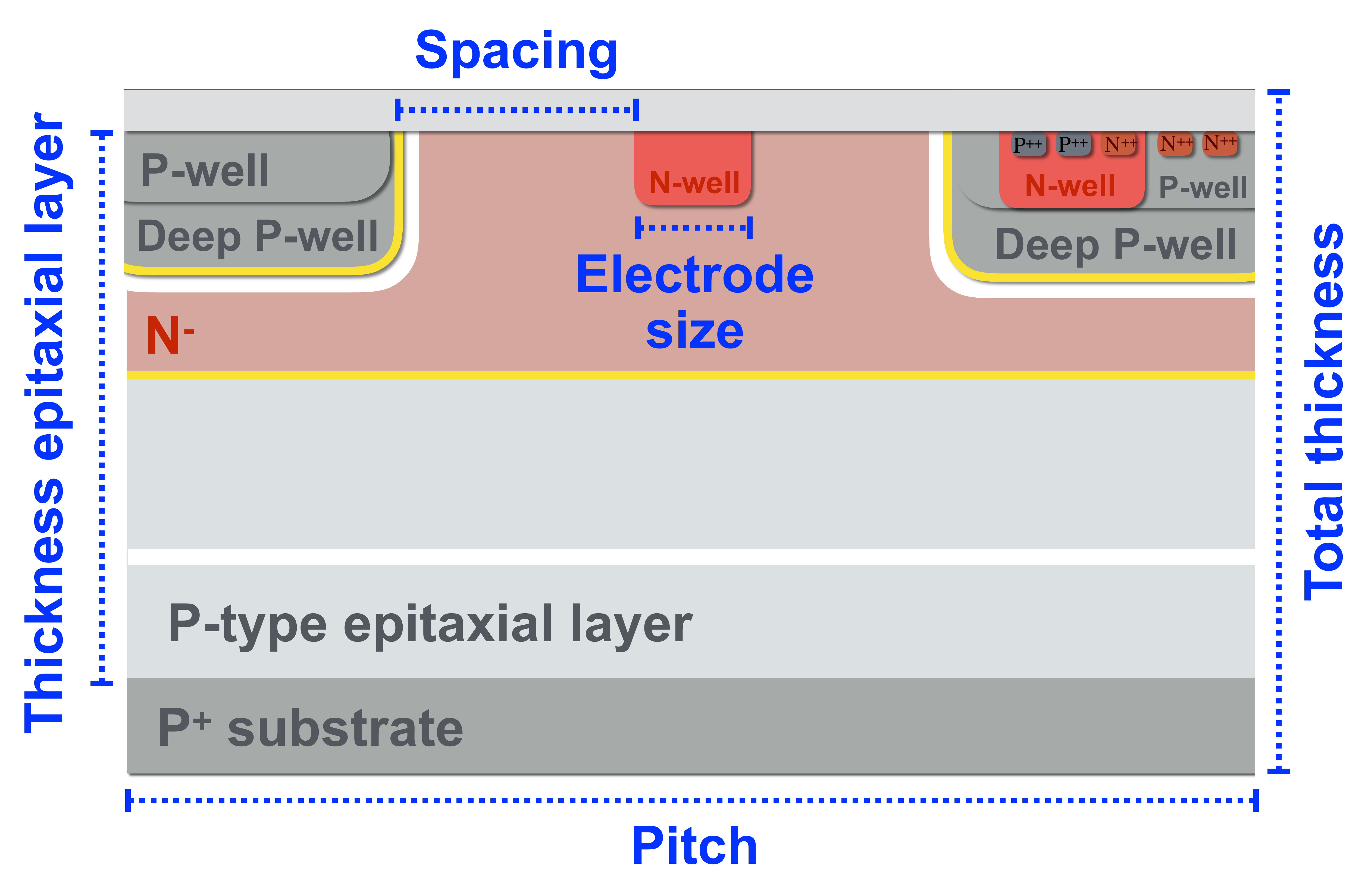}
		\caption{}\label{fig:towerjazz_modified}
	\end{subfigure}
	\caption{Schematic cross-section of a pixel in \subref{fig:towerjazz_standard} the standard \SI{180}{\nm} CMOS imaging sensor process, and \subref{fig:towerjazz_modified} a process modification to obtain full depletion of the epitaxial layer~\cite{Snoeys_modified_process} (not to scale).
	In both cases, a deep p-well shields the n-wells with circuitry from the sensor and allows for full CMOS circuitry in the pixel. The yellow lines mark the junction areas and the white line the edges of the depleted regions.}
	\label{fig:towerjazz}
\end{figure}

\subsection{The ALICE Investigator analogue test-chip} 
The standard imaging process has been chosen by the ALICE collaboration for a monolithic pixel detector chip (ALPIDE) for use in the Inner Tracking System (ITS) upgrade of the ALICE detector~\cite{alice_tdr}.
In the context of the optimisation studies for the ALPIDE chip, a small analogue test-chip, called Investigator, was fabricated to study the analogue performance of the technology in detail.
The Investigator chip contains several small pixel matrices with \num{10x10} pixels, with design variations of the geometry and circuitry of the pixels.
The geometrical parameters of the pixel layout are highlighted by blue dashed lines in~\cref{fig:towerjazz}.
Each pixel contains a source follower and the outputs of the inner \num{8x8} pixel sub-matrix are buffered and routed in parallel off the chip, where they are sampled at 65~MHz and digitised using a 14-bit sampling ADC.
Furthermore, the Investigator chip has also been fabricated in the modified process including the deep n-implant.

\subsection{Test-beam results}

The Investigator chip has been integrated in the Timepix3 based test-beam setup (see~\cref{sec:timepix3_telescope}).
The analogue performance of both process variants has been assessed and evaluated with respect to the requirements for the CLIC tracker and vertex detector for a mini-matrix with the geometrical parameters summarised in~\cref{tab:nom_par}.
\begin{table}[ht!]
\centering
\caption{Geometrical parameters of the studied Investigator chip and mini-matrix.}
\begin{tabular}{ l  r }
\toprule
Parameter &  Value [\si{\micro\metre}]\\
\midrule
Pixel pitch  &  28 \\
Spacing between electrode and deep p-well & 3 \\
Collection electrode size & 2\\
Thickness epitaxial layer & 25\\
Total thickness & 100\\
\bottomrule
\end{tabular}
\label{tab:nom_par}
\end{table}
Studies of different pixel layouts have shown that a spacing between electrode and deep p-well of \SI{3}{\micro \meter} is favourable as a compromise between an improved spatial resolution, for smaller spacing, and a larger efficient operation window at higher thresholds and faster charge collection, for larger spacing.
More details including a study of different bias voltages, as well as a detailed description of the test-beam setup, reconstruction and analysis can be found elsewhere~\cite{ThesisMagdalena,investigator_pub}.

Comparing both process variants, the main performance difference is caused by the non-depleted regions near the pixel edges in the standard process, where charge transport by diffusion dominates.
The cluster size as a function of the track impact positions within the pixel cell is presented in~\cref{fig:cspixeltot_std}.
Significantly more charge sharing can be observed for the standard process compared to the modified process, especially at the pixel edges and corners, which are not fully depleted in the standard process.
\begin{figure}[ht]
	\centering
	\begin{subfigure}[]{.48\linewidth}
		\begin{overpic}[width=1\textwidth]{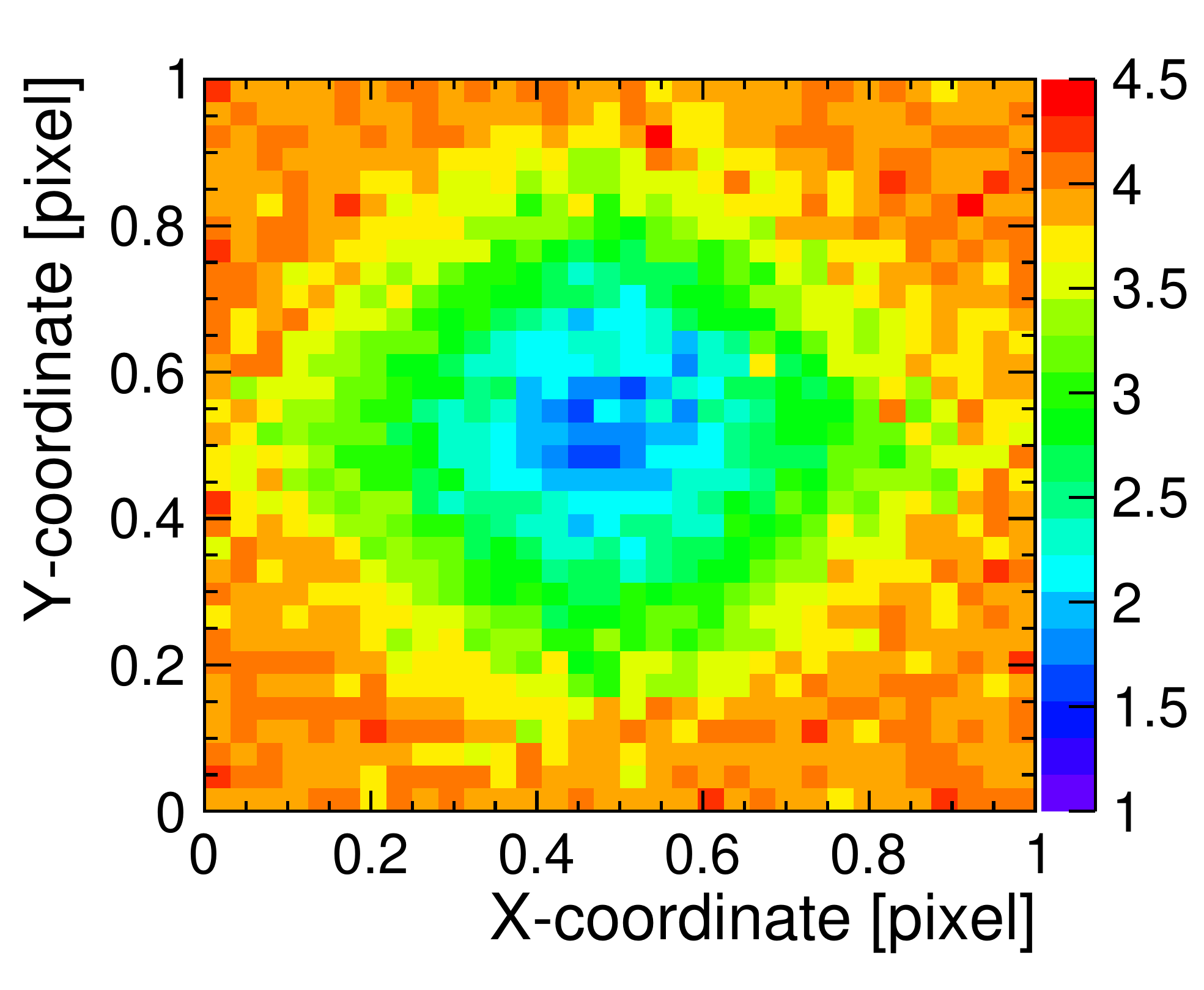}
			\put(100,76){\rotatebox{-90}{\sffamily Total cluster size}}
      \put(74,77){\sffamily\scriptsize CLICdp}
		\end{overpic}
		\caption{}
		\label{fig:cspixel_old}
	\end{subfigure}
	\hspace{0.3cm}
	\begin{subfigure}[]{.48\linewidth}
		\begin{overpic}[width=1\textwidth]{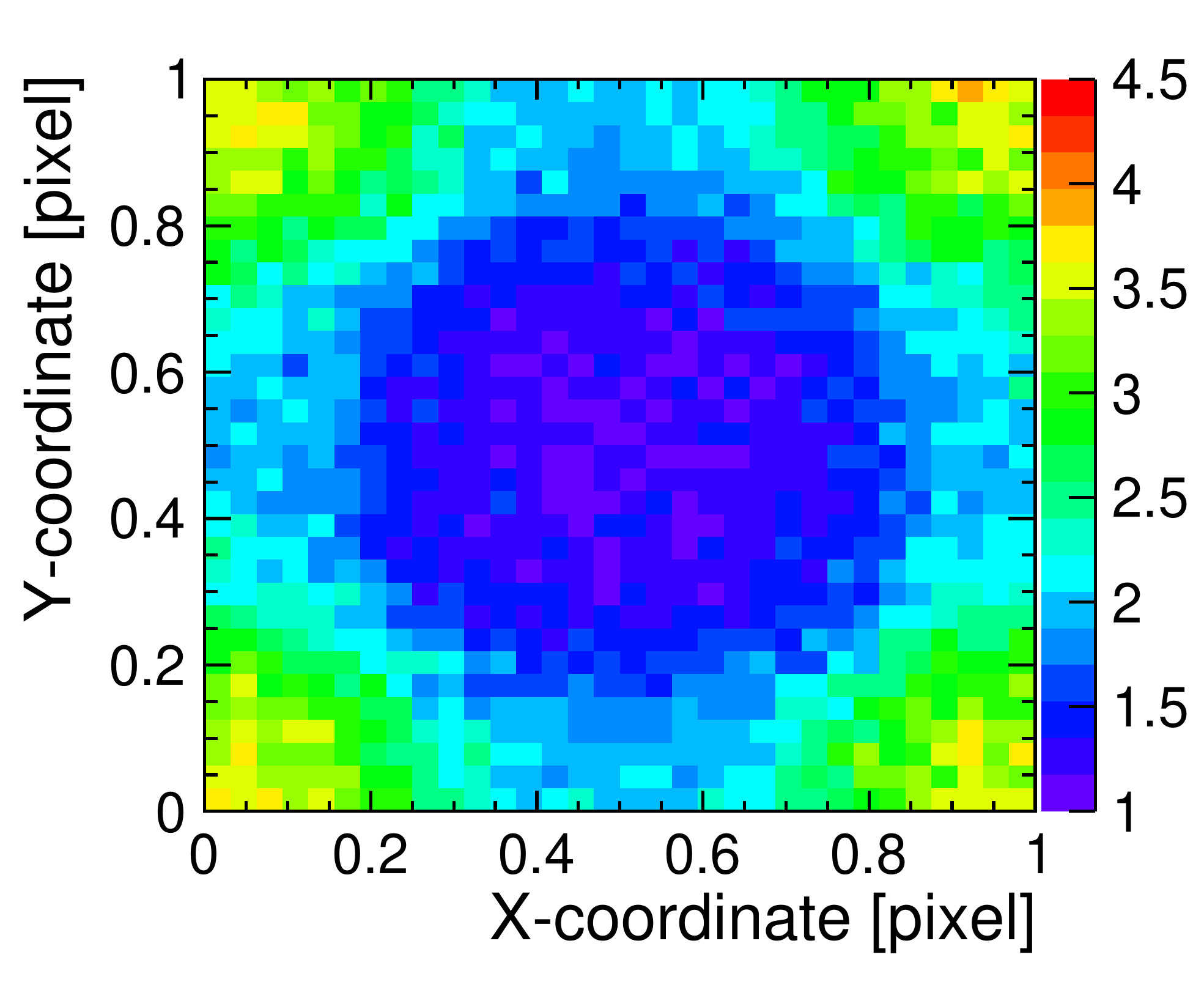}
			\put(100,76){\rotatebox{-90}{\sffamily Total cluster size}}
      \put(74,77){\sffamily\scriptsize CLICdp}
		\end{overpic}
		\caption{}
		\label{fig:cspixel_new}
	\end{subfigure}
	\caption{Average total cluster size as a function of the track impact position within the pixel cell for \subref{fig:cspixel_old} the standard process at a threshold of \SI{41}{\Pem{}} and for \subref{fig:cspixel_new} the modified process at a threshold of \SI{50}{\Pem{}}.
	The different threshold values can be attributed to the slightly higher pixel input capacitance and noise for the modified compared to the standard process~\cite{ThesisMagdalena}.}
	\label{fig:cspixeltot_std}
\end{figure}


The increased charge sharing in the standard process impacts the spatial resolution, as presented in~\cref{fig:res_thr}.
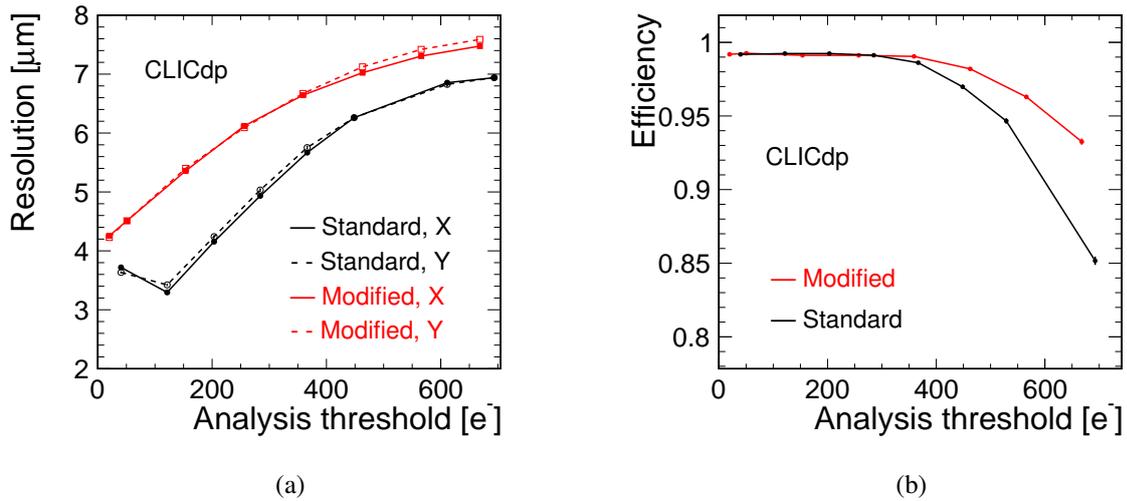
\begin{figure}
	\centering
	\begin{subfigure}[t]{0.48\textwidth}
		\begin{overpic}[width=1\textwidth]{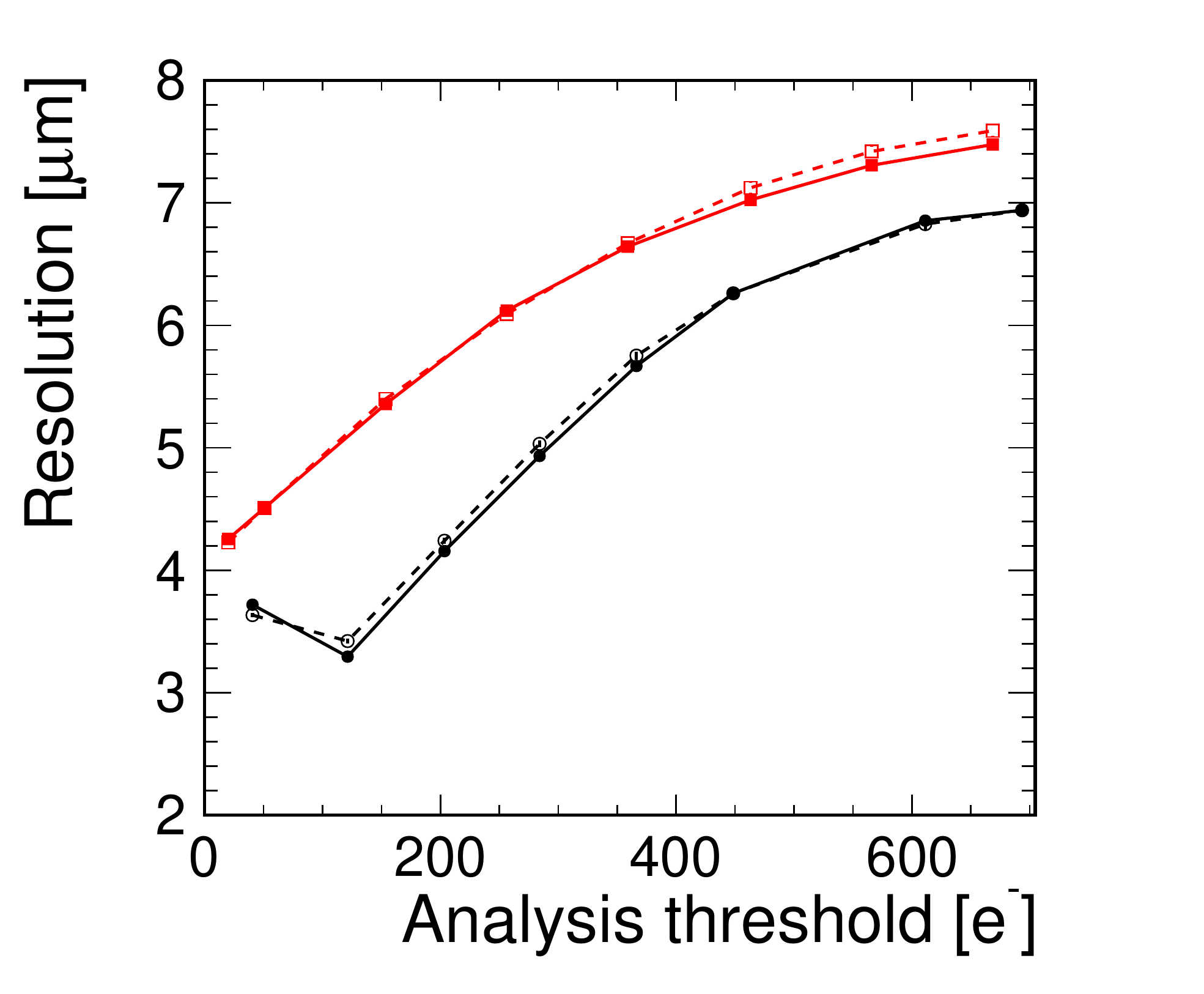}
		\put(50,32){\small \sffamily \textcolor{black}{- - Standard, Y}}
		\put(50,38){\small \sffamily \textcolor{black}{--- Standard, X}}
		\put(50,20){\small \sffamily \textcolor{red}{- - Modified, Y}}
		\put(50,26){\small \sffamily \textcolor{red}{--- Modified, X}}
    \put(25,65){\sffamily\small CLICdp}
		\end{overpic}
		\caption{}
		\label{fig:res_thr}
	\end{subfigure}
	\hspace{0.3cm}
	\begin{subfigure}[t]{0.48\textwidth}
		\begin{overpic}[width=1\textwidth]{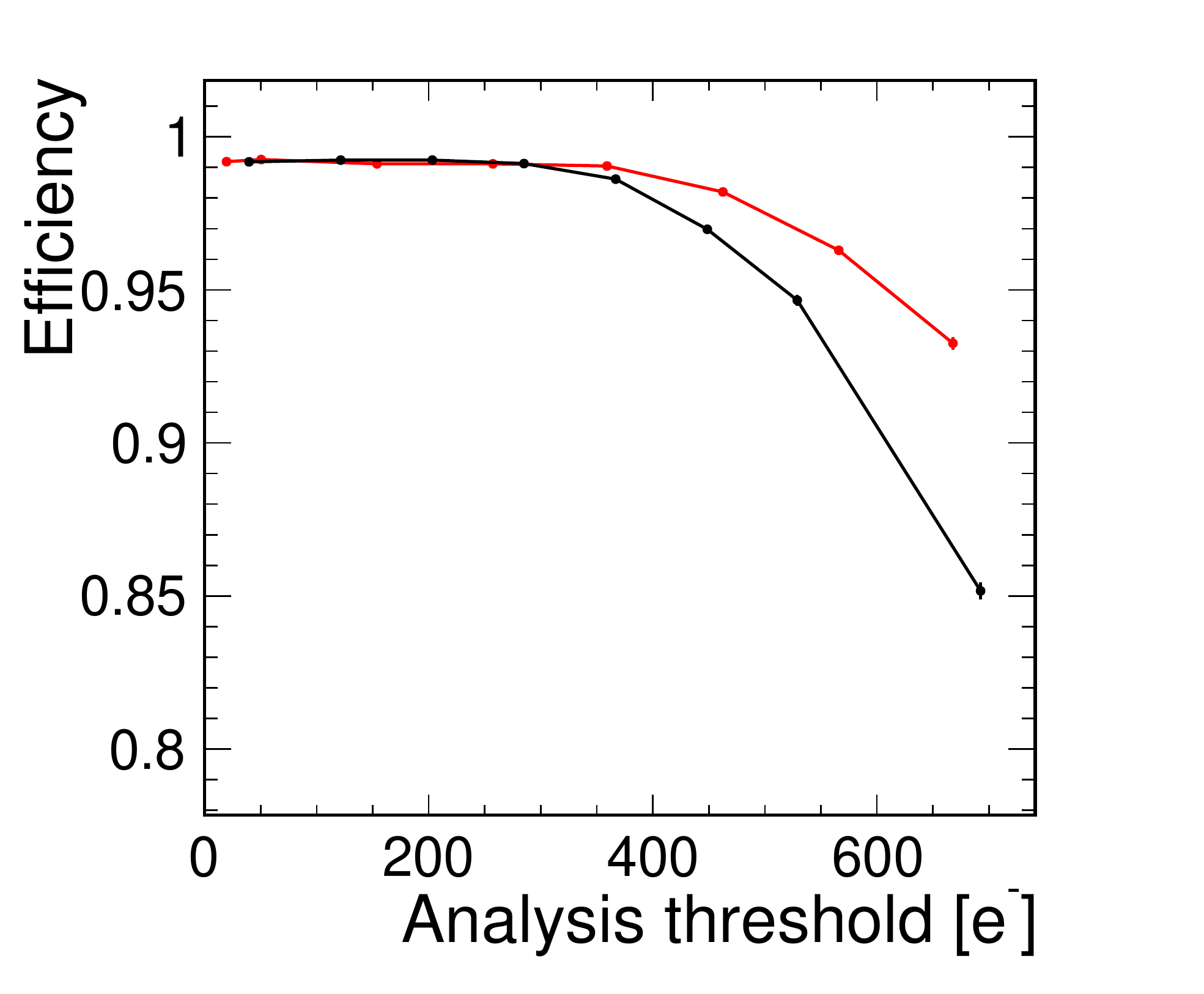}
			\put(26,29){\small \sffamily \textcolor{red}{--- Modified}}
			\put(26,22){\small \sffamily \textcolor{black}{--- Standard}}
      \put(25,50){\sffamily\small CLICdp}
		\end{overpic}
		\caption{}
		\label{fig:eff_thr}
	\end{subfigure}
	\caption{Performance of the standard and modified process for different threshold values: \subref{fig:res_thr} spatial resolution and \subref{fig:eff_thr} efficiency.}
	\label{fig:res_eff}
\end{figure}
The spatial residual is defined as the difference between the calculated cluster position and the track position interpolated on the Investigator reference plane.
The telescope resolution is unfolded from the observed RMS99.7 to obtain the spatial resolution.
The resolution versus threshold (\cref{fig:res_thr}) shows significantly lower values for the standard process over the full threshold range.

As visible in~\cref{fig:eff_thr}, the increased charge sharing in the standard process also impacts the efficient operation window at higher threshold values, resulting in a reduced single pixel signal that causes a drop of the efficiency at higher threshold values.
Even at low threshold values the measured efficiency of \SI[separate-uncertainty = true]{99.2(1)}{\percent} for both process variants shows a discrepancy from \SI{100}{\percent} that can not be explained by the stated statistical uncertainty.
Since this deviation is visible for both process variants, it is assumed to be caused by the data taking setup~\cite{ThesisMagdalena}.

The timing residual is defined as the difference in time between the measured arrival time of the first pixel of a reconstructed Investigator hit and the mean time of all hits on the telescope track. The resulting residual distribution is presented in~\cref{fig:restime_mod} for both process variants.
A Gaussian function is fitted to the timing residual within a range that contains $\mathrm{99.7 \, \%}$ of all statistics.
The timing resolution $\mathrm{\sigma_{time}}$ is calculated unfolding the timing resolution of the telescope of $\mathrm{\sigma_{time,tele}\,= \,1.1 \,ns}$~\cite{calib_timing_tpx3} from the width $\mathrm{\sigma_{Gauss}}$ of this fit.
\begin{figure}[ht]
	\begin{subfigure}[t]{0.48\textwidth}
		\begin{overpic}[width=1\textwidth]{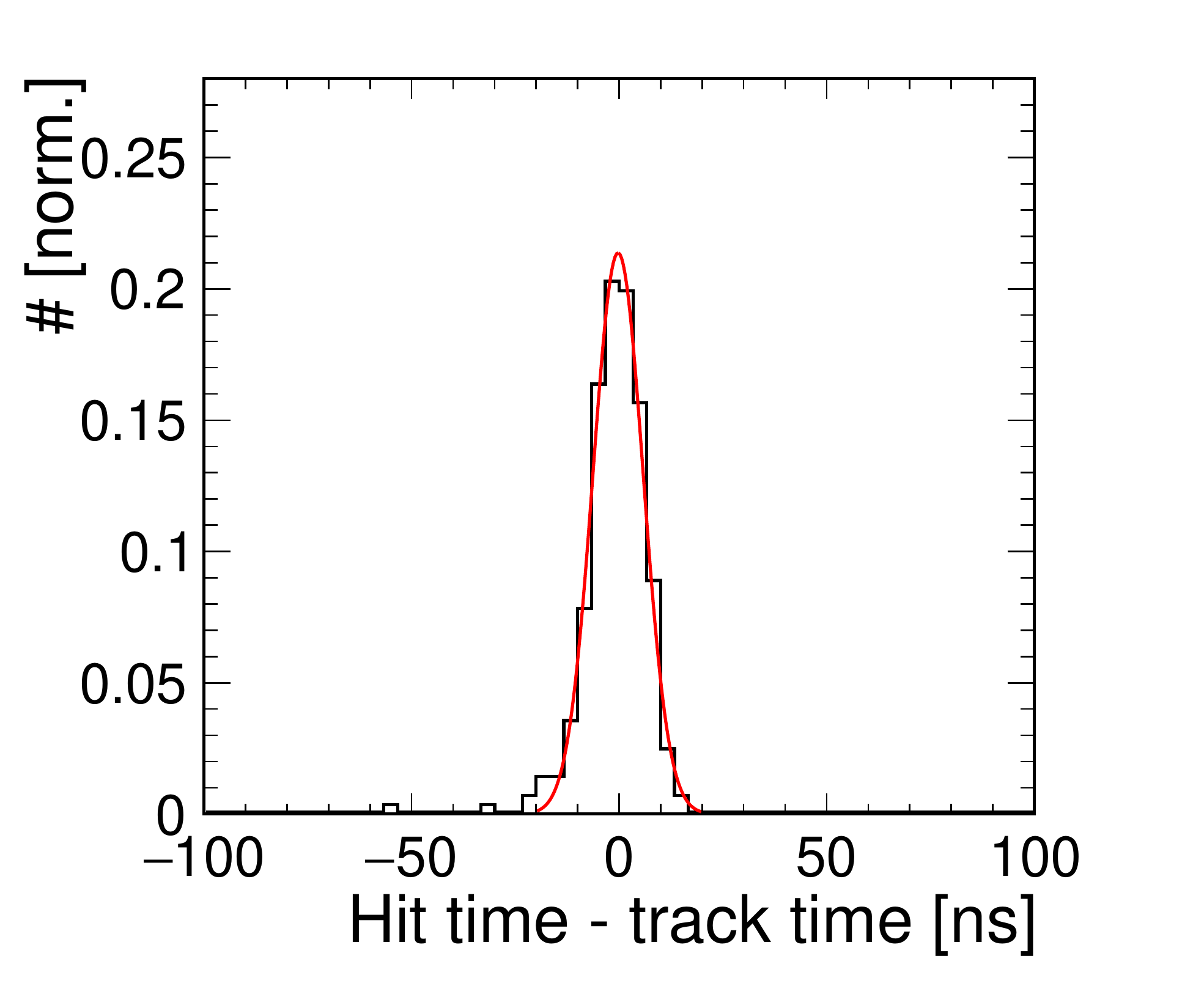}
			\put(21,70){$\color{red}{\sigma = \SI[separate-uncertainty = true]{6.1(5)}{\ns}}$}
      \put(25,50){\sffamily\small CLICdp}
		\end{overpic}
		\caption{}
		\label{fig:time_std}
	\end{subfigure}
	\hspace{0.3cm}
	\begin{subfigure}[t]{0.48\textwidth}
		\begin{overpic}[width=1\textwidth]{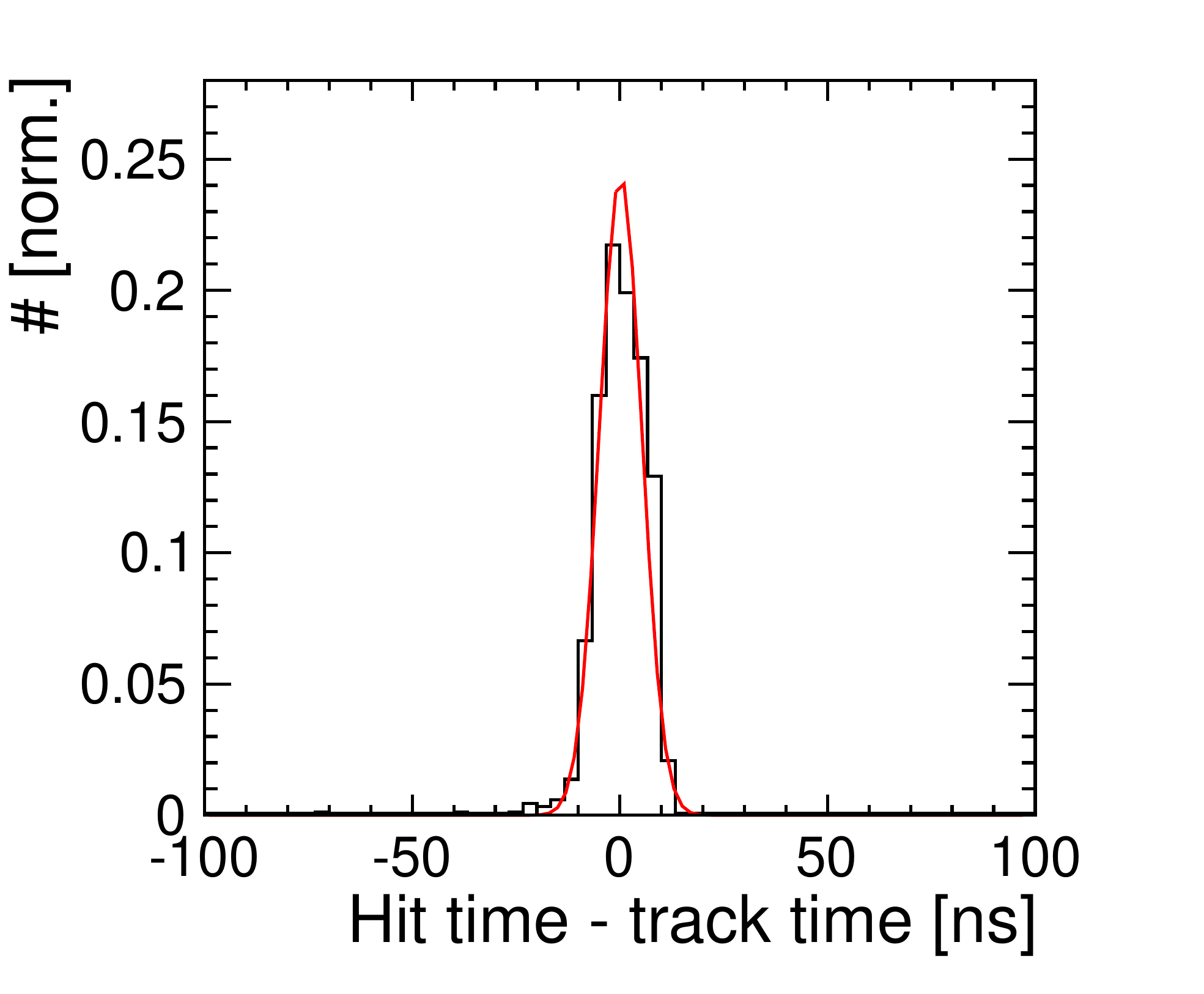}
			\put(21,70){$\color{red}{\sigma = \SI[separate-uncertainty = true]{5.1(3)}{\ns}}$}
      \put(25,50){\sffamily\small CLICdp}
		\end{overpic}
		\caption{}
		\label{fig:time_mod}
	\end{subfigure}
	\caption{Timing residual distribution for \subref{fig:time_std} the standard process and \subref{fig:time_mod} the modified process. Gaussian fits in a range corresponding to 99.7\% of all events are shown in red. The quoted timing resolutions are obtained by unfolding the telescope resolution from the width of the Gaussian fits, and the uncertainty corresponds to the statistical uncertainty on the fit values.}
	\label{fig:restime_mod}
\end{figure}
A slightly more precise analogue timing resolution of $\mathrm{ (5.1 \, \pm \, 0.3) \, ns}$ has been measured for the modified compared to the standard process where an analogue timing resolution of $\mathrm{ (6.1 \, \pm \,0.5)\, ns}$ has been measured.
This is expected from the higher electric field in the sensor of the modified process~\cite{ThesisMagdalena}.
However, the measurement precision is limited by the $\mathrm{65 \, MHz}$ sampling rate of the external ADCs on the readout board~\cite{ThesisMagdalena}.

The time the signal of the seed pixel needs to rise from \SIrange{10}{90}{\percent} of its total amplitude ($T_{10 \, - \, 90}$) is presented in~\cref{fig:risetime}, showing a significantly slower charge collection of approximately \SI{30}{ns} for the standard process compared to approximately \SI{15}{ns} for the modified process, consistent with the expected slower charge collection in the non-depleted regions.

\begin{figure}[ht]
	\centering
	\begin{minipage}[t]{0.48\textwidth}
		\begin{overpic}[width=1\textwidth]{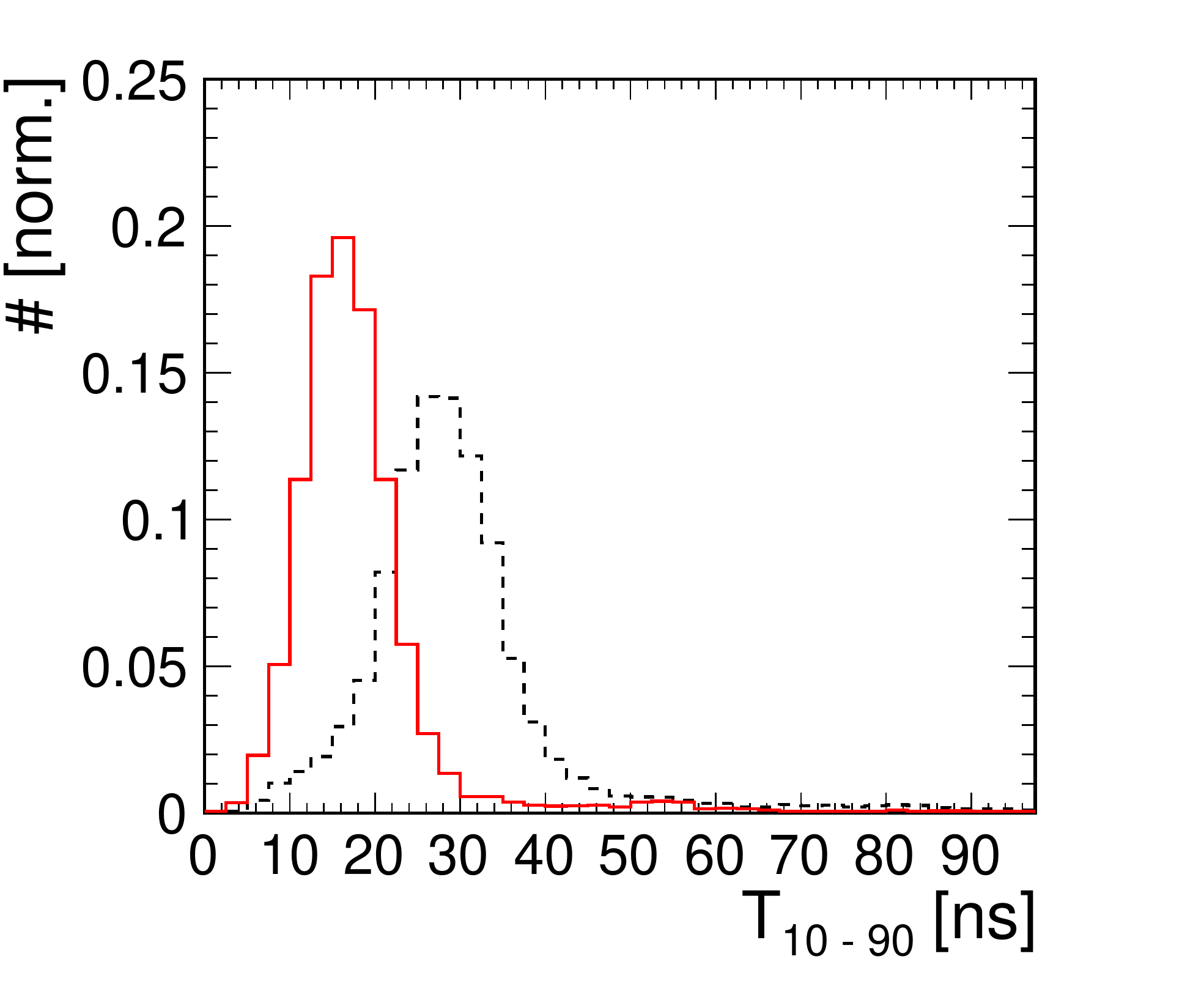}
			\put(53,56){\small \sffamily \textcolor{red}{--- Modified}}
			\put(53,63){\small \sffamily \textcolor{black}{- - Standard}}
      \put(60,40){\sffamily\small CLICdp}
		\end{overpic}
	\end{minipage}
	\caption{Distribution of the time the signal needs to rise from \SIrange{10}{90}{\percent} of its amplitude for both process variants.}
	\label{fig:risetime}
\end{figure}

Overall, the study of the analogue performance of the CMOS sensors with a small collection electrode shows that the standard process offers a better spatial resolution for the same pixel geometry, while the modified process offers better efficiency at high threshold values, and faster signal collection.
Both process variants meet the \SI{7}{\micro \meter} spatial resolution requirement for the tracker.
For the standard process a spatial resolution close to the value of \SI{3}{\micro \meter} required for the vertex detector has been measured.
While the hit time resolution of \SI{\sim 6}{ns} is promising in view of the \SI{5}{ns} time resolution requirement for the tracking system, more refined measurements are needed with a fully monolithic chip in this technology.
The lower rise time can affect the accuracy of the arrival time measurement in future fully integrated chips with an on-chip-threshold and needs to be investigated further.
A noise below \SI{10}{\Pem{}}~\cite{ThesisMagdalena} can be reached with the studied CMOS technologies with a small collection electrode.
This allows for efficient operation for very thin active sensor volumes (epitaxial layer thickness of \SI{25}{\micro \meter}).
The fabricated chips can be thinned down to an overall thickness of \SI{50}{\micro \meter}~\cite{alice_tdr}, making this technology an interesting candidate for achieving a low material budget simultaneously with an excellent spatial resolution and fast timing, as required in particular for the vertex detector.

\subsection{Simulation results}
\label{sec:hrcmos-simulation-results}

Various simulations with different pixel geometries and process parameters have been performed and compared with test-beam measurements. The simulations lead to a better understanding of the technology and are used to optimise the sensor layout for future designs.

Fully self-consistent finite element TCAD simulations have first been performed in two dimensions~\cite{ThesisMagdalena} and later extended to three dimensions.
The electric field and the edge of the depleted region in the simulated three-dimensional single pixel cell are presented in~\cref{fig:efield} for the standard and modified process.
\begin{figure}[ht]
	\centering
	\begin{subfigure}[t]{0.48\textwidth}
		\begin{overpic}[width=1\textwidth]{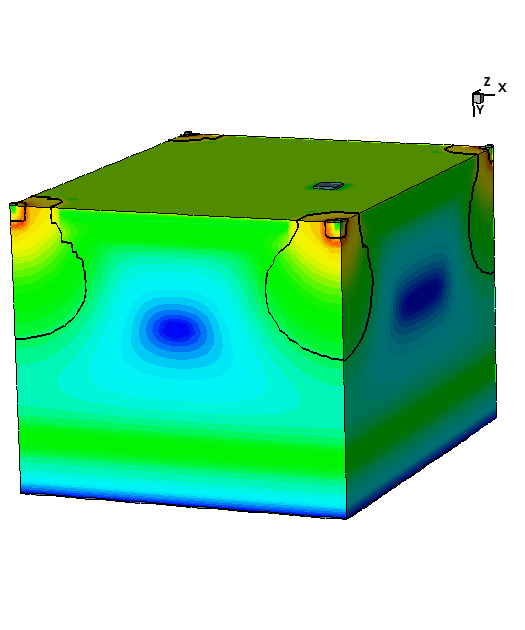}
		\end{overpic}
		\caption{ }
		\label{fig:efield_std}
	\end{subfigure}
	\begin{subfigure}[t]{0.48\textwidth}
		\begin{overpic}[width=1\textwidth]{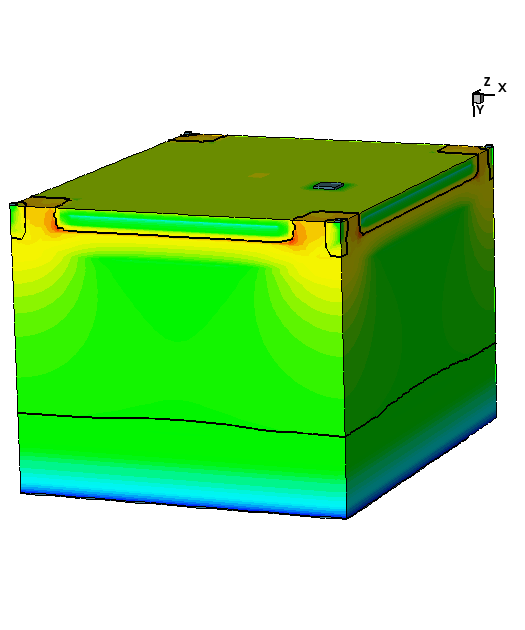}
		\end{overpic}
		\caption{ }
		\label{fig:efield_mod}
	\end{subfigure}
	\caption{Electric field distributions from a three dimensional TCAD simulation of a single pixel cell of \subref{fig:efield_std} the standard process and \subref{fig:efield_mod} the modified process for a pixel pitch of \SI{30x37.5}{\micron}. The collection electrodes are placed at the top corners of the simulated structures. The black lines indicate the edge of the depleted regions.}
	\label{fig:efield}
\end{figure}
Consistent with the test-beam results, differences between both process variants are visible especially at the pixel corner:
in the standard process the low electric field regions are much more prominent in the pixel corners, for which the depleted region does not extend over the full lateral size of the pixel.
The simulations confirm that for both process variants the regions with lowest electric field are located at the pixel corners. 
To push the charges faster out of this low-field regions and speed up the charge collection in the sensor, the modified process has been further optimised using three-dimensional transient TCAD simulations~\cite{pixel_magdalena}.
Both a gap in the deep n-doped implant and an additional deep p-type implant at the pixel border were considered. 
These modifications however reduce charge sharing between pixels and thereby deteriorate the spatial resolution. A compromise was found where a gap in the n-layer is only implemented in one dimension, such that the charge sharing and spatial resolution in the other dimension are not deteriorated. This design is suitable for the CLIC tracking detector, where high spatial resolution is only required in the $\mathrm{r \phi}$ dimension.

To benchmark the worst case scenario in respect of charge collection time, a single MIP has been simulated traversing the pixel corner.
The charge versus integration time from this simulation is compared in~\cref{fig:pulse} for the standard process, the modified process, and the modified process with the gap in the deep n-doped implant in one dimension.
Approximately \SI{20}{\percent} of the charge is collected after \SI{13}{ns} for the standard process and \SI{5}{ns} for the modified process.
For the modified process with the additional gap in the deep n-doped implant in one dimension the charge collection is further accelerated such that \SI{20}{\percent} of the charge is collected before \SI{3}{ns}.
Though no front-end simulation has been included, the results indicate that for the simulated pixel pitch of \SI{30x37.5}{\micron} the much slower charge collection in the pixel corners of the standard process would lead to a significant deterioration of the timing performance. 

The electric field distribution from the electrostatic TCAD simulations of the standard process have been combined with the Monte Carlo simulation framework \apsq (see Section ~\ref{sec:apx_2}) 
to enable high-statistics simulations including stochastic effects while retaining an exact description of the highly non-homogeneous electric field in the sensor ~\cite{katharina_report}.

\begin{figure}[ht]
	\centering
	\begin{minipage}[t]{0.525\textwidth}
		\begin{overpic}[width=1\textwidth]{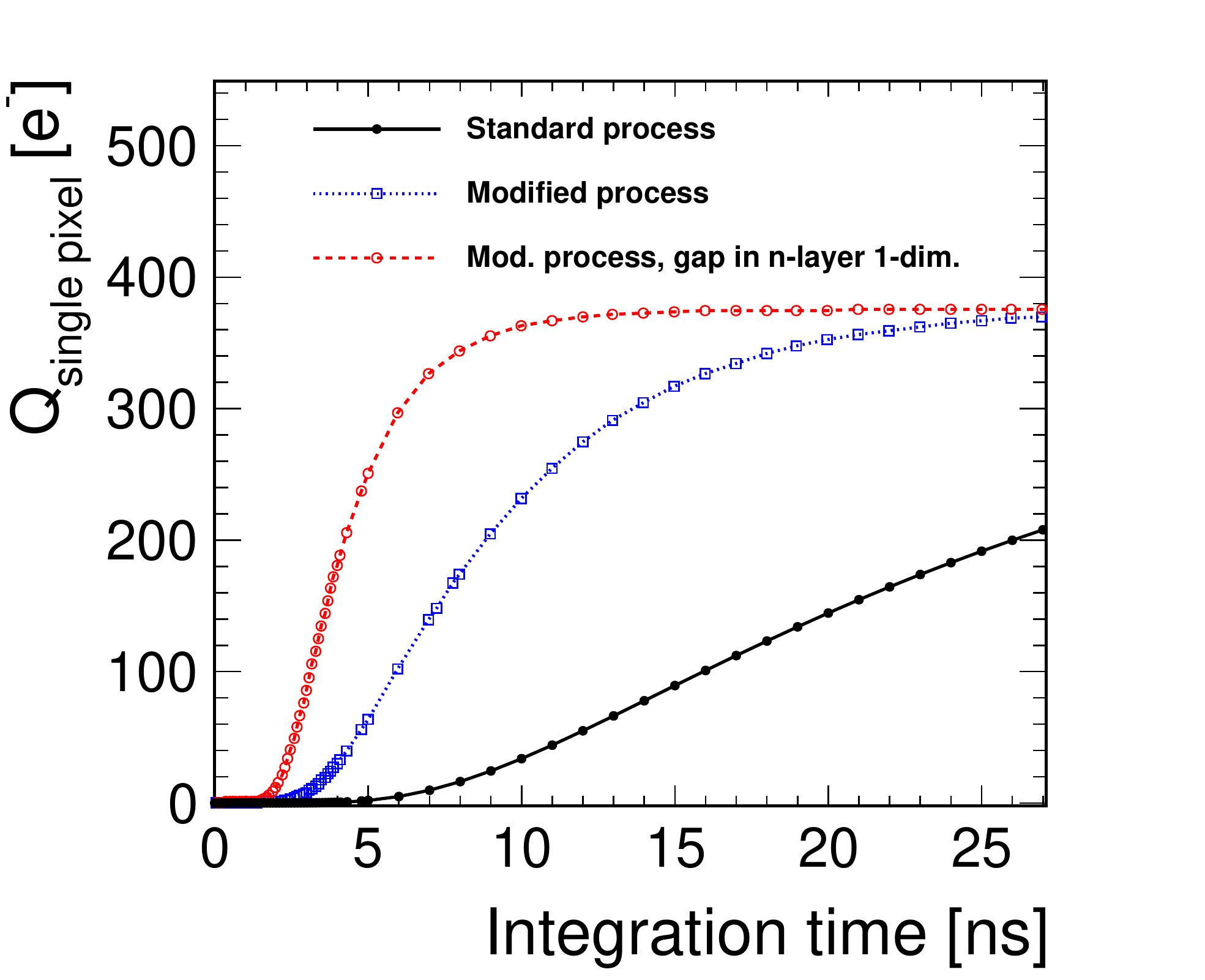}
      \put(65,40){\sffamily\small CLICdp}
		\end{overpic}
    \caption{Collected charge versus integration time for a MIP traversing the corner of a pixel with a size of \SI{30x37.5}{\micron}, comparing the standard process with two variants of the modified process.}
	\label{fig:pulse}
	\end{minipage}
	\hspace{0.3cm}
	\begin{minipage}[t]{0.425\textwidth}
		\begin{overpic}[width=1\textwidth]{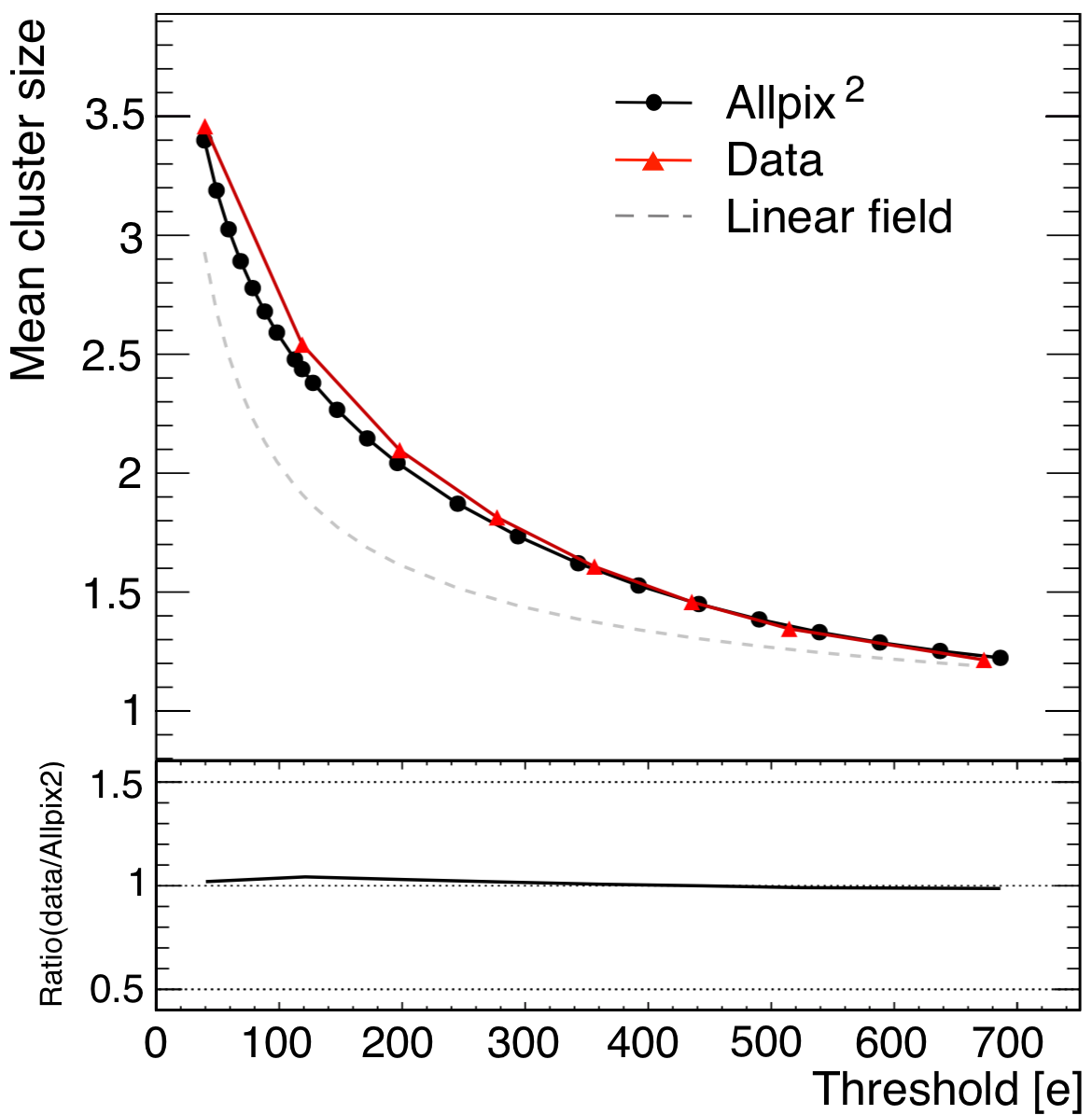}
      \put(65,60){\sffamily\small CLICdp}
		\end{overpic}
		\caption{Mean cluster size versus threshold for data and a combined TCAD+Monte Carlo simulation for the standard process for a pixel pitch of \SI{28x28}{\micron}.
		}
		\label{fig:apx0}
	\end{minipage}
\end{figure}

A simplified approach has been used, calculating the charges that arrive within $\mathrm{20 \, ns}$ at the collection electrode instead of calculating the induced current.
Example results and comparison with test-beam data are depicted in~\cref{fig:apx0}, where the mean cluster size is shown as a function of the detection threshold for simulation and data.
An excellent agreement has been found for various detection thresholds.
Moreover, the comparison with the simulation using a linear electric field (see grey dashed line) in the sensor shows the importance of modelling the electric field in TCAD for such sensors.

\subsection{The CLICTD monolithic tracker test chip}
\label{sec:clictd}

The promising results for the analogue performance of the studied small collection-electrode CMOS sensors led to the design of the fully monolithic CLICTD (CLIC Tracker Detector) chip in the same 180~nm HR-CMOS process, targeting the CLIC tracker requirements~\cite{iraklis_twepp}.

The top-level layout of the CLICTD design is shown in~\cref{fig:clictd_layout} and the main sensor design parameters are summarised in \cref{tab:clictd-parameters}.

\begin{figure}[ht]
	\centering
	\begin{minipage}[t]{0.4\textwidth}
		\begin{overpic}[width=1\textwidth]{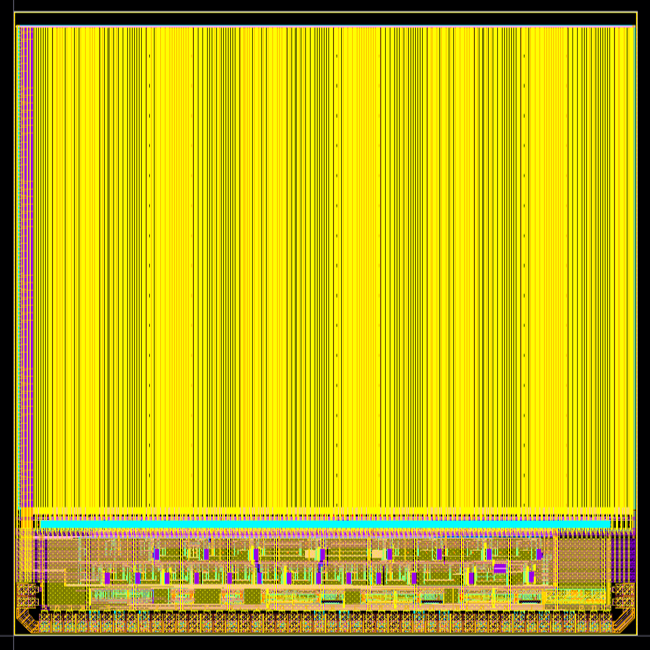}
		\end{overpic}
	\end{minipage}
	\caption{CLICTD mask set after placement and routing. The overall dimensions of the ASIC are \SI{5.0x5.0}{\mm}.}
	\label{fig:clictd_layout}
\end{figure}

\begin{table}[ht]
  \centering
  \begin{threeparttable}
  \caption{Main design parameters of the CLICTD monolithic CMOS sensor.}\label{tab:clictd-parameters}
  \begin{tabular}{l c}
    \toprule
Parameter  & Value \\ \midrule
Process technology & \SI{180}{\nm} HR-CMOS \\
ASIC size & \SI{5.0x5.0}{\mm} \\
Sensitive area & \SI{3.84x4.8}{\mm}  \\
Matrix size & \num{128x16} pixels \\
Pixel pitch & \SI{30x300}{\micron} \\
Analogue sub-pixel pitch & \SI{30x37.5}{\micron} \\
Gain & \SI{550}{\milli\volt / \kilo \Pem{}} \\
Noise RMS (simulated) & \SI{14}{\Pem}\\
Minimum threshold (simulated) & \SI{93}{\Pem} \\
Readout modes & 8-bit ToA + 5 bit ToT / 13 bit ToA / 13 bit photon counting \\
ToA bin size & \SI{10}{\ns} \\
Data compression & Zero suppression per pixel \\
Readout scheme & shutter-based \\
Data output clock & 40~MHz\\
Power pulsing scheme & analogue low-power mode and clock gating \\
Power consumption (w/o power pulsing)\tnote{a} &  210~mW/cm$^2$ + 70~mW periphery\\
Power consumption (after power pulsing)\tnote{b} & 5~mW/cm$^2$ + 70~mW periphery \\
\bottomrule
  \end{tabular}
  \begin{tablenotes} \footnotesize
  \item[a] Average for continuous operation at \SI{2.5}{\micro\second} shutter duration and 3\% occupancy.
  \item[b] Average for CLIC duty cycle and 3\% occupancy.
  \end{tablenotes}
  \end{threeparttable}
\end{table}

The CLICTD matrix has a size of \SI{3.84 x 4.8}{\mm} and comprises a sensitive area of 128 rows and 16 columns. 
The detector unit cell consists of channels with dimensions of \SI{30x300}{\micron}.
The smaller dimension along the r$\phi$ coordinate of the detector is constrained by the spatial resolution requirement of \SI{7}{\micron} in this coordinate.
To ensure prompt charge collection in the sensor diodes, while limiting the amount of digital circuitry, each channel is segmented in sub-pixels.
Each sub-pixel is equipped with a collection electrode and an analogue front end.
The analogue output of each sub-pixel is routed along the long channel dimension to the common digital part.
Thus, the number of sub-pixels per channel and the length of each channel is constrained by the space available within the \SI{30}{\micron} pixel pitch in the short dimension.
Motivated by three-dimensional transient TCAD simulations discussed in \cref{sec:hrcmos-simulation-results}, a geometry with 8 sub-pixels and a sub-pixel pitch of \SI{30 x 37.5}{\micron} has been chosen, as illustrated in \cref{fig:clictd_layout_sketch}.

\begin{figure}[ht]
  \centering
		\includegraphics[width=.8\linewidth]{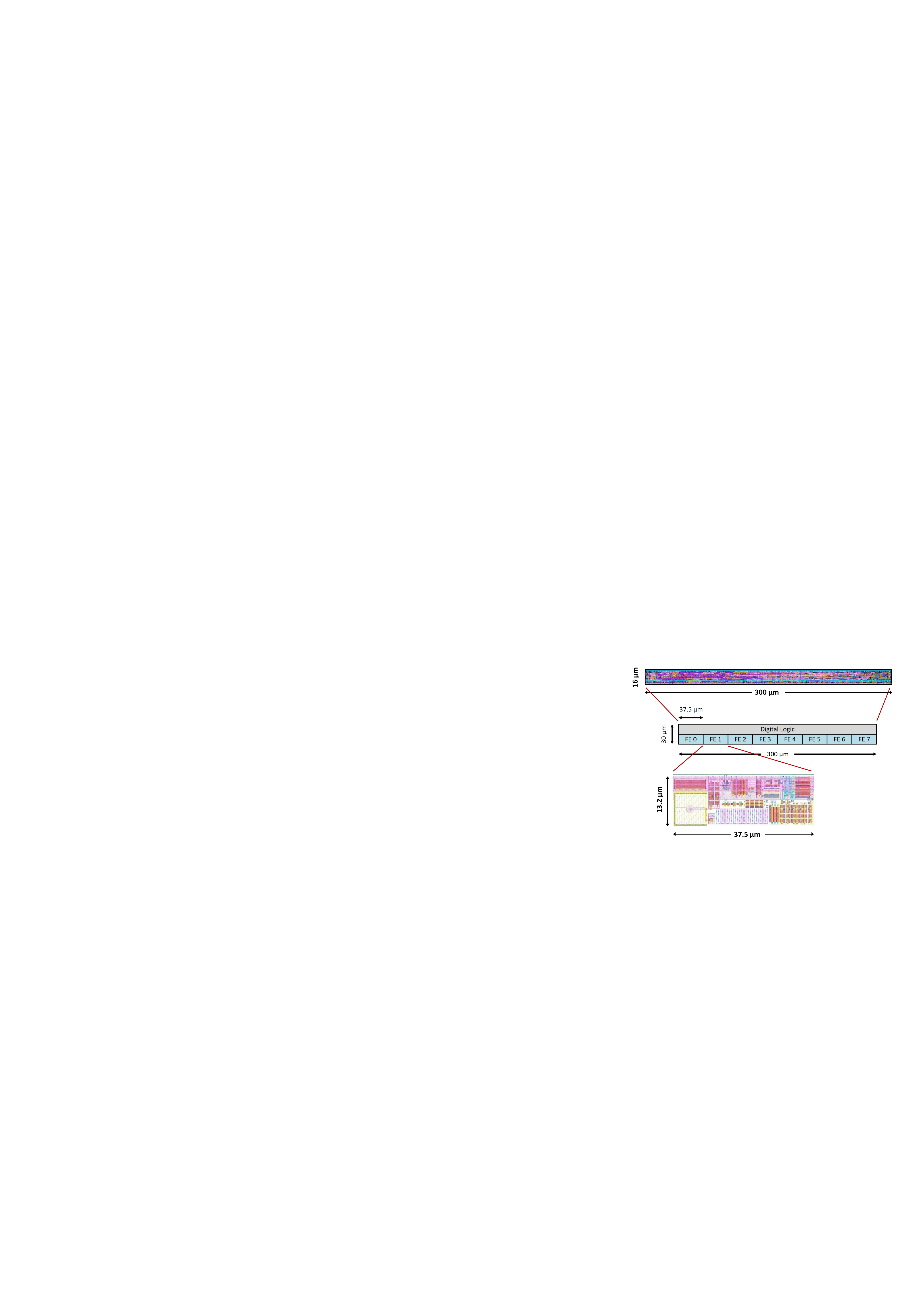}
		\caption{CLICTD pixel layout: The pixel is divided into 8 sub-pixels with separate analogue front-ends, that share a common digital block.}\label{fig:clictd_layout_sketch}
\end{figure}

\cref{fig:clictd} shows a block diagram of the CLICTD channel design. 
The charge collected in each sub-pixel is integrated by a charge-sensitive amplifier.
A global detection threshold is applied to the amplified signal by a threshold discriminator in each sub-pixel, which can be tuned individually with a 3-bit threshold adjustment DAC.
A gain of \SI{550}{\milli\volt / \kilo \Pem{}}, a noise of \SI{14}{\Pem{}} RMS and a minimum detectable charge of \SI{93}{\Pem{}} have been obtained from simulations of the analogue front-end.
Each of the sub-pixels can be masked individually.

\begin{figure}[ht]
	\centering
		\includegraphics[width=.9\textwidth]{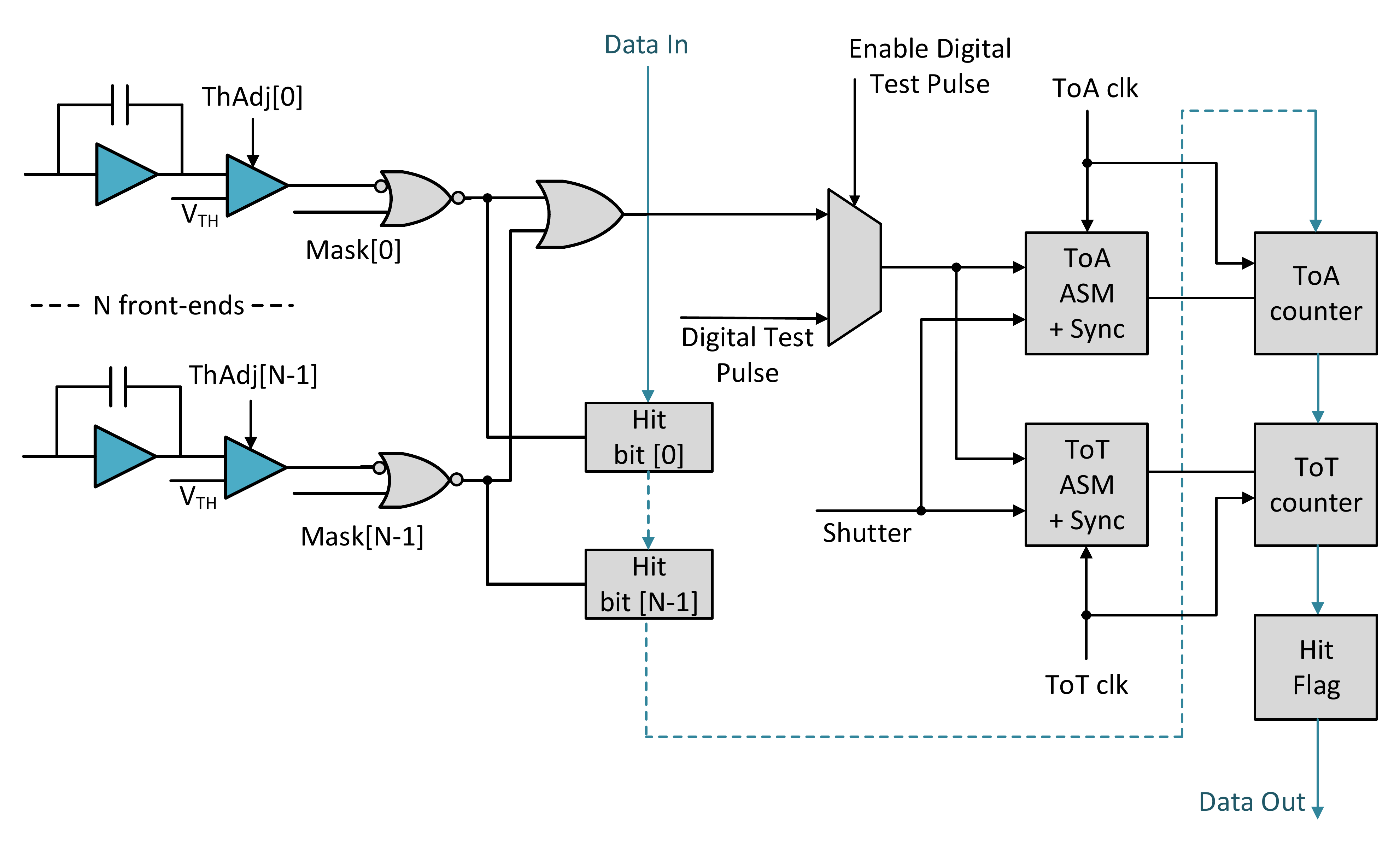}
		\caption{Block diagram of the CLICTD channel design.}
		\label{fig:clictd}
\end{figure}

A shutter-based read out scheme is implemented, optimised for the expected low pixel occupancies in the CLIC tracker resulting in mostly single sub-pixel hits per bunch train.
The discriminator output of all sub-pixels is combined by means of an OR-gate and sent to two asynchronous state machines in each pixel, measuring the ToA and ToT, respectively.
The ToT measurement uses a \SI{100}{\mega\hertz} clock with a programmable clock division factor between 2 and 16 to adjust the dynamic range.
The ToA is determined by a measurement of the clock cycle for which the first sub-pixel signal crosses the threshold, using the shutter signal as a time reference and a \SI{100}{\mega\hertz} clock to achieve the required \SI{10}{\ns} precision.

The address of all sub-pixels detecting a signal above threshold is stored in a bit pattern for each channel.
Besides increasing the spatial hit resolution in the long dimension, this allows for detailed studies of the sensor performance as a function of the pixel occupancy and charge-sharing between sub-pixels.

Different modes can be selected to read out the ToA and ToT information of the combined sub-pixel outputs.
In the nominal configuration a simultaneous 8-bit ToA and a 5-bit ToT measurement is performed.
Alternatively, the chip can be operated in a long counter mode, reading out only the time stamp of the first sub-pixel hit with the combined 13-bit precision, such that the timing performance can be studied in detail.
In addition, the chip also allows for a readout in photon counting mode, where the accumulated number of sub-pixel threshold crossings is counted. 

For threshold tuning and calibration purposes, test pulses can be injected to selected front-ends, as well as directly to the input of the digital logic.

A serial readout is proposed to shift the data off chip at a clock frequency of \SI{40}{\mega\hertz}.
To reduce the amount of data, a compression scheme is implemented, using the first bit of the readout data as a flag for zero suppression:
if a signal above threshold has been recorded in at least one sub-pixel of the channel, this hit-flag is set to a value of one and the following 21 bits of ToA, ToT as well as the bits storing the hit pattern are readout.
Otherwise, only the hit-flag, set to a value of zero, is sent off the chip.

For continuous operation at \SI{2.5}{\micro\second} shutter duration and 3\% occupancy the average power consumption of the matrix was simulated to amount to approximately \SI{210}{\milli\watt\per\cm\squared}. The analogue and digital circuitry both contribute similar amounts to this total power. The periphery blocks, which have not been optimised for low power consumption, contribute an additional \SI{70}{\milli\watt}.
To reduce the average power consumption of the chip to the required level of less than \SI{150}{\milli\watt\per\cm\squared}, a power-pulsing scheme is implemented. It sets the main power consuming parts of the analogue electronics to a low-power mode between subsequent shutters. The average power in the digital domain is minimised by means of clock gating: the clock in the channel logic is stopped when it is not acquiring or shifting out data. For the CLIC duty cycle, an average power consumption in the matrix of approximately \SI{5}{\milli\watt\per\cm\squared} was simulated. The periphery blocks are not power pulsed.

The CLICTD chip has been submitted for production in the modified process with the additional deep n-doped implant (see \cref{sec:towerjazz-process}).
Using a process split, additional wafers are produced with a segmentation of the deep n-doped implant in the long dimension of the pixel, to reduce the charge collection time (see \cref{sec:hrcmos-simulation-results}).

\section{Monolithic SOI sensors} \label{sec:soi}
\subsection{SOI process for pixel detectors}
Silicon-On-Insulator (SOI) wafers implement a layer of $\text{SiO}_{2}$ insulator oxide (Buried Oxide, BOX) between a thick high-resistivity sensor wafer and a thin low-resistivity electronics wafer. SOI wafers allow the fabrication of monolithic pixel detectors, by implementing a sensor matrix with collection electrodes in the high-resistivity substrate and readout electronics on the low-resistivity layer~\cite{SOI-IEE-TNS-2010}. The separation allows for a full depletion of the sensor layer while offering full standard CMOS circuits on the electronics wafer. The insulation layer physically separates the electronics from the substrate wafer, and thus reduces the parasitic capacitances. \cref{fig:soi_scheme} shows a schematic cross section through an SOI pixel-detector. For applications with higher ionising radiation levels than CLIC, process variants including an additional buried silicon layer (Double SOI) can be exploited~\cite{SOIPIX-paper}. 
The additional layer can be connected to an external bias voltage, compensating for threshold voltage shifts in the transistors, and it acts as an additional shielding of the electronics layer from the high potential needed to deplete the sensor layer.

\begin{figure}[ht]
  \centering
  \includegraphics[width=.49\linewidth]{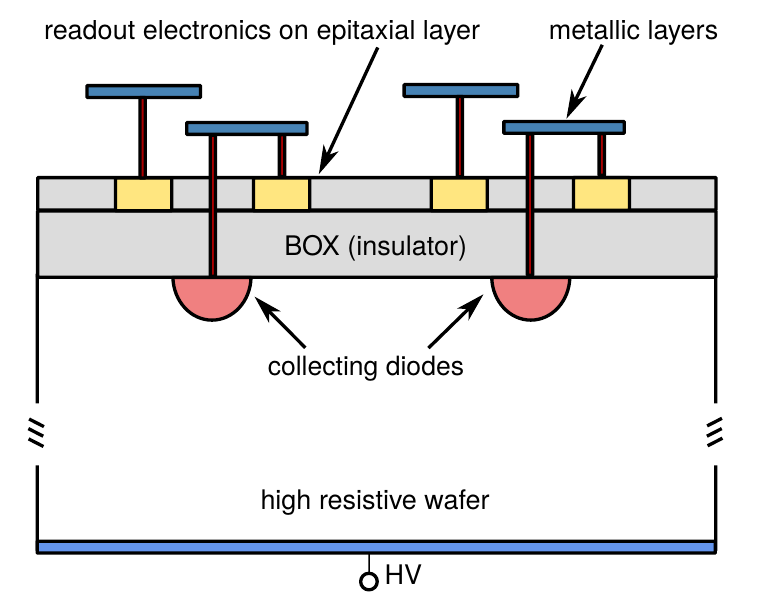}
  \caption{Schematics of an SOI CMOS structure working as a particle detector~\cite{SOI_paper}. A high-resistivity sensing layer and a low resistive electronics layer are separated by a buried oxide layer. Single or double oxide layers can be implemented to separate the sensing layer from the electronics layer.}\label{fig:soi_scheme}
\end{figure}

\subsection{Technology demonstrators}
A technology demonstrator pixel detector has been fabricated in the \SI{200}{\nm} Lapis SOI process~\cite{lapis_soi}. This technology allows for a fabrication of thin sensors, down to \SI{50}{\micron}. The demonstrator contains two different pixel architectures, schematically shown in \cref{fig:SOI-matrix-layout}. One is based on a source-follower (SF) design, the other on a charge-sensitive preamplifier. The source-follower based matrix has \num{8x36} integrating pixels, with a pixel pitch of \SI{30}{\micron}. The matrix with charge-sensitive preamplifier (CSA) is subdivided into two smaller matrices, one with small sensing diodes, the other with larger sensing diodes, each containing \num{8x18} pixels of size \SI{30x30}{\micron}. A rolling shutter readout of the full matrix is implemented. In the first demonstrator, the source-follower matrix was subdivided into eleven smaller sub-matrices (\num{4x6} or \num{8x6} pixels) with the same pixel architecture but slightly different sensor implant layouts and transistor sizes. Test results for the first generation of source follower matrix can be found in~\cite{SOI_paper}. Results presented here are obtained with the second iteration of the design, where the source-follower matrix was designed with uniform geometry and a more homogeneous response.
The test sensors have been fabricated on different silicon-on-insulator materials with different thicknesses and resistivities. The results presented in the following were obtained with \SI{500}{\micron} thick n-doped float-zone silicon and \SI{300}{\micron} thick p-doped double-SOI. Thinning to \SI{100}{\micron} is foreseen for future sensor productions, in order to meet the CLIC vertex-detector requirements in terms of material content. 

\begin{figure}[ht]
  \centering
  \includegraphics[width=0.3\linewidth]{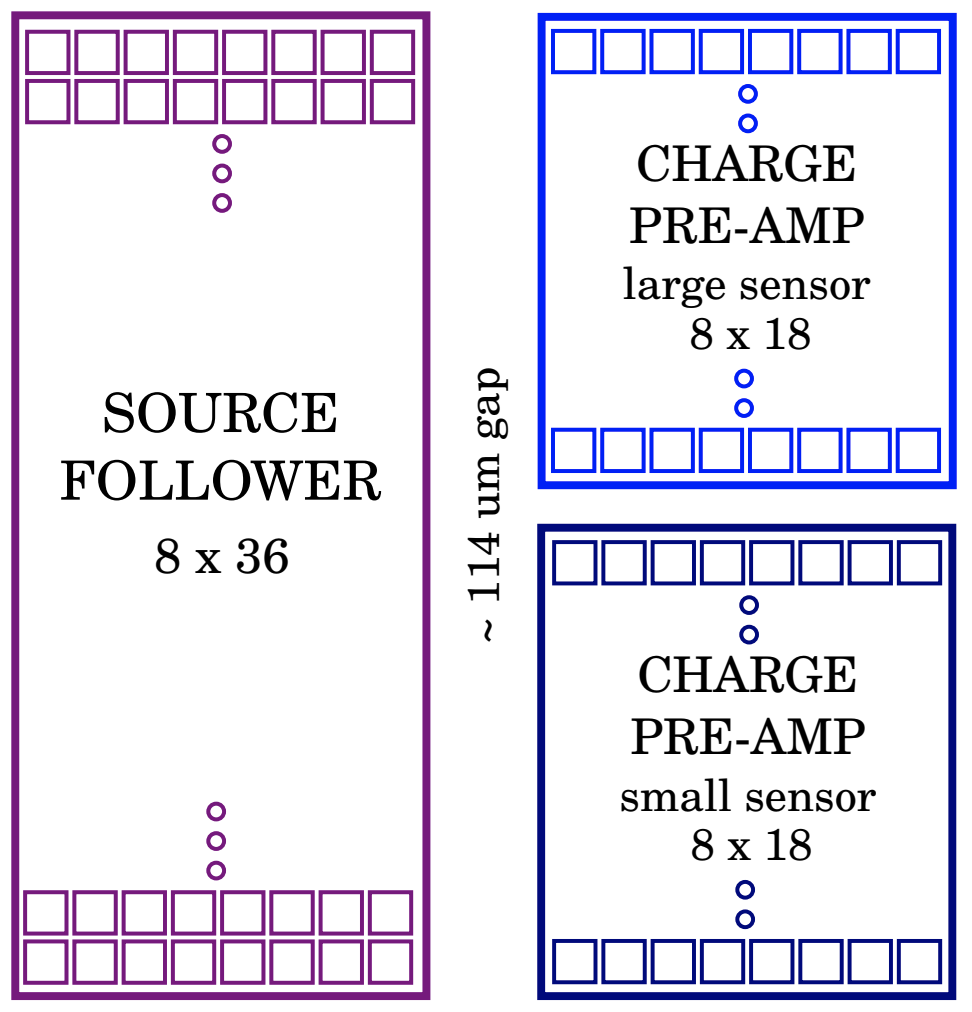}
  \caption{Matrix layout of the SOI technology demonstrator with three sub-matrices, as described in the text.}\label{fig:SOI-matrix-layout}
\end{figure}

\subsection{Characterisation}
The performance of multiple SOI demonstrator detectors has been evaluated in several beam tests using a \SI{120}{\giga\electronvolt} pion beam at CERN. The detector was placed on a mezzanine board that was assembled on a dedicated readout board containing a 12-bit external ADC. This system was connected to a Genesys Virtex-5 FPGA Board for readout. Details on the reconstruction and analysis flow can be found elsewhere~\cite{SOI_paper}. The main focus was the characterisation of a \SI{500}{\micron} thick float-zone n-type detector and a \SI{300}{\micron} thick double-SOI p-type detector.

\subsubsection{Signal-to-noise ratio}
The signal-to-noise ratio obtained in both detectors as a function of the applied bias voltage is displayed in \cref{fig:soi_snr}. For the float-zone n-type detector, all three pixel matrices show an excellent signal-to-noise ratio, as expected for the thick sensing layer. Full depletion of the detection layer is reached around \SI{70}{\volt}. From that point on, the signal-to-noise ratio of the source-follower matrix is stable above \num{350}. The charge-sensitive preamplifier matrices show a maximal signal-to-noise ratio of about \num{250}, with a slightly decreasing trend above full depletion. The origin is suspected to be the limited dynamic range of the preamplifier. The extracted resistivity of the \SI{500}{\micron} float-zone sensor wafer is around \SI{12}{\kilo\ohm\cm}.

Due to a relatively high leakage current, the maximum operation voltage for the double-SOI detector was limited to \SI{70}{\volt}, corresponding to a depletion depth of approximately \SI{130}{\micron}. Therefore, there are two effects reducing signal-to-noise ratio. The first one is directly connected to the lower depletion depth with respect to float-zone detector, affecting all three pixel architectures equally. The second one is caused by the much larger pixel capacitance introduced by an additional coupling to the middle silicon layer, that is affecting mainly the source-follower architecture. The matrices with charge-sensitive preamplifier front-end reach a signal-to-noise ratio around \num{90}, whereas the source-follower based pixels only reach \num{35}.
Since the full depletion is not reached, the resistivity of the double-SOI sensor wafer cannot be extracted from the signal measurements.

\begin{figure}[htb]
  \centering
    \begin{subfigure}[T]{.49\linewidth}
  \begin{tikzpicture}
  \begin{axis}[
    /pgf/number format/1000 sep={},
    axis line style={black,thick},
    ticklabel style={black,font={\sansmath\sffamily\large}},
    xlabel= Bias voltage / \si{\volt},
    ylabel = SNR, 
    ylabel near ticks,
    xlabel near ticks,
		label style={black,font={\sansmath\sffamily\Large}},
    xlabel style={at={(rel axis cs:1,0)}, anchor=north east,yshift=-.4cm},
    ylabel style={at={(rel axis cs:0,1)}, anchor=south east,yshift=.85cm},
    width=7.5cm,
    height=6.5cm,
    xmin=-8,
    ymin=0,
    ymax=400,
    minor y tick num={4},
    minor x tick num={4},
    every tick/.style={black,thick},
    every axis/.append style={font=\footnotesize},
    legend style={draw=none,text=black},
    legend style={text=black},
    legend style={at={(0.95,0.05)},anchor=south east,nodes=right,font={\sansmath\sffamily\footnotesize}},
    ]

    \addplot[mark=o,thick,color=black] plot[] table[x=Voltage,y=CPA_small]{sections/VertexTracking/figures/soi-plots/snr_fzn.dat}; 
    \addlegendentry{CPA small};

    \addplot[mark=square,thick,color=red!75!white] plot[] table[x=Voltage,y=CPA_large]{sections/VertexTracking/figures/soi-plots/snr_fzn.dat}; 
    \addlegendentry{CPA large};
    
    \addplot[mark=triangle,thick,color=blue!75!white] plot[] table[x=Voltage,y=SF]{sections/VertexTracking/figures/soi-plots/snr_fzn.dat}; 
    \addlegendentry{SF};
  
   \node at (rel axis cs:0.25,0.25) {CLICdp};

  \end{axis}
\end{tikzpicture}
  \caption{}\label{fig:soi_snr_fzn}
  \end{subfigure}
  \hfill
  \begin{subfigure}[T]{.49\linewidth}
\begin{tikzpicture}
  \begin{axis}[
    /pgf/number format/1000 sep={},
    axis line style={black,thick},
    ticklabel style={black,font={\sansmath\sffamily\large}},
    xlabel= Bias voltage / \si{\volt},
    ylabel = SNR, 
    ylabel near ticks,
    xlabel near ticks,
		label style={black,font={\sansmath\sffamily\Large}},
    xlabel style={at={(rel axis cs:1,0)}, anchor=north east,yshift=-.4cm},
    ylabel style={at={(rel axis cs:0,1)}, anchor=south east,yshift=.85cm},
    width=7.5cm,
    height=6.5cm,
    xmin=-3,
    xmax=90,
    ymin=-5,
    ymax=105,
    minor y tick num={4},
    minor x tick num={4},
    every tick/.style={black,thick},
    every axis/.append style={font=\footnotesize},
    legend style={draw=none,text=black, fill=none},
    legend style={text=black},
    legend style={at={(0.95,0.05)},anchor=south east,nodes=right,font={\sansmath\sffamily\footnotesize}},
    ]

    \addplot[mark=o,thick,color=black] plot[] table[x=Voltage,y=CPA_small]{sections/VertexTracking/figures/soi-plots/snr_dsoi.dat}; 
    \addlegendentry{CPA small};

    \addplot[mark=square,thick,color=red!75!white] plot[] table[x=Voltage,y=CPA_large]{sections/VertexTracking/figures/soi-plots/snr_dsoi.dat}; 
    \addlegendentry{CPA large};
    
    \addplot[mark=triangle,thick,color=blue!75!white] plot[] table[x=Voltage,y=SF]{sections/VertexTracking/figures/soi-plots/snr_dsoi.dat}; 
    \addlegendentry{SF};
  
    \node at (rel axis cs:0.2,0.8) {CLICdp};

  \end{axis}
\end{tikzpicture}
\caption{}\label{fig:soi_snr_dsoi}
\end{subfigure}
  \caption{Signal-to-noise ratio as a function of the applied reverse bias voltage for the three sub-matrices of \subref{fig:soi_snr_fzn} the \SI{500}{\micron} thick float-zone n-type SOI detector and of \subref{fig:soi_snr_dsoi} the \SI{300}{\micron} thick double SOI p-type detector .}\label{fig:soi_snr}
\end{figure}
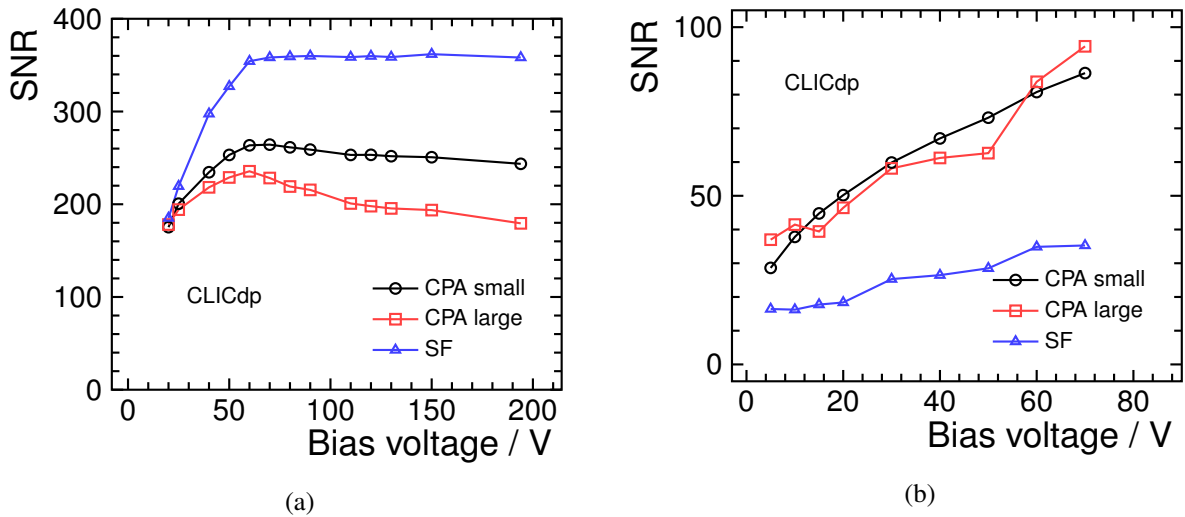

\subsubsection{Efficiency}
\cref{fig:soi_efficiency} shows the detection efficiency of the \SI{500}{\micron} thick float-zone n-type SOI detector as a function of the applied bias voltage at perpendicular particle incidence. The obtained efficiency reaches a plateau in the order of \SI{98}{\percent{}} already for voltages above \SI{20}{\volt}, as expected for the high resistive substrate and the large detector thickness. The slight systematic inefficiency of the detector is expected to be caused by the rolling shutter readout scheme and the inevitable inefficiency of the pixel row that is currently read out and reset. Systematic studies, correlating the track impact point to the rolling shutter position on the matrix are ongoing.

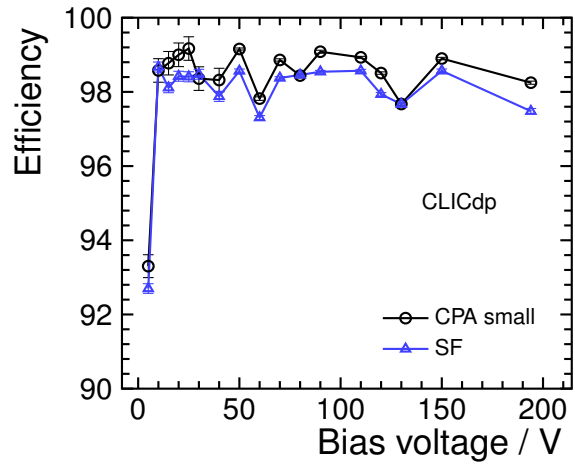
\begin{figure}[htb]
  \centering
  \begin{tikzpicture}
  \begin{axis}[
    /pgf/number format/1000 sep={},
    axis line style={black,thick},
    ticklabel style={black,font={\sansmath\sffamily\large}},
    xlabel= Bias voltage / \si{\volt},
    ylabel = Efficiency, 
    ylabel near ticks,
    xlabel near ticks,
		label style={black,font={\sansmath\sffamily\Large}},
    xlabel style={at={(rel axis cs:1,0)}, anchor=north east,yshift=-.4cm},
    ylabel style={at={(rel axis cs:0,1)}, anchor=south east,yshift=.85cm},
    width=7.5cm,
    height=6.5cm,
    xmin=-8,
    ymin=90,
    ymax=100,
    minor y tick num={4},
    minor x tick num={4},
    every tick/.style={black,thick},
    every axis/.append style={font=\footnotesize},
    legend style={draw=none,text=black},
    legend style={text=black},
    legend style={at={(0.95,0.05)},anchor=south east,nodes=right,font={\sansmath\sffamily\footnotesize}},
    ]

    \addplot[mark=o,thick,color=black] plot[error bars/.cd, y dir = both, y explicit] table[x=Voltage,y=efficiency, y error=error]{sections/VertexTracking/figures/soi-plots/efficiency_fzn_cpa_small_wafer2_12MHz.dat}; 
    \addlegendentry{CPA small};

    \addplot[mark=triangle,thick,color=blue!75!white] plot[error bars/.cd, y dir = both, y explicit] table[x=Voltage,y=efficiency,y error=error]{sections/VertexTracking/figures/soi-plots/efficiency_fzn_sf_wafer12_12MHz.dat}; 
    \addlegendentry{SF};
  
    \node at (rel axis cs:0.75,0.5) {CLICdp};
    
  \end{axis}
\end{tikzpicture}
  \caption{Efficiency as a function of the applied reverse bias voltage for the \SI{500}{\micron} thick FZ-n SOI detector. Both, charge-sensitive preamplifier and source follower pixels are almost fully efficient above \SI{20}{\volt}.}\label{fig:soi_efficiency}
\end{figure}

\subsubsection{Spatial resolution}
The large sensor thickness in combination with the small pixel pitch leads to an excellent spatial resolution of the detector, as summarised in \cref{fig:soi_resolution_summary} for the resolutions in the y-coordinate. The cluster positions for the SOI sensor were obtained after correcting for the non-linear charge sharing between the pixels in the cluster. The resolution values were obtained by quadratically subtracting the telescope pointing resolution of \SI{2}{\micron} from the width of a Gaussian fit to the residuals between the cluster positions and the predicted impact points. Above full depletion the \SI{500}{\micron} thick float-zone n-type detector reaches \SI{3}{\micron} for all pixel types. The performance of the source-follower matrix is slightly superior to the charge sensitive preamplifier due to the better signal-to-noise ratio, as shown in \cref{fig:soi_snr}, and reaches a resolution of about \SI{2.2}{\micron} at high voltages. Due to the reduced signal-to-noise ratio, the slightly thinner double-SOI detector only reaches a resolution of the order of \SIrange{4}{5}{\micron}. For this detector, the charge-sensitive preamplifier matrices with the larger signal-to-noise ratio show the better resolution.

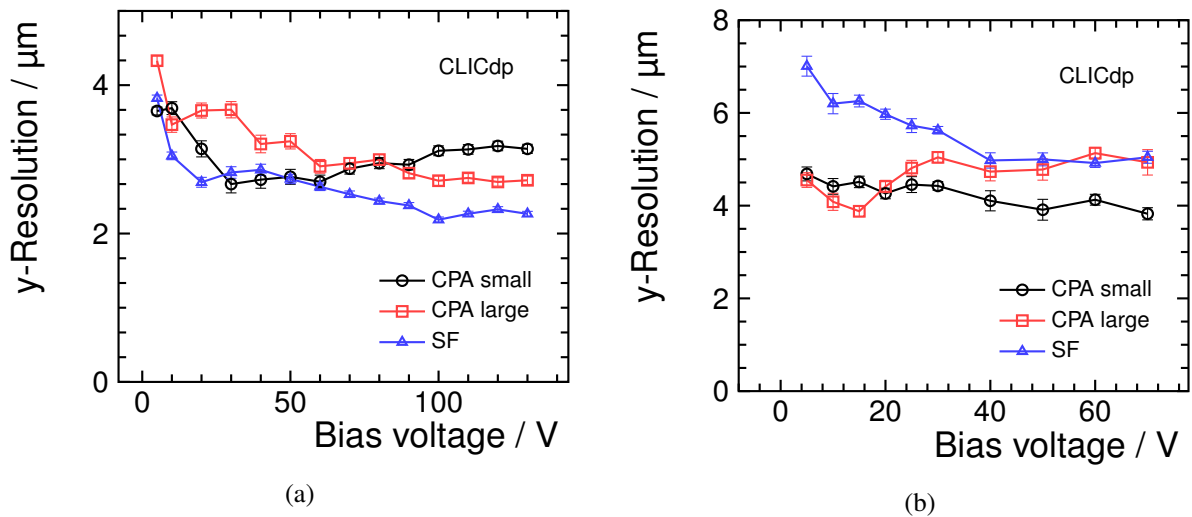
\begin{figure}[htb]
  \centering
  \begin{subfigure}[T]{.49\linewidth}
  \begin{tikzpicture}
  \begin{axis}[
    /pgf/number format/1000 sep={},
    axis line style={black,thick},
    ticklabel style={black,font={\sansmath\sffamily\large}},
    xlabel= Bias voltage / \si{\volt},
    ylabel = y-Resolution / \si{\micron},
    ylabel near ticks,
    xlabel near ticks,
		label style={black,font={\sansmath\sffamily\Large}},
    xlabel style={at={(rel axis cs:1,0)}, anchor=north east,yshift=-.4cm},
    ylabel style={at={(rel axis cs:0,1)}, anchor=south east,yshift=.85cm},
    width=7.5cm,
    height=6.5cm,
    xmin=-8,
    ymin=0,
    ymax=5,
    minor y tick num={5},
    minor x tick num={4},
    every tick/.style={black,thick},
    every axis/.append style={font=\footnotesize},
    legend style={draw=none,text=black},
    legend style={text=black},
    legend style={at={(0.95,0.05)},anchor=south east,nodes=right,font={\sansmath\sffamily\footnotesize}},
    ]

    \addplot[mark=o,thick,color=black] plot[error bars/.cd, y dir = both, y explicit] table[x=Voltage,y=res_cpa_small, y error=error_cpa_small]{sections/VertexTracking/figures/soi-plots/res_fzn.dat}; 
    \addlegendentry{CPA small};
    
    \addplot[mark=square,thick,color=red!75!white] plot[error bars/.cd, y dir = both, y explicit] table[x=Voltage,y=res_cpa_large, y error=error_cpa_large]{sections/VertexTracking/figures/soi-plots/res_fzn.dat}; 
    \addlegendentry{CPA large};
    
    \addplot[mark=triangle,thick,color=blue!75!white] plot[error bars/.cd, y dir = both, y explicit] table[x=Voltage,y=res_sf, y error=error_sf]{sections/VertexTracking/figures/soi-plots/res_fzn.dat}; 
    \addlegendentry{SF};
  
\node[anchor=north east] at (rel axis cs:0.9,0.9) {CLICdp};

  \end{axis}
\end{tikzpicture}
  \caption{}\label{fig:soi_resolution_fzn}
\end{subfigure}
\hfill
\begin{subfigure}[T]{.49\linewidth}
\begin{tikzpicture}
  \begin{axis}[
    /pgf/number format/1000 sep={},
    axis line style={black,thick},
    ticklabel style={black,font={\sansmath\sffamily\large}},
    xlabel= Bias voltage / \si{\volt},
    ylabel = y-Resolution / \si{\micron},
    ylabel near ticks,
    xlabel near ticks,
		label style={black,font={\sansmath\sffamily\Large}},
    xlabel style={at={(rel axis cs:1,0)}, anchor=north east,yshift=-.4cm},
    ylabel style={at={(rel axis cs:0,1)}, anchor=south east,yshift=.85cm},
    width=7.5cm,
    height=6.5cm,
    xmin=-8,
    ymin=0,
    ymax=8,
    minor y tick num={3},
    minor x tick num={4},
    every tick/.style={black,thick},
    every axis/.append style={font=\footnotesize},
    legend style={draw=none,text=black},
    legend style={text=black},
    legend style={at={(0.95,0.05)},anchor=south east,nodes=right,font={\sansmath\sffamily\footnotesize}},
    ]

    \addplot[mark=o,thick,color=black] plot[error bars/.cd, y dir = both, y explicit] table[x=Voltage,y=res_cpa_small, y error=error_cpa_small]{sections/VertexTracking/figures/soi-plots/res_dsoi.dat}; 
    \addlegendentry{CPA small};

    \addplot[mark=square,thick,color=red!75!white] plot[error bars/.cd, y dir = both, y explicit] table[x=Voltage,y=res_cpa_large, y error=error_cpa_large]{sections/VertexTracking/figures/soi-plots/res_dsoi.dat}; 
    \addlegendentry{CPA large};
    
    \addplot[mark=triangle,thick,color=blue!75!white] plot[error bars/.cd, y dir = both, y explicit] table[x=Voltage,y=res_sf, y error=error_sf]{sections/VertexTracking/figures/soi-plots/res_dsoi.dat}; 
    \addlegendentry{SF};
  
    \node[anchor=north east] at (rel axis cs:0.9,0.9) {CLICdp};

  \end{axis}
\end{tikzpicture}
\caption{}\label{fig:soi_resolution_dsoi}
\end{subfigure}
  \caption{Spatial resolution in the y-coordinate as a function of the applied reverse bias voltage for the three sub-matrices \subref{fig:soi_resolution_fzn} of the \SI{500}{\micron} thick FZ-n SOI detector and \subref{fig:soi_resolution_dsoi} of the \SI{300}{\micron} thick double-SOI p-type detector .}\label{fig:soi_resolution_summary}
\end{figure}

\subsection{The CLIPS vertex-detector test chip} \label{sec:clips}
The CLIPS (short for CLIc Pixel Soi) detector was designed to fulfil the CLIC vertex-detector requirements in terms of spatial and timing resolution and material content~\cite{clips-design-review-presentation}. The fully monolithic chip consists of three pixel matrices with \num{64x64} pixels each, with a pixel size of \SI{20x20}{\micron}, and an additional set of test structures. The first matrix M1 is implemented without n-stops between the pixels, the second matrix M2 includes n-stops. The pixel layout in the third matrix M3 is similar to M2, but the feedback capacitance is formed only on parasitic capacitances instead of a dedicated metal-to-metal capacitor implemented in the M2 matrix. The total size of the chip is \SI{4.4x4.4}{\mm}, as illustrated in \cref{fig:clips_layout}.

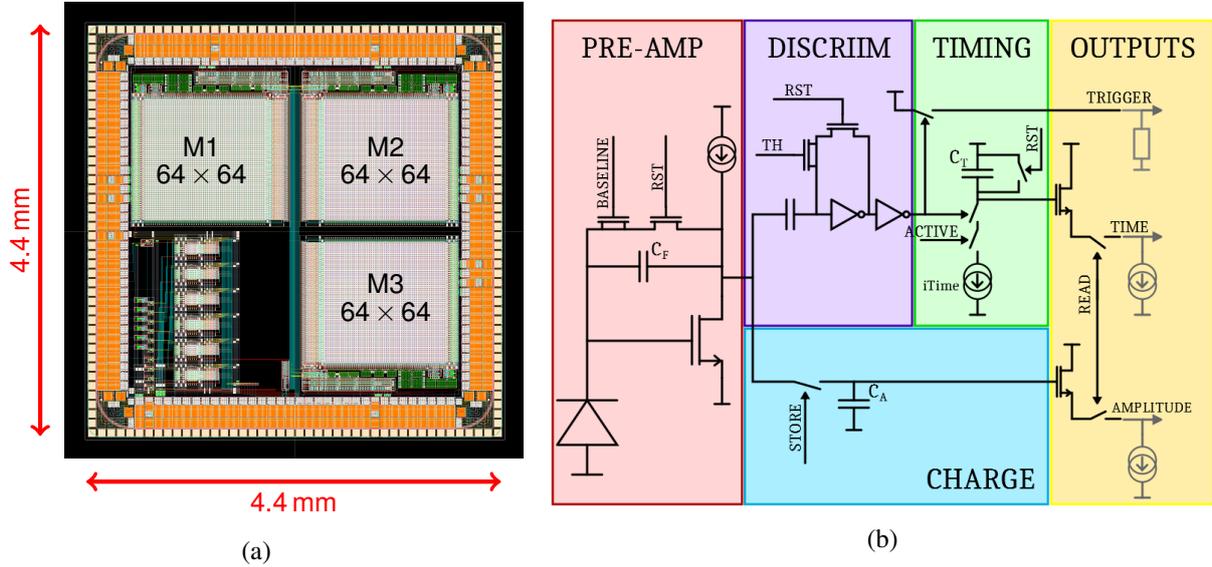
\begin{figure}[!htb]
  \begin{subfigure}[c]{0.42\linewidth}
    \begin{tikzpicture}
    \node[anchor=south west,inner sep=0] at (0,0)(image){\includegraphics[width=.9\linewidth]{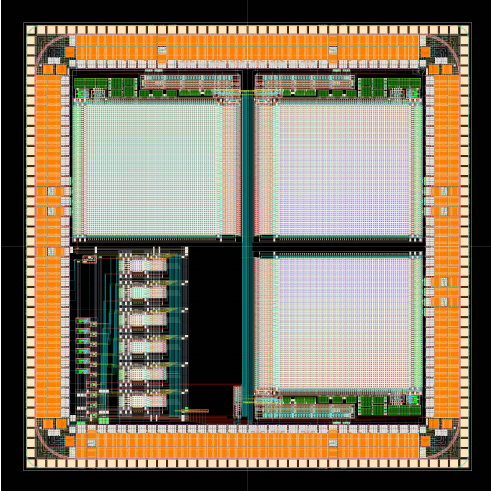}};
    \begin{scope}[x={(image.south east)},y={(image.north west)}]
      \node at (0.3,0.65){\shortstack{M1\\\num{64x64}}};
      \node at (0.7,0.65){\shortstack{M2\\\num{64x64}}};
      \node at (0.7,0.35){\shortstack{M3\\\num{64x64}}};
      \draw[<->,red,ultra thick](0.05,-0.05)--(0.95,-0.05)node[pos=0.5,below]{\SI{4.4}{\mm}};
      \draw[<->,red,ultra thick](-0.05,0.05)--(-0.05,0.95)node[pos=0.5,above,rotate=90]{\SI{4.4}{\mm}};
    \end{scope}
    \end{tikzpicture}
    \caption{}\label{fig:clips_layout}
  \end{subfigure}
  \hfill
  \begin{subfigure}[T]{.55\linewidth}
    \includegraphics[width=\linewidth]{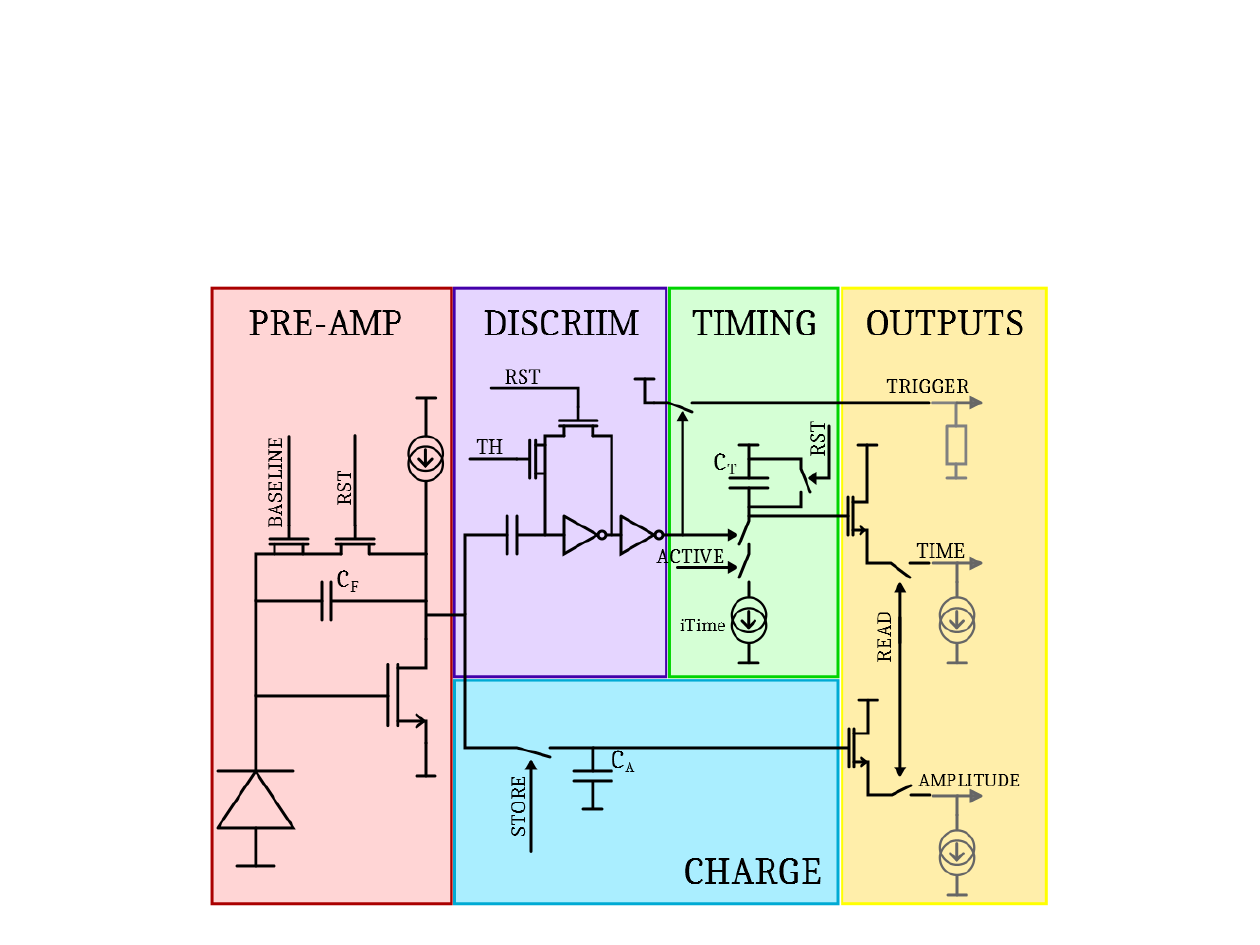}
    \caption{}\label{fig:clips_blockdiagram}
  \end{subfigure}
  \caption{CLIC Pixel SOI chip: \subref{fig:clips_layout} chip layout containing three independent pixel matrices with design variations and \subref{fig:clips_blockdiagram} block diagram of the pixel front-end. }\label{fig:clips}
\end{figure}

The pixel electronics is divided into several functional blocks, as depicted in \cref{fig:clips_blockdiagram}. For all three matrices, the pixel front-end is based on a charge sensitive pre-amplifier with common-source architecture and a non-resistive feedback. The simulated front-end response is linear up to \SI{1}{\femto\coulomb} of input charge, which is approximately the expected ionisation signal in a \SI{100}{\micron} thick detector. Above that, the output starts to saturate. A discriminating stage is implemented for hit detection.

The readout of charge and time information is based on a snapshot concept, well adapted to the low duty cycle of Linear Colliders. The amplified ionisation charge is stored on a storage capacitor, which is sampling the analogue output of the pre-amplifier once a particle hit is detected in the pixel. The Time-of-Arrival of the particle hit is extracted in a similar way. Once a particle hit is detected, a storage capacitor is discharged by a constant current source triggered by the discriminator output signal, thus the charge and voltage on the capacitor decrease linearly with time. The earlier the particle hit appears, the lower the voltage on the capacitor will be at the end of the periodic active detection cycle, which can be synchronised to the bunch-train structure of the Linear Collider. In this way, hit timing is measured during the gaps between colliding bunches, without the need to distribute a clock tree into the pixel matrix. This change in architecture allows to reduce the pixel size below \SI{25}{\micron}. In the first CLIPS demonstrator, analogue-to-digital conversion of the charge and time information for pixels with detected hits (signalled by a trigger output of the discriminator in each pixel) will be performed off-chip by the readout system.

The active time window of the chip is adjustable between \SI{100}{\ns} and \SI{300}{\micro\second}. The discharge current has to be adopted such that the ToA measurement covers the full active time, and thus the achievable timing resolution depends on the active time window. Simulations have been performed for an active window of \SI{1}{\micro\second}, covering well the duration of the bunch trains at CLIC. The extracted non-linearity of the time measurement is below \SI{0.5}{\percent{}} corresponding to less than \SI{5}{\ns} over the full ToA range.

Time-walk of the discriminating stage is an additional source of uncertainty on the ToA measurement. From simulations, the expected time-walk is below \SI{10}{\ns} for signals above \SI{0.3}{\femto\coulomb} at a threshold of \SI{0.1}{\femto\coulomb}. Since both charge and time information are available, a time-walk correction can be applied during the event reconstruction.

The first version of CLIPS has recently been produced. Chips fabricated on \SI{500}{\micron} thick \mbox{FZ-n} wafers have been received. Thinning of selected wafers to \SI{100}{\micron} is foreseen at a later stage. The development of test systems is currently ongoing.

\section{Detector integration}\label{sec:vtx-trk-integration}
The detector performance requirements, in particular the very low material budget for the vertex and tracking detectors, lead to challenging constraints for the mechanical and electrical integration of the detector components and their cooling systems. An integrated approach is followed, addressing simultaneously the critical R\&D issues in the domains of Through-Silicon Via (TSV) interconnects, power-delivery and power pulsing, support structures, detector assembly, cabling and cooling. The work is focused on conceptual designs and validation through Finite-Element-Analysis (FEA) simulations and technology demonstrators of critical components.

\subsection{Back-end processing (TSV)}
The size of pixel readout ASICs is usually limited by the size of the reticle used in the CMOS process. In most current pixel detectors, the ASICs are designed to be abutted on three sides, while the fourth side is used for wire-bond connection.
Multiple ASICs are then flip-chip bonded onto a large sensor to form a module. To minimise inefficient areas, the sensor pixels in the border region of the readout ASICs have to be enlarged, at the cost of degraded spatial resolution. The guard ring and ASIC periphery regions contribute to the material budget as insensitive material.

In order to achieve a full coverage of large areas with hybrid pixel detectors, sensor and readout ASICs should ideally be seamlessly tileable along all four sides. Active edge sensors, as presented in \cref{sec:active_edge_sensors}, can help in reducing inactive material on the sensor side of hybrid pixel detectors and allow neighbouring sensors to be brought close together without an inactive area on the sensor edge. Through-Silicon Via (TSV) vertical interconnects remove the need for wire-bonding connections on the ASIC side of the assembly. Instead of being connected on the ASIC periphery, Input/Output (IO) signals are routed through the silicon die to the ASIC back side, where they are redistributed to Ball Grid Array (BGA) IO pads.

A post-processing "via last" TSV process, developed by CEA-Leti~\cite{cea-leti} and the Medipix3 collaboration~\cite{medipix3}, has demonstrated the feasibility of TSVs on functional detector ASICs from the Medipix/Timepix ASIC family~\cite{tsv_ieee,tsv_campbell}. The process includes thinning of the ASIC wafers to \SI{120}{\micron} and results in vias of \SI{60}{\micron} diameter connecting the IO signals through the silicon die to the redistribution layer on the back side, as shown in \cref{fig:tsv-details}.

\begin{figure}[ht]
  \centering
    \begin{subfigure}[T]{.49\linewidth}
    \begin{tikzpicture}
    \node[anchor=south west,inner sep=0] at (0,0)(image){	\includegraphics[width=\linewidth]{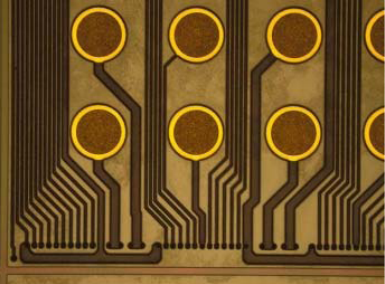}};
    \begin{scope}[x={(image.south east)},y={(image.north west)}]
     \draw[|-|, thick, yellow](0.045,0.08)--(0.25,0.08) node [pos=0.5,below]{\small \SI{1}{\mm}}; 
    \end{scope}
    \end{tikzpicture}
    \caption{}\label{fig:tsv-backside}
    \end{subfigure}
    \hfill
    \begin{subfigure}[T]{.49\linewidth}
      \includegraphics[width=\linewidth]{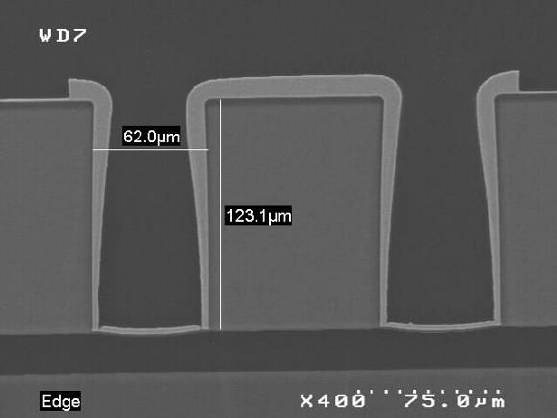}
      \caption{}\label{fig:tsv-cutview}
    \end{subfigure}
  \caption{\subref{fig:tsv-backside} Medipix3RX redistribution layer on the back side of the ASIC for connecting the TSVs on the bottom to BGA bonding pads in the centre. \subref{fig:tsv-cutview} Cut through a TSV in a \SI{120}{\micron} thick Medipix3RX ASIC. Image credit: CEA-Leti.}
  \label{fig:tsv-details}
\end{figure}

Three separate projects have been undertaken with CEA-Leti in order to develop the TSV processing on the IO pads of Medipix3 and Timepix3 wafers. The first one demonstrated the general feasibility of the processing of Medipix3/Timepix3 wafers and established the process steps. The aim of the second project was to evaluate the yield of the TSV processing. The achieved TSV yield was good for most of the wafers (\SIrange{70}{80}{\percent}). The majority of the observed yield loss was restricted to components close to the edge of the wafer, as was expected from equipment limitations. A small number of samples were flip-chip connected to standard silicon sensors and underwent further testing, demonstrating the full functionality of the TSV processed ASICs. From this project it was concluded that the TSV yield is adequate for a small-scale production. In a third project, even thinner wafers down to \SI{50}{\micron} thickness have been processed successfully, aiming mostly at applications in vertex detectors of high-energy physics experiments. The dies are mechanically flat and robust, and electrical testing results are compatible with the one for \SI{120}{\micron} thick ASICs from previous runs.

\subsection{Power delivery and power pulsing for the vertex detector}
\label{sec:power}
In the following a current-based power-delivery and power-pulsing concept for the CLIC vertex detector is described, which has been verified experimentally for a dummy-load setup corresponding to a barrel half ladder~\cite{CLICdp-Note-2015-004}. It targets an average power consumption of less than 50\,mW/cm$^2$ as well as a low material-budget contribution of 0.05\%$X_0$ per single detection layer for cables and local energy storage across the ladder.

\subsubsection{Powering requirements}
Different powering states during the CLIC accelerator cycle can be defined according to the nominal power-consumption values of the CLICpix ASIC. This is shown schematically in \cref{fig:VertexPowerConsumption} for a half ladder consisting of 12 readout ASICs based on the CLICpix architecture (see \cref{sec:CLICpix}) and with a size of 1~cm$^2$ each.

The analogue peak power of \SI{24}{\watt} is only needed during a short acquisition time window of approximately 20\,$\upmu$s around the time of the collisions and can be switched off entirely during the 20\,ms gaps between bunch trains. The average analogue power consumption can thus be reduced by a factor of approximately 1000 to 24\,mW.

The digital part of the CLICpix ASICs has a peak power of 1.2\,W per half ladder during the acquisition time. Afterwards the most power consuming digital blocks are disabled, resulting in an idle state with less than 100\,mW power consumption per half ladder. The data readout is assumed to consume 360~mW per ASIC and will be performed sequentially ASIC-by-ASIC. The readout time per ASIC scales with the pixel occupancy in each ASIC, with an expected maximum of approximately 300\,$\upmu$s per ASIC at 3\% occupancy. The readout of the entire half ladder will be spread over the duration of the 20\,ms gap between bunch trains. The average digital power consumption in this scenario amounts to approximately 160~mW per half ladder, corresponding to a reduction by a factor of 8 with respect to the digital peak power.

Critical voltage levels are regulated inside the readout ASICs. Voltage variations of approximately $\pm10\%$ during load transitions can therefore be tolerated for the supply voltages.

\begin{figure}[ht]
  \centering
  \includegraphics[width=0.9\textwidth]{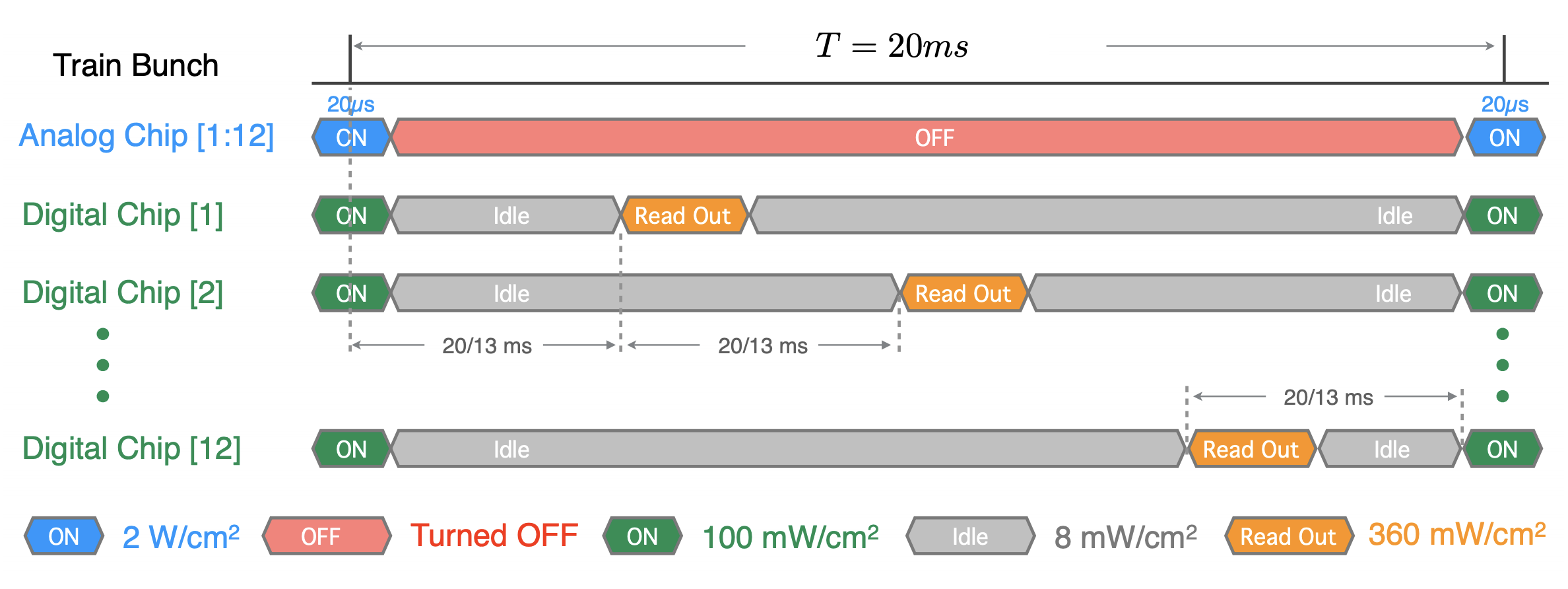}
  \caption{Analogue and digital power consumption states in a vertex-detector half ladder (12 ASICs of 1\,cm$^2$ each) during the CLIC accelerator cycle~\cite{CLICdp-Note-2015-004}.}
  \label{fig:VertexPowerConsumption}
\end{figure}

\subsubsection{Powering concept}
An electrical circuit optimised for low material budget and providing the required powering states for a vertex detector half ladder is shown schematically in~\cref{fig:PowerSchemaVertex}. 
It is based on an FPGA-controlled current source that is located outside of the main detector and that provides a low continuous current.
The current is brought through cables to the detector front end, where it charges up on-detector silicon capacitors for local energy storage at each ASIC.
Low-mass aluminium cables are chosen for the cable sections located inside the detector volume.
The locally stored energy is then used to deliver the power to the readout ASICs in the vertex detector half-ladder.
On-detector low-dropout (LDO) regulators provide the necessary stability of the input voltage for the analogue and digital parts of the readout ASICs. 
The voltage of the capacitors is sensed at the back-end and used to regulate the magnitude of the storage-capacitor currents.
The local energy storage concept results in a largely reduced current from the current source to the vertex-detector region and across the ladder, which reduces the needed cable mass accordingly. 
Separate current sources, cables, storage capacitors and LDOs are used for the analogue and digital parts, such that both parts can be optimised independently. 
 
\begin{figure}[ht]
  \centering
  \includegraphics[width=\textwidth]{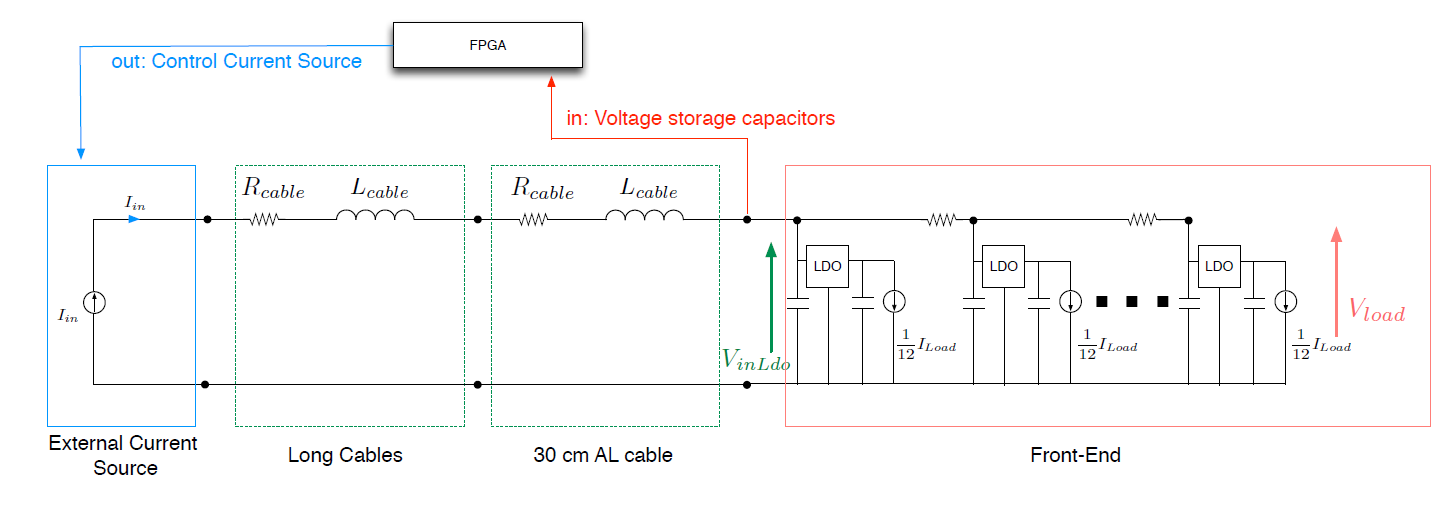}
  \caption{Diagram of a controlled current source powering a half-ladder~\cite{CLICdp-Note-2015-004}.}
  \label{fig:PowerSchemaVertex}
\end{figure}

\subsubsection{Experimental verification}
An experimental setup implementing the proposed circuit for the analogue and digital power pulsing was developed. Silicon capacitors, voltage regulators and dummy loads representing the 12 ASICs of a half ladder are implemented on a printed circuit board, as shown in \cref{fig:TestSetup} for the analogue power-pulsing setup. The board is connected through low-mass aluminium flex cables and copper back-end cables to the controlled current source setup consisting of laboratory power supplies, an FPGA control board and a custom made controlled current-source board.

\begin{figure}[t]
  \centering
  \begin{tikzpicture}
   \node[anchor=south west,inner sep=0] (image) at (0,0){\includegraphics[width=0.8\textwidth]{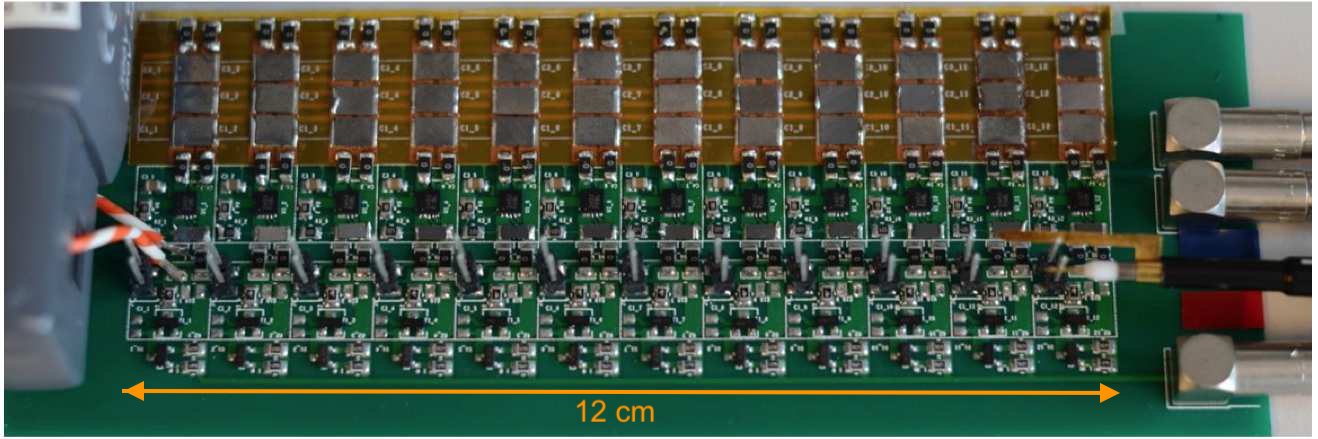}};
     \begin{scope}[x={(image.south east)},y={(image.north west)}]
	\draw[<-,thick, yellow] (0.275,0.8)--(0.325,0.8)node[pos=1, above, anchor=west]{\textcolor{yellow}{\bfseries Silicon capacitors}};
	\draw[<-,thick, yellow] (0.275,0.55)--(0.325,0.55)node[pos=1, above, anchor=west]{\textcolor{yellow}{\bfseries Low-Dropout Regulators}};
	\draw[<-,thick, yellow] (0.275,0.25)--(0.325,0.25)node[pos=1, above, anchor=west]{\textcolor{yellow}{\bfseries Loads}};
     \end{scope}
  \end{tikzpicture}
  \caption{
Printed circuit board of a dummy half ladder for analogue power-pulsing, equipped with storage capacitors, ISL80112 LDOs and dummy loads representing 12 readout ASICs~\cite{CLICdp-Note-2015-004}.}
  \label{fig:TestSetup}
\end{figure}
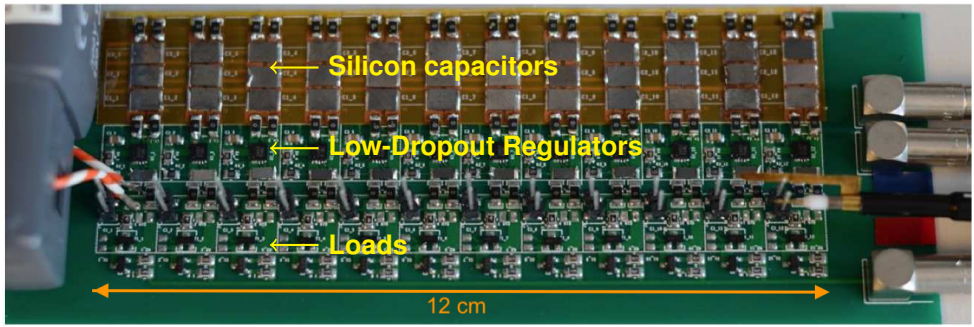

\begin{figure}[t]
  \centering
    \begin{subfigure}[T]{.49\linewidth}
      \includegraphics[width=\linewidth]{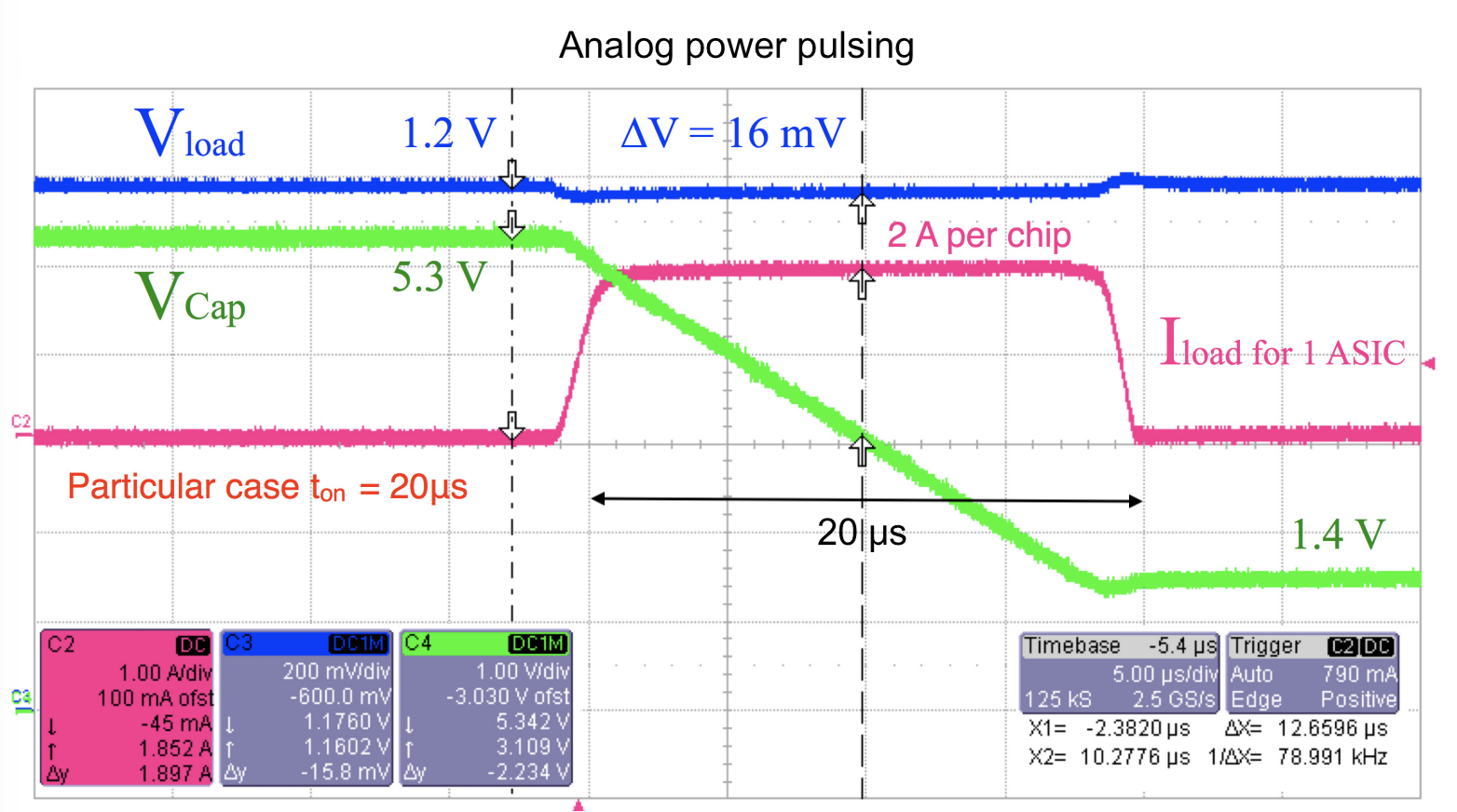}
      \caption{}\label{fig:an-pp-results}
    \end{subfigure}
    \hfill
    \begin{subfigure}[T]{.49\linewidth}
      \includegraphics[width=\linewidth]{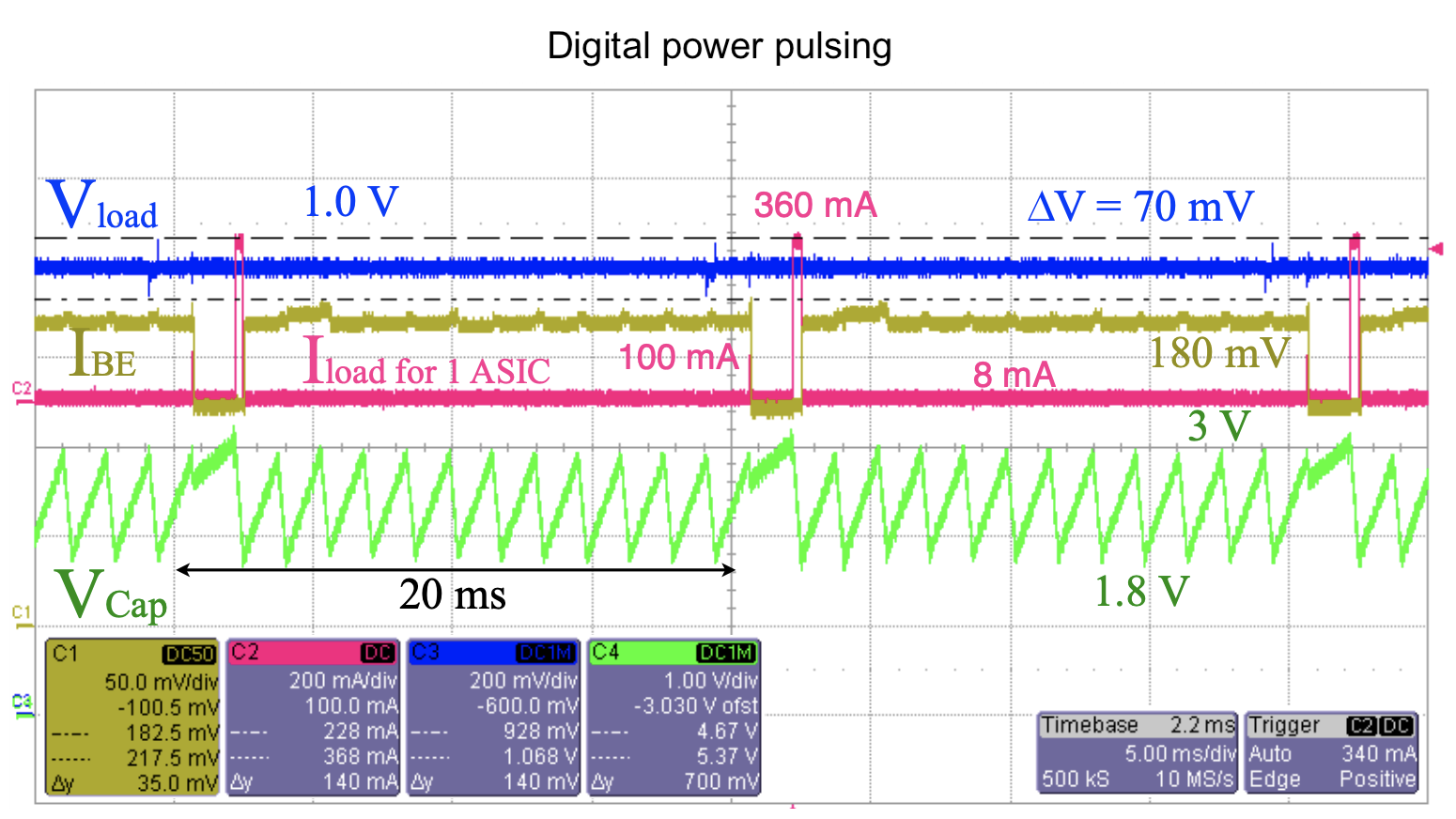}
      \caption{}\label{fig:dig-pp-results}
    \end{subfigure}
  \caption{Measurement results for \subref{fig:an-pp-results} analogue and \subref{fig:dig-pp-results} digital power pulsing~\cite{CLICdp-Note-2015-004}. Shown are the storage capacitor voltage $V_\text{cap}$, the load voltage $V_\text{load}$ and the load current $I_\text{load}$ for the dummy load located furthest away from the current source. For the digital power pulsing also the charge current from the back-end $I_\text{BE}$ is shown.}
  \label{fig:PowerPulsingResults}
\end{figure}

Measurement results of an analogue and a digital power-pulsing sequence are presented in \cref{fig:PowerPulsingResults}.
The voltage of the storage capacitors for the analogue power pulsing decreases during the 20\,$\upmu$s of 2\,A peak current consumption from 5.3\,V to 1.4\,V, before the capacitor gets recharged during the 20\,ms gap before the next on-time. The output voltage for the analogue part is set to approximately 1.2\,V. The voltage drop during the entire on-time is less than $\Delta V \approx 20$\,mV. This level of voltage stability was achieved using capacitors at the input and output of the LDOs of 10\,$\upmu$F and 1\,$\upmu$F, respectively. 

For the digital power pulsing, three load current states are visible, corresponding to the peak current during data acquisition (100 mA), the current during readout (360 mA) and the idle current (8 mA). One readout period of 300\,$\upmu$s is emulated for each of the 12 dummy loads, equally spread over the 20\,ms between consecutive bunch trains. All 12 input storage capacitors of 6.6\,$\upmu$F each are connected in parallel and hence the voltage at each storage capacitor is affected by the load conditions in the entire half ladder. This leads to the observed ramps of each storage capacitor voltage between 1.8\,V and 3\,V. 
The voltage at the LDO output is set to approximately 1.0\,V and variations of this voltage of up to $\Delta V \approx 70$\,mV are observed during load transition periods. A capacitor of 1\,$\upmu$F is placed at the LDO output to reach this level of voltage stability. 

The measured average power dissipation is smaller than $10$\,mW/cm$^2$ for the analogue part and smaller than $35$\,mW/cm$^2$ for the digital part.
The combined power dissipation stays therefore below the anticipated vertex detector limit of $50$\,mW/cm$^2$.
It should be noted that the observed power dissipation is dominated by the dissipation in the LDOs due to the large necessary over-voltages in the storage capacitors for this implementation, which could be further optimised in future implementations of the system, once silicon capacitors with larger energy density become available.

The averaged contributions of the aluminium cable, the LDOs and the capacitors to the material budget in the vertex detector ladder corresponds to \SI{0.1}{\percent{}X_0}. Approximately \SI{80}{\percent{}} of this material is in the silicon capacitors and the rest is shared equally between the cables and LDOs. 
The use of thinner conductors and future improvements of the silicon-capacitor technology are expected to decrease the material budget contribution to the required \SI{0.05}{\percent{}X_0}.

The power-pulsing test setup was operated in a magnetic field of up to \SI{1.5}{\tesla}, showing no degradation of the electrical performance and no measurable mechanical effects.

\subsection{Lightweight support structures}\label{sec:support_structures}
The low overall material budget for the inner detector region results in target values of about \SI{0.05}{\percent{}X_0} for the contribution from the stave supports of each vertex-detector double layer and of less than \SI{1}{\percent{}X_0} for the contributions from supports and cooling infrastructure per single tracker layer. The stiffness and the vibration modes have to be compatible with air-flow cooling in the vertex-detector region and with the large lever arms for the suspension systems in the tracker region. To reach these ambitious goals, new lightweight support structures have been designed, simulated, produced and experimentally evaluated. A variety of Carbon-Fibre-Reinforced-Polymer (CFRP) support structures and a vertex-detector barrel ladder dummy with realistic material composition have been designed, simulated, built and thermo-mechanically characterised. 

\subsubsection{Vertex detector support structures}
Test samples of low-mass supports suitable for the vertex-detector barrel layers based on CFRP have been produced at different manufacturers and at the CERN EP Composite Lab. While earlier studies had evaluated also cross-braced designs~\cite{stave-note-2014}, all recent productions have focussed on full-sandwich staves~\cite{stave-note-2016}. \cref{fig:vtx-stave-support-structures} shows two examples of sandwich staves with very thin CFRP prepreg skins (\SIrange{30}{50}{\micron}) and either Rohacell or Nomex honeycomb cores.
The produced samples have a surface area of 280~mm~$\times$~26~mm, a thickness of approximately 2~mm and a weight between 1.5~g and 2.1~g.
The material budget for each stave was estimated based on calculations and on a photon attenuation source measurement. Values of approximately 0.06\%$X_0$ were obtained, close to the target of 0.05\%$X_0$. 

The flatness on both sides of the staves was measured using a laser sensor and a reference granite surface. Values of approximately \SI{0.6}{\mm} were obtained for a first batch of samples. For a second batch produced with an improved manufacturing procedure, a flatness of approximately \SI{0.1}{\mm} was achieved.
Flexural-stiffness measurements have been performed using a three-point bending test setup (\cref{fig:vtx-stave-support-3point-bending}) following the ASTM D790-02 standard. Stiffness values of \SIrange{0.2}{1.4}{\newton\per\mm} were obtained, in agreement with calculations and ANSYS finite-element simulations (\cref{fig:vtx-stave-support-ansys-simulation}). Structures based on a Rohacell core were found to have higher stiffness and lower deformation than those made with a Nomex honeycomb core.

\begin{figure}[ht]
  \centering
  \begin{subfigure}[T]{.44\linewidth}
    \includegraphics[width=\linewidth]{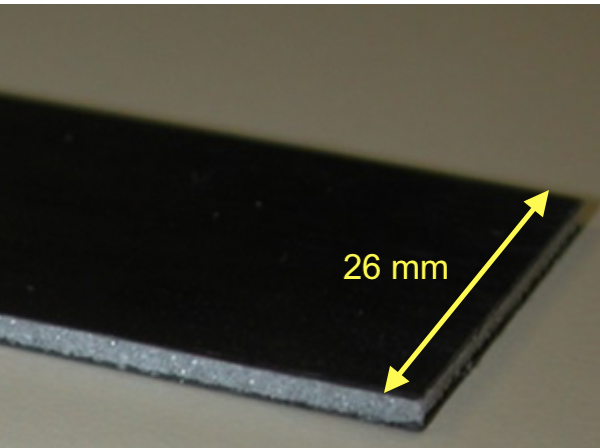}
    \caption{}\label{fig:stave-rohacell}
  \end{subfigure}
  \hfill
  \begin{subfigure}[T]{.42\linewidth}
    \includegraphics[width=\linewidth]{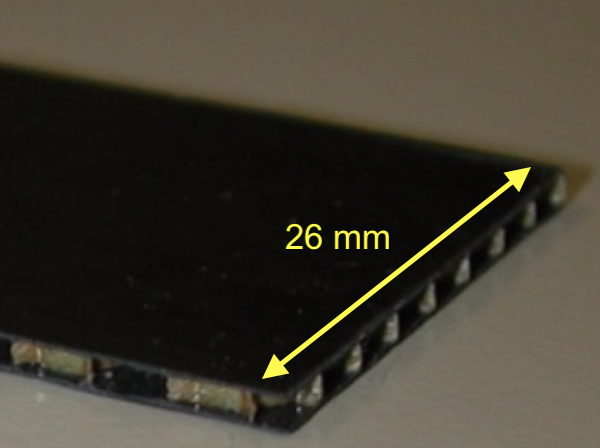}
    \caption{}\label{fig:stave-nomex-honeycomb}
  \end{subfigure}
  \caption{Example CFRP stave supports made from thin CFRP prepregs and a core of \subref{fig:stave-rohacell} Rohacell or \subref{fig:stave-nomex-honeycomb} a Nomex honeycomb structure~\cite{stave-note-2016}.}\label{fig:vtx-stave-support-structures}
\end{figure}
\begin{figure}[ht]
  \centering
    \begin{subfigure}[T]{.49\linewidth}
      \includegraphics[width=\linewidth]{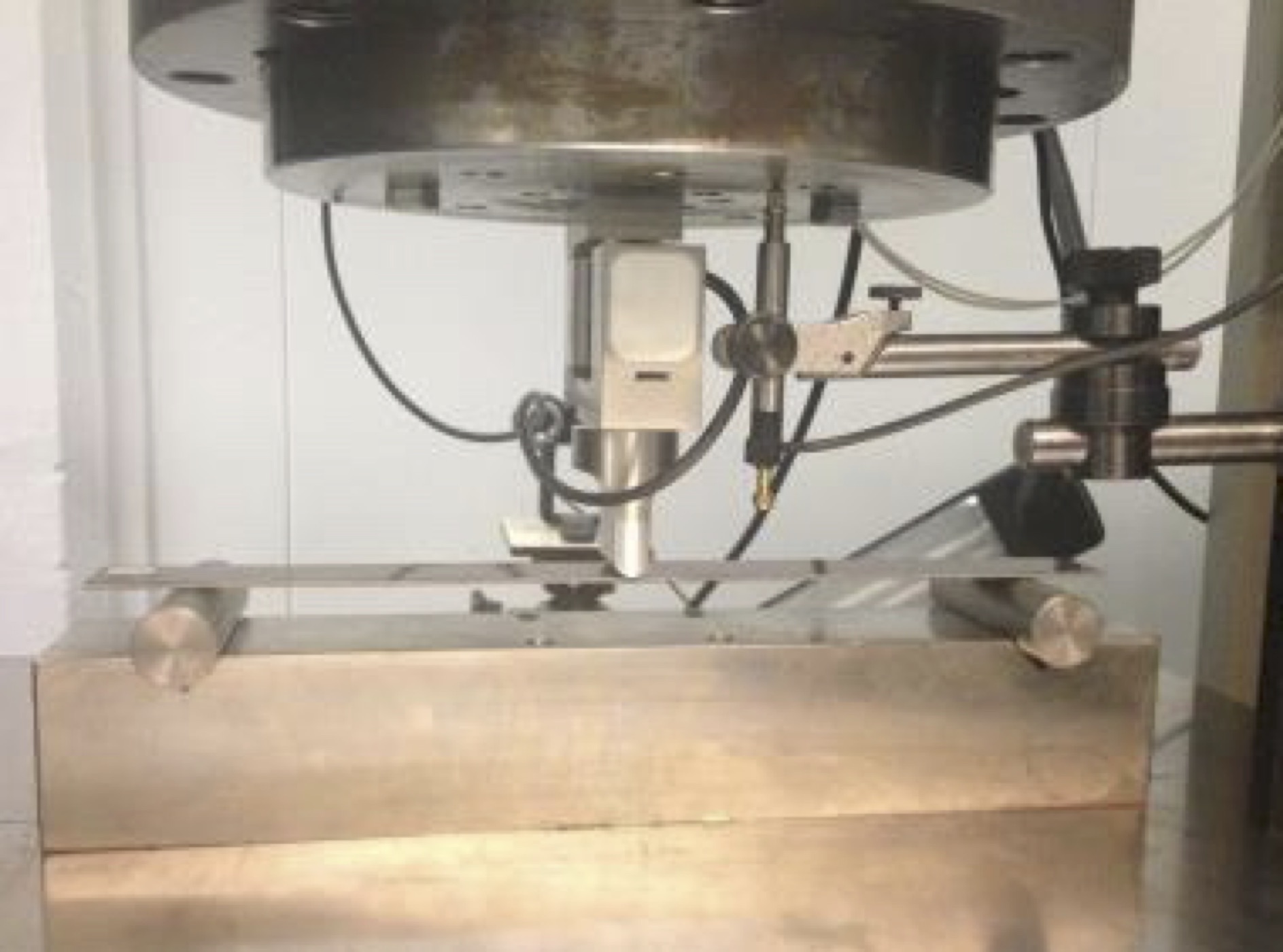}
      \caption{}\label{fig:vtx-stave-support-3point-bending}
    \end{subfigure}
    \hfill
    \begin{subfigure}[T]{.47\linewidth}
      \includegraphics[width=\linewidth]{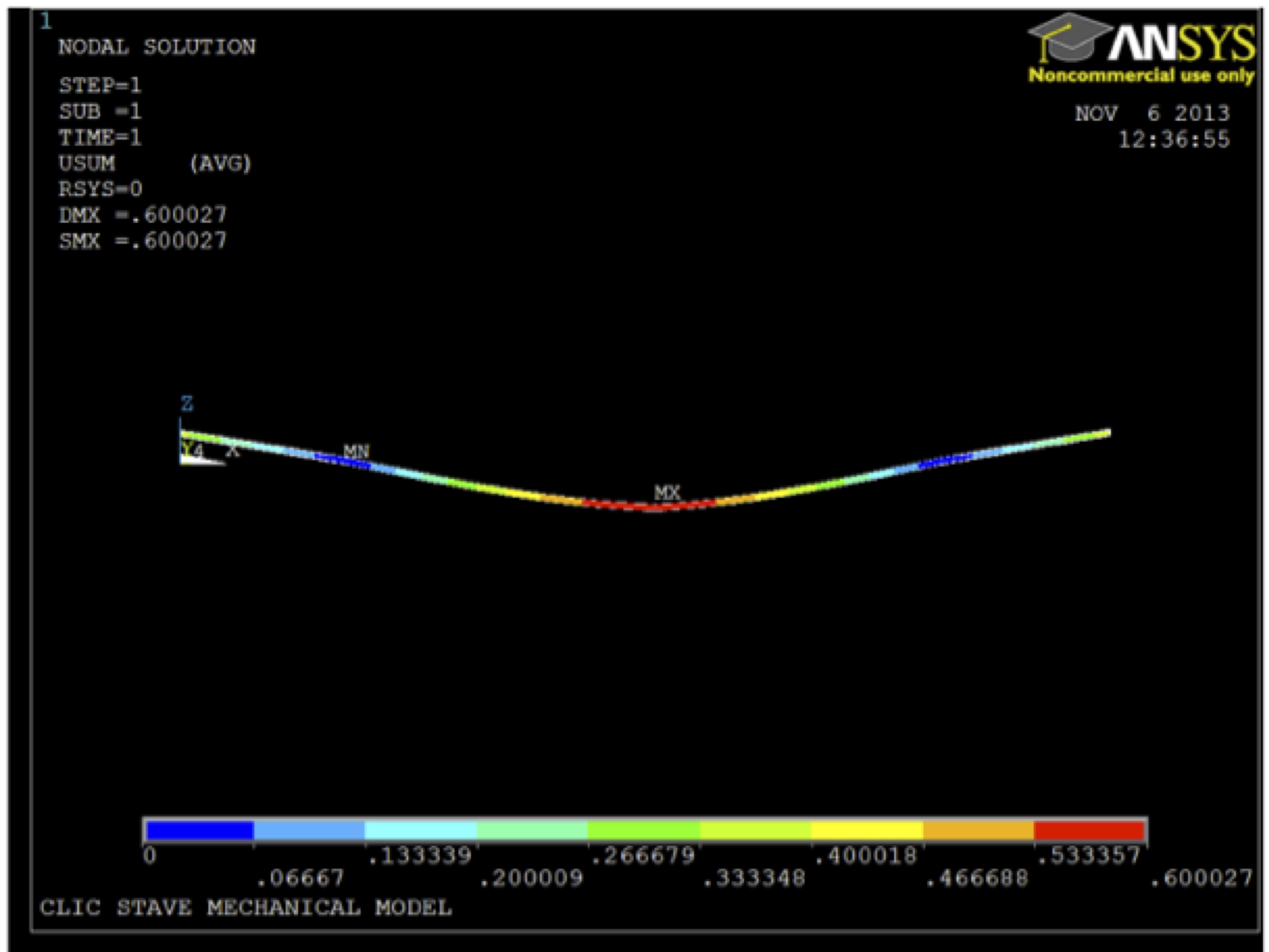}
      \caption{}\label{fig:vtx-stave-support-ansys-simulation}
    \end{subfigure}
  \caption{\subref{fig:vtx-stave-support-3point-bending} Three-point bending test setup for flexural stiffness measurements and \subref{fig:vtx-stave-support-ansys-simulation} corresponding ANSYS finite-element simulation.}
  \label{fig:stave-bending}
\end{figure}

The air-flow cooling setup described in \cref{sec:cooling} was used to measure the vibrations induced on stave supports, for fixation mechanisms of different rigidity and for different air velocities and angles of incidence~\cite{DuarteRamos:2138963}. In all cases, the out-of-plane vibrations were found to be at an acceptable level of less than 2.5~\micron RMS amplitude at nominal flow conditions (5~m/s) and for the worst case of perpendicular incidence of the air flow on the stave. Example measurement results for a bare Rohacell stave with a material budget of approximately 0.06\%$X_0$ are displayed in \cref{fig:vtx-stave-support-vibration-measurements}.  

\begin{figure}[t]
  \centering
  \begin{subfigure}[T]{.49\linewidth}
    \centering
    \begin{tikzpicture}
  \begin{axis}[
    /pgf/number format/1000 sep={},
    axis line style={black,thick},
    ticklabel style={black,font={\sansmath\sffamily\large}},
    xlabel= Air velocity / \si{\meter\per\second},
    ylabel = RMS amp. @ center / \si{\mm},
    ylabel near ticks,
    xlabel near ticks,
		label style={black,font={\sansmath\sffamily\Large}},
    xlabel style={at={(rel axis cs:1,0)}, anchor=north east,yshift=-.4cm},
    ylabel style={at={(rel axis cs:0,1)}, anchor=south east,yshift=.85cm},
    width=7.5cm,
    height=6.5cm,
    ymin=0,
    minor y tick num={3},
    minor x tick num={4},
    every tick/.style={black,thick},
    every axis/.append style={font=\scriptsize},
    legend style={draw=none,text=black, fill=none},
    legend style={text=black},
    legend style={at={(0.05,0.95)},anchor=north west,nodes=right,font={\sansmath\sffamily\footnotesize}},
    ]

    \addplot[mark=o,thick,color=black] plot[error bars/.cd, y dir = both, y explicit] table[x=x,y=0deg]{sections/VertexTracking/figures/cooling/vtx-stave-support-vibration-measurements-vs-velocity.csv}; 
    \addlegendentry{0deg};

    \addplot[mark=square,thick,color=red!75!white] plot[error bars/.cd, y dir = both, y explicit] table[x=x,y=45deg]{sections/VertexTracking/figures/cooling/vtx-stave-support-vibration-measurements-vs-velocity.csv};
    \addlegendentry{45deg};

    \addplot[mark=triangle,thick,color=blue!75!white] plot[error bars/.cd, y dir = both, y explicit] table[x=x,y=90deg]{sections/VertexTracking/figures/cooling/vtx-stave-support-vibration-measurements-vs-velocity.csv};
    \addlegendentry{90deg};

    \draw[black,dashed] ({axis cs:5,0}|-{rel axis cs:0,1}) -- ({axis cs:5,0.003}|-{rel axis cs:0,0});
    \node[anchor=north ] at (rel axis cs:0.6,0.9) {CLICdp};
  \end{axis}
\end{tikzpicture}
    \caption{}\label{fig:vtx-stave-support-vibration-measurement-vs-velocity}
  \end{subfigure}
  \begin{subfigure}[T]{.49\linewidth}
    \centering
    \begin{tikzpicture}
  \begin{axis}[
    /pgf/number format/1000 sep={},
    axis line style={black,thick},
    ticklabel style={black,font={\sansmath\sffamily\large}},
    xlabel= Position along stave / \si{\mm},
    ylabel = RMS amplitude / \si{\mm},
    ylabel near ticks,
    xlabel near ticks,
		label style={black,font={\sansmath\sffamily\Large}},
    xlabel style={at={(rel axis cs:1,0)}, anchor=north east,yshift=-.4cm},
    ylabel style={at={(rel axis cs:0,1)}, anchor=south east,yshift=.85cm},
    width=7.5cm,
    height=6.5cm,
    xmin=-125,
    xmax=125,
    ymin=0,
    ymax=0.00175,
    minor y tick num={3},
    minor x tick num={4},
    every tick/.style={black,thick},
    every axis/.append style={font=\scriptsize},
    legend style={draw=none,text=black, fill=none},
    legend style={text=black},
    legend style={at={(0.95,0.95)},anchor=north east,nodes=right,font={\sansmath\sffamily\footnotesize}},
    ]

    \addplot[mark=o,thick,color=black] plot[error bars/.cd, y dir = both, y explicit] table[x=x,y=0deg]{sections/VertexTracking/figures/cooling/vtx-stave-support-vibration-measurements-vs-position.csv}; 
    \addlegendentry{0deg};

    \addplot[mark=square,thick,color=red!75!white] plot[error bars/.cd, y dir = both, y explicit] table[x=x,y=45deg]{sections/VertexTracking/figures/cooling/vtx-stave-support-vibration-measurements-vs-position.csv};
    \addlegendentry{45deg};

    \addplot[mark=triangle,thick,color=blue!75!white] plot[error bars/.cd, y dir = both, y explicit] table[x=x,y=90deg]{sections/VertexTracking/figures/cooling/vtx-stave-support-vibration-measurements-vs-position.csv};
    \addlegendentry{90deg};

    \node at (rel axis cs:0.25,0.8) {\large \textcolor{black}{\sansmath\sffamily\SI{5}{\meter\per\second}}};
    \node[anchor=north west] at (rel axis cs:0.05,0.95) {CLICdp};
  \end{axis}
\end{tikzpicture}
    \caption{}\label{fig:vtx-stave-support-vibration-measurement-vs-position}
  \end{subfigure}
  \caption{Measurements of the vibration amplitude for a bare CFRP Rohacell stave support \subref{fig:vtx-stave-support-vibration-measurement-vs-velocity} for different angles of incidence as a function of the air velocity and \subref{fig:vtx-stave-support-vibration-measurement-vs-position} as a function of the position along the stave.}\label{fig:vtx-stave-support-vibration-measurements}
\end{figure}

 \subsubsection{Thermo-mechanical vertex-detector stave dummy}\label{sec:thermo-mech-dummy}
 A low-material stave design for the double detector layers in the vertex barrel part has been developed and two dummy staves with silicon dummy layers and a realistic support have been assembled, in order to evaluate the thermo-mechanical characteristics of the design.
 
\cref{fig:thermal_stave} illustrates the layer stack of the demonstrator stave. The structure is based on a carbon-fibre Rohacell sandwich structure, which gives mechanical stability to the stave and provides the \SI{2}{\mm} separation between the two active layers. Ten silicon dummy sensors and a flex circuit are glued with adhesive tape to either side of the support. The dummy sensors, produced at MPG-HLL~\cite{mpg-hll}, are made of an aluminium layer on top of \SI{100}{\micron} non-processed silicon, which mimics the \SI{50}{\micron} active sensor and \SI{50}{\micron} readout ASIC. The aluminium traces allow for resistive heating of the dummy sensor, producing a realistic heat load on to the structure. The flex-PCB on top contains 24 digital temperature probes and the necessary traces for heating of the dummy sensors and for the temperature readout. \cref{fig:dummy_stave} depicts one of the staves during the assembly sequence. On the left side, the flex circuit is already glued to the structure, while on the right half, the silicon sensor tiles are still visible.
The total stave size is \SI{280x26}{\mm}. Besides gaining experience on the assembly procedure of such a structure, the assembled staves have been integrated into the detector mock-up for cooling tests, as outlined in more detail in \cref{sec:cooling}.

\begin{figure}[t]
  \includegraphics[width=\linewidth]{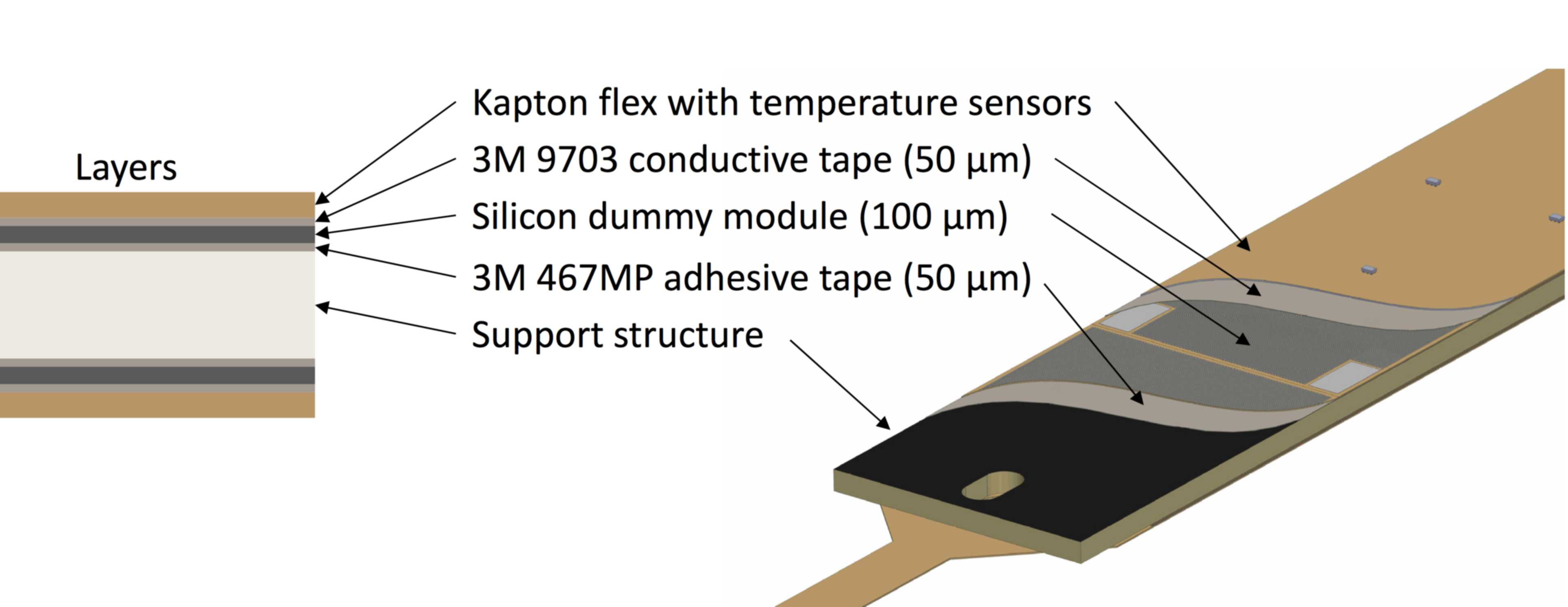}
  \caption{Layer stack-up of a double-sided thermal dummy stave for the vertex barrel detector.}\label{fig:thermal_stave}
\end{figure}
\begin{figure}
\begin{tikzpicture}
\node[anchor=south west,inner sep=0] at (0,0)(image){\includegraphics[width=\linewidth]{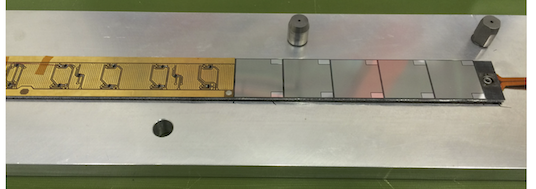}};
\begin{scope}[x={(image.south east)},y={(image.north west)}]
\draw[<->,white,ultra thick](0.23,0.45)--(0.25,0.7)node[pos=0.625,anchor=east]{\SI{26}{\mm}};
\draw[<->,white,ultra thick](0,0.4)--(0.95,0.4)node[pos=0.5,below]{\SI{280}{\mm}};
\end{scope}
\end{tikzpicture}
\caption{Vertex barrel stave (thermal dummy).}\label{fig:dummy_stave}
\end{figure}

\subsubsection{Support structures for the tracker outer barrel}
\cref{fig:tracker_barrel_figure} illustrates the support structure concept for the outer tracker, with one large barrel structure hosting the outer two layers, and two wheel structures on each side of the detector, hosting the four endcap layers per detector side.
The structure is based on a 3D space frame concept. Specially designed and optimised light-weight, custom made node pieces connect commercial off-the-shelf carbon fibre tubes to form the frame. The estimated material content of the inner and outer barrel support structures ranges from \SI{0.13}{\percent X_0} to \SI{0.37}{\percent X_0} per detection layer. A subsection corresponding to 1/6 of one outer barrel ring section has been built at 1:1 scale, as shown in \cref{fig:tracker_barrel_mockup}. The subsection is over \SI{1}{\meter} in size, and the total weight is \SI{926}{\g}. After assembly, the precision of the final structure has been evaluated by means of mechanical and photogrammetric metrology. Overall, the deviation of the node points from their ideal position was below \SI{300}{\micron}, with a standard deviation of less than \SI{150}{\micron}. Static load tests were used to validate the FEA simulation model. With a \SI{50}{\kg} test load placed on the structure, the deformation has been below \SI{500}{\micron}. This confirms the simulation results of the mechanical stability of the design. The estimated sag of the whole structure for the outermost tracker layers is \SI{70}{\micron} for an assumed total load of \SI{150}{\kg}.

\begin{figure}
  \begin{subfigure}[T]{.49\linewidth}
    \includegraphics[width=\linewidth]{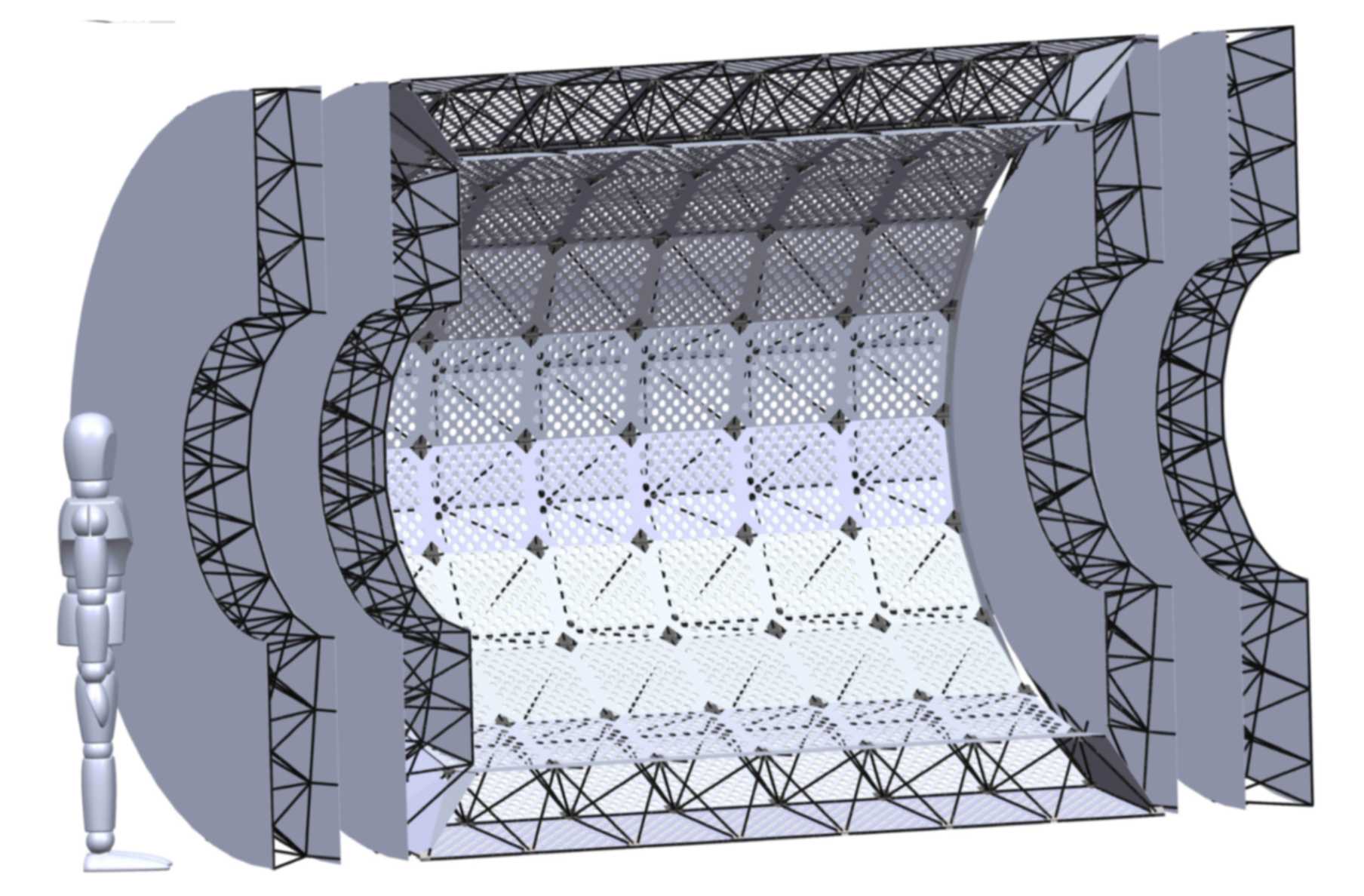}
    \caption{}\label{fig:tracker_barrel_figure}
  \end{subfigure}
  \hfill
  \begin{subfigure}[T]{.49\linewidth}
    \includegraphics[width=\linewidth]{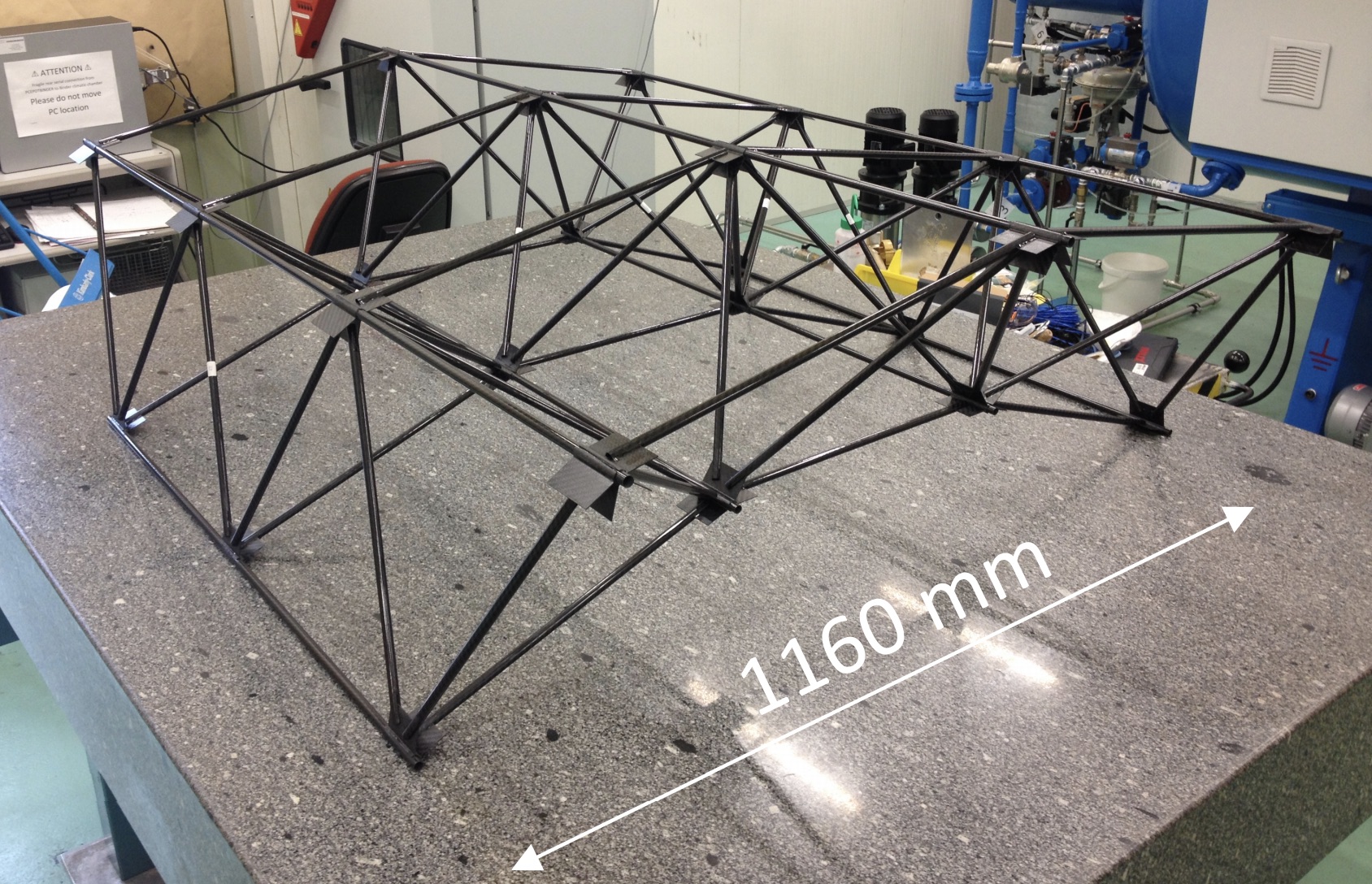}
    \caption{}\label{fig:tracker_barrel_mockup}
  \end{subfigure}
  \caption{\subref{fig:tracker_barrel_figure} Support structure concept for the outer tracker and \subref{fig:tracker_barrel_mockup} full-scale subsection of the tracker barrel support structure.}
\end{figure}

\subsection{Beam pipe design}\label{sec:beam-pipe-design}
A layout for the beam pipe of the CLIC detector at 3~TeV centre-of-mass energy has been developed, consisting of a thin central beryllium cylinder connected to stainless-steel conical sections on either side. While the design was originally worked out for the \clicild detector concept~\cite{VillarejoBermudez:1982810}, small adaptations to the geometry were made~\cite{AlipourTehrani:2254048}, in order to match the inner detector layout of the current CLICdet concept and to maximise the forward acceptance of the detector. The resulting design, shown in \cref{fig:tracker} and \cref{fig:beampipe-engineering-layout-clicdet}, meets the requirements from physics, mechanics and cooling:
\begin{enumerate}
\item To limit the background occupancies in the innermost detector layers to a few percent, the beam pipe needs to be placed outside the cone-shaped region exposed to very high rates of beam-induced background particles.
\item The material content before the first detection layer should stay below 0.2\%$X_0$, to ensure a precise measurement of event vertices. 
\item The beam pipe should provide shielding against back-scattered particles produced in the forward detector region, while maintaining sufficient angular coverage for the forward tracking discs.  
\item A path for the flow of cooling air into the vertex-detector region needs to be provided.
\item The beam pipe has to withstand the loads induced by the vacuum and the loads transmitted from the conical portions to the beryllium cylinder.
\end{enumerate}
To address the first two requirements, the 616~mm long central beryllium part of the beam pipe is foreseen to have an inner radius of 29.4~mm and a wall thickness of 0.6~mm. The third and fourth requirements lead to a double-wall design of the two conical sections with an inner part made of 4.8~mm thick stainless steel and an additional thin ($<1$~mm) CFRP layer around it. The gap of 5~mm between the two walls acts as an air in- and outlet. A spiralling structure made of CFRP is placed in the gap, adding a rotational component to the air flow and thereby maximising the cooling efficiency in the vertex-detector region. The CFRP layer points to the interaction point at an angle of 6.6~degrees, which sets the angular acceptance limit for the forward tracking discs. The last requirement is addressed by a system of support collars and spokes connecting the beam pipe to the main support cylinder for the inner tracking system. The overall length of the beam pipe consisting of the two conical sections and the central cylindrical part is 5.1 m, resulting in a total weight of approximately 100~kg.

\begin{figure}[ht] 
  \centering
  \includegraphics[scale=0.35]{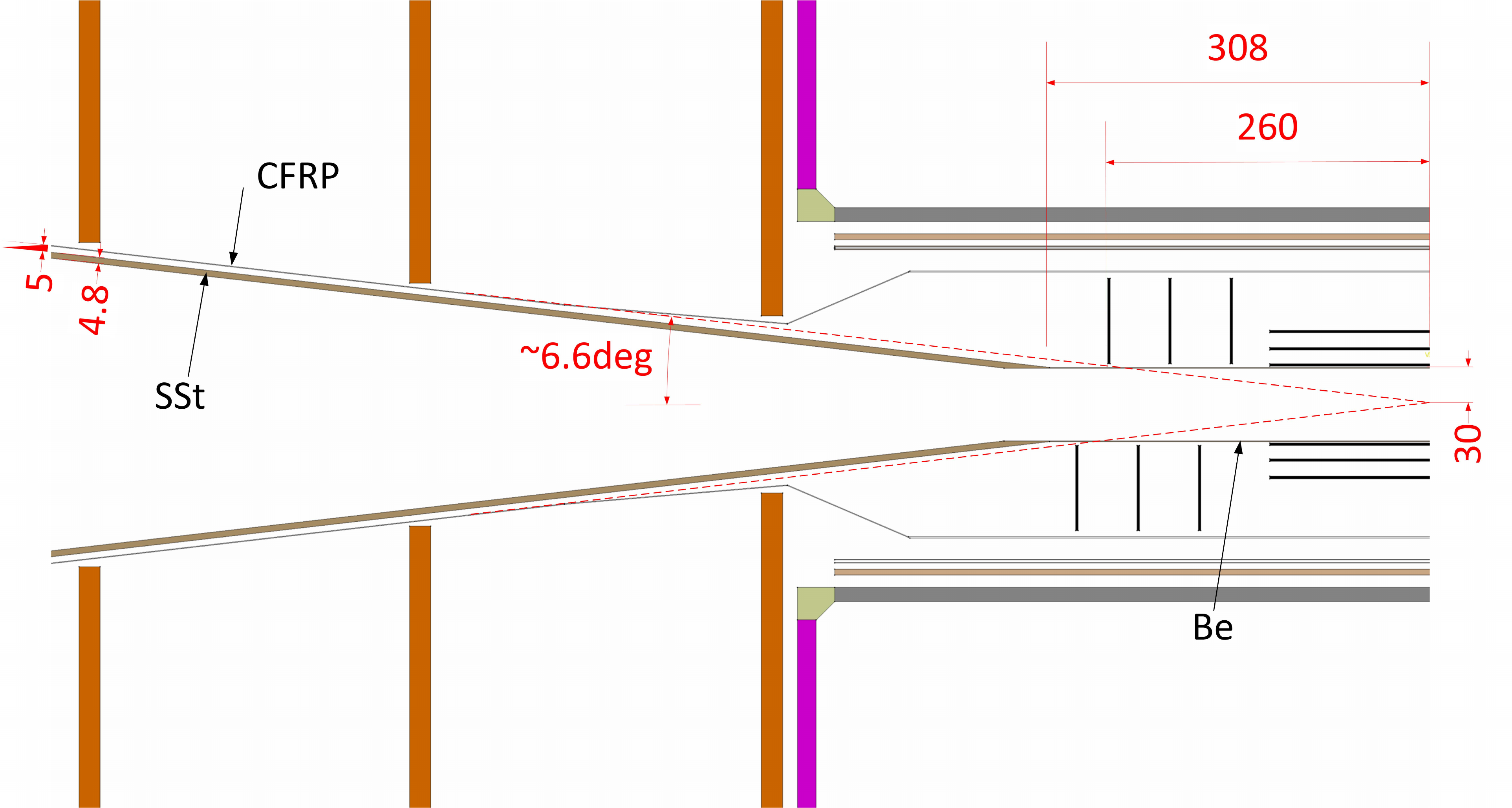}
  \caption{Engineering layout for the innermost region of the detector (from~\cite{AlipourTehrani:2254048}), indicating the cylindrical beryllium (Be) beam-pipe section, the conical beam-pipe section (stainless steel -- SSt), and the outer shell (carbon-fibre-reinforced polymer -- CFRP) used to create an air-channel for the vertex detector cooling. The positions of the vertex barrel and petal layers as well as the innermost tracker barrel layers and discs are indicated. The purple structure is the support of the innermost tracker disk, to which the barrel support structure (shown in grey) is attached. A cylinder keeping the air flow in the close vicinity of the vertex detector is also shown.}\label{fig:beampipe-engineering-layout-clicdet}
\end{figure}

\subsection{Detector assembly}
The CFRP main support cylinder provides the mechanical stiffness and the positioning for all detectors in the inner region as well as the beam pipe, and it directly supports the elements of the outermost inner tracker barrel layer (ITB3). Detailed studies on the assembly procedure of the inner detector have been performed only for the previous detector model CLIC\_ILD~\cite{VillarejoBermudez:1982810}. The overall concept of having a removable support cylinder holding the inner detectors and the beam pipe remains unchanged in the current CLICdet model, but due to the larger radius and the fact that the full inner tracker is supported inside the cylinder, changes to the routing of services are necessary. These changes need to be reflected in the assembly sequence, as well. In the following, the assembly concept developed for CLIC\_ILD is outlined, and changes to adopt to the CLICdet detector model are mentioned where appropriate.

In a first step, both half-shells of the main support cylinder are populated. Starting with ITB3, which is placed directly onto the support cylinder, the inner tracker barrel is assembled and the services are routed along the inner wall of the support cylinder. Subsequently, the support structure for the vertex detector is installed. 
First, the air-flow guiding cylinder is inserted along with the supports of the outermost vertex barrel double layer. The detector layers are then placed onto the support and connected to the services, which are routed along the inner wall of the air-flow cylinder. The procedure is repeated for the remaining two barrel double layers.
The final step is the insertion of each vertex detector petal and its services, starting from the innermost one and going through each petal following their relative placement in the z-direction.
The completed sub-assembly for the CLIC\_ILD concept is shown in \cref{fig:assembly_4}, and a detailed view of the vertex detector region is shown in \cref{fig:assembly_3}. The procedure is repeated for the other half-shell.

For CLIC\_ILD, the services of the vertex detector have been routed along the support cylinder. Due to the increased radius of the support tube in the CLICdet detector layout, the services for the vertex barrel are routed along the conical section of the beam pipe, and thus outside of the tracker acceptance. Eventually, this makes it necessary to insert also the inner endcap discs of the tracking detector at this stage of the assembly procedure, before the vertex detector services can be put in place. 

\begin{figure}[htb]
  
  \begin{subfigure}[T]{.60\linewidth}
    \includegraphics[width=\linewidth,clip,trim=3cm 0 2.5cm 0]{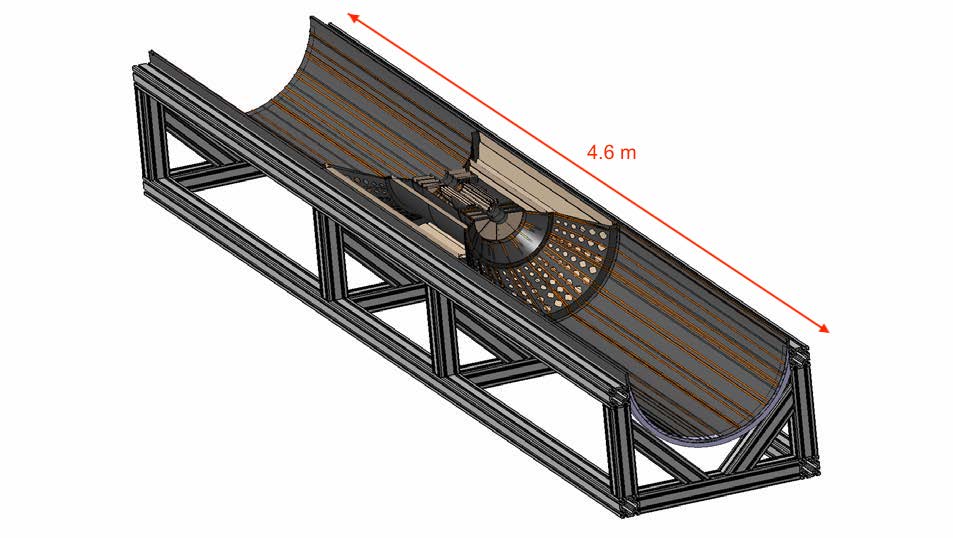}
    \caption{}\label{fig:assembly_3}
  \end{subfigure}
  \hfill
  \begin{subfigure}[T]{.39\linewidth}
    \includegraphics[width=\linewidth,clip,trim=1cm 0 1cm 0]{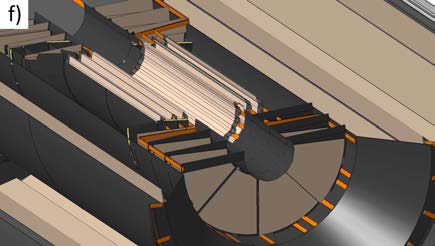}
    \caption{}\label{fig:assembly_4}
  \end{subfigure}
  \caption{Sub-assembly of the vertex detector and inner tracking layers for CLIC\_ILD~\cite{VillarejoBermudez:1982810}: \subref{fig:assembly_3} the complete sub-assembly of the vertex detector and inner tracking layer and \subref{fig:assembly_4} a closeup of the vertex detector barrel layers and petals assembled in the inner detector region.}
\end{figure}

As outlined in \cref{sec:beam-pipe-design}, the beam pipe is composed of a central beryllium cylinder and two stainless steel cones on either side. The beam pipe is supported from the main support cylinder by a system of collars and spokes. The preparation of the beam pipe for the final assembly process is shown in~\cref{fig:assembly_5}.

\begin{figure}[htb]
  \centering
  \includegraphics[width=.66\linewidth]{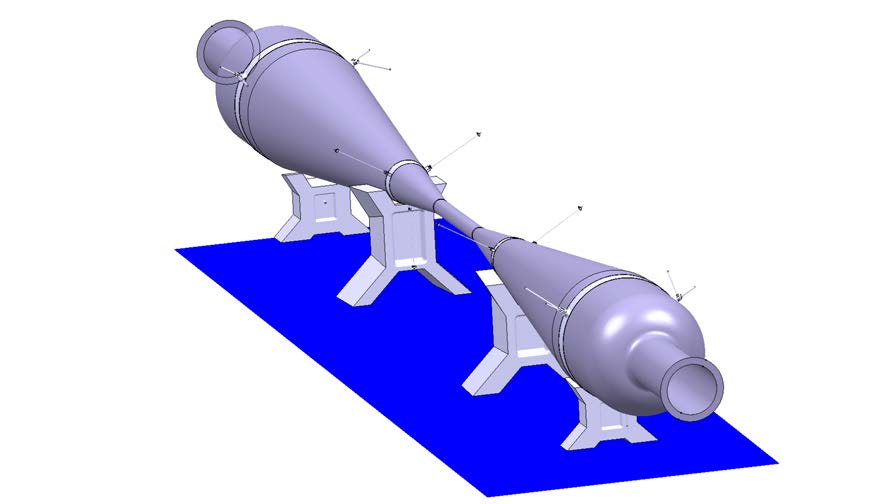}
  \caption{Preparation of the beam pipe for the final assembly of the inner region~\cite{VillarejoBermudez:1982810}. The beam pipe is resting on temporary supports and the collars and their support spokes are added to it.}\label{fig:assembly_5}
\end{figure}

The CFRP layers of the two conical beam-pipe sections will serve as support for the services to the vertex detector and the inner endcap tracking discs. Using a dedicated temporary support, shown in \cref{fig:assembly_6}, two half conical shields are supported from their inner surfaces while the endcap discs are placed starting with the innermost disc closest to the interaction region. Finally, the collars to connect the conical shields to the main support cylinder are attached. The completed subassembly for the CLIC\_ILD model is shown in \cref{fig:assembly_7}.

\begin{figure}[htb]
  \begin{subfigure}[T]{.49\linewidth}
    \begin{tikzpicture}
    \node[anchor=south west,inner sep=0] at (0,0)(image){  \includegraphics[width=\linewidth]{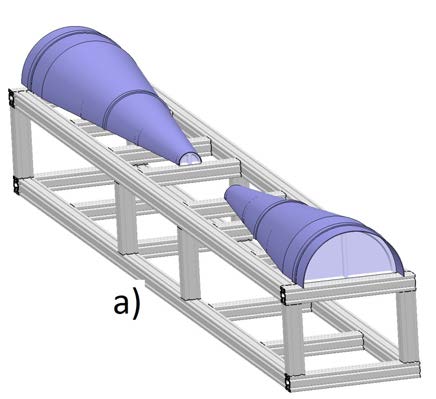}};
    \begin{scope}[x={(image.south east)},y={(image.north west)}]
    \fill[white] (0.22,0.18) rectangle (0.33,0.3);
    \end{scope}
    \end{tikzpicture}
    \caption{}\label{fig:assembly_6}
  \end{subfigure}
  \hfill
  \begin{subfigure}[T]{.49\linewidth}
    \begin{tikzpicture}
    \node[anchor=south west,inner sep=0] at (0,0)(image){  \includegraphics[width=\linewidth]{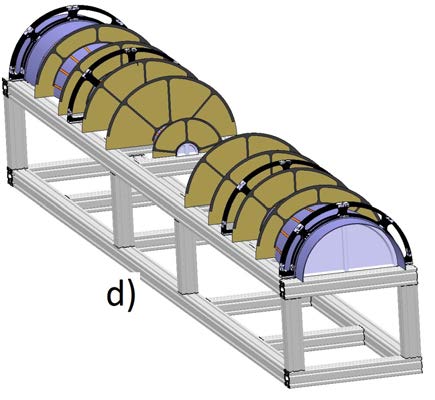}};
    \begin{scope}[x={(image.south east)},y={(image.north west)}]
    \fill[white] (0.24,0.2) rectangle (0.32,0.32);
    \end{scope}
    \end{tikzpicture}
    \caption{}\label{fig:assembly_7}
  \end{subfigure}
  \caption{Sub-assembly of the inner tracking layers and stainless steel conical shields~\cite{VillarejoBermudez:1982810}.}
\end{figure}

With all different sub-assemblies in place, the assembly of the inner detector can proceed by lowering and connecting the sub-assemblies into each other. Once the complete support tube is ready, it can be inserted as one single piece into the detector. 

A similar sequence will be applied to the structural parts of the outer tracker, before they are inserted into the electromagnetic calorimeter.

\subsection{Cables}
Low-mass flexible cables are presently proposed to power the front-end electronics of the vertex and tracking detectors~\cite{VillarejoBermudez:1982810}. Their conductive material is aluminium, which reduces the material contribution by a factor of 4 compared to copper cables with the same total electrical resistance. These cables are composed of 6 layers of different material with a total thickness of \SI{140}{\micron}. Three different materials are used: two aluminium layers to conduct electricity to power the sensors, three kapton layers to provide the desired mechanical properties and electric isolation and one glue layer to glue the intermediate layers. \cref{fig:cabling} shows the different layers and their thickness in micron as well as a photograph of such a cable.

\begin{figure}
  \centering
  \begin{subfigure}[T]{.4\linewidth}
    \includegraphics[width=\linewidth,clip,trim=0 0 9cm 0.5cm]{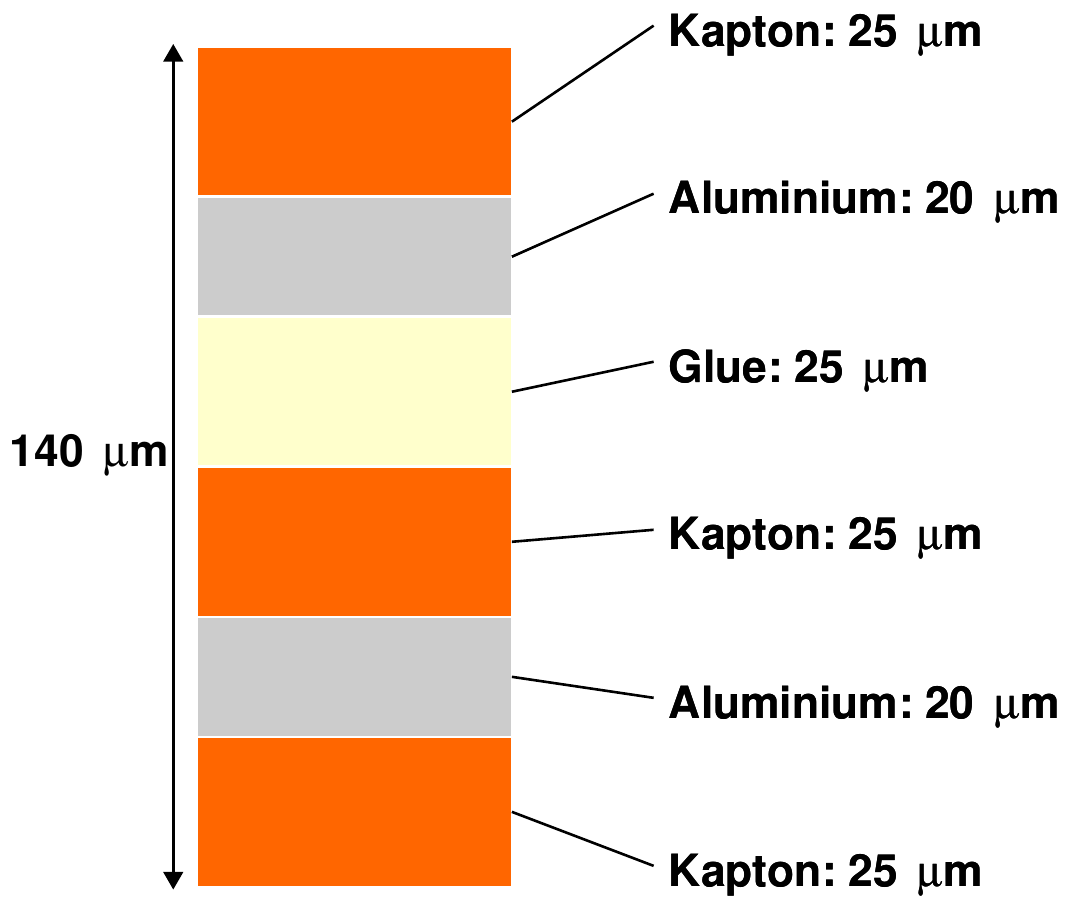}
    \caption{}\label{fig:cabling_layers}
  \end{subfigure}
  \hfill
  \begin{subfigure}[T]{.55\linewidth}
    \includegraphics[width=\linewidth]{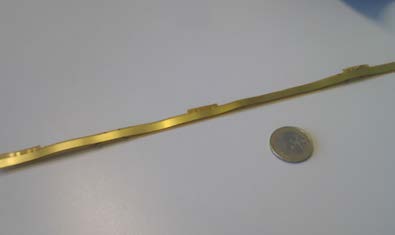}
    \caption{}\label{fig:cabling_photo}
  \end{subfigure}
  \caption{\subref{fig:cabling_layers} Schematic cross section of the proposed front-end power cables, showing the different material layers and their thickness in \si{\micron}. \subref{fig:cabling_photo} Photograph of a \SI{10}{\mm} wide flex cable~\cite{VillarejoBermudez:1982810}.}\label{fig:cabling}
\end{figure}

Cable bending, in particular in the innermost regions of a detector, is an important issue because of the space limitations around the subdetectors and their mechanical support structure. For an initial assessment of the potential issues that might occur when bending layered flat cables, numerical bending simulations have been performed, assessing the stresses on the cables during installation in the detector~\cite{VillarejoBermudez:1700515}. The results show that the proposed cables would withstand the bending and the forces expected during assembly of the detector components. 

\subsection{Cooling}\label{sec:cooling}
Due to the low radiation exposure of less than \SI[per-mode=reciprocal]{e11}{\neq\per\cm\squared\per\year}, the CLIC vertex detector can be operated at room temperature. Still, the dissipated power has to be removed from sensors and readout electronics, to maintain a decent temperature within the detector. The current presumption foresees a temperature below \SI{40}{\celsius} on the active elements. The use of conventional liquid/two-phase cooling solutions in the CLIC vertex detector would result in a significant increase in material budget from both the cooling medium and its tubing, which would exceed the available material budget. Therefore, the use of a dry gas (air or $\text{N}_\text{2}$) as a coolant has been proposed for the inner detector region as a means to achieve the specified material budget~\cite{DuarteRamos:2138963}.

The initial studies on the feasibility of this cooling solution have been performed using Computational Fluid Dynamics (CFD) models and the results are documented in~\cite{ClicVertexCoolingSimulation}. Experimental studies have been conducted in order to validate the results of the CFD simulations.

A bench-top wind tunnel was built as a first step in the validation of the air cooling solution, to better understand how a single vertex detector stave behaves under the action of an air cooling stream. Both thermal studies and vibration test were performed. The CFD thermal simulations reproduced the measured data to within a few degrees. Stave vibrations induced by the air flow were found to be at an acceptable level, as discussed in \cref{sec:support_structures}.

\begin{figure}[ht]
  \begin{subfigure}[T]{.66\linewidth}
 \begin{tikzpicture}
 \node[anchor=south west,inner sep=0] at (0,0)(image){\includegraphics[width=\linewidth,clip,trim=0 0.33cm 0 0.33cm]{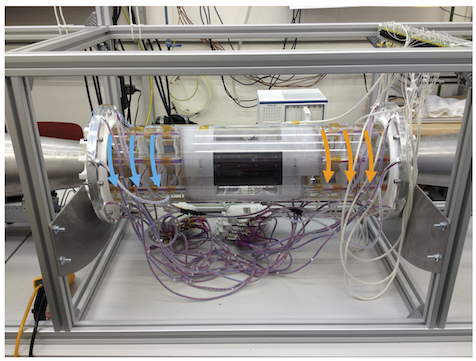}};
 \begin{scope}[x={(image.south east)},y={(image.north west)}]
 \draw[<->,ultra thick,white](0.2,0.2)-- (0.825,0.2)node[pos=0.5,below]{\SI{56}{\cm}};
 \end{scope}
 \end{tikzpicture}
 \caption{}\label{fig:vertex_mockup}
\end{subfigure}
\hfill
 \begin{subfigure}[T]{.33\linewidth}
   \includegraphics[width=\linewidth]{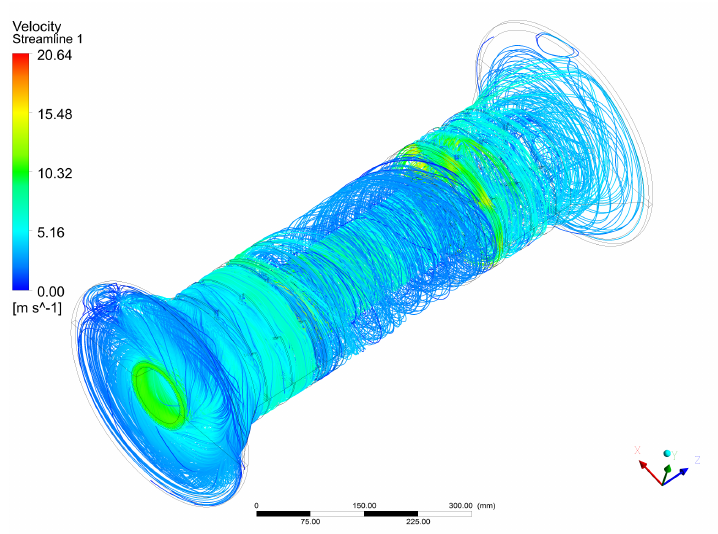}
   \caption{}\label{fig:vertex_airflow_simulation}
 \end{subfigure}
\caption{\subref{fig:vertex_mockup} Photograph of the 1:1 scale mock-up of the vertex detector to study the performance of the air-flow cooling concept. \subref{fig:vertex_airflow_simulation} illustration of a computational fluid dynamics simulation of the air cooling in the CLIC\_ILD vertex-detector region, showing the velocity streamlines of the forced air-flow in the detector volume~\cite{ClicVertexCoolingSimulation}.}
\end{figure}

To have a more realistic test set-up, a 1:1 scale thermal mock-up of the vertex detector was built, as shown in \cref{fig:vertex_mockup}. In a first step, the cooling performance on the barrel staves was studied. To represent the staves in the vertex detector barrel, the mock-up uses double-sided PCBs with a thickness of \SI{1.6}{\mm}. Copper traces embedded on each side of the PCBs were used to apply the same heat load as that generated by the electronics and silicon sensors as expected in a real detector. A total of 176 surface mount PT100 temperature sensors were used to determine the temperature distribution along both sides of each stave. The measurements have shown that at nominal working conditions (average power dissipation of \SI{50}{\milli\watt\per\cm\squared} and a volumetric flow of about \SI{17}{\liter\per\second}, corresponding to an air velocity of \SI{5}{\meter\per\second}), there are significant temperature gradients along the axial and circumferential directions, as illustrated in \cref{fig:cooling_temp_along_stave}.

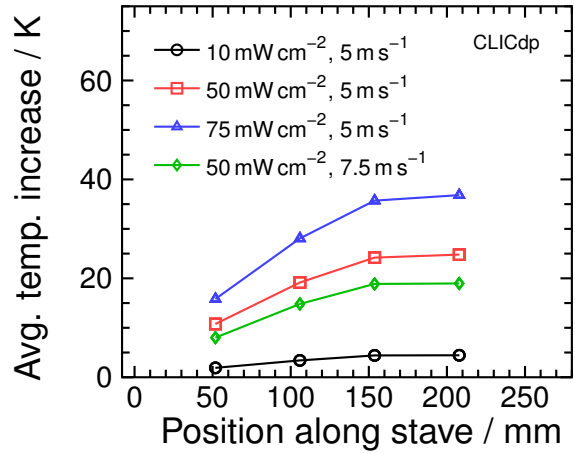
\begin{figure}[ht]
  \centering
  \begin{tikzpicture}
  \begin{axis}[
    /pgf/number format/1000 sep={},
    axis line style={black,thick},
    ticklabel style={black,font={\sansmath\sffamily\large}},
    xlabel= Position along stave / \si{\mm},
    ylabel = Avg. temp. increase / \si{\kelvin},
    ylabel near ticks,
    xlabel near ticks,
		label style={black,font={\sansmath\sffamily\Large}},
    xlabel style={at={(rel axis cs:1,0)}, anchor=north east,yshift=-.4cm},
    ylabel style={at={(rel axis cs:0,1)}, anchor=south east,yshift=.85cm},
    width=7.5cm,
    height=6.5cm,
    xmin=-8,
    xmax=280,
    ymin=0,
    ymax=75,
    minor y tick num={3},
    minor x tick num={4},
    every tick/.style={black,thick},
    every axis/.append style={font=\scriptsize},
    legend style={draw=none,text=black, fill=none},
    legend style={text=black},
    legend style={at={(0.05,0.95)},anchor=north west,nodes=right,font={\sansmath\sffamily\footnotesize}},
    ]

    \addplot[mark=o,thick,color=black] plot[error bars/.cd, y dir = both, y explicit] table[x=x,y=10mW]{sections/VertexTracking/figures/cooling/cooling_temp_along_stave.csv}; 
    \addlegendentry{\SI{10}{\milli\watt\per\cm\squared}, \SI{5}{\meter\per\second}};

    \addplot[mark=square,thick,color=red!75!white] plot[error bars/.cd, y dir = both, y explicit] table[x=x,y=50mW]{sections/VertexTracking/figures/cooling/cooling_temp_along_stave.csv};
    \addlegendentry{\SI{50}{\milli\watt\per\cm\squared}, \SI{5}{\meter\per\second}};

    \addplot[mark=triangle,thick,color=blue!75!white] plot[error bars/.cd, y dir = both, y explicit] table[x=x,y=75mW]{sections/VertexTracking/figures/cooling/cooling_temp_along_stave.csv};
    \addlegendentry{\SI{75}{\milli\watt\per\cm\squared}, \SI{5}{\meter\per\second}};

    \addplot[mark=diamond,thick,color=green!75!black] plot[error bars/.cd, y dir = both, y explicit] table[x=x,y=50mWhighflux]{sections/VertexTracking/figures/cooling/cooling_temp_along_stave.csv};
    \addlegendentry{\SI{50}{\milli\watt\per\cm\squared}, \SI{7.5}{\meter\per\second}};

    \node[anchor=north east] at (rel axis cs:0.95,0.95) {CLICdp};
    
  \end{axis}
\end{tikzpicture}
  \caption{Average temperature increase of the dummy staves with respect to the inlet air temperature as a function of the position along the length of the stave, data from~\cite{DuarteRamos:2138963}.}\label{fig:cooling_temp_along_stave}
\end{figure}

An example result of the computational fluid dynamics simulation of the air cooling in the vertex-detector region is shown in \cref{fig:vertex_airflow_simulation}. The CFD simulation qualitatively reproduces the measured temperature distributions of the middle and outermost barrel layers in the mock-up, and can be used to further optimise the detector geometry. The gradients along the axial direction (up to \SI{35}{\celsius} over the stave length) will be difficult to eliminate without a reduction of the average power dissipation in the staves. Increasing the volumetric flow, as an attempt to reduce these gradients, has been experimentally tested (see \cref{fig:cooling_temp_along_stave}) and has proven to be inefficient~\cite{DuarteRamos:2138963}.

In a later step, the two realistic dummy staves presented in \cref{sec:thermo-mech-dummy} were integrated into the mock-up. One of the PCB-based staves in the middle and one in the outer layer were replaced. The new staves were each equipped with 48 digital temperature sensors, allowing for a better resolution of the temperature gradient along the stave. The realistic stave design had no impact on the cooling performance; similar temperature distributions were achieved.

For the large main tracker (137~m$^2$ surface area), air-flow cooling is not considered to be a viable option owing to the large volume. First conceptual simulation and engineering studies demonstrate the feasibility of liquid cooling, based on e.g. water, in terms of material budget and cooling power for room temperature operation.
Owing to the similarities in the general support structure concept, operation temperature and power dissipation with respect to the upgrade project of the ALICE inner tracking system~\cite{alice_tdr}, no further CLIC tracker specific cooling studies have been performed so far.

\section{Summary and outlook}\label{sec:vtx-trk-summary}
The results presented in this chapter demonstrate that significant progress has been made since the CDR in all areas of the vertex and tracker R\&D towards reaching the challenging detector requirements at CLIC. The CLIC-specific developments largely profit from synergies with other detector development projects and from the rapid progress in the semiconductor industry.

Advanced \textbf{simulation} and \textbf{characterisation tools} have been developed and are used extensively to aid the various design and testing activities.

\textbf{Hybrid pixel detector} assemblies with thin \textbf{planar sensors} (down to \SI{50}{\micron}) have shown excellent detection efficiency ($>99\%$) and timing performance (few nanoseconds resolution). The use of active-edge sensor technology and Through-Silicon Vias (TSV) has enabled detector layouts with minimised inactive areas. Dedicated readout ASICs with small pitch (\SI{25}{\micron}) have been developed in \SI{65}{\nm} process technology and successfully hybridised to planar sensors using intricate single-chip solder bump-bonding techniques. The achievable position resolution in thin sensors was found to be limited by the low amount of charge sharing. Approximately \SI{7}{\micron} were achieved for a pitch of \SI{25}{\micron} and an active thickness of \SI{50}{\micron}. Sensor designs with enhanced lateral drift are under development with the aim to improve the position resolution. Further performance improvements require smaller pixel pitches, which can be achieved in the future using advanced ASIC process technologies (for example \SI{28}{\nm} feature size). Alternative fine-pitch interconnect processes, such as the use of Anisotropic Conductive Film (ACF), could in the future reduce the complexity and cost of the hybridisation process, making the use of hybrid pixel detectors feasible also for the large-area tracker.

\textbf{Capacitively coupled hybrid pixel detectors} with active HV-CMOS sensors have been produced and tested successfully. Non-uniformities observed in the response of the various assemblies were found to be related to the complex hybridisation and calibration procedures. Further refined gluing techniques and a precise characterisation of the signal transfer chain would be needed for maintaining a good performance over larger detector areas.  

Promising results have been obtained from test assemblies implemented in various \textbf{monolithic depleted CMOS pixel sensor} technologies with \SIrange{180}{200}{\nm} feature size. These technologies offer the potential for achieving high performance over large surface areas, with a reduced material budget and at a lower cost than hybrid solutions.
The performance of the fully integrated ATLASpix High-Voltage CMOS test chip with elongated pixels has been found to be close to meeting the requirements of the CLIC tracker. A new version with adapted pixel geometry targeting the CLIC tracker requirements is currently under design. The ALICE Investigator analogue test chip implemented in a High-Resistivity CMOS process with minimised collection electrodes has been tested in view of the CLIC tracker requirements. The excellent results obtained with this test chip in terms of spatial and temporal resolution have led to the design of the fully integrated CLICTD tracker test chip with segmented macro pixels, which is currently in production. Test assemblies with Silicon-On-Insulator technology have demonstrated the potential of this concept for achieving very good spatial resolution. A dedicated demonstrator chip for the CLIC vertex detector, CLIPS, has been designed and is currently in production.

The feasibility of \textbf{power pulsing} with local energy storage optimised for the CLIC duty cycle was demonstrated with a low-mass powering demonstrator setup, including tests in a magnetic field. Power pulsing was also implemented successfully in hybrid ASICs and HV-CMOS sensors for CLIC. Average power consumption levels below \SI{50}{\milli\watt\per\cm\squared} have been achieved.

\textbf{Detector integration} studies have been performed to demonstrate the feasibility of the CLIC vertex and tracking detector concepts in terms of mechanical supports, assembly, cabling and cooling. Demonstrators of \textbf{low-mass support structures} for the vertex and tracker detector have been produced and tested in respect of their mechanical properties. \textbf{Air cooling} for the vertex detector was tested in simulation studies and in air-flow tests using a full-scale CLIC vertex-detector mock-up with realistic heat loads. Further integration studies including larger demonstrator assemblies with functional readout ASICs will be required in the next phase of the project.


\chapter{Electromagnetic and hadronic calorimeters}\label{chap:calorimeters}
This chapter presents the requirements and concepts for the electromagnetic and hadronic calorimeters at CLIC and gives an overview of the R\&D for imaging calorimetry, which is performed within the CALICE collaboration.

\section{Introduction}
The physics at future high-energy lepton colliders, with its requirement for a jet energy reconstruction with unprecedented precision, is one of the primary motivations for the development of highly granular calorimeters. Within the CALICE collaboration~\cite{Adloff:2012dla}, several such concepts for the electromagnetic calorimeter (ECAL) as well as the hadronic calorimeter (HCAL) are being developed. For the CLIC detector a sampling ECAL based on tungsten absorbers and silicon sensors as active material is envisaged, while the HCAL will consist of a steel absorber structure with small scintillator tiles read out individually by Silicon Photomultipliers (SiPMs).

In order to demonstrate the performance of the concepts, in the years 2005 to 2012 the CALICE collaboration built {\em physics prototypes} and exposed them to test beams of various particle types. An overview of the results can be found in~\cite{ref:PfacaloTests}. Since the time of the CLIC CDR~\cite{cdrvol2} several new results have been published, which will be discussed in this chapter. The next step is to demonstrate that the concepts are scalable to a hermetic collider detector, fulfilling realistic requirements on space and power consumption of the readout electronics. Concepts for production and quality control as well as for a cooling system are also mandatory. These points are addressed by {\em technological prototypes}. For the SiPM-on-tile HCAL (\emph{Analogue HCAL}, AHCAL) a large prototype with 38 layers has been built and recently exposed to particle beams, while for the \emph{Silicon ECAL} (SiECAL) up to now a 10-layer prototype has been studied. The completion of the ECAL technological prototype with 30 layers is an important future goal.

\section{Requirements for calorimetry at CLIC}
The CLIC physics programme requires an efficient reconstruction of multi-jet final states and a
jet-energy resolution for light-quark jets of $\sigma_E/E \leq \SI{3.5}{\percent}$ for jet energies in the range \SI{100}{\GeV} to \SI{1}{\TeV} ($\leq \SI{5}{\percent}$ at \SI{50}{\GeV})~\cite{cdrvol2}.
In particular the separation of \PW and \PZ bosons  on an event-by-event basis profits from highly granular (imaging) calorimeters optimised for Particle Flow Analysis (PFA) reconstruction.
A high granularity in the calorimeters is also essential for separating interesting physics events from beam-induced background particles, which lead to large energy deposits in the sensitive volumes of the main calorimeters of up to \SI{12}{\TeV} per bunch train at \SI{3}{\TeV} (see \cref{sec:beam_induced_backgrounds}). Hit-time tagging at the 1-nanosecond level combined with PFA reconstruction techniques are required to efficiently suppress these out-of-time background particles in the reconstructed data. 

In general the use of PFA reconstruction algorithms reduces the influence of the instrinsic calorimeter energy resolution on the jet reconstruction. Nevertheless, the intrinsic energy resolution plays an important role at low jet energies (up to roughly \SI{~100}{\GeV}), for the matching of tracks with calorimeter energy depositions for charged particles, and for the reconstruction of photon final states. In the usual parametrisation of the energy resolution for single particles, $\sigma_{E}/E = a/\sqrt{E(\SI{}{\GeV})} \oplus b$, the \emph{stochastic} term $a$ reflects statistical fluctuations in the shower evolution and measurement, while the \emph{constant} term $b$ arises from imperfections in detector homogeneity, stability and calibration. At higher energies, there are additional contributions from the fluctuations of non-contained energy, or \emph{leakage}.
Photon final states require a good intrinsic resolution in the ECAL of approximately $15\%/\sqrt{E}$, which implies a high sampling fraction, and sufficient containment to limit electromagnetic leakage into the HCAL. An ECAL depth of \SI{22}{X_0} can fulfil this requirement~\cite{AlipourTehrani:2254048}.
A total depth of \SI{7.5}{$\lambda$_I} in the HCAL, in addition to the \SI{1}{$\lambda$_I} in the ECAL, was found to be sufficient to reduce the effect of hadronic leakage on the jet-energy resolution to an acceptable level~\cite{cdrvol2}. The goal for the constant term is \SI{1}{\percent} for the ECAL and \SIrange{2}{3}{\percent} for the HCAL~\cite{ref:PfacaloTests}. In prototype tests very often the constant term cannot be determined precisely for several reasons: small prototypes exhibit leakage, the covered energy range is not wide enough to constrain the fit of the energy resolution, or the data does not follow the simple parametrisation. The constant term can also be affected by limited statistics in calibration samples. In addition, prototypes usually do not have the expected production quality of the components, concerning sample uniformity, stability and sensitivity to environmental conditions.  Therefore, the constant term determined from beam tests of calorimeter prototypes often is larger than what is expected for the full-scale detector.

\section{Calorimeter implementation in CLICdet}
The overall layout of the calorimeters for the CLIC detector is shown in \cref{fig:clicdet}. The ECAL and HCAL barrels are arranged in dodecagons around the tracker volume. The endcap calorimeters are arranged to provide good coverage in the transition region, and maximum coverage to small polar angles. \cref{fig:calo-implementation} shows the implementation of ECAL and HCAL in the current simulation model.

\begin{figure}[ht]
  \centering
  \includegraphics[width=0.6\linewidth]{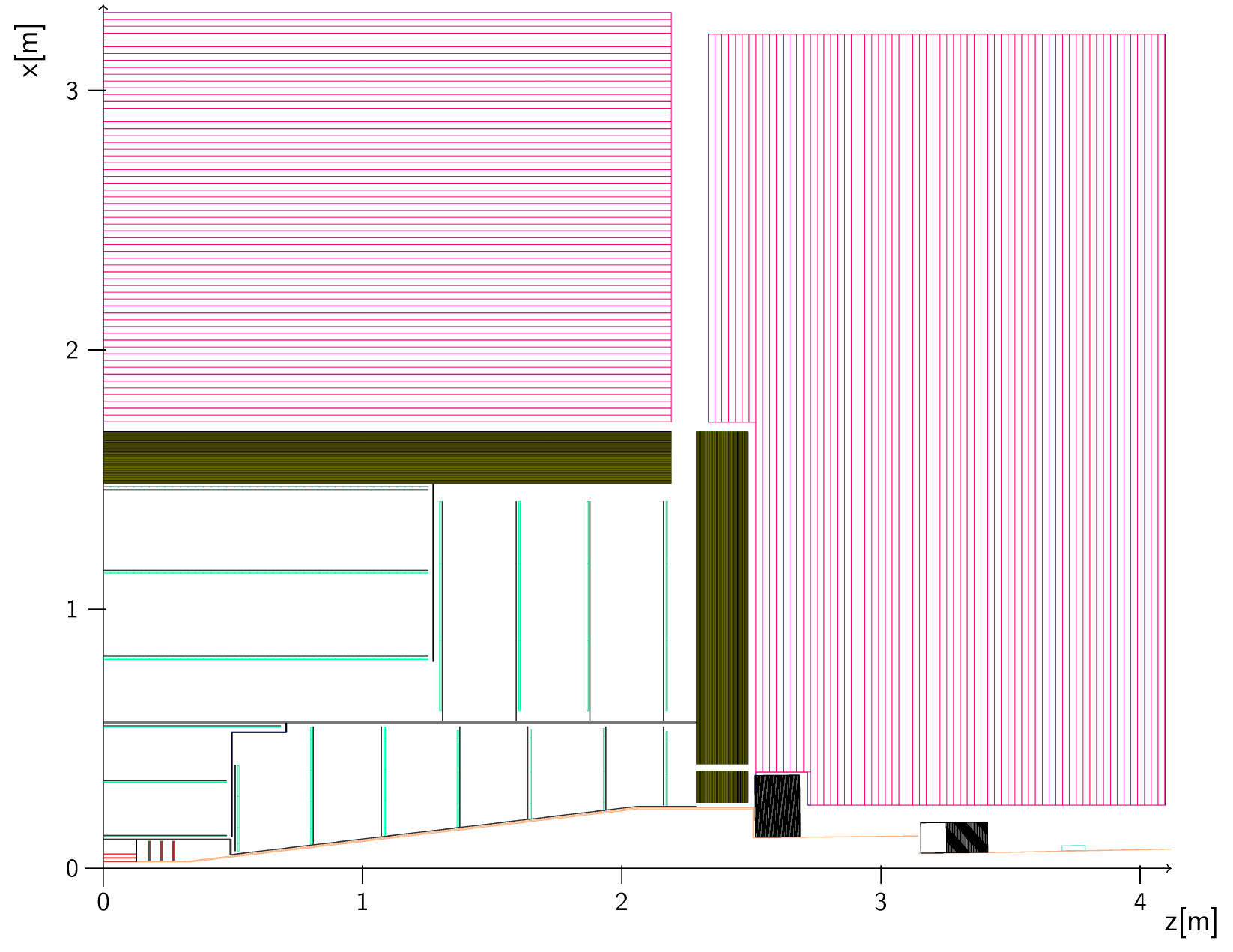}
  \caption{Side view of a quarter of the CLIC detector. The ECAL is shown in dark green; the HCAL in red. In the barrel region, both ECAL and HCAL layers are parallel to the beam, while they are perpendicular to the beam in the endcap. The part of the ECAL closest to the beam pipe is the ECAL plug (see text).}		
  \label{fig:calo-implementation}
\end{figure}

\subsection{ECAL}
The basic ECAL structure of the CLIC detector is a silicon-tungsten sampling calorimeter with \SI[product-units=power]{5x5}{\mm} silicon detector cells. To achieve a good photon energy resolution, in particular at high energy, the ECAL consists of \num{40} layers, with tungsten absorber plates of \SI{1.9}{\mm} thickness and a distance of \SI{3.15}{\mm} between plates~\cite{AlipourTehrani:2254048}. The tungsten plates reduce the size of electromagnetic showers in the calorimeter, due to the small Moli{\`e}re radius (\SI{9}{\mm}) and short radiation length (\SI{3.5}{\mm}) of tungsten. This helps to avoid overlapping of showers produced by nearby particles. The ECAL barrel extends from \SI{150}{\cm} to \SI{170}{\cm} in radius, corresponding to a total thickness of about \SI{22}{X_0} or \SI{1}{$\lambda$_I}. To allow easy access to the detector for maintenance or repairs implies that the ECAL endcap has to be built in two main parts, with a small so-called ECAL plug remaining in place attached to the support structures, while the bulk of the endcap is retracted. The necessary empty space between ECAL endcap and ECAL plug is presently assumed to be shaped as a ring, with a width of \SI{30}{\mm}. The ECAL has in total 101 million readout cells, corresponding to a silicon-detector surface of \SI{2500}{\meter\squared}.

\subsection{HCAL}
The proposed hadronic calorimeter of the CLIC detector consists of \num{60} steel absorber plates, each of them \SI{19}{\mm} thick, interleaved with scintillator tiles, similar to the CALICE calorimeter design for the ILD detector. The gap for the sensitive layers and their cassette is \SI{7.5}{\mm}, including two cassette walls made of \SI{0.5}{\mm} steel each.  The polystyrene scintillator in the cassette is \SI{3}{\mm} thick with a tile size of \SI[product-units=power]{30x30}{\mm}. Analogue readout of the tiles with SiPMs is envisaged. The HCAL barrel extends from \SI{174}{\cm} to \SI{333}{\cm} in radius. Both the endcap and the barrel HCAL are around \SI{7.5}{$\lambda$_I} deep, which brings the combined thickness of ECAL and HCAL to \SI{8.5}{$\lambda$_I}. The part of the HCAL endcap which surrounds the ECAL endcap is treated as a separate entity called the HCAL ring. The HCAL has in total 10 million readout cells, corresponding to a scintillator surface of \SI{8700}{\meter\squared}.

\section{CALICE Silicon Electromagnetic Calorimeter}\label{sec:ECAL}
The basic unit of the active elements of the CALICE SiECAL is the Active Sensor Unit (ASU) with a lateral size of \SI[product-units=power]{18x18}{\cm}. An ASU is the assembly of PCB, ASIC and silicon wafers. A PCB is equipped on one side with \num{16} SKIROC ASICs and on the other with four wafers. Each wafer is subdivided into \num{256} pads of \SI[product-units=power]{5.5x5.5}{\mm}, yielding a total number of \num{1024} pads per ASU. Two versions of ASUs are under development: a more conservative one with the ASICs in ball grid array (BGA) packages, and a more aggressive one with ASICs directly bonded to the PCB, allowing for a smaller layer thickness (see~\cref{fig:ecal-layer}).
A readout layer of the ECAL is a chain of up to \num{12} interlinked ASUs. Each layer is read out individually with an interface board.
 
The CALICE Collaboration is currently developing a technological prototype of the SiECAL. The final prototype will have up to \num{30} layers of one ASU each and a total of about \num{30000} readout channels. Up to now a maximum of \num{10} layers with a total of \num{10240} readout channels have been operated in test beams. The program comprises in particular also the production and tests of long layers with up to \num{12} ASUs.
 
\subsection{SiECAL sensor and ASIC R\&D}
The sensors in the SiECAL technological prototype consist of \num{256} P-I-N diodes and are produced from \SI{6}{inch} silicon wafers. Several designs with and without guard rings have been tested. So far sensors with a thickness of \SI{320}{\micron} and \SI{650}{\micron} have been used. Studies involving \SI{8}{inch} wafers with \SI{725}{\micron} thickness are ongoing.

The data are read out with SKIROC2 ASICs~\cite{ref:skiroc2}, a dedicated 64-channel dual-gain ASIC with a \num{12} bit ADC, designed to cover charges from \SI{2}{fC} to \SI{10}{pC}, equivalent to a dynamic range of about \SI{2500}{MIP}. It provides a cell-by-cell auto trigger down to \SI{0.5}{MIP}. It also offers time tagging with a \num{12} bit TDC.

A key element of the readout electronics is the capability for power-pulsed operation to reduce the average power consumption and thus the need for active cooling. Power pulsing makes use of the low duty cycle in the linear-collider beam time structure. The current SKIROC generation (SKIROC2 and SKIROC2A) is designed for the bunch structure of the ILC with bunch trains of \SI{1}{\ms} length every \SI{200}{\ms}, resulting in a duty cycle of \SI{0.5}{\percent}. Power pulsing can also be applied at CLIC. However, the higher bunch train repetition frequency of \SI{50}{\Hz} and the shorter train duration of \SIrange{156}{176}{\ns} will require design changes in a dedicated readout ASIC for CLIC, which also needs to provide hit time tagging with a precision of approximately \SI{1}{\nano\second}.

\begin{figure}[ht]
                \begin{subfigure}[T]{0.45\textwidth}
\includegraphics[width=\linewidth]{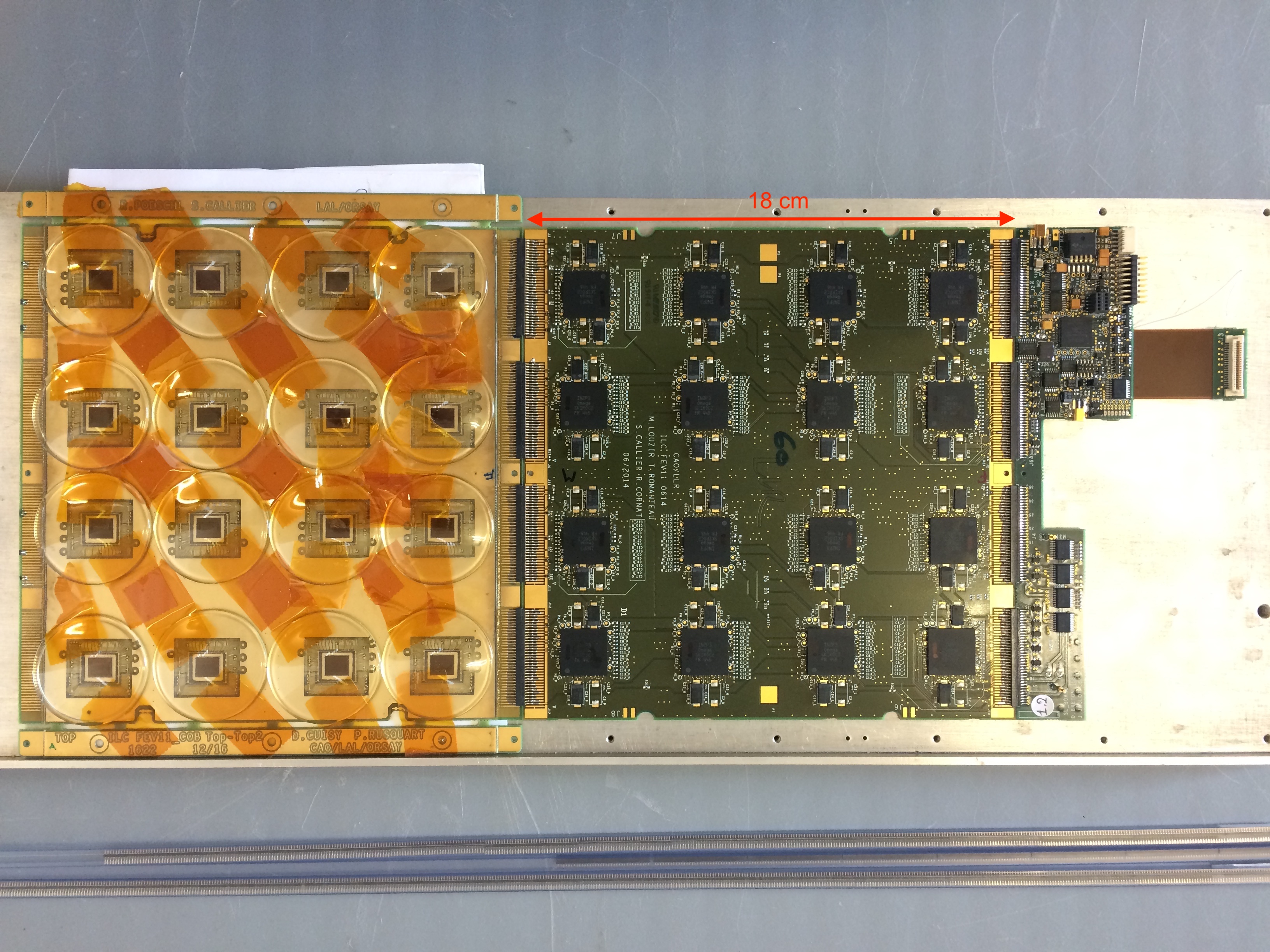}
                         \caption{}\label{fig:ecal-asu}
        \end{subfigure}%
	\hfill
                \begin{subfigure}[T]{0.54\textwidth}	\includegraphics[width=\linewidth]{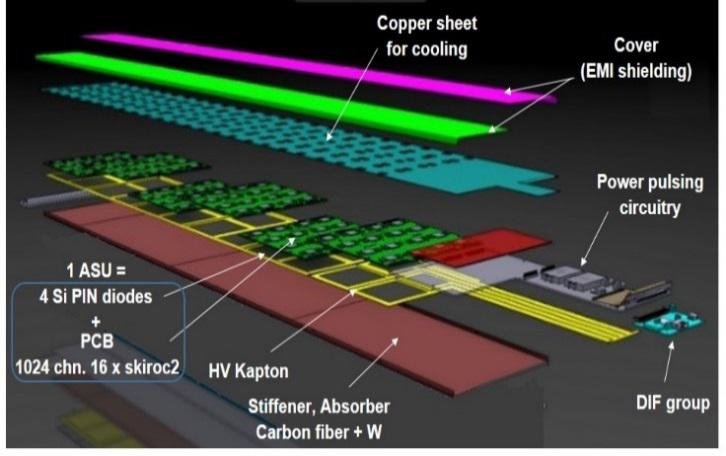}
                         \caption{}\label{fig:ecal-slab}
        \end{subfigure}%
                \caption{CALICE SiECAL technological prototype: \subref{fig:ecal-asu} setup with flat ASU with directly bonded ASICs, ASU with BGA-packaged ASICs, and compact interface card (SL board). \subref{fig:ecal-slab} Exploded drawing of a SiECAL layer.}
    \label{fig:ecal-layer}
\end{figure}

\subsection{SiECAL assembly and mass-production}
A detector designed for Particle Flow techniques requires of the order of \num{100000} highly integrated detection units for the SiECAL alone. Therefore the scalability of the production was included in the design from the start.
In recent years a demonstrator of an assembly and quality assurance chain has been realised~\cite{AIDA2020_SiECAL_assembly} (see~\cref{fig:siecal-benches}). This chain features two test benches and two assembly installations. The first bench characterises the large silicon diodes with the readout electronics and the second one measures the mechanical tolerances of the PCBs with high precision to ensure the sub-millimeter specifications are fulfilled. A semi-automated precision gluing table is used to mount and electrically connect the wafers onto the PCB. Finally, the instrumented ASUs are connected to their back-end interface and assembled in their mechanical support.
These can then be interfaced to the standard data acquisition chain for full validation with cosmic rays, radioactive sources or in beam tests.
A first batch of \num{10} short active layers (1 ASU long) has been produced in 2016. Several critical points in the production process have been identified and corrected, e.g. the precision of the distribution of the glue has been improved, and the alignment procedure for the components of the active layers has been revised to reach the required precision. In addition, the inter-connection between the interface card and the first ASU, as well as between ASUs, has been re-designed to be scalable to a large production. 

\begin{figure}[t]
   \begin{subfigure}[T]{0.49\textwidth} 
      \includegraphics[width=\linewidth]{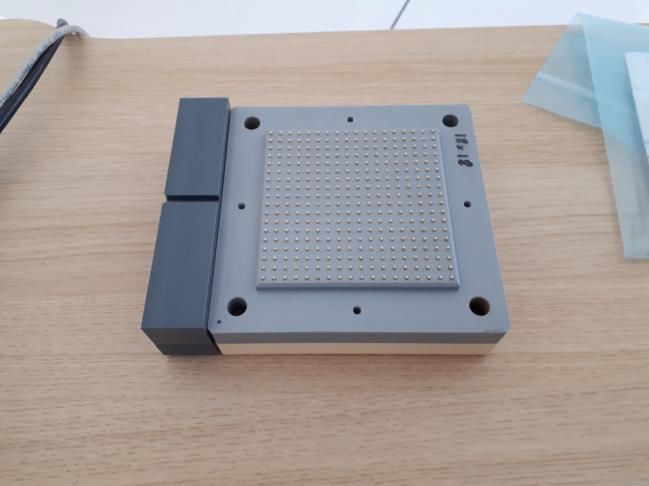}
      \caption{contact springs}\label{fig:siecal-springcontact}
   \end{subfigure}   
   \hfill
   \begin{subfigure}[T]{0.49\textwidth} 
      \includegraphics[width=\linewidth]{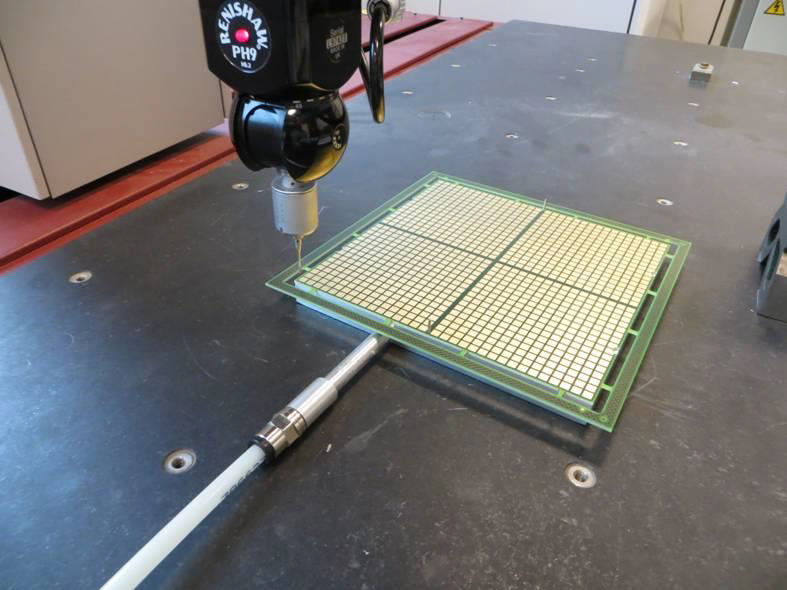}
      \caption{metrology bench}\label{fig:siecal-metrology}
   \end{subfigure}
   
   \begin{subfigure}[T]{0.49\textwidth}	         
      \includegraphics[width=\linewidth]{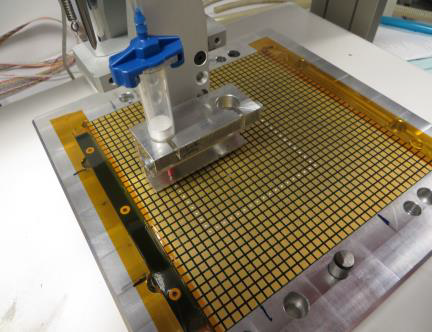}
      \caption{glueing bench}\label{fig:siecal-glueing}
   \end{subfigure}
   \hfill
   \begin{subfigure}[T]{0.49\textwidth} 
      \includegraphics[width=\linewidth]{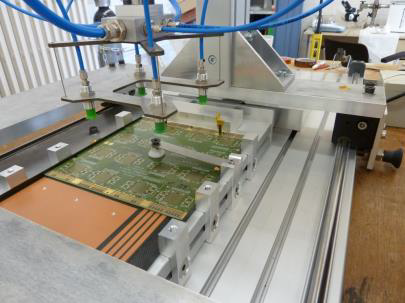}
      \caption{assembly bench}\label{fig:siecal-assembly}
   \end{subfigure}
   \caption{Test and assembly benches for the CALICE SiECAL technological prototype: \subref{fig:siecal-springcontact} contact springs for electrical tests, \subref{fig:siecal-metrology} metrology of the raw PCB of an ASU, \subref{fig:siecal-glueing} conductive glue distribution on a PCB in the glueing bench and ~\subref{fig:siecal-assembly} positioning of an ASU in the assembly bench~\cite{AIDA2020_SiECAL_assembly}.}
    \label{fig:siecal-benches}
\end{figure}

\subsection{SiECAL mechanics, cooling and integration}
The mechanical structure of the SiECAL consists of a carbon reinforced epoxy composite structure (\emph{alveolar structure}) embedding half of the tungsten absorber plates. The other tungsten plates, carrying active readout elements on both sides (a \emph{slab}), are inserted into the alveoli between two tungsten plates. \cref{fig:ecal-slab}} shows a drawing of a SiECAL layer.

Inside the readout layers, the power dissipation is distributed along the detector slab with hot spots at the location of the readout ASICs. The residual heat dissipated by the ASICs along the ASU chains is conducted by a copper drain to a cold block (heat exchanger) located at the end of a slab. This system includes also copper drains adapted to the cooling of the interface cards. The heat exchangers are connected to a leakless water cooling system with small water speed (\SI[per-mode = symbol]{<1}{\metre\per\second}), a small temperature gradient (\SI{<5}{\celsius}) and a very limited risk of water leakage, as the part of the cooling loop inside the detector is below atmospheric pressure. More details and results from prototype tests can be found in~\cite{AIDA2020_calo_cooling}.

\begin{figure}[ht]
                \begin{subfigure}[T]{0.67\textwidth}
                  \includegraphics[width=\linewidth, clip, trim = 0 0 1.5cm 0 ]{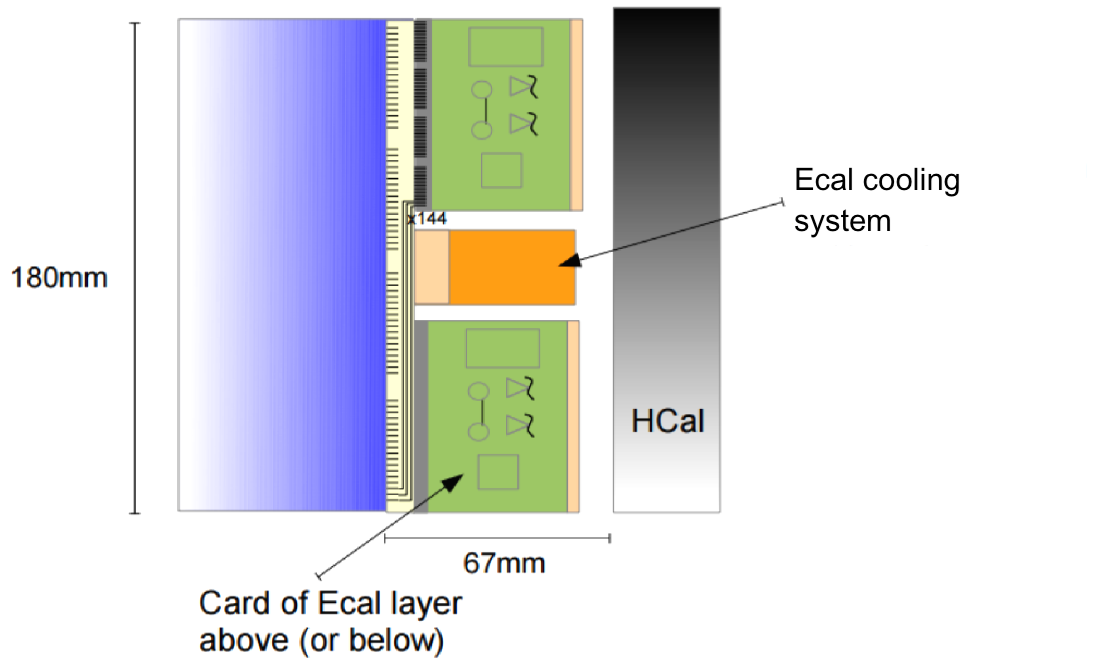}
                         \caption{}\label{fig:ecal-interfaces-sketch}
        \end{subfigure}%
	\hfill
                \begin{subfigure}[T]{0.32\textwidth}	\includegraphics[width=\linewidth]{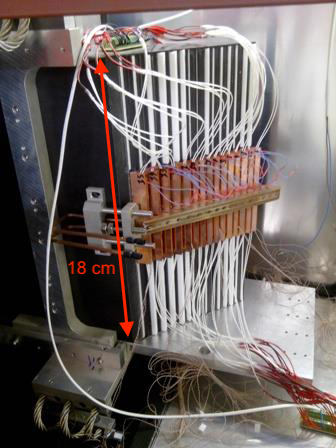}
                         \caption{}\label{fig:ecal-interfaces-cooling}
        \end{subfigure}%
          \caption{SiECAL interface region: \subref{fig:ecal-interfaces-sketch} sketch of the top view of the interface region. The end of the ECAL slabs is on the left (blue) and the adjacent HCAL module is on the right (grey). The cooling interface (orange) is in the middle of each slab, while the position of the DAQ interfaces (SL boards, green) alternates between layers. \subref{fig:ecal-interfaces-cooling} Cooling interfaces during a test of the SiECAL cooling prototype.}
    \label{fig:ecal-interfaces}
\end{figure}

In the SiECAL design for ILD, the slabs in the barrel are oriented perpendicular to the beam direction. This has the advantage of having no gaps in the instrumented volume pointing to the interaction vertex, and requiring shorter slabs. On the other hand, slabs running parallel to the beam direction have the advantage of easier access to the interface boards a the end of the slab, for connecting them to the cooling and the DAQ system, and also for possible maintenance. For the CLIC detector both options are considered. For both options, the space available for interface boards is very limited. A new SiECAL interface board (\emph{SLab interface board}, SL board) fulfilling the ILD constraints has been produced (see~\cref{fig:ecal-asu}), and first tests are encouraging.
A sketch of the SiECAL interface region and a photo of the cooling interface from the prototype test are shown in~\cref{fig:ecal-interfaces}. 

\subsection{SiECAL power pulsing}
The large peak currents (several Ampere per layer) present during power-pulsed operation can have a negative impact on the power supply, necessitating decoupling by large capacitors.

For several years the SiECAL prototype layers are operated regularly in power-pulsing mode. On the testbench the power consumption of the SKIROC2 is \SI{25}{\uW} per channel. A first analysis of test-beam data using simplified layers shows that power pulsing has only a minor influence on the data quality~\cite{rouene:thesis}.
However, a ramp-up time of about \SI{600}{\us} has to be taken into account, which adds to the overall power budget. In earlier versions of the layers the power needed by the ASICs was provided by a large \SI{400}{\milli\F} decoupling capacitor located at the end of a layer. In recent versions of the ASUs this large capacitor has been replaced by smaller capacitors (O(\SI{40}{\milli\F})) that are integrated into the ASUs. This reduces the peak current of about \SI{12}{\A} expected for a long layer to about \SI{1}{\A}. A thorough analysis of peak currents and power consumption at the different stages of the supply and readout chain is still outstanding.

\subsection{SiECAL test-beam results}
Several configurations of short layers of the SiECAL technological prototype have been operated in beam tests, most of them using sensors with \SI{320}{\micron} thickness.
An analysis of a beam test at DESY with low-energy positrons (\SIrange[range-phrase = { to }]{1}{6}{\GeV}) is presented in~\cite{ref:siecal_commissioning}. The largest setup consisted of \num{10} layers (about \num{10000} channels), including \num{4} layers with \SI{650}{\micron} thick sensors. It was operated recently at the CERN SPS, where muon tracks and electron and pion shower data between \num{10} and \SI{150}{\GeV} were recorded. Due to the small signals of the silicon sensors and the very compact electronics design, noise is of particular concern. Several hardware effects were observed for the first ASUs, such as high noise levels due to the routing of the PCB, re-triggering of the ASUs and so called square events caused by capacitive coupling of the channels at the edge of a wafer. In recent tests, these problems were much reduced. Around $4\%$ of the channels had to be masked~\cite{LCWS2018_talk_AIrles}. For the remaining channels a signal-to-noise ratio for MIPs in the trigger branch of \num{11.6 \pm 0.7} has been measured. This value is compatible with the value of \num{12.9 \pm 3.4} obtained on a testbench~\cite{ref:siecal_commissioning}. For the charge measurement after digitisation a signal-to-noise ratio of around \num{20} is reported~\cite{ref:siecal_commissioning}.

Recently, a slab of \num{8} ASUs was tested for the first time. After small changes to the power and clock distribution, MIP signals could be recorded in all ASUs~\cite{LCWS2018_talk_VBoudry}. For the necessary extension to \num{12} ASUs further studies are ongoing.

In the following, relevant results obtained with the SiECAL physics prototype will be discussed. This prototype had 30 active layers with a lateral size of \SI[product-units=power]{18x18}{\cm} segmented into pads of \SI[product-units=power]{1x1}{\cm}. The tungsten absorber layers varied in thickness from \SI{0.4}{X_0} to \SI{1.2}{X_0}, resulting in a total thickness of \SI{20}{\cm}, corresponding to \SI{24}{X_0}. The prototype had \num{9720} readout channels.
The expected performance of the technological prototype is similar or better.

\subsubsection{SiECAL energy resolution}
The energy resolution for positron showers obtained for the SiECAL physics prototype is $(16.51 \pm 0.35)\%/\sqrt{E} \oplus (1.9 \pm 0.23)\%$~\cite{CAN-046}, confirming results obtained previously for electrons~\cite{ref:SiECALresolution}. This corresponds to expectations, and fulfils the CLIC detector requirement of approximately $15\%/\sqrt{E}$. However, one should keep in mind that these results were obtained for positrons (electrons) with beam energies up to \SI{20}{\GeV} (\SI{45}{\GeV}) only, such that leakage is not important. At CLIC much higher electron and photon energies are expected, and the constant term in the energy resolution, which is difficult to determine precisely from low energy data, plays a significant role. Therefore dedicated measurements at higher beam energies with a prototype providing sufficient containment would be useful.

\subsubsection{SiECAL particle separation}
\begin{figure}[ht]
                \begin{subfigure}[T]{0.49\textwidth} \includegraphics[width=\linewidth]{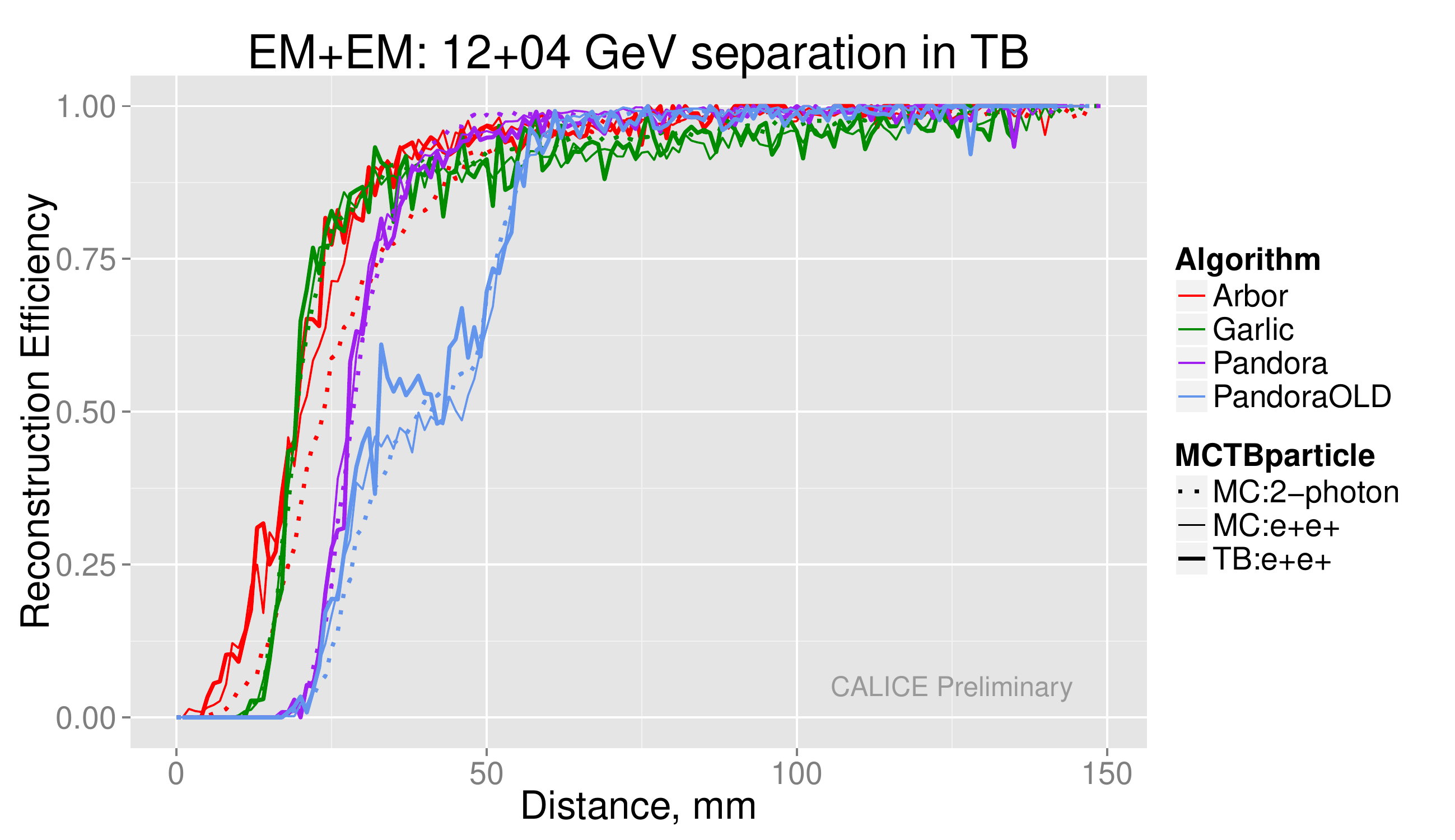}
                         \caption{}\label{fig:separation-em-em}
        \end{subfigure}%
	\hfill
                \begin{subfigure}[T]{0.49\textwidth}	\includegraphics[width=\linewidth]{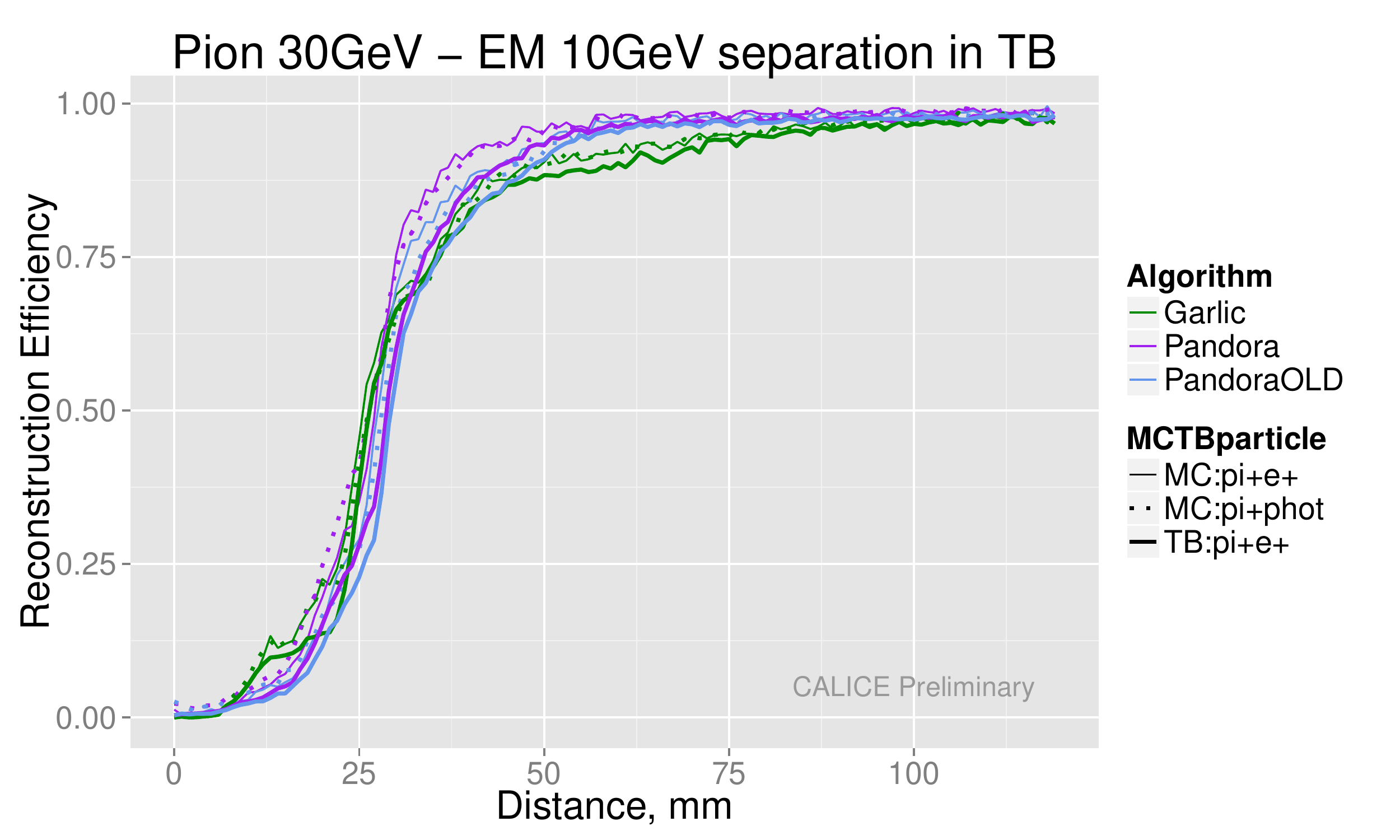}
                         \caption{}\label{fig:separation-pi-em}
        \end{subfigure}%
          \caption{Reconstruction efficiency of two particles as a function of the distance between them: \subref{fig:separation-em-em} of a \SI{12}{\GeV} and a \SI{4}{\GeV} electromagnetic cluster and \subref{fig:separation-pi-em} of a \SI{30}{\GeV} pion and a \SI{10}{\GeV} electromagnetic cluster. Results for various reconstruction algorithms are shown in different colours and the data (thick solid lines) are compared to Monte Carlo simulations containing either positron showers (thin solid lines) or photon showers (thin dashed lines)~\cite{CAN-057}.}
    \label{fig:separation}
\end{figure}

At large jet energies, the \emph{confusion term} in particle-flow reconstruction, caused by the overlap of showers of neutral and charged particles, dominates the jet energy resolution~\cite{thomson:pandora}. The correct separation of nearby showers is therefore an important performance parameter for the calorimeter system of a linear collider detector. The separation of two electromagnetic showers and the separation of an electromagnetic and a hadronic shower have been studied by overlapping shower data from the CALICE SiECAL and AHCAL physics prototypes~\cite{CAN-057}. Depending on particle type and reconstruction algorithm, a reconstruction efficiency of \SI{75}{\percent} or better can be reached for distances between the clusters as small as \SI{2}{\cm}. Above \SI{5}{\cm} the efficiency is typically \SI{90}{\percent} or better (see~\cref{fig:separation}). These results are obtained with a SiECAL cell size of \SI[product-units=power]{10x10}{\mm}. Without re-tuning of the algorithms, a finer granularity usually does not improve the separation efficiency~\cite{CAN-057}. So far no re-tuning has been done.

\section{CALICE SiPM-on-tile Hadronic Calorimeter}\label{sec:HCAL}
The basic unit of the active elements of the CALICE AHCAL is the HCAL Base Unit (HBU)~\cite{ref:HBU}, with a size of \SI[product-units=power]{36x36}{\cm}, holding \num{144} SiPMs controlled and read out by four SPIROC2E ASICs (see~\cref{fig:HBU}).
The AHCAL is designed such that readout layers consist of up to \num{3} slabs along the beam-direction, and each slab consists of up to \num{6} HBUs. An LED system for calibration with one LED per scintillator tile is integrated in the readout layers. Each layer has one set of interface boards providing power, signals for the LED system, and connection to the DAQ.

In 2017 and 2018, the CALICE Collaboration built a large AHCAL technological prototype~\cite{ref:AHCAL_tech_prototype}. It consists of a non-magnetic stainless steel absorber structure with \num{38} active layers of \num{2} slabs of \num{2} HBUs each, and has in total \num{21888} channels.

\subsection{AHCAL sensor and ASIC R\&D}
The dynamic range requirements for the SiPM-tile system are driven by the need to distinguish single SiPM pixel peaks for the LED calibration at very small amplitudes, and the need to detect the highest expected amplitudes in physics events of several hundred MIPs. For operation in self-triggered mode, SiPMs with low noise and small cross talk between pixels are essential. The design has been optimised for a signal of about \num{15} pixels for a MIP crossing the tile. This value was chosen such that a trigger threshold at \num{0.25} to \num{0.5} MIP, which then corresponds to about \num{4} to \num{7} pixels including safety margins for channel-to-channel variations, leads to a tolerable noise rate above the trigger threshold for most SiPMs. The low trigger threshold at \num{0.5} MIP or less is necessary to reach an efficiency of \SI{95}{\percent} or better for detecting a MIP crossing the tile, ensuring a good imaging performance and the ability to use tracks within showers for calibration. Recently, SiPMs with trenches between pixels, leading to low noise (typically of the order of \SI[per-mode=symbol]{50}{\kHz\per\mm\squared}) and low cross talk (typically a few percent), have become available from several vendors.
 
\begin{figure}[t]
  \begin{minipage}{\textwidth}
  \begin{minipage}[][9.5cm][t]{0.38\textwidth} 
    \centering
    \begin{subfigure}[t]{0.6\textwidth} 
      \begin{tikzpicture}
         \node[anchor=south west,inner sep=0] at (0,0)(image){ \includegraphics[width=\linewidth]{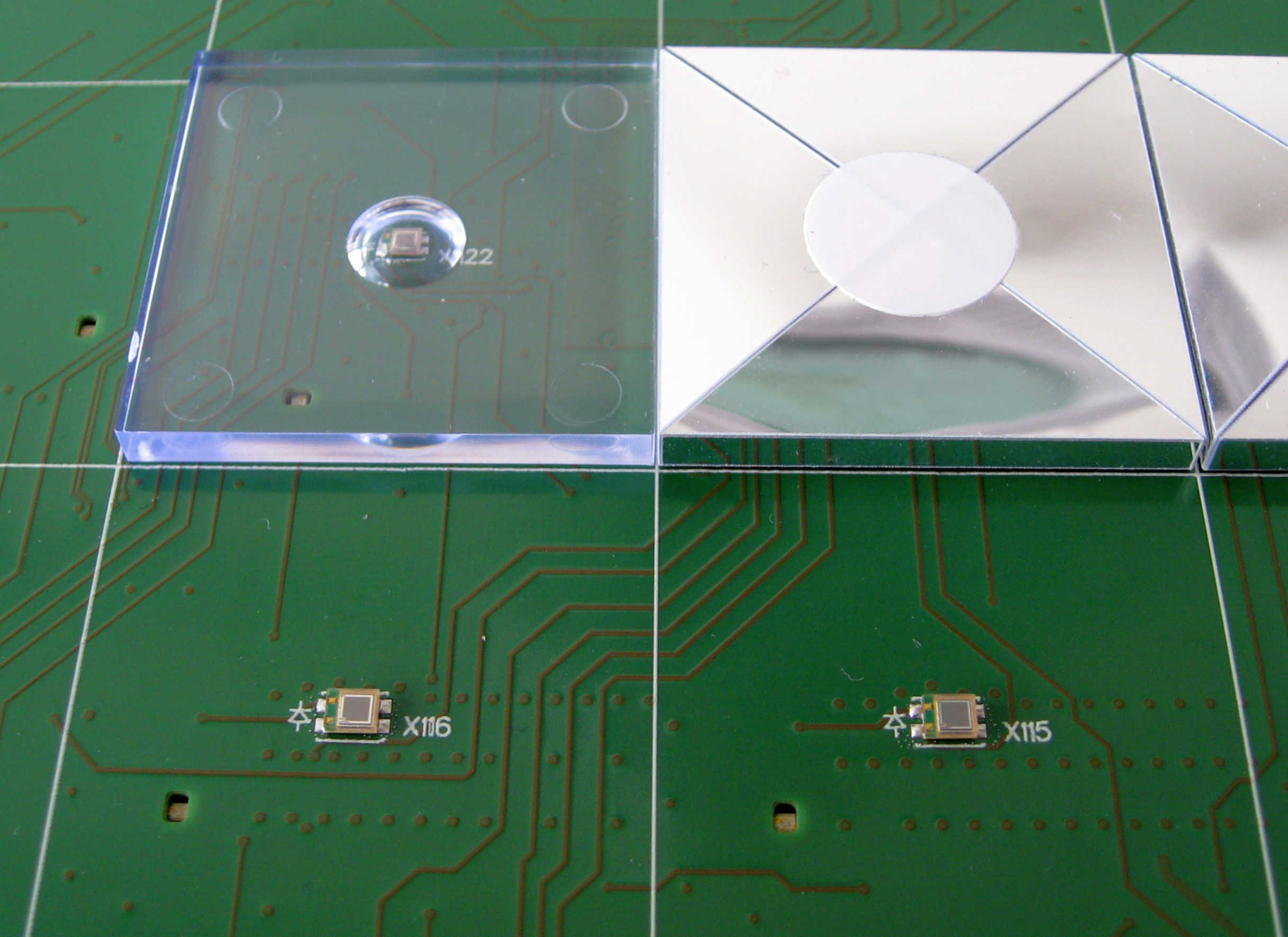}};
         \begin{scope}[x={(image.south east)},y={(image.north west)}]
            \draw[<->,yellow,thick](0.1,0.45)--+(0.41,0) node[below,pos=0.5]{$\SI{30}{\mm}$};
         \end{scope}
      \end{tikzpicture}
      \caption{SiPM and tiles}\label{fig:hcal-tiles}
   \end{subfigure}
   
   \vfill
   \begin{subfigure}[b]{0.95\textwidth} 
      \begin{tikzpicture}
         \node[anchor=south west,inner sep=0] at (0,0)(image){ \includegraphics[width=\linewidth]{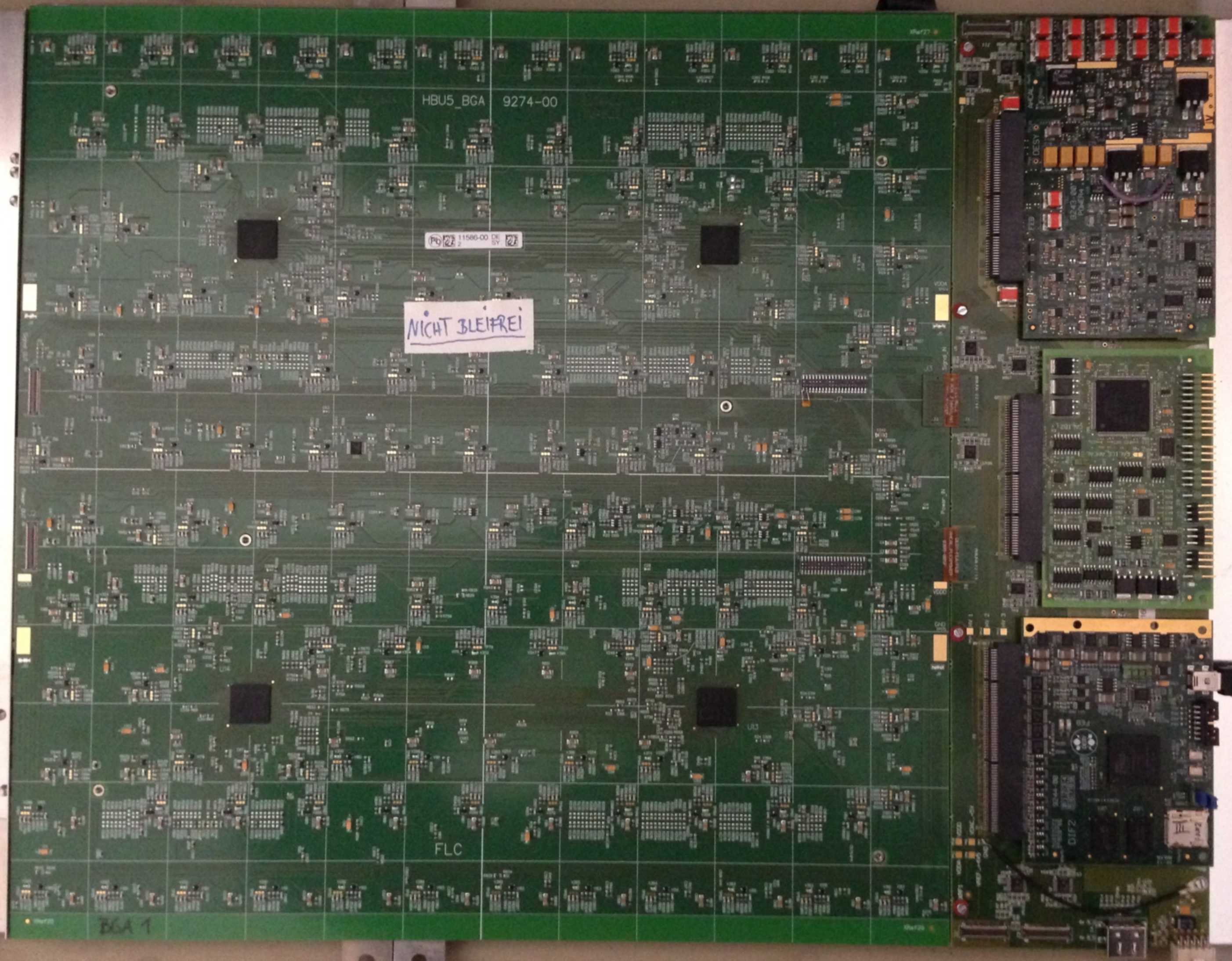}};
         \begin{scope}[x={(image.south east)},y={(image.north west)}]
            \draw[<->,thick](0.01,-0.03)--+(0.77,0) node[below,pos=0.5]{$\SI{36}{\cm}$};
         \end{scope}
      \end{tikzpicture}

                         \caption{HBU}\label{fig:HBU}
   \end{subfigure}
 \end{minipage} 
 \hfill
 \begin{minipage}[][9.5cm][c]{0.6\textwidth} 
   \begin{subfigure}[]{\textwidth}	         
      \begin{tikzpicture}
         \node[anchor=south west,inner sep=0] at (0,0)(image){ \includegraphics[width=\linewidth]{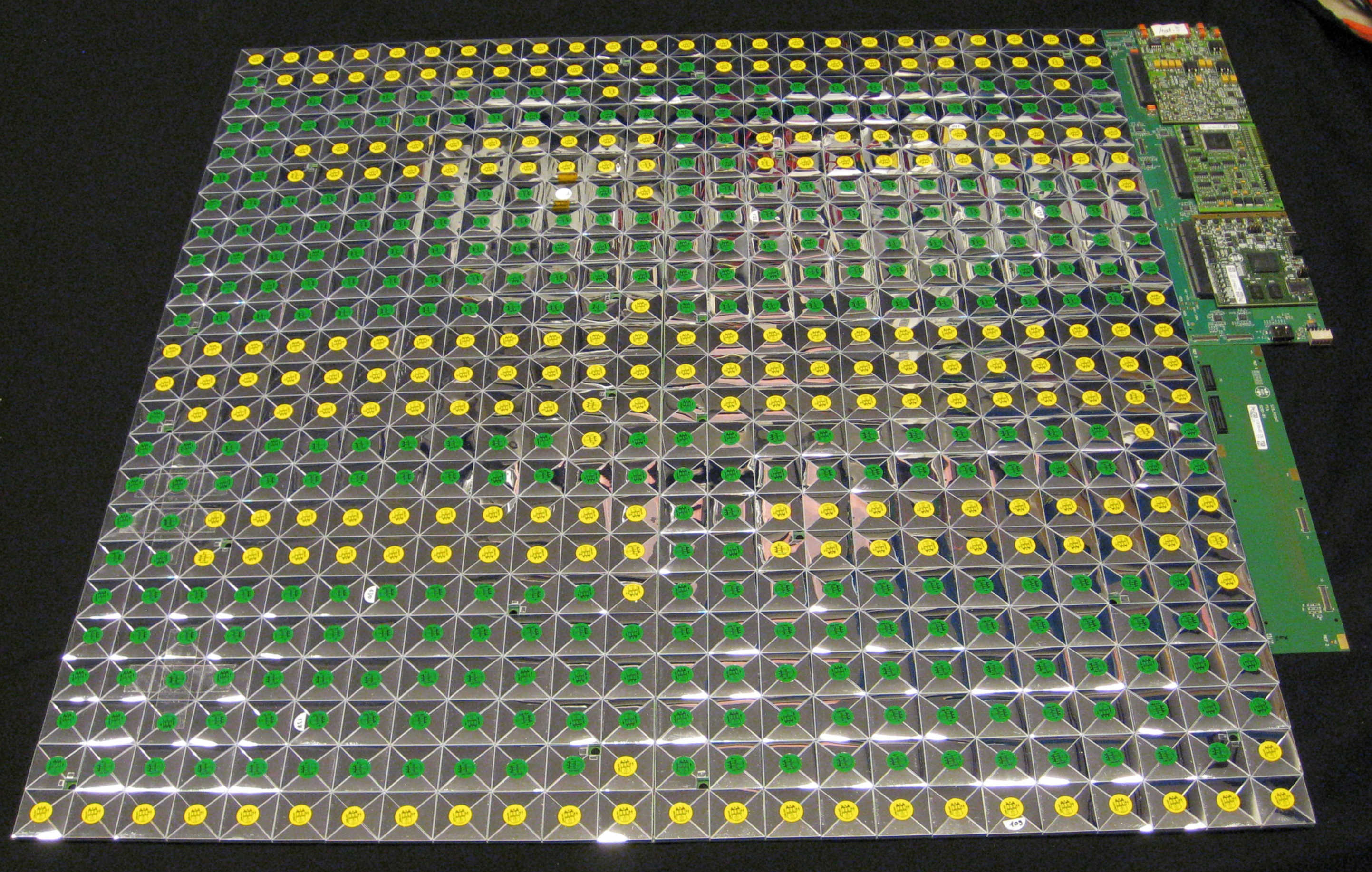}};
         \begin{scope}[x={(image.south east)},y={(image.north west)}]
             \draw[<->,thick](0.0,-0.03)--+(0.96,0) node[below,pos=0.5]{$\SI{72}{\cm}$};
         \end{scope}
      \end{tikzpicture}
      \caption{AHCAL test-beam prototype layer}\label{fig:ahcal-layer-backside}
   \end{subfigure}
 \end{minipage} 
 \end{minipage}
   \caption{CALICE AHCAL technological prototype: \subref{fig:hcal-tiles} scintillator tiles with central dimple, wrapped and unwrapped, mounted on an HBU with SiPMs,~\subref{fig:HBU} electronics side of an HBU and \subref{fig:ahcal-layer-backside} view of the scintillator tiles from a full test-beam prototype layer.}
    \label{fig:tiles}
\end{figure}
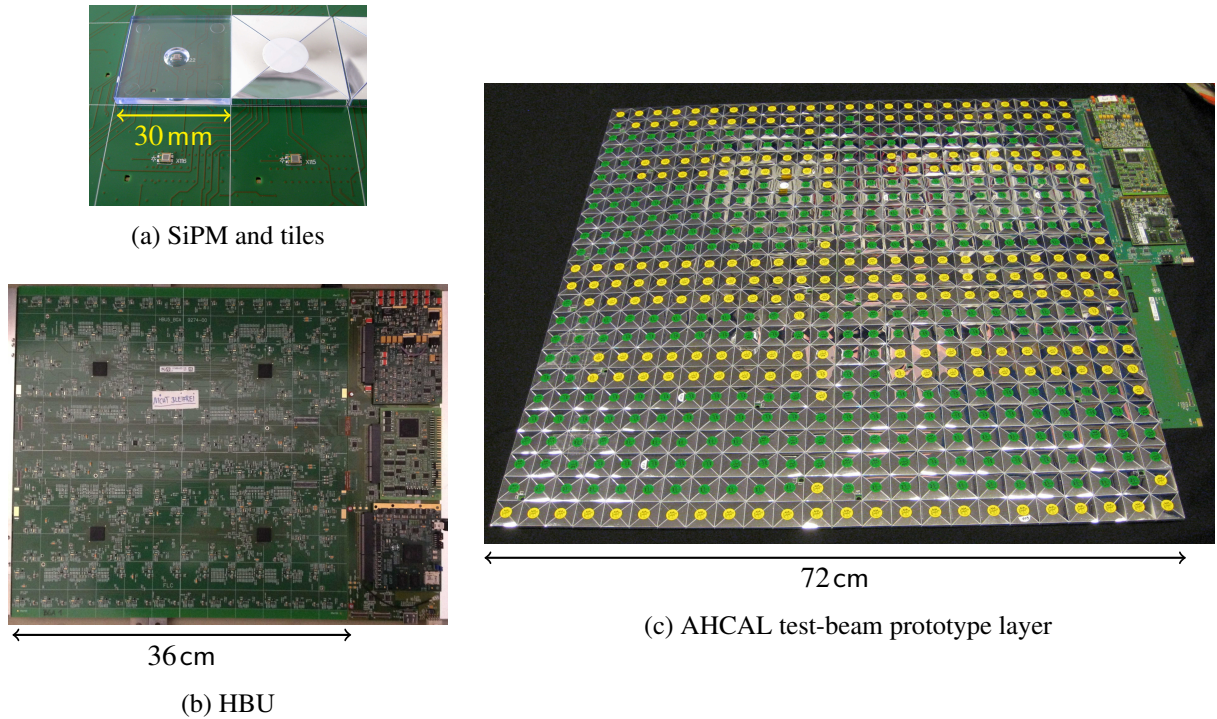

The new AHCAL prototype uses Hamamatsu MPPC S13360-1325PE photon sensors~\cite{Hamamatsu:MPPC}. They have a size of \SI[product-units=power]{1.3x1.3}{\mm} with \num{2668} pixels with a pitch of \SI{25}{\um}. Compared to the SiPMs previously used in the physics prototypes, their probability for after-pulses and cross talk between the pixels is about an order of magnitude lower.
The tiles are injection-moulded polystyrene scintillator tiles with a central dimple for optimal light collection, wrapped individually in reflector foil, as shown in~\cref{fig:tiles}.

The SiPMs were delivered in lots of \num{600} pieces with a uniform break-down voltage for each lot within \SI{\pm 100}{\mV}. The gain of the SiPMs was found to be uniform within \SI{2.4}{\percent} when operated at a common over-voltage. 

The SiPMs are read out by SPIROC2E ASICs~\cite{ref:spiroc}, a dedicated 36-channel ASIC providing a \num{12} bit dual-gain energy measurement. The input is coupled capacitively via an \num{8} bit DAC, allowing a cell-by-cell tuning of the SiPM bias voltage in a range of up to \SI{4.5}{\V}. For the AHCAL technological prototype this feature is not used at the moment. Instead all SiPMs connected to one ASIC are operated with the same bias voltage. The ASIC features a cell-by-cell auto trigger and time stamping at the few ns level in test-beam operations. In operating conditions with shorter data-taking windows similar to the bunch train structure of linear colliders, a time resolution of a \SI{}{\ns} or better is expected.

Like the SKIROC2, also the SPIROC2 is designed for operation in power-pulsed mode. Similar to SKIROC, the different operating conditions at CLIC is likely to require changes in the ASIC design, which up to now is optimised for the ILC bunch structure.

\subsection{AHCAL assembly and mass production}
As a full linear collider calorimeter AHCAL will consist of about \num{10} million channels, the design is optimised for mass production and automatic assembly. Quality control and quality assurance take place at several steps of the assembly process~\cite{AIDA2020_calotestbenches}.

\begin{figure}[ht]
   \centering
\includegraphics[width=0.6\linewidth]{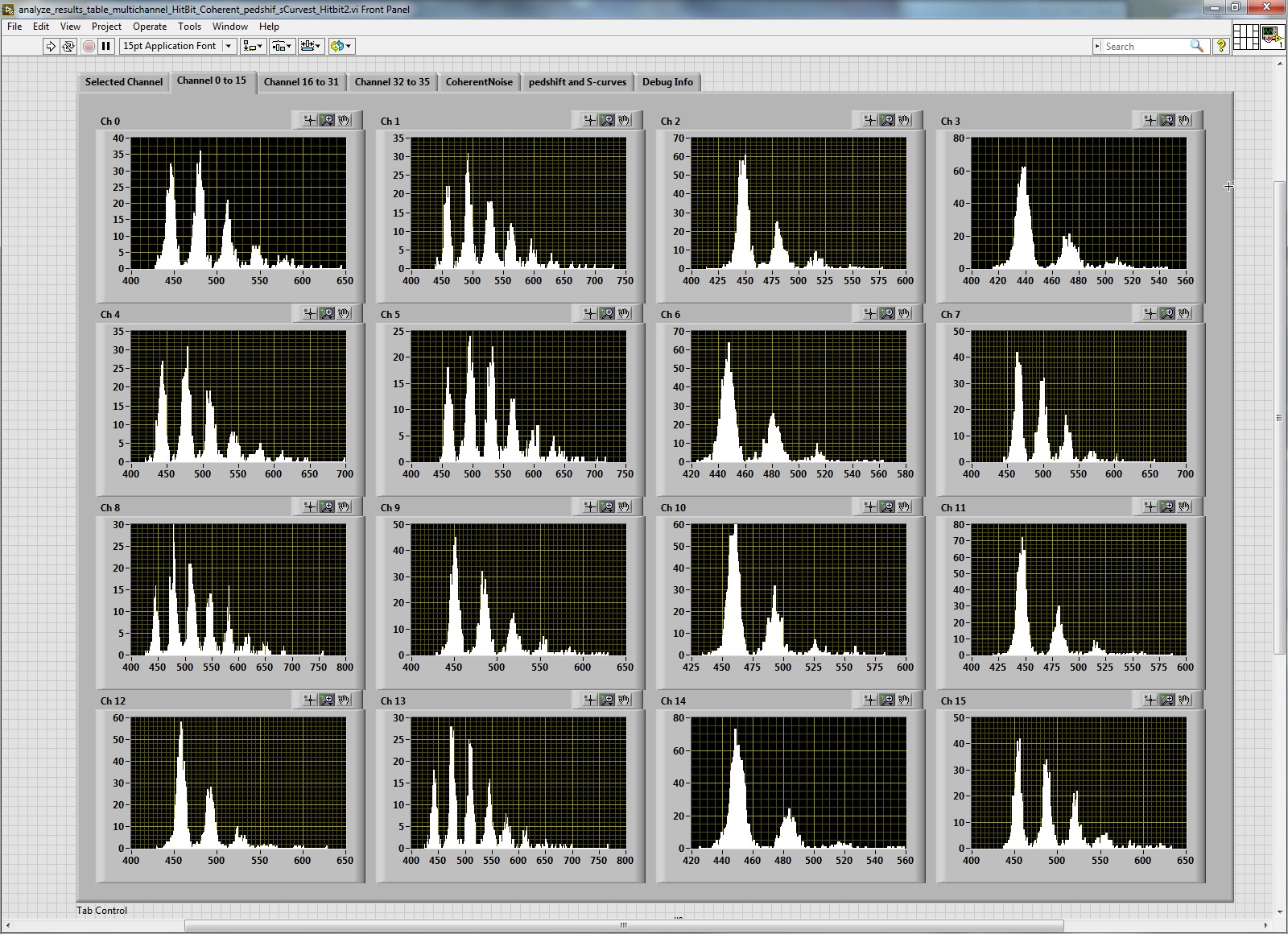}
   \caption{Example of LED amplitude spectra recorded during tests with an HBU before tile assembly. Peaks corresponding to integer numbers of pixels are clearly visible for all channels.}
   \label{fig:ahcal-HBUqa}
\end{figure}

Spot-samples of all SiPM lots, and all of the ASICs, undergo semi-automatic testing procedures before soldering the HBUs. The HBUs are equipped with all components, including ASICs and surface-mount SiPMs, in a standard pick-and-place process. At this step a first test of the electronics is done by placing the HBU on a reflective surface. The LED system is used to record signals from all channels (see~\cref{fig:ahcal-HBUqa}), and resoldering or replacement of components is possible. 

The polystyrene scintillator tiles are injection moulded. Without any further surface treatment, they are wrapped in laser-cut reflective foil by a robotic procedure and mounted on the HBUs using a pick-and-place machine, after glue dispensing with a screen printer. Up to \num{30} HBUs are then stacked above each other in a cosmic ray test stand, where their performance parameters (gain, light yield) are determined.

The HBUs together with the interface boards are then integrated into cassettes. \cref{fig:ahcal-layer-backside} shows an active layer, with the scintillator tiles visible. For the new large prototype, an initial MIP calibration has been determined for all layers at the DESY beam test facility, by exposing them without absorbers to a \SI{3}{\GeV} electron beam. The calibration confirmed the results for the single HBUs from the cosmic ray test stand.

\subsection{AHCAL mechanics, cooling, integration and temperature compensation}
\begin{figure}[ht]
   \centering
\includegraphics[width=0.7\linewidth]{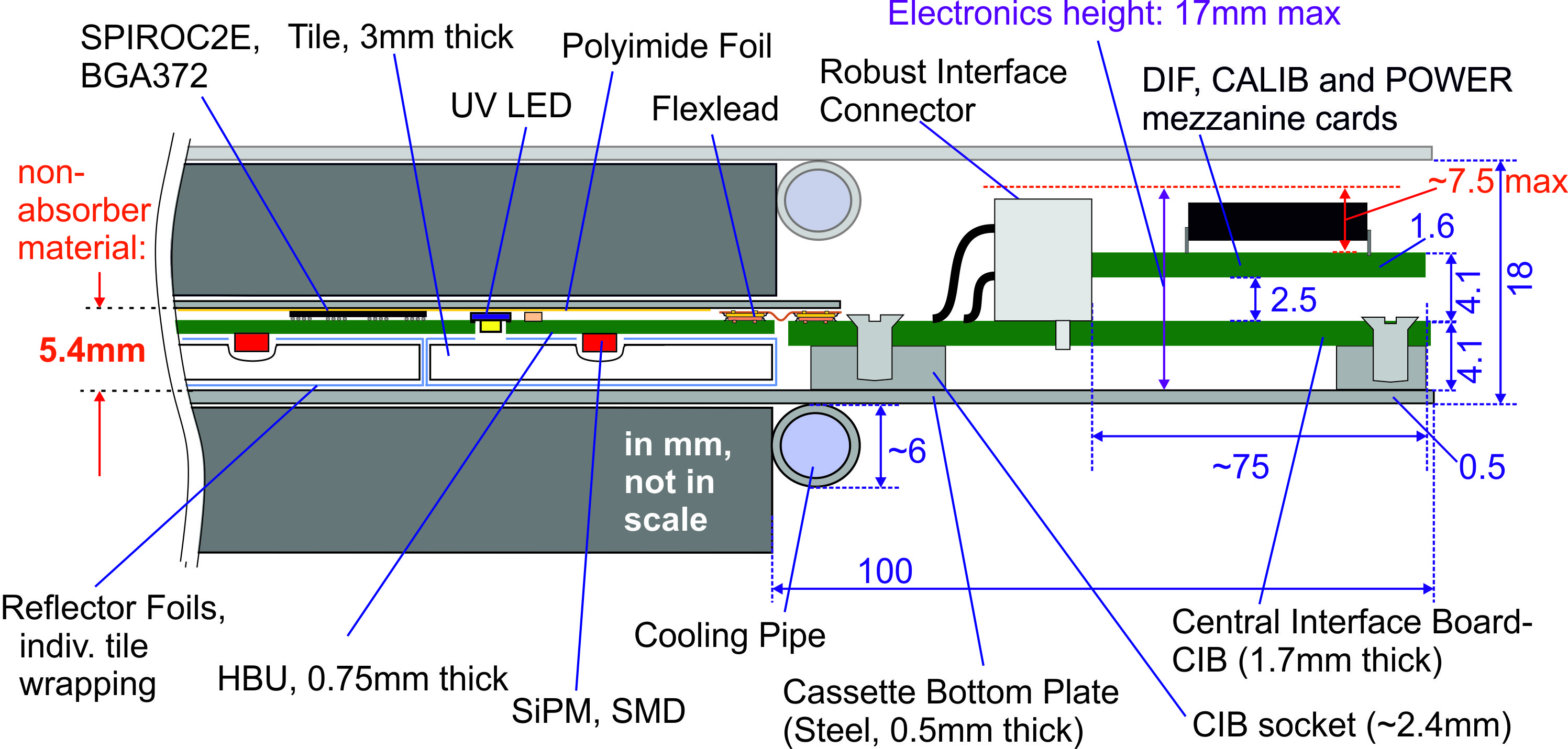}
          \caption{Structure of one layer of the AHCAL technological prototype.}
    \label{fig:ahcal-layerstructure}
\end{figure}

The AHCAL absorber is a self-supporting steel structure, into which the readout layers are inserted. The cassettes are made from \SI{0.5}{\mm} thick steel and have bolts to hold the HBUs in place. The thin cassette walls are flexible enough to follow the gravitational sag of the absorber plates, and minimize the non-structural mass. The total thickness of the non-absorber material in a layer is only \SI{5.4}{\mm}, including \SI{3}{\mm} for the scintillator tile. It is kept as small as possible to minimize the total thickness of the calorimeter, and therefore the inner diameter of the magnet.
The interface boards are outside of the absorber such that they can easily be accessed for replacement or repair. A sketch of the side view of an AHCAL layer including the interface board region is shown in \cref{fig:ahcal-layerstructure}.

The power consumption of the SiPMs and of the readout ASICs in the active layers, which is reduced by power pulsing of the ASICs, is small enough to be extracted through the steel absorber structure. However, the interface boards need cooling. Like the SiECAL, the AHCAL is envisaged to have a leakless water cooling system, operated below atmospheric pressure~\cite{AIDA2020_calo_cooling}. For the AHCAL test-beam prototype, an overpressure cooling system has been used, but with diameters of all pipes as foreseen for the leakless system. The system operated reliably, keeping the temperatures stable and low enough for a safe operation of the detector.

The gain as well as the photon detection efficiency of a SiPM depend on the temperature. In order to keep these quantities stable, the interface boards of the AHCAL technological prototype offer the possibility of adjusting the bias voltage of the SiPMs to obtain a stable overvoltage. This feature has been used routinely in recent beam tests, resulting in very stable running and gain variations of typically within \SI{\pm 1}{\percent}.

\subsection{AHCAL power pulsing}\label{sec:caloPP}
The AHCAL technological prototype allows operation with and without power pulsing, and both modes have been used in beam tests. In addition to the power pulsing mode foreseen for ILC with a duty cycle of about \SI{0.5}{\percent}, the SPIROC2 can also be operated with a reduced or without sleep time between acquisition cycles. This mode improves the data taking efficiency in beam tests, but reduces the power consumption less than the ILC mode, and leads to more frequent power switching. This power-pulsing mode without sleep time was also tested successfully with the AHCAL technological prototype operated in beam tests.

The signal quality and power consumption for a full AHCAL slab of \num{6} HBUs with SPIROC2B was studied in the laboratory~\cite{ref:AHCALpowerpulsing}. With a ramp-up time of about \SI{150}{\us}, the effect of power pulsing on the data is found to be negligible. Recent studies with a layer of \num{2} full slabs of \num{6} HBUs equipped with SPIROC2E showed a power consumption of about \SI{150}{\uW} per channel for a duty cycle of \SI{1}{\percent}, depending slightly on the number of events that need to be read out. In order to reach the goal of \SI{25}{\uW} per channel, further optimisation is required and possible. 
A setup with several layers of the full size of \num{3x6} HBUs, which will allow a system-wide test in realistic conditions, is in preparation.

\subsection{AHCAL test-beam results}
The large AHCAL technological prototype has been operated in beam tests at the CERN SPS. Several $10^7$ events with muon tracks as well as electron and pion showers have been collected. The energies range from
\SIrange[range-phrase = { to }]{10}{100}{\GeV} for electrons and from \SIrange[range-phrase = { to }]{10}{200}{\GeV} for hadrons. The data taking rate averaged over the spills, of about \SI{5}{\s} duration, was up to \num{400} events per second. The operation of the detector was stable, and less than \SI{0.1}{\percent} non-working channels were observed. The analysis of the data is ongoing, and more results are expected in the near future.

In the following sections, relevant results obtained with the AHCAL physics prototype, based on data recorded in 2007 to 2011, will be discussed. This prototype had up to 38 active layers with a lateral size of \SI[product-units=power]{90x90}{\cm} consisting of scintillator tiles with sizes from \SI[product-units=power]{3x3}{\cm} to \SI[product-units=power]{12x12}{\cm}. The scintillator tiles were read out by wavelength-shifting fibres coupled to SiPMs. Together with the cassettes of the active layers, the absorber thickness was \SI{21.4}{\mm} of steel per layer, resulting in a total HCAL thickness of \SI{120}{\cm}, corresponding to \SI{5.26}{$\lambda$_I}. The prototype had \num{7608} readout channels. 
The expected performance of the technological prototype is similar or better, and it offers channel-by-channel hit time information at the few-nanosecond level in addition.

The AHCAL test-beam data provide information of unprecedented granularity on hadronic shower structures and responses, thereby allowing for detailed comparisons with \geant simulations. For example, a multitude of shower-shape parameters are well described by the FTFP\_BERT and QGSP\_BERT physics lists (see  \cref{sec:shower-shapes}). This is essential for achieving reliable \geant-based predictions of the overall CLICdet detector responses. Predictions of the CLICdet performance for particle identification and jet final states moreover rely on particle-flow analysis with PandoraPFA. To this end a study was performed in which two pion-induced test-beam events were superimposed, with one event having its incoming track removed to simulate a neutral particle~\cite{ref:PFAtest}. Distances between the particles were varied in the study. The results show that the probability of PandoraPFA correctly resolving the situation is well described by simulations based on the QGSP\_BERT list in \geant.

\subsubsection{AHCAL energy resolution}
\begin{figure}[ht]
                \begin{subfigure}[T]{0.49\textwidth} \includegraphics[width=\linewidth]{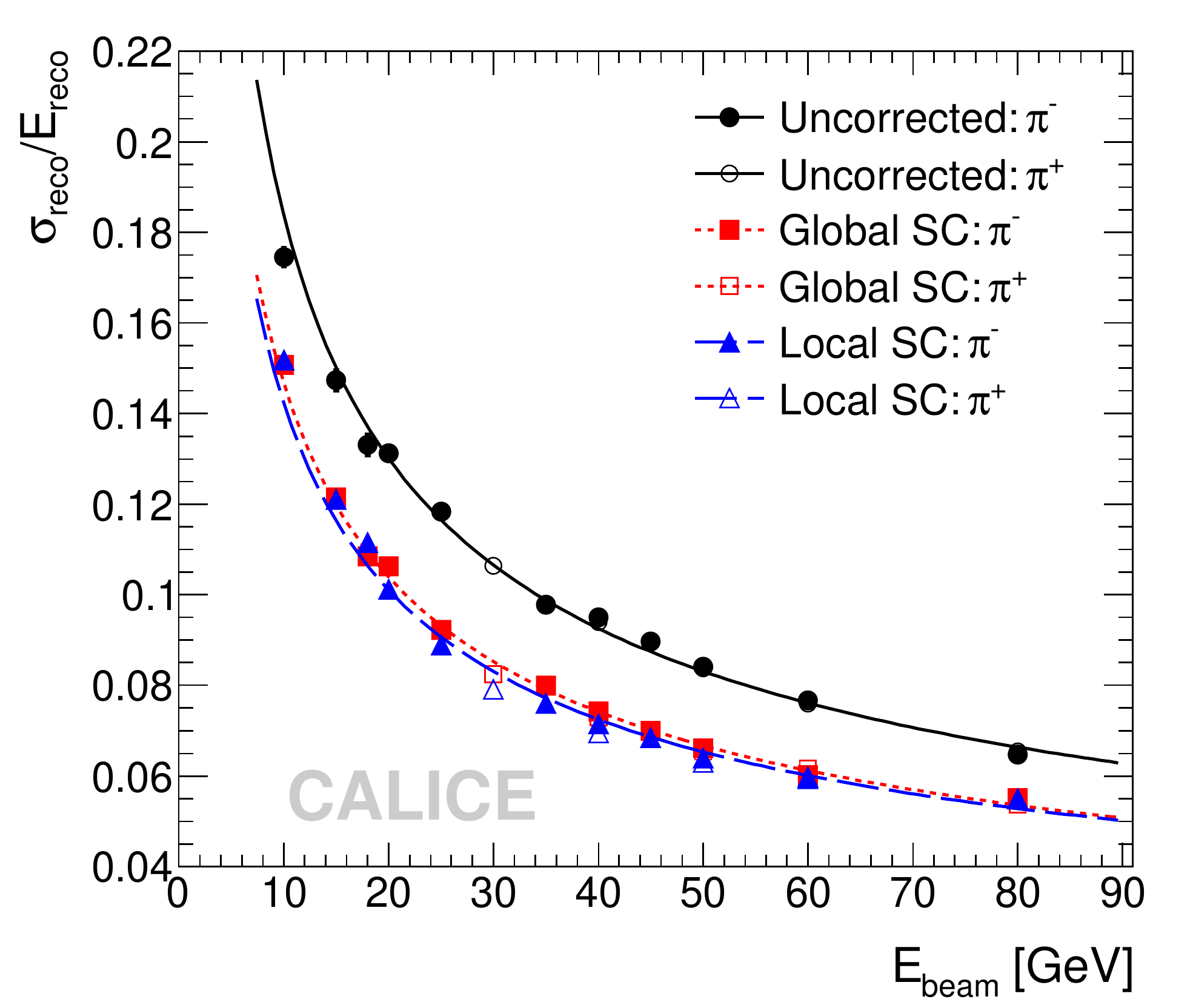}
                         \caption{}\label{fig:ahcal-resolution}
        \end{subfigure}%
	\hfill
                \begin{subfigure}[T]{0.49\textwidth}	\includegraphics[width=\linewidth]{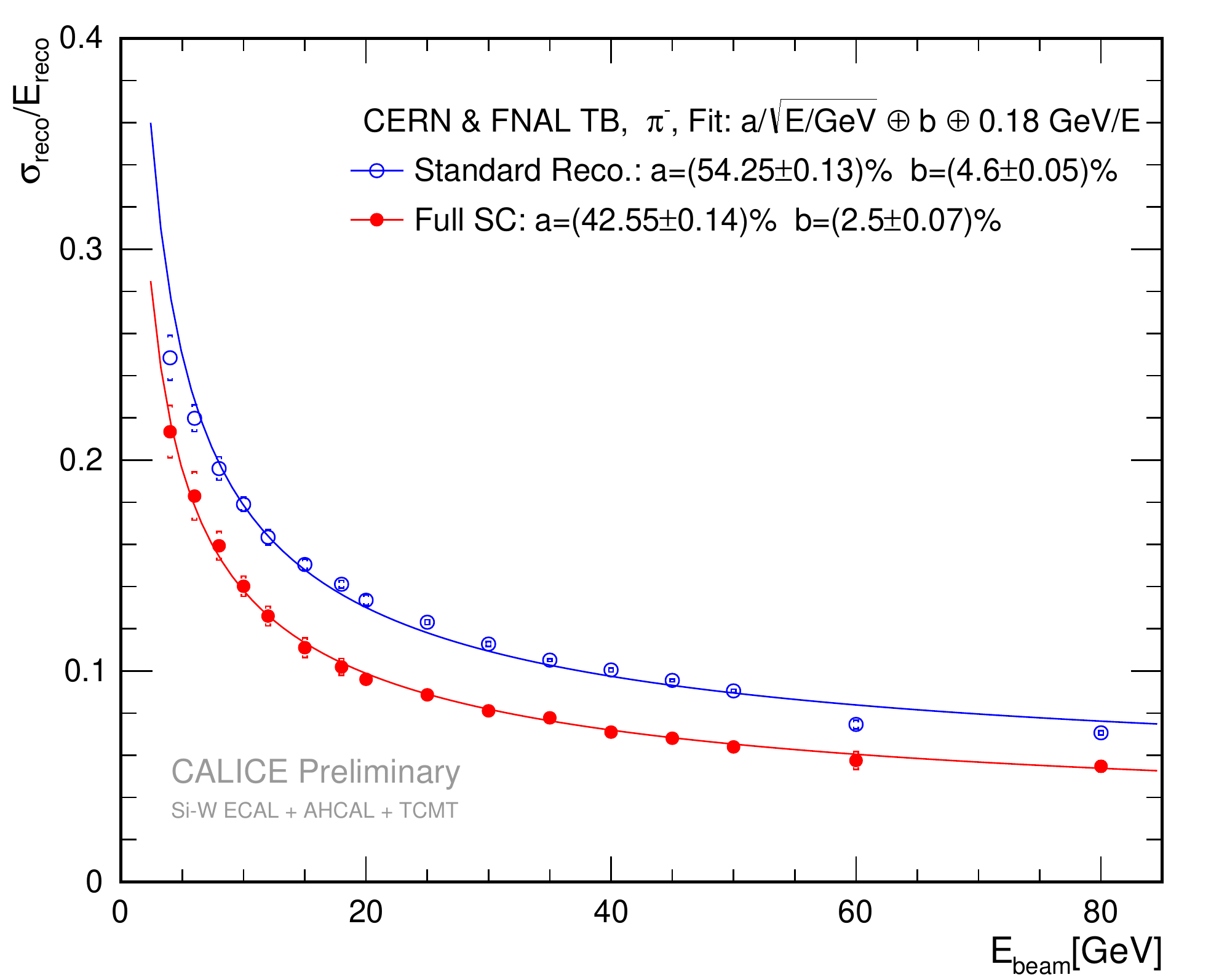}
                         \caption{}\label{fig:combined-siecal-ahcal-resolution}
        \end{subfigure}%
          \caption{Energy resolution for pion showers: \subref{fig:ahcal-resolution} measured in the AHCAL physics prototype~\cite{ref:AHCALresolution} and \subref{fig:combined-siecal-ahcal-resolution} measured in the combined setup of the SiECAL and AHCAL physics prototype~\cite{CAN-058}. For both setups results with and without software compensation are shown.}
    \label{fig:hadenergyresolution}
\end{figure}

The energy resolution for pions obtained by summing the energy deposits in the AHCAL physics prototype is $(57.6 \pm 0.4)\%/\sqrt{E} \oplus (1.6 \pm 0.3)\% \oplus 0.18\,\mathrm{GeV}/E$~\cite{ref:AHCALresolution}. Since the AHCAL is non-compensating, the energy resolution is affected by fluctuations in the electromagnetic fraction of hadronic showers. Due to the high granularity of the calorimeter, it is possible to apply individual weighting of the shower components, in order to compensate for differences between the hadronic and electromagnetic response as well as for the invisible energy depositions. This approach, known as {\em software compensation}, yields a significant improvement in the fitted combined resolution. Two approaches to software compensation, a global and a local one, have been explored~\cite{ref:AHCALresolution}, both leading to a similar improvement of the energy resolution to approximately $(45\%)/\sqrt{E} \oplus (1.7\%) \oplus 0.18\,\mathrm{GeV}/E$, see \cref{fig:ahcal-resolution}.

The readout layers of the AHCAL physics prototype have also been exposed to test-beam in an absorber structure made of \SI{1}{\cm} thick tungsten plates~\cite{ref:WAHCAL1, ref:WAHCAL2}, as was considered previously for a detector at CLIC. Because of the very small radiation length of tungsten, this layer thickness corresponds to \SI{2.8}{X_0}, which leads to a rather coarse sampling of electromagnetic showers or shower components. The energy resolution for electrons is significantly worse than with steel absorber, but for pions a rather similar resolution of $(57.9 \pm 1.1)\%/\sqrt{E} \oplus (4.6 \pm 0.4)\% \oplus 0.065\,\mathrm{GeV}/E$ is obtained~\cite{ref:WAHCAL2}.
In the measured range of energies up to \SI{60}{\GeV}, the W-AHCAL is found to be approximately compensating, such that the improvement by software compensation is much smaller than for steel absorber~\cite{CAN-062}.
The simulations based on the physics models implemented in the \geant physics lists QGSP\_BERT\_HP and FTFP\_BERT\_HP reproduce average hadron shower properties at the percent level, and spatial shower profiles at the \SI{15}{\percent} level or better~\cite{ref:WAHCAL2}. For tungsten absorber and scintillator as active material the data-driven Neutron High Precision (HP) Models and Cross Sections in \geant are important for the agreement of simulation with data for all physics lists.
The results confirm the validity of the use of tungsten absorber in a hadron calorimeter, and show that \geant can be used to predict the detector performance also in this case.

\subsubsection{AHCAL timing}
\label{sec:ahcal_timing}
The measurement of hit times at the nanosecond level in the hadronic calorimeter can be used for the association to a bunch crossing, for background rejection, and in the study of the intrinsic time structure of hadronic showers. The CALICE T3B experiment has demonstrated this with a small setup of \num{15} plastic scintillator tiles read out with SiPMs. The data were recorded with a fast USB oscilloscope with a sampling frequency of \SI{1.25}{GS/s} and \SI{8}{bit} ADC resolution. With this setup a time resolution between
\SIrange[range-phrase = { and }]{0.7}{1.07}{\ns} has been reached~\cite{ref:T3B}, which includes contributions from the readout system, the data analysis procedure and the time jitter of the trigger. The analysis showed that the late shower component is more pronounced for tungsten than for steel absorber.
 
The dedicated readout with a fast oscilloscope is not scalable to a large collider detector. In the SPIROC2 readout ASIC of the AHCAL, the hit time measurement is performed with a \SI{12}{bit} TDC that encodes the time when a signal crosses the trigger threshold. The resolution of this kind of measurement depends on the ramp speed of the TDC, and therefore the bunch clock. Up to now this measurement has only been performed with a previous version of the ASIC, SPIROC2B, and a bunch clock of \SI{250}{\kHz} optimised for test-beam campaigns. With this setup, a hit time resolution for muons of about \SI{5}{\ns} has been obtained, while the resolution degrades for showers due to electronics effects~\cite{ThesisEldwan}.
The SPIROC2 ASIC is designed to reach sub-ns time resolution when operated with a bunch clock of \SI{5}{\MHz}, as foreseen for ILC. This still needs to be demonstrated in beam tests.
A readout based on the sampling of the pulse height at regular intervals and a fit of the pulse shape, as foreseen for CLIC (see \cref{chap:electronics-daq}), might lead to a significantly improved resolution for the hit times.

\subsection{Optimisation of the AHCAL readout granularity}
One possibility for reducing the number of readout channels and thus the complexity of the calorimeter is a coarser granularity in the rear layers of the detector. For ILD the energy reconstruction of jets up to \SI{250}{\GeV} for an AHCAL with \SI[product-units=power]{30x30}{\mm} tiles in the front part and \SI[product-units=power]{60x60}{\mm} tiles in the rear part has been studied in simulation~\cite{ILDCaloGranularity}. For a detector with the rear half of the AHCAL layers equipped with \SI[product-units=power]{60x60}{\mm} tiles, the degradation was found to be negligible for jet energies up to \SI{180}{\GeV}, while \SI{250}{\GeV} jets show a slightly worse resolution. First practical experience has been gained with one layer equipped with \SI[product-units=power]{60x60}{\mm} tiles in the recent beam tests of the AHCAL technological prototype. Further studies for the higher particle and jet energies expected at CLIC are needed to evaluate whether a coarser granularity in parts of the calorimeter is also a viable option here. 

\section{Combined performance of ECAL and HCAL}
In a collider experiment, the energy resolution for hadrons will be given by the combination of the measurements in the ECAL and the HCAL. The energy resolution of the combined setup of the SiECAL and the AHCAL physics prototypes for pions has been found to be $(54.25 \pm 0.13)\%/\sqrt{E} \oplus (4.60 \pm 0.05)\% \oplus 0.18\,\mathrm{GeV}/E$ for the standard reconstruction, while software compensation improves the resolution to $(42.55 \pm 0.14)\%/\sqrt{E} \oplus (2.50 \pm 0.07)\% \oplus 0.18\,\mathrm{GeV}/E$~\cite{CAN-058}, see \cref{fig:combined-siecal-ahcal-resolution}.

Consistent results, showing that the single-pion energy resolution of the combined ECAL+HCAL system is similar to that of the AHCAL alone for both reconstruction methods, have been obtained with an ECAL based on tungsten absorber and scintillator strips read out by SiPMs~\cite{ref:CombinedSciResolution}.

\subsection{Shower shapes}\label{sec:shower-shapes}
The physics prototypes of the SiECAL and the AHCAL have also been used to study observables sensitive to the shapes and structure of hadron showers, like the transverse shower profile or track segments within a shower. The prediction of the particle separation capabilities relies on these distributions being modelled correctly by the simulation. Studying these variables directly is more sensitive to differences in the simulation models, and offers interesting insights into the evolution of hadronic showers.

In the AHCAL with steel absorber, the agreement between data and simulation for shower-shape observables is usually better than \SI{10}{\percent} for the FTFP\_BERT and QGSP\_BERT physics lists~\cite{ref:AHCALG4Validation, ref:AHCALShowerShapes, ref:AHCALShowerDecomposition}. This good agreement is also found for track segments within pion showers~\cite{ref:AHCALTrackSegments}.
For pion showers in the AHCAL with tungsten absorber, deviations between data and simulation of up to \SI{20}{\percent} are observed for both FTFP\_BERT\_HP and QGSP\_BERT\_HP~\cite{ref:WAHCAL2}.
In the radial shower profile, both physics lists predict too much energy in the shower core. At beam energies up to \SI{10}{\GeV}, in general the deviations between simulation and data for the shower shapes are below \SI{10}{\percent}, with QGSP\_BERT\_HP performing remarkably well for both pions and protons, with deviations within \SI{3}{\percent} or better for most of the studied variables~\cite{ref:WAHCAL1}.

Due to the very high granularity, the SiECAL can resolve even finer structures of hadronic showers than the AHCAL, but only for the first part of the shower close to the first hard interaction of the incoming particle. For particle energies between \num{2} and \SI{10}{\GeV}, none of the studied physics lists describe the entire set of data, but overall the simulations are within \SI{20}{\percent} of the data and for most observables much closer~\cite{ref:SiECALHadShowers}.
For tracks originating from the first hard interaction, in most cases data and simulation agree within \SI{10}{\percent} without revealing a clear preference for one of the studied physics lists~\cite{CAN-055}. The largest source of discrepancy between data and simulation is the energy and radius of the interaction region. The measured energy deposition in the interaction region is up to \SI{20}{\percent} higher than predicted by the simulation.

For most variables describing the shapes of pion or proton showers, simulations based on the FTFP\_BERT(\_HP) and QGSP\_BERT(\_HP) \geant physics lists agree with the data to better than \SI{10}{\percent} for both SiECAL and AHCAL and steel or tungsten absorber. The good agreement allows a reliable prediction of the PFA performance of the CLIC detector by the simulation.

\section{Applications other than at linear colliders}
Based on the positive experience of the CALICE prototypes, the CMS experiment has chosen a high granularity calorimeter with silicon sensors and scintillator tiles read out by SiPMs for the upgrade of their calorimeter endcaps~\cite{HGCALTDR} for the High-Luminosity LHC. This environment poses much stronger requirements on radiation hardness as well as data readout and trigger rates, and the detector will have to be operated at \SI{-30}{\celsius}. On the other hand, the power budget is much larger, and the space requirements are less stringent than for an electron--positron collider detector. For this upgrade, a new readout ASIC is being developed that is based on the SKIROC, but implements a sampling of the signal pulse height with \SI{40}{\MHz}, similar to the scheme planned for the electronics of the CLIC detector. Two versions of the ASIC are planned, one for the readout of silicon sensors and one for SiPM readout.
Intense prototyping for the CMS calorimeter endcaps is currently ongoing, as they have to be built and installed for the start of the High-Luminosity LHC in 2026. The similarity of the concepts creates large synergies between the CMS calorimeter upgrade and the developments for an electron--positron collider detector.

\section{Summary and outlook}

In recent years, the CALICE collaboration has demonstrated successfully the concepts of a highly granular ECAL based on silicon sensors and tungsten absorber as well as of a highly granular HCAL based on scintillator tiles read out by SiPMs and steel or tungsten absorber. Beam tests of the physics prototypes of both concepts have shown that the requirements for the CLIC detector can be met, and that \geant simulations describe the data sufficiently well to have confidence in predictions of the overall detector performance also for jet final states. 

The next generation of prototypes, the technological prototypes, address the scalability of the calorimeter concepts to a full collider detector. For both concepts technological prototypes have been built, exercising mass assembly and quality assurance procedures. For the SiECAL, a 10-layer prototype has been built, and tested in particle beams. In the coming years, an extension to a total number of \num{30} layers -- as planned for the ILD detector concept -- is foreseen. This will  allow cross-checking of the performance under realistic collider detector conditions. A 38-layer AHCAL technological prototype has recently been operated in beam tests with muon, electron and pion beams, and the data will be used to measure the energy and time reconstruction capabilities as well as to perform shower-shape studies. A combined beam test of SiECAL and AHCAL is important to evaluate the performance of the calorimeter system.
For both concepts, the operation of layers of the full size, as foreseen in a collider detector, has to be demonstrated. First steps have been made, and results are expected soon. Both technological prototypes have dedicated readout ASICs designed for very low power consumption, which is achieved by power pulsing according to the ILC beam structure. This has been used successfully in beam tests, but not all aspects have been tested so far.

For the CLIC detector, a test-beam measurement of the energy resolutions for high energy electrons would be important to ensure that also these requirements are fulfilled. No fundamental show-stoppers have been identified to achieve the required \SI{1}{\nano\second} time resolution for either silicon or scintillator and SiPM active active layers. Nevertheless, dedicated adaptions to the readout ASIC designs have to be implemented in order to demonstrate this in real prototypes. At the same time those readout ASICs need to include the implementation of CLIC-compatible power pulsing. The large occupancy expected in the ECAL and HCAL cells close to the beam pipe might require overall design changes with improved shielding against back-scattering particles, smaller readout cells and shorter integration times (see \cref{sec:detector-optimisation-high-occupancy}).

\chapter{Very forward calorimeters}\label{chap:fcal}
This chapter presents the concepts for the forward calorimeters at CLIC and gives an overview of the corresponding detector developments carried out by the FCAL collaboration. The motivation for forward calorimetry at linear colliders is introduced in \cref{sec:fcal-motivation}, followed by a description of the forward-calorimeter concepts for CLIC. Ongoing hardware developments are discussed in \cref{sec:fcal-activities}. Results of beam tests with multi-layer prototypes and with irradiated sensors are described in \cref{sec:fcal-testbeam}. The chapter concludes with a brief description of other applications and future plans in \cref{sec:fcal-future}.

\section{Motivation for very forward calorimetry at linear collider detectors}\label{sec:fcal-motivation}
Very forward calorimeters at linear collider detectors are indispensable for precisely measuring the luminosity, the key quantity needed to convert measured count rates into cross sections.  The required accuracy for the absolute luminosity measurement is a few $10^{-3}$~\cite{Abramowicz:2016zbo,Robson:2018zje,deBlas:2018mhx}.
The gauge process used is low-angle Bhabha scattering, which can be calculated with high precision in QED. Very forward calorimeters are also foreseen to assist in beam tuning, e.g. in a fast feedback system to maximise the luminosity during collider operation. Finally, yet importantly, very forward calorimeters increase the angular coverage of the detector. This leads to improved measurements of missing energy, which is important e.g. to detect BSM phenomena which result in forward-going particles in their decay chain.

\section{Very forward calorimeters at CLIC}\label{sec:fcal-at-clic}
       
\subsection{Introduction}
\begin{figure}[ht]
\begin{center}
    \includegraphics[width=0.65\columnwidth]{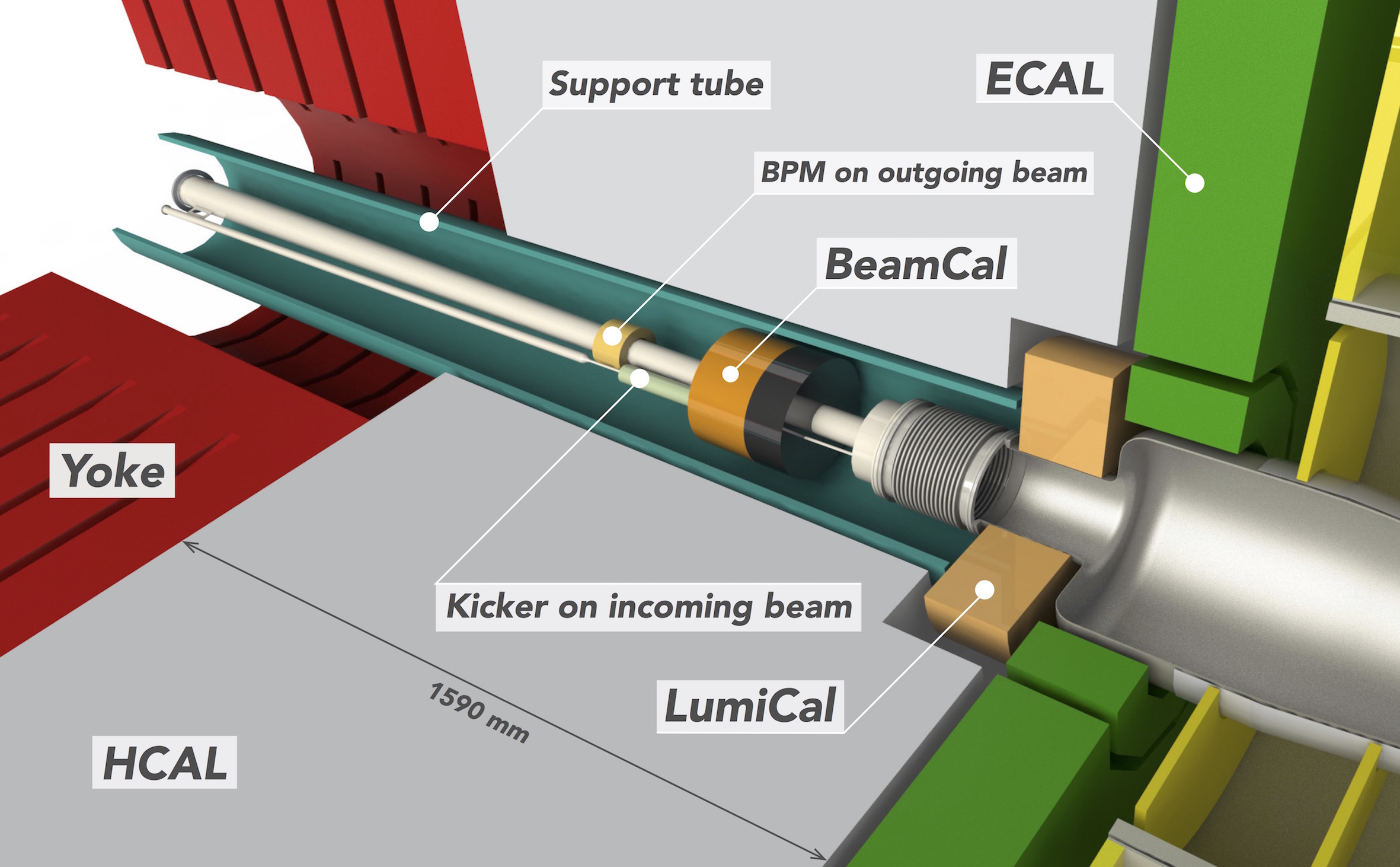}
  \end{center}
          \caption{Isometric view of the forward region of the CLIC detector~\cite{AlipourTehrani:2254048}. 
LumiCal is located immediately downstream of the ECAL endcap. BeamCal and a graphite cylinder to prevent back-scattering into the vertex region are located inside the cylindrical support tube.
A kicker and a beam position monitor (BPM) are part of the intra-train feedback system.}
    \label{fig:forward}
\end{figure}

Two small dedicated electromagnetic sampling calorimeters, called LumiCal and BeamCal, are to be installed in the very forward region of the CLIC detector. Detailed Monte Carlo studies have been performed to optimise the design and location of these calorimeters, estimate the background from physics processes and understand the impact of backgrounds from beam-beam interactions on the luminosity measurement~\cite{cdrvol2,LCD-Note-2009-002}.
A schematic layout of the forward region of CLIC is shown in~\cref{fig:forward}.

To ensure a high efficiency and accuracy for the detection of high-energy electrons in a large beamstrahlung background, compact calorimeters with a small Moli\'ere radius are needed. In order to cope with high occupancies, LumiCal and BeamCal need a dedicated fast readout. In addition, the lower polar angle range of BeamCal is exposed to a large flux of particles, leading to energy depositions of up to one MGy per year. Hence, radiation-hard sensors are required.

\subsection{Calorimeter implementation in CLICdet}
LumiCal is primarily designed with an accurate luminosity measurement in mind.
It is located in a cut-out of the HCAL endcap, at 2.5\,m from the interaction point, covering polar angles of 39 to 134\,mrad.
The calorimeter has 40 layers on both sides of the interaction point.
Each layer comprises a 3.5\,mm thick tungsten absorber and a 320\,$\upmu$m thick silicon sensor.
The silicon area is subdivided into 64 radial and 48 azimuthal segmentations, resulting in pad sizes ranging from $3.75 \times 13\,\text{mm}^2$ to  $3.75 \times 44\,\text{mm}^2$.

BeamCal, at 3.2\,m from the IP and located in the cylindrical support tube, completes the angular coverage for forward-electron tagging (from 10 to 46\,mrad).
BeamCal is implemented in the CLICdet simulation model with 40 layers on both sides of the interaction point, each comprising a 3.5\,mm thick tungsten absorber and a 300\,$\upmu$m diamond sensors.
In addition to diamond, GaAs sensors that are also radiation hard are under consideration for the active layers.
The sensor area is subdivided into regular cells, independent of the radius, of $8 \times 8\,\text{mm}^2$.
In order to reduce the flux of secondary particles scattering back into the central detector, a 100\,mm thick graphite layer is placed in front of BeamCal.

\subsection{Concept of the calorimeters}
\begin{figure}[ht]
\begin{center}
    \includegraphics[width=0.6\columnwidth]{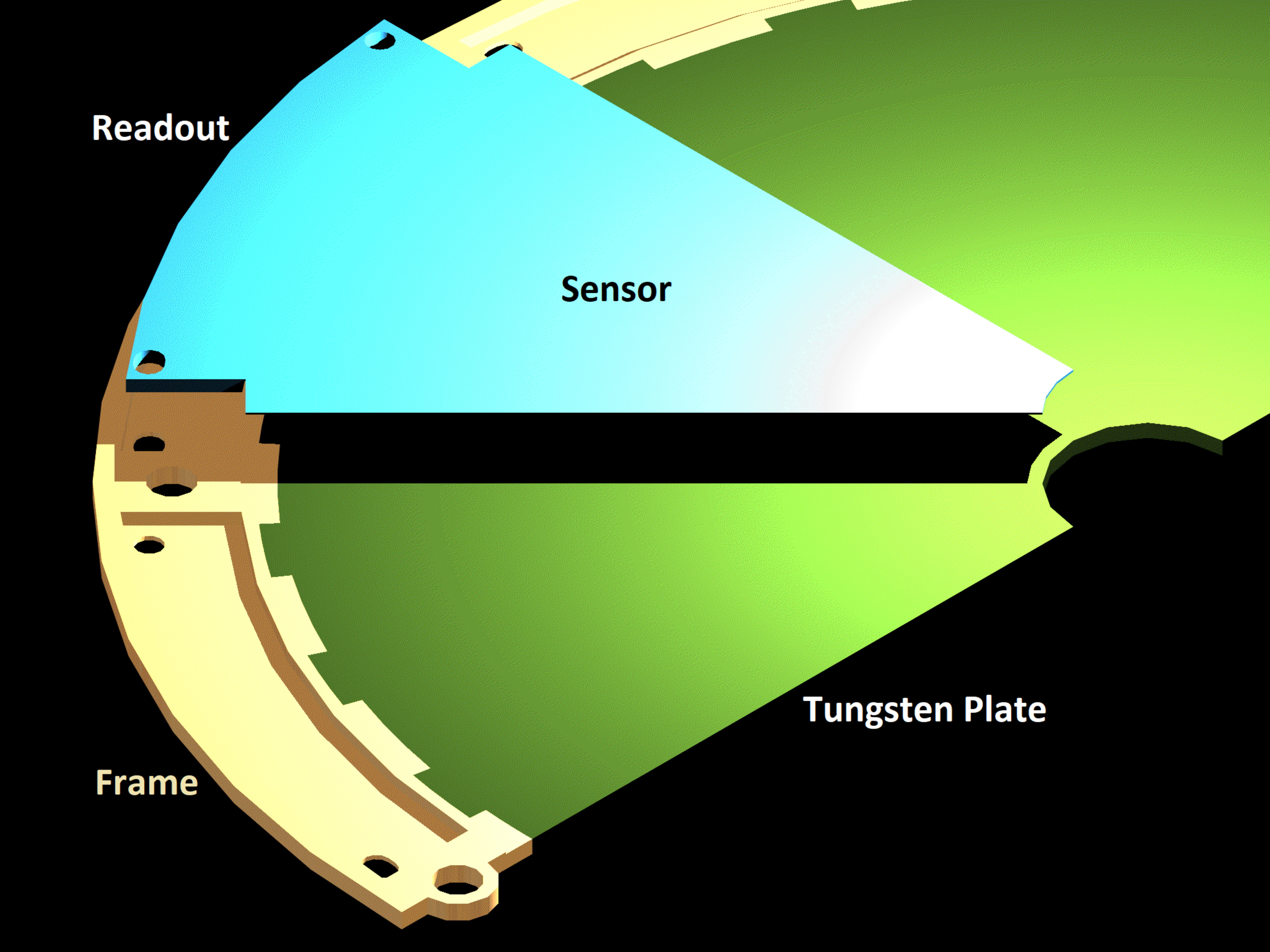}
  \end{center}
          \caption{Schematic view of a half-layer of BeamCal, showing the tungsten absorber, the sensor with the readout electronics at its periphery, and the steel frame.}
    \label{fig:layer}
\end{figure}

The conceptual design, prototyping and testing in beam of LumiCal and BeamCal elements are being pursued by the FCAL collaboration~\cite{FCAL_website}. 
While the focus in FCAL is primarily on the very forward calorimeters for the ILC, many of the concepts developed, and the simulation tools used, can be applied for CLIC studies.
The small Moli\`ere radius in LumiCal and BeamCal is achieved by a compact design of cylindrical sandwich calorimeters: tungsten absorber discs of about 1\,\xo
thickness (3.5\,mm) are interspersed with segmented silicon (for LumiCal) or GaAs sensor planes (for BeamCal), as sketched in~\cref{fig:layer}. 

At CLIC, with centre-of-mass energies of up to 3\,TeV, 40 such sandwich layers are required in both LumiCal and BeamCal. In order to achieve a small distance between the tungsten plates (1\,mm were recently achieved~\cite{VETA:2018}, 0.5\,mm is the ultimate goal), the readout electronics is located outside of the calorimeter structure. Given the time structure of the CLIC beam, with bunch trains colliding at 50\,Hz, power pulsing of the calorimeters allows the heat dissipation to be strongly reduced. Some cooling is needed for the electronics, on the periphery of LumiCal and BeamCal, where space is available.

The Bhabha cross section has a very strong polar-angle dependence. Given the required absolute accuracy of the luminosity measurement (few $10^{-3}$), this strong polar-angle dependence translates into tight mechanical tolerances for the construction and alignment of LumiCal. The inner radius of LumiCal has to be controlled with a precision of about 10\,$\upmu$m, and the distance between the LumiCal modules on both sides of the IP to better than 100\,$\upmu$m.  While this has been shown to be feasible, and that monitoring these positions with lasers can be achieved, the integration and detailed design in the CLIC detector is still pending.

\section{Present FCAL activities}\label{sec:fcal-activities}
       
\subsection{Flexible mechanical infrastructure for beam tests}
\begin{figure}[ht]
  \centering
  \vspace{5mm}
  \begin{subfigure}[T]{.48\linewidth}
    \centering
    \includegraphics[height=1.25\linewidth]{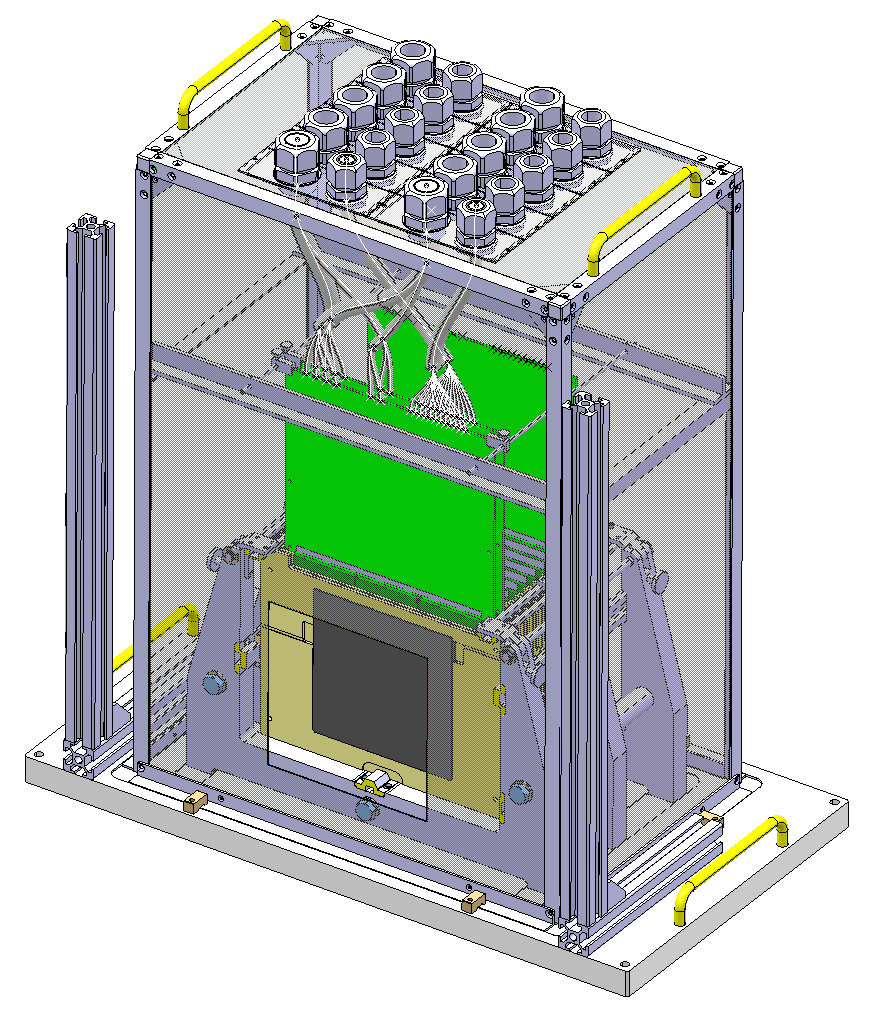}
    \caption{}\label{fig:infra_sketch}
  \end{subfigure}\hspace{5mm}
  \begin{subfigure}[T]{.48\linewidth}
    \centering
    \includegraphics[height=1.35\linewidth]{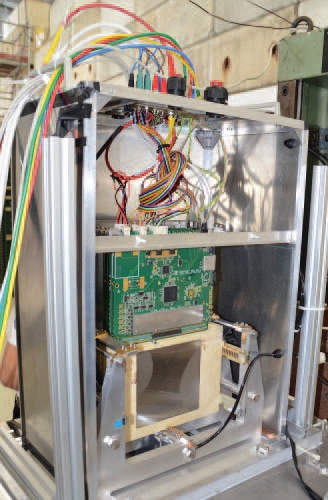}
    \caption{}\label{fig:infra_photo}
  \end{subfigure}
  \caption{The flexible mechanical structure developed for FCAL prototype tests as described in~\cite{2017EPJ}.
Tungsten plates of 14$\times$14\,cm, glued on permaglass frames, are inserted into slots of the precision combs. Sensor planes connected to PCBs (in green) are placed in between the tungsten plates.
\subref{fig:infra_sketch} A drawing of the structure~\cite{2017EPJ}.
\subref{fig:infra_photo}~A picture of the structure during beam tests~\cite{Idzik:2016xpn}.}
\label{fig:infra}
\end{figure}

A mechanical structure, precise but flexible to accommodate different LumiCal or BeamCal prototype systems, has been built and is shown in~\cref{fig:infra}. Up to 30 tungsten absorber plates of 3.5\,mm thickness, glued to permaglass frames, can be very precisely positioned on a set of assembly \emph{combs}, with exactly 1\,mm distance between the tungsten plates. Sensor planes can be inserted between the tungsten plates, with flexibility to choose the layout of each prototype (e.g. 2\,W plates / 1 \,sensor / 2 \,W plates etc.). Precision machining of tungsten plates is challenging: the flatness of the four best tungsten plates used is better than 40\,$\upmu$m, their roughness is less than 10\,$\upmu$m -- a set of plates from a second manufacturer fails to meet these specifications by a factor of 3 to 5. The mechanical structure is housed in a light-tight box, equipped with feed-throughs for signals and for nitrogen gas-flow cooling.
            
\subsection{Additional high quality tungsten absorber plates}
Partners in industry have produced a batch of 25 additional W absorber plates. The detailed composition of the material, given by the manufacturer, is W 92.5\%, Ni 5.25\% and Cu 2.25\%, similar to the plates purchased elsewhere earlier. The nominal thickness of the plates is 3.5\,mm. When measured with a Zeiss 3D coordinate measurement system (precision 2.5\,$\upmu$m), variations of the average thickness from 3.43 to 3.62 mm were found, with the exception of one outlier. On average, the density of the plates is found to be 17.47$\pm$ 0.02\,g/cm$^3$, in agreement with expectations for the type of alloy used. The flatness of the plates varies from better than 50\,$\upmu$m to around 100\,$\upmu$m, depending on the plate. The plates will be mounted into permaglass frames, and will be used in the mechanical structure for future LumiCal prototype test-beam campaigns (see~\cref{subsec:next_prototype}).

\subsection{Currently used sensors and ASICs}\label{sec:FcalAsics}
\begin{figure}[ht]
  \centering
  \begin{subfigure}[T]{.45\linewidth}
    \includegraphics[width=\linewidth]{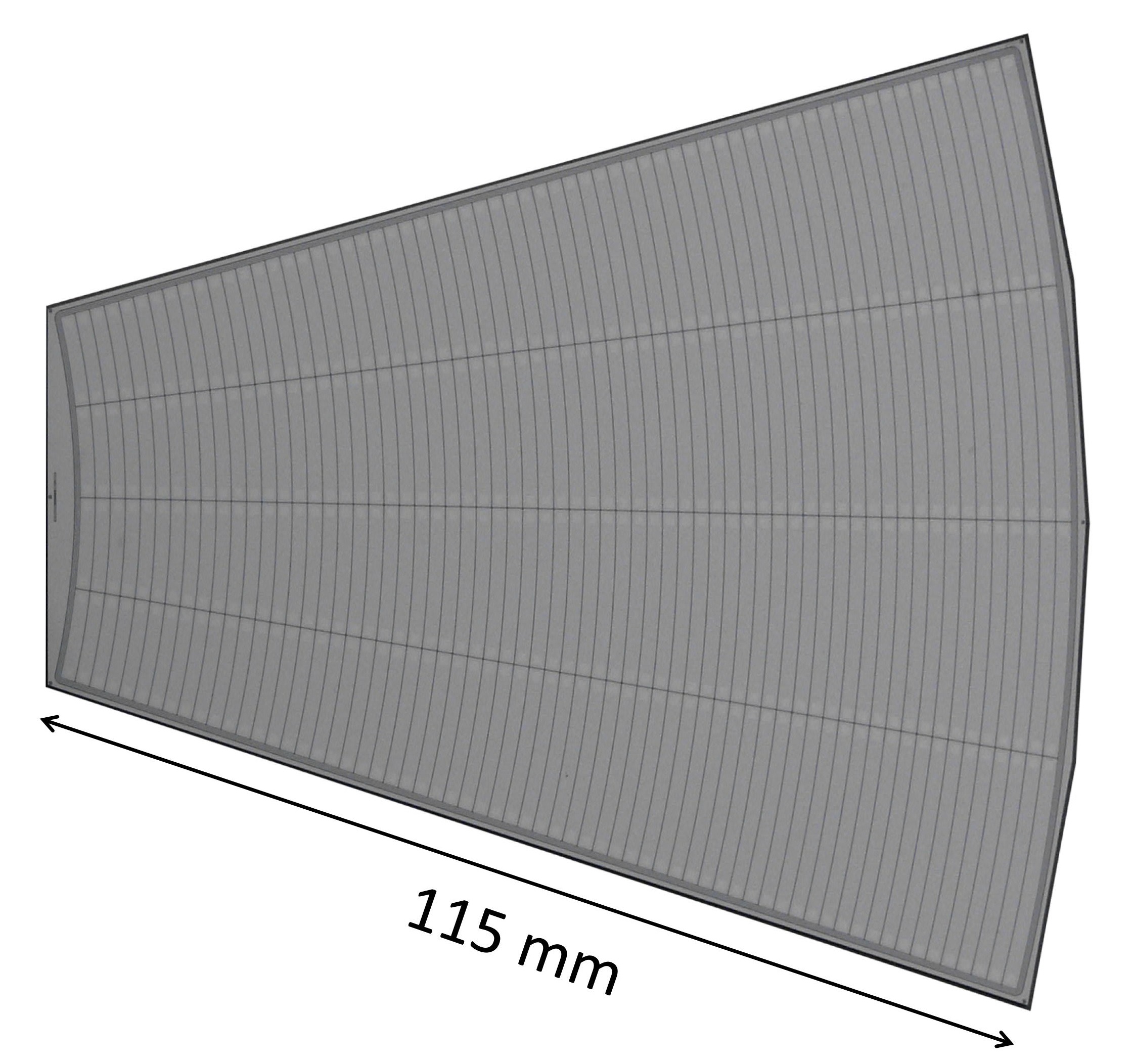}
    \caption{}\label{fig:lumical_sensor}
  \end{subfigure}
  \begin{subfigure}[T]{.45\linewidth}
    \includegraphics[width=\linewidth]{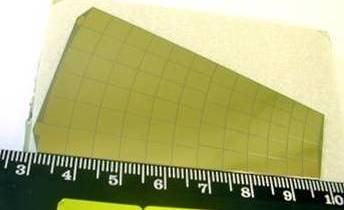}
    \caption{}\label{fig:beamcal_sensor}
  \end{subfigure}
  \caption{\subref{fig:lumical_sensor} LumiCal silicon pad sensor, and \subref{fig:beamcal_sensor} GaAs sensor for BeamCal.}
\label{fig:sensors}
\end{figure}

Prototypes of LumiCal sensors for the ILC detectors have been manufactured in industry. These n-type silicon sensors are 320\,$\upmu$m thick. Each sensor covers a 30$^\circ$ ring segment (see~\cref{fig:sensors}, left), with a radial pitch of 1.8 mm pads (gaps of 0.1 mm between pads). The bias voltage for full depletion ranges from 39 to 45\,V, and the leakage currents per pad are below 5 nA above full depletion~\cite{eudet_sensors}.
     
Large area prototype GaAs sensors for BeamCal, shown in~\cref{fig:sensors} (right), were developed for the ILC detectors and produced with partners in industry. The Liquid Encapsulated Czochralski (LEC) method was used. The sensors were doped by a shallow donor (Sn or Te), then compensated with Chromium.  This results in a semi-insulating GaAs material with a resistivity of about 10$^7$\,\si{\ohm}m. The sensors are 500\,$\upmu$m thick, with pads of several mm$^2$ area. The operation voltage is about 100\,V with leakage current per pad of less than 500\,nA.

A dedicated ASIC for LumiCal, containing 8 front-end channels, was designed choosing an architecture comprising a charge-sensitive amplifier and a shaper. Such an ASIC was produced in 0.35\,$\upmu$m CMOS technology~\cite{4600902}. In addition, a low-power, small-area multichannel ADC ASIC was developed~\cite{6156491}. It contains eight 10-bit power and frequency scalable pipeline ADCs, and the necessary auxiliary equipment. The 32-channel readout system, based on four sets of the front-end and ADC ASICs, was developed and used successfully in several test-beam campaigns~\cite{sesnorplaneperformance}.
 
\subsection{Development of the FLAME readout ASIC}
\begin{figure}[ht]
\begin{center}
    \includegraphics[width=0.9\columnwidth]{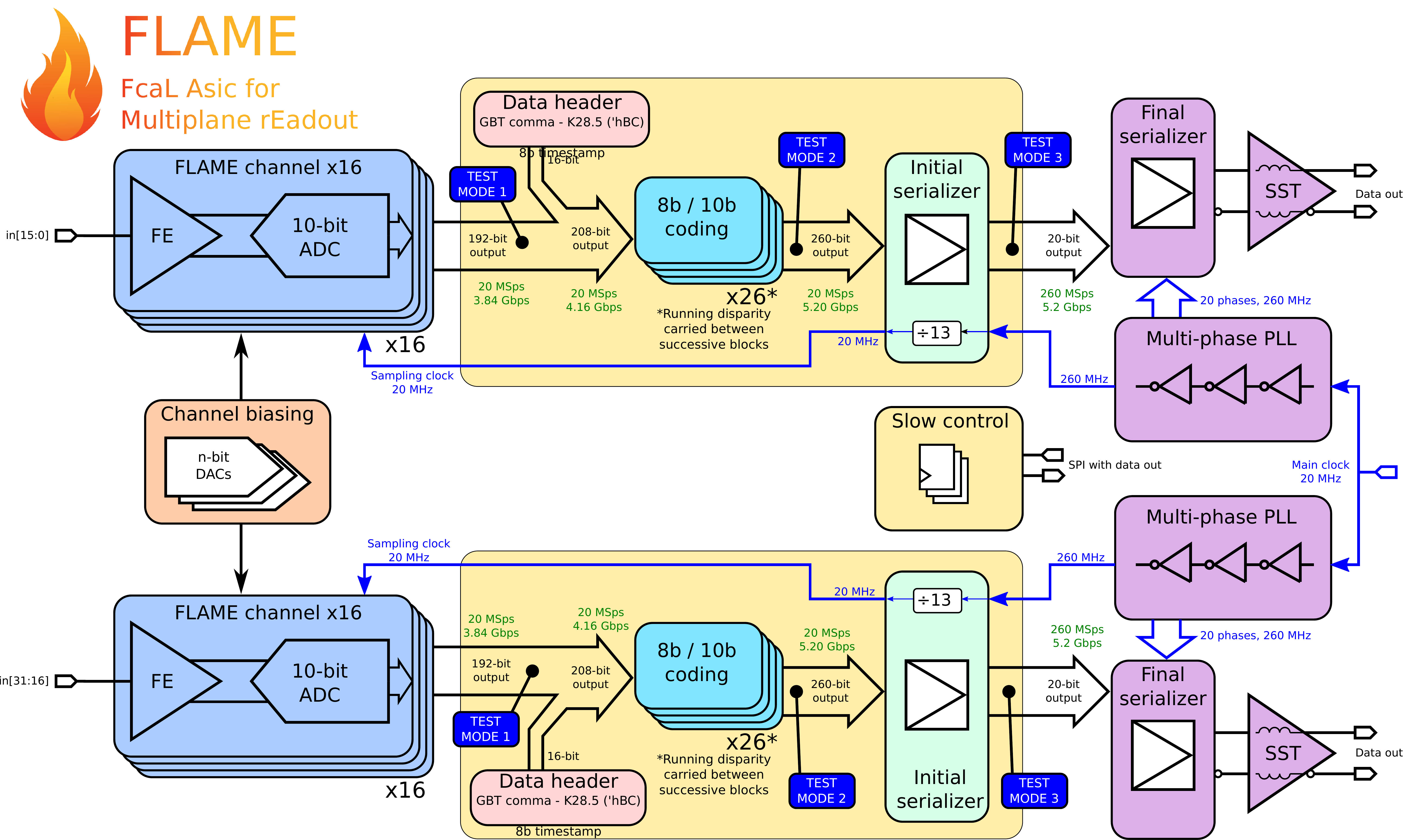}
  \end{center}
          \caption{Block diagram of the FLAME readout ASIC.}
    \label{fig:FLAME}
\end{figure}

The existing readout is limited in the number of channels, which allows building only small (32 readout channels) prototypes of detector modules. For this reason a new development of a LumiCal readout ASIC called FLAME (FcaL Asic for Multiplane rEadout) was started~\cite{Idzik:2016xpn}. The block diagram of FLAME is shown in~\cref{fig:FLAME}. 
FLAME is based on the same architecture as the previous readout, with an analogue front-end and a 10-bit ADC in each channel.
It is developed in a smaller feature-size TSMC 130\,nm CMOS technology. This choice allows to obtain a large reduction of the power consumption and much better radiation hardness. A System on Chip (SoC) architecture was chosen, comprising all functionality (analogue front-end, ADC, data serialisation and transmission) in one ASIC. This will simplify the architecture of the overall readout system, minimising the number of its components. FLAME is designed as a 32-channel ASIC. By designing a readout board with 8 ASICs one can build a detector module that reads the whole LumiCal sensor tile, which contains 256 channels. The chip is built of two identical 16-channel blocks. The data from each block is sent out by a very fast (5.2 {Gbps}) serialiser and a serial data transmission block. The output data are coded and formatted and can be received directly by fast FPGA links.

The development of FLAME is in an advanced stage. Two prototypes of critical blocks, i.e. the 8-channel front-end plus ADC ASIC and the serialiser and data transmission ASIC, were produced and tested. Presently the integration into a complete FLAME ASIC is ongoing.

\section{Test-beam results}\label{sec:fcal-testbeam}
       
\subsection{Performance of prototype calorimeters}
\label{subsec:prototype_calo}

\begin{figure}[ht]
  \centering
  \begin{subfigure}[T]{.45\linewidth}
    \includegraphics[width=\linewidth]{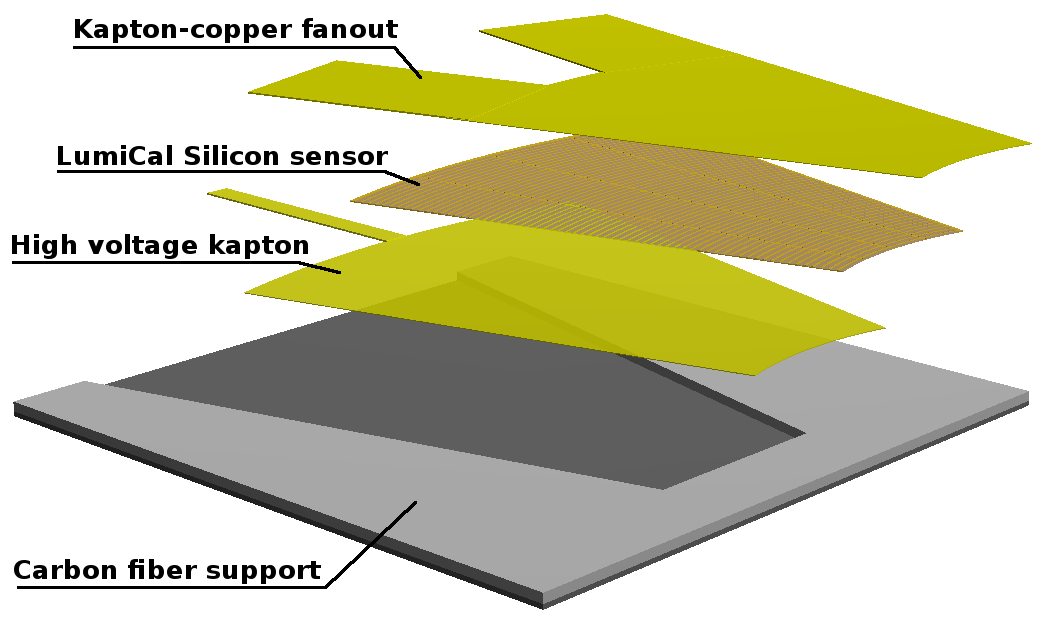}
    \caption{}\label{fig:thin_sketch}
  \end{subfigure}
  \begin{subfigure}[T]{.45\linewidth}
    \includegraphics[width=\linewidth]{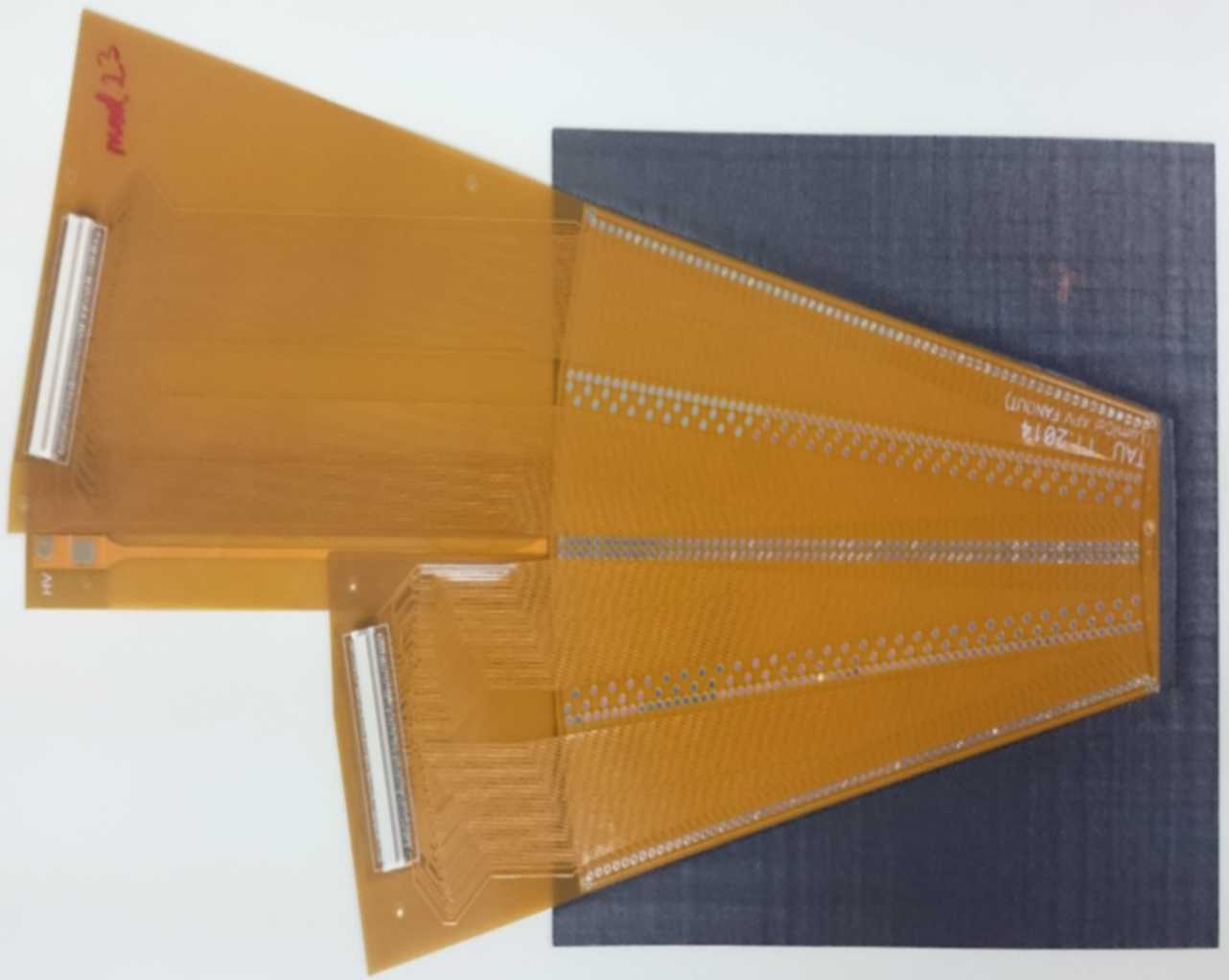}
    \caption{}\label{fig:thin_sensor}
  \end{subfigure}
  \caption{\subref{fig:thin_sketch} Schematic and \subref{fig:thin_sensor} photograph of a thin LumiCal module assembly. The thickness of the adhesive layers (not shown) between the components amounts to 10 -- 15\,$\upmu$m. The total thickness of the assembly in the sensor/fan-out region is 650\,$\upmu$m, while the outer frame of the carbon fibre support is less than 750\,$\upmu$m.}
\label{fig:thin}
\end{figure}

Prototypes of detector planes using LumiCal and BeamCal sensors, equipped with FE and ADC ASICs, were built and successfully operated in a test-beam~\cite{sesnorplaneperformance}. In a next step, fully equipped detector planes using LumiCal sensors were used to study the shower development in a stack of up to 11 tungsten plates interleaved with four sensor planes, inserted in different positions for a sequence of measurements. Multilayer operation was demonstrated and lateral and longitudinal shower development was measured and found to be in good agreement with Monte Carlo simulation~\cite{2017EPJ}. The thickness of the readout system imposed a distance between tungsten plates of 4.5\,mm -- correspondingly, the shower spread laterally much more than required in LumiCal and BeamCal, and an effective Moli\`ere radius of 24.0$\pm$1.7\,mm was found.
For a further test-beam campaign, much thinner sensor/readout planes were developed, allowing a 1\,mm spacing between the tungsten plates in the stack. A schematic view and a photograph of such a sensor plane assembly is shown in~\cref{fig:thin}.

Six such sensor layers were interspersed by tungsten absorbers of 1\,\xo. 
The sensors were read out using the APV25 chip hybrid board~\cite{APV_ieee, APV_nima}. Charge dividers were used to enlarge the dynamic range of the APV25 chip. The powering circuits and the fan-out part which connects silicon sensor pads with the FE board inputs were made from flexible Kapton-copper foils with thickness of 70\,$\upmu$m for the high voltage one, applied to the back n-side of the sensor, and about 120\,$\upmu$m for the fan-out. Ultrasonic wire bonding was used to connect conductive traces on the fan-out to the sensor pads. A support structure made of carbon fibre composite with a thickness of 100\,$\upmu$m in the sensor-gluing area provides mechanical stability. The ultrasonic wire bonding proved to provide good electrical performance, but for a module thinner than 1\,mm, the wire loops, which are typically 100$-$200\,$\upmu$m high, cause a serious problem when the module needs to be installed in a 1\,mm gap between absorber plates. The bonding machine was tuned to make the loop as low as possible and technically acceptable. The sampling based measurements, which were done using a confocal laser scanning microscope, show that the loop height is in the range from 50\,$\upmu$m  to 100\,$\upmu$m. The total thickness of the sensor plane was 650\,$\upmu$m.

\begin{figure}[ht]
\begin{center}
    \includegraphics[width=0.6\columnwidth]{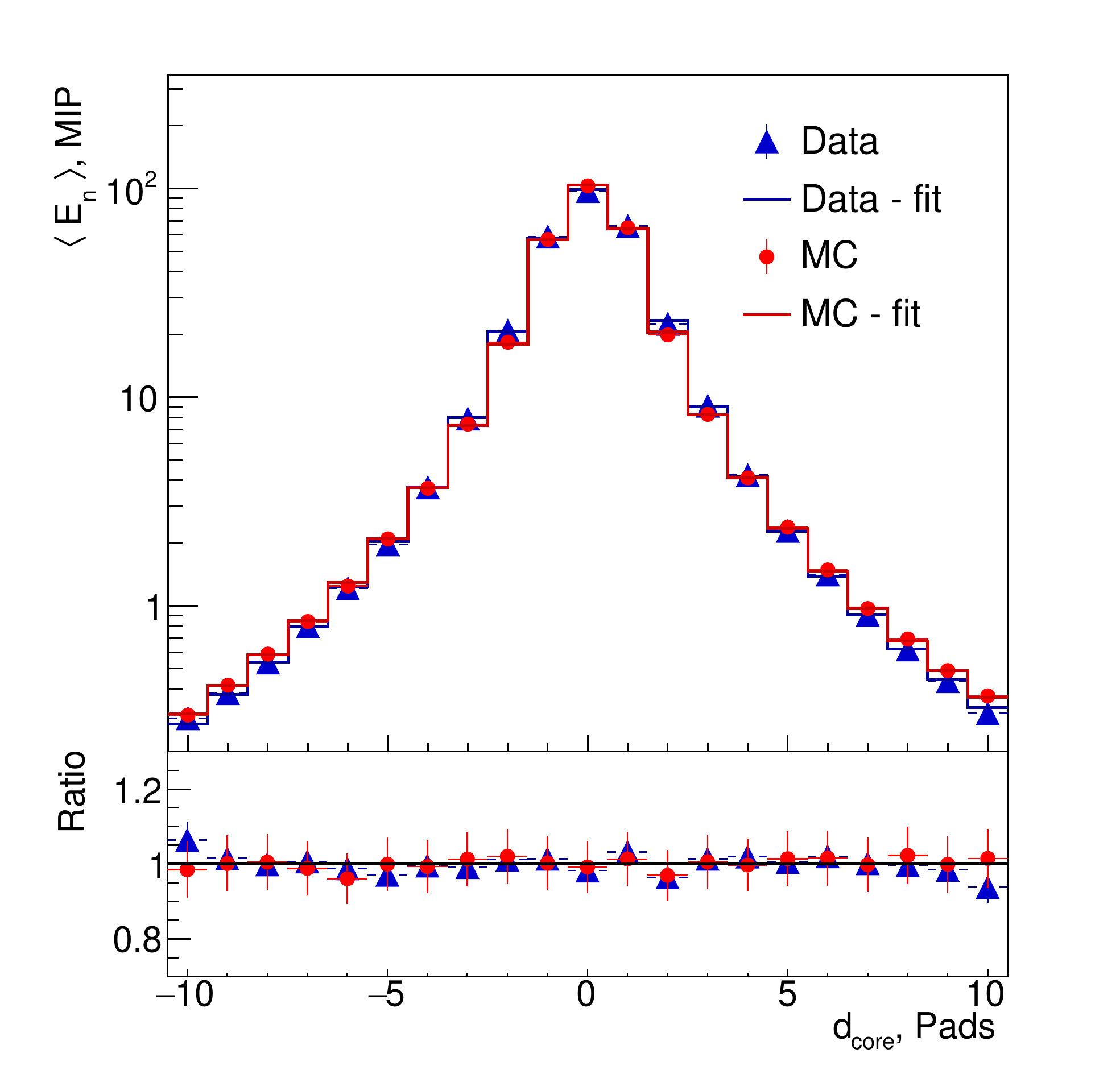}
  \end{center}
          \caption{ The average transverse shower profile, $\langle E_\text{n} \rangle$, as a function of the distance from the core, $d_\text{core}$, in units of 1.8 mm wide pads, for data (blue triangles) and MC simulation (red circles). The histograms are the results of fits to data and MC using a parameterisation of the shower shape. The lower part of the figure shows the ratio of the distributions to the fitted function, for the data (blue) and the MC (red).}
    \label{fig:MR}
\end{figure}

The calorimeter prototype was studied in a 1 to 5 GeV electron beam at DESY. The shower position was reconstructed with a resolution of (440$\pm$20)\,$\upmu$m. The average transverse shower profile for 5 GeV electrons is shown in~\cref{fig:MR} for data and Monte Carlo simulation. The effective Moli\`ere radius for this setup is determined to be 8.1$\pm$0.3 mm in data, in agreement with the 8.4$\pm$0.1 mm obtained in Monte Carlo simulations~\cite{new_FCAL_paper_2019}.

\subsection{Radiation damage studies}
Two studies of the radiation tolerance of potential BeamCal sensors have been carried out. 
In the first study, the radiation tolerance of prototype GaAs sensors has been explored by exposing the sensors to direct radiation from an electron beam of about 10 MeV~\cite{sdalinac}. 
It was found that the sensors can be operated at room temperature up to approximately 1\,MGy with only a factor of two increase in the leakage current~\cite{1748-0221-7-11-P11022}; however, 
a significant loss in the response to ionising particles was observed. This loss can be partially compensated by increasing the bias voltage. 
In addition, a new series of GaAs:Fe prototypes is under study: these include a small dopant concentration of iron, which is expected to mitigate radiation damage effects.  

In the second study~\cite{2017arXiv1703.05429}, different types of solid-state sensors were exposed to varying levels of radiation induced by electrons from the SLAC End Station A Test Beam (ESTB). 
For this study, the beam, with energies varying between 3 and 15\,GeV, was directed into a tungsten beam dump. This dump was split at the depth of the shower maximum and the sensor inserted, 
leading to an exposure incorporating the full spectrum of particle species that will irradiate the BeamCal sensors. 
GaAs and silicon sensors were exposed, as well as samples of sapphire and SiC.
For GaAs, the observed charge collection loss was similar to that of the first study, although a high room-temperature leakage current of order 10\,$\upmu$A/cm$^2$ was observed for 1\,MGy-scale doses for a sensor bias of 600\,V.
SiC sensors and sapphire sensors were irradiated to 0.8\,MGy and\,3 MGy, respectively, leading to a charge-collection loss of 50\% and 75\%.
For silicon diode sensors irradiated to 6\,MGy, the charge-collection loss measured was 40\%, but a significant leakage current was observed,
which could however be reduced after annealing and by lowering the operating temperature to -30$^\circ$\,C.
Radiation damage studies for FCAL are continuing at different facilities, and are also being complemented by FLUKA Monte Carlo studies~\cite{FLUKA2}.

\section{Applications and future plans}\label{sec:fcal-future}

\subsection{Applications other than at Linear Colliders}
The expertise acquired within the FCAL collaboration for radiation-hard sensors and fast front-end electronics was used to build, commission and operate fast beam-condition monitors at the CMS experiment at LHC.
Radiation-hard sensors developed by FCAL are used as beam-loss monitors with nano-second time resolution at FLASH, European XFEL and LHC~\cite{Ignatenko:2012zz,BELL2010433}.
In addition, front-end ASICs using design features similar to the ones developed for FCAL are being produced for the upgrade of the LHCb tracker.
       	
\subsection{Next LumiCal prototype}
\label{subsec:next_prototype}
FCAL is preparing a full-depth LumiCal prototype for ILC with at least 20 sensor planes, to be tested with beam. The existing mechanical structure will be equipped with 3.5 mm thick W plates and sensors mounted in the recently developed, very compact assembly (see~\cref{subsec:prototype_calo}). FLAME ASICs and FPGAs for data concentration are being prepared. In addition, a pixel tracker in front of the LumiCal prototype will improve the impact position resolution and allow separating electrons from photons.

\subsection{Pixelated LumiCal and BeamCal for CLIC}\label{sec:pixelated-lumical}
Given the high occupancies in LumiCal and BeamCal, shown in \cref{chap:electronics-daq}, sensors with much smaller pads than currently used in beam tests are desirable. Ongoing R\&D on fully integrated CMOS silicon sensors with small pixel sizes (see \cref{sec:clictd}) open perspectives towards pixelated (analogue or digital) LumiCal sensor layers, offering compactness as well as large dynamic range for the measurement of high-energy electromagnetic showers.


\chapter{Readout electronics and data acquisition system}
\label{chap:electronics-daq}
This chapter covers subdetector readout implementations and the resulting data rates at CLIC. Selected example readout schemes anticipated for the different subdetectors of CLICdet are summarised briefly. Based on these example implementations, an estimation of the data rates at CLIC is given. Finally, scenarios for data transfer from the detector to the computing infrastructure are discussed.

\section{Subdetector implementation}\label{sec:implementation}
\paragraph*{Vertex detector and first tracker disc:}
Silicon pixel detectors with a cell size of $25\times25\,\upmu\text{m}^2$ are considered for the vertex detector as well as the first tracker disc.
Using the readout implementation as proposed in CLICpix2 (see~\cref{sec:CLICpix2}), the charge (5\,bits) as well as the arrival time (8\,bits) is recorded for one hit readout per bunch train (total of 13\,bits).
Zero suppression is applied as well as address encoding.

\paragraph*{Main tracker:}
A pixel tracker implementation based on the CLICTD monolithic sensor (see~\cref{sec:clictd}) using $30\times300\,\upmu$m$^2$ super-pixels is under consideration for the main tracker.
Each super-pixel is furthermore subdivided into 8 pixels of $30\times37.5\,\upmu$m$^2$ surface.
The charge (5\,bits) as well as the arrival time (8\,bits) of one super-pixel is recorded for one hit readout per bunch train. 
An additional 8\,bits provide a hit map within the super-pixel (total of 21\,bits).
Zero suppression is applied as well as address encoding.

\paragraph*{ECAL and HCAL:}
Silicon pad sensors with a cell size of $5 \times 5\,\text{mm}^2$ are considered for the active layers of the ECAL. For HCAL, scintillator tiles with a size of $30 \times 30\,\text{mm}^2$ each coupled to a silicon photomultiplier are considered.
For the ECAL and HCAL readout introduced in~\cref{sec:ECAL} and~\cref{sec:HCAL}, sampling the calorimeter-cell pulse height (16\,bits) at regular intervals (40\,MHz, corresponding to a sampling period of 25\,ns) is foreseen for operation at CLIC.
The sampling time window for the calorimeters is 200\,ns containing the full bunch-train duration (156\,ns at 3\,TeV) and additional time for the shower development, resulting in 8 time samples.
The required time resolution of the calorimeter is then achieved by offline signal shape analysis of the recorded pulse heights.
Zero suppression is applied as well as address encoding.

For low-occupancy regions in the calorimeters, the pulse-height sampling may be replaced with one hit readout per bunch train combined with individual hit time stamping, as is planned for ILC operation as discussed in~\cref{sec:ahcal_timing}.
For the data rate estimation discussed below, continuous sampling provides an upper limit.

\paragraph*{LumiCal and BeamCal:} 
Silicon pad sensors with a cell size that increases as a function of the radius from $3.75 \times 13\,\text{mm}^2$ to $3.75 \times 44\,\text{mm}^2$ are considered for the active layers of LumiCal.
For BeamCal, GaAs or diamond sensors with a cell size $8 \times 8\,\text{mm}^2$ are foreseen.
For the LumiCal and BeamCal readout discussed in \cref{sec:FcalAsics}, sampling the pulse height (16\,bits) of the calorimeter cells with 100\,MHz (corresponding to a sampling period of 10\,ns) throughout the full bunch train and shower development period is assumed, resulting in 20 time samples.
Also here, the required time resolution is achieved by offline signal shape analysis.
It is anticipated that the readout solution for the very forward detectors accommodates up to 5 subsequent hits per bunch train to cope with the high occupancies.
When including safety factors, this value is currently exceeded in both LumiCal and BeamCal.
Further detector optimisation is foreseen in the next project phase in order to reduce the high cell occupancies.
For LumiCal zero suppression is applied as well as address encoding.
Due to the high occupancies in the current BeamCal design, no zero suppression is considered, and therefore serial readout without address encoding can be performed.

\paragraph*{Muon detectors:}
For the muon detectors with $30\times30\,$mm$^2$ readout cells, a binary readout with a multi-hit TDC (24\,bits) is foreseen for one hit readout per bunch train.
Zero suppression is applied as well as address encoding.
The background occupancies for the muon detectors are obtained with additional shielding against secondary particles~\cite{ArominskiEDMS}.

\section{Estimation of CLIC detector data rates}
\label{sec:datarate}

\subsection{Readout parameters for CLICdet subdetectors}
\cref{tab:DataRateCLIC3TeV} summarises different readout aspects of the CLICdet subdetectors.
It includes the timing requirements of the CLICdet subdetectors, the foreseen sizes of the cells and the corresponding number of cells for the implementation examples discussed before.
In addition, average and maximal background occupancies per bunch train are included~\cite{ArominskiEDMS}, which can be approximated for low-occupancy regions (see~\cite{SailerOccupancy}) as
\begin{equation}
 \text{Occupancy/train} = \sum_{\text{process}} n_\text{hits, process}/ (\text{mm}^2 \cdot \text{BX}) \times n_{\text{bunches}} \times w \times l \times cs \times sf_{\text{proc}}.
\label{eq:occ}
\end{equation}
Here, the number of hits ($\text{n}_{\text{hits}}$) for two background processes, incoherent $\Pep\Pem$ pairs and $\PGg\PGg\to$ hadrons events, are included.
These are the dominating background processes for the detector region (see~\cref{chap:overview-backgrounds}).
The number of hits per bunch crossing (BX) per mm$^2$ are multiplied by the number of bunch crossings per train ($n_{\text{bunches}}$, e.g.\ 312 at 3\,TeV).
The cell width ($w$) and the cell length ($l$), as well as the average cluster size ($cs$) are taken into account.
The cluster size corresponds to the estimated number of readout cells per hit
(3 for the tracking detector, 5 for the vertex detector, 1 for all other detector regions).
Safety factors ($sf_\text{proc}$, 2 for $\PGg\PGg\to$ hadrons, 5 for incoherent pairs) to account for the uncertainty of the background-process simulation are applied.

For high-occupancy regions like the calorimeter endcaps and the very forward calorimeters, the low-occupancy approximation used in \cref{eq:occ} is not valid.
For these regions, safety factors as well as occupancies from different sources need to be combined probabilistically using the method described in \cite{SailerOccupancy}.

Furthermore, \cref{tab:DataRateCLIC3TeV} lists the number of bits to be read out per hit cell. 
These are, in case of the calorimeters, multiplied by the number of time samples corresponding to the bunch train duration and shower development period.

\begin{table}[hbtp]
\caption{Readout parameters for an example implementation of CLICdet at 3\,TeV, including zero suppression and address encoding.
The address encoding is implemented per detector region, assuming non-ambiguous addressing within each row in the table.
The average and maximal occupancies from beam-induced backgrounds and the resulting data volumes include safety factors as described in the text.
BeamCal is fully saturated and the data volume corresponds to a full subdetector readout.
}
\label{tab:DataRateCLIC3TeV}
\footnotesize
 \centering
     \begin{tabular}{l  c c r @{$\times$} l r r @{ / } l c r}
      \toprule
Detector region        & time      & hit         & \multicolumn{2}{c}{}        & number   & \multicolumn{2}{c}{average to} & number   & data    \\
                       & sampling  & time        & \multicolumn{2}{c}{cell}    & of       & \multicolumn{2}{c}{max.\ train}& of bits  & volume   \\
                       & period    & resolution  & \multicolumn{2}{c}{size}    & channels & \multicolumn{2}{c}{occupancy}  & per cell & per train  \\
                       & [ns]      & [ns]        & \multicolumn{2}{c}{[mm$^2$]}& [10$^6$] & \multicolumn{2}{c}{[\%]}       & [bits]   & [MByte] \\
\midrule
Vertex barrel 1        & 10        & $\sim 5$    & 0.025           & 0.025     & 89       &   2.1    & 2.8                 & 13       & 9.5     \\
Vertex barrel 2        & 10        & $\sim 5$    & 0.025           & 0.025     & 95       &   1.6    & 2.1                 & 13       & 7.7     \\
Vertex barrel 3        & 10        & $\sim 5$    & 0.025           & 0.025     & 126      &   0.62   & 0.74                & 13       & 3.9     \\
Vertex barrel 4        & 10        & $\sim 5$    & 0.025           & 0.025     & 132      &   0.52   & 0.61                & 13       & 3.4     \\
Vertex barrel 5        & 10        & $\sim 5$    & 0.025           & 0.025     & 166      &   0.24   & 0.31                & 13       & 2.1     \\
Vertex barrel 6        & 10        & $\sim 5$    & 0.025           & 0.025     & 172      &   0.21   & 0.25                & 13       & 1.8     \\
\midrule
Vertex disc 1          & 10        & $\sim 5$    & 0.025           & 0.025     & 93       &   0.39   & 2.1                 & 13       & 1.8     \\
Vertex disc 2          & 10        & $\sim 5$    & 0.025           & 0.025     & 93       &   0.39   & 2.1                 & 13       & 1.8     \\
Vertex disc 3          & 10        & $\sim 5$    & 0.025           & 0.025     & 93       &   0.43   & 2.4                 & 13       & 2.0     \\
Vertex disc 4          & 10        & $\sim 5$    & 0.025           & 0.025     & 93       &   0.43   & 2.4                 & 13       & 2.0     \\
Vertex disc 5          & 10        & $\sim 5$    & 0.025           & 0.025     & 93       &   0.47   & 2.7                 & 13       & 2.2     \\
Vertex disc 6          & 10        & $\sim 5$    & 0.025           & 0.025     & 93       &   0.47   & 2.8                 & 13       & 2.2     \\
\midrule
Inner tracker barrel 1 & 10        & $\sim 5$    & 0.03            & 0.3       & 88       &  0.28    &  0.36               & 21       & 1.5     \\
Inner tracker barrel 2 & 10        & $\sim 5$    & 0.03            & 0.3       & 244      &  0.096   &  0.13               & 21       & 1.4     \\
Inner tracker barrel 3 & 10        & $\sim 5$    & 0.03            & 0.3       & 580      &  0.040   &  0.051              & 21       & 1.5     \\

Outer tracker barrel 1 & 10        & $\sim 5$    & 0.03            & 0.3       & 1589     &  0.014   &  0.018              & 21       & 1.5     \\
Outer tracker barrel 2 & 10        & $\sim 5$    & 0.03            & 0.3       & 2258     &  0.0080  &  0.011              & 21       & 1.2     \\
Outer tracker barrel 3 & 10        & $\sim 5$    & 0.03            & 0.3       & 2893     &  0.0054  &  0.0070             & 21       & 1.0     \\
\midrule 
Inner tracker disc 1   & 10        & $\sim 5$    & 0.025           & 0.025     & 2000     &  0.019   & 0.18                & 13       & 2.1     \\
Inner tracker disc 2   & 10        & $\sim 5$    & 0.03            & 0.3       & 252      &  0.098   & 0.71                & 21       & 1.5     \\
Inner tracker disc 3   & 10        & $\sim 5$    & 0.03            & 0.3       & 244      &  0.090   & 0.42                & 21       & 1.3     \\
Inner tracker disc 4   & 10        & $\sim 5$    & 0.03            & 0.3       & 228      &  0.084   & 0.37                & 21       & 1.2     \\
Inner tracker disc 5   & 10        & $\sim 5$    & 0.03            & 0.3       & 218      &  0.070   & 0.36                & 21       & 0.93    \\
Inner tracker disc 6   & 10        & $\sim 5$    & 0.03            & 0.3       & 208      &  0.050   & 0.14                & 21       & 0.64    \\
Inner tracker disc 7   & 10        & $\sim 5$    & 0.03            & 0.3       & 202      &  0.046   & 0.073               & 21       & 0.57    \\

Outer tracker disc 1   & 10        & $\sim 5$    & 0.03            & 0.3       & 1546     &  0.0070  & 0.025               & 21       & 0.70    \\
Outer tracker disc 2   & 10        & $\sim 5$    & 0.03            & 0.3       & 1546     &  0.0072  & 0.029               & 21       & 0.73    \\
Outer tracker disc 3   & 10        & $\sim 5$    & 0.03            & 0.3       & 1546     &  0.0066  & 0.026               & 21       & 0.67    \\
Outer tracker disc 4   & 10        & $\sim 5$    & 0.03            & 0.3       & 1546     &  0.0071  & 0.026               & 21       & 0.71    \\
\midrule
ECAL barrel            & 25        & 1           & 5               & 5         & 72       &   0.36   &  2.4                & $16\times8$  & 5.0  \\
ECAL endcap            & 25        & 1           & 5               & 5         & 29       &   1.1    &  33                 & $16\times8$  & 6.1   \\
HCAL barrel            & 25        & 1           & 30              & 30        & 4.8      &   0.11   &  0.86               & $16\times8$  & 0.10 \\
HCAL endcap            & 25        & 1           & 30              & 30        & 4.5      &   5.4    &  100                & $16\times8$  & 4.6   \\
HCAL rings             & 25        & 1           & 30              & 30        & 0.4      &   0.010  &  1.4                & $16\times8$  & 0.00075 \\
LumiCal                & 10        & 5           & 3.75            & 13--44    & 0.25     &   90     &  100                & $16\times20$ & 9.3  \\
BeamCal                & 10        & 5           & 8               & 8         & 0.093    &   \multicolumn{2}{c}{100}      & $16\times20$ & 3.7  \\
\midrule
MUON barrel            & 25        & 8           & 30              & 30        & 2.4      &  0.00020 &  0.084              & 24           & 0.000028\\
MUON endcap            & 25        & 8           & 30              & 30        & 1.7      &  0.043   &  9.2                & 24           & 0.0041  \\
\bottomrule
Total (rounded)        &           &             &  \multicolumn{2}{c}{}       & \num{18700}&   \multicolumn{2}{c}{}       &              &  88 \\
\bottomrule
\end{tabular}
\end{table}

\subsection{Event sizes}
\label{sec:address}
Based on \cref{tab:DataRateCLIC3TeV}, upper limits for the data volume of the CLIC detector are estimated including zero suppression and address encoding (except for BeamCal).
Address encoding is implemented per detector region as indicated in~\cref{tab:DataRateCLIC3TeV}.
Each cell in a detector region has a unique identifier, adding $n$\,bits to each data word for a hit cell in a subdetector region with $2^n$ channels, with $n$ being an integer. For the purpose of the address encoding, each detector region with a dedicated address space corresponds to one layer in the vertex/tracking detectors and one entire detector section for the calorimeters and muon detectors. The actual implementation of the address encoding and zero suppression can be further optimised in the next project phase.

As the beam-induced backgrounds are the dominating source of detector occupancy, the average occupancy per bunch train is taken as a basis to estimate the data volume per bunch train.
After each bunch train, the data recorded by the CLIC subdetectors are sent off the detector, without the use of triggers.

The CLIC detector with the subdetectors shown in~\cref{tab:DataRateCLIC3TeV} has in total 18.7~billion readout channels.
The data volume is 88\,MByte per bunch train at 3\,TeV, dominated by the data from the vertex barrel detector and the calorimeters.
For 380\,GeV, the beam-induced backgrounds are smaller, resulting in a data volume per bunch train of 29\,MByte.

\subsection{Alternative tracker readout concept}
The CLIC detector implementation presented here has 18.7~billion channels of which 15.2~billion channels are in the CLICTD-based pixel tracker (excluding the first tracker discs).
A second main tracker option based on silicon strips was therefore studied in view of occupancies and data rates.
The strip dimensions are chosen such that the strip occupancies do not exceed a few percent across the full tracker volume.
Strips of 50\,$\upmu$m pitch and 1 to 10\,mm strip lengths are used.
As in the CLICTD-based implementation example discussed above, the first tracker discs are equipped with $25 \times 25\,\upmu\text{m}^2$ pixels for this implementation.
Such a main tracker would have in total approximately 560~million strip detector cells resulting in 4~billion readout channels for the full CLIC detector.
In addition, for each strip detector cell only the charge (5\,bits) and the arrival time (8\,bits) would be recorded (total of 13\,bits instead of the 21\,bits for the CLICTD-based implementation including the additional 8\,bits for the super-pixel hit map).
The combined effect of the higher occupancies in this configuration, the lower number of bits read out per cell and the lower number of bits needed for the address encoding of the strip tracker would result in an approximately 25\% lower data rate of the strip layers with respect to the CLICTD-based layers.
Furthermore it would result in an overall reduction of the CLICdet data rate by approximately 5\%.

The first tracker discs are currently considered to use the same technology as the vertex detector, with $25 \times 25\,\upmu\text{m}^2$ pixels over an area of $1.2\,\text{m}^2$, in order to reduce the confusion for the track reconstruction in the forward tracker region.
In view of the occupancy shown in \cref{tab:DataRateCLIC3TeV}, the first tracker discs could be equipped with larger pixels.
Taking into account both the occupancy and the pattern-recognition needs, an optimisation of the granularity of the first tracker discs is foreseen, following a similar approach as described in~\cite{Aplin:2013oca}.

\subsection{Alternative implementation of zero suppression}
In addition to address encoding, an alternative, serial zero-suppressed implementation was evaluated, using 1\,bit for cells that are not hit and $1+n$\,bits for hit cells.
$n$ is here for instance 13 for the vertex detector and the first tracker discs and 21 for the remaining part of the  tracker.
No bits are dedicated to the address; the cell address instead can be found from the cell position in the data stream.
Due to the large number of readout channels and the low average occupancy, the data rates using this scheme would be dominated by the empty channels resulting in data rates that would be significantly larger than those using the address encoding described in \cref{sec:address}.
The data volume at 3\,TeV for the full CLICdet detector would add up to approximately \num{2380}\,MByte instead of approximately 88\,MByte using address encoding.
This is dominated by the main tracker which would have a data volume in this serial readout scheme of \num{2150}\,MByte instead of 19\,MByte.
Additional zero suppression for fully unoccupied modules could however greatly improve this serial zero-suppression method.

\subsection{Output data rates}
The data volume presented in \cref{tab:DataRateCLIC3TeV} combined with a bunch-train repetition rate at CLIC of \SI{50}{\hertz} results in an average data rate of 4.40\,GByte/s at 3\,TeV.
The corresponding estimate for the background conditions at 380\,GeV results in an average data rate of 1.45\,GByte/s.
Continuous data readout in the gaps between colliding bunches is assumed.
\cref{tab:DataVolumeComparison} compares these data rates to those of the LHC experiments.
The data rates at CLIC without the use of triggers are far below those at the LHC.

\begin{table}[ht]
\caption{Data rates of CLICdet. The numbers are compared to those of the LHC experiments~\cite{Albrecht:2018iur}.}
\label{tab:DataVolumeComparison}
\centering
\begin{tabular}{lrrlr}
\toprule
Detector               &  \multicolumn{1}{c}{Event size}   &  \multicolumn{2}{c}{Repetition rate} & \multicolumn{1}{c}{Data rate} \\
                       &  \multicolumn{1}{c}{[MByte]}      &  \multicolumn{2}{c}{[Hz]           } & \multicolumn{1}{c}{[GByte/s]}\\
\midrule
CLICdet @ 3\,TeV       &   88           &   50             & (train)           & 4.40 \\
CLICdet @ 380\,GeV     &   29           &   50             & (train)           & 1.45 \\
\midrule
ALICE (2017)           &  100           &   300            & (central Pb--Pb)  & 30 \\
ATLAS (2017)           &  1             &   100\,000       & (level-1 trig.)   & 100 \\
CMS (2017)             &  1.5           &   100\,000       & (level-1 trig.)   & 150 \\
LHCb (2017)            &  0.07          &   1\,000\,000    & (level-0 trig.)   & 70 \\
\midrule
ALICE (Run 3)          &  60            &   50\,000        & (no trig.)        & 3\,000 \\
ATLAS (HL-LHC)         &  5             &   1\,000\,000    & (level-1 trig.)   & 5\,000 \\
CMS (HL-LHC)           &  4.2--4.6      &   750\,000       & (level-1 trig.)   & 3\,150--3\,450 \\
LHCb (Run 3--4)        &  0.13          &   30\,000\,000   & (no trig.)        & 3900 \\
\bottomrule
\end{tabular}
\end{table}

The CMS permanent storage rate after the first trigger level and further online data treatment is currently $\sim5$\,GByte/s and therefore at the same level as the CLIC data rate without the use of triggers.
Therefore, one can conclude that the data-storage needs for the running CLIC experiment will not exceed those of a current multi-purpose LHC detector.

During calibration runs of CLICdet, no zero suppression and address encoding would be used.
In this configuration, one full detector readout would produce a data volume of 47\,GByte for CLICdet using a CLICTD-based main tracker or 8\,GByte for CLICdet using a strip tracker.
These numbers include approximately 1.8\,GByte for the full calorimeter system including multiple time samples as listed above.
The larger data volume per calibration event would need to be compensated by repetition frequencies well below the bunch train repetition rate or by data links and computing infrastructure optimised for the larger throughput during calibration runs. The required additional data-transfer and storage volume for calibration data could however be significantly reduced by implementing on-chip calibration features for the tracking detectors and calorimeters.

\section{Data links}
Based on the data rates discussed in \cref{sec:datarate}, the number of optical data links needed for CLICdet readout can be estimated.
In view of the ongoing development of radiation-tolerant optical data links, a bandwidth of O(20\,Gbit/s) per link is predicted for the time beyond 2025~\cite{CERN-EP-detectorRD-programme}.

The CLIC detector readout during 3\,TeV operation would be achievable with a few 20\,Gbit/s links, assuming continuous readout in the gaps between colliding bunches.
Taking into account the geometrical distribution of the detector elements, more links would be needed.

The CLIC detector readout during calibration runs with event sizes of up to 47\,GByte could be achieved with approximately 100 data links for 5\,Hz or 1000 data links for 50\,Hz.
Data taking in calibration runs is then limited by the computing infrastructure, namely the availability of storage space and the rate of writing the data to storage.
The implementation of on-chip calibration features for the tracking detectors and calorimeters would, on the other hand, significantly reduce the optical data-link bandwidth for calibration data taking.

\section{Detector readout optimisation for high-occupancy regions}
\label{sec:detector-optimisation-high-occupancy}
The currently foreseen configurations of BeamCal, LumiCal and the HCAL endcap see train occupancies well above a few \% at 3\,TeV.
Three options are considered to reduce the occupancies in the main calorimeter-endcaps and forward-calorimeter regions~\cite{vanDam:1751528,CLICdet-background}.

\begin{figure}[t]
\begin{subfigure}[T]{.49\linewidth}
\includegraphics[width=\linewidth]{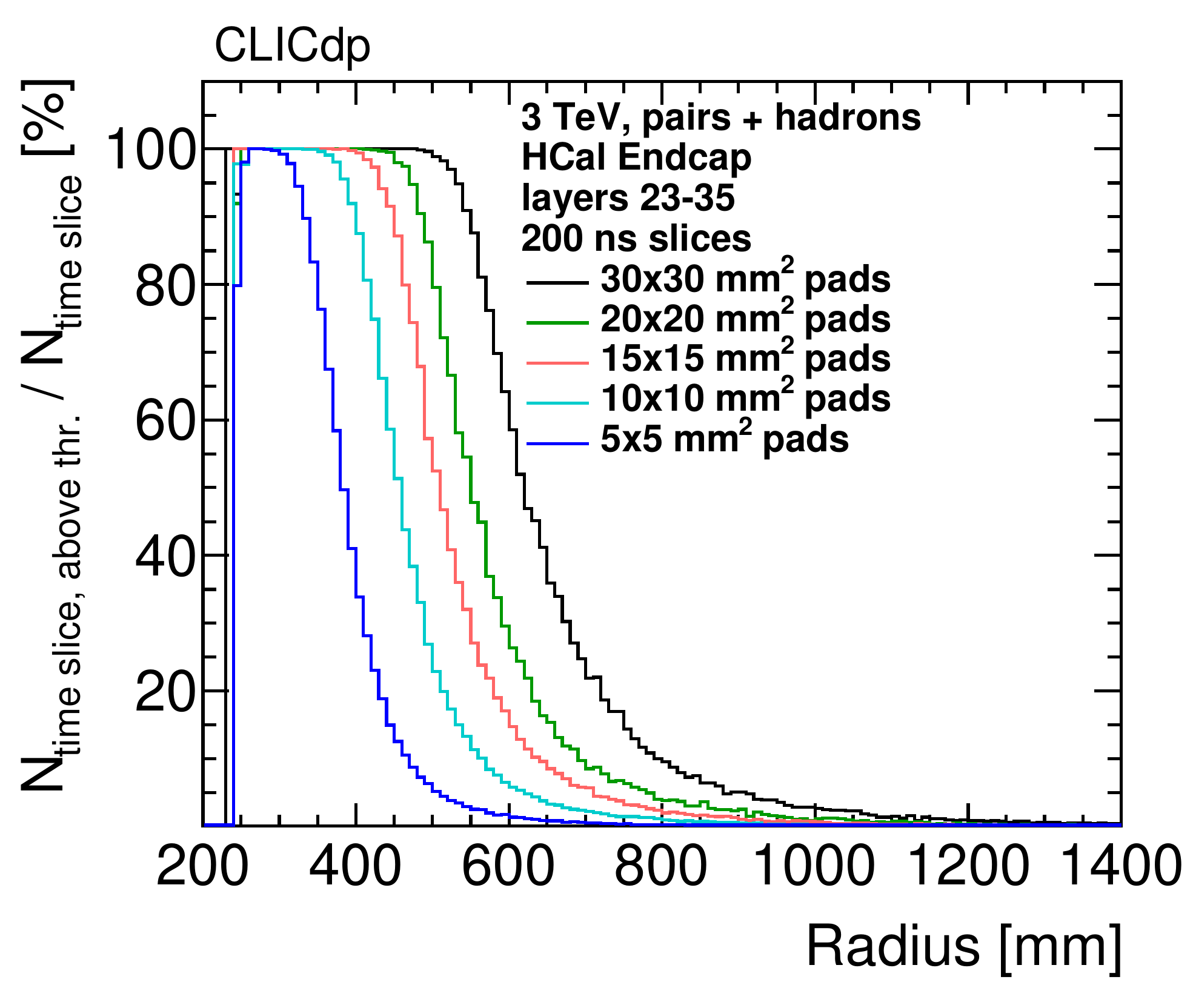}
\caption{}\label{fig:reduce_cellsize}
\end{subfigure}
\hfill
\begin{subfigure}[T]{.49\linewidth}
\includegraphics[width=\linewidth]{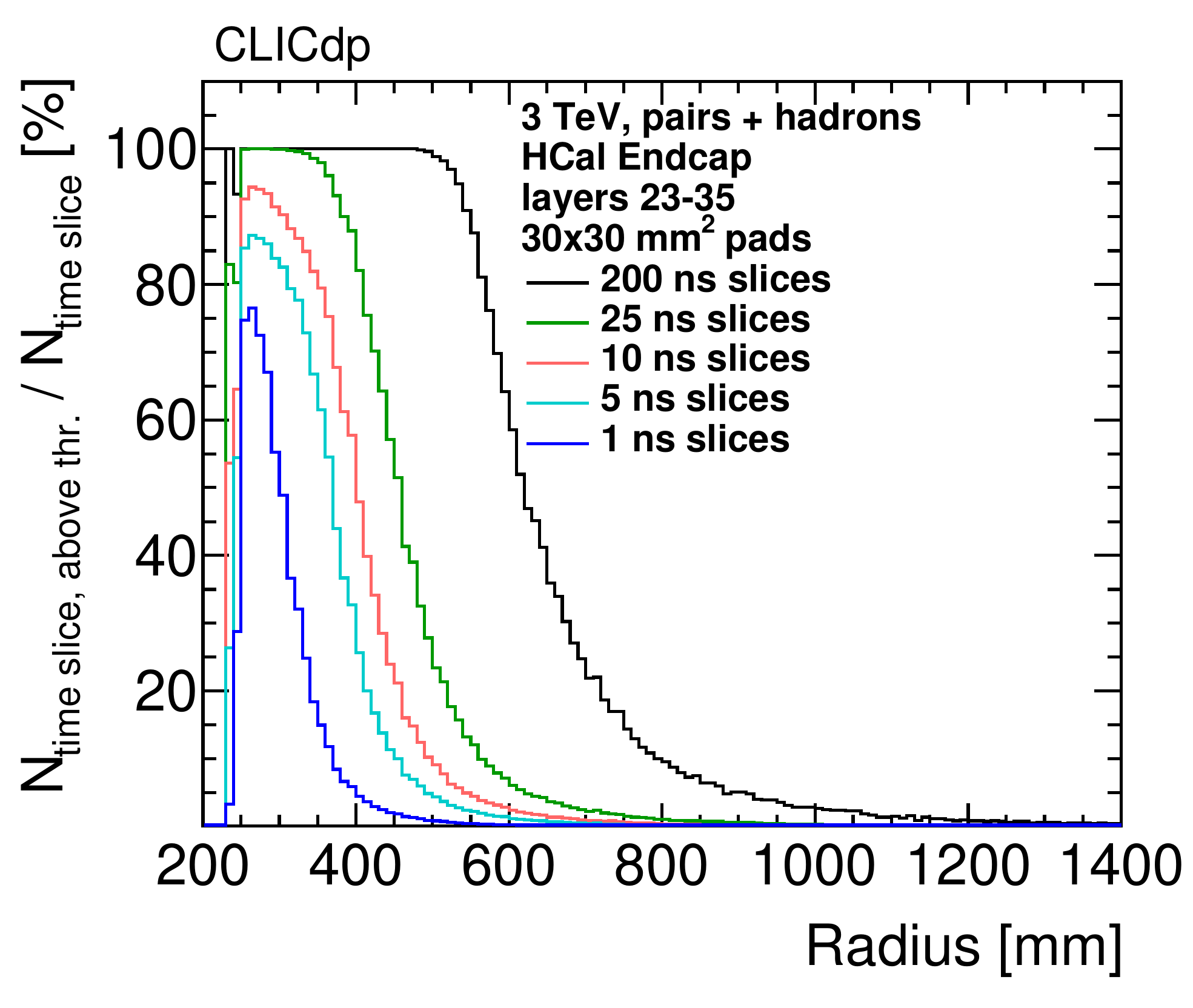}
\caption{}\label{fig:increase_sampling}
\end{subfigure}
\caption{Radial distributions of the fraction of time slices per train with at least one hit above threshold.
The distributions are shown for \subref{fig:reduce_cellsize} different cell sizes and \subref{fig:increase_sampling} different time slices.
The distributions are shown for layers 23-35 of the HCAL endcap at 3\,TeV, including the $\PGg\PGg\to$ hadron and incoherent pair backgrounds as well as safety factors.
}
\label{fig:modification}
\end{figure}

The first option is based on an increase in the spatial granularity.
Its effect on the occupancy is shown in \cref{fig:reduce_cellsize} for layers 23-35 of the HCAL endcaps.
Reducing the cell size by a factor of 36 ($5 \times 5\,\text{mm}^2$) still leaves cells at the lowest radii of the HCAL endcap 100\% occupied. 

The second option uses an increased temporal granularity.
As discussed in \cref{sec:implementation}, it is foreseen to sample the calorimeter cell pulse height throughout the train and shower-development period, in order to achieve the required cluster time resolution in the CLICdet calorimeters.
This allows for a detection of multiple hits per train, depending on the signal shaping time and the sampling rate.
The assumed sampling period is 25\,ns for ECAL and HCAL and 10\,ns for LumiCal and BeamCal.
The effect of reducing the sampling period is shown in \cref{fig:increase_sampling} for layers 23-35 of the HCAL endcaps.
The signal-development times in the sensors and the shaping and deconvolution of overlapping signals are not taken into account in the simulation.
Therefore the calculation of the time-slice occupancies for short sampling periods ($<10$\,ns) should be considered as lower limits and are shown solely to illustrate the trend.
Using the default cell size of $30 \times 30\,\text{mm}^2$ and reducing the time slice duration achieves a reduction of the overall occupancy to values lower than 95\% at 10\,ns, 87\% at 5\,ns and 77\% at 1\,ns.

\begin{figure}[t]
\centering
\includegraphics[width=0.5\linewidth]{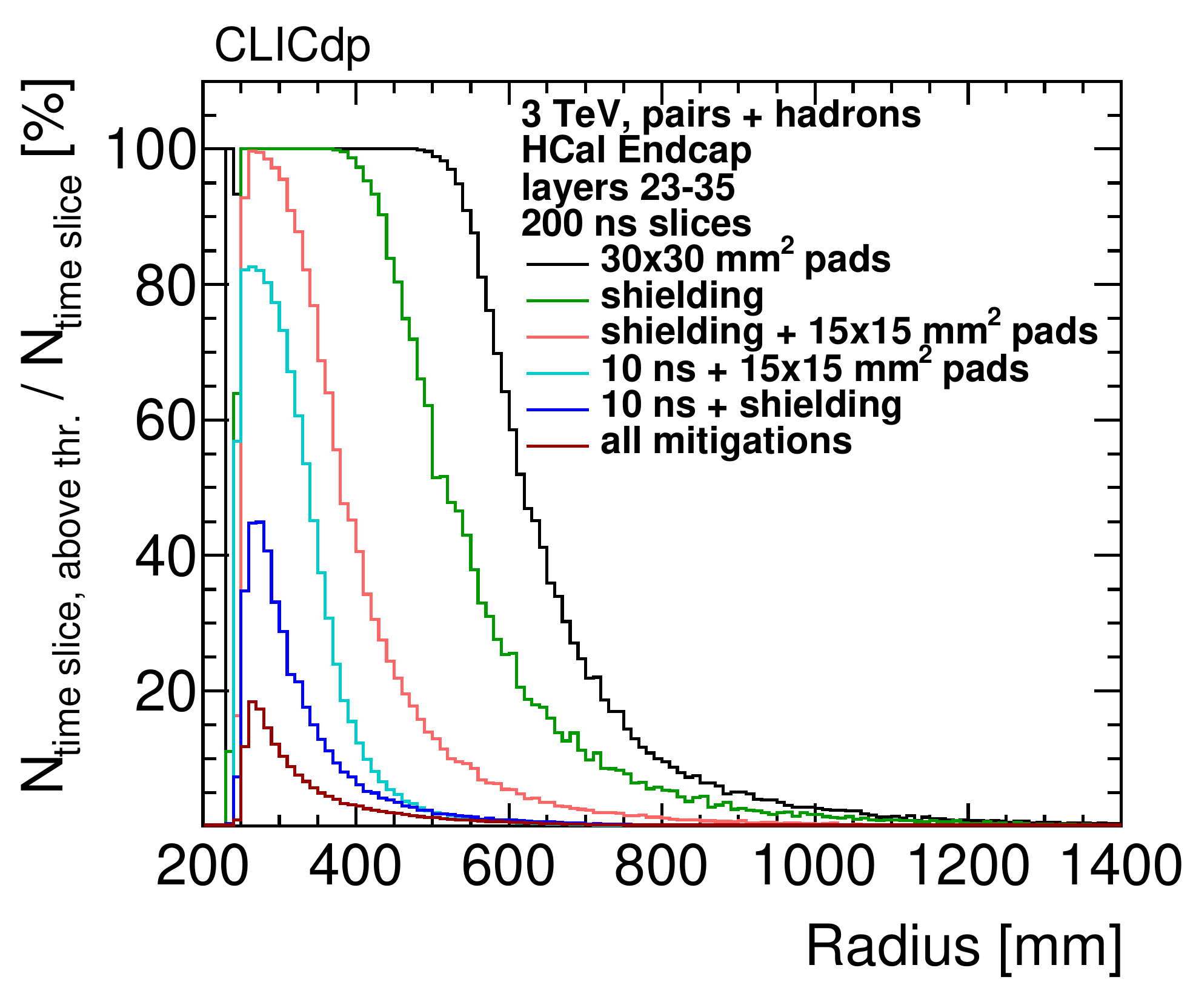}
\caption{Radial distributions of the fraction of time slices per train with at least one hit above threshold for different shielding, layout and readout configurations.
The distribution is shown for layers 23-35 of the HCAL endcap at 3\,TeV, including the $\PGg\PGg\to$ hadron and incoherent pair backgrounds as well as safety factors.
}
\label{fig:add_shielding}
\end{figure}

As a third option, the use of additional shielding material is considered for reducing the occupancies from secondaries~\cite{vanDam:1751528,CLICdet-background}.
A proposed shielding concept consists of 50\,mm of tungsten and 30\,mm of SWX (95\% polyethylene, 5\% boron carbide) which is added to the inside of the cylindrical support tube hosting the beam pipes and the BeamCal (see~\cref{fig:forward}).
The shielding needs to be optimised such that it absorbs large fractions of the secondaries while not reducing the acceptance of the detectors.
The effect of the shielding proposal is shown in \cref{fig:add_shielding} for layers 23-35 of the HCAL endcaps.
Furthermore, \cref{fig:add_shielding} shows the combined effect of using shielding, reducing the spatial granularity and reducing the time slice duration.
The occupancy could be reduced to values lower than 20\% when using the described shielding, a cell size of $15 \times 15\,\text{mm}^2$ and a time slice duration of 10\,ns.

For the BeamCal and LumiCal, pixels with a pitch in the order of few 10 to \SI{100}{\micro\metre} are considered as discussed in~\cref{sec:pixelated-lumical}.
The exact layout adaptation is a subject for further studies.

While the increase in spatial and temporal granularity would result in an increase in the data volume per detector, additional shielding would reduce the data volume.

\section{Summary and outlook}
Detector occupancies, data volumes and data output rates have been estimated for all CLIC subdetectors, based on example readout schemes and on the expected experimental conditions.
Acceptable cell occupancies are obtained for most detector regions, except for the inner regions of the endcap calorimeters and the very forward calorimeters LumiCal and BeamCal.
Upper limits on the total data output rate of 1.45\,GByte/s at 380\,GeV and 4.40\,GByte/s at 3\,TeV have been obtained, dominated by beam-induced background hits.
The resulting data storage volume is of the same order as that for the current LHC experiments.
Therefore dedicated studies of the data-link and computing infrastructure have been deferred to a later stage.

An optimisation of the tracker granularity taking into account occupancy and pattern-recognition constraints could lead to a largely reduced number of cells.
This is desired in view of a reduced power budget and readout complexity.

Ongoing optimisation studies for the readout of the main calorimeter endcaps show that increased spatial and temporal granularity in combination with improved shielding result in acceptable cell occupancies in these regions.

Planned future work in the data-acquisition domain include an optimisation of the LumiCal and BeamCal readout schemes based on pixelated sensors (see \cref{sec:pixelated-lumical}), as well as the development of detector-calibration concepts optimised for low bandwidth.


\chapter{Conclusions and future developments}\label{chap:conclusions}
The CLIC detector technology studies summarised in this report cover the time from the publication of the CLIC CDR in 2012 until the launch of the European Strategy Upgrade process at the end of 2018. Over this period significant progress has been made in all critical areas of detector R\&D and design. In most cases the development targets consist of validated \emph{technology demonstrators} of the most challenging components, rather than full prototypes of final detectors. Extensive test-beam programmes complement the laboratory characterisation studies for various detector technologies. Future work until the time of project approval will continue this technology exploration, while simultaneously shifting the focus towards building larger demonstrators and performing full design studies. Once the project is approved, detector development efforts will have to increase significantly and be prioritised according to the engineering constraints and construction schedule. 

The optimised CLICdet detector design is driven by the CLIC physics programme and the experimental conditions provided by the accelerator, and takes into account constraints and results from the hardware and simulation studies. Future improvements to the detector design and to the full-detector simulation and reconstruction algorithms will profit from the detailed performance assessments and simulations performed for the various technology demonstrators.

CLIC-specific detector R\&D is mainly performed for the vertex and tracker region, for which the requirements from physics and accelerator are the most challenging. In particular the sensor and readout developments benefit from the rapid progress in the silicon industry and from synergies with detector upgrade projects for the High-Luminosity LHC and other ongoing detector projects. They also rely on dedicated simulation and characterisation tools, which are being developed with contributions from the world-wide silicon detector community. The R\&D for the CLIC vertex-detector region has to meet the most challenging combination of requirements. Therefore initial studies have focused on hybrid pixel detector technologies, optimising separately the design of small-feature size high-performance CMOS readout ASICs and ultra-thin sensors with the smallest feasible pixel pitch. While each of the requirements for the vertex detector could be met individually, the combination of very low material budget and high spatial resolution still remains to be achieved. Future developments will profit from the availability of more advanced CMOS process technologies and will pursue new sensor and interconnect technologies suitable for even smaller pixel pitch. A variety of promising monolithic depleted CMOS sensor technologies are under study, which have the potential to deliver high performance over large detector areas, as required for the outer tracker. CLIC-specific monolithic designs are in preparation, targeting both the tracker and the vertex detector. Future designs are expected to profit from the availability of new monolithic CMOS technologies with smaller feature size, which will allow for smaller pixel pitch and increased performance. Powering and detector-integration studies have demonstrated the feasibility of the major detector design features, such as power pulsing, air-flow cooling, low-mass support structures and cabling. Future work in this domain will shift towards building larger combined demonstrators, profiting from the experience gained with ongoing LHC detector upgrade projects. 

R\&D for the main calorimeters is well covered by the generic Linear Collider calorimeter activities within the CALICE collaboration, complemented by dedicated studies targeting the most important CLIC-specific requirements such as containment for higher centre-of-mass energies and sub-nanosecond timing capability for reconstructed objects. The feasibility of highly granular imaging calorimetry and the validity of the performance assumptions in the full-detector simulation models has been demonstrated both for the electromagnetic and hadronic calorimeters through a large-scale test-beam programme with prototypes of increasing size and complexity. The most recent generation of prototypes addresses the scalability of the imaging calorimeter technology to a full collider detector. Plans for future studies include the construction and testing of prototypes with additional layers and combined ECAL and HCAL beam tests. Dedicated tests with very high-energy electrons will be needed to ensure that the CLIC resolution requirements for electromagnetic objects are met. The need for precise hit timing at CLIC will profit from the developments in the context of the ongoing collaboration between CALICE and the High-Granularity Calorimeter HL-LHC Upgrade Project of the CMS collaboration (CMS-HGCAL). In particular the high-occupancy endcap regions at CLIC will require adaptations to the currently developed CALICE readout concepts. Initial power-pulsing studies for ILC conditions will have to be adapted to the duty cycle at CLIC. 

The very forward calorimetry for luminosity measurement as well as electron and photon tagging is conducted within the FCAL collaboration. The measurements in this detector region require a very compact detector setup and a large dynamic range, and are complicated by very high rates of beam-induced background particles and consequently high radiation levels. Developments for BeamCal focus mainly on radiation-hardness studies for various sensor materials. Very compact and highly granular technology demonstrators for LumiCal have been constructed and validated successfully in beam tests. Initial concepts for achieving the required high alignment accuracy for BeamCal and LumiCal have been devised. Future FCAL plans include tests of a large-scale prototype with custom-designed FCAL electronics and performance assessments of radiation-hard BeamCal sensor material. The expected high occupancies at CLIC will require reduced pad sizes and further improvements to the readout electronics. Monolithic CMOS sensor technologies studied in the context of the tracker developments may allow for very compact pixelated LumiCal sensor layers with a large dynamic measurement range.

In summary, a targeted R\&D programme for the CLIC detector has explored and advanced the state-of-the art in many areas of detector technology, and is on track to meet the challenging requirements with a variety of innovative concepts. Future work will focus on the most critical development goals, benefiting from anticipated continuing technological progress, and prepare for detector construction in line with the progression of the CLIC project.


\appendix



\chapter*{Acknowledgements}

We acknowledge the support from CERN as the host laboratory for the CLICdp collaboration.
We would like to gratefully acknowledge CERN, DESY and FNAL and their accelerator staff for the reliable and efficient test-beam operation.
This work benefited from services provided by the ILC Virtual Organisation, supported by the national resource providers of the EGI Federation. This research was done using resources provided by the Open Science Grid, which is supported by the National Science Foundation and the U.S. Department of Energy's Office of Science.
This work was supported by
the European Union's Horizon 2020 Research and Innovation programme under Grant Agreement No.\,654168 (AIDA-2020);
the National Commission for Scientific and Technological Research (CONICYT), Chile;
the DFG cluster of excellence ``Origin and Structure of the Universe'', Germany;
the Federal Ministry of Education and Research (BMBF), Germany under Grant Agreement No.\,05H18VKRD1; 
the Israel Science Foundation (ISF);
the I-CORE Program, Israel;
the Israel Academy of Sciences;
the Research Council of Norway;
the National Science Centre, Poland, HARMONIA project under contract UMO-2015/18/M/ST2/00518 and OPUS project under contract UMO-2017/25/B/ST2/00496;
the Polish Ministry of Science and Higher Education under contract No.\,3501/H2020/2016/2 and 3812/H2020/2017/2; 
the Spanish Ministry of Economy, Industry and Competitiveness under projects MINEICO/FEDER-UE, FPA2015-65652-C4-3-R, FPA2015-71292-C2-1-P and FPA2015-71956-REDT; the Generalitat Valenciana under grant PROMETEO/2018/060; the IFIC, IFCA, IFT and CIEMAT grants under the Centro de Excelencia Severo Ochoa and Maria de Maeztu programs, SEV-2014-0398, MDM-2017-0765, SEV-2016-059, MDM-2015-0509, Spain; 
the UK Science and Technology Facilities Council (STFC), United Kingdom; 
and the U.S. Department of Energy, Office of Science under contract DE-AC02-06CH11357.

\printbibliography[title=References]


\end{document}